\RequirePackage{lineno}\newdimen\linenumbersep\linenumbersep=2pt
\documentclass[aps,amsmath,amssymb,floatfix,preprintnumbers,showkeys,superscriptaddress,twocolumn]{revtex4-2} 
\usepackage{array, amsmath, amssymb, booktabs, color, eurosym, graphicx, longtable, pbox, siunitx, url, verbatim}
\usepackage{hyperref}
\hypersetup{colorlinks,allcolors=blue}

\graphicspath{{./figures/}}

\newcommand{\1}[1]{\, \mathrm{#1}} 
\newcommand{\n}[1]{\mathrm{#1}}    

\newcommand{\tonneyear}{\ensuremath{\mathrm{t}\times\mathrm{y}}}

\begin{document}

\title{A Next-Generation Liquid Xenon Observatory for Dark Matter and Neutrino Physics}



\newcommand{\alabama}{\affiliation{University of Alabama, Department of Physics \& Astronomy, Tuscaloosa, AL 34587, USA}}
\newcommand{\albany}{\affiliation{Department of Physics, The University at Albany, The State University of New York, Albany, NY 12222, USA}}
\newcommand{\apc}{\affiliation{Universit\'e de Paris, CNRS, Astroparticule et Cosmologie, F-75013 Paris, France}}
\newcommand{\argonne}{\affiliation{High Energy Physics Theory Group, Argonne National Laboratory Argonne, IL 60439, USA}}
\newcommand{\arizonastate}{\affiliation{Department of Physics, Arizona State University, Tempe, AZ 85281, USA}}
\newcommand{\banjaluka}{\affiliation{Faculty of Architecture, Civil Engineering and Geodesy, University of Banja Luka, Bulevar vojvode Petra Bojovica 1a, 78000 Banja Luka, Bosnia and Hercegovina}}
\newcommand{\barcelona}{\affiliation{Department of Quantum Physics and Astrophysics and Institute of Cosmos Sciences, University of Barcelona, 08028 Barcelona, Spain}}
\newcommand{\beihang}{\affiliation{School of Physics, Beihang University, Beijing, 100083, P.R.China}}
\newcommand{\belgrade}{\affiliation{Vinca Institute of Nuclear Science, University of Belgrade, Mihajla Petrovica Alasa 12-14, Belgrade, Serbia}}
\newcommand{\bern}{\affiliation{Albert Einstein Center for Fundamental Physics, Institute for Theoretical Physics, University of Bern, Sidlerstrasse 5, 3012 Bern, Switzerland}}
\newcommand{\blackhills}{\affiliation{Black Hills State University, School of Natural Sciences, Spearfish, SD 57799, USA}}
\newcommand{\bltp}{\affiliation{Bogoliubov Laboratory of Theoretical Physics, Joint Institute for Nuclear Research, 141980 Dubna, Russia}}
\newcommand{\bologna}{\affiliation{Department of Physics and Astronomy, University of Bologna and INFN-Bologna, 40126 Bologna, Italy}}
\newcommand{\brandeis}{\affiliation{Department of Physics, Brandeis University, 415 South Street, Waltham, MA 02453, USA}}
\newcommand{\bristol}{\affiliation{University of Bristol, H.H. Wills Physics Laboratory, Bristol, BS8 1TL, UK}}
\newcommand{\brookhavennl}{\affiliation{Brookhaven National Laboratory (BNL), Upton, NY 11973, USA}}
\newcommand{\brown}{\affiliation{Department of Physics, Brown University, 182 Hope Street, Providence, RI 02912, USA}}
\newcommand{\budker}{\affiliation{Budker Institute of Nuclear Physics SB RAS, Lavrentiev avenue 11, 630090 Novosibirsk, Russia}}
\newcommand{\caltech}{\affiliation{Division of Physics, Mathematics, \& Astronomy, California Institute of Technology, Pasadena, CA 91125, USA}}
\newcommand{\cern}{\affiliation{Theoretical Physics Department, CERN, 1211 Geneva 23, Switzerland}}
\newcommand{\chalmers}{\affiliation{Chalmers University of Technology, Department of Physics, SE-412 96 G\"oteborg, Sweden}}
\newcommand{\chicagophysics}{\affiliation{Department of Physics \& Kavli Institute for Cosmological Physics, The University of Chicago, Chicago, IL 60637, USA}}
\newcommand{\chicagoastro}{\affiliation{Department of Astronomy and Astrophysics \& Kavli Institute for Cosmological Physics, The University of Chicago, Chicago, IL 60637, USA}}
\newcommand{\cincinnati}{\affiliation{Department of Physics, University of Cincinnati, Cincinnati, Ohio 45221, USA}}
\newcommand{\coimbralib}{\affiliation{LIBPhys, Department of Physics, University of Coimbra, 3004-516 Coimbra, Portugal}}
\newcommand{\coimbralip}{\affiliation{LIP-Coimbra, Department of Physics, University of Coimbra, 3004-516 Coimbra, Portugal}}
\newcommand{\columbia}{\affiliation{Physics Department, Columbia University, New York, NY 10027, USA}}
\newcommand{\comenius}{\affiliation{Department of Nuclear Physics and Biophysics, Comenius University, Mlynsk\'a dolina F1, SK–842 15 Bratislava, Slovakia}}
\newcommand{\copenhagen}{\affiliation{Niels Bohr International Academy and DARK, Niels Bohr Institute, University of Copenhagen, Blegdamsvej 17, 2100, Copenhagen, Denmark}}
\newcommand{\daejeon}{\affiliation{IBS Center for Underground Physics (CUP), Yuseong-gu, Daejeon, Republic of Korea}}
\newcommand{\desy}{\affiliation{Deutsches Elektronen-Synchrotron DESY, Notkestr. 85, 22607 Hamburg, Germany}}
\newcommand{\dresden}{\affiliation{Technische Universit\"at Dresden, 01069 Dresden, Germany}}
\newcommand{\dublin}{\affiliation{Dublin Institute for Advanced Studies, Dublin, D02 XF86, Ireland}}
\newcommand{\durham}{\affiliation{Institute for Computational Cosmology, Department of Physics, University of Durham, South Road, Durham, DH1 3LE, UK}}
\newcommand{\edinburgh}{\affiliation{SUPA, School of Physics and Astronomy, University of Edinburgh, Edinburgh, EH9 3FD, UK}}
\newcommand{\fermilab}{\affiliation{Fermi National Accelerator Laboratory, Batavia, IL 60510, USA}}
\newcommand{\ferrara}{\affiliation{Department of Physics and Earth Sciences, University of Ferrara and INFN-Ferrara, 44122, Italy}}
\newcommand{\fiorentino}{\affiliation{INFN, Sezione di Firenze Via G. Sansone 1, 50019 Sesto Fiorentino, Italy}}
\newcommand{\freiburg}{\affiliation{Physikalisches Institut, Universit\"at Freiburg, 79104 Freiburg, Germany}}
\newcommand{\harvard}{\affiliation{Department of Physics \& Laboratory for Particle Physics and Cosmology, Harvard University, Cambridge, MA 02138, USA}}
\newcommand{\hawaii}{\affiliation{Department of Physics and Astronomy, University of Hawai'i, Honolulu, HI 96822, USA}}
\newcommand{\heidelbergmpi}{\affiliation{Max-Planck-Institut f\"ur Kernphysik, 69117 Heidelberg, Germany}}
\newcommand{\heidelberguni}{\affiliation{Physikalisches Institut, Ruprecht-Karls-Universit\"at Heidelberg, Heidelberg, Germany}}
\newcommand{\helmholtzdarmstadt}{\affiliation{ExtreMe Matter Institute EMMI, GSI Helmholtzzentrum f\"ur Schwerionenforschung GmbH, 64291 Darmstadt, Germany}}
\newcommand{\ias}{\affiliation{School of Natural Sciences, Institute for Advanced Study, Princeton, NJ 08540, USA}}
\newcommand{\icrr}{\affiliation{Kamioka Observatory, Institute for Cosmic Ray Research, the University of Tokyo, Higashi-Mozumi, Kamioka, Hida, Gifu, 506-1205, Japan}}
\newcommand{\ictp}{\affiliation{International Centre for Theoretical Physics, Strada Costiera 11, 34151, Trieste, Italy}}
\newcommand{\ieap}{\affiliation{Institute of Experimental and Applied Physics, Czech Technical University, 128 00 Prague, Czech Republic}}
\newcommand{\iitg}{\affiliation{Department of Physics, Indian Institute of Technology - Guwahati, Guwahati 781039, India}}
\newcommand{\imperial}{\affiliation{Department of Physics, Blackett Laboratory, Imperial College London, London SW7 2BW, UK}}
\newcommand{\inr}{\affiliation{Institute for Nuclear Research, Moscow, 117312, Russia}}
\newcommand{\ipmu}{\affiliation{Kavli Institute for the Physics and Mathematics of the Universe (WPI), the University of Tokyo, Kashiwa, Chiba, 277-8582, Japan}}
\newcommand{\ippp}{\affiliation{Institute for Particle Physics Phenomenology, Durham University, South Road, Durham, United Kingdom}}
\newcommand{\itp}{\affiliation{Institute for Theoretical Physics, University of Heidelberg, Philosophenweg 16, D-69120 Heidelberg, Germany}}
\newcommand{\kcl}{\affiliation{Physics, King's College London, Strand, London WC2R 2LS, UK}}
\newcommand{\kias}{\affiliation{School of Physics, KIAS, 85 Hoegiro, Seoul 02455, Republic of Korea}}
\newcommand{\kiasquc}{\affiliation{Quantum Universe Center, KIAS, 85 Hoegiro, Seoul 02455, Republic of Korea}}
\newcommand{\kitiap}{\affiliation{Institute for Astroparticle Physics, Karlsruhe Institute of Technology, Karlsruhe, Germany}}
\newcommand{\kitetp}{\affiliation{Institute of Experimental Particle Physics, Karlsruhe Institute of Technology, Karlsruhe, Germany}}
\newcommand{\kitipe}{\affiliation{Institute for Data Processing and Electronics, Karlsruhe Institute of Technology, Karlsruhe, Germany}}
\newcommand{\kobe}{\affiliation{Department of Physics, Kobe University, Kobe, Hyogo 657-8501, Japan}}
\newcommand{\korea}{\affiliation{Department of Physics, Korea University, Anam-ro 145, Sungbuk-gu, Seoul 02841, Korea}}
\newcommand{\lal}{\affiliation{LAL, Universit\'e Paris-Sud, CNRS/IN2P3, Universit\'e Paris-Saclay, F-91405 Orsay, France}}
\newcommand{\laquila}{\affiliation{Department of Physics and Chemistry, University of L’Aquila, 67100 L’Aquila, Italy}}
\newcommand{\lbnl}{\affiliation{Lawrence Berkeley National Laboratory, Berkeley, CA 94720, USA}}
\newcommand{\liverpoollodge}{\affiliation{Oliver Lodge Laboratory, University of Liverpool, Liverpool, UK}}
\newcommand{\liverpool}{\affiliation{University of Liverpool, Department of Physics, Liverpool L69 7ZE, UK}}
\newcommand{\liverpoolmath}{\affiliation{University of Liverpool, Department of Mathematical Sciences, Liverpool L69 3BX, UK}}
\newcommand{\llnl}{\affiliation{Lawrence Livermore National Laboratory, Livermore, CA 94550, USA}}
\newcommand{\lngs}{\affiliation{INFN-Laboratori Nazionali del Gran Sasso and Gran Sasso Science Institute, 67100 L'Aquila, Italy}}
\newcommand{\lpnhe}{\affiliation{LPNHE, Sorbonne Universit\'{e}, CNRS/IN2P3, 75005 Paris, France}}
\newcommand{\mainz}{\affiliation{Institut f\"ur Physik \& Exzellenzcluster PRISMA, Johannes Gutenberg-Universit\"at Mainz, 55099 Mainz, Germany}}
\newcommand{\maryland}{\affiliation{Department of Physics, University of Maryland, College Park, MD 20742, USA}}
\newcommand{\massachusetts}{\affiliation{Department of Physics, University of Massachusetts, Amherst, MA 01003, USA}}
\newcommand{\melbourne}{\affiliation{ARC Centre of Excellence for Dark Matter Particle Physics, School of Physics, The University of Melbourne, VIC 3010, Australia}}
\newcommand{\mephi}{\affiliation{National Research Nuclear University “MEPhI” (Moscow Engineering Physics Institute), Moscow, 115409, Russia}}
\newcommand{\michigan}{\affiliation{University of Michigan, Randall Laboratory of Physics, Ann Arbor, MI 48109, USA}}
\newcommand{\minnesota}{\affiliation{William I. Fine Theoretical Physics Institute, School of Physics and Astronomy, University of Minnesota, Minneapolis, MN 55455, USA}}
\newcommand{\mitctp}{\affiliation{Center for Theoretical Physics, Massachusetts Institute of Technology, Cambridge, MA 02139, USA.}}
\newcommand{\munster}{\affiliation{Institut f\"ur Kernphysik, Westf\"alische Wilhelms-Universit\"at M\"unster, 48149 M\"unster, Germany}}
\newcommand{\nagoya}{\affiliation{Kobayashi-Maskawa Institute for the Origin of Particles and the Universe, and Institute for Space-Earth Environmental Research, Nagoya University, Aichi 464-8602, Japan}}
\newcommand{\nanjing}{\affiliation{School of Physics, Southeast University, Nanjing 211189, China}}
\newcommand{\nankai}{\affiliation{School of Physics, Nankai University, Tianjin 300071, China}}
\newcommand{\naples}{\affiliation{Department of Physics ``Ettore Pancini'', University of Napoli and INFN-Napoli, 80126 Napoli, Italy}}
\newcommand{\nikhef}{\affiliation{Nikhef and the University of Amsterdam, Science Park, 1098XG Amsterdam, Netherlands}}
\newcommand{\northwestern}{\affiliation{Department of Physics \& Astronomy, Northwestern University, Evanston, IL 60208, USA}}
\newcommand{\novosibirsk}{\affiliation{Novosibirsk State University, Pirogov street 2, 630090 Novosibirsk, Russia}}
\newcommand{\ntu}{\affiliation{Department of Physics, National Taiwan University, Taipei, Taiwan 10617, ROC}}
\newcommand{\nyuad}{\affiliation{Division of Science, New York University Abu Dhabi, Saadiyat Island, Abu Dhabi, United Arab Emirates}}
\newcommand{\oregon}{\affiliation{Department of Physics and Institute for Fundamental Science, University of Oregon, Eugene, OR 97403, USA}}
\newcommand{\origins}{\affiliation{Origins Project Foundation, Phoenix, AZ 85020, USA}}
\newcommand{\oxford}{\affiliation{Department of Physics, University of Oxford, Keble Rd, Oxford OX1 3RH, UK}}
\newcommand{\pennstate}{\affiliation{Department of Physics, Pennsylvania State University, 104 Davey Lab, University Park, PA 16802, USA}}
\newcommand{\pisasns}{\affiliation{Scuola Normale Superiore, Piazza dei Cavalieri 7, 56126 Pisa, Italy}}
\newcommand{\pisainfn}{\affiliation{INFN, Sezione di Pisa, Largo Bruno Pontecorvo 3, I-56127 Pisa, Italy}}
\newcommand{\pisafermi}{\affiliation{Dipartimento di Fisica E. Fermi, Università di Pisa, Largo B. Pontecorvo 3, I-56127 Pisa, Italy}}
\newcommand{\prague}{\affiliation{Institute of Physics, Czech Academy of Sciences, 182 00 Prague, Czech Republic}}
\newcommand{\princeton}{\affiliation{Department of Physics, Princeton University, Princeton, NJ 08544, USA}}
\newcommand{\purdue}{\affiliation{Department of Physics and Astronomy, Purdue University, West Lafayette, IN 47907, USA}}
\newcommand{\queens}{\affiliation{Department of Physics,~Engineering Physics and Astronomy, Queen's University, Kingston ON K7L 3N6, Canada}}
\newcommand{\rensselear}{\affiliation{Department of Physics, Applied Physics and Astronomy, Rensselaer Polytechnic Institute, Troy, NY 12180, USA}}
\newcommand{\rice}{\affiliation{Department of Physics and Astronomy, Rice University, Houston, TX 77005, USA}}
\newcommand{\rkmveri}{\affiliation{Ramakrishna Mission Vivekananda Educational and Research Institute, Belur Math, Howrah 711202, India}}
\newcommand{\roma}{\affiliation{Universit\`a degli Studi and INFN Roma Tre, Via della Vasca Navale 84, I-00146, Rome}}
\newcommand{\royalholloway}{\affiliation{Royal Holloway, University of London, Department of Physics, Egham, TW20 0EX, UK}}
\newcommand{\rutherford}{\affiliation{STFC Rutherford Appleton Laboratory (RAL), Didcot, OX11 0QX, UK}}
\newcommand{\sanfordlab}{\affiliation{South Dakota Science and Technology Authority (SDSTA), Sanford Underground Research Facility, Lead, SD 57754, USA}}
\newcommand{\schoolofmines}{\affiliation{South Dakota School of Mines and Technology, Rapid City, SD 57701, USA}}
\newcommand{\shanghai}{\affiliation{School of Physics and Astronomy, Shanghai Jiao Tong University, Shanghai, 200240, P.R.China}}
\newcommand{\shandong}{\affiliation{School of Physics, Shandong University, Jinan, 250100, P.R.China}}
\newcommand{\sheffield}{\affiliation{University of Sheffield, Department of Physics and Astronomy, Sheffield S3 7RH, UK}}
\newcommand{\shsu}{\affiliation{Department of Physics, Sam Houston State University, Huntsville, TX 77341, USA}}
\newcommand{\sinp}{\affiliation{Saha Institute of Nuclear Physics, HBNI, 1/AF Bidhannagar, Kolkata 700064, India}}
\newcommand{\sissa}{\affiliation{SISSA, Theoretical and Scientific Data Science group, Via Bonomea 265, 34136 Trieste, Italy}}
\newcommand{\skutek}{\affiliation{SkuTek Instrumentation, West Henrietta, NY 14586, USA}}
\newcommand{\slac}{\affiliation{SLAC National Accelerator Laboratory, Menlo Park, CA 94025, USA}}
\newcommand{\stanfordu}{\affiliation{Physics Department, Stanford University, Stanford, CA 94305, USA}}
\newcommand{\stanfordkavli}{\affiliation{Kavli Institute for Particle Astrophysics and Cosmology, Stanford University, Stanford, CA 94305, USA}}
\newcommand{\stockholm}{\affiliation{Oskar Klein Centre, Department of Physics, Stockholm University, AlbaNova, Stockholm SE-10691, Sweden}}
\newcommand{\strasbourg}{\affiliation{IPHC, CNRS, 67037 Strasbourg, France}}
\newcommand{\subatech}{\affiliation{SUBATECH, IMT Atlantique, Universit\'e de Nantes, CNRS/IN2P3, Nantes 44307, France}}
\newcommand{\tamu}{\affiliation{Mitchell Institute for Fundamental Physics and Astronomy, Texas A\&M University, College Station, TX 77843, USA}}
\newcommand{\tokyo}{\affiliation{Center for Nuclear Study, The University of Tokyo, 113-0033 Tokyo, Japan}}
\newcommand{\torino}{\affiliation{INAF-Astrophysical Observatory of Torino, Department of Physics, University of Torino and  INFN-Torino, 10125 Torino, Italy}}
\newcommand{\triumf}{\affiliation{TRIUMF, 4004 Wesbrook Mall, Vancouver, BC V7A 4N4, Canada}}
\newcommand{\tsinghua}{\affiliation{Department of Physics \& Center for High Energy Physics, Tsinghua University, Beijing 100084, China}}
\newcommand{\tudarmstadt}{\affiliation{Department of Physics, Technische Universit\"at Darmstadt, 64289 Darmstadt, Germany}}
\newcommand{\ucberkeley}{\affiliation{Department of Physics, University of California Berkeley, Berkeley, CA 94720, USA}}
\newcommand{\ucdavis}{\affiliation{University of California Davis, Department of Physics, One Shields Ave., Davis, CA 95616, USA}}
\newcommand{\ucirvine}{\affiliation{Department of Physics and Astronomy, University of California, Irvine, CA 92697, USA}}
\newcommand{\ucla}{\affiliation{Physics \& Astronomy Department, University of California, Los Angeles, CA 90095, USA}}
\newcommand{\uclondon}{\affiliation{Department of Physics and Astronomy, University College London, London WC1E 6BT, United Kingdom}}
\newcommand{\ucriverside}{\affiliation{Department of Physics and Astronomy, University of California, Riverside, CA 92521, USA}}
\newcommand{\ucsantabarbara}{\affiliation{Department of Physics, University of California, Santa Barbara, Santa Barbara, CA 93106, USA}}
\newcommand{\ucsandiego}{\affiliation{Department of Physics, University of California San Diego, La Jolla, CA 92093, USA}}
\newcommand{\ucsbkitp}{\affiliation{Kavli Institute for Theoretical Physics, University of California, Santa Barbara, CA 93106, USA}}
\newcommand{\rochester}{\affiliation{Department of Physics and Astronomy, The University of Rochester, Rochester, NY 14627, USA}}
\newcommand{\sydney}{\affiliation{School of Physics, The University of Sydney, NSW 2006 Camperdown, Sydney, Australia}}
\newcommand{\ustc}{\affiliation{Department of Physics, University of Science and Technology of China, Hefei, Anhui, P.R.China}}
\newcommand{\vatican}{\affiliation{Vatican Observatory, Castel Gandolfo, V-00120, Vatican City State}}
\newcommand{\wisconsin}{\affiliation{Department of Physics, University of Wisconsin-Madison, Madison, WI 53706, USA}}
\newcommand{\washington}{\affiliation{Institute for Nuclear Theory, University of Washington, Seattle, WA 98195, USA}}
\newcommand{\weizmann}{\affiliation{Department of Particle Physics and Astrophysics, Weizmann Institute of Science, Rehovot 7610001, Israel}}
\newcommand{\wspc}{\affiliation{Imperial College Press, World Scientific Publishing, London SW7 2AZ, United Kingdom}}
\newcommand{\yokohama}{\affiliation{Department of Physics, Faculty of Engineering, Yokohama National University, Yokohama, Kanagawa 240-8501, Japan}}
\newcommand{\zurich}{\affiliation{Physik-Institut, University of Zurich, 8057  Zurich, Switzerland}}

\author{J.~Aalbers}\slac\stanfordkavli 
\author{K.~Abe}\icrr\ipmu 
\author{V.~Aerne}\zurich 
\author{F.~Agostini}\bologna 
\author{S.~Ahmed Maouloud}\lpnhe 
\author{D.S.~Akerib}\slac\stanfordkavli 
\author{D.Yu.~Akimov}\mephi 
\author{J.~Akshat}\purdue 
\author{A.K.~Al Musalhi}\oxford 
\author{F.~Alder}\uclondon 
% ...

\begin{abstract}
\noindent The nature of dark matter and properties of neutrinos are among the most pressing issues in contemporary particle physics. The dual-phase xenon time-projection chamber is the leading technology to cover the available parameter space for Weakly Interacting Massive Particles (WIMPs), while featuring extensive sensitivity to many alternative dark matter candidates. These detectors can also study neutrinos through neutrinoless double-beta decay and through a variety of astrophysical sources. A next-generation xenon-based detector will therefore be a true multi-purpose observatory to significantly advance particle physics, nuclear physics, astrophysics, solar physics, and cosmology. This review article presents the science cases for such a detector.
\end{abstract}


\maketitle

\phantom{LatexCanBeAnnoying}\newpage
\phantom{LatexCanBeAnnoying}\newpage

\tableofcontents

\clearpage\newpage

\section{Introduction}\label{sec:introduction}

\subsection{An Observatory for Rare Events}

\begin{figure}[!htbp]
\begin{center}
\includegraphics[width=.9\columnwidth,clip,trim=140 460 230 120]{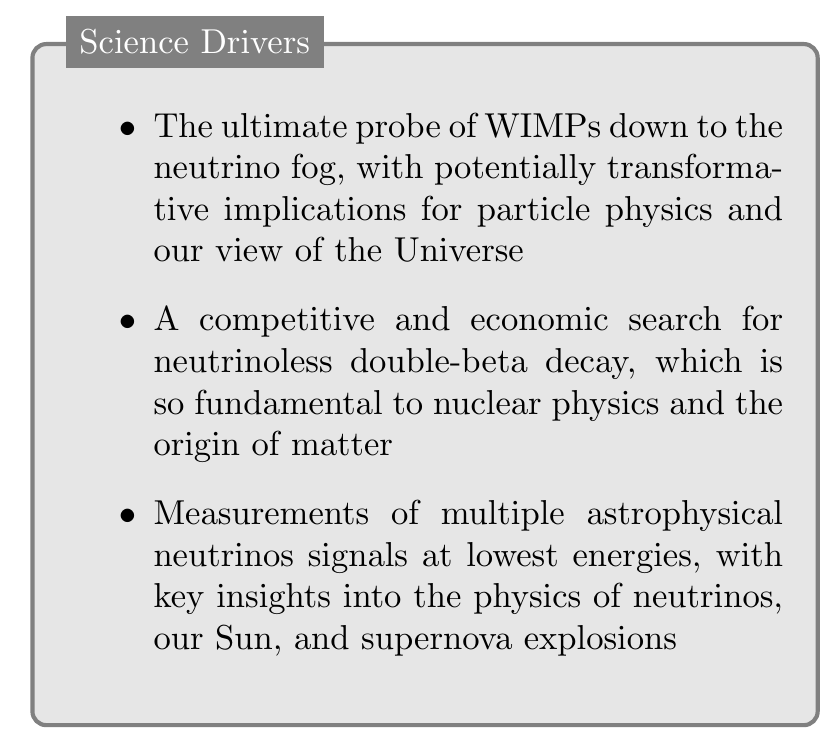}
\caption{Main science drivers for the next-generation liquid xenon observatory.}\label{fig:mainpoints}
\end{center}
\end{figure}

\begin{figure*}[!htbp]
\begin{center}
\includegraphics[width=2\columnwidth,clip,trim=2 294 0 44]{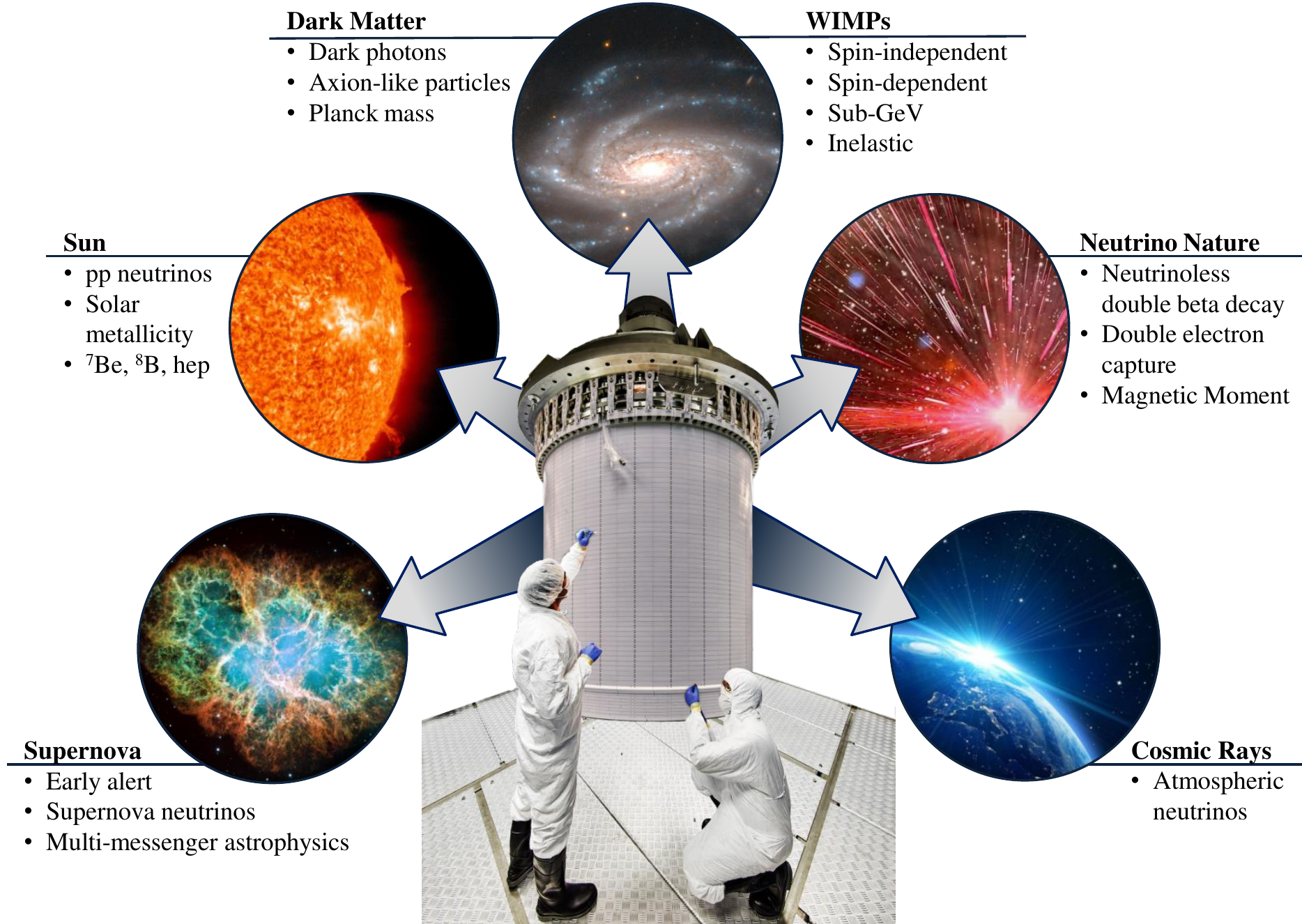}
\caption{The science channels of a next-generation liquid xenon observatory for rare events spans many areas and is of interest to particle physics, nuclear physics, astrophysics, solar physics, and cosmology.}\label{fig:sciencechannels}
\end{center}
\end{figure*}

Identifying the true nature of dark matter is one of the most important questions in physics today. As we show in this review, liquid xenon time projection chambers (TPCs) are the leading technology in searches for a large variety of dark matter particle candidates. Following two decades of evolution of this technology, now is the time to design the ultimate next-generation dark matter experiment in order to probe the widest possible range of dark matter candidates. A possible realization of such a detector has been proposed by the DARWIN collaboration~\cite{Aalbers:2016jon}. This experiment will also have competitive sensitivity to search for neutrinoless double-beta decay and other rare events. Furthermore, we show in this review that such an experiment serves as a versatile astroparticle physics observatory that is sensitive to neutrinos from our Sun, the atmosphere, and Galactic supernovae. Figs.~\ref{fig:mainpoints} and \ref{fig:sciencechannels} illustrate these topics. 

\subsection{Evidence for Dark Matter}

Strong evidence on astronomical and cosmological scales suggests gravitational interaction between baryonic matter and an unknown type of non-luminous matter, called dark matter~\cite{Bertone:2004pz,Bertone:2016nfn}. In 1922, the Dutch astronomer Jacobus Kapteyn was among the first to use the word ``dark matter'' to refer to invisible matter whose existence is inferred only from its gravitational effects. At the time, it was thought that such dark matter consisted of obscure stars; these might have been extinct and dark or simply not bright enough to be seen. While Kapteyn found that the amount of dark matter in the Milky Way ``cannot be excessive''~\cite{Kapteyn:1922zz}, later that same year the British astronomer James Jeans came to another conclusion~\cite{Jeans:1922}. In fact, by reanalysing the data, Jeans realized that ``there must be about three dark stars in the universe to every bright star.'' The next decade, Jan Oort used the vertical kinematics of Milky Way stars to constrain the local dark matter content~\cite{Oort:1932}, while Fritz Zwicky became the first to use the virial theorem to infer the presence of dark matter within the Coma Cluster~\cite{Zwicky:1933gu}. Further evidence for dark matter in galaxies came in 1978, when Vera Rubin and collaborators established that the rotation velocities of stars in spiral galaxies consistently differ from the distribution expected given the amount of baryonic matter~\cite{Rubin:1970zza}. 

Evidence for dark matter has now been found across all time and length scales~\cite{Lisanti:2016jxe}, spanning from the Big Bang to today, and from the cosmos as a whole down to individual galaxies~\cite{Kolb:1990vq}. Gravitational effects of dark matter can be observed in the cosmic microwave background, e.g.~with the Planck satellite~\cite{Akrami:2018vks}. Current estimates put the dark matter mass-energy density at five times that of baryonic matter in the observable universe~\cite{Aghanim:2018eyx}. Increased understanding of large-scale structure formation points to the existence of non-relativistic (cold) dark matter~\cite{Springel:2006vs,Knobel:2012wa,Coil:2012vw}. Gravitational lensing strongly suggests the presence of a significant amount of non-baryonic matter with no observable electromagnetic interaction~\cite{Bartelmann:1999yn}. 

The critical role of dark matter in the formation of galaxies~\cite{Silk:2012ra} such as our own Milky Way~\cite{Read:2014qva} underlines its significance to humanity. Galactic rotation curves~\cite{Salucci:1996bf,Richards:2015gla} and dynamics~\cite{Sofue:2000jx,Foster:2010ri} provide evidence for the existence of a uniformly-distributed halo of dark matter around most galaxies. A precise determination of the local dark matter halo density is fundamental for interpreting results from direct detection experiments; however, density estimates are model-dependent and may vary depending on the method used~\cite{Read:2014qva,deSalas:2019rdi}. Methods used to determine the local dark matter density can be broadly classified into local methods and global methods. Local methods focus on a small volume around the solar system, while global methods analyse data over a much bigger volume and attempt to determine the local dark matter density based on our position within the halo. Based on global methods using Gaia data release~2~\cite{Brown:2018dum}, the local dark matter density has been determined to be in the range $\rho = (0.3-0.4)$~$\1{GeV/cm^3}$~\cite{deSalas:2019pee}, while local methods based on the same data have produced a wider range of $\rho = (0.4-1.5)$~$\1{GeV/cm^3}$~\cite{Hagen18,Buch:2018qdr,Widmark:2018ylf} with some tendency towards higher values~\cite{Wu:2019nhd}. When presenting results from direct dark matter searches, it is common to assume $\rho = 0.3$~$\1{GeV/cm^3}$~\cite{Baxter:2021pqo}, which results from mass modelling of the Milky Way using parameters in agreement with observational data~\cite{Green:2011bv}.

While the existence of dark matter is thus well established, its physical characteristics remain elusive. Astrophysical observations indicate that dark matter could take the form of a new elementary particle outside the current Standard Model (SM) of particle physics~\cite{Tanabashi:2018oca}. The nature of this non-baryonic component is still unknown: its existence would be one of the strongest pieces of evidence that the current theory of fundamental particles and forces, summarized in the SM, is incomplete. A number of proposed candidates have been put forward over time, with some of the most popular candidates discussed in sections~\ref{sec:wimps} and~\ref{sec:broaderdarkmatter}.

\subsection{Dark Matter Direct Detection}

Since the 1980s, there have been large efforts to develop experiments on Earth that are able to directly search for dark matter, particularly for Weakly Interacting Massive Particles (WIMPs)~\cite{Drukier:1983gj,Smith:1983jj,Goodman:1984dc,Drukier:1986tm}, one popular dark matter candidate. Given the low expected interaction strength, the probability of multiple collisions within a detector is negligible, resulting in a recoil spectrum of single scattering events. 

A possible dark matter signature would be an annual modulation of the interaction count rate due to the motion of the Earth around the Sun~\cite{Goodman:1984dc,Drukier:1986tm,Freese:2012xd}. The relative velocity of dark matter particles in the Milky Way halo with respect to the detector on Earth depends on the time of year; therefore, the measured count rate is expected to exhibit a sinusoidal dependence with time, where the amplitude and phase of modulation will depend on the dark matter distribution within the halo~\cite{Copi:2002hm}.
While there is general consensus on standard values to be used to calculate expectations for direct experiments~\cite{Baxter:2021pqo}, this scenario can be modified in a number of possible astrophysical scenarios such as the presence of dark matter streams~\cite{Savage:2006qr,Lang:2010cd}, halo substructure~\cite{Kuhlen:2009vh,Schneider:2010jr,OHare:2019qxc}, a dark disk~\cite{Ling:2009eh} or local captured populations of WIMPs resulting from interactions in the Sun~\cite{Damour:1998vg} and Earth~\cite{Krauss:1988qm}.
    
In the effort to directly detect dark matter, many technologies have been developed to measure dark matter interactions with target nuclei. Complementary searches with different targets, discussed further in \autoref{sec:complementarity}, are essential to unveil the nature of dark matter. In the most common approach, experiments attempt to measure the nuclear recoil energy produced by collisions between dark matter candidates and target nuclei in the detector. The recoiling nucleus can deposit energy in the form of ionization, heat, and/or light that is subsequently detected. Different technologies have been explored so far to achieve this goal~\cite{Schumann:2019eaa}. Successful targets include solid state crystals~\cite{Aalseth:2014eft,Agnese:2014aze,Armengaud:2016cvl,Shields:2015wka,Aguilar-Arevalo:2016ndq,CRESST:2019jnq,Crisler:2018gci,SuperCDMS:2020ymb,SuperCDMS:2020aus}, metastable fluids~\cite{Behnke:2016lsk,Szydagis:2018wjp}, and noble liquids~\cite{Aalseth:2018gq,Akerib:2016vxi,Aprile:2017aty,Cui:2017nnn,Kobayashi:2018jky,Amole:2017dex}. 

\subsection{An Evolution of Scales}\label{sec:evolution}

\begin{figure}[!htbp]
\begin{center}
\includegraphics[width=0.99\columnwidth]{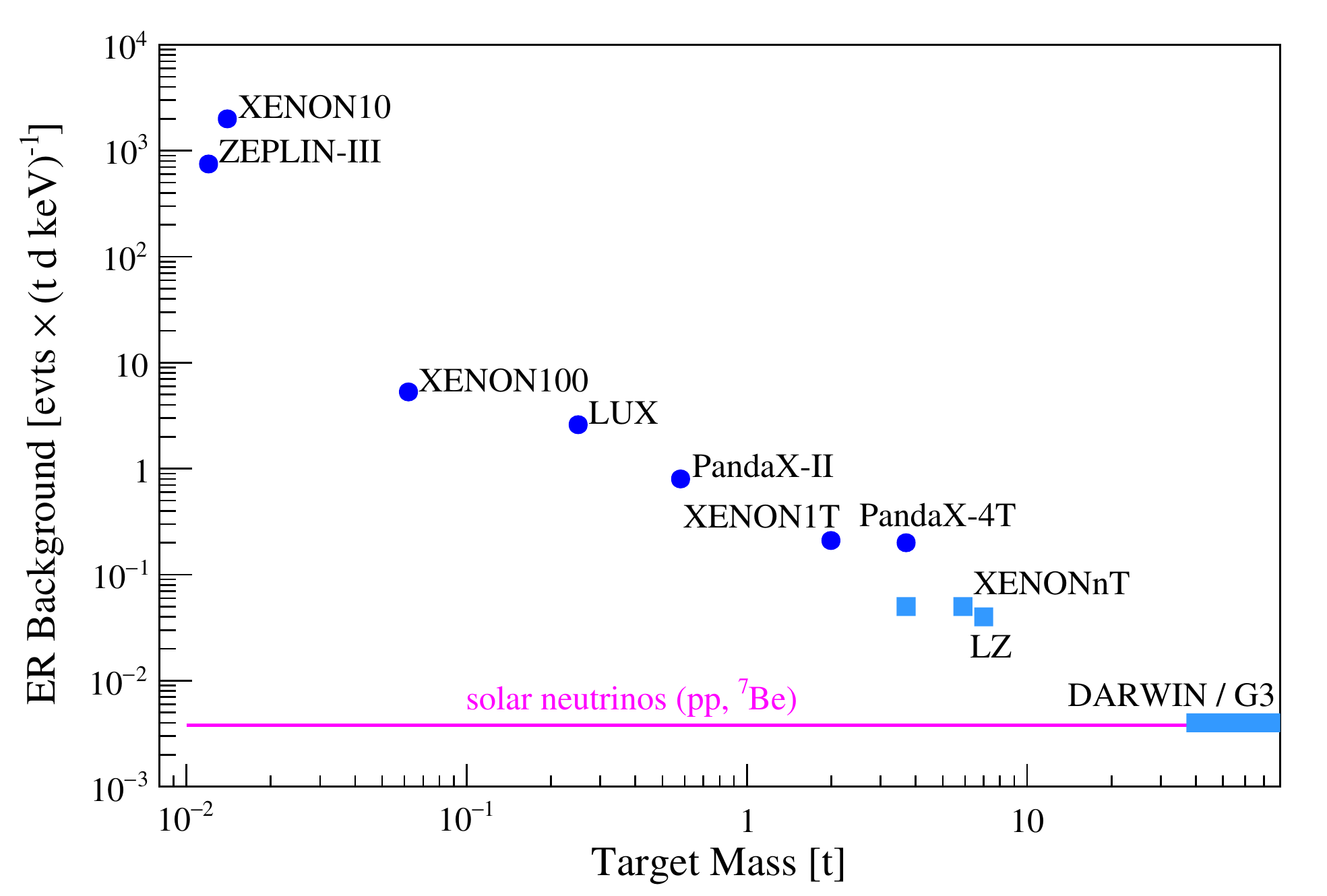}
\caption{The background rates in liquid xenon TPCs (before discrimination) have decreased exponentially over the years. This has been a key accomplishment that has enabled an exponential gain in sensitivity with ever-larger detectors. Solid dots are the best achieved limits, open squares the expected sensitivities. The experiment discussed here is labeled DARWIN/G3 and will at low energies be dominated by the signal from solar neutrinos. See text for references.}\label{fig:evolution}
\end{center}
\end{figure}

Liquid xenon TPCs in particular have demonstrated their exceptional capabilities for rare event detection as a result of an intense, decade-long development. The interested reader is referred to~\cite{Bolozdynya:2010, hatano_applications_2010, Akimov:2021} for detailed discussions of this technique. The two-phase (or dual-phase) emission detector that underlies liquid xenon TPCs was proposed a half-century ago~\cite{Dolgoshein:1970}. Its use for the detection of dark matter particles and neutrinos was proposed in 1995~\cite{Bolozdynya:1995}, with more mature conceptional designs developed around the turn of the millennium~\cite{Cline:2000, Aprile:2002ef}. Evolving out of ZEPLIN-I~\cite{Alner:2005pa}, the ZEPLIN-II~\cite{Alner:2007ja} detector was the first two-phase xenon dark matter experiment, with both experiments setting competitive limits on WIMP interactions at that time. This technology was further advanced in ZEPLIN-III~\cite{Lebedenko:2008gb,Araujo:2011as} and with XENON10~\cite{Angle:2007uj} saw the first leading limits on WIMP interactions. While XMASS provided an impressive demonstration of fiducialization in liquid xenon even in a single phase detector~\cite{XMASS:2018bid}, further evolution progressed through successively larger, cleaner, and thus more sensitive detectors: from XENON100~\cite{Aprile:2011dd,Aprile:2012nq}, LUX~\cite{Akerib:2016vxi}, PandaX-I~\cite{Xiao:2014xyn} and PandaX-II~\cite{Wang:2020coa} to XENON1T~\cite{Aprile:2018dbl} and the current generation PandaX-4T~\cite{Zhang:2018xdp}, XENONnT~\cite{Aprile:2020vtw}, and LZ~\cite{Akerib:2018lyp} (\autoref{fig:evolution}). The next-generation experiment discussed here is labeled DARWIN/G3 in \autoref{fig:evolution}~\cite{Schumann:2015cpa} and represents a natural continuation of this evolution towards larger xenon exposures, as presented in the sensitivity studies shown below. Yet, despite even its sensitivity to low-energy neutrino interactions, such a next-generation experiment will remain compact, with height and diameter only $\sim3\1{m}$.

\subsection{The Liquid Xenon Time Projection Chamber}

\begin{figure}[!htbp]
\begin{center}
\includegraphics[width=0.99\columnwidth]{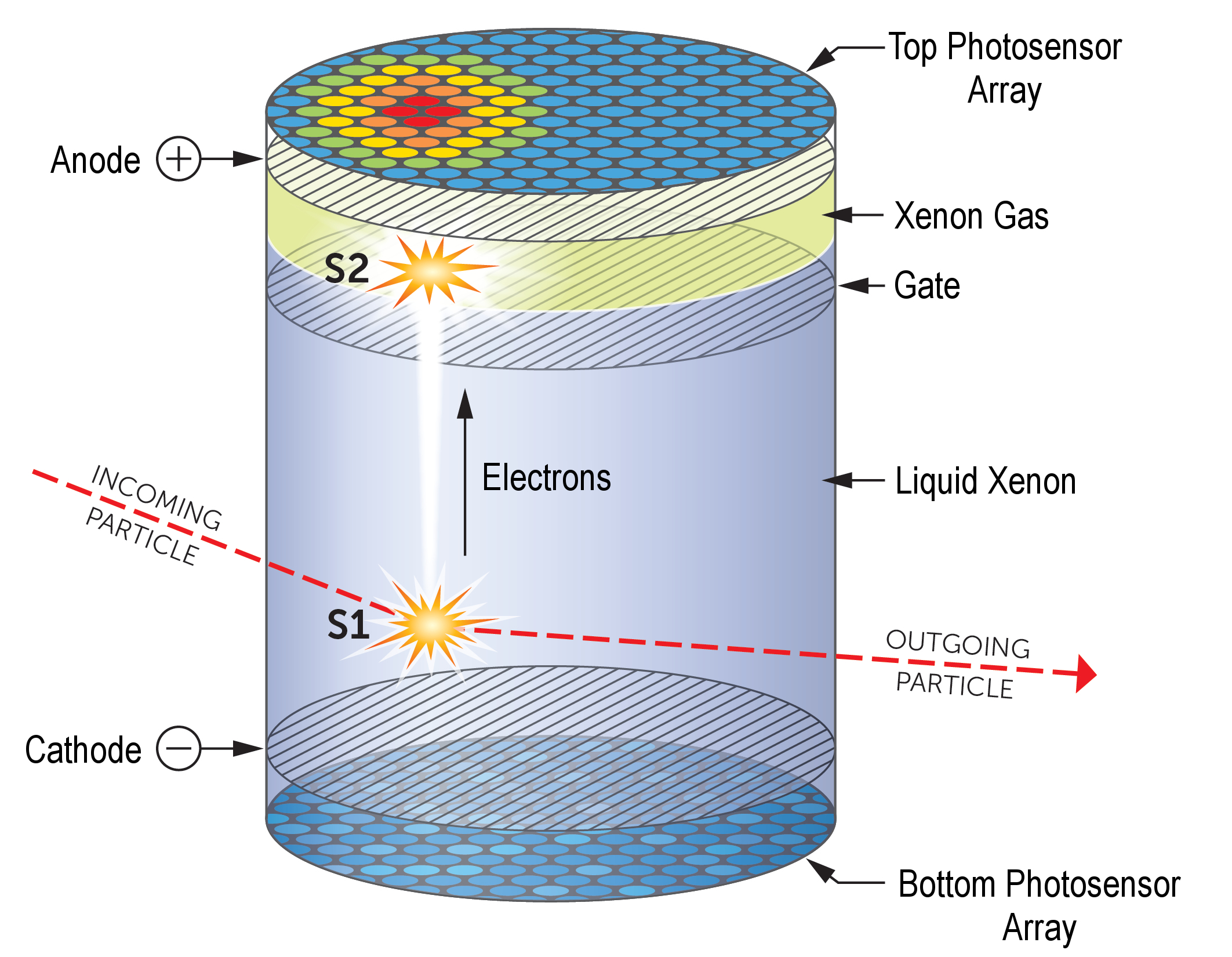}
\caption{Principle of a dual-phase liquid xenon time projection chamber. Energy from a particle interaction within the active liquid xenon volume produces prompt scintillation light (S1) and a delayed signal (S2) from electroluminescence (proportional scintillation) in the gaseous xenon layer. The localization of the S2 signal and the time difference between S1 and S2 allows for determination of the original vertex location.}\label{fig:tpcsketch}
\end{center}
\end{figure}

Conventionally, a next generation liquid xenon TPC will consist of a central liquid xenon volume surrounded by light reflectors for vacuum ultra-violet (VUV) light, allowing maximum light detection~\cite{Aprile:2004ey}. Two arrays of light sensors, such as photomultiplier tubes (PMTs)~\cite{Akerib:2012da,XENON:2015ara} or silicon photomultipliers (SiPMs)~\cite{Aprile:2005qq,Baudis:2020nwe}, are arranged on the top and bottom part of the TPC to detect light signals, see \autoref{fig:tpcsketch}. 

A particle incident on the liquid xenon target deposits energy and produces both prompt scintillation light and ionization electrons. The scintillation signal is immediately detected by the photosensors as the S1 signal. The \textit{active} liquid xenon volume is defined by a cathode and a gate electrode, separated by $\sim$3~meters to provide a drift field for the electrons. These drifting ionization electrons are then extracted into the gas phase above the liquid xenon, where they produce electroluminescent light~\cite{Lansiart:1976}. Typical dual phase detectors operate at $\sim1.5\1{bar}$, where $5\1{kV/cm}$ for the extration field is sufficient to create proportional scintillation. This electroluminescence is then detected by the same photosensors and is known as the S2 signal~\cite{Yamashita:2003rc,Aprile:2004ey, Mount:2017qzi}.

The time delay between S1 and S2 in addition to the localization of the S2 light pattern on the top photosensor array~\cite{PANDA-X:2021jua,Liang:2021nsz,LUX:2017lif} allows precise reconstruction of the three-dimensional interaction vertex~\cite{Angle:2006rj}. Fiducialization in event selection enables an effective way to suppress external gamma and neutron background for all rare event searches and to minimize effects from imperfections of the TPC near its surface. The ratio of S2 and S1 signals further allows for discrimination between different types of interaction in the liquid xenon TPC: nuclear recoils (NR) and electronic recoils (ER). Nuclear recoils are most notably induced by WIMPs, through Coherent Elastic Neutrino-Nucleus Scattering (CE$\nu$NS), and by neutrons; whereas electronic recoils are produced by gamma rays and internal beta decays~\cite{Aprile:2005mz,Akerib:2020lkv}. Nuclear recoils exhibit a lower S2/S1 ratio and can therefore be distinguished from electronic recoils~\cite{Aprile:2006kx,Akerib:2020lkv}. Excellent energy resolution further helps to differentiate various signals from relevant background~\cite{XENON:2020iwh}. As we explain in \autoref{sec:s2only}, the scientific reach of these TPCs can be extended towards lower energies by dropping the requirement of the presence of an S1; this results in sensitivity to single electrons.

\subsection{Xenon as a Detector Medium}\label{sec:detector_medium}

Xenon as a detection medium exhibits several desirable features~\cite{Doke:2002oab,Aprile:2008bga}, giving it a significant advantage as a target material. Assuming a W-value of 11.5\,eV~\cite{Baudis:2021dsq}, averaged over scintillation and ionization, leads to yields as high as $\sim$65 photons per keV for gamma rays that are of order $\sim$100~keV~\cite{Chepel:2012sj,Lenardo:2014cva}, similar to other excellent scintillators such as NaI and CsI. For nuclear recoils, the yields are still $\sim$10\% of that, even below $\sim$10~keV~\cite{Aprile:2018jvg,PandaX-II:2021jmq}. Energy resolutions better than 1\% ($\sigma/E$) have been achieved at MeV scales~\cite{XENON:2020iwh} and mm-level position resolution can routinely be achieved~\cite{Solovov:2011aa,Aprile:2012vw,Baudis:2020nwe}. 

Liquid xenon is a well-understood and well-characterized detector medium. Based on world data, its response can be accurately modeled using the Noble Element Simulation Technique (NEST), a code package to simulate interactions in liquid xenon (and argon) and their detection in a TPC~\cite{Szydagis:2011tk, Szydagis:2013sih, Lenardo:2014cva, szydagis_m_2022_6028483}. This includes various interactions of interest, such as electronic recoils induced by gamma and beta rays, nuclear recoils, energy deposits by alphas, and more complex decays such as that from $^{83\text{m}}$Kr. These models have been demonstrated to correctly reproduce the mean yields and their widths across a wide range of detector parameters and energies even down to $300\1{eV}$, making this simulation package a mature, comprehensive tool for liquid xenon experiments. As a result of this and other efforts~\cite{Sorensen:2008ec,Dahl:2009nta,Bezrukov:2010qa,Sorensen:2010hq,Mu:2013dga,Mu:2013pja,Wang:2016obw}, the light and charge yields can now be accurately described and predicted, with good comparisons to existing calibration data sets, as a function of energy, stopping power, drift electric field, extraction electric field, particle and interaction type, and in some cases even concerning density, phase, and angle of the recoil relative to the drift field. 

Liquid xenon may be naturally contaminated with radioactive isotopes that could produce a dark matter background, such as $^{37}$Ar, $^{85}$Kr or $^{222}$Rn. However, purification to very high levels has already been demonstrated in dark matter~\cite{Abe:2008py,Akerib:2016vxi,Aprile:2016swn,Aprile:2018dbl} and neutrinoless double-beta decay experiments~\cite{Albert:2014awa}. Cosmogenically-produced $^{37}$Ar decays away quickly~\cite{LUX-ZEPLIN:2022sad}, and purity levels achieved for $^{85}$Kr are already sufficient for the next-generation detector discussed here. Further advantages of xenon are obtained through its high charge number $Z$, mass number $A$, and density; these allow for self-shielding of gamma-ray and neutron backgrounds, which will multiply-scatter, especially in the outer limits of the fiducial volume (\autoref{sec:backgrounds}). Xenon also contains odd-neutron isotopes for spin-dependent neutron coupling (\autoref{sec:sd}), and enough residual spin-dependent proton sensitivity to produce competitive results for that science channel. In addition, natural xenon possesses promising isotopes for the search for neutrinoless double-beta decay (\autoref{sec:0nubb}) and double electron capture (\autoref{sec:dec}). Finally, the mass of the xenon nucleus makes it kinematically ideal for WIMPs in the mass range above $\sim 10\1{GeV/c^2}$.

In the following sections, we highlight the science case for a large, next-generation liquid xenon TPC detector for astroparticle physics. In \autoref{sec:wimps}, we overview various WIMP dark matter models, and the expected sensitivity when probing such models. In \autoref{sec:broaderdarkmatter}, we discuss other dark matter models that a next-generation liquid xenon TPC is sensitive to. In \autoref{sec:doublebeta}, we review double-beta processes to probe physics beyond the Standard Model. In \autoref{sec:neutrinos}, we discuss the science channels using neutrinos for astroparticle physics and particle physics. In \autoref{sec:otherstuff}, we collect physics channels that are not part of the above categories. Mitigation and rejection of detector backgrounds is sketched out in \autoref{sec:backgrounds}. The relation of a next-generation liquid xenon TPC to other future experimental efforts is discussed in \autoref{sec:complementarity}. Finally, we review some of the already-documented support for such a detector in the particle physics community in \autoref{sec:priority} before concluding in \autoref{sec:summary}.

\section{Dark Matter WIMPs}\label{sec:wimps}

\subsection{WIMP Direct Detection}
A well-motivated candidate for particle dark matter is the Weakly Interacting Massive Particle (WIMP)~\cite{Jungman:1995df,Gelmini:2010zh}. While the list of possible dark matter candidates is now quite long, the WIMP model remains a leading scenario, with large regions of well-motivated yet unprobed parameter space~\cite{Arcadi:2017kky}. The hierarchy problem~\cite{Witten:1981nf}, specifically the surprisingly and unnaturally low mass of the Higgs particle, continues to strongly motivate searches for new physics and new particles at the $\sim$100\,GeV scale of the electroweak force~\cite{Tanabashi:2018oca}. This alone would motivate the WIMP hypothesis and WIMP-focused searches, but there is also an additional motivation from what has been termed the `WIMP miracle': if a new stable particle existed in this mass range, and if it interacted with Standard Model particles via some force also at the electroweak scale, then a very simple thermal freeze-out process in the early universe would result in the dark matter density we see in the universe today~\cite{Bertone:2010zza,Steigman:2012nb}. It is this surprising connection of particle physics at the Weak scale and the evolution of the macroscopic density in the early universe that continues to motivate searches for WIMP dark matter. Few other models can point to as clear a convergence.

The assumption of a massive (electroweak scale) mediator implies a lower bound on the WIMP mass, the so-called Lee-Weinberg limit~\cite{Kolb:1985nn}. A heavy mediator will suppress the dark matter annihilation cross section. Thus, for dark matter with a mass of less than $\sim2$\,GeV$/c^2$, the thermal freeze-out process of the early universe ends too early and results in a dark matter density that is too high and inconsistent with observation.

Because WIMPs are so well-motivated, searches for particles satisfying these criteria are underway in parallel following three general and complementary strategies~\cite{Balazs:2014rsa,Roszkowski:2017nbc} (see also \autoref{sec:complementarity}): 1) WIMP production at high energy colliders such as the Large Hadron Collider (LHC)~\cite{Kahlhoefer:2017dnp}; 2) indirectly via WIMP annihilation in dense astrophysical environments that produce astrophysical signals in various Standard Model particle channels~\cite{Gaskins:2016cha}; and 3) directly via observation of nuclear recoils produced by dark matter scattering as proposed here~\cite{Goodman:1984dc,Undagoitia:2015gya}. The next-generation liquid xenon-based experiment discussed here is complementary to the next generation of astrophysical searches~\cite{Ong:2017ihp} and the high-luminosity LHC~\cite{Arduini:2016xsb} at a similar time scale.

Direct detection experiments are particularly interesting for a variety of reasons. As scattering interactions happen at energies far below the electroweak scale itself, the interaction mechanism can be simplified and described as a contact interaction. A diverse set of high-energy models will therefore appear nearly identical at these low energies, distinguished almost exclusively by the characteristic scattering cross section. This generality of direct detection via low-energy scattering is a significant advantage to this detection approach. Also, for a large set of WIMP models and a wide range of WIMP masses, direct dark matter experiments depend only linearly on the local dark matter density, which makes results robust against astrophysical uncertainties. Further, for relics produced by the freeze-out process, such as WIMPs, the relic density is inversely related to the thermal annihilation cross section, such that a dimensional argument can be made that the expected scattering rate in a detector (which goes like density times cross section) should be roughly independent of details of the theory. Another expression of the generality of the direct detection approach is its sensitivity to a large mass range~\cite{Smith:1986ms}. The kinematics for non-relativistic scattering remain unchanged once the WIMP mass is much larger than the target mass, rendering these experiments sensitive to dark matter masses well beyond 100~TeV (and in principle even up to the Planck mass~\cite{Bramante:2018qbc}, see \autoref{sec:planck}). Thus, a single experiment can probe many orders of magnitude of the allowed dark matter mass parameter space.

Xenon in particular is expected to couple well to WIMP dark matter for several reasons~\cite{Newstead:2013pea}. First, in the low momentum-transfer regime of direct detection, a generic spin-independent scattering (\autoref{sec:si}) will interact with the nucleus as a whole as a many-nucleon object, and this coherence provides a significant boost to the corresponding scattering cross section~\cite{Smith:1983jj,Goodman:1984dc}, scaling roughly as the square of the number of nucleons. Therefore, a heavy nucleus like xenon is significantly favored over lighter options. A second advantage is the large number of common natural xenon isotopes, resulting in a diversity of nuclear properties. This variety of isotopes gives xenon significant sensitivity to other interaction models as well, such as spin-dependent (\autoref{sec:sd}) or various effective couplings (\autoref{sec:eft}).

\subsection{WIMP Sensitivity Projections: Method}
\begin{figure}[!htbp] 
	\centering
    \includegraphics[width=\columnwidth,clip]{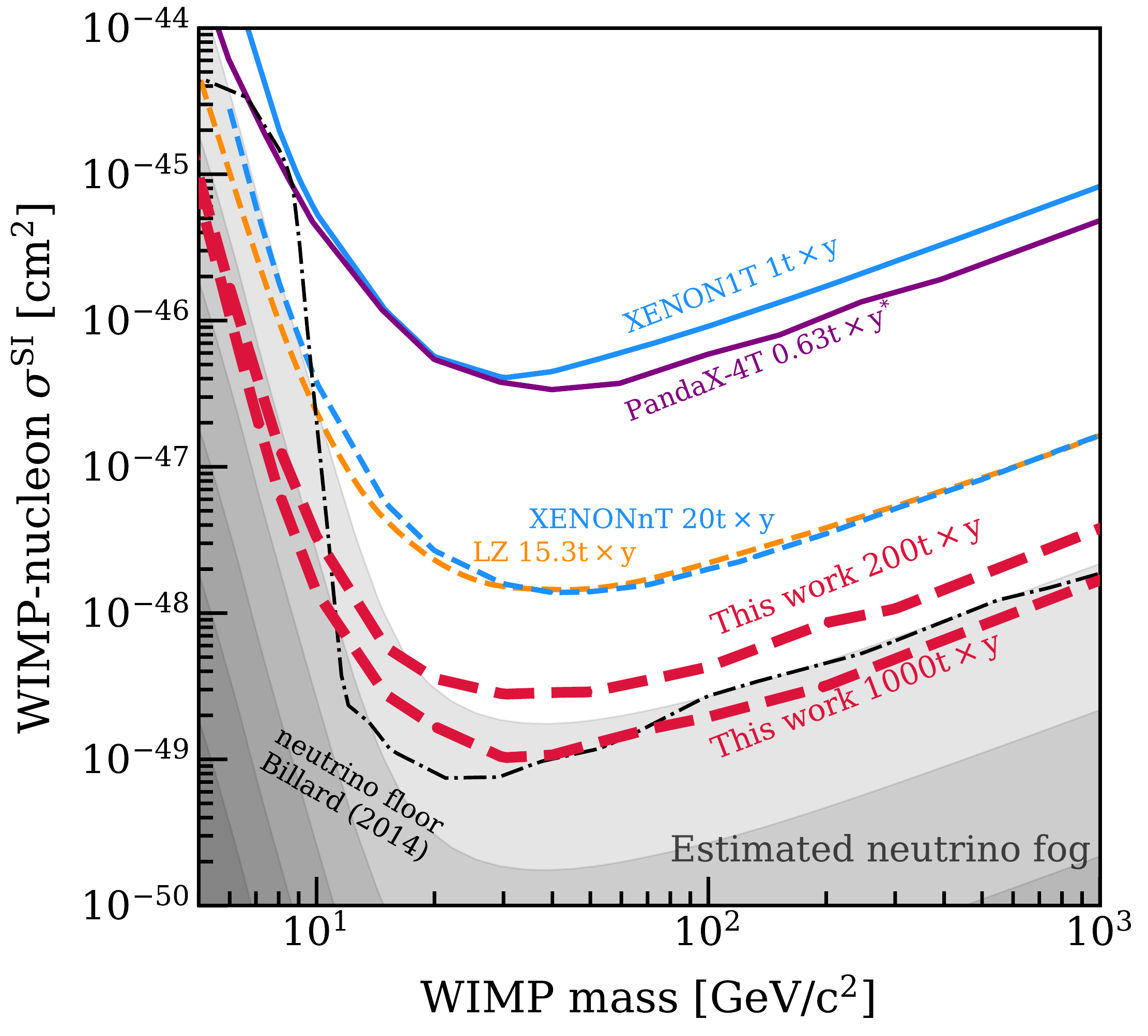}
    \caption{Projections for the next-generation experiment discussed here, together with projected and current leading $90\%$ upper limits, on the spin-independent WIMP-nucleon cross section. Blue and purple solid lines show the current limits from XENON1T~\cite{Aprile:2018dbl} and PandaX-4T~\cite{PandaX-4T:2021bab} (non-blind*). Dashed blue and orange lines indicate sensitivity projections from LZ~\cite{Akerib:2018lyp} ($15.3\,\tonneyear$, one-sided) and XENONnT~\cite{Aprile:2020vtw} ($20\,\tonneyear$). Projected median upper limits for exposures of $200\,\tonneyear$ and $1000\,\tonneyear$ are plotted in dashed red. The dashed line shows one definition of the ``neutrino floor''~\cite{Billard:2013qya}, the shaded gray area indicates the ``neutrino fog'', specifically where more than one, 10, 100, etc. neutrino events are expected in the $50\%$ most signal-like S1/S2 region. Calculations follow Refs.~\cite{wimprates,Lewin:1995rx}.}
\label{fig:si_sensitivity}
\end{figure}

\autoref{fig:si_sensitivity} on page~\pageref{fig:si_sensitivity} and \autoref{fig:SD_sensitivity} on page~\pageref{fig:SD_sensitivity} show sensitivity estimates for a liquid xenon TPC with only neutrino-induced backgrounds and the double beta decay of ${}^{136}$Xe considered. We use a binned likelihood in position-corrected $cS1$ and $cS2$ (see e.g.~\cite{Akerib:2017vbi,Aprile:2019bbb}), and assume the log-likelihood ratio test statistic is asymptotically distributed~\cite{Wilks:1938dza}. 

Particle yields and the detector response to electronic and nuclear recoils is simulated using v2.1.0~\cite{szydagis_m_2022_6028483} with the \texttt{LUXrun03} detector model, roughly corresponding to the model presented in~\cite{LUX:2019ius} for the first science run of LUX. The modelled detector response is similar to the models assumed for the sensitivity projections for LZ~\cite{Akerib:2018lyp} and XENONnT~\cite{Aprile:2020vtw}. For S1 scintillation signals, the detector model assumes a $g_1$ value of $0.12~\mathrm{phd/photon}$  and a 2-fold photon hit threshold for S1s. For S2 ionization signals, the $g_2$ value depends primarily on the photon yield $g^\text{gas}_1$ and field strength in the gas gap, which are $0.1~\mathrm{phd/photon}$ and $6.4~\mathrm{kV/cm}$, respectively; more details can be found in ~\cite{LUX:2019ius}. This corresponds roughly to a $g_2=13.8~\mathrm{phd/e}$. The S2 threshold is assumed as $165~\mathrm{phd}$. The spatially varying drift field for this simulation is between $\sim90~\mathrm{V/cm}$ and $300~\mathrm{V/cm}$, and the electron lifetime assumed is $\sim16$~times the maximum drift length. We leave a detailed parametric investigation of the sensitivity of such a next-generation detector to a future study.

The background model is made up of the intrinsic electronic and nuclear recoil backgrounds. The expectation value is dominated by electronic recoils from naturally occurring $^{136}$Xe and solar (mostly $pp$) neutrinos scattering off electrons. Electronic recoil events can be distinguished from a WIMP signal using the ionization signal. 

Nuclear recoil events from neutrons scattering in the detector volume can be separated to some degree from a WIMP signal based on the recoil energy spectrum and their tendency to scatter multiple times. Further, neutrons can be tagged surrounding the detector with an active neutron veto. We thus only include nuclear recoil backgrounds from $^8$B, HEP, diffuse supernovae and atmospheric neutrinos. These neutrino signals, while being an interesting signal in their own right (Sec.~\ref{sec:neutrinos}), may significantly affect the sensitivity to dark matter as they are becoming the dominant background (Sec.~\ref{sec:nufloor}). 

The neutrino recoil spectra, as well as flux uncertainties on the different components, are taken to be the same as in~\cite{Aprile:2020vtw}, with spectra from Ref.~\cite{KotilaIachelloWebSite,Chen:2016eab,Billard:2013qya,Newstead:2020fie}. WIMP recoil spectra are computed using the \texttt{wimprates} package~\cite{wimprates}, with spin-independent computations from Ref.~\cite{Lewin:1995rx}, spin-dependent computations from Ref.~\cite{Klos:2013rwa}, and WIMP-pion recoil spectra from Ref.~\cite{Aprile:2018cxk,Hoferichter:2018acd}. We use the background and signal distributions to construct signal regions for each WIMP interaction and mass as the $50\%$ most signal-like region in S1 and S2, ordered by signal-to-background ratio. We indicate the region at which neutrinos become an appreciable background as the cross section where the WIMP and neutrino expectation in the signal region are equal. Levels where the neutrino signal is equal, 10 times, 100 times etc.~of the WIMP signal are indicated by the shared gray regions labeled ``neutrino fog'' in Figures~\ref{fig:si_sensitivity} and~\ref{fig:SD_sensitivity}. Estimates of where experimental sensitivity will improve only very slowly with exposure depend crucially on the uncertainty on the neutrino signal and detector response. Attempts to quantify the ``neutrino floor``, such as~\cite{Billard:2014yka,OHare:2016pjy}  (the former is included as a dashed line in figure~\ref{fig:si_sensitivity}) often assume e.g.~very low experimental energy thresholds in order to reflect the ultimate limit. Further discussion of the neutrino background may be found in section~\ref{sec:nufloor}.

\subsection{Spin-Independent WIMPs}\label{sec:si}

The next-generation detector proposed here can be thought of as the `ultimate' WIMP dark matter detector in two senses: exposure and energy threshold. Traditionally, WIMP detection has been limited primarily by the experiment's exposure (expressed in mass $\times$ time), and sensitivity has progressed proportionally to that exposure. This linear scaling will hold as long as contamination by any non-WIMP recoils remains small. This next-generation WIMP detector will be the last to benefit from this proportional scaling over much of its operating time. Any larger experiment would face a rate of coherent elastic neutrino-nucleus scattering from astrophysical sources~\cite{Monroe:2007xp,Billard:2013qya}. While that is an interesting signal in its own right (\autoref{sec:neutrinos}), neutrinos present an unavoidable background to WIMP sensitivity. 

The energy threshold of this search is also important. A recoil threshold of $\sim$keV is required in order to efficiently test WIMP hypotheses down to the Lee-Weinberg limit of few GeV$/c^2$ mass. The goal for an ultimate WIMP dark matter detector, then, can be described as testing the entire WIMP mass range ($\sim$2~GeV$/c^2$ -- $\sim$100~TeV$/c^2$) down to cross sections limited by neutrino scattering. Such a detector also has sensitivity to many theoretically interesting and yet unexplored dark matter candidates (\autoref{sec:broaderdarkmatter}) and probes the coupling of dark matter to the Higgs boson~\cite{Feng:2014vea}.

\begin{figure}[!htbp] 
	\centering
    \includegraphics[width=\columnwidth]{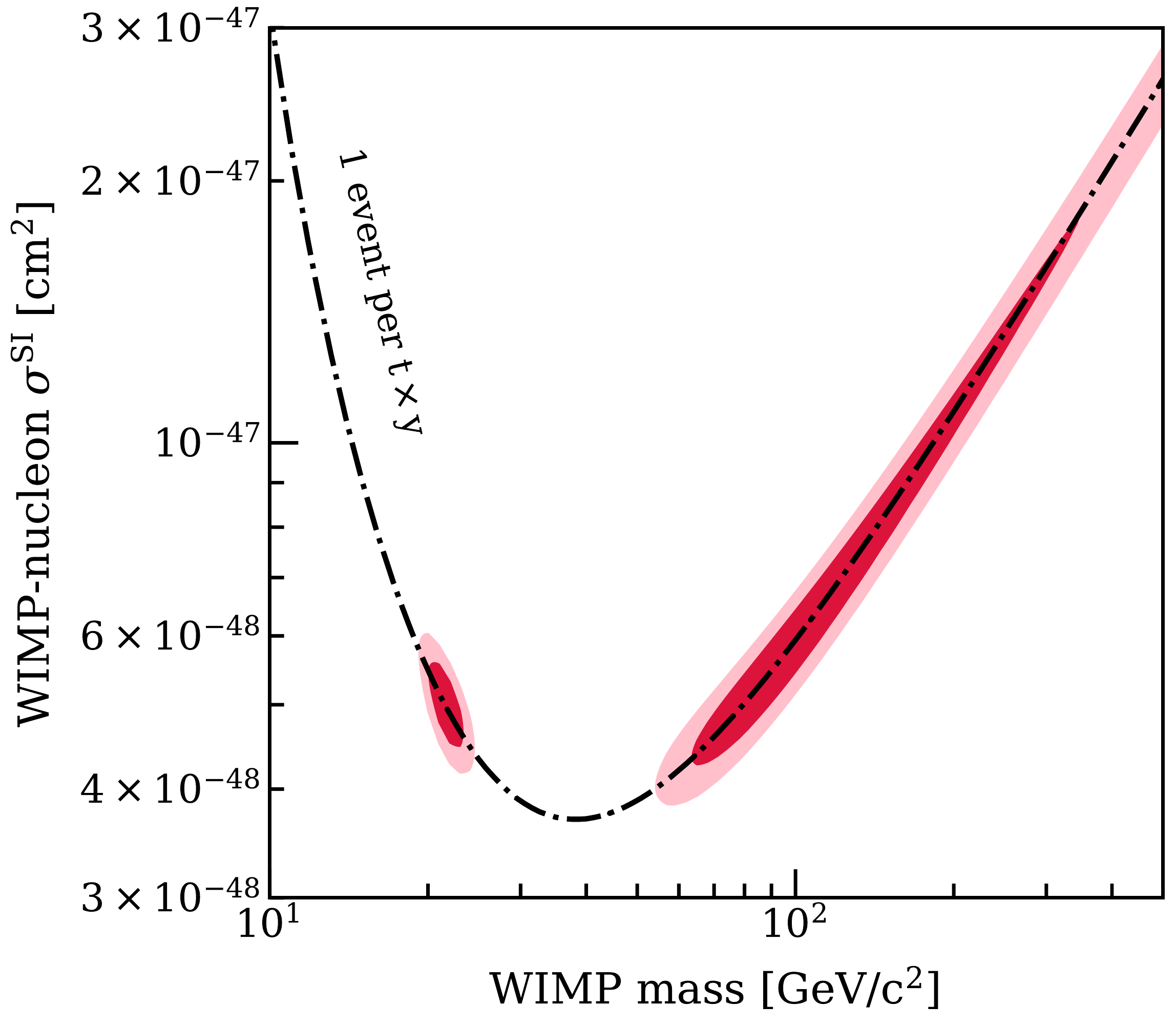}
    \caption{Illustration of 1-~and 2-sigma (dark and light red) confidence intervals on spin-independent WIMP signals with a $1000\,\tonneyear$ exposure and WIMP masses of either 20 or $100\1{GeV/c^2}$. The signal expectation for the excesses is $1/\tonneyear$, indicated by the black dash-dotted line.}
\label{fig:excess_contour}
\end{figure}

To indicate the WIMP mass and cross section resolution expected for a signal from WIMPs roughly one order of magnitude below current constraints (one event per tonne-year), \autoref{fig:excess_contour} shows confidence intervals for spin-independent WIMP signals at 20 and $100\1{GeV/c^2}$. At high masses, spin-independent WIMP spectra are degenerate in WIMP mass (as the kinematics only depend on the reduced mass). This leads to poor mass resolution, which can be significantly improved using additional, different target materials~\cite{Pato:2010zk}. An excess for intermediate and low masses will be well-constrained both in mass and cross section using a xenon target alone.

A simple variation of the vanilla spin-independent WIMP scenario is to allow the interaction strength to depend on the nucleon type (proton or neutron) with non-trivial coupling strengths $f_p, f_n$~\cite{Kamionkowski:1994rm}. The deviation of the ratio $f_p/f_n$ from $1$ will then depend on the specific dark matter model. If for a given nuclear isotope, $f_p/f_n = (Z-A)/Z$, then this isotope would give no constraint. Fortunately, the mixture of multiple isotopes in xenon detectors provides sensitivity to even the most difficult case of $f_p/f_n\simeq -1.4$~\cite{Chang:2010yk,Feng:2011vu,Frandsen:2011ts}, providing yet another benefit of xenon as a target material.

\subsection{Spin-Dependent Scattering}\label{sec:sd}

\begin{figure}[!htbp] 
	\centering
    \includegraphics[width=\columnwidth,clip]{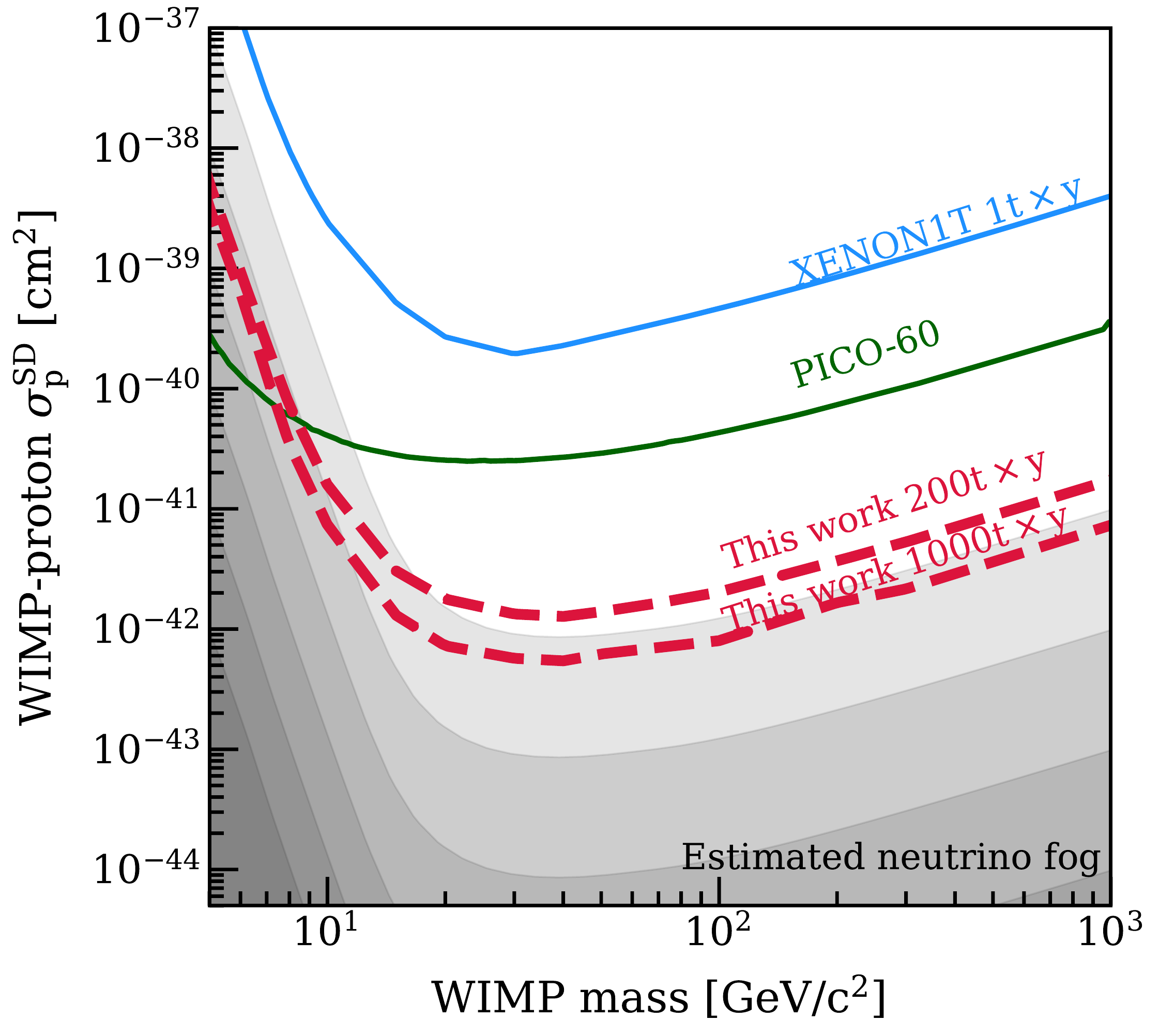}
    \includegraphics[width=\columnwidth,clip]{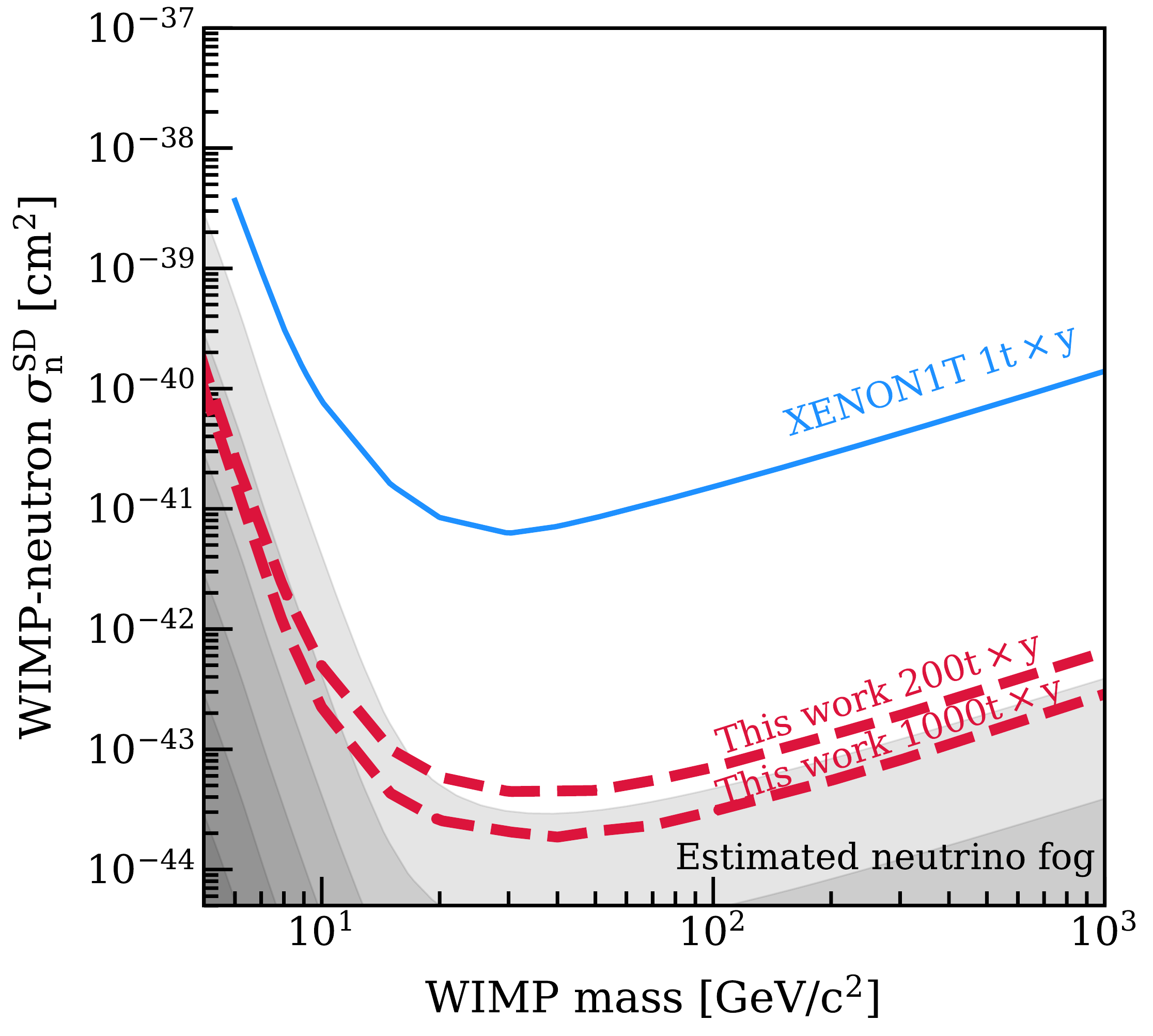}
    \caption{Projections and current leading $90\%$ upper limits on the spin-dependent WIMP-nucleon cross section, assuming that the WIMP couples only to proton spins (top) or neutron spins (bottom). Green and blue solid lines show the current leading limits by PICO-60~\cite{Amole:2017dex} and XENON1T~\cite{Aprile:2018dbl,Aprile:2019xxb}. Projected median upper limits for exposures of $200\,\tonneyear$ and $1000\,\tonneyear$ are plotted in red. The shaded gray areas indicate the ``neutrino fog'' with the lightest area showing the WIMP cross section where more than one neutrino event is expected in the $50\%$ most signal-like $S1,S2$ region. Subsequent shaded areas indicate tenfold increases of the neutrino expectation. Calculations follow Refs.~\cite{wimprates,Klos:2013rwa}.}
\label{fig:SD_sensitivity}
\end{figure}

The simplest deviation from the spin-independent scattering to a more complicated coupling can be modeled by allowing the WIMP to interact solely with the nuclear spin but with different couplings $a_p,a_n$ to protons and neutrons. This scenario is typically referred to as spin-dependent scattering~\cite{Iachello:1990ut,Engel:1992bf,Tovey:2000mm}.  If one simplifies this picture by assuming that one coupling vanishes, then the derivation of a differential rate of scattering events by WIMPs depends on the spins and nuclear structure (mostly of the unpaired nucleon) of the nuclei in the target. Contributions from two-nucleon currents improve the sensitivity to the spin-dependent WIMP-proton coupling in xenon, see \autoref{sec:cheft}.

For xenon detectors, the two naturally occurring isotopes $^{129}$Xe (spin-1/2) and $^{131}$Xe (spin-3/2), with natural abundances of 26.4\% and 21.2\%, respectively, are most relevant for this spin-dependent coupling. Both have an unpaired neutron, making xenon also an ideal target for detecting the spin-dependent WIMP-neutron cross section. The projected sensitivity for a next-generation liquid xenon TPC is shown in \autoref{fig:SD_sensitivity}, calculated using the same assumptions and method as in \autoref{fig:si_sensitivity}.

\subsection{Effective Field Theory}\label{sec:eft}

The spin-independent and spin-dependent scattering discussed in the previous sections~\ref{sec:si} and~\ref{sec:sd} are the more frequently studied interactions of the WIMP with Standard Model fields. Their motivation dates back to dark matter candidates in supersymmetric theories~\cite{Engel:1992bf} defining the leading responses related to the nuclear density (therefore scaling coherently with the number of nucleons $A$, spin-independent interactions) or to the nuclear spin (spin-dependent interactions). A more systematic picture covering more general WIMP-nucleus interactions beyond standard spin-independent and spin-dependent scattering has been worked out recently using effective field theories (EFTs). This includes both a nonrelativistic framework, see \autoref{sec:nreft}, as well as chiral effective field theory, see \autoref{sec:cheft}, which incorporates the constraints from QCD at low energies.

\subsubsection{Nonrelativistic Effective Field Theory}\label{sec:nreft}

The nonrelativistic EFT (NREFT)~\cite{Fan:2010gt,Fitzpatrick:2012ix,Anand:2013yka} integrates out all degrees of freedom except for nucleons and the WIMP. The effective operators that describe the coupling of the WIMP to nucleons are constructed imposing Galilean invariance in terms of the momentum transfer $q$, the WIMP transverse relative velocity $v^\perp$, and the spins of the nucleon and the WIMP~\cite{Fan:2010gt, Fitzpatrick:2012ix,Anand:2013yka}. At lowest (zeroth) order in $q$ and $v^\perp$, the only contributions correspond to the leading operators considered for spin-independent and spin-dependent scattering. Up to second order in $q$ and first order in $v^\perp$, 14~operators appear at the single-nucleon level for spin-1/2 dark matter, each with different isoscalar and isovector (or, equivalently, proton and neutron) couplings~\cite{Anand:2013yka}. The corresponding coefficients, usually considered to be independent, have been constrained from several experiments~\cite{Schneck:2015eqa,Aprile:2017aas,Xia:2018qgs,Angloher:2018fcs}. With few exceptions, the best constraints are given by experiments using a xenon target. For an extension of NREFT to dark matter of spin 1 or higher, see~\cite{Dent:2015zpa,Catena:2019hzw,Gondolo:2020wge}. Even given the significant uncertainties in the WIMP halo phase space distribution, NREFT coefficients could nevertheless be constrained by a single next-generation direct detection experiment, if some dozen events would be detected~\cite{Krauss:2018pvg}.   

Since the NREFT is limited to nucleons as degrees of freedom, additional matching steps are required to constrain particular WIMP models from experimental limits. This is because the NREFT coefficients contain information on the underlying WIMP-quark or WIMP-gluon operators, but also on hadronic matrix elements (\autoref{sec:structure}). In addition, there is a~priori no hierarchy among the various NREFT operators apart from their scaling in $q$ and $v^\perp$. In that sense, the NREFT can be considered minimal, as even constraints from QCD are not imposed. In addition, the NREFT formalism has also been used to represent contributions beyond the applicability of the strict EFT. For example, long-range effects due to pion exchange (as occurs in chiral EFT) or electromagnetic interactions (such as dipole operators) can be expressed in terms of $q$-dependent NREFT Wilson coefficients. For a complete description, however, the corresponding degrees of freedom need to be included in the EFT.  

\subsubsection{Chiral Effective Field Theory}\label{sec:cheft}

Chiral EFT~\cite{Epelbaum:2008ga,Machleidt:2011zz,Hammer:2012id} classifies the possible interactions of the WIMP with nucleons according to their chiral scaling, i.e., the scaling with momenta and quark masses, with nucleons and WIMPs but also pions (and, in SU(3), kaons and $\eta$-mesons) as active degrees of freedom. In this way, the constraints from the chiral symmetry of QCD are automatically included. The chiral regime is appropriate to study WIMP-nucleus scattering because the typical momentum transfer $q$ is of the order of the mass of the pion, the pseudo-Nambu-Goldstone boson resulting from the spontaneous breakdown of chiral symmetry. This is also the relevant scale for the typical momenta in heavier nuclei, such as xenon, used for direct detection experiments.

At the single-nucleon level, the chiral analysis can be mapped onto the NREFT operator basis~\cite{Hoferichter:2015ipa,Bishara:2016hek,Bishara:2017pfq}. This provides a prediction for an additional hierarchy of the NREFT operators based on their chiral scaling, which significantly simplifies the number of one-body operators relevant for WIMP-nucleus scattering. A second important advantage of the chiral EFT framework is that it predicts subleading multi-nucleon effects. For example, contributions in which the WIMP couples to a virtual pion exchanged between the nucleons inside the nucleus (\autoref{sec:wimp_pion}) appear at subleading order in the chiral expansion. Such meson-exchange currents (or two-body currents) provide subleading contributions to generalized spin-independent and spin-dependent scattering and have been studied in a number of papers~\cite{Prezeau:2003sv,Cirigliano:2012pq,Menendez:2012tm,Klos:2013rwa,Cirigliano:2013zta,Hoferichter:2015ipa,Hoferichter:2016nvd,Korber:2017ery,Hoferichter:2017olk,Andreoli:2018etf,Hoferichter:2018acd,Hu:2021awl}. For a xenon target, two-nucleon currents improve the sensitivity to spin-dependent WIMP-proton scattering by more than an order of magnitude.

\subsubsection{WIMP-Pion Coupling}\label{sec:wimp_pion}

A novel contribution to WIMP-nucleus scattering that emerged from the chiral EFT analysis (\autoref{sec:cheft}) concerns meson-exchange currents. In the simplest case, the WIMP couples to a virtual pion exchanged between two nucleons within the nucleus. Interestingly, meson-exchange contributions, which enter at subleading order in chiral EFT, can scale coherently with the number~$A$ of nucleons. The combination of the nuclear and chiral hierarchies defines a scaling that lies between the spin-independent and spin-dependent responses, coherent but suppressed in the chiral counting. Chiral EFT also predicts the leading meson-exchange currents to dominate over all other NREFT operators except the standard spin-independent one.

\begin{figure}[!htbp] 
	\centering
    \includegraphics[width=\columnwidth,clip]{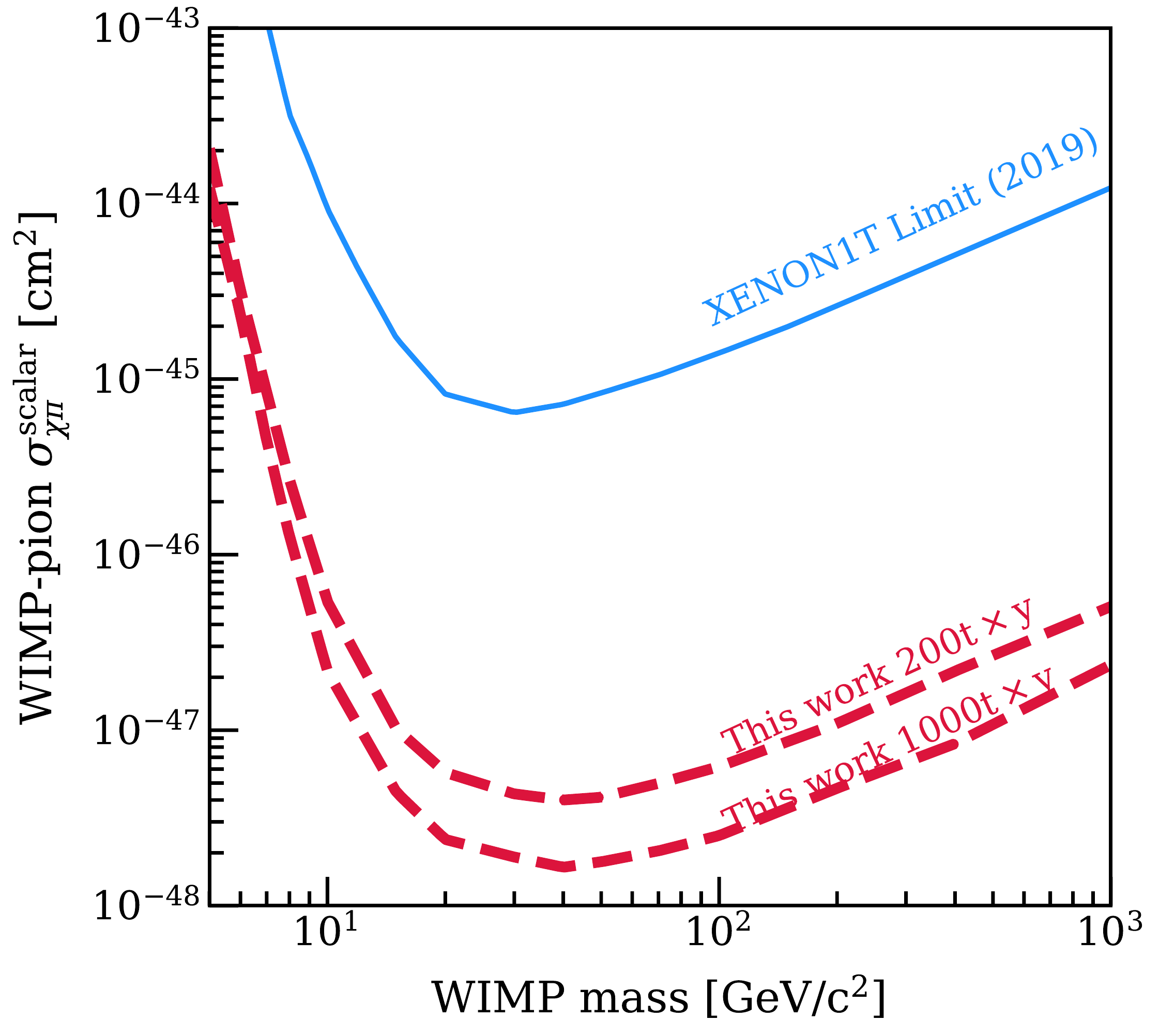}
    \caption{Projections and current leading $90\%$ upper limits on the scalar WIMP-pion interaction cross section. Blue solid lines show the current leading limits by XENON1T~\cite{Aprile:2018cxk}. Projected median upper limits for exposures of $200\,\tonneyear$ and $1000\,\tonneyear$ are plotted in red. Calculations follow Refs.~\cite{wimprates,Hoferichter:2018acd}.}
\label{fig:wimppion_sensitivity}
\end{figure}

As observed in~\cite{Aprile:2018cxk}, once an underlying quark-level operator is specified, the resulting limits can be expressed in terms of a WIMP-pion cross section, in close analogy to the spin-independent and spin-dependent WIMP-nucleon cross sections. The proposed next-generation liquid xenon experiment will improve this result by a similar factor as the standard spin-independent limit, shown in \autoref{fig:wimppion_sensitivity}.

\subsubsection{Three-Flavor EFT and the UV}\label{sec:quark-level}

From a particle physics point of view, the most immediate parameterization of dark matter interactions at low energies, $\mu\simeq 2$ GeV, is in terms of three-flavor dark matter EFT that has been studied extensively~\cite{Goodman:2010yf,Goodman:2010ku,Fox:2011pm,Crivellin:2014qxa,DEramo:2014nmf,Hill:2014yxa,Hill:2014yka,Bishara:2016hek,Bishara:2017pfq,Bishara:2017nnn,Brod:2017bsw,Hoferichter:2015ipa,Hoferichter:2016nvd,Hoferichter:2018acd}. This model has as degrees of freedom the dark matter particle, the lightest three flavors of quarks ($u,d,s$), the gluon, and the photon. Dark matter interactions are organized in terms of dimensions of the operators, so that the effective Lagrangian takes the form ${\cal L}_{\rm DM~EFT}= \sum_{d,a} {\cal C}_a^{(d)} {\mathcal Q}_a^{(d)} / \Lambda^{d-4} $, where ${\cal C}_a^{(d)}$ are dimensionless Wilson coefficients and $\Lambda$ the typical scale of the UV theory for dark matter. The sum is over different types, $a$, and dimensions, $d$, of the operators ${\mathcal Q}_a^{(d)}$. An example of a dimension-6 operator for fermionic dark matter is $(\bar \chi \gamma^\mu \chi)(\bar q\gamma_\mu q)$ for dark matter-quark vector interactions, or a dimension-7 operator $m_q(\bar \chi \chi)(\bar q q)$ for scalar interactions, both of which lead to spin-independent scattering. The full basis of up to and including dimension-7 operators in the limit $\Lambda \gg m_\chi \sim m_W$ can be found in~\cite{Brod:2017bsw}. The case of the heavy dark matter limit is discussed in~\cite{Hill:2011be,Hill:2014yka,Hill:2014yxa,Chen:2019gtm}). The chiral EFT of \autoref{sec:cheft} then gives the hadronization of the three-flavor dark matter EFT and the nuclear response.

The three-flavor dark matter EFT is a valid description for dark sector mediators that are heavier than a few hundred MeV. In this case, the mediators are heavier than the typical momenta exchanged in the dark matter scattering on nuclei and can be integrated out. The Wilson coefficients, ${\cal C}_a^{(d)}$, are constants that contain all the UV dark matter physics. In the absence of a complete UV theory of dark matter they can be freely varied when comparing the results of dark matter direct detection experiments. For $\Lambda$ well above the nuclear scale, the higher dimension operators are suppressed, making the framework predictive. For instance, for $\Lambda \gg m_\chi$,  a truncation at dimension~7 is expected to capture most new physics models.

The connection between the three-flavor dark matter EFT at $\mu =2\1{GeV}$ and the UV theory at $\mu \simeq \Lambda$ is achieved by going through a series of appropriate EFTs and performing the matchings at each threshold~\cite{Crivellin:2014qxa,Brod:2018ust,Bishara:2018vix,DEramo:2014nmf,Crivellin:2014gpa}. In this way the results of indirect dark matter searches and the dark matter searches at the LHC can be reliably compared with the direct detection results. From the perspective of direct detection experiments, the UV physics is encoded in the values of Wilson coefficients ${\cal C}_a^{(d)}$. One can then compare the constraints on ${\cal C}_a^{(d)}$ obtained from direct detection experiments with the constraints imposed by the LHC and indirect dark matter searches on either complete dark matter models or on simplified models by going through the above series of matchings and renormalization group evolutions. 

\subsection{Nuclear Structure Factors}\label{sec:structure}

The WIMP-nucleus cross section is proportional to the nuclear structure factors, which encode the relevant information of the structure of the target nuclei. The EFT operators at the WIMP-nucleon level generate, at the nuclear level, six different nuclear one-body responses analogous to semileptonic weak interactions~\cite{Serot:1978vj,Donnelly:1979ezn,Serot:1979yk}. In addition, chiral EFT predicts two-nucleon nuclear responses associated with meson-exchange currents. The corresponding nuclear structure factors are obtained from the one-body nuclear responses $\mathcal{F}^M$, $\mathcal{F}^{\Phi ''}$, $\mathcal{F}^{\Sigma '}$, $\mathcal{F}^{\Sigma ''}$, $\mathcal{F}^{\tilde{\Phi} '}$, and $\mathcal{F}^{\Delta}$~\cite{Engel:1992bf,Anand:2013yka}, and the two-body nuclear responses $\mathcal{F}_{\pi}$, $\mathcal{F}_{\rm b}$~\cite{Hoferichter:2016nvd,Hoferichter:2018acd}. The one-body structure factors decompose into isoscalar and isovector (or, equivalently, proton and neutron) contributions, e.g., $\mathcal{F}^M_\pm$. $\mathcal{F}^M$ and the two-body $\mathcal{F}_{\pi}$, $\mathcal{F}_{\rm b}$ can receive coherent contributions from all $A$ nucleons in the nucleus while about one in five nucleons contributes coherently to $\mathcal{F}^{\Phi ''}$. These responses dominate spin-independent ($\mathcal{F}^M$) and scalar WIMP-pion ($\mathcal{F}_{\pi}$) scattering. $\mathcal{F}^{\Sigma '}$ and $\mathcal{F}^{\Sigma ''}$ are usually rewritten in terms of the more common $S_{00}$, $S_{01}$, $S_{11}$ or $S_p$, $S_n$ in spin-dependent analyses. They are related to the spin distribution in the nucleus and are not coherent because the nuclear pairing interaction couples nucleons in spin-zero pairs.

The nuclear structure factors allow one to factorize the nuclear response from the hadronic matrix elements and the couplings of the WIMP. Schematically, the WIMP-nucleus cross section decomposes as
\begin{equation}
 \frac{{\rm d}\sigma_{\chi\mathcal{N}}}{{\rm d}q^2}\propto \sum_i \big|c_i \mathcal{F}_i(q^2)\big|^2 + {\rm interference\ terms},
\end{equation}
where the $\mathcal{F}_i(q^2)$ denote the nuclear structure factors and the $c_i$ involve a convolution of hadronic matrix elements and WIMP couplings. Thus, in the case of scalar operators, the role of the pion-nucleon $\sigma$ term is well known in the literature~\cite{Bottino:1999ei,Bottino:2001dj,Ellis:2008hf,Crivellin:2013ipa,Hoferichter:2015dsa,Gupta:2021ahb}. In special cases, the WIMP-nucleus cross section can be expressed in terms of single-particle cross sections: (i) if only $c^M_+$ is non-vanishing it can be expressed by the spin-independent isoscalar WIMP-nucleon cross section (\autoref{sec:si}); (ii) if only the coefficients of $S_p$ or $S_n$ are non-vanishing, by the spin-dependent WIMP-proton or WIMP-neutron cross section (\autoref{sec:sd}); and (iii) if only $c_\pi$ is non-vanishing, by the scalar WIMP-pion cross section (\autoref{sec:wimp_pion}).

\begin{figure}[!htbp] 
\centering
	\includegraphics[width=\columnwidth,clip]{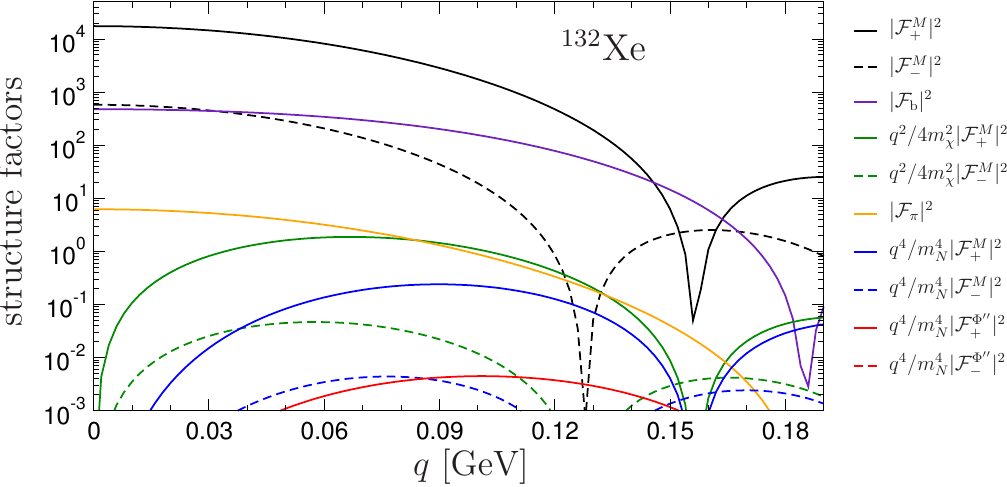}
	\caption{Structure factors for $^{132}$Xe from one- and two-body contributions (without interference terms). Solid lines show isoscalar and two-body contributions while dashed lines indicate isovector couplings. Figure taken from Ref.~\cite{Hoferichter:2018acd}.}
	\label{fig:structure_factors_xe132}
\end{figure}

In general, reliable nuclear structure factors for any nuclear response require a good description of the nuclear target. The only exception is the leading $\mathcal{F}^M_+$ structure factor in spin-independent scattering, for which the purely phenomenological Helm form factor~\cite{Helm:1956zz,Lewin:1995rx} is a common and good description~\cite{Vietze:2014vsa}. For heavy targets such as xenon, structure factors need to be calculated from nuclear theory. The nuclear shell model is presently the method of choice with significant progress in recent years. The shell model solves the many-body problem in a relatively small configuration space (one or two harmonic oscillator shells near the Fermi surface) with a phenomenological nuclear interaction adapted to the configuration space~\cite{Caurier:2004gf}. The description of excitation energies, charge radii, and electromagnetic properties of medium- and heavy-mass nuclei, including all stable xenon isotopes, is already very good~\cite{Hoferichter:2018acd,Vietze:2014vsa,Klos:2013rwa}. State-of-the-art nuclear structure factors are easily available in dedicated notebooks documented in~\cite{Hoferichter:2018acd,Anand:2013yka}. \autoref{fig:structure_factors_xe132} shows nuclear structure factors for a general coherent WIMP scattering off $^{132}$Xe (26.9\% natural abundance) with the hierarchy given by chiral EFT.

More advanced nuclear structure {\it ab initio} calculations treat explicitly all nucleons in the nucleus (see e.g.~\cite{Kamada:2001tv,Carlson:2014vla,Navratil:2016ycn,Hagen:2013nca,Hergert:2015awm,Stroberg:2019mxo}). They can use nuclear interactions based on chiral EFT, thus consistently providing the nuclear states and WIMP-nucleon operators that enter the calculation of the structure factors. This will allow one to estimate theoretical uncertainties. While nuclear structure factors obtained with {\it ab initio} many-body techniques have historically been limited to light nuclei with $A \leq 6$~\cite{Gazda:2016mrp,Andreoli:2018etf,Korber:2017ery}, recent progress has been significant and {\it ab initio} spin-dependent structure factors have been calculated very recently~\cite{Hu:2021awl} with the valence-space in-medium similarity renormalization group method.

\subsection{Inelastic Scattering}\label{sec:inelastic}

The recoil energy spectrum resulting from spin-dependent interactions is similar to the one expected from spin-independent interactions. Using different target materials with other experiments can help to break that degeneracy, as would be a different mixture of isotopes of xenon in a target. In addition, liquid xenon TPCs can even differentiate between these two interaction channels with one and the same exposure, as WIMPs might alternatively scatter inelastically off nuclei that possess low-lying excited states up to $\sim$100 keV~\cite{Ellis:1988nb}, including $^{129}$Xe and $^{131}$Xe~\cite{Baudis:2013bba,Aprile:2017ngb,Aprile:2020sfu}. This inelastic scattering in the nuclear sector is not to be confused with dark matter models in which the WIMP can be excited, as is discussed in the context of the inelastic Dark Matter (iDM) model in \autoref{sec:inelasticdarkmatter}. 

Inelastic scattering is always non-coherent, because of the different initial and final nuclear states. This would allow one to narrow the nature of the underlying WIMP-nucleon interaction, testing the spin-dependent case upon detection in the simplest scenario. In addition to the nuclear recoil, a prompt electronic recoil is caused by the de-excitation of the up-scattered xenon nucleus. Such interactions thus suffer from the larger background of electronic recoils. However, since they would only be expected for non-coherent spin-dependent interactions, given sufficient statistics, a single xenon detector would be able to extract information about dark matter that is inaccessible to the elastic channel alone. 

Observation of such inelastic scattering would provide a range of further insights to the nature of dark matter: Each unique nuclear excitation is sensitive to a distinct portion of the WIMP halo, so that multiple contributions from the inelastic channel could be combined with that of the elastic channel to constrain the WIMP velocity distribution~\cite{Baudis:2013bba}. In addition, the range of observed recoil energies as well as the energy at which the inelastic channel begins to overtake the elastic one would indicate the mass of the incident WIMP. Finally, in contrast to the elastic channel, the inelastic event rate may be enhanced or suppressed with the enrichment or depletion of $^{129}$Xe and $^{131}$Xe. This flexibility would allow one to optimize data acquisition in a xenon detector. The most stringent limit on inelastic WIMP-nucleon scattering currently comes from the XMASS detector~\cite{Suzuki:2019ine}. Prospects for a future detection of dark matter detection with inelastic xenon transitions are further discussed in~\cite{McCabe:2015eia}.

\subsection{Discriminating Between WIMP-Nucleus Responses}

Given the number of different nuclear responses (\autoref{sec:structure}), a key question is how they could, in the event of a detection, be distinguished in order to extract information on the nature of the WIMP~\cite{Krauss:2018pvg}. One possible strategy concerns the study of inelastic scattering into low-lying excited states of the xenon target, discussed in the previous \autoref{sec:inelastic}. The detection of the inelastic channel, in addition to the elastic scattering would primarily point to the non-coherent character of the WIMP-nucleus interaction, suggesting a spin-dependent interaction as the prime choice.

\begin{figure}[!htbp] 
	\centering
	\includegraphics[width=\columnwidth]{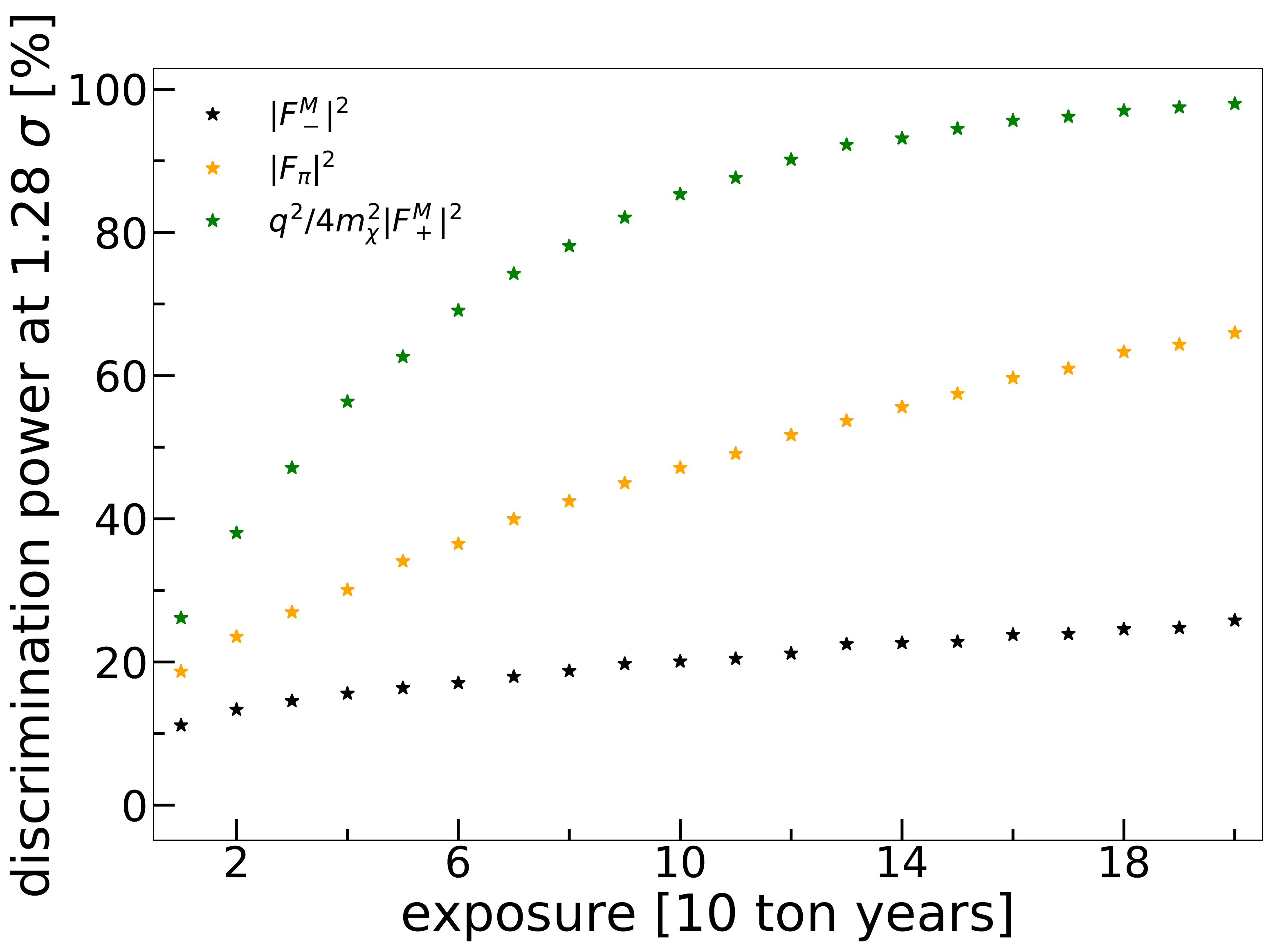}
	\caption{Discrimination power against $|\mathcal{F}_+^M|^2$ vs.\ exposure for three selected structure factors, $|\mathcal{F}_-^M|^2$ (black), $|\mathcal{F}_\pi|^2$ (orange), and $q^2/4m_\chi^2 |\mathcal{F}_-^M|^2$ (green). 
	The detector setting is like the one discussed here, for a WIMP mass of $m_\chi =100$ GeV$/c^2$ and interaction strength $\sigma_0=10^{-47}\,{\rm cm}^2$. Figure taken from Ref.~\cite{Fieguth:2018vob}.} 
	\label{fig:results_example}
\end{figure}

A second handle to discriminate the nuclear responses exploits their different dependence on the momentum transfer, see \autoref{fig:structure_factors_xe132}. The feasibility of this approach has been explored for several WIMP-nucleon interactions~\cite{Fieguth:2018vob,Rogers:2016jrx}. In particular, Ref.~\cite{Fieguth:2018vob} considered realistic detector settings, including projections for a next-generation experiment like the one proposed here, see \autoref{fig:results_example}. As with inelastic scattering, for most responses a discrimination becomes possible with sufficient statistics. However, due to the similarities in the $q$-dependence, a separation of isoscalar and isovector responses will be difficult.

Finally, the nature of the WIMP-nuclear response and in particular its spin-dependent character can be tested by varying the enrichment or depletion on the isotopes with odd $A$ $^{129}$Xe and $^{131}$Xe that is possible with a liquid xenon target. In this case it will be most powerful to combine the results of the proposed experiment with searches using spinless nuclear targets, such as argon, to further test the spin-dependent hypothesis. Likewise, to discriminate between isoscalar and isovector responses the most promising strategy takes advantage of the different proton to neutron ratios in different target nuclei. In this sense, xenon isotopes exhibit the smallest proton to neutron ratios, in contrast to the largest ones which are found e.g.~in fluorine or argon.

\subsection{Scattering at High Momentum Transfer}

Traditional momentum- and velocity-independent dark matter models used to drive experimental developments already starting in the 1990s. Those lead to the well-known low-energy recoil spectra, resembling simple distributions exponentially falling with energy~\cite{Lewin:1995rx}. Consequently, significant experimental effort went into lowering the energy threshold, the calibration for nuclear recoils in this energy regime, and improved understanding of relevant background sources. 

\begin{figure}[!htbp]
\centering
\includegraphics[width=0.47\textwidth,clip,trim=0 0 40 30]{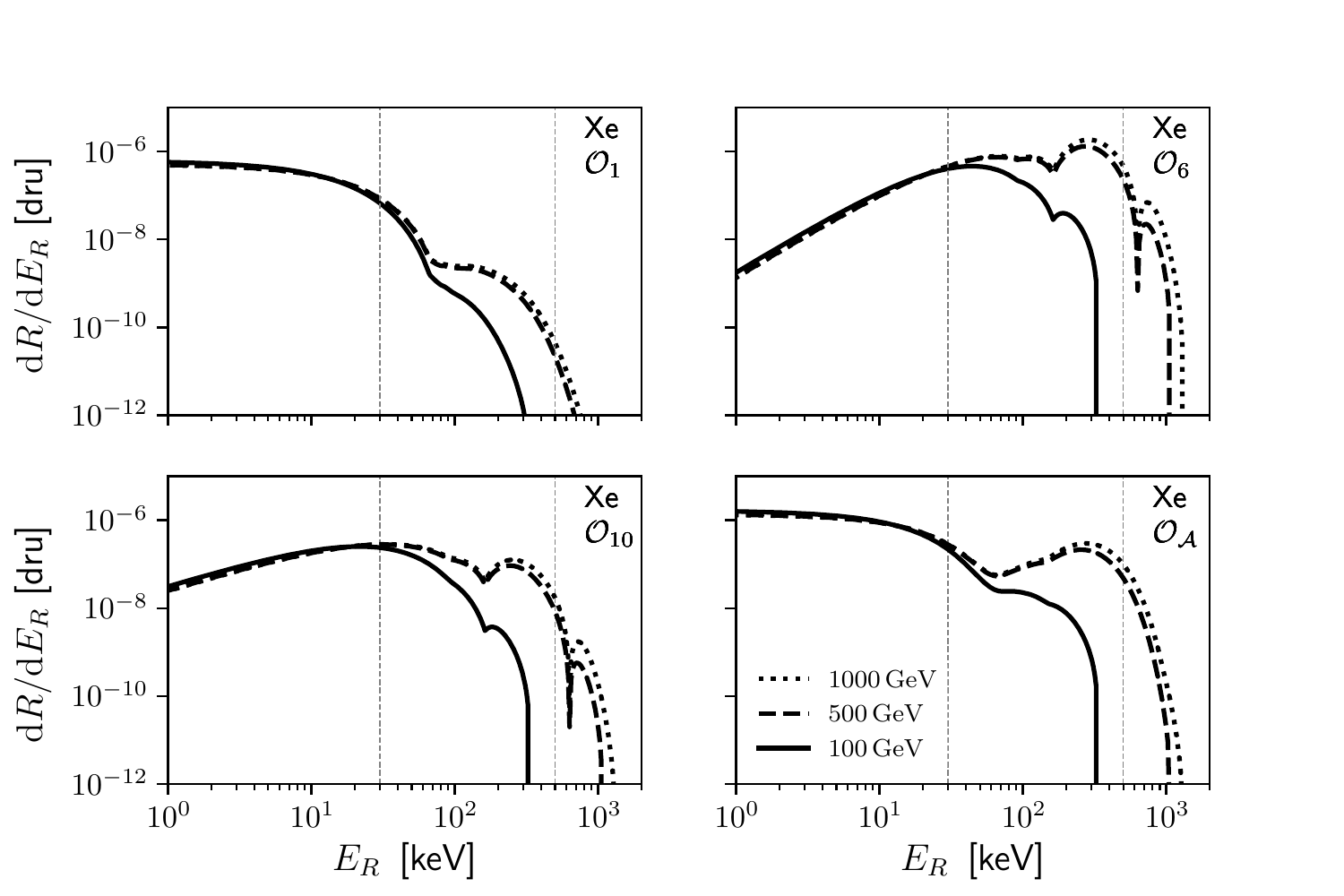}
\caption{The expected recoil spectrum for EFT operators, $O(1)$ (top left panel), $O(6)$ (top right panel), $O(10)$ (bottom left panel), and for anapole interactions (bottom right panel) in a xenon experiment. The dark matter particle mass is chosen to be $m_{\chi}=100$~GeV/$c^2$ (solid), 500~GeV$/c^2$ (dashed), and 1000~GeV$/c^2$ (dotted). The vertical dashed lines represent $E_{\mathrm{max}}=30$~keV and 500~keV. The coupling for each operator has been fixed to produce 100 events in the energy range $[3,\,30]$~keV~\cite{Bozorgnia:2018jep}.}
\label{fig:high_nr_spectra}
\end{figure}

However, many models, such as momentum-dependent effective models or non-trivial mixtures of interactions, result in a more complex nuclear recoil signature with characteristic peaks in the higher nuclear energy regime. This includes many of the models discussed in the following, such as inelastic, composite, exothermic, and magnetic dark matter~\cite{TuckerSmith:2001hy,Graham:2010ca,Dienes:2014via,Hardy:2015boa,Schneck:2015eqa,Bramante:2016rdh} but also the well-known EFT operators for elastic scattering~\cite{Gluscevic:2015sqa, Aprile:2017aas,Gelmini:2018ogy}. These effects manifest themselves often outside the traditionally analyzed energy ranges. The fact that most particles in the Standard Model adhere to such more complex interactions provides strong motivation to explore this important higher-energy parameter space. \autoref{fig:high_nr_spectra} show possible recoil spectra for selected interactions, taken from Ref.~\cite{Bozorgnia:2018jep}. Further motivation to also probe higher recoil energies stems from the presence of Galactic streams that may result in higher recoil energies than from the customarily assumed isothermal halo~\cite{Freese:2003tt,Helmi:2004id,Vogelsberger:2007ny,Kuhlen:2009vh,Lang:2010cd,Purcell:2012sh,Buckley:2019skk}. 

In case of discovery, features in the higher nuclear recoil tails of recoil energy spectra might be used to determine the property of the dark matter-matter interaction. Further, the high-energy nuclear recoil tails of the recoil spectrum are especially sensitive to astrophysical parameters that describe the dark matter velocity distribution, such as the maximum velocity, Galactic escape speed~\cite{McCabe:2010zh,Wu:2019nhd}, or the presence of tidal streams~\cite{OHare:2018trr,OHare:2019qxc,Adhikari:2020gxw}. By employing multiple targets, it is possible to significantly reduce astrophysical uncertainties. For example, the complementarity between argon and xenon-based targets aids to determine the properties of the dark matter particle~\cite{Bertone:2007xj, Pato:2010zk, Newstead:2013pea, Cerdeno:2013gqa, Peter:2013aha, Edwards:2018lsl}.

\subsection{Simplified Models} 

Despite the fact that dark matter-nucleus scattering is characterized by low-energy processes (for which EFTs could provide swift analyses and some general conclusions), further exploration of the internal structures in the interactions between dark matter and Standard Model particles would involve high-energy processes such as those probed at colliders and in the early Universe. At sufficiently high energies, the EFT treatment will break down, as the internal mediators generating the effective dark matter-Standard Model couplings become on-shell.

Simplified models of dark matter can provide a predictable framework to remedy the aforementioned problem, while keeping the number of free parameters manageable, see e.g.~\cite{Hisano:2015bma, DeSimone:2016fbz,Abdallah:2015ter,DiFranzo:2013vra,Abercrombie:2015wmb} and references therein for review and~\cite{Arina:2014yna,Hisano:2018bpz,Balazs:2017hxh, Jacques:2016dqz,Buckley:2014fba,Berlin:2014tja} for some specific studies. In the simplest scenario, only the dark matter mass, mediator mass and a few couplings (depending on the specific models) connecting the dark sector to ours are assumed. This can readily build the interplay among dark matter signals in direct detection, high-energy colliders and astrophysical/cosmological evolution. In light of these complementary approaches, it should be noted that a next-generation xenon experiment is particularly well-suited to probe most of the remaining parameter space in some broad classes of simplified models, e.g., $Z^\prime$ mediated WIMP models~\cite{Blanco:2019hah}. In some realizations of simplified models, the tree-level dark matter-nucleon scattering cross section could exhibit either velocity-suppressed or spin-dependent features to pass the current strong constraints from the existing liquid xenon limits. Examples are a pseudo-scalar or axial-vector current in the interactions of the mediator with the Standard Model quarks and/or the dark sector. It is also worth mentioning that the tree-level interactions between the dark matter and the Standard Model in simplified models can generate loop processes which may still induce detectable signals. These can play important roles in future of direct detection experiments such as the one proposed here~\cite{Drees:1993bu, Hisano:2010ct, Baek:2016lnv, Baek:2017ykw, Arcadi:2017wqi, Li:2018qip,Abe:2018emu, Li:2019fnn,Mohan:2019zrk, Ertas:2019dew, Giacchino:2015hvk, Giacchino:2014moa, Ibarra:2014qma, Colucci:2018vxz, Colucci:2018qml, Chao:2019lhb,LaFontaine:2021cin}.

\subsection{Electroweak Multiplet Dark Matter}

One particularly simple case among WIMP candidates is the dark matter particle as the lightest member of an electroweak multiplet. This is in essence the original WIMP model, sometimes also called the ``minimal dark matter" scenario~\cite{Cirelli:2005uq,Cirelli:2009uv,DiLuzio:2018jwd}. Where "WIMP" refers to particles interacting through the Weak force, this WIMP is the same an as electroweak multiplet, by definition. The interaction between the dark matter and the Standard Model particles are therefore mediated by the Standard Model gauge bosons and the Higgs boson, without the need to introduce additional mediators. Since the interactions are governed by the Standard Model gauge invariance, this is a very predictive scenario and serves as an example of a simple and elegant WIMP dark matter model that is still largely unexplored by experimental searches.

In this model, the fermionic multiplets only have gauge interactions at the renormalizable level. In general, we could consider multiplets $(1, n, Y)$ under the Standard Model gauge group SU(3)$_{\rm C} \times$SU(2)$_{\rm L} \times$U(1)$_{\rm Y}$. The mass scale of the electroweak multiplet is set by a vector-like mass parameter. After electroweak symmetry breaking, the mass spectrum of the multiplet is not exactly degenerate. Minimally, the degeneracy will be lifted by electroweak loop corrections~\cite{Thomas:1998wy,Buckley:2009kv,Cirelli:2005uq,Cirelli:2009uv,Ibe:2012sx,Bottaro:2021snn}. For a large multiplet $n > 7$, the Landau pole will be about one order of magnitude above the mass of electroweak multiplet~\cite{DiLuzio:2015oha}, which makes the model contrived; conversely, new physics below the scale of the Landau pole may lead to an asymptotically safe scenario~\cite{Pelaggi:2017abg}. Ultimately, perturbative unitarity of the annihilation cross section provides a limit of $n < 14$~\cite{Smirnov:2019ngs,Bottaro:2021snn}.

\begin{figure}[!htbp] 
	\centering
	\includegraphics[width=\columnwidth]{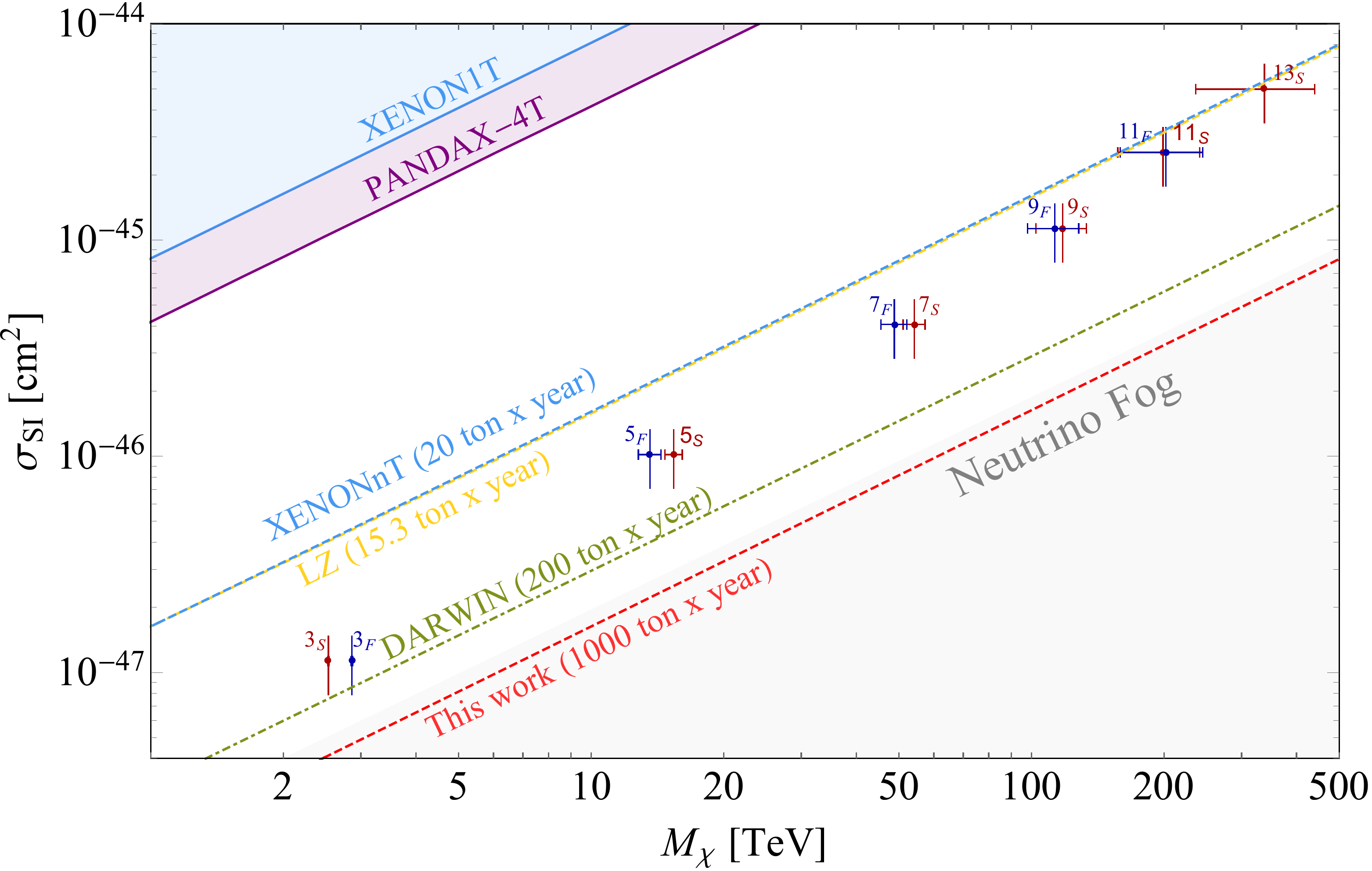}
	\caption{Expected spin-independent scattering cross-section for Majorana multiplets (red) and for real scalar multiplets (blue), assuming the Higgs portal coupling $\lambda H = 0$). Vertical errors correspond to LQCD uncertainties on the elastic cross-section, horizontal errors indicate uncertainties from the determination of the WIMP freeze out mass. The next-generation experiment discussed here will fully probe these classes of highly motivated WIMP dark matter models. Figure adopted from~\cite{Bottaro:2021snn}.} 
	\label{fig:bottaro}
\end{figure}

Sommerfeld enhancement~\cite{Belotsky:2005dk,Hisano:2006nn,Cirelli:2007xd,March-Russell:2008lng,ArkaniHamed:2008qn,Cassel:2009wt} and bound state effects~\cite{March-Russell:2008klu,vonHarling:2014kha,An:2016gad,Mitridate:2017izz,Binder:2021vfo} need to be included in accurate calculations of predictions. Target masses of the electroweak multiplet dark matter are in the range of~1 to 30~TeV~\cite{DiLuzio:2018jwd,DelNobile:2015bqo,Mitridate:2017izz} for $n<7$, but can approach the unitarity bound for larger multiplets, which saturates at $n=13$~\cite{Smirnov:2019ngs,Bottaro:2021snn}. These masses are beyond the reach of the Large Hadron Collider~\cite{Low:2014cba,Cirelli:2014dsa,Han:2018wus,CidVidal:2018eel} and would require one of the proposed future high energy colliders~\cite{Strategy:2019vxc,Han:2020uak,Capdevilla:2021fmj,Bottaro:2021snn}. In contrast, the direct detection of the electroweak multiplet dark matter is through 1-loop processes involving the Standard Model W, Z, and Higgs bosons. The spin-independent cross sections have been computed to be around $10^{-47}\1{cm^2}$ for the Majorana triplet (wino)~\cite{Hisano:2015rsa} and $10^{-48}\1{cm^2}$ for the Dirac doublet (Higgsino)~\cite{Hill:2013hoa}. The other cases are expected to be within the same order~\cite{Smirnov:2019ngs}. As shown in \autoref{fig:bottaro}, this level of spin-independent cross section is well within the reach of the next-generation liquid xenon detector discussed here~\cite{He:2016mls,Thornberry:2021ych,Bottaro:2021snn}. To avoid confusion, note that the LZ line in~\cite{Bottaro:2021snn} corresponds to the sensitivity from the LZ Design Reports~\cite{LZ:2015kxe,Mount:2017qzi} instead of the goals shown Ref.~\cite{Akerib:2018lyp}.

\subsection{Implications for Supersymmetry} 

One classic WIMP dark matter model is the lightest supersymmetric partner (LSP). Supersymmetric models, such as the Minimal Supersymmetric Standard Model (MSSM), with an exact R-parity, predicts that a stable electrically neutral LSP could be a cold dark matter candidate~\cite{Jungman:1995df}. There are three possibilities for a stable neutral LSP: sneutrino, gravitino and neutralino. Among them, the most attractive scenario for direct detection is neutralino dark matter. For a general review on supersymmetry and its low-energy phenomenology, see~\cite{Martin:1997ns}. 

In the MSSM, two neutral higgsinos and two neutral gauginos could mix with each other after electroweak symmetry breaking to form four mass eigenstates called neutralinos. Current direct detection is sensitive to the scattering of WIMPs off nuclei through tree-level Higgs exchange. Thus, existing data has ruled out a significant part of the parameter space of the ``well-tempered" neutralino scenario~\cite{ArkaniHamed:2006mb}, in which the LSP is a mixed neutralino (e.g., mixed bino and higgsino) with the right thermal relic abundance and couplings to the nucleus through the Higgs boson. 

Yet, there are large regions of parameter space unprobed by current experiments. In the MSSM, the reason is that for an LSP that is predominantly a bino, there is a general reduction of the spin-independent direct detection cross section for negative values of the higgsino mass parameter $\mu$. This reduction is induced by a decrease of the coupling of the bino to the Higgs boson~\cite{Cheung:2012qy}, as well as by a destructive interference between the contributions of the standard Higgs with the ones of non-standard Higgs bosons~\cite{Huang:2014xua,Huang:2017kdh}.  The same happens in other minimal supersymmetric extensions, like the NMSSM, but for a singlino dark matter candidate, the reduction occurs for positive values of $\mu$~\cite{Baum:2017enm}. Moreover, there are regions of parameter space, called blind spots, in which the scattering amplitude is drastically reduced~\cite{Cheung:2012qy,Huang:2014xua,Baum:2017enm,Cabrera:2019gaq}. The precise parameter space associated with these blind spots is slightly modified by loop corrections~\cite{Han:2018gej}. Quite generally, for the appropriate signs of $\mu$, the spin-independent scattering cross section can easily be below $10^{-47}\1{cm^2}$~\cite{Baum:2017enm,Carena:2018nlf,Cao:2019qng,Wang:2020xta}. This range of cross sections are out of the reach of current experimental searches but can be probed by next generation direct detection experiments like the one discussed here.

In addition to the well-tempered neutralino at the blind spot, nearly pure wino or higgsino dark matter can scatter off nuclei elastically at one-loop level with a small cross section~\cite{Hisano:2011cs, Hill:2011be}. The pure wino scenario has been strongly constrained by indirect detection of gamma rays from the Galactic center~\cite{Cohen:2013ama,Fan:2013faa} and local spheroidal satellite galaxies~\cite{Ackermann:2013yva,Bhattacherjee:2014dya}, although the former is subject to large uncertainty from the dark matter profile. The spin-independent pure wino-nucleon cross section is around $2\times10^{-47}\1{cm^2}$~\cite{Hisano:2015rsa}, which can be probed by next-generation direct detection experiments. The elastic scattering cross section of the higgsino is found to be below $10^{-48}\1{cm^2}$ with a large theoretical uncertainty~\cite{Hill:2013hoa}. Depending on the mass splitting between neutral higgsinos, the inelastic scattering of higgsino dark matter could be potentially probed with such a future experiment~\cite{Bramante:2016rdh}. 

It is also possible that dark matter could have multiple components such as a combination of very light QCD axions and neutralinos in a supersymmetric theory that solves the strong CP problem~\cite{Baer:2011hx}. In this scenario, direct detection experiments probe the dark matter fraction of the WIMP times its scattering cross section. The next-generation experiment is thus also motivated as pushing its sensitivity to lower cross section enhances the sensitivity to smaller fractions of dark matter in a multi-component scenario~\cite{Zurek:2008qg,Profumo:2009tb,Kajiyama:2013rla,Herrero-Garcia:2017vrl,Herrero-Garcia:2018qnz,Scaffidi:2020wpa}.

\subsection{Inelastic Dark Matter}\label{sec:inelasticdarkmatter}

Inelastic Dark Matter (iDM) was originally proposed~\cite{TuckerSmith:2001hy} to resolve the tension between results published by the DAMA/LIBRA collaboration~\cite{Bernabei:2008yh,Bernabei:2010mq,Bernabei:2018yyw} and other direct and indirect observations~\cite{Ullio:2000bv,TuckerSmith:2004jv,Chang:2008gd}. Multiple particle candidates have since been proposed as inelastic dark matter~\cite{ArkaniHamed:2008qn,Cui:2009xq}, mostly motivated by the measured DAMA/LIBRA spectrum and the constraints for other experiments. Although ultimately this model failed given later exclusions from XENON100~\cite{Aprile:2011ts}, inelastic dark matter has sparked significant theory development and has remained as an interesting and well-studied family of dark matter models. A common feature is a dark matter particle that scatters off Standard Model particles through an excited state of the dark matter particle itself. The mass difference of the excited state $\delta$ imposes a threshold on the energy transfer of the interaction, below which interactions are suppressed. This threshold on the energy transfer E$_{\rm nr}$ limits the population of dark matter that can interact with a given target to those with a minimum velocity $\beta_{\rm min}$ expressed by

\begin{equation}
      \beta_{\rm min} = \sqrt{\frac{1}{2M_N E_{\rm nr}}} \left(\frac{M_N E_{\rm nr}}{\mu} + \delta\right)  
\end{equation}
where $M_N$ is the nucleus mass and $\mu$ is the reduced mass of the dark matter and target particles. Enforcing this constraint alters the spectrum of the expected interaction and can result in peaked recoil spectra~\cite{TuckerSmith:2001hy,Lang:2010cd}, strong dependencies on the particular target material~\cite{Chang:2010pr}, or halo distributions with differing high velocity behavior~\cite{March-Russell:2008rkh}. Note that number-changing interactions that involve multiple dark matter and one standard model particle (Co-SIMPs) lead to similar effects, since rest mass is converted to kinetic energy~\cite{Smirnov:2020zwf}. Calculating this spectrum for a given detector can be done in a model-independent way; software packages have been developed~\cite{Barello:2014uda} to perform these calculations in a consistent manner. Dedicated searches for inelastic dark matter are thus required and have been carried out in XENON100~\cite{Aprile:2017aas}, PandaX-II~\cite{Chen:2017prd}, and LUX~\cite{LUX:2021ksq}.

\subsection{Self-Interacting Dark Matter}\label{sec:selfinteracting}

\begin{figure}[!htbp] 
	\centering
	\includegraphics[width=\columnwidth]{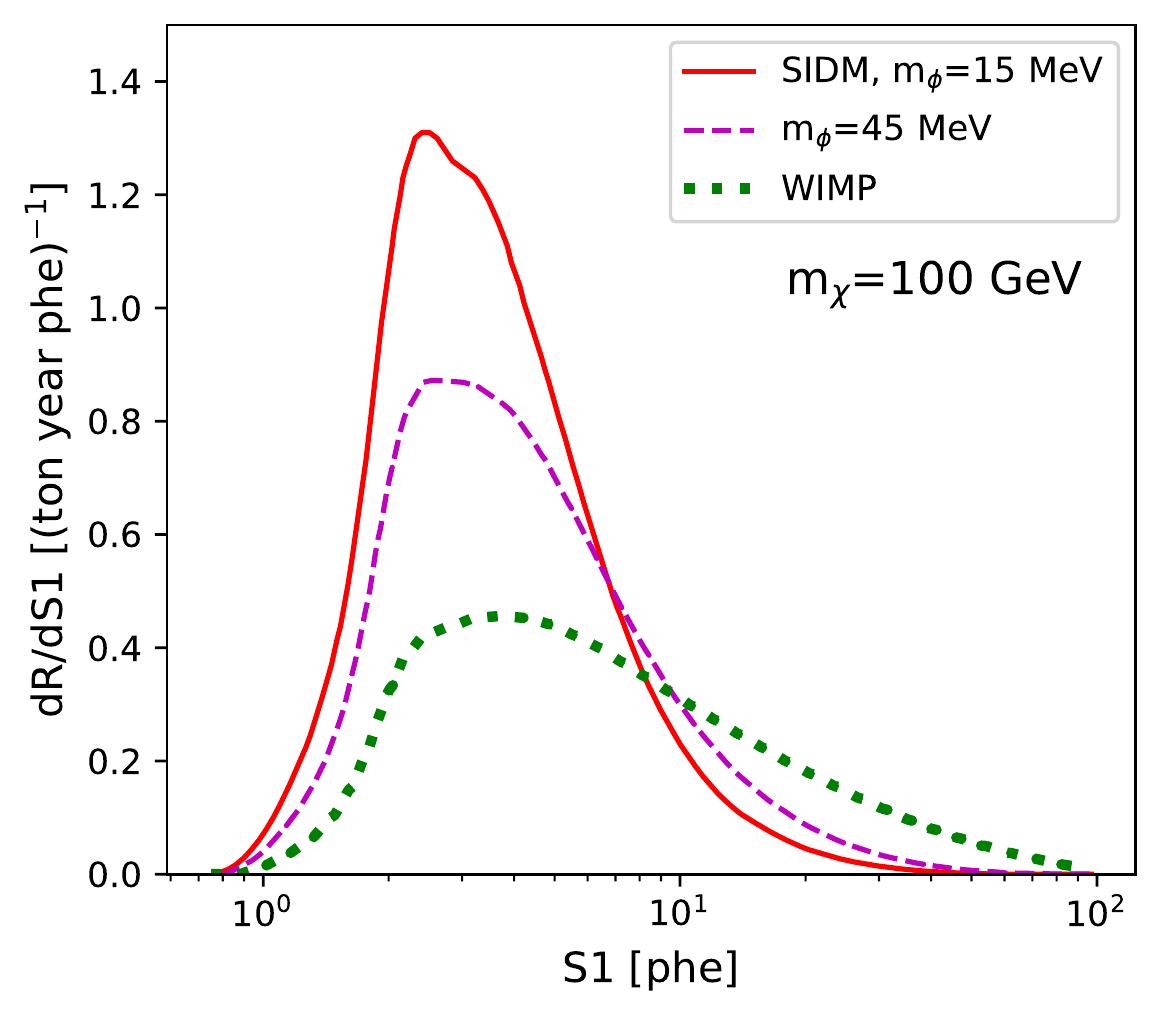}
	\caption{Predicted event rates at a xenon-based experiment for a self-interacting dark matter model with a light mediator (solid red), a model with three times the mediator mass (dashed magenta), and the vanilla WIMP model with contact interaction (dotted green). The spectra are normalized to have the same number of total events within the signal range. See~\cite{DelNobile:2015uua} for details.} 
	\label{fig:s1_sidmspectrum}
\end{figure}

Self-interacting dark matter (SIDM)~\cite{Spergel:1999mh,Kaplinghat:2015aga} is a leading candidate that can resolve both long-standing and more recent tensions between small-scale structure observations and prevailing cold dark matter predictions, see~\cite{Tulin:2017ara} for a review. SIDM phenomenology is also expected if the dark matter is composite (\autoref{sec:compositedm}) or arises from a Mirror World scenario (\autoref{sec:mirrordm}). Dark matter self-interactions, analogous to the nuclear interactions, can thermalize the inner Galactic halo in the presence of the stellar component and tie dark matter and baryon distributions in accord with observations~\cite{Kaplinghat:2013xca,Kamada:2016euw,Creasey:2016jaq,Ren:2018jpt}. In many particle physics realizations of SIDM, there exists a light force carrier that mediates dark matter self-interactions~\cite{Feng:2009hw, Buckley:2009in,Loeb:2010gj,March-Russell:2011ang,Aarssen:2012fx,Tulin:2013teo,GarciaGarcia:2015pnn,Kahlhoefer:2017umn,Chu:2018faw}. When the mediator couples to Standard Model particles, it may generate dark matter signals in direct detection experiments~\cite{Kaplinghat:2013yxa}. For a typical SIDM model, the mediator mass is comparable to or less than the momentum transfer in nuclear recoils. Compared to WIMPs with a contact interaction, the SIDM signal spectrum is then more peaked towards low recoil energies~\cite{DelNobile:2015uua,Kahlhoefer:2017ddj}, see \autoref{fig:s1_sidmspectrum}. Thus the detection of such a spectrum can be an indication of the self-scattering nature of dark matter. Even a null result can put a stringent constraint on the coupling constant between the two sectors. Recently, the PandaX-II collaboration analyzed their data based on an SIDM model with a dark photon mediator and derived an upper bound of $\sim10^{-10}$~\cite{Ren:2018gyx,PandaX-II:2021lap} on the kinematic mixing parameter between the dark and visible photons. Limits from liquid xenon experiments set the strongest constraints also on light SIDM models~\cite{Tsai:2020vpi}. The next-generation liquid xenon experiment discussed in this review will further test SIDM models, and dark matter models with a light mediator in general. 

\subsection{Leptophilic Interactions}

While past efforts in direct dark matter detection have mostly focused on WIMP couplings to nucleons, it is also possible that dark matter would couple preferentially to leptons. Such ``leptophilic'' dark matter candidates have been discussed extensively in the context of the cosmic ray positron excess observed by PAMELA~\cite{Adriani:2013uda} and AMS-02~\cite{Aguilar:2019owu}, as well as the bright $511\1{keV}$ X-ray signal from the Galactic Center~\cite{Knodlseder:2005yq}, and the high-energy cosmic ray electron data from DAMPE~\cite{Ambrosi:2017wek}. Leptophilic dark matter is easily realized in concrete models. This is the case, for instance, if dark matter interactions with the Standard Model particles are mediated by a new gauge boson that couples predominantly to leptons~\cite{Fox:2008kb,Bell:2014tta}. Another example are dark matter interactions mediated by new scalar particles carrying lepton number, such as the sleptons in supersymmetric models~\cite{Chun:2009zx,Bringmann:2012vr,Agrawal:2014ufa,Kopp:2014tsa,Fukushima:2014yia}.

Even if the tree-level interactions of dark matter are leptophilic, couplings to nucleons can be induced at loop level. In that case the WIMP--nucleon scattering is the more promising detection channel, despite loop-suppression, as long as the WIMP is much heavier than the electron~\cite{Kopp:2009et,Schmidt:2012yg,Kopp:2014tsa,Chang:2014tea,Bai:2014osa,Bell:2014tta,Kile:2014jea,Roberts:2016xfw,DEramo:2017zqw}. The reason for this is the more favorable kinematics: the scattering of a heavy WIMP ($\gg \mathrm{MeV}$) on an electron leads to a very small momentum transfer, mostly invisible to a typical direct detection experiment.

However, there are scenarios in which WIMP couplings to nucleons are absent even at the loop level. This can happen for instance if WIMP--lepton interactions are mediated by a new axial vector boson. In this case, the dominant direct detection signal is dark matter scattering on electrons~\cite{Kopp:2009et,Roberts:2015lga,Roberts:2016xfw}, and searches for this process have been carried out by many experiments, including XENON100~\cite{Aprile:2015ade,Aprile:2015ibr,Aprile:2017yea} and LUX~\cite{Akerib:2018zoq}. Scattering on electrons is particularly efficient for sub-GeV dark matter~\cite{Essig:2011nj}, making it the primary detection channel in that mass range. 

Scattering on electrons is also the most efficient channel for dark matter capture in the Sun~\cite{Kopp:2009et,Garani:2017jcj,Liang:2018cjn}. Therefore, if dark matter annihilates into a final state including high-energy neutrinos, searches for these neutrinos from the Sun leads to highly competitive and complementary limits. On the collider side, strong limits on leptophilic dark matter are obtained from LEP data~\cite{Fox:2011fx}. Future lepton collider would lead to further improvements~\cite{Dreiner:2012xm,Freitas:2014jla,Rawat:2017fak}. Progress with these experiments will be complementary to advances from the experiment proposed here.

\subsection{Modulation Searches}

As the Earth revolves around the Sun, a sinusoidal annual modulation should be observable in the dark matter flux hitting direct detection experiments underground~\cite{Drukier:1986tm,Freese:1987wu}, with details depending on the phase space distribution of the halo~\cite{Copi:1999pw,Copi:2000tv}. The DAMA/LIBRA collaboration upholds a long-standing claimed observation of an annually modulating event rate~\cite{Bernabei:2018yyw} with a statistical significance in excess of 9$\sigma$. However, most interpretations of this signal in terms of WIMPs have been ruled out by numerous other experiments. A substantial level of particle model fine-tuning is now required to reconcile the DAMA/LIBRA observation with other null results~\cite{Copi:2002hm,Baum:2018ekm}. Moreover, experiments such as ANAIS~\cite{Amare:2019jul,Amare:2021yyu}, COSINE-100~\cite{Adhikari:2019off,Adhikari:2021szr} and SABRE~\cite{SABRE:2018lfp} attempt to replicate DAMA/LIBRA with an identical sodium iodide target but have not found any evidence of modulation.

A next-generation liquid xenon experiment will be robustly constructed using long-term infrastructure that is made to last multiple years or even decades. Combined with the extremely low background and large target mass, a next-generation experiment may be the ideal experiment to perform an annual modulation search. An annual modulation analysis thus is an integral part of the primary dark matter data analysis, with a sensitivity enhanced by the long data taking time spanning many annual cycles.

A diurnal modulation is guaranteed for most dark matter candidates due to the varying speed of the Earth relative to the dark matter wind as the Earth rotates, though this will be around two orders of magnitude smaller than the annual modulation. However, if dark matter interacts more strongly inside the Earth, then there may be a much larger diurnal modulation effect as the Earth's ``shadow'' eclipses the dark matter wind from the perspective of an experiment~\cite{Collar:1993ss,Hasenbalg:1997hs,Kavanagh:2016pyr,Emken:2017qmp}. Such shadowing effect also provides additional sensitivity to cosmic-ray boosted dark matter (CRDM) with mass lower than around 1~GeV~\cite{PROSPECT:2021awi}.  Many models within the scope of a future xenon experiment will exhibit such a modulation (\autoref{sec:broaderdarkmatter}).

Experimentally, the challenge for detecting diurnal modulations remains to understand sub-1\% variations in detector parameters on a daily basis rather than from weekly or monthly calibrations. In a massive next-generation detector, spatial variation of quantities such as light collection efficiencies may be inherently greater, but there is no reason to assume that temporal variation will be worse than in contemporary detectors. These experiments can teach us how to better control variation, through existing logging of temperature and pressure data as function of time, and excellent handles for temporal systematic uncertainties~\cite{Mount:2017qzi,Aprile:2017aty}, especially when coupled to frequent calibrations using fast-decaying radioisotopes such as $^{83\textrm{m}}$Kr~\cite{Kastens:2009pa, Manalaysay:2009yq}. 

\subsection{Confronting the Neutrino Fog}\label{sec:nufloor}

As the size and sensitivity of direct detection experiments improves, the detectable signal of dark matter will become so small that it will reach a level similar to the strength of the coherent elastic neutrino-nucleus scattering signal of astrophysical neutrinos~\cite{Monroe:2007xp,Strigari:2009bq,Billard:2013qya}. While there is a substantial science case for the detection of astrophysical neutrinos in their own right (\autoref{sec:neutrinos}), for dark matter searches they are a critical background.

When searching for a signal that is mimicked by a background, discovery is only possible when an excess in events is larger than the expected statistical fluctuations and systematic uncertainties of that background. For the neutrino background, the systematic uncertainties on the flux normalizations dominate, which range from 1\%--50\%. Many of the particle models discussed here will eventually be limited in some way by the neutrino background, in both the electronic~\cite{Wyenberg:2018eyv,Essig:2018tss} and nuclear recoil channels~\cite{Billard:2013qya,Baudis:2013qla,Dent:2016iht,Dent:2016wor,Dent:2019krz}. This background is often referred to as the ``neutrino floor’’, or more accurately, the ``neutrino fog'', as it represents a gradual worsening of sensitivity and a dependence on the systematics of the neutrino flux. Various definitions of this neutrino fog have been put forward~\cite{OHare:2021utq}. Just like any generic limit on dark matter, the shape of a neutrino fog is dependent on nuclear~\cite{Papoulias:2018uzy}, astrophysical~\cite{OHare:2016pjy} and particle model~\cite{Dent:2016iht,Dent:2016wor,Gelmini:2018ogy} inputs for the dark matter signal. Given non-standard neutrino-nucleus interactions, these could be further modified~\cite{AristizabalSierra:2017joc,Gonzalez-Garcia:2018dep} and even raised by several orders of magnitude~\cite{Boehm:2018sux}. 

Unlike many other backgrounds, neutrinos cannot be shielded, so they must be dealt with statistically, or by searching for some discriminating features. Techniques that have been discussed in the past include exploiting the differing annual modulation signatures~\cite{Davis:2014ama}, or the complementarity between different target nuclei~\cite{Ruppin:2014bra}. However, only direction-dependence provides enough of a discriminant to fully subtract the background~\cite{Grothaus:2014hja,O'Hare:2015mda,Mayet:2016zxu,Franarin:2016ppr,OHare:2017rag}, but measuring this in any large-scale experiment is extremely challenging. 

\begin{figure}[!htbp]
\begin{center}
\includegraphics[width=0.98\columnwidth]{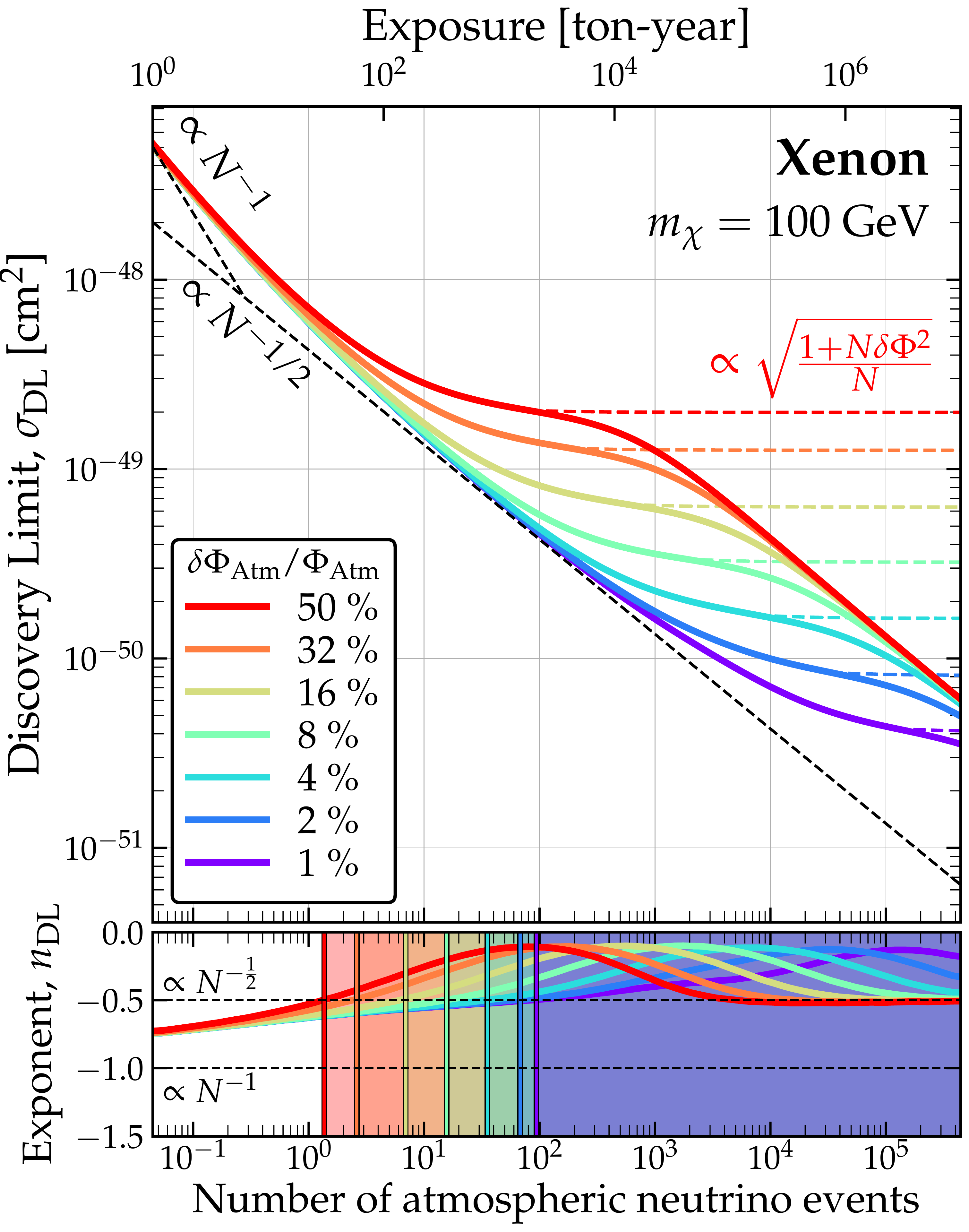}
\caption{Spin-independent discovery limits at $m_\chi = 100$~GeV as a function of the expected number of atmospheric CE$\nu$NS events $N$, and the fractional uncertainty on the atmospheric neutrino flux, $\delta \Phi_{\rm Atm}/\Phi_{\rm Atm}$ from Ref.~\cite{OHare:2016pjy}. Three scaling regimes as a function of $N$ are shown with dashed lines: 1) ``background-free'' $\sigma \sim N^{-1}$, 2) Poissonian $\sigma \sim N^{-1/2}$, and 3) Saturation $\sigma \sim \sqrt{(1+\delta\Phi^2 N)/N}$. The bottom panels in each case show the logarithmic scaling exponent defined as: $n_{\rm DL} \equiv \textrm{d} \ln\sigma_{\rm DL}/\textrm{d} \ln N$. This figure shows the importance of the neutrino flux systematic uncertainty in extending the dark matter physics reach below the neutrino fog.}
\label{fig:fig_nufloor_100GeV}
\end{center} 
\end{figure}

Fortunately for the next-generation xenon experiment, extending the dark matter physics reach below the neutrino fog will be facilitated by complementary measurements made by neutrino experiments. Taking the example of standard WIMP-nucleon cross sections, the most important backgrounds will be $^8$B solar neutrinos for WIMP masses below $\sim$10\,GeV/$c^2$, and atmospheric neutrinos above that. The $^8$B flux is measured at the 2\% level from Solar neutrino data~\cite{Bergstrom:2016cbh}. The atmospheric flux on the other hand, is difficult to measure and to theoretically predict at the relevant sub-100~MeV energies, so it still has a $\sim$20\% uncertainty (\autoref{sec:atmnu} and~\cite{Newstead:2020fie}). Any reduction in these uncertainties will, in effect, ``lower’’ the neutrino fog. Indeed, gradual improvements in neutrino flux measurements are expected independent of the experiment under discussion here. For example, experiments like SNO+~\cite{Caden:2017htb}, JUNO~\cite{Jinping:2016iiq} and DUNE~\cite{Abi:2018dnh,Kelly:2019itm} will be either operating or under construction over a similar timescale to the next-generation xenon experiment.

In \autoref{fig:fig_nufloor_100GeV} we show how the minimum discoverable spin-independent cross section for a $100\1{GeV}$ WIMP evolves with increasing exposure in a xenon experiment. The brief plateau in the discovery limit is the impact of the atmospheric neutrino background. However, in the limit of high statistics, the number of observed background events will eventually be large enough to account for the finite uncertainty. At this point, the discovery limit breaks past the neutrino fog and smaller cross sections can be accessed. Comparing the different lines, we see clearly the importance of the systematic uncertainty. A future improvement down to $\sim$4\% would be enough to extend the reach of a 1000~tonne-year xenon experiment almost an order of magnitude into the neutrino fog at high masses~\cite{OHare:2020lva}. This is where much of the remaining supersymmetric WIMP candidates~\cite{Roszkowski:2014iqa,Athron:2017qdc,Hisano:2011cs,Kobakhidze:2018vuy}, as well as many alternative WIMP models~\cite{Arcadi:2017wqi,Baker:2019ndr,Arina:2019tib} lie.

\section{Broadening the Dark Matter Reach}\label{sec:broaderdarkmatter}

Liquid xenon experiments have already demonstrated that they are versatile detectors with significant sensitivity to a variety of non-WIMP dark matter models. Traditionally, WIMPs are searched-for using analyses that exploit the electronic/nuclear recoil discrimination capability of liquid xenon and achieve the lowest nuclear recoil background of any dark matter direct detection technology. To broaden this reach, a number of different analyses and technologies are available as presented in this section. This in turn enables liquid xenon experiments to achieve competitive sensitivity to a number of dark matter models that are also described here. In particular, subsections~\ref{sec:dpe}--\ref{sec:hydrodoping} describe dedicated analyses and technologies to lower the energy threshold of liquid xenon TPCs. Subsections~\ref{sec:upscattdm}--\ref{sec:asymmdm} describe models that especially profit from such lower thresholds, and subsections~\ref{sec:bosonicwimp}--\ref{sec:lumidm} models where the signal can be in the electronic recoil band. Subsections~\ref{sec:maginelasticdm} and~\ref{sec:planck} describe two models that require dedicated analyses to increase the reach of liquid xenon TPCs to complex interactions and up to Planck mass dark matter, respectively.

With the WIMP model being probed extensively by experiment, the community is in parallel starting to work on detector concepts that can probe dark matter over a much wider mass range~\cite{Battaglieri:2017aum}, in particular covering thermal relic particles in the $\n{MeV/c^2}-\n{GeV/c^2}$ mass range~\cite{Essig:2011nj,Essig:2013lka,Hochberg:2014dra,Kuflik:2015isi,Alexander:2016aln,Battaglieri:2017aum}. Searches in this lower mass range were pioneered with liquid xenon detectors~\cite{Essig:2012yx}. While many experiments optimized for very low-energy recoils now exist~\cite{Agnese:2015nto,Angloher:2015ewa,Petricca:2017zdp,Angloher:2017sxg,NEWS-G:2017pxg}, liquid argon~\cite{Agnes:2018ves,Agnes:2018oej} and xenon~\cite{Essig:2017kqs,Akerib:2018hck,Aprile:2019xxb} TPCs still remain the leading technologies even for sub-GeV masses. There is thus significant interest in achieving the lowest-possible energy threshold in a next-generation liquid xenon detector.

\begin{figure}[!htbp]  
\begin{center}
\includegraphics[width=0.99\columnwidth]{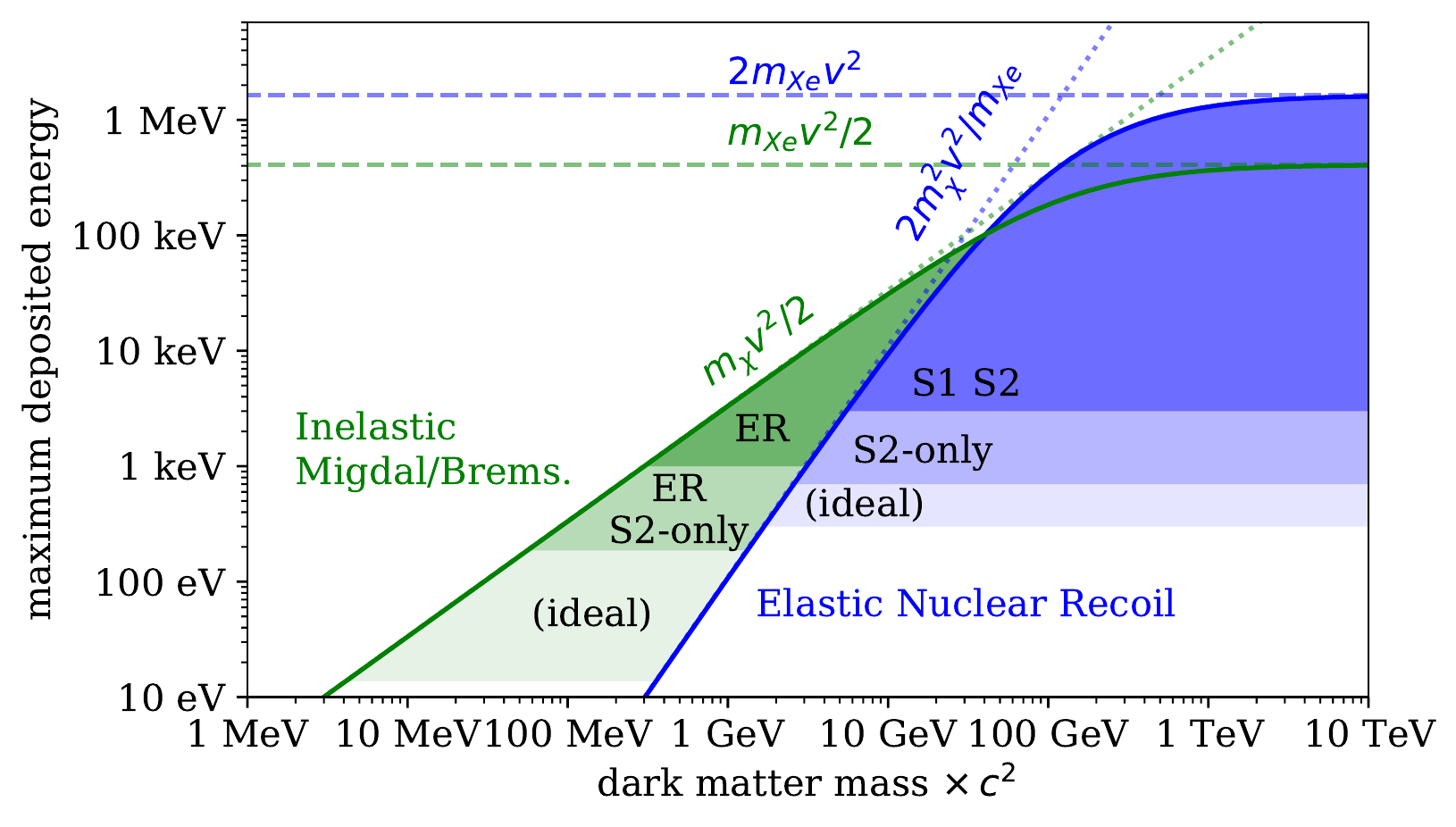}
\caption{Maximum recoil energy transferred in elastic dark matter interactions to a xenon nucleus (blue) or in inelastic dark matter interactions to an electron (green). Currently-achieved energy thresholds are indicated for both the traditional S1$+$S2 analysis~\cite{Aprile:2018dbl,Aprile:2020tmw} as well as a S2-only analysis~\cite{Aprile:2019xxb}. The ultimate thresholds for an ideal detector are also shown ($13.7\1{eV}$ for inelastic scatters~\cite{Baudis:2021dsq} and $0.3\1{keV}$ for elastic nuclear recoils~\cite{Lenardo:2019fcn}).}
\label{fig:kinematics}
\end{center} 
\end{figure}

\autoref{fig:kinematics} visualizes the relevant dark matter scattering kinematics. For a maximum-velocity dark matter particle ($v=v_{\mathrm{esc}}+v_{\mathrm{Earth}}$) and a head-on dark matter-nucleus collision, it shows the maximum recoil energy for either elastic scatters resulting in a nuclear xenon recoil, or inelastic scatters resulting in electronic recoils. For a given energy threshold, this translate into a minimum mass for the dark matter particle to be able to leave a signal in the xenon target. As can be seen, lowering the threshold increases the dark matter mass range that the detector is sensitive to. Further, inelastic scatters as discussed below can be used to probe drastically lighter dark matter candidates (see e.g. Ref.~\cite{Kouvaris:2016afs}).  

\subsection{Double Photoelectron Emission}\label{sec:dpe}

In the traditional analysis where both primary scintillation (S1) and ionization (S2) signals are read out, the energy threshold of two-phase liquid xenon TPCs is set by the smallest scintillation signal that can be confidently discriminated from background sources. Typically, dark matter experiments require an $n$-fold coincidence of PMTs within a short time window for a pulse to be classified as an S1. The optimal value of $n$ (typically in the range 2--4) is a compromise between signal efficiency and the rejection of fake S1 pulses, caused by random coincidences of PMT dark counts~\cite{Aprile:2020thb}.

This methodology makes no attempt to otherwise discriminate dark count background pulses from actual photon-induced pulses. However, it is known that, for some PMT photocathodes, the energy of the liquid xenon scintillation photons (175\,nm or 7\,eV~\cite{Fuji:2015}) is enough to produce two photoelectrons on the PMT photocathode a fraction of the time, resulting in pulses that are on average twice as large as a single photoelectron pulse.

\begin{figure}[!htbp]  
\begin{center}
\includegraphics[width=0.99\columnwidth]{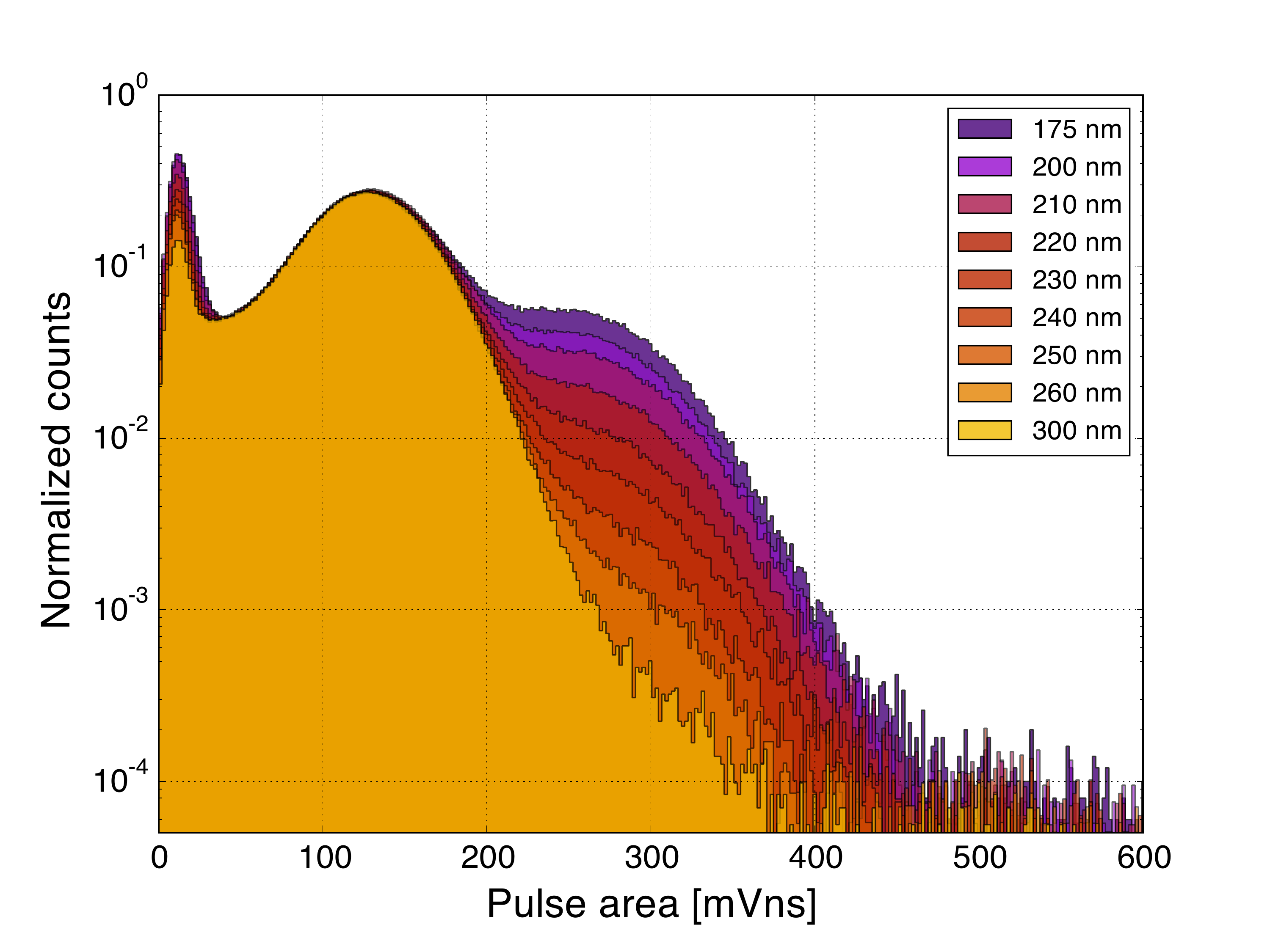}
\caption{Superposition of single photon pulse area spectra of a R11410 PMT for different wavelengths. Each spectrum is normalized by the integral in the region between 50--120~mV~ns in order to show the effect more clearly. Figure from Ref.~\cite{Faham:2015kqa}.}
\label{fig:R11410_rainbow_plot}
\end{center} 
\end{figure}

This so-called Double Photoelectron Emission (DPE) effect can therefore be exploited to increase the signal efficiency beyond the standard $n$-fold optimisation, provided that the DPE fraction and efficiency gain can be properly calibrated. This requires the precise determination of the PMT DPE probability, which depends strongly on the wavelength of the impinging light, as well as on the composition and thickness of the photocathode. For the widely-used Hamamatsu R11410 PMT model, a wavelength scan was performed with single photons down to the VUV range on one unit~\cite{Faham:2015kqa} (see \autoref{fig:R11410_rainbow_plot}). The inter-PMT variability due to the photocathode manufacturing process has also been measured at low temperature with a batch of 35 R11410-22 PMTs~\cite{Paredes:2018hxp}. Measuring the DPE probability is also crucial for pulse area calibration. A pulse area in `photoelectrons' does not represent the number of photon hits detected, but can be understood and calibrated if the DPE probability is known. Experiments have reported average values of their PMT DPE probability of around $\sim$10--20\% given liquid xenon scintillation light~\cite{Araujo:2020rwg, Akerib:2019zrt}.

\begin{figure}[!htbp]  
\begin{center}
\includegraphics[width=0.99\columnwidth]{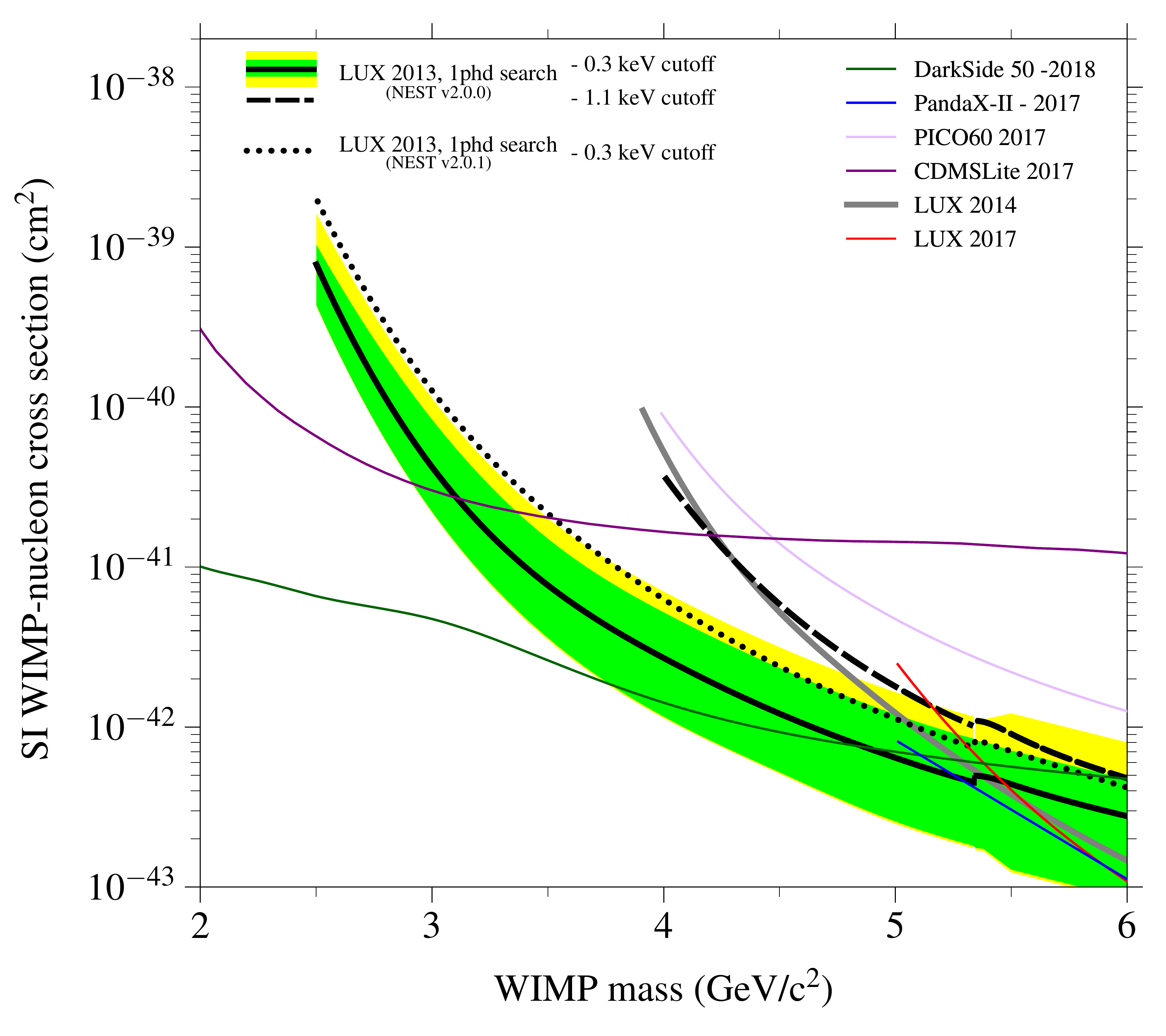}
\caption{90\%~CL upper limits on the spin-independent WIMP-nucleon cross section obtained using the single-photon population producing Double Photoelectron Emission in the LUX 2013 WIMP search. The observed limit with a 0.3~keV NR energy cut-off is shown in solid black, with 1$\sigma$ and 2$\sigma$ sensitivity bands shown in green and yellow. The dashed black line is derived from the same analysis but with a model cut-off at 1.1~keV. Both of these results correspond to the NEST v.2.0.0 model. The upper limit using a 0.3~keV NR energy cut-off with the newer NEST v.2.0.1 model is shown using a dotted black line. Also shown are other results current at the time, namely from the LUX 2013 search~\cite{Akerib:2015rjg} (gray), the LUX complete exposure~\cite{Akerib:2016vxi} (red), DarkSide-50~\cite{Agnes:2018ves} (green), PandaX-II~\cite{Cui:2017nnn} (blue), PICO60~\cite{Amole:2017dex} (lilac) and CDMSLite~\cite{Agnese:2015nto} (purple). Figure from Ref.~\cite{Akerib:2019zrt}.}
\label{fig:lux_DPE_limits}
\end{center} 
\end{figure}

The LUX experiment exploited this DPE to lower the coincidence condition from two PMTs to just a single PMT with an S1 pulse consistent with DPE~\cite{Akerib:2019zrt} (\autoref{fig:lux_DPE_limits}). In general, an experiment may lower its $n$-fold condition by requiring that a subset of PMT hits are consistent with DPE. A PMT with a low dark count rate and high DPE probability might enhance the low-energy reach of a next-generation dark matter experiment with a straightforward extension of the analysis~\cite{Akerib:2021pfd}.

\subsection{Charge-Only Analysis}\label{sec:s2only}

Interactions from WIMP candidates below $\sim\n{GeV/c^2}$ would produce scintillation (S1) signals close to or below the typical low-energy threshold of liquid xenon TPCs. This loss of efficiency can be bypassed by removing the requirement that the S1 signal be detected at all, and leveraging the inherent gain in the S2 signal~\cite{Aprile:2016wwo, Agnes:2018ves, Aprile:2019xxb, Akerib:2021pfd, XENON:2021myl, PandaX-II:2021nsg}. Relaxing the requirement of an observed S1 allows events which resulted in even a single extracted electron to be analyzed. This increased sensitivity to low-mass dark matter candidates comes at the expense of recoiling particle discrimination (usually from the S2/S1 ratio) and accurate determination of the z-coordinate (usually from the delay between the S1 and S2 signals). While sometimes these analyses still make use of S1 pulses to reject background events, when they do not require an S1 to be present, they are commonly referred to as `charge-only' or `S2-only' analyses.

\begin{figure}[!htbp] 
\begin{center}
\includegraphics[width=0.99\columnwidth]{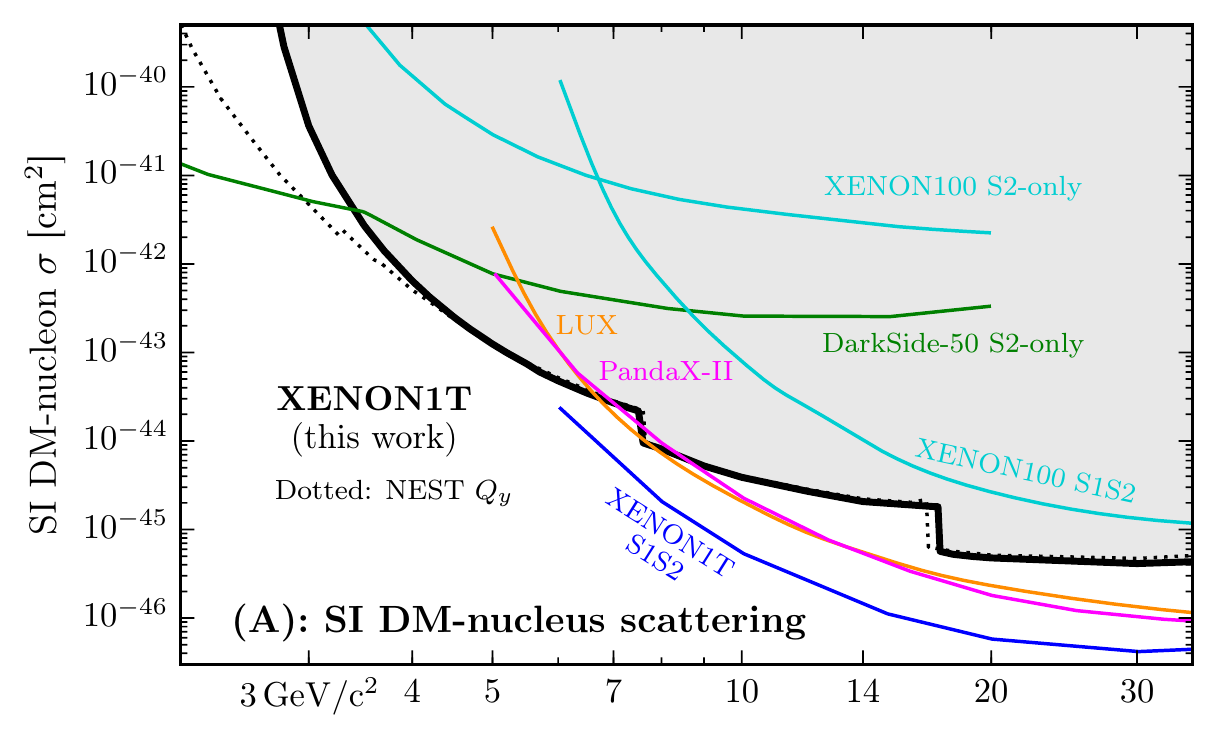}
\caption{Shown are 90\% confidence level upper limits (black lines with gray shading above) on spin-independent dark matter-nucleus scattering with the dark matter mass, $m_\chi$, on the horizontal axis. The thick black line is the result from the XENON1T charge-only analysis. Other results are shown from XENON1T in blue~\cite{Aprile:2018dbl}, LUX in orange~\cite{Akerib:2016vxi}, PandaX-II in magenta~\cite{Ren:2018gyx}, DarkSide-50 in green~\cite{Agnes:2018ves}, XENON100 in turquoise~\cite{Aprile:2016wwo, Aprile:2016swn}. Dotted lines show the XENON1T limit when assuming the $Q_y$ from NEST v2.0.1~\cite{szydagis_m_2019_3357973} cut off below 0.3~keV. Figure taken from Ref.~\cite{Aprile:2019xxb}.}
\label{fig:sub_GeV}
\end{center} 
\end{figure}

Thus far, charge-only analyses have been background-limited due to large single- and few-electron backgrounds, which have yet to be reliably quantified and mitigated~\cite{Aprile:2013blg,Akimov:2016rbs,Sorensen:2017kpl,Sorensen:2017ymt,Akerib:2020jud,Kopec:2021ccm,Bodnia:2021flk,XENON:2021myl, PandaX-II:2021nsg}. The extended drift volume of a next-generation detector may be subject to stronger electron lifetime effects, but will also provide improved identification of S2s originating from the bottom of the detector because of increased electron diffusion (resulting in wider S2 pulses). Additionally, xenon contamination from out-gassing or surface detachment of impurities will benefit from the relative scaling of volume and surface area. Despite being background limited, charge-only analyses have been used to set leading limits on dark matter interaction rates, see \autoref{fig:sub_GeV}. The sensitivity of liquid xenon TPCs to signals at the level of single electrons results in leading sensitivity to sub-GeV WIMPs as well as other particle models. Specifically, charge-only analyses are especially good for detecting electronic recoil signals, as their S1 is much smaller than for a nuclear recoil of the same S2 size. A charge-only analysis in a next-generation detector will further improve this sensitivity over the current generation of xenon TPCs (see also \autoref{fig:kinematics}). 

\subsection{General Dark Matter-Induced Atomic Responses}\label{sec:genatomresp}

A dark matter particle with mass in the MeV--GeV range can deposit enough energy in the collision with an electron in a xenon atom to ionise the target and produce a detectable S2 signal in a TPC detector~\cite{Kopp:2009et,Essig:2011nj}. This charge-only analysis has mainly been performed with models where the interaction between dark matter and electrons is mediated by a new hypothetical force carrier such as the dark photon~\cite{Agnes:2018oej,Aprile:2019xxb, XENON:2021myl}. To avoid confusion, here the dark photon acts as force carrier, as opposed to the analysis described in \autoref{sec:dark_photon}, where the dark photon itself is the dark matter candidate. In this framework, the total ionisation rate for a given xenon orbital can be expressed in terms of a single target-dependent ionisation form factor, which is a function of the initial and final state electron wave functions~\cite{Kopp:2009et,Essig:2011nj}.

Xenon detectors can also probe more complex models, such as those where the amplitude for dark matter scattering by a free electron, $\mathcal{M}$, depends on the initial electron momentum~\cite{Catena:2019gfa}. These include models where dark matter couples to electrons via magnetic dipole or anapole interactions. By expanding $\mathcal{M}$ using effective theory methods similar to the ones previously discussed in the context of dark matter-nucleon interactions (see \autoref{sec:nreft}), reference~\cite{Catena:2019gfa} found that the most general form for the total ionisation rate of a given xenon orbital is a linear combination of four target-dependent atomic responses, which are defined in terms of initial and final state electron wave function overlap integrals. Assuming that dark matter is made of fermions with mass in the MeV-GeV range and interactions dominated by electromagnetic moments of higher order, such as the electric and magnetic dipoles or the anapole moment, reference~\cite{Catena:2020tbv} showed that liquid xenon TPCs can shed light on whether dark matter is a Dirac or Majorana particle. By using Monte Carlo simulations and a non-trivial extension of the likelihood ratio test to the case where one of the hypotheses lies on the boundary of the parameter space, only about $45-610$ signal events are required to reject Majorana dark matter in favour of Dirac dark matter at 3~sigma confidence level. 

\subsection{Migdal Effect and Bremsstrahlung}\label{sec:migdal}

\begin{figure}[!htbp] 
\begin{center}
\includegraphics[width=0.99\columnwidth]{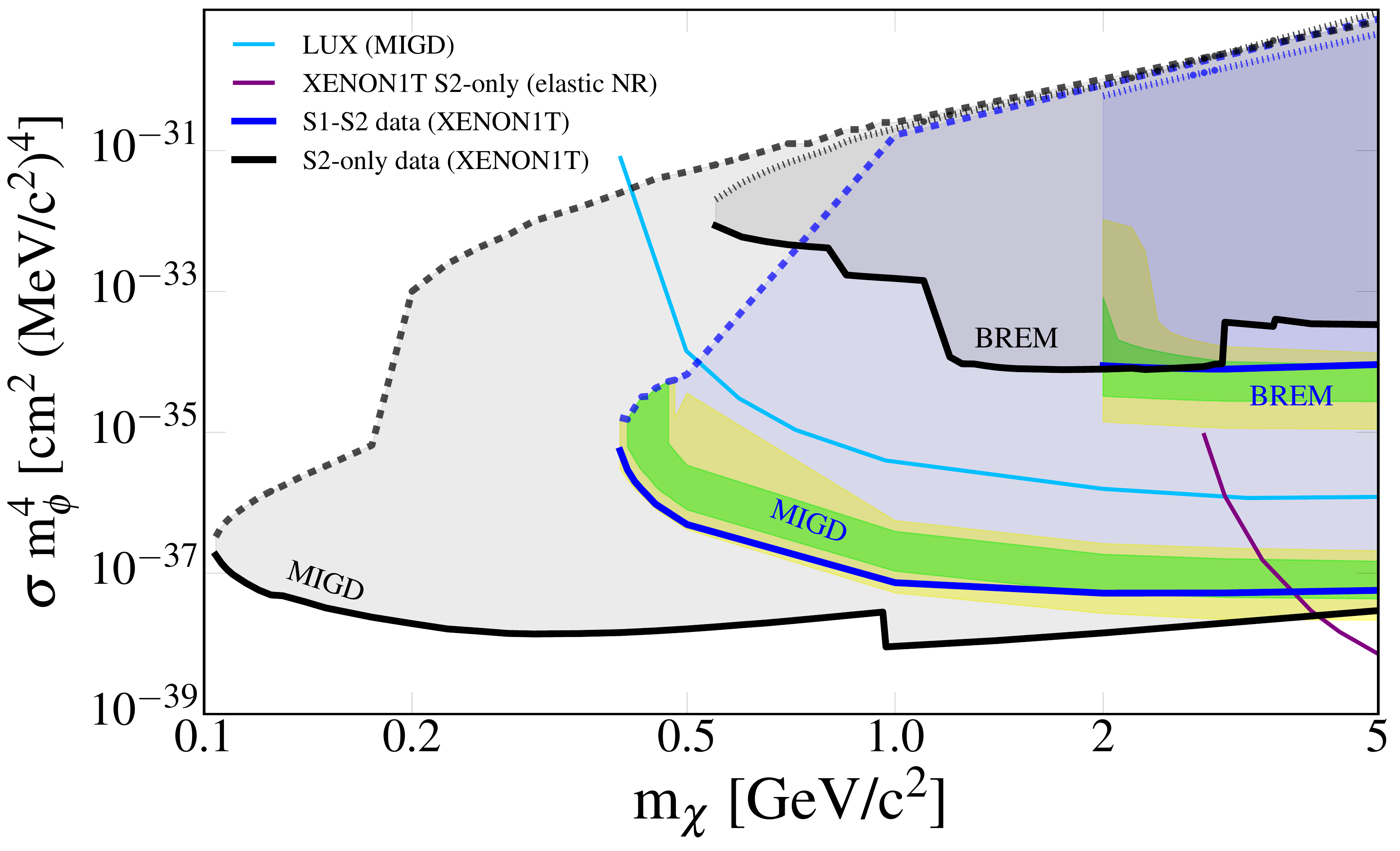}
\caption{Limits on the spin-independent light mediator dark matter-nucleon interaction cross section at 90\% confidence level using signal models from the Migdal effect and Bremsstrahlung in the XENON1T experiment with the S1-S2 data (blue contours and lines) and charge-only data (black contours and lines). The solid and dashed (dotted) lines represent the lower boundaries (also referred to as upper limits) and Midgal (Bremsstrahlung) upper boundaries of the excluded parameter regions. Green and yellow shaded regions give the 1 and 2$\sigma$ sensitivity contours for upper limits derived using the S1-S2 data, respectively. The upper limits on the spin-independent dark matter-nucleon interaction cross sections from LUX~\cite{Akerib:2018hck} and XENON1T charge-only (elastic nuclear recoil results)~\cite{Aprile:2019xxb} are also shown. Figure taken from Ref.~\cite{Aprile:2019jmx}.}
\label{fig:migdal}
\end{center} 
\end{figure}

\begin{figure}[!htbp]
    \includegraphics[width=0.99\columnwidth]{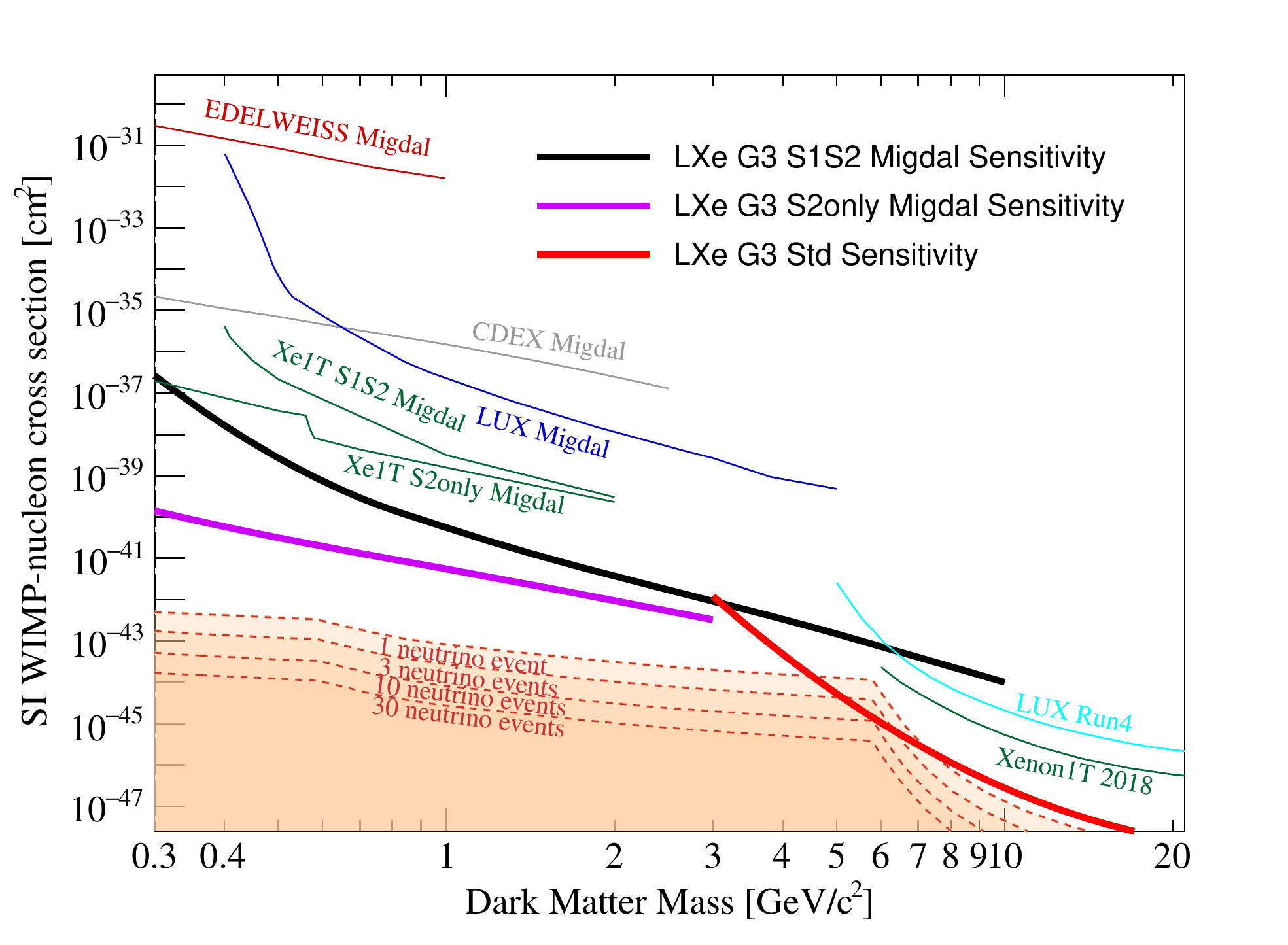}
    \caption{Spin-independent sensitivity for the electronic recoil-inducing Migdal effect for the case of a heavy scalar mediator. The S1-S2 sensitivity (black, solid) and the charge-only sensitivity (violet, solid) are shown. The charge-only analysis improves the sensitivity by more than two orders of magnitude with respect to the standard S1-S2 analysis (red, solid). Experimental limits from similar analyses in LUX (blue, solid)~\cite{Akerib:2018hck}, XENON1T (green, solid)~\cite{Aprile:2019xxb} and CDEX (gray, solid)~\cite{Liu:2019kzq} are also shown.\label{fig:migdalProjection}}
\end{figure}

The progressive loss of the scintillation (S1) response with decreasing nuclear recoil energy impedes the ability of liquid xenon TPCs to reach sensitivity for sub-GeV dark matter masses. However, the standard dark matter-nucleus interaction can also induce an inelastic atomic scattering signal. 
The Migdal effect~\cite{Migdal:1941} predicts the ionization of the atom with some (small) probability due to the sudden nuclear acceleration caused by the dark matter collision, resulting in excitation and ionization processes from the electrons~\cite{Ibe:2017yqa}. Since electronic recoils produce a more detectable signal than nuclear recoils, this channel enables liquid xenon detectors to reach dark matter masses of order $\sim100\1{MeV/c^2}$~\cite{Dolan:2017xbu,Akerib:2018hck,Aprile:2019jmx,Wang:2021oha}, see \autoref{fig:migdal}. The sensitivity of liquid xenon detectors to sub-GeV dark matter achieved using the Migdal effect is competitive with other detectors that are dedicated to searches of light dark matter~\cite{NEWS-G:2017pxg,CRESST:2019jnq,SuperCDMS:2020ymb}. \autoref{fig:migdalProjection} shows a conservative projected Migdal sensitivity for a next-generation detector assuming LZ detector parameters with an extended exposure of 300~tonne-years, equivalent to e.g.~a 56~tonne fiducial mass and 5.4~live-years. However, the Migdal effect has not yet been observed directly in dark matter targets. A dedicated calibration could be performed using a low-energy neutron beam~\cite{Bell:2021ihi}. This could provide a direct test of the theoretical predictions of the Migdal effect.

Similar to the Migdal effect, nuclear Bremsstrahlung searches leverage the fact that in liquid xenon at low energies, electronic recoils produce a stronger signal than nuclear recoils~\cite{Kouvaris:2016afs}. Bremsstrahlung searches consider the emission of a photon from the recoiling atomic nucleus. In the atomic picture this can be viewed as the dipole emission of a photon from a xenon atom that has been polarized in the dark matter-nucleus scattering. In xenon, the emission of the Bremsstrahlung photon is more heavily suppressed compared to the Migdal effect and hence results in a weaker signal for all interaction types~\cite{Bell:2019egg}. The theoretical motivation and event rates for Bremsstrahlung have been derived in~\cite{Kouvaris:2016afs} and searches using liquid xenon detectors have been published in~\cite{Akerib:2018hck,Kobayashi:2018jky,McCabe:2017rln}.

\subsection{Hydrogen Doping}\label{sec:hydrodoping}

Kinematically, the large xenon nucleus (average mass 122~GeV/$c^2$) is not well suited for an efficient transfer of energy from Galactic dark matter with mass $\lesssim$1~GeV/$c^2$. As a result, nearly all of the resulting xenon nuclear recoils fall below the energy threshold for detection. A possible solution for enhancing the sub-GeV sensitivity of liquid xenon TPCs is to dissolve a lighter species in the liquid xenon bulk~\cite{Akerib:2021hydrox}. In this configuration, the lighter nucleus becomes the dark matter target, and the xenon becomes the sensing medium.

This strategy exploits two of the primary advantages of the liquid xenon medium. First, the high atomic number and density of liquid xenon provides excellent self-shielding of external backgrounds from the central volume of the detector. Such a suppression would not be possible in a similarly-sized detector comprised of the light species alone. Second, the high yield of detectable quanta (electrons and photons) resulting from low-energy particle interactions makes xenon an ideal sensor for the recoiling light nuclei.

Having the lightest nucleus of any element, hydrogen is kinematically the best candidate species for detecting interactions from sub-GeV dark matter and astrophysical neutrinos~\cite{Beacom:2002hs}. The lone proton comprising hydrogen's nucleus additionally provides unique sensitivity to the spin-dependent dark matter coupling to protons. Likewise, doping the xenon target with deuterium would provide similar sensitivity to the neutron-only couplings. There are still significant open questions concerning the actual feasibility of adding H$_2$ to a liquid xenon TPC. Drifting electrons in the detector's gas space will be cooled down by the hydrogen and therefore the electric field strength needed to extract quasi-free electrons out of the liquid and into the gas space will be increased~\cite{Bolozdynya:1991}. Furthermore, the light yield of xenon electroluminescence will be suppressed. Molecular species within the liquid space are also known to quench the S1 light production. S1 as well as ionization signals for H$_2$ mole fractions up to $5.7\,$\% have been observed in a 26~atm gaseous xenon TPC from $^{241}$Am $5.5\,$MeV alpha particles~\cite{Tezuka:2004gza}, although there is a reported loss of about half of the S1 and electron signals for an H$_2$ mole fraction of $1.1\,$\%. To broaden the understanding of the exact properties of such an H$_2$-doped liquid xenon TPC, further measurements are in progress. Helium is also a viable option as the light mass target species, as it would not have the signal quenching properties of H$_2$, but its spin-dependent sensitivity would be comparatively poor. Introduction of helium into the detector might not be suitable if PMTs are used as a photosensor due to its ability to diffuse into and degrade the PMT vacuum. However, helium could be considered as a dopant if silicon photomultipliers (SiPMs) are used to detect light signals instead of PMTs.

\subsection{Upscattered Dark Matter}\label{sec:upscattdm}

The sensitivity reach of liquid xenon detectors for sub-GeV dark matter is significantly enhanced if some dark matter particles receive a kinematic boost from up-scattering with cosmic rays~\cite{Bringmann:2018cvk,Alvey:2019zaa,Cappiello:2019qsw,Dent:2019krz,Bondarenko:2019vrb,Wang:2019jtk,Dent:2020syp}. This process, often denoted Cosmic Ray Dark Matter (CRDM), will create a small population of fast or even relativistic dark matter particles. This in turn provides sensitivity to liquid xenon experiments to dark matter with masses several orders of magnitude below 1~GeV. Cosmic ray up-scattered dark matter is only a fraction of the Galactic dark matter population, with an abundance and flux that depends on the dark matter-cosmic ray scattering cross section, the local dark matter density, and the local interstellar spectrum of cosmic rays. Relatively large cross section values of e.g. $\sigma_{\chi N}\gtrsim 10^{-31}{\rm{cm}}^2$ for $m_\chi = 1$~GeV are required to have a notable impact on the sensitivity of a typical liquid xenon detector. For spin-independent interactions, liquid xenon experiments are competitive with and complementary to existing neutrino experiments~\cite{Bringmann:2018cvk, Cappiello:2019qsw,  PROSPECT:2021awi}, which have sensitivity in a similar region of parameter space. Additionally, cosmic-ray upscattering extends the sensitivity of liquid xenon detectors to inelastic dark matter models with mass splittings up to $\sim 100$~MeV~\cite{Bell:2021xff}.

Another upscattering mechanism extending the sensitivity of liquid xenon detectors to dark matter masses down to keV scales is a process called ``solar reflection''~\cite{An:2017ojc,Emken:2017hnp,Liang:2021zkg,An:2021qdl,Emken:2021lgc}. This is based on the observation that scatterings on thermal electrons and nuclei within the Sun can accelerate light dark matter particles. This could give rise to an observable flux of highly energetic particles in a liquid xenon TPC detector.

\subsection{Dark Matter Annihilation Products}\label{sec:dmproducts}

In several models of so-called ``neutrino portal" dark matter, Galactic dark matter self-annihilates into neutrinos~\cite{Cherry:2014xra,GonzalezMacias:2015xbi,Becker:2018rve,Lamprea:2019qet,Patel:2019zky}. These in turn may be detected at direct detection experiments with high rates via coherent elastic neutrino-nucleus scattering (see \autoref{sec:cevns}). A next-generation liquid xenon detector would be sensitive to the flux of these neutrinos for dark matter mass (respective neutrino energy) of $[0.01-1]\1{GeV/c^2}$~\cite{McKeen:2018pbb}. The sensitivity to this neutrino flux would complement neutrino detectors such as Super-Kamiokande. Similarly, dark matter that annihilates to a second component of nucleophilic, ``boosted" dark matter could also be discovered. A next-generation liquid xenon TPC will be sensitive to the effective baryonic coupling of thermal boosted dark matter that is as low as the weak interaction~\cite{McKeen:2018pbb}.

Alternatively, dark matter may annihilate or decay within the target volume of future liquid xenon detectors~\cite{Undagoitia:2021tza}. Considering deposited energies up to a few MeV, the relevant final states include annihilation into $\gamma\gamma$ and $e^-e^+$, and decays into $\gamma\gamma$, $\gamma\nu$ and $e^-e^+$. Although the sensitivity obtained is not as high as the current limits from cosmological considerations~\cite{Slatyer:2015jla} and X-ray measurements~\cite{Ng:2019gch}, this is a complementary approach in a well-understood background environment and free of the large uncertainties typically present in indirect detection experiments.

\subsection{FIMPs and Super-WIMPs}\label{sec:bosonicwimp}

Broad classes of non-WIMP dark matter candidates are Feebly Interacting Massive Particles (FIMPs), which are produced by the thermal freeze-in mechanism~\cite{McDonald:2001vt,Hall:2009bx}, as well as Super Weakly Interacting Massive Particles (super-WIMPs), which are produced by the decay of a freeze out-produced state to a lighter state which is secluded from the Standard Model~\cite{Feng:2003xh}. Both classes share the feature that they couple to Standard Model particles with cross sections far smaller than the Weak scale. These include fermions such as sterile neutrinos~\cite{Dodelson:1993je} and gravitinos, both of which only couple to the Standard Model gravitationally and thus are impossible to observe in a typical direct detection experiment. However, both multi-GeV bosonic and fermionic FIMPs and keV-scale bosonic FIMPs or super-WIMPs can couple to light Standard Model particles in such way to be observed with low-background experiments~\cite{Pospelov:2008jk,Hambye:2018dpi,Belanger:2020npe}. We note here that dark sectors with non-trivial dynamics, for instance an early universe thermal phase transition, 
allow freeze-in production of dark matter in the mass range $10\1{keV/c^2} \lesssim m \lesssim 100\1{MeV/c^2}$ with relatively large scattering cross section with nucleons and/or electrons, which blurrs the distinction between FIMPs and WIMPs~\cite{Elor:2021swj}. In the following, we give two examples of possible keV-scale candidates, namely dark photons and axion-like particles, and discuss related signals.

\subsubsection{Dark Photons}\label{sec:dark_photon}

Dark photons, more properly dark $Z'$ vector bosons, are 
a possible FIMP or super-WIMP candidate if they are stable over cosmological time scales~\cite{Pospelov:2008jk}, and even if unstable, they can act as the mediator of dark matter--Standard Model interactions~\cite{Essig:2013lka}. Like axions, they are well motivated in many UV constructions, and have the advantage over some other candidates of having a mass which is naturally protected from large corrections. In addition to production related to the usual thermal freeze-in or freeze-out, there is also an attractive universal inflationary fluctuation mechanism that gives the observed relic density depending only on the vector-boson mass and inflationary scale~\cite{Graham:2015rva}. A well-studied interaction of dark photons with the Standard Model is via kinetic mixing~\cite{Holdom:1985ag} with hypercharge, and thus with both the Standard Model photon and the Z-boson~\cite{Babu:1996vt,Babu:1997st}. As a consequence, dark photons can be absorbed in a detector with a cross section proportional to the photoelectric cross section. The expected signal is therefore a mono-energetic electronic recoil peak at the dark photon mass. Direct detection experiments have set competitive constraints on kinetic mixing parameter $\kappa$ of dark photons, in a mass range from several to hundreds of keV~\cite{An:2014twa,Aprile:2019xxb,Aprile:2020tmw}. A next-generation detector such as the one discussed here will have improved sensitivity to this mixing parameter $\kappa$. Further, dedicated low-energy calibrations, for example using $^{37}$Ar diluted in the liquid xenon, will help to improve the search in the keV mass range and reduce the relevant detector-specific systematics to negligible levels for a discovery experiment. Using a low-energy charge-only analysis (\autoref{sec:s2only}), the sensitive mass range can be extended to the sub-keV level, see e.g.~\cite{LZ:2021xov}. In addition, XENON1T data already results in the current-best limits on the solar emission of dark photons for some masses~\cite{An:2020bxd}. These channels offer significant room for improvement with a next-generation detector.

\subsubsection{Axions and Axion-Like Particles}

The QCD axion is a pseudoscalar Nambu-Goldstone boson originally proposed as a solution to the strong CP problem of QCD~\cite{PecceiQuinn_1977,Weinberg:1977ma,Wilczek:1977pj}. The QCD axion is also an excellent dark matter candidate~\cite{Preskill:1982cy,Abbott:1982af,Dine:1982ah,Krauss:1985ww,Irastorza:2021tdu}: As they acquire their mass via non-perturbative QCD effects, they are stable on cosmological timescales. Further, QCD axions offer a variety of well-motivated production mechanisms. They couple to the Standard Model very weakly, with couplings suppressed by a high energy scale $f_a$. In terms of this unknown (but constrained) scale, QCD axions are predicted to have a mass $m_a\simeq 5.7 \mu {\rm eV} (10^{12} {\rm GeV}/f_a)$~\cite{GrillidiCortona:2015jxo}. The strict lower bound $f_a\gtrsim {\rm few}\times 10^7 {\rm GeV}$ (see \cite{Zyla:2020zbs}) arising from astrophysical constraints~\cite{Raffelt:2006cw,Sikivie:2006ni,Ayala:2014pea,Chang:2018rso,Capozzi:2020cbu} and the solar axion helioscope CAST~\cite{Anastassopoulos:2017ftl} implies that QCD axion dark matter is beyond the reach of detectors like the one discussed here, and requires dedicated experiments. However, axions produced in the Sun would have thermal spectra with keV energies, and could be detected with a xenon TPC as discussed in \autoref{sec:solar_axions}.

Axion-like particles (ALPs) are a generalization of the QCD axion in that they share many of the same properties, except that the strict relationship between the mass and the scale $f_a$ is relaxed and that the various possible couplings of ALPs to the Standard Model vary over greater ranges than the QCD axion. In particular ALPs can be both much lighter than the QCD axion or much heavier. These particles do not solve the strong CP problem, but nevertheless are good dark matter candidates and can show up abundantly in theories for physics beyond the Standard Model, in particular String Theory~\cite{Witten:1983ar,Svrcek:2006yi,Conlon:2006tq,Arvanitaki:2009fg,Cicoli:2012sz}. In a similar way to dark photons, ALPs can be detected via an analogous process to the photoelectric effect~\cite{Derevianko:2010kz}. For dark matter ALPs, the resulting signal is again a mono-energetic spectrum of electronic recoils at the ALP mass, with an event rate proportional to the square of the dimensionless axion-electron coupling~$g_{ae}$. 

Due to their ultra-low electronic recoil background levels, liquid xenon TPCs have placed the strongest constraints to date on keV ALP dark matter~\cite{Aprile:2014eoa, Abe:2014zcd,Akerib:2017uem, Fu:2017lfc,Abe:2018owy,Aprile:2019xxb,Aprile:2020tmw,Zhou:2020bvf, XENON:2021myl,LZ:2021xov}. Next-generation detectors including the one discussed here will continue to set world-leading constraints~\cite{Aalbers:2016jon}.

\subsubsection{Solar Axions, Dark Matter, and Baryon Asymmetry}

As discussed in \autoref{sec:solar_axions}, a liquid xenon TPC can detect the QCD axion and axion-like particles from the Sun for sufficiently large couplings of the axion with electrons or photons. Such relatively large couplings correspond to a small decay constant, $f_a$, of the axion. In this case, the cosmological abundance of axions produced by conventional mechanisms~\cite{Abbott:1982af,Dine:1982ah,Preskill:1982cy,Davis:1986xc} is too small for the axion to explain the observed dark matter. However, the axion abundance can be large enough to be dark matter in the various scenarios proposed in~\cite{Sikivie:1982qv,Visinelli:2009kt,Hiramatsu:2010yn,Co:2017mop,Co:2018mho,Co:2019jts,Hook:2019hdk}, with couplings that are sufficiently large to be detected in the proposed detector. 

In one such cosmological scenario for example~\cite{Co:2019jts}, a non-zero field velocity delays the onset of field oscillations, enhancing the axion abundance relative to the conventional misalignment mechanism. For axion-like particles, this field velocity can simultaneously explain the baryon asymmetry of the universe~\cite{Co:2020xlh}. Fitting the ratio of dark matter to baryon abundances predicts the axion coupling in terms of its mass $m_a$
\begin{equation}
g_{a \gamma} \simeq 2\times 10^{-11}c_\gamma~{\rm GeV}^{-1} \left(\frac{m_a}{\rm meV}\right)^{1/2},
\end{equation}
where $c_\gamma$ is a model-dependent constant of order unity. For any axion mass, this is much larger than the photon coupling of the QCD axion. A next-generation liquid xenon TPC with a 1000 ton-year exposure can probe this coupling down to $g_{a\gamma}\sim 3 \times 10^{-11} {\rm GeV}^{-1}$~\cite{Dent:2020jhf}, corresponding to $m_a \sim$~meV.

\subsection{Asymmetric Dark Matter}\label{sec:asymmdm}

Asymmetric Dark Matter (ADM) is one of the most motivated non-WIMP dark matter candidate. Similar to the physics that sets the Standard Model cosmic baryon abundance, Asymmetric Dark Matter posits that the dark matter relic density is determined by a dark-sector particle-antiparticle asymmetry associated to a new conserved quantity. In the most attractive scenarios where the Standard Model is linked to the dark sector by connector interactions with non-trivial Standard Model global quantum numbers, this dark asymmetry may be thought of as dark ``baryon" or ``lepton" number, though more complicated cases involving, e.g. flavor are also possible~\cite{Nussinov:1985xr,Gelmini:1986zz,Kaplan:1991ah,Hooper:2004dc,Kitano:2004sv,Cosme:2005sb,Kaplan:2009ag,March-Russell:2009vla,Frandsen:2010yj,Frandsen:2011kt,Petraki:2013wwa,Zurek:2013wia}. Then, similarly to protons and baryon number, the lightest symmetry-carrying state in the dark sector is cosmologically stable. Unlike WIMPs, this dark matter particle is necessarily non-self-conjugate. If $\eta_B$ and $\eta_{dm}$ are, respectively, the baryon and dark matter asymmetries, and if like for baryons and anti-baryons the process of particle-antiparticle annihilation is efficient in the dark sector, then the ratio of the dark matter to the baryon densities is given by 
\begin{equation}
\frac{\Omega_{dm}}{\Omega_B}= \left|\frac{\eta_{dm}}{\eta_B}\right|\frac{m_{dm}}{m_B}~.
\end{equation}
Here, $m_{dm}$ is the mass of the lightest asymmetry-carrying dark matter particle. 

In many beyond the Standard Model theories, the asymmetries
are naturally equal and opposite, up to a computable coefficient which is of order $\sim 1$. Thus, in this case we have an explanation of why the observed baryon and dark matter densities are so close, $\Omega_{dm}\simeq 5 \Omega_B$, if $m_{dm}\sim {\rm few}~{\rm GeV}/c^2$. This strongly motivates searches for dark matter particles in the $\sim 1-10~{\rm GeV}/c^2$ mass range. Though less minimal, this mass range can be extended in more involved models with an $\eta_{dm}/\eta_B$ ratio much different than~1. Since the freeze-out mechanism is no longer setting the dark matter density, the scattering cross section of the dark matter with the Standard Model can be smaller (or larger) than that for WIMPs. Moreover, it is noteworthy that independently-motivated theories, such as those based on ``neutral naturalness" Twin Higgs~\cite{Chacko:2005pe} or composite Higgs explanations of the LHC-data driven little hierarchy problem for the Weak Scale also give non-self-conjugate dark-sector states at the $1-10 {\rm GeV}/c^2$ scale~\cite{GarciaGarcia:2015fol,Craig:2015xla,GarciaGarcia:2015pnn,Cacciapaglia:2021aex}.  

In theories such as Twin Higgs the leading interactions of the individual ADM particles with the Standard Model are due to Higgs-portal and/or kinetic mixing with hypercharge, so often the direct-detection phenomenology is similar to the case of WIMPs with these interactions, though without a necessary link to
a crossed-channel freeze-out process. In other cases the leading higher-dimensional interaction of the individual ADM particles arises from the ``connector" interaction that determines the relation between $\eta_{dm}$ and $\eta_B$~\cite{Hall:2010jx,Buckley:2010ui,Buckley:2011kk,March-Russell:2011ang,March-Russell:2012elz}, or can sometimes be related to a freeze-out process involving heavier states~\cite{Cui:2020dly}.    
\subsection{Composite Dark Matter}\label{sec:compositedm}

Similar to the very rich set of cosmologically stable composite states that exist in the Standard Model sector, all or part of the dark matter density might be in the form of bound states of individual dark matter particles. This is a natural possibility, especially in asymmetric dark matter case, if the dark matter is part of a dark sector as is in turn often so in explicit constructions of physics beyond the Standard Model, or if the dark matter is self interacting. These composites may be strongly-bound ``dark nuclei" or ``dark nuggets"~\cite{Detmold:2014qqa,Hardy:2014mqa,Detmold:2014kba,Wise:2014ola,Hardy:2015boa,Gresham:2017zqi,Bai:2018dxf} of possibly extremely large dark nucleon number, or they could be ``dark atoms"~\cite{Kaplan:2011yj,Petraki:2014uza,GarciaGarcia:2015pnn} made of dissimilar stable dark matter particles, or they could be weakly bound two-body ``dark-onium" states of identical or conjugate particles~\cite{March-Russell:2008klu,Wise:2014jva,Petraki:2015hla,Laha:2015yoa}. Q-balls~\cite{Coleman:1985ki} can also be thought of as a form of composite state carrying a conserved quantum number, the Q-ball description sometimes applying to bound states of bosonic dark matter particles. 

As well as implications for cosmological and indirect detection observations, such composite states typically give rise to a wide variety of new or modified signatures in direct detection experiments.  One of the most studied cases is that of large dark nuclei in which case the recoil spectrum is modified by a characteristic quasi-universal dark sector form factor  \cite{Hardy:2015boa,Butcher:2016hic,Grabowska:2018lnd,Coskuner:2018are}. Since there is often an $N^2$ ($N\gg 1$) dark nucleon number coherent enhancement in the direct detection scattering cross-section which is not present in the collider production of (pairs of) individual dark matter particles, the usual collider bounds on direct detection cross sections can be much weakened~\cite{Hardy:2015boa}. In addition since many of these composite states have low lying single particle or collective excitations of parametrically small energy splitting above the ground state, leading to a very rich set of possible inelastic signatures depending on the exact model (\autoref{sec:inelasticdarkmatter}). It is also possible to have an enhanced diurnal modulation signal as a consequence of the dissipative dynamics that is naturally associated with such composite dark matter states~\cite{Foot:2014osa}.

\subsection{Mirror Dark Matter}\label{sec:mirrordm} 

Closely related to both asymmetric and composite dark matter is the idea of Mirror Dark Matter~\cite{Hodges:1993yb,Berezhiani:2003xm,Okun:2006eb,Foot:2007nn}. This is the intriguing idea that an exact copy of the Standard Model in the dark sector is invoked with an unbroken symmetry between the two. Like some other asymmetric and composite dark matter models, Mirror Dark Matter can generate signatures both in nuclear recoils similar to those expected from $\sim 7\1{GeV}/c^2$ WIMPs with a cross section around $10^{-44}\1{cm^2}$, as well as in electronic recoils~\cite{Foot:2010hu,Foot:2014mia,Foot:2018jpo}. The strongest direct detection constraint on the kinetic mixing currently comes from LUX~\cite{LUX:2019gwa}, with a factor of $\sim$2 better projected sensitivity for the current-generation detectors~\cite{LZ:2021xov}. Mirror dark matter as a hypothesis is potentially entirely falsifiable by the next-generation liquid xenon experiment discussed here~\cite{Clarke:2016eac}.

\subsection{Luminous Dark Matter}\label{sec:lumidm}

It is possible to construct models where the dominant signal of the dark matter originates from photons. These photons could be observed in direct detection experiments as a monoenergetic line produced by dark matter decay from an excited state~\cite{Feldstein:2010su,Pospelov:2013nea}. The excited state could be populated through upscattering in or near the detector (and have short lifetimes $\sim1\,\mu s$) or in the Earth (lifetimes $\sim(1-10)$\,s). For this scenario to work, the elastic cross section needs to be small relative to the inelastic cross section. A simple way to achieve this is with a magnetic dipole operator which couples two distinct Majorana fermions that have a small, keV-order mass splitting. Such ``Luminous dark matter'' was first proposed as a potential explanation of the DAMA/LIBRA modulation~\cite{Feldstein:2010su} and has since been used to explain the recent XENON1T excess~\cite{Bell:2020bes}. While the former scenario is now strongly constrained, the latter scenario will be confirmed or ruled out by a next-generation liquid xenon experiment.

\subsection{Magnetic Inelastic Dark Matter}\label{sec:maginelasticdm}

A natural scenario where dark matter dominantly scatters off nuclei through an inelastic transition in the dark sector (\autoref{sec:inelasticdarkmatter}) is the case of a magnetic dipole interaction~\cite{Chang:2010en}. This model relies on the fact that fermionic dipole operators vanish for Majorana fermions. Thus, if a dark matter Dirac fermion state is split into two nearly degenerate Majorana fermions, an elastic dipole transition is forbidden, leaving the leading dark matter interaction to be an inelastic magnetic dipole transition. Many examples of dark matter models that can realize this scenario have been studied, see e.g.~\cite{Kumar:2011iy,Patra:2011aa,Weiner:2012gm, Pierce:2014spa}.   

For magnetic inelastic dark matter, the sensitivity of a direct detection experiment is modified by both the kinematic constraints of inelastic transitions and by the dependence on the charge and magnetic dipole moment of the target nuclei~\cite{Chang:2010en}. Depending on the dark matter mass splitting and the size of the dipole moment, the excited dark matter state can also decay in the detector, thus yielding both a nuclear recoil from the initial scatter and an electronic recoil from the decay shortly thereafter. Searching for the photons produced by this decay can be an additional handle on uncovering such a scenario~\cite{Lin:2010sb, Pospelov:2013nea}. A dedicated search has been performed by XENON100~\cite{Aprile:2017kek}, with the proposed experiment providing significantly improved sensitivity not only due to the lower background and longer exposure, but also due to the larger size of the detector which translates into a sensitivity to longer decay times.

\subsection{Dark Matter around the Planck Mass} \label{sec:planck}

\begin{figure}[!htbp]  
\begin{center}
\includegraphics[width=0.99\columnwidth]{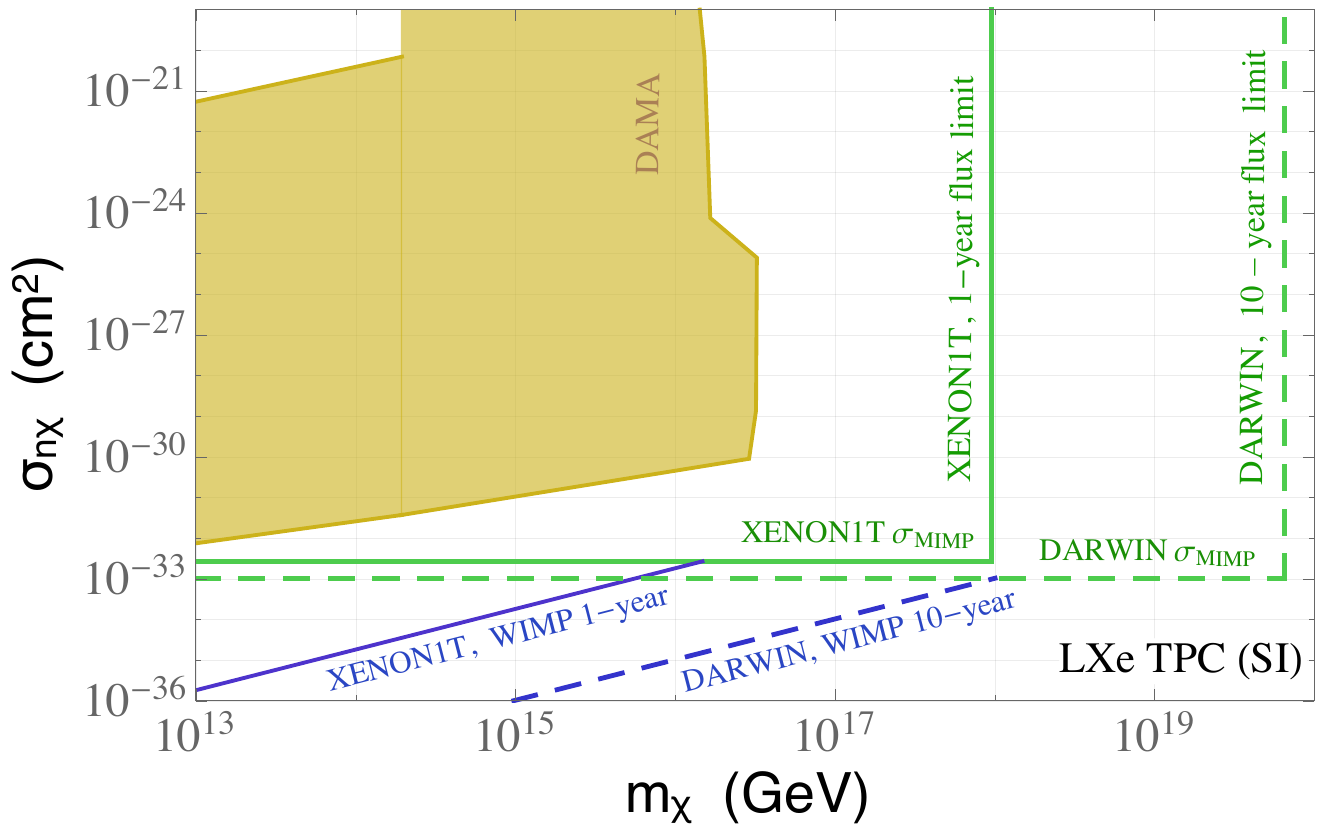}
\caption{Per-nucleon spin-independent scattering cross sections and dark matter masses that can be probed by liquid xenon dark matter detectors via dedicated searches for multi-scatter signals. For cross sections above $\sigma_{\rm MIMP}$ (horizontal green lines) one expects dark matter to scatter multiple times in the detector while transiting. The maximum mass reachable (vertical green lines) is limited by the total integrated flux of dark matter in the detector over the run-time of the experiment. Masses up to and beyond the Planck mass $\simeq 10^{19}\1{GeV/c^2}$ may be probed with a next-generation detector. Only smaller cross-sections and smaller masses are probed by the standard single-scatter analyses (blue lines). Figure taken from Ref.~\cite{Bramante:2018qbc}.}
\label{fig:multiscatter_SI}
\end{center} 
\end{figure}

The observed local dark matter mass density could be made up of few but very massive dark matter particles with masses around the Planck mass $\simeq 10^{19}\1{GeV/c^2}$, as opposed to numerous lighter particles. Super-massive species are motivated by supersymmetric and grand unified theories~\cite{Kolb:2017jvz}, or production by Hawking evaporation of early universe primordial black holes~\cite{Fujita:2014hha,Lennon:2017tqq}. In extensions of the WIMP scenario, with two thermal relics, one of which has a finite lifetime, super-massive dark matter particles with stronger interactions than typical electroweak WIMPs are naturally expected~\cite{Asadi:2021bxp}, and provide excellent targets. The detection of any such particles could help determine parameters of the early universe such as inflation~\cite{Chung:1998zb}, or an epoch of early matter domination~\cite{Hamdan:2017psw,Asadi:2021bxp} leading to efficient primordial black hole production. Their existence could also imply new light mediators beyond the Standard Model~\cite{Davoudiasl:2018wxz}. 

Due to the small number density, the flux of these particles through a given detector would be very low, and thus any detection would both imply and require a very high scattering cross section, such that almost all particles impinging on the detector prompt a signal. When the cross section becomes high enough that these particles would interact more than once in the detector, discovery requires a dedicated analysis looking for multiple-scatter events. Such events are typically discarded in WIMP-like dark matter analyses, leaving many orders of magnitude of unexplored parameter space, see \autoref{fig:multiscatter_SI}. In this multiple scattering regime, a next-generation liquid xenon experiment would be capable of probing dark matter masses up to and beyond the Planck mass $\simeq 10^{19}$~GeV~\cite{Bramante:2018qbc}, in a complementary way to the range that could be probed using dedicated neutrino experiments~\cite{Bramante:2018tos,Bramante:2019yss,Clark:2020mna}. A dedicated search using the DEAP-3600 liquid argon detector has already been published~\cite{Adhikari:2021fum}.
Clusters of dark matter formed through self-attraction~\cite{Hardy:2014mqa,Butcher:2016hic,Acevedo:2021kly}, such as ``dark blobs'' or ``dark nuggets''~\cite{Hardy:2014mqa,Hardy:2015boa,Grabowska:2018lnd,Coskuner:2018are} could exist at these high masses and create tracks if they have sufficiently large cross-sections.

\section{Double Beta Processes}\label{sec:doublebeta}

\subsection{Neutrinoless Double Beta Decay of \texorpdfstring{$^{136}$Xe}{Xenon-136}} \label{sec:0nubb}

Among the main intellectual challenges facing the nuclear and particle physics communities today are the neutrino-mass generation mechanism, the absolute neutrino-mass scale, and the neutrino-mass spectrum. One of the best ways to address these fundamental questions is to search for neutrinoless double beta decay ($0\nu\beta\beta$)~\cite{Doi:1992dm,Avignone:2007fu, Dolinski:2019nrj,Agostini:2022zub}. The observation of this rare nuclear decay process, forbidden in the Standard Model, would imply that the lepton number is violated by two units and confirm the Majorana nature of the neutrinos. It would also provide invaluable information about the dominance of matter over antimatter in the universe, because two matter particles---electrons---are emitted in the decay without the balance of the corresponding antiparticles. Double beta decay can occur in the two xenon isotopes $^{134}$Xe~\cite{LZ:2021rff} and $^{136}$Xe, with the latter offering a larger sensitivity to the $0\nu\beta\beta$ half-life ($T^{0\nu}_{1/2}$). The best experimental constraint on the $^{136}$Xe $0\nu\beta\beta$ half-life, $T^{0\nu}_{1/2} > 1.07 \times 10^{26}\1{years}$ (90\% CL), is set by the KamLAND-Zen collaboration using $^{136}$Xe dissolved in a liquid scintillator~\cite{KamLAND-Zen:2016pfg}. Among double-beta experiments, only $^{76}$Ge, $^{82}$Se, $^{100}$Mo and $^{130}$Te offer $0\nu\beta\beta$ half-life limits comparable to $^{136}$Xe~\cite{CUORE:2021gpk,GERDA:2020xhi,CUPID:2020aow,Majorana:2019nbd,EXO-200:2019rkq,CUPID:2019gpc}. The EXO-200 collaboration demonstrated that better energy resolution and background rejection can be achieved with a liquid xenon TPC~\cite{Albert:2017owj}, and the PandaX-II collaboration conducted a first search using a dual-phase natural xenon detector~\cite{Ni:2019kms}. XENON1T recently demonstrated that energy resolutions below $\sigma/\mu=1$\% at $Q_{\beta\beta}$ can be achieved in liquid xenon TPCs used for dark matter searches~\cite{XENON:2020iwh}.

\begin{figure}[!htbp]
\begin{center}
\includegraphics[width=0.99\columnwidth]{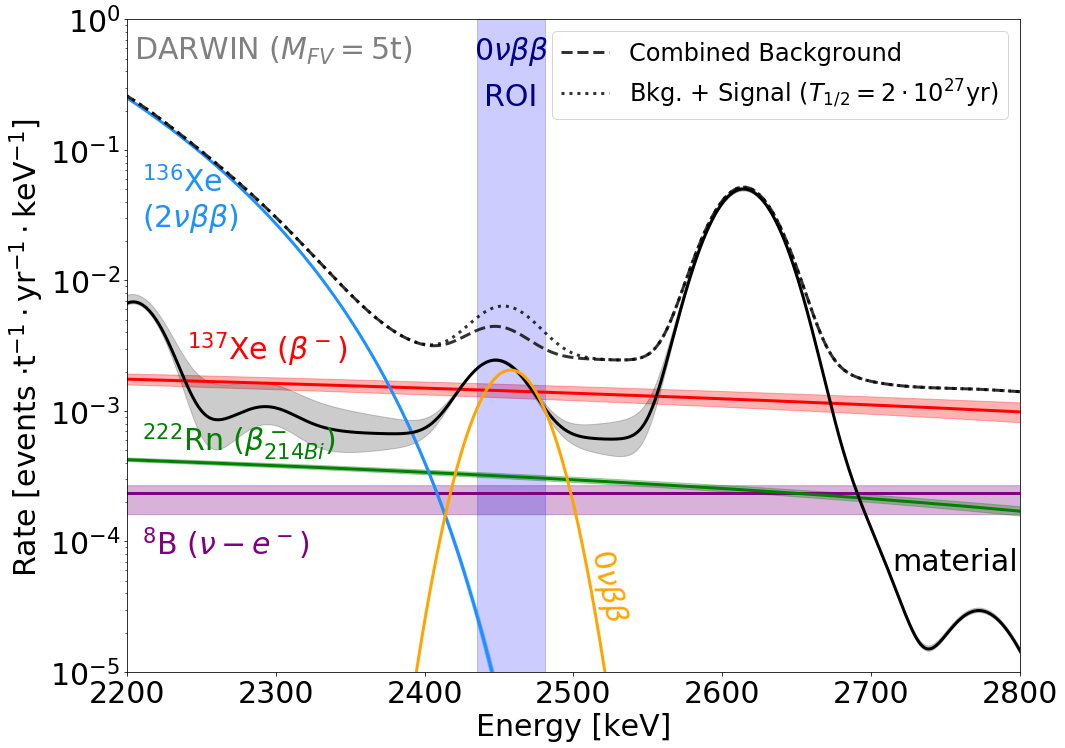}
\caption{Predicted  background  spectrum around the $0\nu\beta\beta$ energy region of interest (ROI) for a proposed next-generation dark matter experiment. Rates are averaged over a fiducial volume (FV) containing $5000\1{kg}$ of liquid xenon with natural isotopic abundance. Bands indicate $\pm~1\sigma$ uncertainties. The orange line represents a hypothetical signal corresponding to $T_{1/2}=2\times 10^{27}\1{years}$. Figure from Ref.~\cite{Agostini:2020adk}.}\label{fig:0vbb_signal}
\end{center}
\end{figure}

A next-generation liquid xenon detector will contain multiple tonnes of the $^{136}$Xe isotope, either at the natural abundance of 8.9\%, or, as a possible upgrade, using enriched xenon. Given a TPC design optimized for WIMP searches, a detector instrumenting $\sim 40,000\1{kg}$ of non-enriched xenon can already improve the sensitivity to $0\nu\beta\beta$ decay by more than one order of magnitude over current limits, without any interference with its primary dark matter science goal. Taking advantage of the excellent self-shielding of liquid xenon, the material-induced gamma ray background can be suppressed below the total intrinsic background rate (\autoref{sec:erfiducialization}). \autoref{fig:0vbb_signal} shows the relevant sources of background with a hypothetical $0\nu\beta\beta$ signal for the innermost $5000\1{kg}$ of natural xenon in the TPC of a proposed next-generation detector~\cite{Agostini:2020adk}. The background from material-induced gamma rays will be further reduced in more massive detectors than the one simulated in \autoref{fig:0vbb_signal}.

\begin{figure}[!htbp]
\begin{center}
\includegraphics[width=0.99\columnwidth]{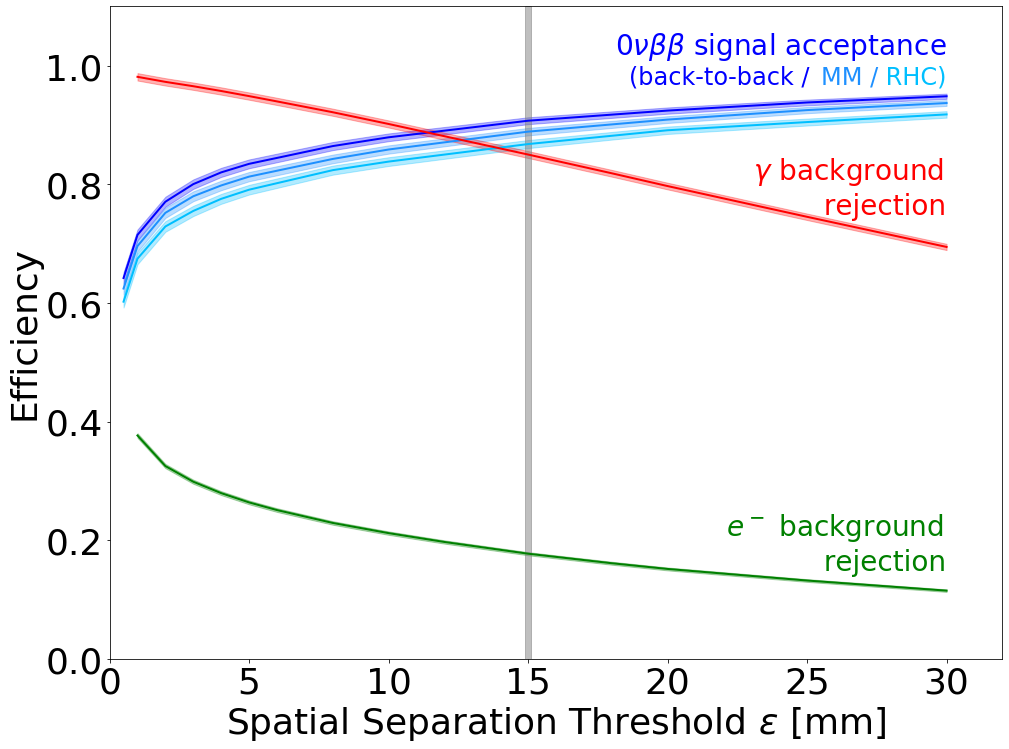}
\caption{Efficiency of $0\nu\beta\beta$ signal acceptance and background rejection as a function of the minimum distance for individual reconstruction of energy depositions. The three signal lines (blue) compare different energy and angular distributions for the $0\nu\beta\beta$ signal based on a back-to-back electron emission, a mass mixing (MM) mechanism and a right-handed current (RHC) model. The background rejection efficiency is shown for $\gamma$s (red) and electrons (green) with $E=Q_{\beta\beta}=2457.8\1{keV}$. The vertical line (gray) corresponds to the value assumed here. Bands indicate $\pm~2\sigma$ uncertainties~\cite{Agostini:2020adk}.}\label{fig:0vbb_acceptance}
\end{center}
\end{figure}

\begin{figure}[!htbp]
\begin{center}
\includegraphics[width=0.99\columnwidth]{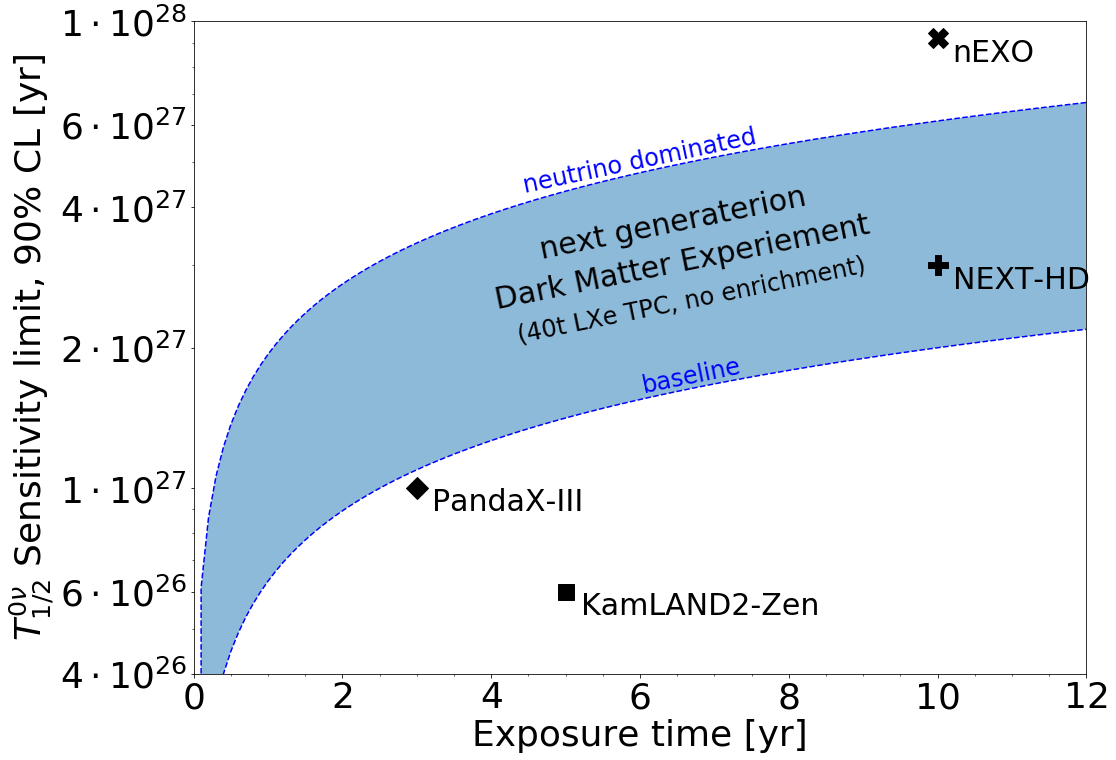}
\caption{Predicted median $T_{1/2}^{0\nu}$ sensitivity at 90\%~CL as a function of the exposure time for a next generation TPC detector containing \SI{40}{t} of liquid xenon with natural isotopic abundance. The band indicates the sensitivity range between a baseline radio purity scenario at a depth of \SI{3500}{m} water equivalent to a scenario with neutrino dominated background. Sensitivity projections for future $^{136}$Xe $0\nu\beta\beta$ experiments~\cite{Agostini:2020adk, Gomez_NEXT:2019, Chen:2016qcd, Albert:2017hjq, Barabash:2015eza} are shown for comparison. Figure based on the one in~\cite{Agostini:2020adk}.}\label{fig:0vbb_sensitivity}
\end{center}
\end{figure}

Selecting ultra-low background materials for construction can further reduce the material contribution as well as the background rate from $^{222}$Rn, which emanates from material surfaces into the target volume. A sufficiently deep laboratory suppresses cosmogenic background sources, such as the in-situ activation of $^{136}$Xe by muon-induced neutrons (producing $^{137}$Xe)~\cite{Cebrian:2020bwn,Rogers:2020npx}, down to the limit set by electron scattering of solar $^{8}$B neutrinos. Optimizing the detector design for an improved spatial resolution would allow to further exploit background rejection, based on the different topology of background and signal events caused by bremsstrahlung radiation, as shown in \autoref{fig:0vbb_acceptance}. Combining these measures, the experimental sensitivity can be further enhanced to make a next-generation dark matter detector competitive to dedicated next-generation, tonne-scale $0\nu\beta\beta$ experiments, as shown in \autoref{fig:0vbb_sensitivity}. Naturally, a larger target mass would allow further gains in self-shielding and even more competitive sensitivity to rival the next generation of dedicated experiments. Isotopic enrichment in $^{136}$Xe would further improve this sensitivity, as it linearly increases the signal, although this also increases the background rate from the two-neutrino double beta decay ($2\nu\beta\beta$) of $^{136}$Xe and $\beta$-decay of $^{137}$Xe.

In addition, there exist mechanisms of $0\nu\beta\beta$~decay where the lepton-number violation necessary for the decay is due to a lepton-number violating mechanism other than the standard scenario of exchange of light neutrinos. A large fraction of these models can be tested using an effective field theory approach~\cite{Pas:1999fc,Pas:2000vn,Prezeau:2003xn,deGouvea:2007qla,Cirigliano:2017djv,Graf:2018ozy,Cirigliano:2018yza,Dekens:2020ttz,Deppisch:2020ztt,Kotila:2021xgw} which in the case of chiral EFT also provides a hierarchy for the relevant nuclear matrix elements~\cite{Menendez:2011qq,Wang:2018htk,Cirigliano:2017tvr,Pastore:2017ofx,Cirigliano:2018hja,Cirigliano:2019vdj,Cirigliano:2020dmx,Cirigliano:2021qko,Jokiniemi:2021qqv,Wirth:2021pij}, along similar lines as described in \autoref{sec:eft} for dark matter. These theoretical models can thus be used as a low-energy test of new physics phenomena that complements high-energy searches at accelerators.

Besides the search for $0\nu\beta\beta$ decay, precision measurements of the $2\nu\beta\beta$ decay can reduce the experimental uncertainty on the $2\nu\beta\beta$ nuclear matrix element~\cite{Barabash:2020nck} and constrain the underlying nuclear theories~\cite{Saakyan:2013yna,Engel:2016xgb,Agostini:2022zub}. This is especially relevant regarding $0\nu\beta\beta$ decay, because predictions of nuclear matrix elements disagree by a factor three or more, severely limiting the interpretation of current limits and the physics reach of future searches, for instance in terms of neutrino masses~\cite{Engel:2016xgb,Agostini:2022zub}. The nuclear many-body methods used to calculate $0\nu\beta\beta$ nuclear matrix elements are generally the same that are also used to obtain the nuclear structure factors in \autoref{sec:structure}. The nuclear shell model among other more phenomenological approaches yields most predictions~\cite{Rodriguez:2010mn,Mustonen:2013zu,Hyvarinen:2015bda,Horoi:2015tkc,Menendez:2017fdf,Song:2017ktj,Simkovic:2018hiq,Fang:2018tui,Deppisch:2020ztt,Coraggio:2020hwx}, complemented by recent {\it ab initio} studies~\cite{Yao:2019rck,Novario:2020dmr,Belley:2020ejd,Wirth:2021pij}. A precise $2\nu\beta\beta$ spectrum shape measurement can provide insights towards reliable $0\nu\beta\beta$ nuclear matrix element calculations~\cite{KamLAND-Zen:2019imh}. In addition, precision measurements of $2\nu\beta\beta$ decay can also be used to probe new physics. For example, right-handed lepton currents affect the angular and energy distributions of the decay \cite{Deppisch:2020mxv}, MeV-scale sterile neutrinos can be searched for through kinks in the $2\nu\beta\beta$ spectrum \cite{Bolton:2020ncv, Agostini:2020cpz}, and neutrino self-interactions can leave an imprint on the spectrum as well \cite{Deppisch:2020sqh}. Because lepton number is not necessarily violated in $2\nu\beta\beta$ decay, this is independent of the neutrino nature and can be used to constrain or pinpoint properties of both Majorana and Dirac neutrinos.

\subsection{Double Electron Capture on \texorpdfstring{$^{124}$Xe}{Xenon-124}}\label{sec:dec}

Similar to double beta decay, double electron capture is a second order Weak Interaction process~\cite{Winter:1955zz,Blaum:2020ogl} with extremely long half-lives. Two electrons are captured from the atomic shell and two protons are converted into neutrons. In the Standard Model decay, two neutrinos carrying virtually the total Q-value are emitted and leave the active volume undetected ($2\nu$ECEC). The measurable signal is constituted by the atomic de-excitation cascade of X-rays and Auger electrons that occurs when the vacancies of the captured electrons are refilled. In a liquid xenon detector, this cascade is measured as a single resolvable signal at $64.3\1{keV}$ for the double K-capture~\cite{Nesterenko:2012xp} as the most likely case~\cite{Doi:1991xf}. The half-life of this decay is of great interest with regard to nuclear matrix element calculations, as it provides a benchmark point from the proton-rich side of the nuclide chart~\cite{Suhonen:2013rca,Pirinen:2015sma,Perez:2018cly}. A precise measurement would help to narrow down uncertainties, which in turn have implications on the neutrino mass scale derived from $0\nu\beta\beta$ as discussed in \autoref{sec:0nubb}. 

Following hints in XMASS~\cite{Abe:2018gyq}, the half-life of the $^{124}$Xe double K-capture has recently been measured by XENON1T~\cite{XENON:2019dti}. At $T_{1/2}^{2\nu\text{KK}} = (1.8 \pm 0.5_\text{stat} \pm 0.1_\text{sys}) \times 10^{22}\1{years}$, it agrees with recent theoretical predictions~\cite{Suhonen:2013rca,Pirinen:2015sma,Perez:2018cly}. For comparison, this is about one order of magnitude slower than the $2\nu\beta\beta$ decay of $^{136}$Xe due to the small overlap of the K-electron with the nucleus. Assuming a natural isotopic abundance similar to XENON1T, a next-generation experiment would record on the order of 10,000 events in its full exposure. This will allow a precision measurement of the double K-capture half-life to the percent level. Additionally, an observation of the KL-capture and LL-capture would be within reach~\cite{Doi:1991xf}. Their measurement would help to decouple the nuclear matrix element from phase-space factors.

Beyond the Standard Model, the double electron capture on $^{124}$Xe without neutrino emission ($0\nu$ECEC) can complement $0\nu\beta\beta$ in addressing fundamental questions about the mass and nature of the neutrino~\cite{Bernabeu:1983yb,Sujkowski:2003mb}. To conserve energy and momentum, $0\nu$ECEC is possible if the $^{124}$Xe decay populates an excited state of the $^{124}$Te daughter nucleus, so that the final energy matches the initial one. This scenario would allow a resonant enhancement of this channel, which would be needed to provide accessible half-lives~\cite{Kotila:2014zya}. A suitable daughter state exists, but current measurements of the Q-value indicate only an approximate match of the $^{124}$Te level, two-hole energy, and Q-value that would only provide a minor enhancement~\cite{Nesterenko:2012xp}. If this decay is realized, the experimental signature contains multiple $\gamma$-rays emitted in a cascade, so coincidence techniques could increase experimental sensitivity \cite{Wittweg:2020fak}.

\subsection{Other Double-Beta Processes}

The $^{124}$Xe $Q$-value of 2857~keV also allows second-order decays involving positrons~\cite{Kim:1982vi}. Examples are the as-yet unobserved Standard Model $2\nu$EC$\beta^+$ and $2\nu\beta^+\beta^+$ decays, as well as the hypothetical $0\nu$EC$\beta^+$ and $0\nu\beta^+\beta^+$ decays. The decay $2\nu$EC$\beta^+$ is predicted to have a half-live one order of magnitude above that of $2\nu$ECEC~\cite{Kotila:2013gea}. Exploiting the coincidence signature of the positron annihilation and the atomic de-excitation cascade, this decay could already be within reach of LZ and XENONnT~\cite{Barros:2014exa, Wittweg:2020fak} and be a sure signal in the next-generation detector. The $2\nu\beta^+\beta^+$ decay is expected to be several orders of magnitude slower~\cite{Kotila:2013gea}. It exhibits a unique signature with five point-like ionization clusters, located in the same plane with the central vertex~\cite{Bolozdynya:1997pdbd}.

On the neutrinoless side, $0\nu$EC$\beta^+$ would be favoured in the absence of resonance enhancement for $0\nu$ECEC, even though the decay rate is expected to be suppressed by about three orders of magnitude with respect to the $0\nu\beta\beta$ decay of $^{136}$Xe~\cite{Rath:2009dr,Kotila:2013gea,Barea:2013wga,Kotila:2012zza}. Here, the current lower limits on the half-lives are on the order of $10^{21}$ years~\cite{Angloher:2016ktr,Lehnert:2016gra,CUORE:2017dbf}. These would be accessible to a large extent in a next-generation liquid xenon experiment when exploiting the coincidence signature of the atomic relaxation, the mono-energetic positron, and the two subsequent back-to-back $\gamma$-rays. Moreover, limits on half-lives of neutrinoless second-order weak decays in $^{124}$Xe could complement $0\nu\beta\beta$ searches in $^{136}$Xe and other nuclei and help to identify the decay mechanism~\cite{Hirsch:1994es, Deppisch:2006hb,Gehman:2007qg,Wittweg:2020fak}. These channels provide an exciting avenue for the next-generation detector discussed here to complement ongoing searches for double-weak processes.

\section{Neutrinos for Astrophysics} \label{sec:neutrinos}

\begin{figure}[!htbp]
\begin{center}
\includegraphics[width=0.99\columnwidth]{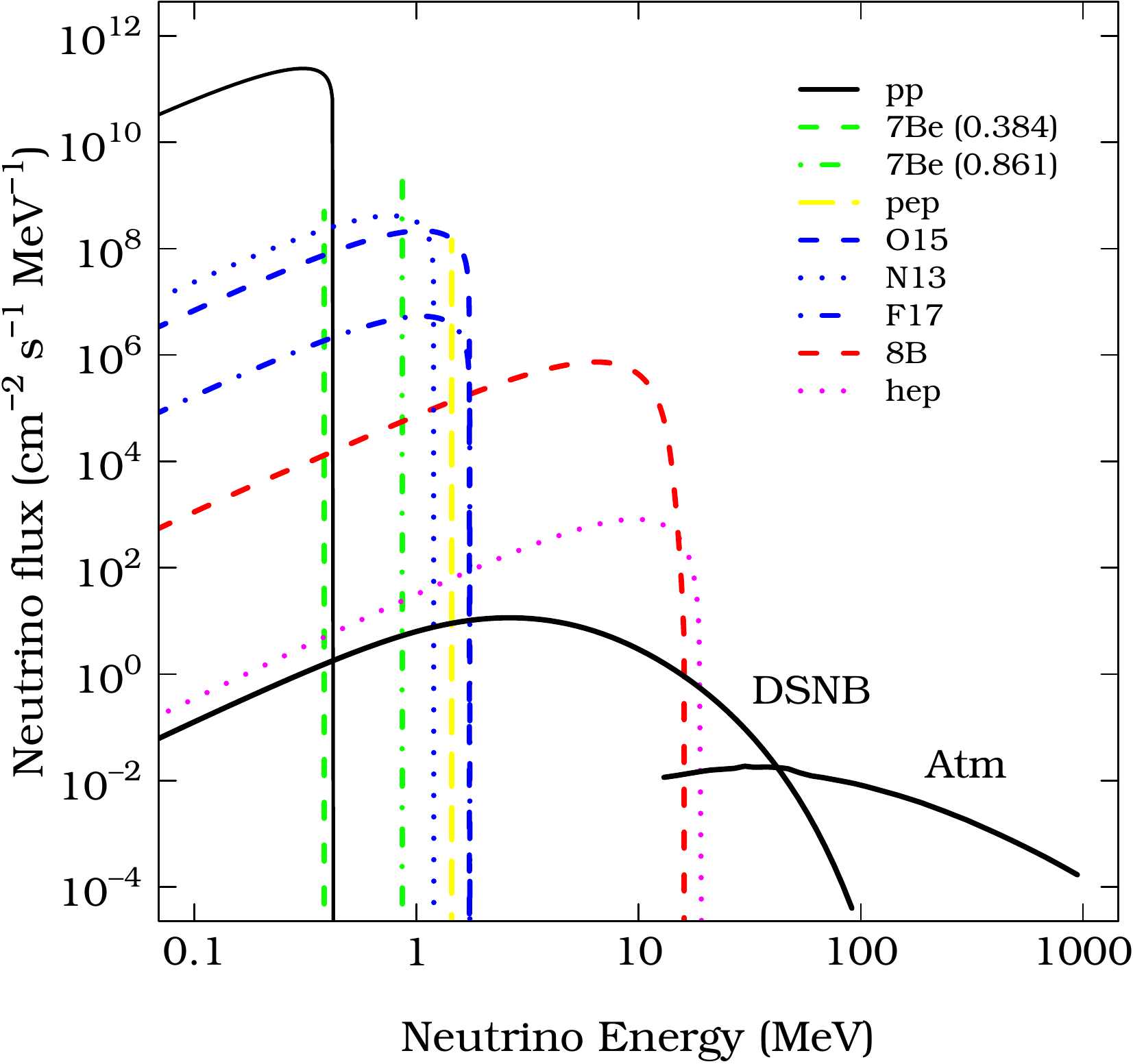}
\caption{Astrophysical neutrino fluxes span many orders of magnitude in flux and energy. This explains the different exposures and energy thresholds required to measure them. Figure adopted from Ref.~\cite{Dutta:2019oaj}.}\label{fig:nufluxes}
\end{center}
\end{figure}

Many sources of astrophysical neutrinos exist~\cite{Vitagliano:2019yzm,Gann:2021ndb}, and those in the relevant energy range for xenon experiments are shown in \autoref{fig:nufluxes}~\cite{Dutta:2019oaj}. Overall, the flux is dominated by pp solar neutrinos, which will be the leading source of low-energy electronic recoils. Atmospheric neutrinos have the highest energy and can induce sizable nuclear recoils of tens of keV through Coherent Elastic Neutrino-Nucleus Scattering; their measurement is a goal of the next-generation liquid xenon experiment. A prominent source is $^8$B solar neutrinos as they lie in a sweet spot: their energy is high enough that nuclear recoils are visible in dedicated xenon TPCs, while their flux is so large that a first measurement can be achieved already with the currently-running generation of liquid xenon experiments.

A next-generation detector will make important advances in neutrino astrophysics, covering low-energy realms that are out of the reach of experiments such as Hyper-K~\cite{Abe:2018uyc} or DUNE~\cite{Acciarri:2016crz}. This section outlines the scientific scope of the next-generation detector, including solar, atmospheric, and supernova neutrinos, and discusses the unique interaction channels that this detector will be sensitive to. 

\subsection{Neutrino Interactions}

Neutrinos can interact with liquid xenon through Charged Current (CC) and/or Neutral Current (NC) interactions to produce detectable electronic and nuclear recoils. The neutrino-induced rate is
\begin{equation}
\frac{{\rm d}R}{{\rm d}T_R} = \mathcal{N} \times \int_{E_{\nu}^{min}}{\phi\left(E_{\nu}\right) \times \frac{{\rm d}\sigma(E_\nu, T_R)}{{\rm d}T_R} {\rm d}E_\nu}
\end{equation}
where $\mathcal{N}$ is the number of target nuclei or electrons per unit of mass of detector material (for nuclear and electronic recoils, respectively), $\phi\left(E_{\nu}\right)$ is the neutrino flux as a function of the neutrino energy as shown in \autoref{fig:nufluxes}~\cite{Dutta:2019oaj}, and $E_{\nu}^{min}$ is the minimum neutrino energy required to generate a recoil at an energy $T_R$. For a nuclear recoil, in the limit where $m_N \gg E_\nu$, the minimum energy is given by 
\begin{equation}
E_{\nu}^{min} = \sqrt{\frac{m_N T_R}{2}},
\end{equation}
whereas in the case of an electronic recoil, it is given by 
\begin{equation}
E_{\nu}^{min} = \frac{1}{2}\left( T_R + \sqrt{T_R\left( T_R + 2m_e \right)}\; \right)
\end{equation}

The differential cross section depends on the nature of the interaction. In the next two sections, we discuss Coherent Elastic Neutrino-Nucleus Scattering and the Electroweak interaction, which constitute the major contributions to the potential detectable signal for liquid xenon detectors.

\subsubsection{Coherent Elastic Neutrino-Nucleus Scattering} \label{sec:cevns}

In the Standard Model, elastic neutrino-nucleon scattering proceeds only through neutral current interaction with the exchange of a $Z$-boson. The resulting differential neutrino-nucleus cross section as a function of the nuclear recoil energy $T_R$ and the incoming neutrino energy $E_\nu$ is
\begin{align}
\label{CEvNS}
\frac{{\rm d}\sigma(E_\nu, T_R)}{{\rm d}T_R} &= \frac{G^2_f}{\pi} m_N  \left( Z g_v^p + N g_v^n \right)^2 \notag\\
&\times \left(1 - \frac{m_NT_R}{2E^2_{\nu}}
\right) F^2(T_R),
\end{align}
where $m_N$ is the target nucleus mass, $G_f$ is the Fermi coupling constant, $N$ the number of neutrons, $Z$ the number of protons, $g_v^n = -1/2$, and $g_v^p = 1/2 - 2 \sin^2 \theta_w$, where $\theta_w$ the Weak mixing angle. Because $\sin^2{\theta_w}\simeq 0.23$, the cross section scales roughly with the number of neutrons squared. The nuclear form factor $F(T_R)$ describes the loss of coherence due to the internal structure of the nucleus. For momentum transfers less than the inverse of the size of the nucleus, the coherence condition is largely satisfied and $F(T_R) \rightarrow 1$. In lieu of experimental data on the neutron distribution in the nucleus, a typical parameterization is the Helm form factor~\cite{Helm:1956zz} that is also commonly used for WIMP direct detection~\cite{Engel:1991wq, Lewin:1995rx}. Note that Eq.~\eqref{CEvNS} gives the leading form of the cross section, keeping only the coherently enhanced part of the vector interaction; Ref.~\cite{Hoferichter:2020osn} has the complete expressions. 

\begin{figure}[!htbp]
\begin{center}
   \includegraphics[width=0.99\columnwidth]{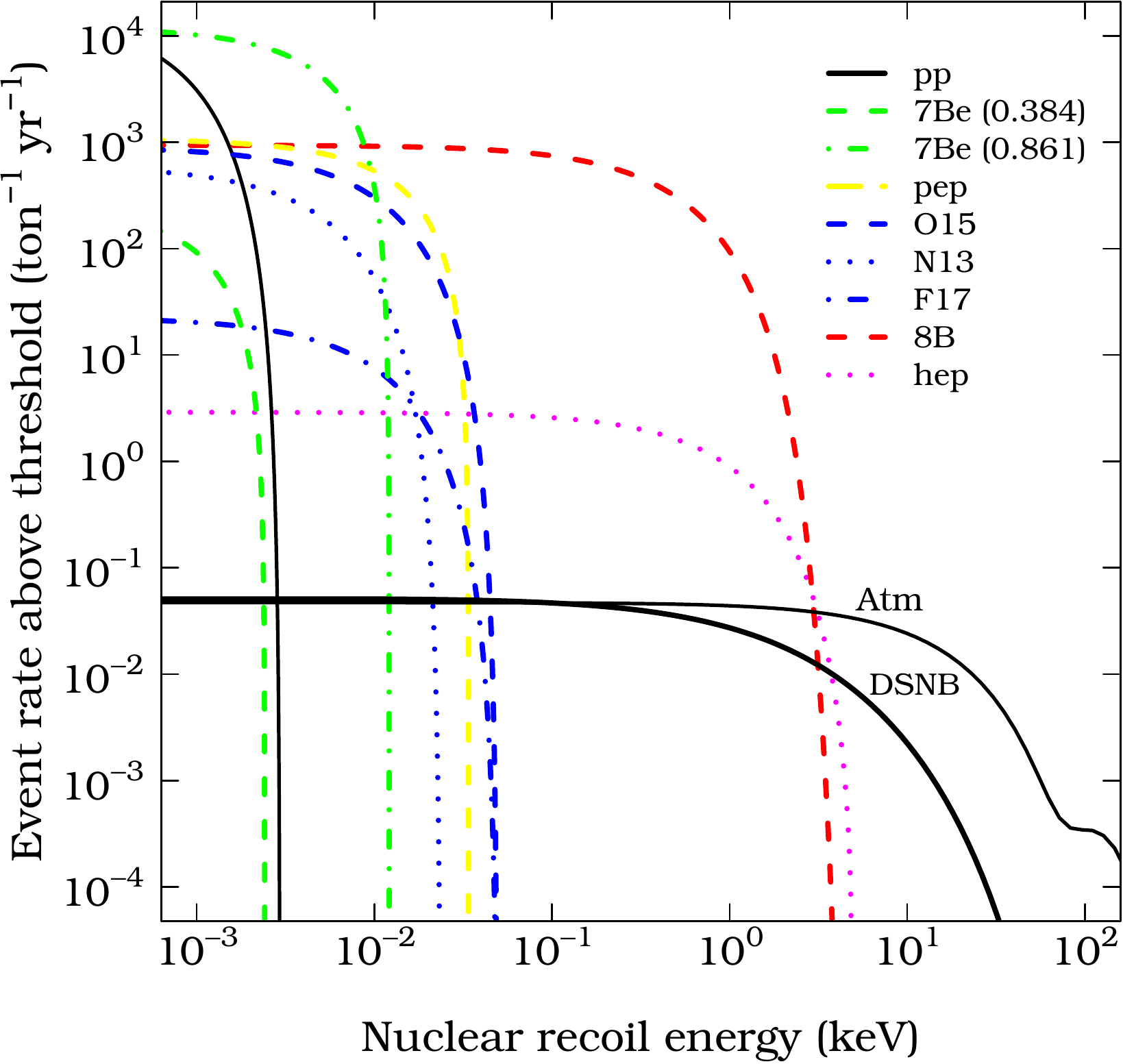}
   \caption{Nuclear recoil event rates from astrophysical neutrinos via CE$\nu$NS. $^8$B solar neutrinos are expected to be measured first in the currently-running generation of experiments. The detector proposed here targets a precision measurement of that flux, and a first measurement of the atmospheric neutrino flux.}\label{fig:nurates}
\end{center}
\end{figure}

In effect, this Coherent Elastic Neutrino-Nucleus Scattering (CE$\nu$NS)~\cite{Akimov:2017ade} increases the cross section for heavy nuclei such as xenon, while pushing the recoil energy spectrum to small energies of keV or less. Given the excellent performance of liquid xenon detectors at such low energies, this channel thus opens the possibility to detect neutrinos from astrophysical sources with a target mass that is modest in comparison to other neutrino detectors, see \autoref{fig:nurates}. In addition to providing the possibility to measure some astrophysical neutrino sources for the first time, the fact that this interaction is flavor-independent provides complementary information for sources that have been measured by other neutrino detectors~\cite{Cabrera:1984rr,Krauss:1985pf}.

\subsubsection{Electroweak Interaction}\label{sec:ewinteraction}

The neutrino-electron electroweak interaction proceeds through both Charged Current ($W$-boson exchange) and Neutral Current ($Z$-boson exchange) interactions. In the free electron approximation, the resulting differential neutrino-electron cross section as a function of the electronic recoil energy $T_R$ and the incoming neutrino energy $E_\nu$ is
\begin{align}
\frac{{\rm d}\sigma(E_\nu, T_R)}{{\rm d}T_R} & = \frac{G^2_f m_e}{2 \pi} \Bigg[ \left( g_v + g_a \right)^2 + \left(g_{a}^{2} - g_{\nu}^{2} \right) \frac{m_e T_R}{E_{\nu}^2}\notag\\
&+ \left( g_v - g_a \right)^2 \left(1 - \frac{T_R}{E_{\nu}}\right)^2\Bigg],
\end{align}
where $m_e$ is the electron mass, $G_f$ is the Fermi coupling constant, $g_v = 2\sin^2{\theta_w} - 1/2$ and $g_a = 1/2$ are respectively the vectorial and axial coupling, and $\theta_w$ is the weak mixing angle. In the context of $\nu_e + e \rightarrow \nu_e + e$ scattering, the interference coming from the addition of the charge current leads to a shift in axial and vectorial couplings such as: $g_v \rightarrow g_v +1$ and  $g_a \rightarrow g_a +1$. This is then contributing to enhance the $\nu_e + e \rightarrow \nu_e + e$ scattering cross section with respect to the $\nu_{\tau, \mu} + e \rightarrow \nu_{\tau, \mu} + e$ cross section by about one order of magnitude. Further, neutrino oscillations also are an important factor that needs to be taken into account to properly calculate neutrino-induced event rates.

\begin{figure}[!htbp]
\begin{center}
\includegraphics[width=0.99\columnwidth]{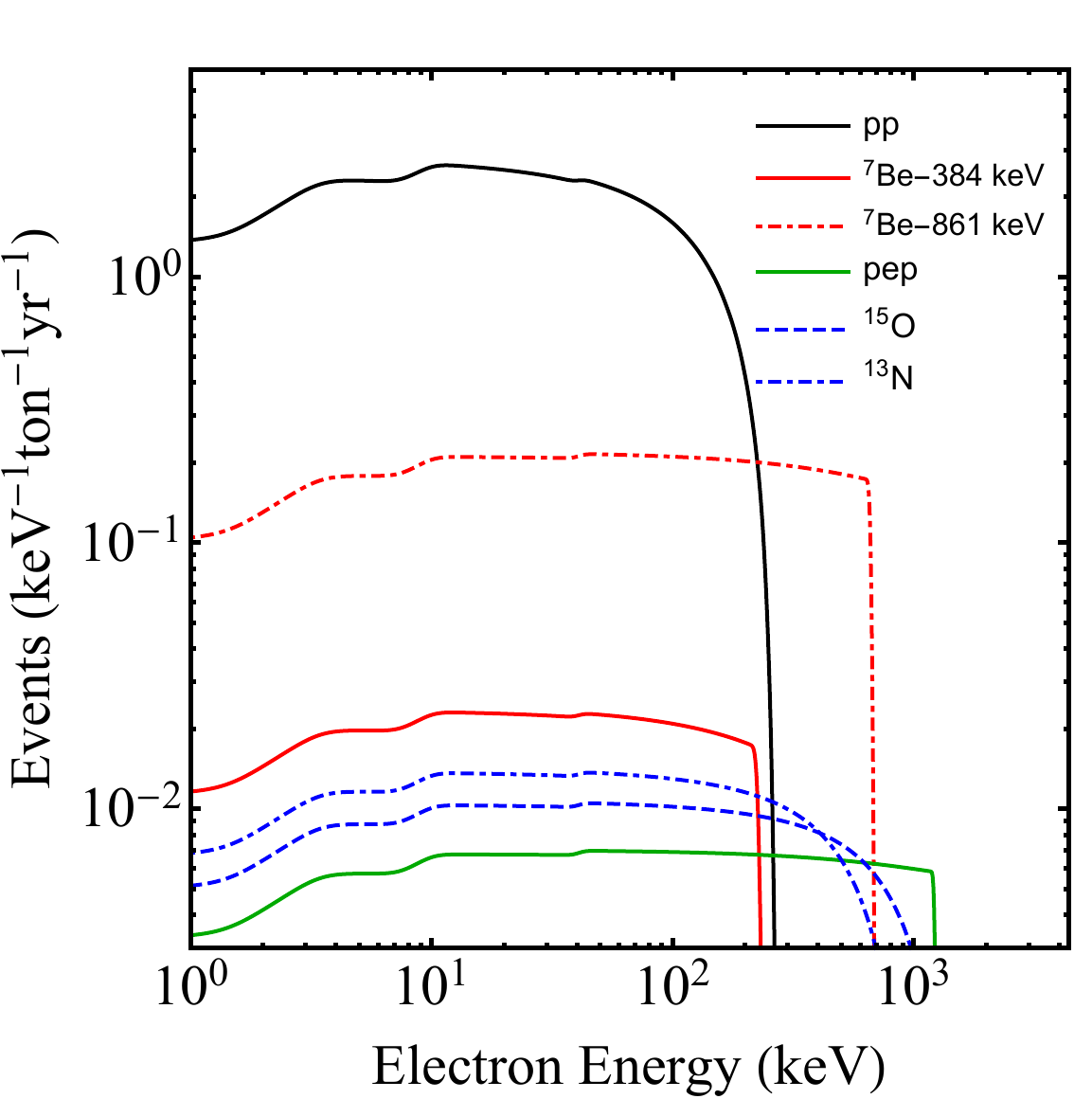}
\caption{Electronic recoil scattering rates from solar neutrinos. The step-wise decrease in event rate towards low energies corresponds to the energy levels of electrons in the xenon atom. Figure from Ref.~\cite{Newstead:2018muu}.}\label{fig:electronrates}
\end{center}\end{figure}

Low-energy electronic recoil starts to deviate from the simple free electron approximation below a few tens of~keV. Therefore, it is important to include corrections for the stepping of atomic shells and atomic binding. This has been included into the calculation by using the Relativistic Random Phase Approximation (RRPA) as presented in~\cite{Chen:2016eab}. The inclusion of these atomic effects result in a reduction of the neutrino-induced electronic recoil event rate below $\sim 5$~keV. Importantly, this reduces the neutrino background in the $[2-10]\1{keV}$ energy range by $\sim$22\%. \autoref{fig:electronrates} shows the electronic recoil event rate for different neutrino flux contributions. The wavy features in the energy spectra are due to the stepping of atomic shells, smoothed by detector resolution effects.

\subsection{Solar Neutrinos}\label{sec:solarneutrinos}

Experimental studies of solar neutrinos date back to over half of a century ago~\cite{Bahcall:1976zz}. The primary goal of these experiments is to measure the different components of the solar neutrino flux, in order to provide an understanding of the physics of the solar interior. Many different types of solar neutrino experiments were operated, and they have evolved in their size and scientific scope since the original experiments~\cite{Robertson:2012ib}. The combination of all solar neutrino data with terrestrial experiments that study neutrinos in the same energy range has led to the LMA-MSW solution to neutrino flavor transformation from the Sun to the Earth~\cite{deHolanda:2002dko}. With this solution, at low energies $\lesssim 5\1{MeV}$, vacuum oscillations describe the neutrino flavor transformation, and the electron neutrino survival probability is $\gtrsim 50\%$. At energies $\gtrsim 5\1{MeV}$, matter-induced transformations describe the flavor transformation, with a corresponding survival probability of $\gtrsim 1/3$. 

However, in spite of all the theoretical and experimental progress in the field of solar neutrino physics over the past several decades, there are still outstanding questions that surround some of the data. For example, three experiments (Super-Kamiokande~\cite{Abe:2016nxk}, SNO~\cite{Aharmim:2011vm}, and Borexino~\cite{Agostini:2018uly}) that are sensitive to electronic recoils from neutrino-electron elastic scattering find that, at electronic recoil energies of a few~MeV, the data are $\sim 2\sigma$ discrepant relative to the prediction of the best-fitting LMA-MSA solution. In addition, the recent measurement of the solar mass-squared difference from solar neutrino data, in particular from the day-night Super-Kamiokande data~\cite{Abe:2016nxk}, is discrepant at the $\sim 2 \sigma$ level relative to that measured by KamLAND~\cite{Gando:2010aa}. Non-standard interactions provide a possible solution to this discrepancy~\cite{Liao:2017awz}. 

Another outstanding question relates to the measured neutrino flux, and how it is able to inform the physics of the solar interior. There is a long-standing problem with standard solar models (SSMs) and predictions of the abundances of heavy elements, or metals, in the Sun. Older abundance calculations~\cite{Grevesse:1998bj} relied on many simplifying assumptions, but nevertheless fit solar observables well, in particular helioseismology data. More recent calculations however~\cite{Asplund:2009fu}, while more sophisticated in construction, were ultimately worse fits to the data~\cite{Asplund:2009fu,Scott:2014lka,Scott:2014mka,Grevesse:2014nka}. These calculations are referred to as high-Z and low-Z models respectively, according to their relative predicted metallicities. Their disagreement is known as the solar abundance problem~\cite{Serenelli:2009yc}, and has not yet been resolved. A global analysis of all solar neutrino fluxes remains inconclusive~\cite{Bergstrom:2016cbh}. A step towards resolving this problem will be to accurately measure the flux of solar neutrinos from the CNO nuclear fusion cycle, first achieved by Borexino~\cite{Agostini:2020mfq}, which is possible with the experiment discussed here (\autoref{sec:cno}).

\subsubsection{Boron-8 Solar Neutrinos (NR)}

Combined with the neutrino-electron scattering data from SNO, Super-Kamiokande and Borexino, precision measurements of $^8$B neutrino induced CE$\nu$NS in a next-generation liquid xenon detector will constrain the $\nu_e$ survival probability in the 5--15 MeV range. A significant deficit from the theoretical prediction can be interpreted as evidence of active-to-sterile neutrino oscillation~\cite{Billard:2014yka}. A next-generation liquid xenon detector will provide an independent measurement of the neutral current component of the solar $^8$B neutrino flux, with an expected event rate of $\sim 90$ events per tonne-year~\cite{Aalbers:2016jon}, measured to be right in between that predicted by the low and high metallicity Standard Solar Model~\cite{Agostini:2017cav,Agostini:2018uly,Aprile:2020thb}.

\subsubsection{Hep Solar Neutrinos (NR)}

A future next-generation detector may detect neutrinos from the minor branch of the pp chain that generates the most energetic neutrinos via the reaction $^{3}$He + p $\rightarrow ^{4}$He + e$^-$ + $\nu_e$. Along with $^{8}$B neutrinos, neutrinos from this hep reaction also undergo adiabatic conversion in the solar interior. Neutrinos from the hep reaction have not been directly identified in solar neutrino experiments; the best upper bound from the SNO experiment is $\sim 4$ times greater than the SSM prediction~\cite{Aharmim:2006wq}. 

\subsubsection{pp Solar Neutrinos (ER)}\label{sec:ppneutrinos}

The possibility to use liquid xenon as a low-energy solar neutrino detector by means of $\nu + e$ scattering was suggested in~\cite{Suzuki:2000ch} but only now is becoming a realistic measurement. A next-generation liquid xenon detector will provide a new, high-precision observation of the electronic recoil energy spectrum induced by elastic scattering of pp neutrinos, see \autoref{fig:electronrates}. This, in turn, will improve measurements of the Sun's (neutrino) luminosity. The pp neutrino flux was first indirectly identified as a component of the Gallium data, and Borexino was the first experiment to make a measurement of the spectral energy distribution of electronic recoil events induced by pp neutrinos~\cite{Smirnov:2015lxy}. The Borexino measurement uncertainty on this component is now down to $\lesssim 10\%$~\cite{Agostini:2018uly}. Further improving upon the measurement of this component will better constrain the ``neutrino luminosity'' of the Sun because pp neutrinos account for 86\% of all solar neutrino emission~\cite{Bahcall:2001pf}. Projections for a next-generation xenon experiment indicate that the pp neutrino flux can be measured to 0.15\% uncertainty with 300 tonne-years of exposure. Combined with a 1\% measurement of the next-largest component, $^7$Be, such a detector could ultimately achieve 0.2\% uncertainty in the neutrino-inferred solar luminosity~\cite{Aalbers:2020gsn}. This will also have the important consequence of constraining alternative sources of energy production in the solar interior~\cite{Newstead:2018muu}. 

\subsubsection{CNO Neutrinos (ER)}\label{sec:cno}

The flux of CNO neutrinos from the Sun makes up less than 1\% of the Sun's total neutrino luminosity but is sensitively dependent on the solar metallicity, with higher metallicity models predicting a higher CNO component. A precise measurement of the CNO flux would provide the necessary information to discriminate between the low and high-Z calculations, thereby resolving the solar abundance problem directly. The very first measurement of CNO neutrinos was achieved recently by Borexino~\cite{Agostini:2020mfq}, though with insufficient statistics to yet conclusively resolve the abundance problem. 

Due to the small CNO luminosity fraction, measuring the CNO flux in a xenon TPC will require large experimental exposures and well controlled backgrounds. A next-generation liquid xenon detector would be capable of measuring the $^{13}$N and $^{15}$O fluxes individually (20-25\%) even in the presence of the $2\nu\beta\beta$ decay background from $^{136}$Xe~\cite{Aalbers:2020gsn}. Significant improvements to the precision of these measurements can be achieved through depletion of the natural xenon target from the $^{136}$Xe isotope~\cite{Newstead:2018muu}, while negating the possibility of a $0\nu\beta\beta$ search (\autoref{sec:0nubb}). Hence, both a natural xenon target and a $^{136}$Xe-depleted target provide exciting physics opportunities for a next-generation liquid xenon detector.

\subsubsection{Neutrino Capture on Xenon-131 and Xenon-136}

Solar neutrinos may also be observed through the neutrino capture process on xenon: $\nu_e + \,^{A}_{54}\textrm{Xe} \to \,^{A}_{55}\textrm{Cs}^{(*)} + e^{-}$~\cite{Georgadze:1997zv}. The isotopes $^{131}$Xe and $^{136}$Xe have sufficiently low reaction thresholds of $Q=355$~keV and $Q=90.3$~keV for capture of all solar neutrino species. The prompt electron gives an electronic recoil with an energy that is offset from that of the captured neutrino as $E_e = E_\nu-Q-E_{\rm ex}$, where $E_{\rm ex}$ is the excitation energy of the resulting Cs nucleus.

The possibility of tagging neutrino capture events which populate excited states in the product Cs nuclei has been explored in~\cite{Haselschwardt:2020ffr}. The emission of $\gamma$-rays and/or conversion electrons during relaxation of the excited nuclear state in conjunction with the primary fast electron provides opportunities for background rejection. 

An especially high suppression of background can be achieved if a delayed coincidence signature in the Cs de-excitation could be exploited. The product isotopes $^{131}$Cs and $^{136}$Cs are unstable with half-lives 9.7~d and 13.0~d, respectively. Detection of the corresponding electron-capture and $\beta$-decay signatures which occur long after the initial capture event may also be possible. With abundances of 21.2\% and 8.9\%, one expects 0.6 and 0.7 neutrino capture events per tonne of natural Xe per year on $^{131}$Xe and $^{136}$Xe, respectively~\cite{Haselschwardt:2020ffr}.

\subsection{Atmospheric Neutrinos (NR)}\label{sec:atmnu}

The collisions of cosmic rays in the atmosphere produce neutrinos over a wide range of energies. A precise determination of this atmospheric neutrino flux depends on several factors, including the cosmic-ray flux at the top of the Earth's atmosphere, the propagation of cosmic rays through the atmosphere, and the decay of the mesons and muons as they propagate though the atmosphere to Earth's surface. Since the flavors of neutrinos that are produced in the decays are known, theoretical models accurately predict the ratio of the flavor components of neutrinos across all energies. However, the normalizations of the fluxes differ depending upon the theoretical input. 

\begin{figure}[!htbp]
    \centering
    \includegraphics[width=\columnwidth]{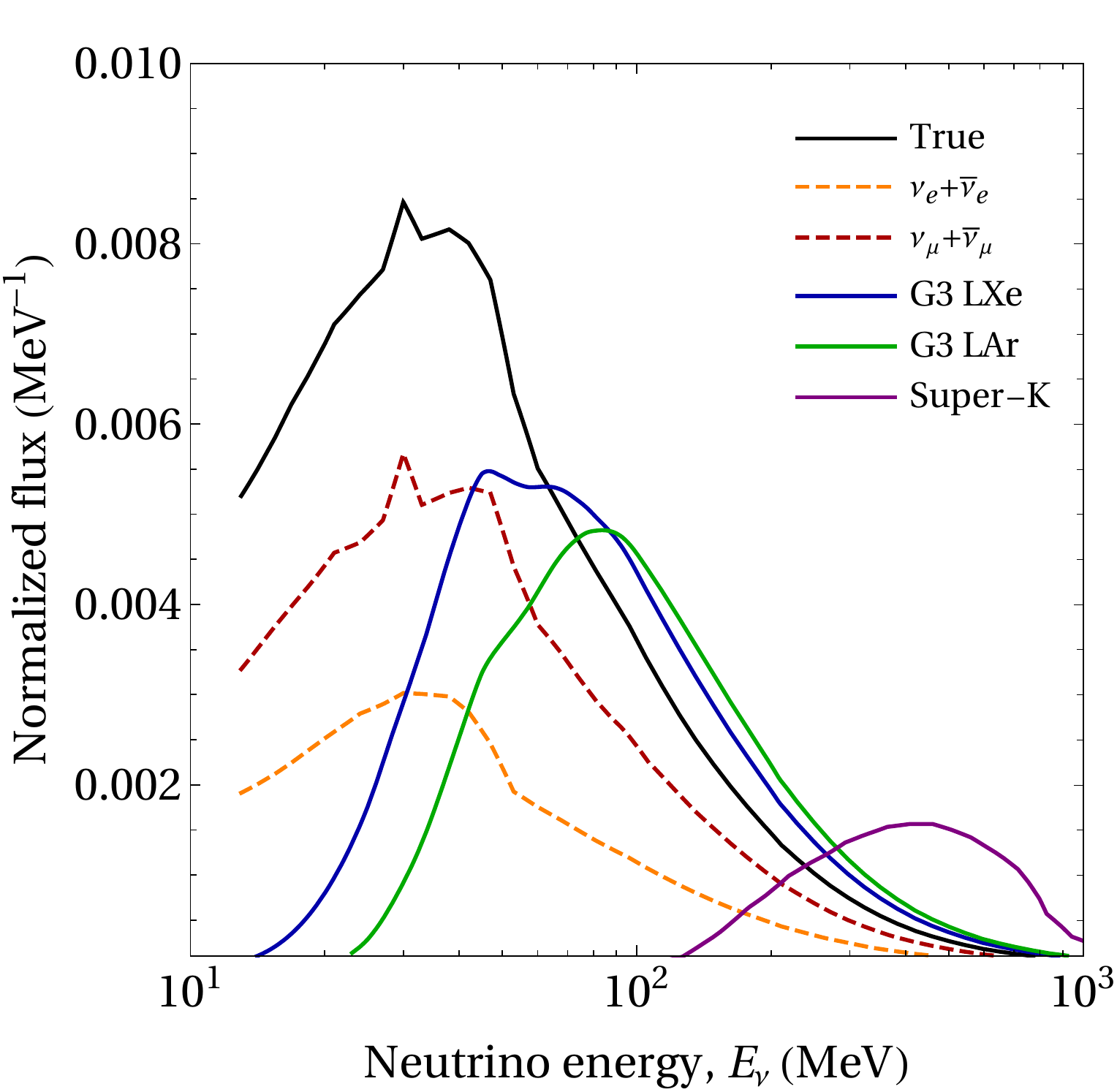}
    \caption{The differential fluxes of atmospheric neutrinos that are accessible by various experiments, normalized such that the area under the curves is equal to unity. The flux accessible to a next-generation xenon experiment (labeled G3 LXe) is shown in blue, and reaches much lower in energy than Super-Kamiokande currently does (shown as solid violet). Figure from Ref.~\cite{Newstead:2020fie}.}
    \label{fig:atmneutrino}
\end{figure}

While the atmospheric neutrino flux for energies $\gtrsim1\1{GeV}$ has been well studied by the aforementioned experiments, the low-energy flux of atmospheric neutrinos, $\lesssim100\1{MeV}$, is difficult to both model and measure~\cite{Battistoni:2005pd}. The resulting energy spectrum of neutrinos corresponds to that from muon and pion decay at rest, but the absolute normalization of the flux is less well constrained, due to uncertainties that arise from several uncertain physical processes. A next-generation dark matter detector will measure this neutrino flux at so-far unexplored low energy, see \autoref{fig:atmneutrino}. The flux at these energies is impacted by the geomagnetic field and modulated by the solar cycle, but the corresponding effects, namely a larger modulation at higher latitudes but an overall smaller flux at lower latitudes~\cite{Zhuang:2021rsg} will be challenging to discern, given the low interaction rates. In fact, measuring atmospheric neutrinos will require an exposure of order 700~tonne-years~\cite{Newstead:2020fie}, thus providing a benchmark target exposure for a next-generation liquid xenon observatory.

\subsection{Supernova Neutrinos (NR)}\label{sec:supernovaneutrinos}

The next supernova event in the Milky Way or in nearby galaxies will provide unprecedented information on the physics of neutrino propagation from the supernova core~\cite{Janka:2006fh,Janka:2012wk}. For example, large water Cherenkov detectors such as Super-Kamiokande will measure thousands of events, mostly through the charged-current inverse beta decay channel, and hundreds of events through various other elastic and inelastic channels~\cite{Krauss:1991xv,Scholberg:2012id, Baxter:2021stl}. Dark matter detectors can play an important role in supernova neutrino astrophysics through their sensitivity to supernova neutrinos via coherent elastic scattering, yielding complementary information for example on the nature of stellar collapse and the explosion energy of the supernova~\cite{Freedman:1977xn, Munoz:2021sad}.

\subsubsection{Galactic Supernova Neutrinos}

Current and future liquid xenon dark matter detectors are uniquely sensitive to neutrinos of all flavors through CE$\nu$NS~\cite{Horowitz:2003cz,XMASS:2016cmy}, whether from core-collapse (Type~II)~\cite{Chakraborty:2013zua,Lang:2016zhv} or thermonuclear runaway fusion (Type~Ia)~\cite{Raj:2019sci}. This provides a calorimetric measurement of the explosion energy going into neutrinos, independent of oscillation effects~\cite{Lang:2016zhv}. The physics available with the statistics collected by a next-generation liquid xenon detector would complement that of larger, dedicated neutrino observatories: in a next-generation detector, there are of order 100 expected events from a core-collapse supernova within 10~kpc of Earth~\cite{Lang:2016zhv}.

CE$\nu$NS is the primary detection interaction from galactic neutrinos in liquid xenon detectors, but charge current reactions are also possible. A supernova within 10~kpc could produce a handful of charge current interactions in a next-generation detector, particularly interacting with the $^{136}$Xe isotope~\cite{Pirinen:2018gsd,Ydrefors2015}. Even the large water Cherenkov veto volumes that typically surround these detectors may record notable supernova neutrino event rates~\cite{Litvinovich:2017smi}. 

Also possible are inelastic charged current interactions of the supernova electron neutrinos with the xenon nuclei. Such interactions, while creating an electron in the final state, leave the post-interaction target nucleus in an excited state. Its subsequent de-excitation produces, among other particles, gamma rays and neutrons~\cite{Bhattacharjee:2020rhs,Bhattacharjee:2020qrj}. The electron and the de-excitation gamma rays give rise to electronic recoils. On the other hand, neutrino-induced neutrons ($\nu$I$n$) from the de-excitation of the final state nucleus can create, through their multiple scattering on the xenon nuclei, additional xenon nuclear recoil events. The rate of $\nu$I$n$ nuclear recoil events is generally low compared to CE$\nu$NS nuclear recoils. However, it may still be possible to identify these $\nu$I$n$ nuclear recoil events using the capability of large liquid xenon detectors to tag neutrons which undergo multiple scatterings, both within the TPC and using the external neutron veto detector. Detection and identification of both electronic and nuclear recoils from supernova $\nu_e$ charged current interactions, together with the nuclear recoil events from neutral current CE$\nu$NS, may thus provide an additional probe of the distribution of the total supernova explosion energy going into different neutrino flavors.

Observations of astrophysical neutrinos are complementary to terrestrial experiments which are sensitive to MeV-scale neutrinos~\cite{Krauss:1991xv}. The recent detection of CE$\nu$NS has provided novel bounds on new physics, for example in the form of kinetic mixing, hidden sector models, flavor models, and sterile neutrinos~\cite{Akimov:2017ade}. Future measurements of supernova neutrinos at dark matter detection experiments can improve on this sensitivity~\cite{Suliga:2020jfa}, providing further information on new physics models (see also \autoref{sec:otherstuff}).

\subsubsection{Pre-Supernova Neutrinos}

\begin{figure}[!htbp]
    \centering
    \includegraphics[width=\columnwidth]{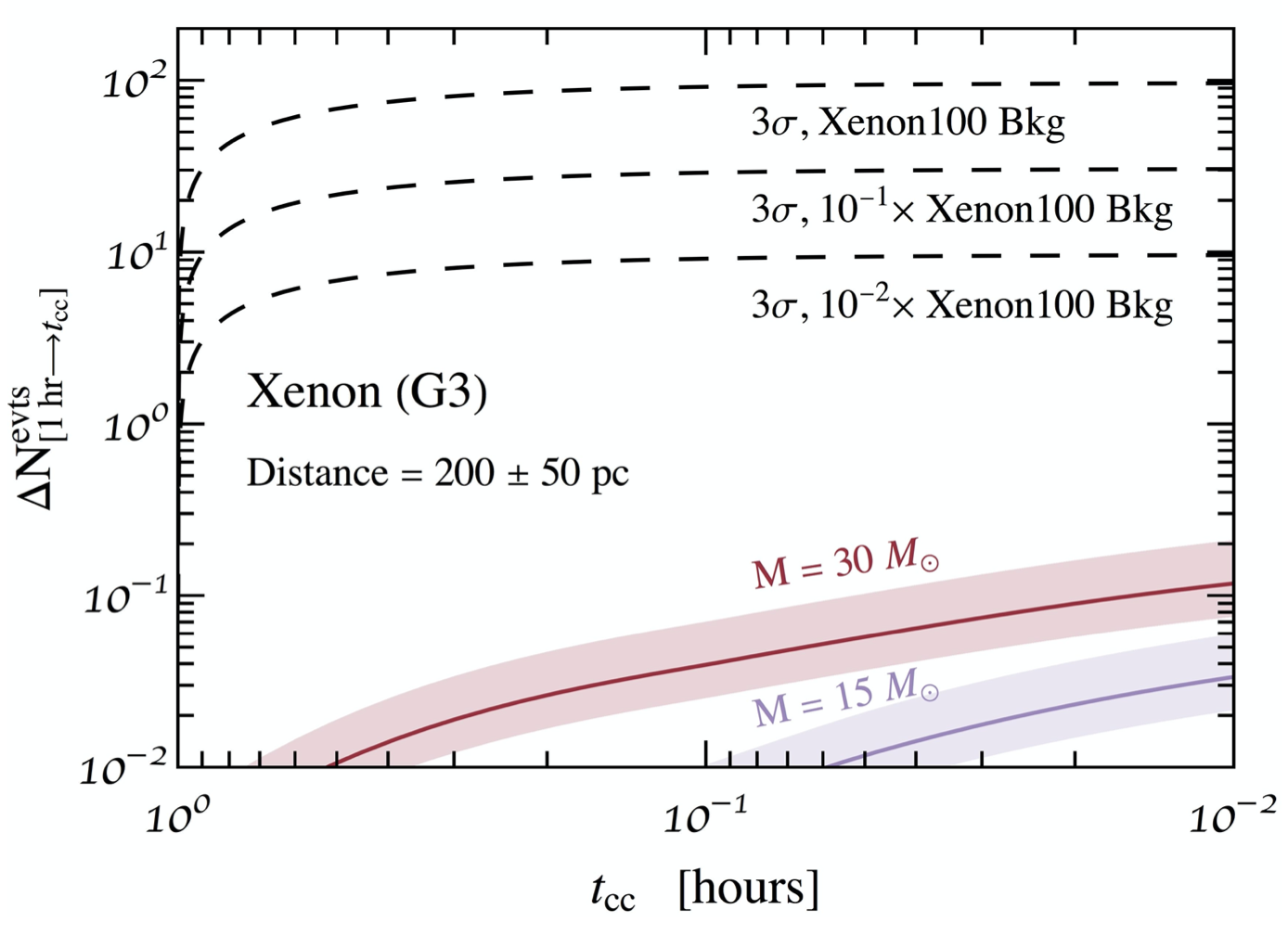}
    \caption{For a next-generation liquid xenon dark matter experiment with an assumed target mass of 50~tonnes, the expected number of pre-supernova neutrinos above the detection threshold is shown as function of time until the core collapse, for two different stellar masses at a distance of 200~pc. Figure from Ref.~\cite{Raj:2019wpy}.}
    \label{fig:pre_sn}
\end{figure}

In the event of a near-Earth ($d <$~kpc) core-collapse supernova, future liquid xenon detectors will also be sensitive to neutrinos of all flavors that are emitted by a massive star in its silicon-burning stage, a few hours \textit{prior to} core collapse~\cite{Odrzywolek:2003vn,Odrzywolek:2004em}. Due to lower stellar temperatures before collapse, these ``pre-supernova" neutrinos are $\mathcal{O}$(10) softer than supernova neutrinos, and therefore require low thresholds for detection~\cite{Kato:2020hlc}. \autoref{fig:pre_sn} indicates that a next-generation liquid xenon experiment operating at 0.1 keV energy threshold would detect, in a 12 hour window prior to collapse, $\mathcal{O}$(100) pre-supernova neutrinos from a massive star 200\,pc away, e.g.~Betelgeuse~\cite{Raj:2019wpy}. Such a detection would constitute the first measurement of the final stages of stellar evolution, and provide a valuable warning before the explosion. Pre-Supernova neutrinos can also help constrain dark photon, axion, and ALPs parameters~\cite{Sieverding:2021jfa, Mori:2021muf, Ge:2020zww}. 

\subsubsection{Supernova Early Warning System} 

In order to be optimally prepared for the next supernova, the Supernova Early Warning System (SNEWS) was developed~\cite{snewsweb}. SNEWS is an inter-experiment network to prepare and provide an early warning for Galactic supernovae: in contrast to the optical signal, neutrinos basically free-stream from the collapsing star and thus reach Earth minutes, hours or even days before the optical counterpart becomes visible. Therefore, by detecting supernova neutrinos, an early alert can be sent to astronomers to facilitate early observations of the Supernova~\cite{Antonioli:2004zb}. SNEWS is in the process of being revamped and amplified to SNEWS2.0 which will have a larger physics reach~\cite{Kharusi:2020ovw,Baxter:2021aaa}. The next-generation detector discussed here will be able to contribute to this network.

\subsubsection{Diffuse Supernova Neutrinos}

In addition to the yield from a Galactic supernova event, an exciting prospect is the detection of the diffuse supernova neutrino background (DSNB)~\cite{Krauss:1983zn,Lunardini:2010ab,Beacom:2010kk}, i.e.~the neutrinos emitted from past supernovae occurring across the universe. Modern predictions put this flux at approximately $6\1{cm^{-2}}\1s^{-1}$~\cite{Horiuchi:2008jz} for neutrino energies above 17.3~MeV, including contributions from all neutrino flavors. In addition to being a probe on supernova physics, the diffuse supernova neutrino background is an independent probe of the local core-collapse supernova and cosmic star formation rate~\cite{Krauss:1983zn,Hopkins:2006bw}. Although this signal has not yet been directly detected, there are strong upper bounds on the $\bar{\nu}_e$ component of the flux from Super-Kamiokande~\cite{Bays:2011si}. The best predictions for the flux of all flavors, with an expected event rate of $\sim 0.05$ events per tonne-year, implies that liquid xenon dark matter detectors with exposures $\sim 1000$~tonne-year may have discovery potential to this signal above known backgrounds~\cite{Strigari:2009bq, Suliga:2021hek}. Besides, these detectors might be the only ones capable of probing the non-electron component of the DSNB, thanks to their excellent sensitivity.		

\subsection{Terrestrial Antineutrinos (ER)}\label{sec:terrestrialantineutrinos}

The Earth is a rich source of antineutrinos with energies in the MeV range, due to radioactive decays in Earth's crust and interior~\cite{Krauss:1983zn}.  A signal from these geoneutrinos has been measured at Kamland~\cite{Araki:2005qa} and Borexino~\cite{Borexino:2010dli}.  Coherent neutrino interactions with xenon will produce recoil energies that are likely below experimental thresholds. However, depending on exposure time, mass, and other backgrounds, several neutrino-electron scattering events may be detectable in a next-generation xenon detector.  

\subsection{Other Neutrino Physics}

\subsubsection{Measuring the Weinberg Angle}

The solar pp flux is very strongly determined by the luminosity constraint on the total neutrino flux, to a precision of $\sim 0.4$\%~\cite{Bergstrom:2016cbh}. The dependence of the neutrino-electron cross section on the Weinberg (weak) angle $\sin^2\theta_W$ thus allows for an independent measurement of this quantity, at energies far below the reach of colliders. Precision determinations of $\sin^2\theta_W$ must be made by running LEP measurements (at $\sim$100\,GeV) down to lower energies. At present, the lowest-energy determination of $\sin^2 \theta_W$ remains above the MeV~scale~\cite{Bouchiat:1983uf}. Electronic recoils from pp neutrinos yield an exchanged momentum on the order of $\sim$~keV, so a detection of the pp~flux via electronic recoils in next-generation xenon experiments will cover new and uncharted territory. The next-generation xenon detector discussed here would be able to constrain $\sin^2 \theta_W$ with (4--5)\% precision~\cite{Aalbers:2020gsn} even without any additional constraints from other experiments. Alternatively, using the solar luminosity condition, a liquid xenon experiment with a 200~tonne-year exposure can already yield a measurement of $\sin^2 \theta_W$ with a precision of 1.5\% at the keV scale~\cite{Cerdeno:2016sfi}. This is complementary to measurements using CE$\nu$NS of pion-decay neutrinos as achieved by the COHERENT collaboration~\cite{Cadeddu:2019eta}. Applying the Relativistic Random Phase Approximation correction (see \autoref{sec:ewinteraction} and~\cite{Chen:2016eab}), the expected electronic recoil rate from solar pp neutrinos is $\sim 90$~counts per 1000~tonne-day in the (0--15)~keV energy range (or 780~counts per 1000~tonne-day in the full energy range). Hence, a 150~tonne-year exposure can reduce the statistical uncertainty in the measurement of $\sin^2 \theta_W$ down to 1.4\% for the energy transfer in the range of (0--15)~keVee. 

\begin{figure}[!htbp]
    \centering
    \includegraphics[width=\columnwidth]{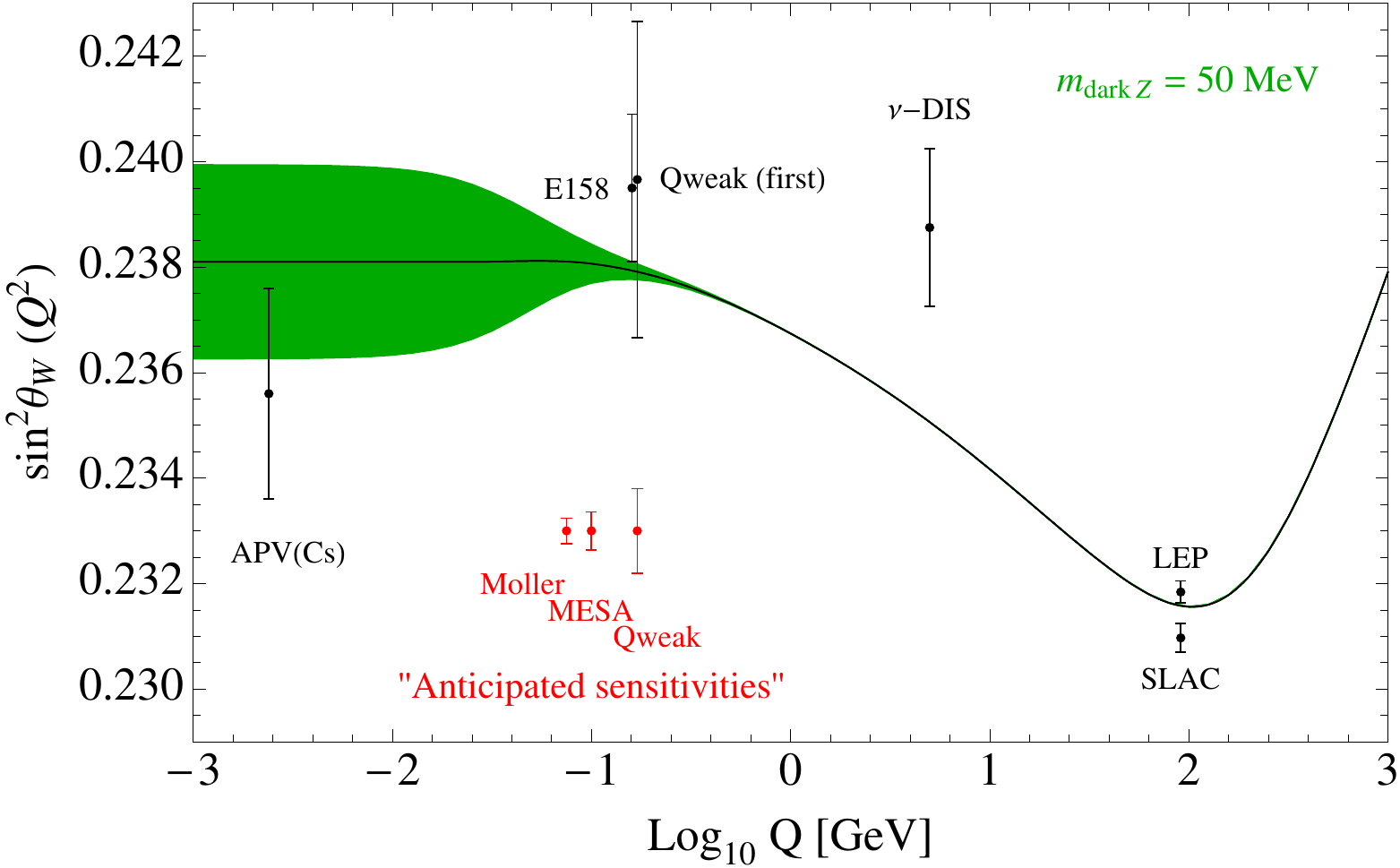}
    \caption{Running of the Weinberg angle $\sin^2 \theta_W$ as a function of momentum scale $Q^2$, along with measured values. A deviation from Standard Model predictions (black line) could indicate the presence of new physics effects. The green band indicates the effect of a new $Z'$ with $m_{Z'} = 50 \,\textrm{MeV}$, where the width of the band is determined by the strength of the kinetic mixing parameter with $U(1)_Y$. An $\mathcal{O}$(tonne-year) xenon dark matter observatory can extend the reach of these measurements down to the keV~scale, significantly to the left of this plot, via the measurement of the pp solar neutrino flux. Figure from Ref.~\cite{Davoudiasl:2014kua}.}
    \label{fig:weinberg}
\end{figure}

A deviation of $\sin^2 \theta_W$ from the computed value at low energies would be an indication of new physics. For example, a new light gauge boson could lead to a different value at low momentum $Q^2$. \autoref{fig:weinberg} shows an example of the variation that could be produced by a 50~MeV $Z'$-mediator, with a coupling in the range required to simultaneously explain the muon $(g - 2)_\mu$ anomaly~\cite{Davoudiasl:2014kua,Aoyama:2020ynm,Muong-2:2021ojo}.

\subsubsection{Electron-Type Neutrino Survival Probability}

\begin{figure}[!htbp]
    \centering
    \includegraphics[width=\columnwidth]{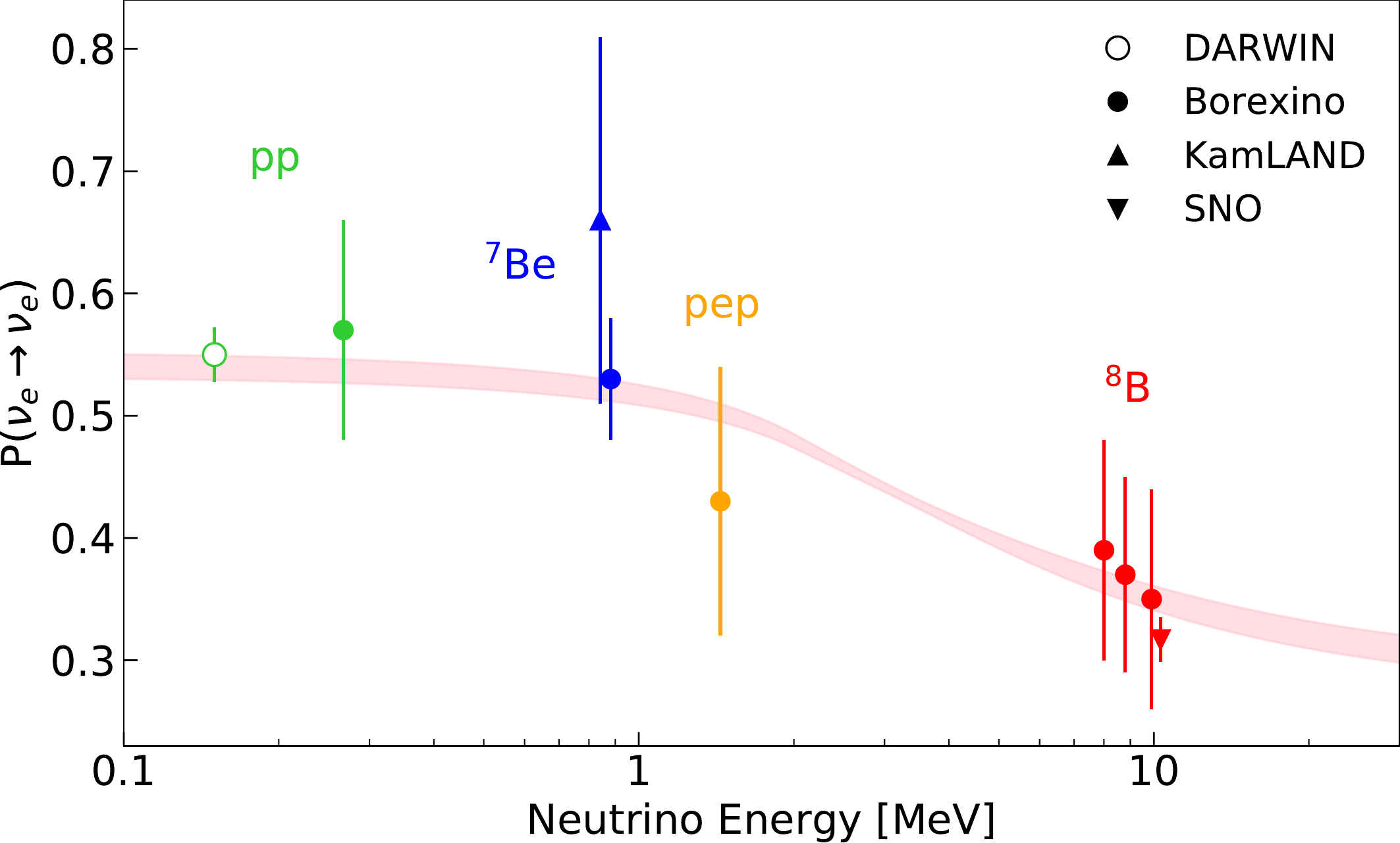}
    \caption{The $\nu_e$ survival probability versus neutrino energy, assuming the high-Z SSM. Dots represent the solar measurements of pp (green), $^7$Be (blue), pep (orange), and $^8$B (red) from Borexino. The upward (downward) triangle shows a measurement of $^7$Be ($^8$B) from KamLAND (SNO). The open point indicates that a next-generation liquid xenon experiment could enhance the precision of the $\nu_e$ survival probability to 0.02 below 200\,keV, using solar pp neutrino events. The pink band represents the 1$\sigma$ prediction of the MSW-LMA solution. Figure from Ref.~\cite{Aalbers:2020gsn}.}
    \label{fig:nusurvival}
\end{figure}

The total electron-neutrino scattering rate receives neutral-current contributions from all three flavors, but charge-current contributions only from the electron-type neutrino. Consequently, a high-statistics observation of solar pp neutrinos enables a liquid xenon experiment to directly measure the oscillation probability of the electron-type neutrinos emitted from the Sun in an energy range that is not accessible to any other experiment. \autoref{fig:nusurvival} shows that with an exposure of 300\,tonnes-years, a liquid xenon detector would measure the low-energy survival probability to 3--4\%~\cite{Aalbers:2020gsn}. Such a measurement would serve as a test of the MSW-LMA (Mikheyev-Smirnov-Wolfenstein Large Mixing Angle) solution of neutrino oscillation and a probe of exotic neutrino properties and non-standard interactions. This can also be used to perform a solar-neutrino-only measurement of the magnitude of the $U_{e3}$ entry of the neutrino mixing matrix, to search for very light sterile neutrinos in currently unexplored regions of parameter space, and one can extend the sensitivity to the hypothesis that neutrinos are pseudo-Dirac fermions by an order of magnitude~\cite{deGouvea:2021ymm}.

\subsubsection{Searching for New Physics of Neutrinos}

A next-generation liquid xenon detector will also be a powerful tool to search for new physics of neutrinos via elastic neutrino-electron scattering. An extensively-studied scenario of new physics of neutrinos is the so-called non-standard interaction (NSI)~\cite{Dev:2019anc}, which might play a potentially important role in future long-baseline experiments such as DUNE~\cite{deGouvea:2015ndi}. In addition, there has also been rising interest in new interactions mediated by light mediators~\cite{Datta:2017ezo,AristizabalSierra:2020edu,Khan:2020vaf,Fernandez-Moroni:2021nap}. A next-generation liquid xenon detector can significantly improve the sensitivity to sterile neutrino mixing parameters, particularly if solar neutrino detection via elastic neutrino-electron scattering (pp component) and the CE$\nu$NS channel (\textsuperscript{8}B component) is combined with reactor neutrino data from JUNO~\cite{Goldhagen:2021kxe}. Notably, the correlation of the mixing angles $\sin^2 \theta_{12}$ and $\sin^2 \theta_{14}$ can be broken by a combined analysis of these complementary data sets.

It has been shown that when combined with a radioactive source, a multi-tonne-scale liquid xenon detector can significantly improve current bounds on leptonic NSIs~\cite{Link:2019pbm} and light mediators, thanks to the high electron density in liquid xenon. In addition, xenon nuclei lie in a range where radiative corrections are particularly sensitive to new weak isospin conserving processes from new physics and are insensitive to isospin violating processes~\cite{Krauss:1991ba}.  Considering solar neutrinos as the source, since the Borexino experiment has demonstrated excellent sensitivities to such new interactions~\cite{Khan:2019jvr, Kamada:2015era}, especially to $\nu_{\tau}$ interactions, it is expected that a next-generation liquid xenon detector will be superior in searching for new physics of neutrinos~\cite{Dutta:2020che}.

\section{Additional Physics Channels}\label{sec:otherstuff}

At the time of writing, an excess of electronic recoil events below $7\1{keV}$ has been reported by XENON1T~\cite{Aprile:2020tmw}. With a statistical significance of about 3\,$\sigma$, this excess has received enormous interest from the community~\cite{Abe:2021ocf,Abellan:2020pmw,Aboubrahim:2020iwb,Alhazmi:2020fju,Alonso-Alvarez:2020cdv,Amaral:2020tga,An:2020bxd,An:2020tcg,Anchordoqui:2020tlp,Arcadi:2020zni,ArguellesDelgado:2021lek,Arias-Aragon:2020qtn,Arias:2020tzl,AristizabalSierra:2020edu,AristizabalSierra:2020zod,Athron:2020maw,Babu:2020ivd,Babu:2021jnu,Baek:2020owl,Baek:2021yos,Bally:2020yid,Baryakhtar:2020rwy,Baym:2020riw,Bell:2020bes,Benakli:2020vng,Bhattacherjee:2020qmv,Bloch:2020uzh,Boehm:2020ltd,Borah:2020jzi,Borah:2020smw,Borah:2021jzu,Bramante:2020zos,Brdar:2020quo,Buch:2020mrg,Budnik:2020nwz,Buttazzo:2020vfs,Cai:2020kfq,Cao:2020bwd,Cao:2020oxq,Chakraborty:2020vec,Chala:2020pbn,Chao:2020yro,Chen:2020gcl,Chen:2021qao,Chen:2021uuw,Chiang:2020hgb,Chigusa:2020bgq,Choi:2020kch,Choi:2020udy,Choi:2020ysq,Choudhury:2020xui,Coloma:2020voz,Croon:2020ehi,Davighi:2020vap,Davoudiasl:2020ypv,DelleRose:2020pbh,Dent:2020jhf,DeRocco:2020xdt,Dessert:2020vxy,Dey:2020sai,DiLuzio:2020jjp,Dror:2020czw,Du:2020ldo,Du:2020ybt,Dutta:2021nsy,Dutta:2021wbn,Ema:2020fit,Escribano:2020wua,Farzan:2020llg,Fayet:2020bmb,Fonseca:2020pjs,Foot:2020ehn,Fornal:2020npv,Gao:2020wer,Gao:2020wfr,Ge:2020jfn,Guo:2020oum,Han:2020dwo,Harigaya:2020ckz,Harnik:2020ugb,Haselschwardt:2020iey,Hayen:2020mod,He:2020sat,He:2020wjs,Hoof:2021mld,Hryczuk:2020jhi,Ibe:2020dly,Ilie:2021iyh,Inan:2020kif,Jaeckel:2020oet,Jeong:2021ivd,Jho:2020sku,Jia:2020omh,Kahlhoefer:2020gkz,Kannike:2020agf,Karmakar:2020rbi,Karozas:2020pun,Keung:2020uew,Khan:2020csx,Khan:2020pso,Khan:2020vaf,Khruschov:2020cnf,Kim:2020aua,Ko:2020gdg,Lee:2020wmh,Li:2020naa,Lin:2020mhx,Lindner:2020kko,Long:2020uyf,McKeen:2020vpf,Miranda:2020kwy,Nakayama:2020ikz,Okada:2020evk,Paz:2020pbc,Robinson:2020gfu,Seymour:2020yle,Shakeri:2020wvk,Shoemaker:2020kji,Smirnov:2020zwf,Straniero:2020iyi,Studenikin:2021fai,Su:2020zny,Sun:2020iim,Szydagis:2020isq,Takahashi:2020bpq,Takahashi:2020uio,Tan:2021nif,Vagnozzi:2021quy,VanDong:2020bkg,Xu:2020qsy,Ye:2021zso,Zhang:2020htl,Zhou:2020bvf,Zioutas:2020cul,Zu:2020bsx,Zu:2020idx}. We refrain here from discussing whether one or the other explanation is more likely and instead mention the various explanations in the respective sections of this review.

\subsection{Solar Axions}\label{sec:solar_axions}

Originally postulated to resolve the strong CP problem in QCD~\cite{PecceiQuinn_1977,Weinberg:1977ma,Wilczek:1977pj}, axions have emerged as a suitable non-baryonic dark matter candidate \cite{Preskill:1982cy,Abbott:1982af,Dine:1982ah,Ipser:1983mw,Duffy:2009ig}. As such, there has been a growing interest in the last few decades to search for axion particles in general, and for axion dark matter in particular~\cite{Krauss:1985ub,Hagmann:1990tj,Sikivie:1999sy,Raffelt:2006rj,Arvanitaki:2009fg,Arik:2011rx,Du:2018uak}. They may be sought in the dark matter galactic halo within which they would cluster~\cite{Hagmann:1998cb} as a cold dark matter axion. 

Independently of being dark matter, if an axion or axion-like particle exists in nature, then it should be produced copiously in the hot solar plasma~\cite{Krauss:1984gm,Dimopoulos:1986kc}. Due to the $\sim$keV temperature of the Sun, solar axions are produced with roughly thermal fluxes in the $1-10\1{keV}$ energy range, and are thus well-suited for detection in xenon experiments. Via their coupling to the photon, $g_{a\gamma}$, the most widely considered process of axion production is Primakoff conversion in which photons convert into axions inside the electromagnetic fields of the electrons and ions of the solar plasma. This flux is dominant in hadronic QCD axion models like the ``KSVZ'' axion~\cite{Kim:1979if,Shifman:1979if}. Another widely considered QCD axion model labelled the ``DFSZ'' axion~\cite{Dine:1981rt,Zhitnitsky:1980tq} possesses a tree-level coupling to electrons, $g_{ae}$, which brings sizable fluxes from the so-called ``ABC'' processes: atomic recombination and deexcitation, Bremsstrahlung, and Compton scattering~\cite{Redondo:2013wwa}. 

The primary way for xenon experiments to measure the axion is through the axioelectric effect~\cite{Derevianko:2010kz}, which allows constraints to be set on $g_{ae}$. Xenon experiments may also constrain $g_{a\gamma}$: both by measuring the Primakoff component~\cite{Pirmakoff:1951pj} of the solar flux (which is dependent only on $g_{a\gamma}$), as well as by exploiting inverse Primakoff conversion inside the detector~\cite{Dent:2020jhf,Gao:2020wer}: $a Z \rightarrow \gamma Z$. In the latter case, the sensitivity to solar axions is boosted, even if the value of $g_{a\gamma}$is small. A final component of the solar axion flux beyond ``ABC'' and Primakoff components is the $^{57}$Fe axion-nucleon interaction, which depends on $g_{an}$~\cite{Moriyama:1995bz}. A next-generation xenon experiment with a $\sim$1000~tonne-year exposure may even be able to out-perform devoted solar axion telescopes~\cite{Dent:2020jhf} such as the planned International Axion Observatory (IAXO)~\cite{Armengaud:2019uso}.

The electronic recoil background level of the detector is the main limiting factor for its sensitivity to solar axions. Liquid xenon TPCs are well known for their very low electronic recoil background levels and are therefore ideal for this search. Amongst underground detectors, liquid xenon TPCs place the strongest constraints to-date on $g_{ae}$ with solar axions~\cite{Ahmed:2009ht,Abe:2012ut,Aprile:2014eoa,Akerib:2017uem,Fu:2017lfc,Aprile:2020tmw}. 

The excess of electronic recoil events seen in XENON1T~\cite{Aprile:2020tmw} has a spectrum that matches the expected solar axion flux. However, the amplitude of the excess would require large couplings that would place the excess in conflict with more stringent astrophysical bounds~\cite{Capozzi:2020cbu,Athron:2020maw,Li:2020naa,Croon:2020ehi}. The proposed next-generation liquid xenon TPC will enable this excess to be robustly tested, should it persist, perhaps leading to the discovery of solar axions.

\subsection{Neutrino Dipole Moments and Light Mediators}

Dark matter searches start to probe various novel neutrino-induced signals, see e.g.~\cite{Billard:2013qya,Link:2019pbm}. Therefore, the interpretation of potential discoveries as coming from new neutrino physics becomes increasingly plausible. As a result, next-generation dark matter detectors will be capable of placing interesting limits on models of new physics in the neutrino sector, often complementary with other experiments~\cite{Dutta:2020che}. 

\begin{figure}[!htbp]
\centering
\includegraphics[width=0.47\textwidth]{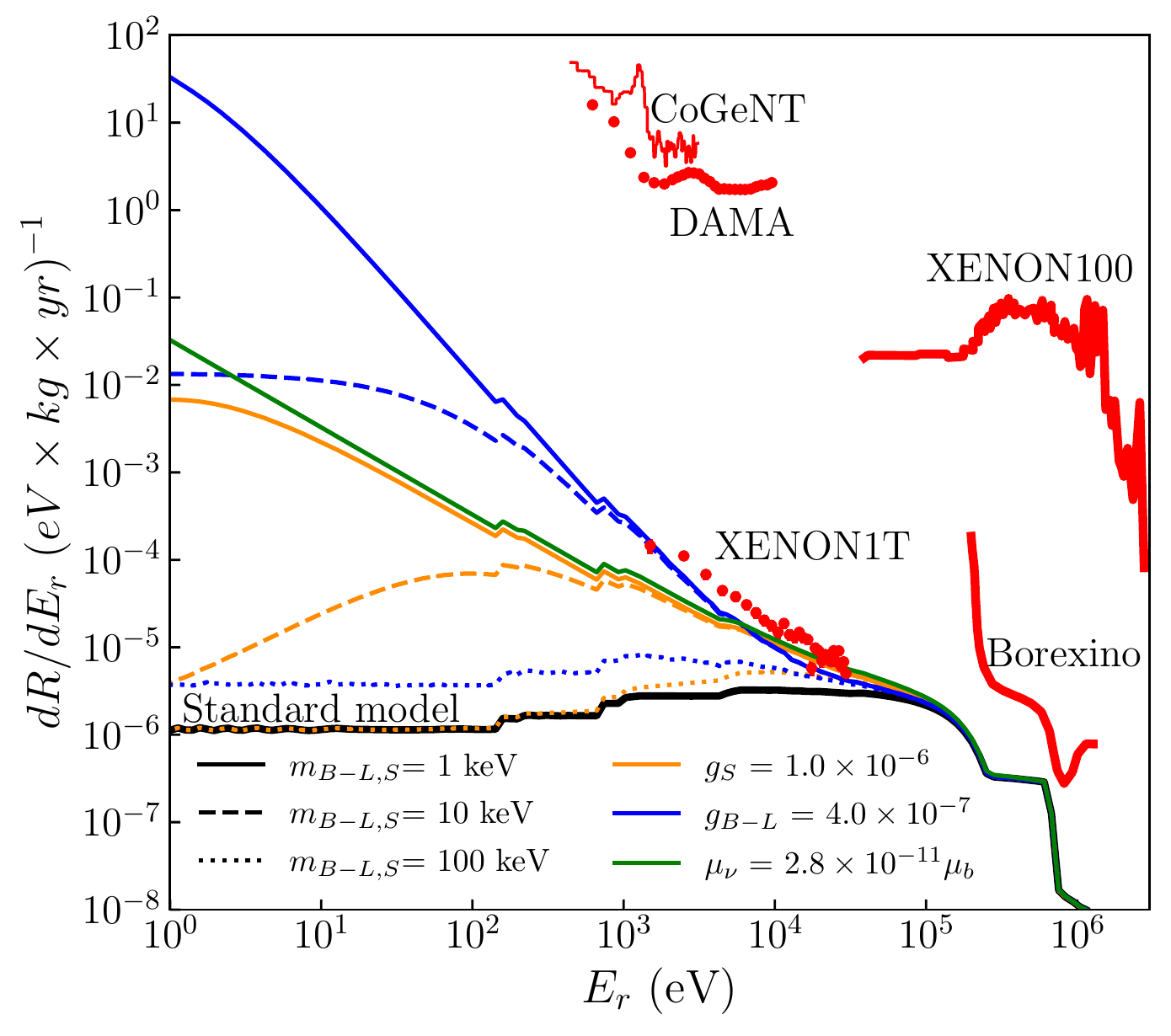}
\caption{\label{fig:NuInDM} Neutrinos may show up in dark matter experiments well above the neutrino fog. Shown in red are the electron recoil spectra in several experiments taken from Ref.~\cite{Harnik:2012ni, Schwemberger:2022det}, with the background level in XENON1T indicated~\cite{Aprile:2017aty,Aprile:2020tmw}. The spectrum expected from Standard Model solar neutrinos is in solid black. The colored curves are the solar neutrino spectra for several new physics models discussed in the text, with line-styles corresponding to various mediator masses. The jagged steps below $\sim5\1{keV}$ are an effect of the electron binding energy as discussed in~\cite{Chen:2017plb}.}
\end{figure}

This is apparent in limits from electronic recoils. In \autoref{fig:NuInDM} we show the observed electronic recoil spectrum observed by several dark matter as well as neutrino experiments, adopted from Refs.~\cite{Harnik:2012ni, Schwemberger:2022det} with the most recent XENON1T measurements~\cite{Aprile:2017aty} included. They represent about two orders of magnitude improvement over the XENON100 background rate. 

The next generation of experiments will have further sensitivity~\cite{Jeong:2021ivd,Babu:2021jnu}. In \autoref{fig:NuInDM} we show several spectra from new physics models which lead to an enhanced scattering rate at low energies. The green curve shows the recoil spectrum in the case that the neutrino possesses a magnetic dipole moment around that which is allowed by current solar neutrino data from Borexino~\cite{Borexino:2017fbd}. In this case the differential cross section is
\begin{equation}
\frac{\mathrm{d}\sigma}{\mathrm{d}E_r}= \mu_\nu^2 \alpha \left(\frac{1}{E_r}-\frac{1}{E_\nu}\right)
\end{equation}
where $\mu_\nu$ is the neutrino dipole moment and $E_r$ is the recoil energy of the electron. At high recoil energies, the dipole-induced scattering is lower than the Standard Model rate and in agreement with the Borexino rate. However, due to the $E_r^{-1}$ falloff, the rate is higher at low recoil energies. Already an analysis of XENON1T~\cite{Aprile:2020tmw} improves the limit in dipole moments to $<3\times 10^{-11}$ times a Bohr Magneton. The next-generation experiment will precisely measure the $pp$ solar neutrino spectrum at low energies and thus further improve this sensitivity. In addition to the neutrino magnetic moment, a new interaction mediated by a scalar propagator will also fall as $E_r^{-1}$ as described in~\cite{Schwemberger:2022det}. An example of such a scalar-mediated interaction is plotted in orange in \autoref{fig:NuInDM}.

One can also consider models with a faster falling spectrum. For example, blue curves of \autoref{fig:NuInDM} are the spectra in a model with a new very light $B-L$ gauge boson which is mediating a new interaction between neutrinos and electrons. The cross section is
\begin{equation}
\frac{\mathrm{d}\sigma}{\mathrm{d}E_r}
=\frac{g_{B-L}^4 m_e}{4\pi (2m_e^2E_r^2 + m_{B-L}^2)^2}\,,
\end{equation}
where $m_{B-L}$ and $g_{B-L}$ are the mass and coupling. Here, we have dropped subleading terms in $E_r/E_\nu$ as well as interference with the SM process which is unimportant at most recoil energies. If the mass of the gauge boson is small, the cross section falls as $E_r^{-2}$. This behavior is due to the $1/(q^2-m_{B-L}^2)$ propagator in the amplitude, with $q^2=2m_e E_r$. Again it can be seen that a next-generation experiment will have significant sensitivity, well beyond that achieved by the Borexino experiment~\cite{Bellini:2011rx}, the GEMMA reactor experiment~\cite{Beda:2009kx}, or the XMASS experiment~\cite{Abe:2020nwr}. In fact, the discussion around the possible excess observed by XENON1T~\cite{Aprile:2020tmw} can already be used to place a constraint on $g_{B-L}<3.6\times 10^{-7}$ for mediators with mass $m_{B-L}<10\1{keV}$. This is already comparable with the constraint from GEMMA~\cite{Boehm:2020ltd}.  A next-generation liquid xenon experiment will be able to strengthen this bound.

\begin{figure}[!htbp]
\centering
\includegraphics[width=0.4\textwidth]{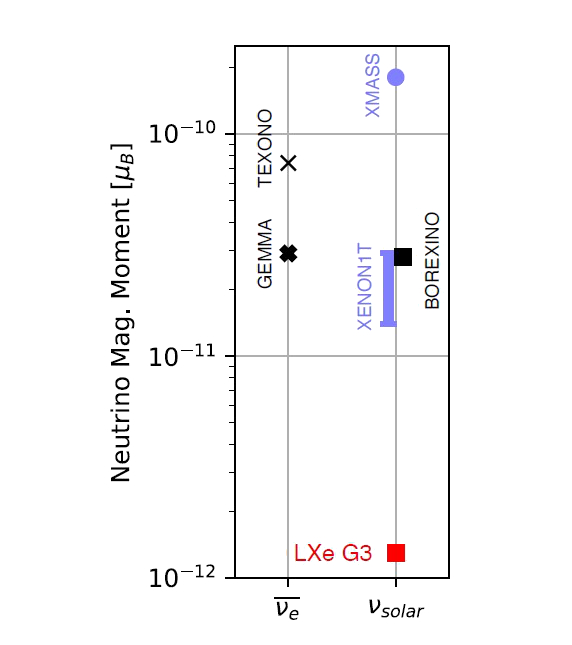}
\caption{\label{fig:NuMagMomSens} Projected neutrino magnetic moment sensitivity (red) along with current limits from reactor-based experiments (left markers) and experiments exposed to a solar flavor mixture (right markers). All the upper limits are reported at 90\% CL, except the XENON1T result, which shows the 10-90\% confidence interval~\cite{Abe:2020nwr,Borexino:2017fbd,Aprile:2020tmw}.}
\end{figure}

It is interesting to consider a scenario in which the next generation of xenon experiments uncovers an excess above the solar neutrino fog. In this case we will immediately entertain both the possibility of dark matter and that of new neutrino physics, as evidenced by the list of papers discussing the excess observed by XENON1T. Fortunately, this degeneracy can be disentangled with reactor neutrino experiments. To those that stand within 100~m of the core, nuclear reactors are a brighter source of neutrinos than the Sun. A low-threshold detector near a reactor, such as GEMMA~\cite{Beda:2009kx}, can thus place strong limits or distinguish whether an excess is coming from dark matter or neutrinos. \autoref{fig:NuMagMomSens} shows the current limits on the neutrino magnetic moment from both large underground detectors and reactor experiments, as well as the projected sensitivity for a next-generation liquid xenon detector with a 750~tonne-year exposure, complementary to dedicated experiments such as CONNIE~\cite{Aguilar-Arevalo:2016khx}. 

In addition to the physics discussed in \autoref{sec:shortbaseline}, a mono-energetic neutrino source in combination with a large xenon detector could set very competitive sensitivities to neutrino magnetic dipole moments and sterile neutrino oscillation~\cite{Coloma:2014hka}. A 50-day run with a 3~MCi $^{51}Cr$ source produces 82~$\nu-e$ elastic scattering events in the XENON1T detector, which corresponds to 9.8 times better signal-to-noise ratio compared to using solar neutrinos~\cite{Coloma:2020voz}. The next-generation experiment discussed here could set much more stringent bounds on the neutrino magnetic moment, if combined with an electron capture source. 

\subsection{Fractionally Charged Particles}

The quantization of the electric charge has been one of the long-standing mysteries stemming from empirical observation. In principle, the Standard Model $U(1)$ allows arbitrarily small real-number charge, but so far, experiments indicate that there is a fundamental unit of electric charge of 1/3~e. This has sparked theoretical explanations including Dirac quantization~\cite{Dirac:1931kp} and provides one of the major motivations for a Grand Unification Theory (GUT)~\cite{Pati:1973uk,Georgi:1974sy}. The search for such fractionally charged or millicharged particles is a test of the paradigm of charge quantization~\cite{Dobroliubov:1989mr,Prinz:1998ua,Davidson:2000hf,Prinz:2001qz,Golowich:1986tj,Babu:1993yh,Gninenko:2006fi,CMS:2012xi,Agnese:2014vxh,Haas:2014dda,Ball:2016zrp,Alvis:2018yte,Magill:2018tbb,Kelly:2018brz,Berlin:2018bsc}. If such a particle was found, its small charge may or may not be the new electric charge unit, but in either case it will inevitably change our understanding of the current charge quantization built on quark charges, and contradict the predictions of certain GUTs.

One can consider the kinetic mixing between the SM $U(1)_Y$ and an additional gauge group $U(1)_D$, with additional matter particles $\xi$ charged under a dark $U(1)_D$. In the limit when the dark $U(1)_D$ vector boson (often called a dark photon, see \autoref{sec:dark_photon}) is massless, the would-be dark sector particles which are charged under $U(1)_D$ become electromagnetically ``fractionally charged'' or ``millicharged''. The level of kinetic mixing is often $\sim 10^{-3}$ or smaller, from loop effects in either QFT or String Theory~\cite{Holdom:1985ag,Dienes:1996zr,Abel:2008ai}, which naturally gives small electric charges. For masses of such new particles below $\sim {\rm MeV}/c^2$, the limits on the kinetic mixing parameter (and thus the fractional charges) are stringent $<10^{-15}$~\cite{Davidson:2000hf}. For heavier states, the limits are weaker~\cite{Dubovsky:2003yn}. Direct detection experiments can possibly observe bound state formation between $q<0$ millicharged particles and nuclei~\cite{Pospelov:2020ktu}. One can also look for millicharged particles without massless gauge bosons in regimes where the dark photon is constrained~\cite{Kelly:2018brz}. A search for such a particle can be a test of GUTs and certain string compactification scenarios~\cite{Shiu:2013wxa}. Further, liquid xenon detectors are sensitive to the possible millicharge of solar neutrinos~\cite{Abe:2020nwr}.

\subsection{Nucleon Decay} 

In the Standard Model, the conservation of baryon number $B$ is an empirically observed symmetry. If $B$ were an exactly conserved quantum number, then protons, being the lightest baryons, would be stable. However, baryon number could be an approximate symmetry of Nature, and violated by small amounts, as predicted for example by many Grand Unified Theories. This could explain the observed matter-antimatter asymmetry of the universe~\cite{Babu:2013jba}. 

Several liquid xenon detectors, such as DAMA-LXe~\cite{Bernabei:2000xp,Bernabei:2006tw} and EXO-200~\cite{Albert:2017qto}, explored the possibility to investigate nucleon decay through so-called invisible decay modes, where the final states (neutrinos, or more exotic particles such as dark fermions) are not detected. One example for an invisible mode is $n \rightarrow \nu\nu\nu$, as proposed in~\cite{Mohapatra:2002ug}. Following such a decay, the daughter nuclei would be left in an excited state, and would emit a detectable signal, such as a $\gamma$-ray, once they de-excite. \autoref{tab_Xe} illustrates the various signatures for two xenon isotopes, $^{129}$Xe and $^{136}$Xe. These decays can be searched-for with a next-generation liquid xenon detector with unprecedented sensitivity.

\begin{table}[!htbp]
\footnotesize\centering
\begin{tabular}{lcll}
        & Invisible &          &  \\
        & decay     &          &  \\
Isotope & mode      & Daughter & Subsequent decays \\
\midrule
$^{129}$Xe & n   & $^{128}$Xe & stable                                                     \\
           & p   & $^{128}$I  & $^{128}$I $\rightarrow[Q = 2.217]{\text{$\beta^{-}$}}$ $^{128}$Xe \\
           &     &            & or $^{128}$I $\rightarrow[Q = 1.258]{\text{EC+$\beta^{+}$}}$ $^{128}$Te     \\
           & nn  & $^{127}$Xe & $^{127}$Xe $\rightarrow[Q = 0.664]{\text{EC}}$ $^{127}$I              \\
           & pn  & $^{127}$I  & stable                                                      \\
           & pp  & $^{127}$Te & $^{127}$Te $\rightarrow[Q = 0.694]{\text{$\beta^{-}$}}$ $^{127}$I      \\
\midrule
$^{136}$Xe & n   & $^{135}$Xe & $^{135}$Xe $\rightarrow[Q = 1.151]{\text{$\beta^{-}$}}$ $^{135}$Cs \\
           & p   & $^{135}$I  & $^{135}$I $\rightarrow[Q = 2.648]{\text{$\beta^{-}$}}$ $^{135}$Xe \\
           &     &            & \phantom{$^{135}$I} $\rightarrow[Q = 1.151]{\text{$\beta^{-}$}}$ $^{135}$Cs  \\
           & nn  & $^{134}$Xe & stable \\
           & np  & $^{134}$I  & $^{134}$I $\rightarrow[Q = 4.175]{\text{$\beta^{-}$}}$ $^{134}$Xe \\
           & pp  & $^{134}$Te & $^{134}$Te $\rightarrow[Q = 1.550]{\text{$\beta^{-}$}}$ $^{134}$I \\
           &     &            & \phantom{$^{134}$Te} $\rightarrow[Q = 4.175]{\text{$\beta^{-}$}}$ $^{134}$Xe  \\
           & nnn & $^{133}$Xe & $^{133}$Xe $\rightarrow[Q = 0.4274]{\text{$\beta^{-}$}}$ $^{133}$Cs \\
           & nnp & $^{133}$I  & $^{133}$I $\rightarrow[Q = 1.770]{\text{$\beta^{-}$}}$ $^{133}$Xe \\
           &     &            & \phantom{$^{133}$I} $\rightarrow[Q = 0.4274]{\text{$\beta^{-}$}}$ $^{133}$Cs \\
           & npp & $^{133}$Te & $^{133}$Te $\rightarrow[Q = 2.920]{\text{$\beta^{-}$}}$ $^{133}$I \\
           &     &            & \phantom{$^{133}$Te} $\rightarrow[Q = 1.770]{\text{$\beta^{-}$}}$ $^{133}$Xe \\
           &     &            & \phantom{$^{133}$Te} $\rightarrow[Q = 0.4274]{\text{$\beta^{-}$}}$ $^{133}$Cs \\
           & ppp & $^{133}$Sb & $^{133}$Sb $\rightarrow[Q = 4.003]{\text{$\beta^{-}$}}$ $^{133}$Te \\
           &     &            & \phantom{$^{133}$Sb} $\rightarrow[Q = 2.920]{\text{$\beta^{-}$}}$ $^{133}$I \\
           &     &            & \phantom{$^{133}$Sb} $\rightarrow[Q = 1.770]{\text{$\beta^{-}$}}$ $^{133}$Xe \\
           &     &            & \phantom{$^{133}$Sb} $\rightarrow[Q = 0.4274]{\text{$\beta^{-}$}}$ $^{133}$Cs \\
\end{tabular}
\caption{The daughter isotopes and their decay modes that follow the invisible mono- and di-nucleon decays of $^{129}$Xe and $^{136}$Xe as well as the tri-nucleon decays of $^{136}$Xe. This table is adapted after \cite{Bernabei:2000xp,Bernabei:2006tw}. The Q-values are reported in MeV.}
\label{tab_Xe}
\normalsize
\end{table}

\subsection{Short-Baseline Oscillations}\label{sec:shortbaseline}

Persistent anomalies in short baseline experiments, including LSND and MiniBooNE, are suggestive of an additional undiscovered neutrino mass eigenstate at the $\sim1\1{eV}$ mass scale~\cite{Abazajian:2017tcc}. However, there is significant tension between different experiments that has yet to be explained~\cite{Diaz:2019fwt}. Given the energy of $^{51}$Cr neutrinos, the oscillation pattern is expected to be within a meter-scale detector. This would make a next-generation liquid xenon TPCs well-suited to conclusively test the existence of sterile neutrinos~\cite{Coloma:2014hka}. In addition, such an experiment would be able to rule out portions of currently allowed parameter space, potentially resolving the existing tension if sterile neutrinos do not exist~\cite{Coloma:2014hka}.

\section{Background Considerations}\label{sec:backgrounds} 

As discussed in this present paper, the proposed next-generation liquid xenon experiment is a versatile observatory for a number of relevant science channels, spanning low-energy nuclear recoils in particular for dark matter, electronic recoils for a number of measurements, and reaching up to high energy events expected from neutrinoless double-beta decay. In order to support this broad physics reach, multiple background sources must be considered. In addition to improved xenon purification, further scrutiny of materials in assays and exploration of discrimination techniques will be necessary to minimize backgrounds for rare event searches. The choice of host facility and detector design are also important considerations that will impact which and how backgrounds manifest. Modelling of detector performance and simulations of background events will be critical in informing the design of the experiment, deriving sensitivities, and ultimately in achieving final science results~\cite{Undagoitia:2015gya}.

\subsection{Underground Laboratories}

\begin{figure}[!htbp]
\begin{center}
\includegraphics[width=0.99\columnwidth,clip,trim=20 10 60 50]{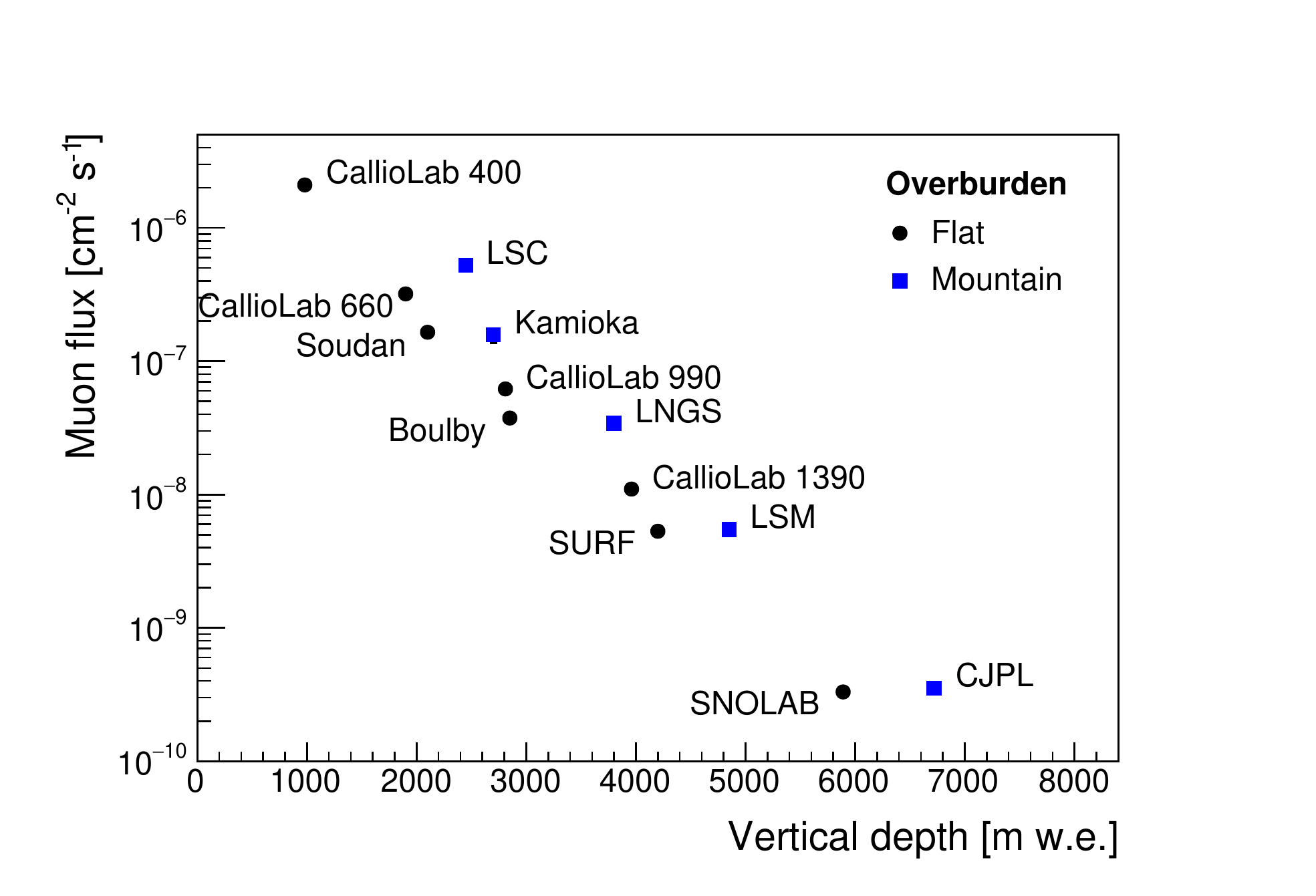}
\caption{Depth-dependent muon flux at various underground laboratories (measured in “meters water equivalent” (m.w.e.). While the depth increases from 620 m.w.e. to 6720 m.w.e, the muon flux decreases by more than four orders of magnitude. The data points represent the following measurements: CallioLab (Pyh\"{a}salmi, Finland) at various depths~\cite{Polaczek-Grelik:2020brg}, LSC (Canfranc, Spain)~\cite{Trzaska:2019kuk} (with depth taken from~\cite{Morales:2005}), Soudan (Minnesota, USA)~\cite{Zhang:2014jsq}, Kamioka (Japan)~\cite{Super-Kamiokande:2015xra} (conversion from muon rate to flux based on simulations from~\cite{Tang:2006uu}), Boulby (UK) at 1,100~m level (2850~m~w.e.)~\cite{Reichhart:2013xkd}, LNGS (Gran Sasso, Italy)~\cite{Borexino:2018pev}, SURF (South Dakota, USA)~\cite{MAJORANA:2016ifg} (depth taken from~\cite{Cherry:1983dp}), LSM (Modane, France)~\cite{FREJUS:1989lko}, SNOLAB (Sudbury, Canada)~\cite{SNO:2009oor}, and Jingping (China)~\cite{JNE:2020bwn}.
}\label{fig:muonflux}
\end{center}
\end{figure}

Muons traversing detectors or surrounding materials will induce primary backgrounds as well as secondary neutrons and cosmogenic backgrounds from activation of materials~\cite{Mei:2005gm,Kudryavtsev:2008fi}. Dark matter detectors are thus deployed in deep underground laboratories, where cosmic-ray muon backgrounds are greatly reduced by the rock overburden~\cite{Bettini:2014tva}. Nevertheless, for high-sensitivity experiments, active muon shielding is still required in order to tag remaining muon-related background events and reduce the muon-induced background to a negligible level compared to other sources. Muon fluxes at the typical underground laboratories range from 1~muon/m$^2$/hour at the Laboratori Nationali del Gran Sasso  (LNGS, 3,100 meters water equivalent deep)~\cite{Mei:2005gm,Votano:2012fr} to about 5~muons/m$^2$/month at China's Jinping underground laboratory (CJPL, 6,720 meters water equivalent deep)~\cite{Yu-Cheng:2013iaa, Cheng:2018lcf} (\autoref{fig:muonflux}).

There are a host of underground laboratories that can be considered as location for the next-generation observatory discussed here. This includes the laboratories hosting the current generation of liquid xenon detectors: LNGS~\cite{Bettini:2007xc} (location of XENONnT), CJPL~\cite{Li:2014rca} (location of PandaX-4T) and SURF~\cite{Heise:2017rpu} (location of LZ). The Boulby Underground Laboratory has a similar muon flux to LNGS with notably low radon levels~\cite{Paling:2015ss}. The Modane Underground Laboratory~\cite{Piquemal:2012fs} includes a facility for radon-free air, and the entire scientific campus of SNOLAB~\cite{Lawson:2012sga} is equipped as a cleanroom with some of the lowest available muon flux. Several of these laboratories entertain feasibility studies for creating additional underground space for scientific use. All in all, there is a favorable outlook that suitable underground space can be made available for the next-generation liquid xenon observatory. While radioactive and muon-induced backgrounds are dominated by the rock composition and overburden, respectively, the varying geomagnetic field at different latitudes has a (modest) impact on the science that can be done with atmospheric neutrinos~\cite{Zhuang:2021rsg}.

\subsection{Fiducialization}\label{sec:erfiducialization}

The dominant background component in liquid xenon detectors at a given energy range has evolved with increasing detector size. In earlier detectors (e.g.~XENON100~\cite{Aprile:2011vb}, LUX and PandaX-I/II), the gamma radiation from radioactive contaminants of detector construction materials contributed significantly to the electronic recoil background for dark matter searches in the keV energy range. A fiducial volume selection is typically applied to reduce these backgrounds, which predominantly appear towards the boundaries of the bulk xenon: with larger detector masses and hence smaller surface-to-volume ratios, fiducialization can preserve a higher proportion of the active volume for a physics search. Gamma-induced electronic recoils will remain a significant background for rare event searches at the MeV scale such as the $0\nu \beta \beta$-decay of $^{136}$Xe (\autoref{sec:0nubb}, \autoref{fig:0vbb_signal}).

Radioactive contaminants of detector materials are also the source of radiogenic neutrons through spontaneous fission or $(\alpha,\mathrm{n})$ reactions. Fiducialization of the liquid xenon target is not quite as effective for neutrons compared to gamma radiation. Therefore, current and future liquid xenon detectors are equipped with dedicated neutron veto systems for efficient mitigation of nuclear recoil background events.

\subsection{Material Selection}

The selection of materials featuring the lowest contamination with radioactive impurities is the most important strategy for background mitigation in current and future liquid xenon experiments \cite{Aprile:2017ilq,Akerib:2020com}. Trace amounts of uranium and thorium can be detected by means of highly sensitive gamma-spectrometers, Inductively-Coupled Plasma Mass Spectrometry (ICPMS) measurements or neutron activation analysis. Radon emanation rates are determined in dedicated setups where the radon which is emanating from the sample accumulates in an ambient carrier gas, before the radon activity in this sample gas is measured, typically using proportional counters or electrostatic radon monitors~\cite{Akerib:2020com,Aprile:2020vmn}. 

Various material samples are measured in intensive screening campaigns in order to pre-select and built radiopure detector components which fit the requirements for a next-generation liquid xenon experiment~\cite{XENON:2015ara}. Once the materials are selected, the screening measurements are used for background modeling by means of simulation. Precise knowledge of emanation sources can further be used to optimize the online purification systems.

\subsection{Intrinsic Background Mitigation}

Sources of so-called intrinsic backgrounds are typically radioactive noble gases which are homogeneously mixed within the liquid xenon target. Thus, any type of shielding remains ineffective. Trace amounts of $^{85}$Kr that stay in the xenon during its distillation from air will cause, if not removed, a low-energy electronic recoil background from its beta decay. Due to the absence of krypton sources within the detector, the $^{85}$Kr contamination is constant over time and scales with the liquid xenon mass. $^{85}$Kr thus needs to be removed through cryogenic separation of the xenon target, as pioneered by the XMASS collaboration~\cite{Abe:2008py}. The purification of xenon from trace amounts of $^{85}$Kr has been successfully demonstrated using both cryogenic distillation~\cite{Abe:2008py,Wang:2014ehv,Aprile:2016xhi} and adsorption~\cite{Bolozdynya:1997,Akerib:2016hcd} as separation techniques. Cryogenic distillation in particular is appropriate to process large amount of xenon gas before being filled into the experiment, but also an online krypton purification at a running detector has been demonstrated. Starting with a Kr-nat contamination of several ppm in commercial xenon, a purification to $(360\pm60)\1{ppq}$ was achieved in XENON1T using the online krypton purification~\cite{XENON:2021fkt}. The lowest concentration to-date was measured in the outlet of the XENON1T distillation system to be below $26\1{ppq}$ (90\% CL)~\cite{Aprile:2016xhi}, using an enhanced Rare Gas Mass Spectrometer (RGMS) with a sensitivity of $8\1{ppq}$~\cite{Lindemann:2013kna}. This level is well below even the requirements of a next-generation liquid xenon experiment.

$^{222}$Rn is not immanent in the xenon gas, but continuously emanates from surfaces of detector materials. Due to its subsequent beta decays, radon is the dominant background source in current liquid xenon detectors. A smaller surface-to-volume ratio will naturally decrease the radon concentration in next-generation liquid xenon detectors. However, further mitigation strategies are needed to achieve a level of about $0.1\1{\mu Bq/kg}$ that is required in order to render radon-induced backgrounds sub-dominant versus the irreducible contributions from neutrino signals. Once a new detector has been built, its emanation rate of $^{222}$Rn is set and expected to be constant over the lifetime of the experiment. A further suppression of the radon induced background can be achieved through continuous purification of the xenon target. The key for an efficient radon removal is a good separation technique and a high purification flow which revolves the entire xenon target fast with respect to the 3.8\,d half-life of radon. Radon removal based on cryogenic distillation has been successfully tested in large scale liquid xenon experiments~\cite{Aprile:2017kop} and will be used also in XENONnT. Radon purification systems designed for small purification flows can also significantly reduce the radon concentration in xenon. Since the dominating radon emanation sources in an experiment are known from screening, dedicated purge flows towards the radon removal system can prevent radon to enter the liquid xenon target~\cite{Akerib:2020com}.

Another intrinsic background source is the decay of $^{137}$Xe. It is naturally created inside the xenon target through activation by muon-induced neutrons. Thus, the $^{137}$Xe induced background strongly depends on the muon rate at the experimental site.

\subsection{Isolated Light and Charge Signals and Accidental Coincidences} 

Due to the large electroluminiscence gain that is exploited in dual-phase liquid xenon TPCs, even a single extracted electron can produce a detectable S2 of tens of photoelectrons in size~\cite{Burenkov:2009zz,Santos:2011ju,Angle:2011th,Aprile:2013blg,Edwards:2007nj,Edwards:2017emx,Xu:2019dqb}. A standalone search with S2s (i.e.~without requiring an accompanying S1 signal, see \autoref{sec:s2only} on page~\pageref{sec:s2only}) can thus lower the energy threshold, improving the reach to low-mass WIMPs, solar axions and solar neutrinos, and other physics that have associated low-energy recoil signatures~\cite{Aprile:2019xxb}. However, this sensitivity to any process that can release even single electrons from their shell brings in additional instrumental backgrounds. Several sources of S2-only, single and few-electron backgrounds are known and being further investigated~\cite{Akerib:2020jud, Kopec:2021ccm, Bodnia:2021flk, Akimov:2016rbs, Aprile:2013blg, Sorensen:2017kpl, Sorensen:2017ymt, Tomas:2018pny, XENON:2021myl}. Photoionization backgrounds caused by large S2s die away within a maximum drift time after the S2~\cite{Aprile:2013blg}. Emission from metal surfaces would be evident in specific locations that could be avoided with positional cuts. The most impactful background appears to be S2s up to five electrons in size that continue for times up to seconds after a large S2. The rates of these correlated small S2s, which appear in the same location as a previous large S2, decrease according to a power law with time after the large S2~\cite{Kopec:2021ccm,Akerib:2020jud,XENON:2021myl}. This background can be mitigated with positional and temporal cuts after large S2s.

\begin{figure}[!htbp]
\begin{center}
\includegraphics[width=0.95\columnwidth]{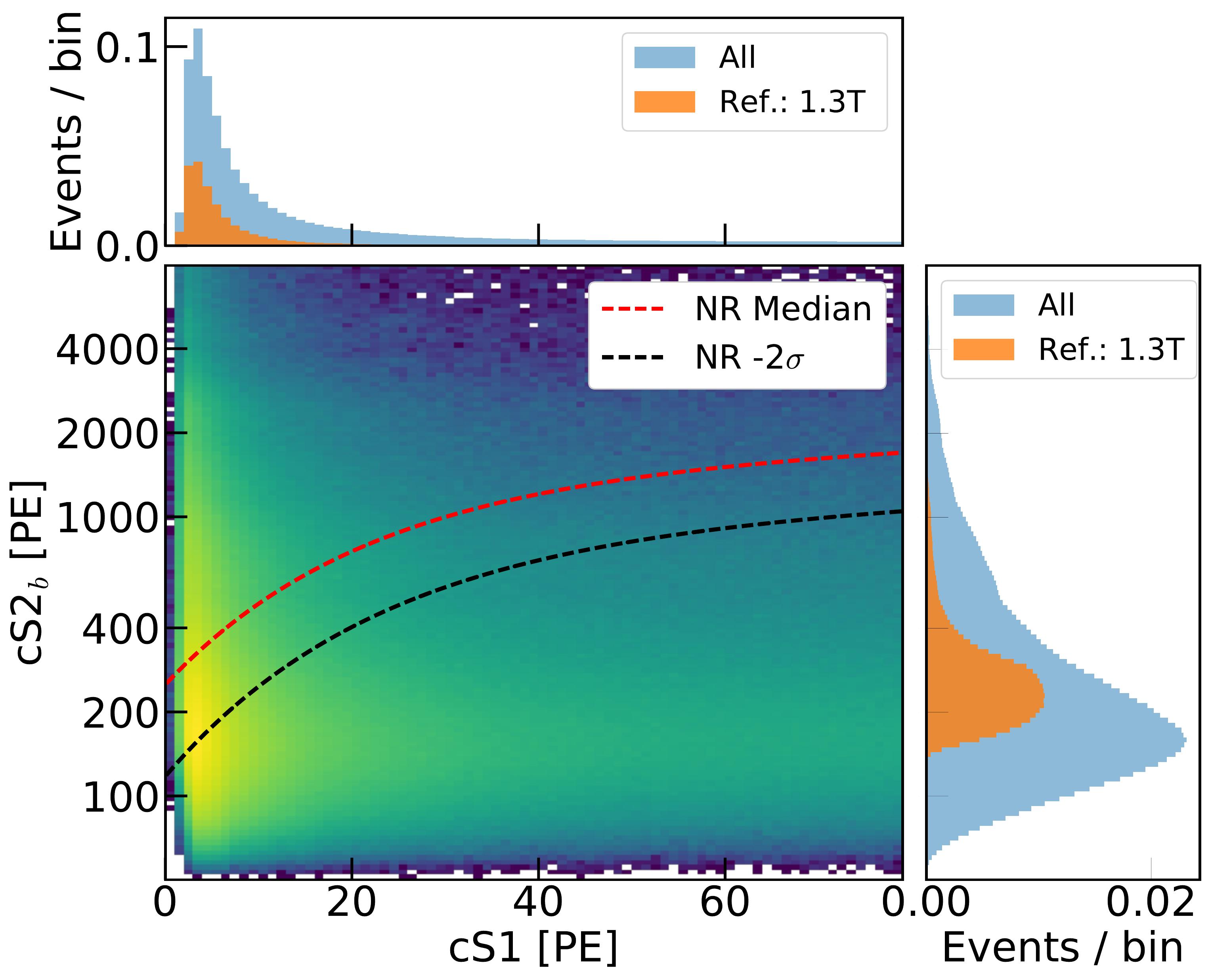}
\caption{Illustration of the accidental coincidence background distribution from XENON1T in cS1 and log10(cS2b), with projections on each axis showing the expected distribution within the entire analysis space (blue), and in the reference region for 1.3~tonne fiducial volume. The reference region lies between the nuclear recoil median and $-2\sigma$ quantile lines, marked by red and black lines, respectively. Figure from Ref.~\cite{Aprile:2019dme}.}\label{fig:accidental}
\end{center}
\end{figure}

S1-only backgrounds also exist due to interactions in areas insensitive to the charge channel. One such origin of lone S1 events is from outside of the main drift field region of the TPC, notably below the cathode. Another origin is from volumes where charges are depleted, or where electrons can not reach the extraction region. Most notably this can be as a result of the field configuration towards the edges of the detector~\cite{Akerib:2016vxi}. Unrelated, isolated S1-only and S2-only signals can be close enough in time to be mis-identified as a single event. Such accidental coincidences of instrumental backgrounds can thus mimic a physics interaction for a conventional search in S1-S2 phase space. As both lone S1s and S2s are more likely to manifest at smaller signal values, these accidental coincidence backgrounds are particularly problematic for the WIMP search region of interest. The absolute incidence of such accidental events should increase in a next-generation detector, as the corresponding surfaces and volumes from which lone S1s and S2s can arise become larger. However, their impact on the WIMP search will decrease with increasing detector size, due to the favorable surface-to-volume ration. Hence, combining a data-driven approach, as adopted for XENON1T (\autoref{fig:accidental}), with detailed detector simulations, should be sufficient to characterise this background for a next-generation experiment.

\subsection{Monte-Carlo Simulation of Backgrounds}

\subsubsection{Background Model} 

To construct a model of expected backgrounds, the activities and normalizations found from material assays and physics estimates must be paired with the corresponding detection efficiencies of the associated events. These efficiencies are determined through Monte Carlo simulations of event primaries, such as daughters from radioactive decay, within a realistic representation of the experiment. Simulations are used to determine the energy depositions from backgrounds within the detector, and in translating them into observed signals.

A framework based on the GEANT4 toolkit~\cite{Agostinelli:2002hh} allows for the tracking of particles within a rendering of the detector geometry. Custom additions can enhance the modelling of various physics processes and phenomena beyond the standard physics lists available, as in the case of modelling neutron captures on Gd using ANNRI~\cite{Hagiwara:2019ptep,Tanaka:2020ptep} or DICEBOX~\cite{Becvar:1998sd} derived outcomes for an improved veto assessment. Bespoke event generators enable the simulation and study of more involved scenarios that are not well-captured in default GEANT4, such as $(\alpha,\mathrm{n})$ reactions accompanied by a varying multiplicity of gammas, or events from atmospheric muons which penetrate the laboratory rock overburden~\cite{Akerib:2021ap}.

Analysis cuts can be applied to remove events with coincident scatters in veto detectors, and restrict to an energy region of interest and/or a fiducial volume. Persistent background counts can then be compared to those anticipated from potential signals. This approach has been used in predicting the background burden of many present-generation experiments, forming the basis of WIMP sensitivity estimates for XENONnT~\cite{Aprile:2020vtw}, LZ~\cite{Akerib:2018lyp} and PandaX-4T~\cite{Zhang:2018xdp}. 

\subsubsection{Generation of S1 and S2 Signals}

\begin{figure*}[!htbp]
\includegraphics[width=2.05\columnwidth]{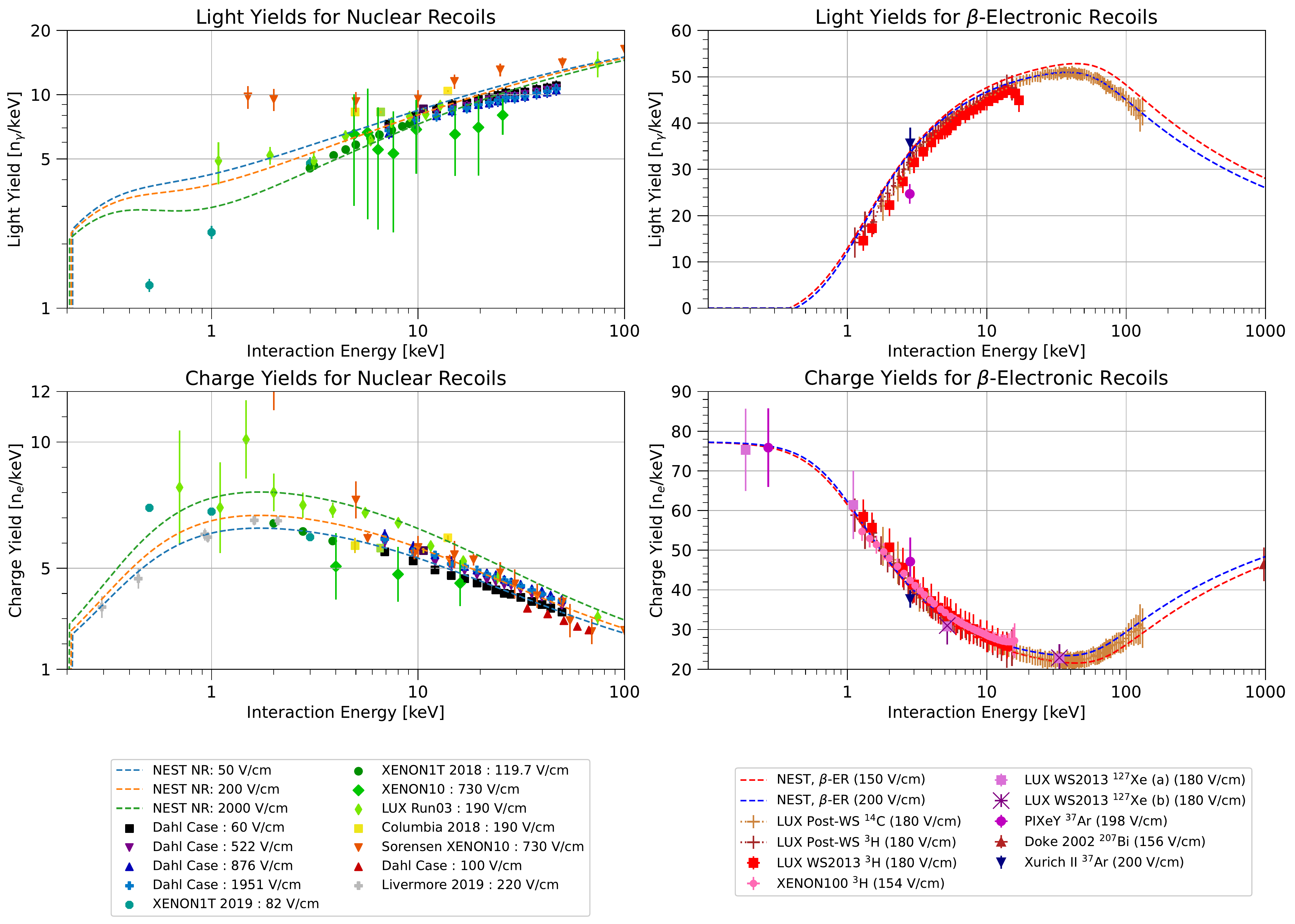}
\caption{The light and charge yields for nuclear and electronic recoils, as measured by various experiments and as modeled by the Noble Element Simulation Technique (NEST) v2.3.5~\cite{szydagis_m_2022_6028483}. The light (charge) yield is defined as the number of photons (electrons) leaving the recoil site after electron-ion recombination, per unit energy. For electronic recoils, NEST has two models for $\beta$-induced and $\gamma$-induced recoils, respectively, and we show the $\beta$ model. Correspondingly, we only show experimental measurements from $\beta$ calibrations or low-energy line sources, which are observed to fit the $\beta$ model better than the $\gamma$ model.
The nuclear recoil data points are from E.~Dahl's thesis~\cite{Dahl:2009nta}, XENON1T~\cite{Aprile:2019dme}, XENON10~\cite{Sorensen:2008ec, Sorensen:2010hq, Angle:2011th, Sorensen:2010hv}, LUX Run~3 (WS2013)~\cite{Akerib:2016mzi}, and dedicated xenon TPCs at Columbia University~\cite{Aprile:2018jvg}, Case Western Reserve University~\cite{Aprile:2006kx}, and Lawrence Livermore National Lab~\cite{Lenardo:2019fcn}. The electronic recoil data points are from LUX~\cite{Akerib:2019jtm, Akerib:2015wdi, Akerib:2017hph, Akerib:2016qlr}, XENON100~\cite{Aprile:2017xxh}, PIXeY~\cite{Boulton:2017hub}, Xurich~II~\cite{Baudis:2020nwe}, and a paper by Doke et~al.~\cite{Doke:2002oab}.
}
\label{fig:NEST_Light_and_Charge_Yields}
\end{figure*}

Generally, energy depositions are converted to observable S1s and S2s to construct PDFs for background components and signal for likelihood-based analysis~\cite{Aprile:2011hx}. The microphysics behind the interactions of particles with the active xenon is captured by the Noble Element Simulation Technique (NEST)~\cite{Szydagis:2011tk, Szydagis:2013sih, Mock:2013ila, Lenardo:2014cva, szydagis_m_2022_6028483}. NEST offers a comprehensive and mature framework to simulate the atomic and nuclear physics of energy deposition and the resulting detector response. Using world data from previous experiments including LUX, XENON, PIXeY~\cite{Singh:2019nrd,Bodnia:2021flk}, neriX~\cite{Plante:2011hw,Goetzke:2016lfg}, ZEPLIN-III~\cite{Araujo:2020rwg}, and Xurich~\cite{Baudis:2017xov,Baudis:2020nwe}, the NEST collaboration has developed models for the light and charge yields of various interactions. These models are semi-empirical and reproduce calibration data from alphas, betas, gammas, nuclear recoils, and exotic interactions like the two-step internal conversion of $^{\text{83m}}$Kr. The excellent agreement between the NEST models and data can be seen in \autoref{fig:NEST_Light_and_Charge_Yields}. Physicists on a next-generation liquid xenon experiment are thus able to take advantage of NEST to accurately simulate the signals induced by both signals and backgrounds, including their S1, S2, position, and pulse timing. 

\subsection{Discrimination}

\begin{figure}[!htbp]
\begin{center}
\includegraphics[width=0.99\columnwidth,clip,trim=100 20 110 40]{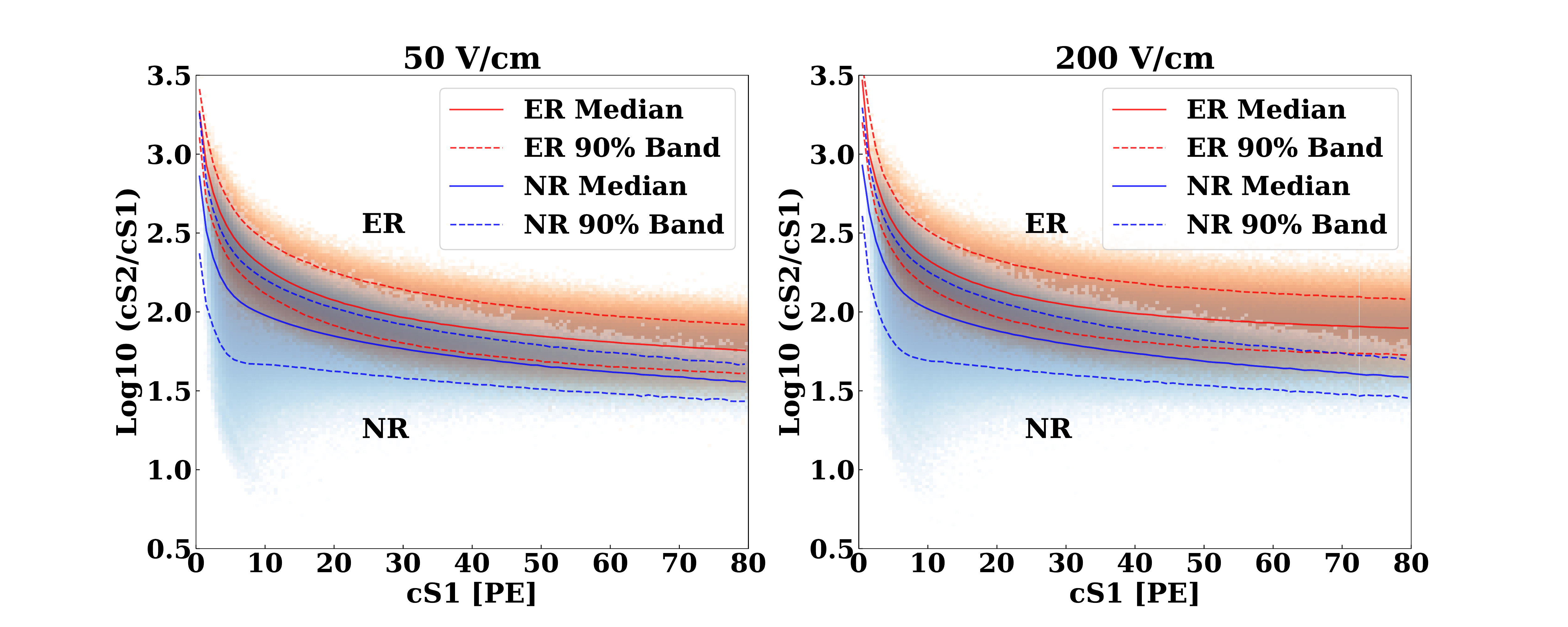}
\caption{Histograms for a flat electronic recoil spectrum and a nuclear recoil spectrum as from a 50~GeV spin-independent WIMP, simulated for different electric fields (using NEST~2.3.5). The top band for each plot is the electronic recoil band, the bottom is the nuclear recoil band. Red lines refer to the median for either band, and the white dotted lines delimit the one-sigma region. Already-demonstrated discrimination is expected to be sufficient for a next-generation detector.}\label{fig:wimperhistos}
\end{center}
\end{figure}

The results of simulating events from a type of source, be it electronic or nuclear recoils, are a list of S1 and S2 values for each event. Binning these events into a 2D~histogram will show how a given type of interaction looks like in terms of the signals received. An example shown in \autoref{fig:wimperhistos} is the histogram of a 50~GeV spin-independent WIMP, which is lower in S2 than that of electronic recoil events. This difference allows for distinction between these two types of interactions and can be measured quantitatively in a few ways. Leakage is the proportion of electronic recoil events per bin that lie below the nuclear recoil median line; rejection is the percentage of background events that are not in the region of interest given by $(1-\mathrm{leakage})$. Various instrumental parameters affect the discrimination capability. For example, the drift field affects the gap between the nuclear and electronic recoil spectra, as does the $g1$~parameter which measures the S1 light yield in the detector. As can be seen from \autoref{fig:wimperhistos}, even moderate drift fields provide satisfactory discrimination. Discrimination is also affected by the atomic structure of xenon, leading to increased leakage from neutrino and Compton scatters on L-shell electrons due to the accompanying atomic de-excitation via Auger electron cascades. This effect is still under study, but available measurements indicate a reduction in rejection by a factor of $6\times$ near the L-shell binding energy ($5.2\1{keV}$) compared to predictions from valence electronic recoils and $\beta$-decays~\cite{Temples:2019taup, Temples:2021prd}. Including this effect for the solar neutrino-induced electronic recoil background results in an 8\% relative increase in leakage from 5.2--8~keV, for 50\% nuclear recoil acceptance.

Importantly, already with the performance of running detectors, discrimination between electronic and nuclear recoils in liquid xenon is sufficient to achieve the various science goals presented in this review, whether they pertain to WIMPs, neutrino-induced signals in both the electronic and nuclear recoil band, or the search for neutrinoless double-beta decay. The accuracy of these simulation results is confirmed \textit{in situ} using dedicated calibration sources, such as dissolved gamma line sources ${}^{83\mathrm{m}}$Kr~\cite{Kastens:2009pa}, dissolved beta-spectrum sources ${}^{220}$Rn~\cite{Lang:2016zde,Aprile:2016pmc}, TH$_3$C~\cite{Akerib:2015wdi}, as well as various neutron sources~\cite{Collar:2013xva,Aprile:2013teh,Akerib:2016mzi}.

\section{Complementarity with Other Experimental Efforts}\label{sec:complementarity}

\subsection{Crossing Symmetry for Freeze-Out Relic Particles}

\begin{figure}[!htbp]
\centering
\includegraphics[width=0.8\columnwidth,clip, trim=1.0in 6.45in 3.75in 2.0in]{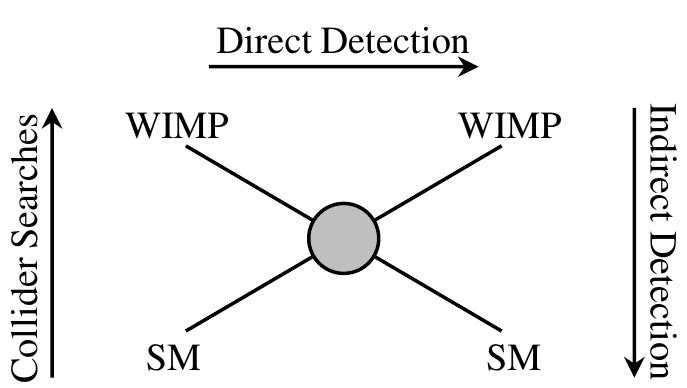}
\caption{Based on the general idea of thermal relic particles (such as WIMPs) interacting with the standard model, three detection techniques are possible: production at colliders, scattering from a target material (direct detection) and annihilation resulting in cosmic rays (indirect detection).}
\label{fig:basicfeyn}
\end{figure}

Production, decay and scattering of dark matter particles are often governed by the same or similar interaction. Relic particles such as WIMPs are a textbook example of this situation, where the production through freeze-out from the thermal equilibrium in the early Universe creates the observed relic density~\cite{Srednicki:1988ce, Bertone:2004pz, Feng:2010gw}. The three principal approaches to discover thermal freeze-out relic particles correspond to the $s$- or $t$-channel processes of dark matter production or scattering, and the time-reversed process to production, which might lead to annihilation of dark matter. This results in the following detection channels, related through crossing symmetry~\cite{Profumo:2013hqa} (see also \autoref{fig:basicfeyn}):

\begin{enumerate}
    \item{\textit{Direct Detection:}} Direct detection of particle dark matter in the Galactic halo in underground experiments as described here;
    \item{\textit{Collider Production:}} Production of dark matter in the laboratory, usually using high energy particle collisions;
    \item{\textit{Indirect Detection}}: Detection of products of dark matter annihilating or decaying in our local universe.
\end{enumerate}

\subsection{Dark Matter at Colliders}

The electroweak energy scale is powerfully probed by the Large Hadron Collider (LHC) at CERN~\cite{Evans:2008zzb}. The freeze-out mechanism requires significant couplings between the dark matter and the standard model, which further motivates searches at a particle collider. Moreover, many `Beyond the Standard Model' (BSM) theories in high energy physics require new particles at the electroweak scale, which are either viable dark matter candidates or might couple to particle dark matter. The most prominent example of such a theory that connects naturally astrophysical and theoretical motivation is supersymmetry (SUSY), which not only remedies many known problems of the Standard Model, such as the hierarchy problem, but also provides an excellent dark matter candidate~\cite{Golfand:1971iw, Clavelli:1970qy, Jungman:1995df, Martin:1997ns}.  

Another motivation for collider searches is the potential to study dark matter in the laboratory. Collider production implies production of the mediator, i.e. the force carrier that connects the dark sector with the visible sector of the Standard Model. Hence, collider dark matter searches are in essence searches for the mediator rather than dark matter. Most collider dark matter searches assume maximal decay of the mediator into dark matter~\cite{Goodman:2010ku, Fox:2011pm}. This is in particular true for 'mono-X' searches, where the main signature is missing momentum in the transverse plane due to the dark matter particle escaping the detector undetected. Constraints placed on the mediator masses are typically about twice as strong as the constraints on the dark matter mass itself. Other analyses attempt to place constraints on the nature of dark matter by looking for deviations in the properties of known particles, for example the Higgs boson, or to search for the mediator directly, such as in dijet searches~\cite{Penning:2017tmb}. Further complementarity stems from searches at colliders for axion-like particles and dark photons, see e.g.~\cite{Bonilla:2022pxu,LHCb:2017trq}. 

\subsection{Indirect Dark Matter Searches}

Dark matter annihilation and decay into standard model particles lead to potential signatures in the Cosmic Microwave Background~\cite{Finkbeiner:2011dx, Galli:2013dna, Slatyer:2015jla} and astrophysical observables such as X-rays~\cite{Boyarsky:2007ge, Yuksel:2007xh, Perez:2016tcq}, gamma rays~\cite{Gunn:1978gr, Stecker:1978du, Berezinsky:1994wva, Bergstrom:1997fj, Gondolo:1999ef, Gehrels:1999ri, Ullio:2000bv, Cesarini:2003nr, Peirani:2004wy, Dodelson:2007gd, Consortium:2010bc, Cholis:2013ena, Ando:2015qda, Ackermann:2015tah, DiMauro:2015tfa, Ajello:2015mfa, Abdallah:2016ygi, Ahnen:2016qkx, Zitzer:2016fvx, Blanco:2017sbc, Archambault:2017wyh, Lisanti:2017qlb, Ahnen:2017pqx, Abeysekara:2017jxs, Blanco:2018esa, Abdalla:2018mve, Abdallah:2018qtu, Blanco:2019eij}, antiprotons~\cite{Aguilar:2016kjl, Cuoco:2016eej, Cui:2016ppb, Cuoco:2017rxb, Cuoco:2017iax, Cui:2018klo}, positrons~\cite{Cholis:2008hb, Bergstrom:2008gr, Zurek:2008qg, Harnik:2008uu, Cirelli:2008jk, Hooper:2008kg}, neutrinos~\cite{Silk:1985ax, Hagelin:1986gv, Freese:1985qw, Krauss:1985aaa, Gaisser:1986ha, Desai:2004pq, PalomaresRuiz:2007ry, Murase:2012xs, Aartsen:2016zhm, Aartsen:2016fep,  Aartsen:2016pfc, Aartsen:2017ulx}, or other particles~\cite{Ellis:1988qp, Donato:1999gy, Fuke:2005it, Donato:2008yx, Ibarra:2013qt, Hryczuk:2014hpa, Carlson:2014ssa, Aramaki:2015laa, Korsmeier:2017xzj, Reinert:2017aga,Arguelles:2019ouk}. Thermal relic dark matter candidates are generically expected to have a thermally averaged annihilation cross section $\langle\sigma v\rangle \simeq 2.2 \times 10^{-26} \mathrm{cm}^{3} / \mathrm{s}$ ~\cite{Kolb:1990vq, Steigman:2012nb}, though other production mechanisms or annihilation channels are known to predict much smaller or larger annihilation cross sections, e.g., Sommerfeld enhancement~\cite{Hisano:2004ds, ArkaniHamed:2008qn}, non-thermal/out-of-equilibrium production~\cite{Gelmini:2006pq, Gelmini:2006pw, Merle:2013wta, Konig:2016dzg}, asymmetric dark matter~\cite{Graesser:2011wi, Lin:2011gj, Iminniyaz:2011yp, Zurek:2013wia}, co-annihilation~\cite{Griest:1990kh, Edsjo:1997bg, Ellis:1998kh}, velocity-dependent annihilation, or non-standard cosmologies~\cite{Fornengo:2002db, Gelmini:2006pq, Gelmini:2006pw,  Boeckel:2009ej, Boeckel:2011yj, Kane:2015jia, Davoudiasl:2015vba, Berlin:2016gtr, Berlin:2016vnh}. The thermal relic cross section, however, provides an important benchmark for indirect detection efforts. While this leads to a characteristic signature of dark matter in the corresponding cosmic ray spectrum~\cite{Gunn:1978gr, Bergstrom:1997fj, Gaskins:2016cha}, the topology, spectral shape and strength of such a signal is rather model dependent and affected by astrophysical foregrounds.

Since the annihilation rate of dark matter is proportional to the square of the dark matter density at the location of annihilation, the brightest signals are expected to come from dense structures. Present searches for annihilation signatures aim at a variety of targets, with the Galactic center and local spheroidal satellite galaxies (dSphs) of the Milky Way being among the most prominent ones. The former is expected to be the brightest source in the sky because of the large dark matter overdensity, but it also has the brightest foregrounds and complex dynamics. The latter are the most extreme dark matter-dominated galaxies known to us, but have a much lower J-factor~\cite{Abdalla:2016olq, Chiappo:2018mlt} compared to the Galactic center (around $10^{17}-10^{19}\,\rm GeV^2 cm^{-5}$ for dSphs and about $10^{22}\,\rm GeV^2 cm^{-5}$ for the Galactic centre, leading to a fainter potential signal. 

In contrast, the flux of particles from decaying dark matter is only proportional to a single power of density, so it is predicted to give rise to less clumpy signals compared to those from annihilating dark matter. Constraints on decaying dark matter come for example from isotropic gamma-ray and neutrino observations~\cite{Murase:2012xs, Blanco:2018esa}.

There is more than just phenomenological support for the hypothesis that dark matter self-annihilates. $N$-body simulations suggest that dark matter halos, in the absence of baryonic effects, follow a density profile which behaves like $\rho \propto r^{-1}$ irrespective of initial conditions. This is referred to as a Navarro-Frenk-White (NFW) profile~\cite{Navarro:1996gj, Navarro:1996bv}. Measured profiles of dwarf spheroidal galaxies appear to follow a shallower density profile $\rho \propto r^0$. This disagreement is referred to as the `core-cusp-problem' and could possibly be resolved by co-annihilating or self-interacting dark matter (\autoref{sec:selfinteracting}).

\subsection{Measurements of Standard Model Parameters}

In some models of dark matter, the dark matter mass and nucleon-dark matter scattering cross section are predicted or bounded as functions of standard model parameters such as the top quark mass and the strong coupling constant. In such scenarios, dark matter searches constrain the standard model parameters, which can be precisely measured by future colliders~\cite{Seidel:2013sqa,Horiguchi:2013wra,Kiyo:2015ooa,Beneke:2015kwa,Gomez-Ceballos:2013zzn} and lattice QCD computations~\cite{Lepage:2014fla}.

For example, in a model that solves the strong CP problem by a space-time parity symmetry~\cite{Dunsky:2019api}, the dark matter mass is proportional to the energy scale at which the standard model Higgs quartic coupling vanishes ($10^9-10^{12}\1{GeV}$), which is sensitive to the standard model parameters. Dark matter couples to a massless dark photon and the dark matter-nucleon scattering arises from unavoidable quantum corrections leading to photon-dark photon mixing. 

\begin{figure}[!htbp]
\centering
\includegraphics[scale=.52]{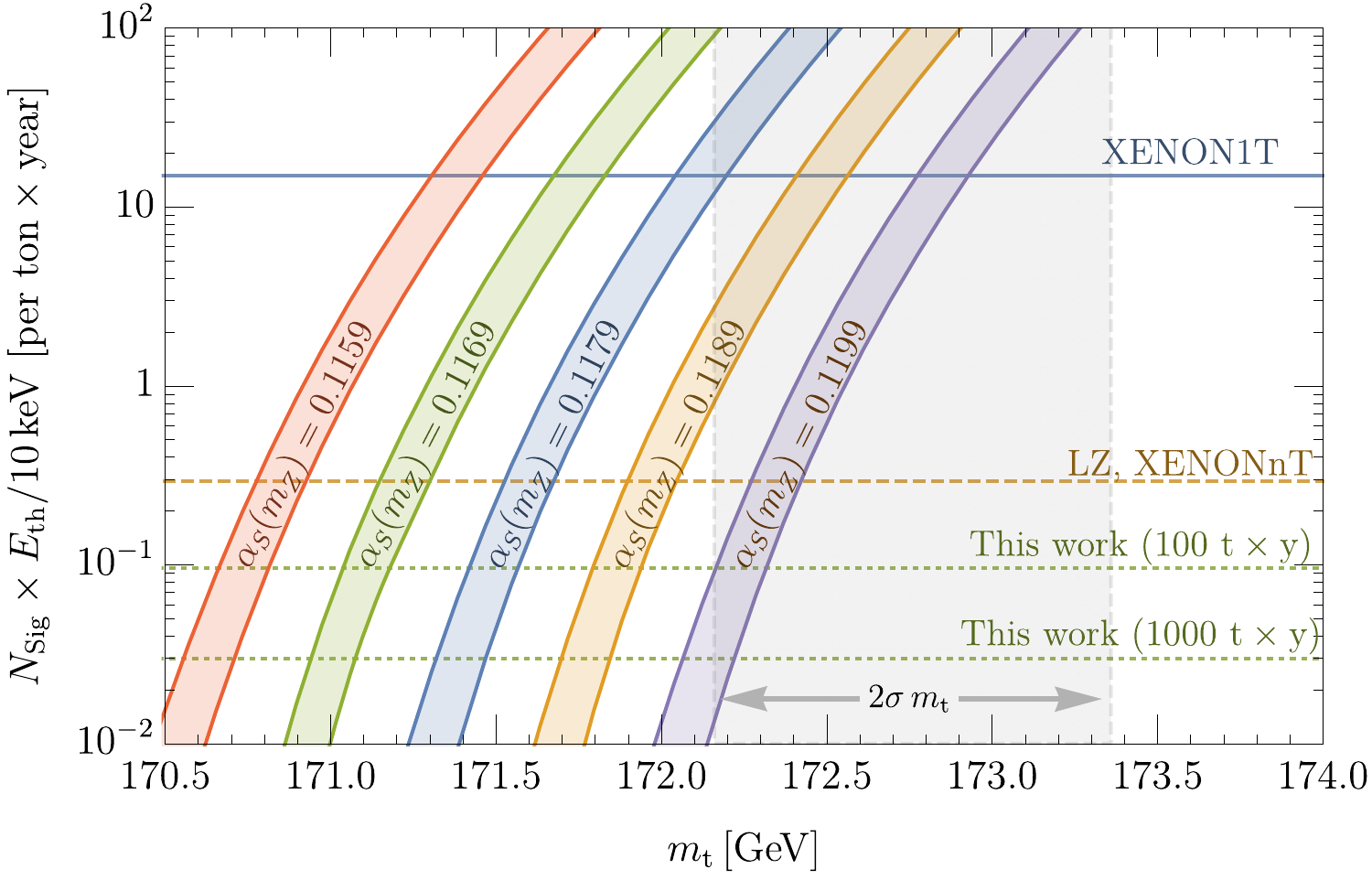}
\caption{The expected number of signals per tonne-year exposure as a function of the top quark mass, $m_t$, and the strong coupling constant evaluated at the $Z$~boson mass scale, $\alpha_s(m_Z)$, in the model described in~\cite{Dunsky:2019api}. The signal count is inversely proportional to the threshold energy $E_{\rm th}$. The thickness of each colored band corresponds to $2\sigma$ uncertainty in the Higgs mass.}
\label{fig:mtop_nsig}
\end{figure}

The resultant correlation between the dark matter signal rate and the standard model parameters that determine the scale where the Higgs quartic coupling vanishes is shown in \autoref{fig:mtop_nsig}. Here, to estimate the projections for next-generation experiments, we scale the limit from XENON1T according to the projections in the high mass region shown in \autoref{fig:si_sensitivity}. Another example is sneutrino or higgsino dark matter in supersymmetric theories, where the dark matter mass is predicted to be smaller than the scale at which the standard model Higgs quartic coupling vanishes~\cite{Dunsky:2020yhv}. Dark matter scatters with nuclei via tree-level $Z$~boson exchange, generating signals detectable in a 1000~tonne-year exposure even for a dark matter mass as large as $10^{12}$~GeV. Detection of or constraints on nucleon-dark matter scattering signals will give an upper bound on the top quark mass and a lower bound on the strong coupling constant.

\subsection{Other Direct Dark Matter Searches}

\begin{figure}[!htbp]
\centering
\includegraphics[width=\columnwidth]{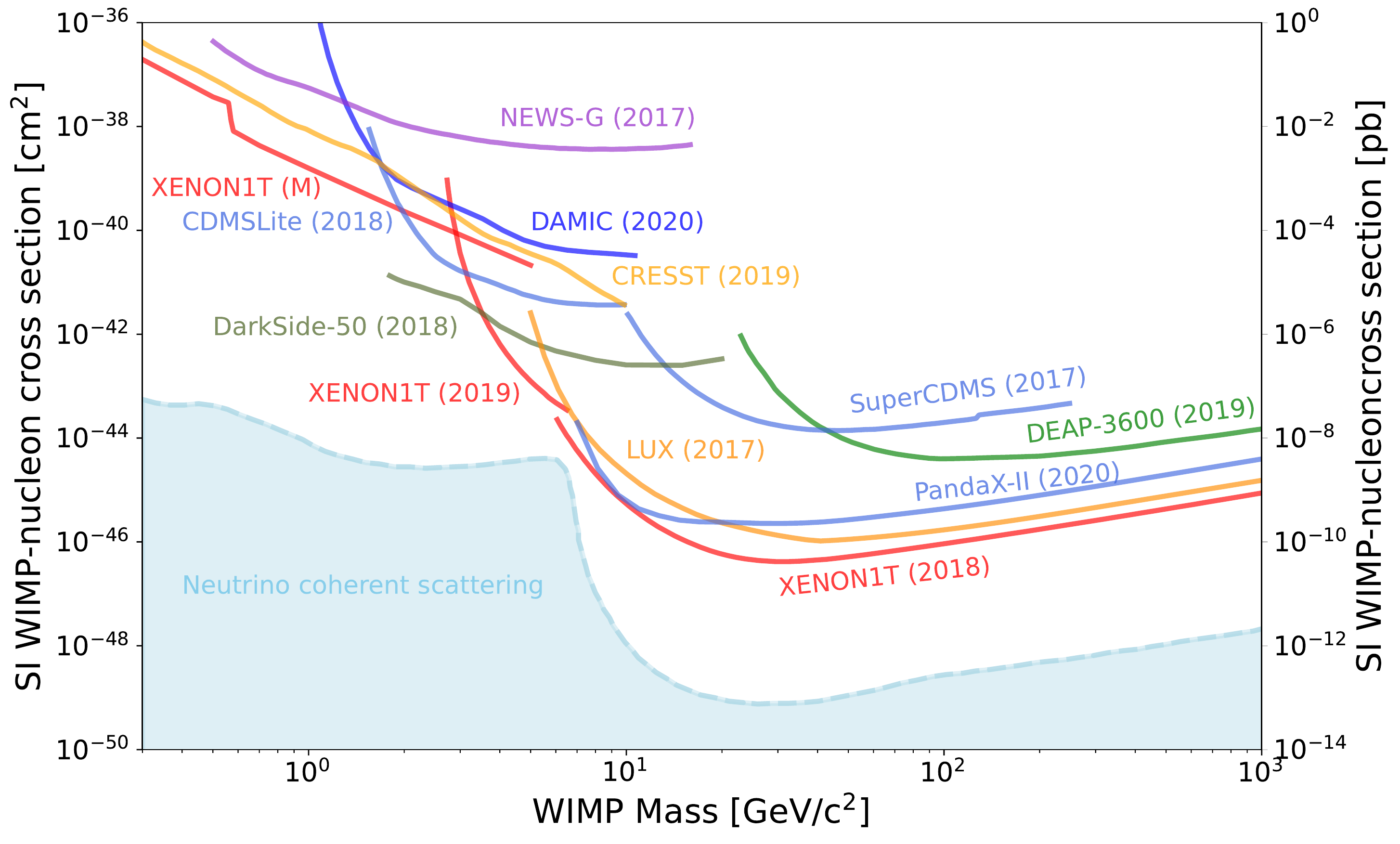}
\caption{Spin-independent dark matter-nuclear scattering limits set by leading direct detection experiments. Complementary experiments with different targets are essential for breaking degeneracies between signals from CE$\nu$NS and WIMP dark matter. Additionally, a variety of targets covers a wider range of potential dark matter masses. Figure adopted from Ref.~\cite{Zyla:2020zbs}.}
\label{fig:directdetectionexp}
\end{figure}

In the search for dark matter directly interacting with a laboratory target, a host of synergistic detectors are required to overcome signal degeneracies, particularly as experiments begin probing the neutrino fog. With different technologies and targets, complimentary experiments can confirm potential dark matter signatures and disentangle them from both neutrino-induced (CE$\nu$NS) signals and instrumental backgrounds. Additionally, a large variety of detectors can probe a wider dark matter mass range, as shown in \autoref{fig:directdetectionexp}. Target materials range from solid state crystals to dense liquids~\cite{Undagoitia:2015gya, Tanabashi:2018oca}. Even within the context of liquid xenon TPCs, larger detectors are required for dark matter nuclear scattering searches, but smaller detectors optimized for single-electron signals might achieve better sensitivity to lower masses and dark matter scattering with electrons~\cite{Bernstein:2020cpc}. 

\subsubsection{Solid State Detectors}

Germanium detectors (HPGe detectors) are a well-understood target for dark matter searches and have been used particularly by CoGeNT~\cite{Aalseth:2014eft} and EDELWEISS~\cite{Shields:2015wka}. DAMIC~\cite{Aguilar-Arevalo:2016ndq} and SENSEI~\cite{Crisler:2018gci} are using silicon CCDs to look for dark matter interactions, in particular through the electronic recoil channel. SuperCDMS~\cite{Agnese:2014aze} measures both phonons and ionization in silicon and germanium crystals cooled to millikelvin temperatures. CRESST~\cite{CRESST:2019jnq} uses calcium tungstate crystals at millikelvin temperatures to read both phonons and scintillation. Sodium iodide is used by ANAIS~\cite{Amare:2021yyu}, SABRE~\cite{Shields:2015wka} as well as DAMA/LIBRA~\cite{Bernabei:2006tw}. Depending on the target and readout, crystals have different sensitivities to different dark matter interaction energies, but largely overlap across dark matter masses in the GeV and sub-GeV range.

\subsubsection{Liquid Target Detectors}

The PICO experiment~\cite{Amole:2017dex} uses Octafluoropropane in a bubble chamber to search for dark matter-induced signals with a particularly strong sensitivity for spin-dependent interactions. Piezo-electric sensors detect bubble formation in the superheated target during an interaction, and cameras record the bubble nucleation. Perhaps the most complementary experiments to liquid xenon TPCs are liquid argon TPCs, such as DarkSide~\cite{Aalseth:2018gq}. The operating principle is identical to the detector described here, but the smaller atomic mass of argon and its effect on collision kinematics makes Argon invaluable for breaking the energy degeneracy of dark matter scatters and coherent elastic neutrino-nucleus scatters, once observed in liquid xenon TPCs. While self-shielding of external backgrounds is better in xenon, the ability to discriminate electronic from nuclear recoils is better in argon. Taken together, the two target elements provide a complementary approach to probing dark matter down to the signal from atmospheric neutrinos. 

\subsection{Neutrinoless Double Beta Decay Experiments}

Experimental searches for $0\nu\beta\beta$ decay~\cite{DellOro:2016tmg} span a variety of isotopes, including $^{76}$Ge, $^{82}$Se, $^{100}$Mo, $^{130}$Te, $^{136}$Xe, and $^{150}$Nd. The choice of isotope is driven by the Q-value of the $2\nu\beta\beta$ mode, the ability to obtain high isotopic abundance, and compatibility with a suitable detection technique. Detection techniques include semiconductor crystals, cryogenic bolometers, time projection chambers, and organic and inorganic scintillators. The choice of detection technique is a balance of the detector energy resolution at the Q-value, the scalability of the technology to large masses, and ability to achieve ultra low backgrounds. The leading experimental efforts to date include MAJORANA~\cite{Majorana:2019nbd}, GERDA~\cite{Agostini:2019hzm}, CUORE~\cite{Adams:2019jhp}, EXO-200~\cite{EXO-200:2019rkq}, and KamLAND-Zen~\cite{KamLAND-Zen:2016pfg, Gando:2020cxo}. The next generation of $0\nu\beta\beta$ searches includes LEGEND\cite{Abgrall:2017syy}, nEXO~\cite{Albert:2017hjq}, NEXT~\cite{Adams:2020cye}, CUPID~\cite{CUPIDInterestGroup:2019inu}, KamLAND2-Zen~\cite{Nakamura:2020szx}, and SNO+~\cite{Andringa:2015tza}. Compared to the experiments using semiconductors and bolometers, a next-generation liquid xenon detector will have significantly larger isotopic mass, even with a natural abundance of $^{136}$Xe, but it will have poorer energy resolution compared to other technologies. The ability to fiducialize the detector means that the backgrounds from radiogenic sources are significantly reduced. A next-generation liquid xenon dark matter detector can have a $0\nu\beta\beta$ sensitivity comparable to those of the next generation of dedicated $0\nu\beta\beta$ experiments (\autoref{sec:0nubb}). 

\subsection{\texorpdfstring{CE$\nu$NS}{CEvNS} Experiments}

The identification of neutrinos in dark matter experiments, in particular through the CE$\nu$NS channel, will be complementary to terrestrial neutrino experiments which operate in a similar energy regime. The COHERENT experiment~\cite{Akimov:2018ghi} uses a stopped-pion source of neutrinos, generated by the Spallation Neutron Source (SNS) at the Oak Ridge National Laboratory. Muon neutrinos with energy $30\1{MeV}$ are produced from charged pion decays, and $\bar{\nu}_\mu$ and $\nu_e$ are produced with a Michel energy spectrum from the subsequent decay of muons at rest. $\bar{\nu}_\mu$ and $\nu_e$ from muon decays are delayed relative to the $30\1{MeV}$ $\nu_\mu$ neutrinos produced from the prompt pion decay. With characteristic energies of tens of MeV, a large sample of the neutrino-nucleus interactions are coherent, which together with the timing structure, permits a measurement of the CE$\nu$NS process.

Using 14.6-kg CsI[Na] scintillator detectors, the COHERENT collaboration announced the first detection of CE$\nu$NS in 2017~\cite{Akimov:2017ade}, with a best-fit count of 134$\pm$22 CE$\nu$NS events, which is 77$\pm$16 percent of the Standard Model prediction. From this initial detection, COHERENT was able to constrain non-standard interactions in a regime of parameter space that had not been possible to probe. In particular, the COHERENT data is sensitive to u- and d-type non-standard interactions for flavor-diagonal muon components, $\epsilon_{\mu \mu}$. The COHERENT detection also set new constraints on exotic solutions to solar neutrino mixing, places novel constraints on new physics that manifests through the neutrino sector (e.g.~\cite{delaVega:2021mhj}), constrains the neutron form factor for CsI~\cite{Ciuffoli:2018qem}, sterile neutrinos~\cite{Coloma:2017egw,Coloma:2017ncl}, and the $g-2$ anomaly~\cite{Liao:2017uzy,Aoyama:2020ynm,Muong-2:2021ojo}. The collaboration has since also measured CE$\nu$NS on Argon~\cite{Akimov:2020pdx}.

Nuclear reactors have been purposed as a copious source of electron anti-neutrinos. The characteristic neutrino energy is $\lesssim 1$ MeV; as such, the coherence condition for the recoil is largely preserved over the entire reactor energy regime. The primary difficulty in detecting CE$\nu$NS using reactors is that detectors have not been able to achieve the low threshold required to identify the CE$\nu$NS nuclear recoil signal. With further improvements in detector technology, several experiments are poised to identify CE$\nu$NS at reactors~\cite{RED:2012hpm,Soma:2014zgm,Aguilar-Arevalo:2016khx,Agnolet:2016zir,Akimov:2017hee,Strauss:2017cuu,Leder:2017lva, Akimov:2019ogx}. The two-phase noble gas detection technique is very promising for this purpose~\cite{Akimov:2019ogx,Ni:2021mwa}, and currently the RED-100 detector (with $\sim 160\1{kg}$ of active liquid xenon) is being tested at the Kalinin NPP site. This is the first CE$\nu$NS experiment using a two-phase emission technique, and other experiments, such as NUXE with liquid xenon and CHILLAX with xenon-doped liquid argon, are currently being developed.

\subsection{Solar Neutrino Experiments}

Direct real-time measurements of solar neutrinos have been observed for the first time by the Borexino experiment~\cite{Arpesella:2007xf, Agostini:2018uly, Agostini:2020mfq}. A next-generation liquid xenon experiments can offer complementary measurements when detecting solar neutrinos through elastic neutrino-electron scattering (\autoref{sec:solarneutrinos}). Neutrinos that contribute to the signal are generated from the $pp$-reaction chain, produced from the electron-capture decay of beryllium-7, and emitted in the carbon, nitrogen, oxygen (CNO) fusion cycle. The measurement of CNO neutrinos in liquid xenon-based detectors is limited by the presence of the two-electron spectrum arising from the $2\nu \beta \beta$-decay of $^{136}$Xe. Depletion of $^{136}$Xe by at least a factor of 100 relative to its natural abundance would be necessary to detect the CNO solar neutrino component in the next-generation xenon-based detector~\cite{Newstead:2018muu}.  

\subsection{Gravitational Wave Searches}

Liquid xenon-based detectors can be utilized to look for neutrinos and gamma-rays released in association with gravitational waves emitted during cataclysmic cosmic events. Such ``simultaneous'' observation is related then to supernova detection and multi-messenger astrophysics, as discussed in \autoref{sec:supernovaneutrinos}.

\subsection{Xenon in Medical Physics}

The great advantages of using liquid xenon for medical imaging have been noticed already back in the 1970s~\cite{Zaklad4326941, Lavoie594289}. Its fast primary scintillation and sufficiently large ionization yield make it especially attractive for Positron Emission Tomography (PET). The two-phase emission detector with condensed xenon as working medium has been tested in the 1980s as a high-quality gamma-camera for nuclear medicine~\cite{Egorov:1983}. The 1990s saw compressed xenon gas being proposed for very effective, collimator-less SPECT systems~\cite{Bolozdynya:1997ecc, Bolozdynya:1997ccc, Rogers:2004cc}. The first liquid xenon detector prototypes for PET scans were built and tested around the same time~\cite{Chepel1993ANL, Chepel790822, Chepel:2002ucm}. Further developments of liquid xenon detectors for PET came recently~\cite{ GallegoManzano:2015hkg,Ferrario:2017sgq,Zhu:2019wbt,Giovagnoli:2021bwc}. In a PET scan, patients are injected with small amounts of chemicals, such as sugar, where molecules have had common stable carbon atoms replaced with positron-emitting isotopes. Position-electron annihilation in the electron shells of atoms in the body leads to back-to-back 511~keV $\gamma$-rays. Cancerous tumors will preferentially absorb more sugar than other parts of the body due to higher rates of metabolic activity~\cite{Kevles:1998}. Liquid xenon is being explored due to its advantageous spatial, temporal, and energy resolutions. Already, PETALO is a full-body PET scanner using the S1 signals in liquid xenon to achieve a high resolution image with reduced patient radioactivity exposure~\cite{Renner:2021vyz}. Many of the developments in liquid xenon detector technology for particle physics thus directly benefit other scientific disciplines, such as medicine, as evidenced here.

Xenon in its gaseous form has also come into use as an alternate anesthetic with minimal dangerous side-effects~\cite{Lynch:2000}. Xenon anesthesia results in a more stable blood pressure, a lower heart rate, and faster emergence from anesthesia than other conventional methods, despite a higher risk of nausea~\cite{Law:2016}.  Most oddly and uniquely, there are some studies that suggest it is useful for treating Traumatic Brain Injuries and Post-Traumatic Stress Disorder (PTSD)~\cite{Campos-Pires,Dobrovolsky}. While in the United States these claims have not been evaluated by the Food and Drug Administration (FDA), in Russia, the inhalation of xenon is used for selective ``deletion'' of traumatic memories, associated with negative emotions~\cite{Dobrovolsky}. All this serves to illustrate the extreme versatility of the element. The study of xenon for particle physics may thus have side-effects spilling over into numerous other fields that seem entirely unrelated.

\subsection{Liquid Xenon TPCs for Nuclear Security}

The neutron is the gold-standard calibration particle of choice for any WIMP detector, since it is supposed to emulate the nuclear recoil generated by dark matter WIMPs. This implies that a WIMP dark matter detector is also an outstanding neutron detector. Liquid xenon TPCs stationed at seaports and airports can allow for non-intrusive inspection (NII) of fissionable materials in cargo, by detecting gamma rays and fast neutrons emitted spontaneously or by stimulation from nuclear materials~\cite{Nikkel:2012zz}. The ability of discriminating between nuclear and electronic recoils allows to better discriminate against activation from other backgrounds. This holds true even through shielding, given the low $\sim$keV energy thresholds achieved recently for dark matter searches. 

Another homeland security application is monitoring of nuclear reactors at power plants for fuel rod theft, which could change the outgoing neutrino (and not just neutron) flux, or rod type replacement, which could change the balance of uranium and plutonium amounts. This concept has been explored by the Nucifer Experiment~\cite{Boireau:2015dda} with a scintillating liquid. Liquid xenon could be ideal for detecting the resulting change in the rate of CE$\nu$NS, already discussed in detail above. Liquid xenon is being considered for detecting CE$\nu$NS by the RED~\cite{Akimov:2019ogx} and NUXE~\cite{Ni:2021mwa} collaborations.

\subsection{Data-Intensive and Computational Sciences}

All of the aforementioned scientific deliverables require the development of cyber-infrastructure such as algorithms, methods, and tools, for the wider benefit of data-intensive sciences. Years of substantial improvements in dark matter detectors means that the field launched into the realm of petabyte data science. The computational science pursued in this field includes, but is not limited to: how ultra-low-energy simulations are performed, including relevant microphysics (\autoref{sec:detector_medium}); how event reconstruction can be performed with the aid of e.g. machine learning~\cite{Khosa:2019qgp}; and how high-throughput/high-level triggers can be deployed on such non-collider experiments.  

This effort requires computational science developments, which can benefit other scientific efforts at both small and large scales. For small scales, there has been an explosion in the number of experiments in recent years. Examples of advancements in this field that are broadly impactful to those having to harness their data are integrating smaller efforts into existing infrastructure using frameworks such as GAUDI~\cite{Barrand:2001ny}; data management systems such as OSG~\cite{Jayatilaka:2017twe}; and demonstrating the effectiveness of columnar compilers in tackling data-intensive applications in high-level descriptive languages. For large scales, overcoming computational science hurdles with novel technologies means serving as a test bed for technologies for Big Science projects such as the HEP Software Project~\cite{Alves:2017she}. Therefore, achieving our physical science goals requires novel computational science and cyber-infrastructure development.

\section{Research Community Priority}\label{sec:priority}

The need for a next-generation liquid xenon TPC is strongly acknowledged throughout the international particle physics community. Studies towards a large-scale liquid xenon dark matter detector started already in 2009 within the EU-ASPERA program, which eventually led to the DARWIN project. The support for DARWIN was strongly recommended in the 2011 update of the ASPERA roadmap~\cite{ASPERA2011}. During the ``Snowmass" process to plan research priorities in 2013, U.S. particle physicists concluded that the discovery goal of liquid xenon dark matter detectors must be to ``search for WIMPs over a wide mass range (1~GeV to 100~TeV)... until we encounter the coherent neutrino scattering signal that will arise from solar, atmospheric and supernova neutrinos."~\cite{Snowmass:2013}. In 2017, the Astro Particle Physics European Consortium (APPEC) devised a European Strategy, which aimed to converge ``with its global partners" on the realization of at least one ``ultimate Dark Matter detector based on xenon"~\cite{APPEC:2017}. 

The Division of Particles and Fields of the American Physical Society defined the next step for the detection of WIMPs to be ``to partner with Europe and Asia on one large international generation-3 detector"~\cite{DPF:2018a} and they note that detector R\&D looks promising for ``the scaling up of liquid noble...detectors to cover the WIMP mass range to the coherent neutrino floor"~\cite{DPF:2018b}. The Chinese community also endorses a next generation deep underground xenon observatory as one of the top priorities in particle astrophysics~\cite{China:2021}. The APPEC Dark Matter Report states that underground dark matter programs with the sensitivity to reach down to the ``neutrino floor at the shortest possible timescale'' should receive enhanced support~\cite{Billard:2021uyg}. Clearly, the main goal is to search for dark matter, but it is understood that such ultimate detector will have other important implications for astrophysics and the quest for the nature of neutrinos. This present paper is a response to the global support for a next-generation liquid xenon TPC, as evidenced here. Progress is also made in assembling a strong global liquid xenon detector community, evidenced for example by the signing in 2020 of a joint Memorandum of Understanding between the members of the LZ and XENON collaborations.

\subsection{Dark Matter}

In the past two decades, the goal of liquid xenon TPCs has been to detect theorized elastic scatters of WIMP dark matter off xenon nuclei. In addition to WIMPs, these detectors have sensitivity to a large host of well-motivated dark matter candidates, as outlined in this work. About 10~different xenon-based dark matter detectors were built over the years, increasing the xenon target mass by almost three orders of magnitude, reducing the electronic recoil background by about four orders of magnitude and improving the sensitivity to WIMP dark matter by more than a factor~1000. After the pioneering work by ZEPLIN-II/III and XENON10, XENON100, LUX and PandaX managed to build a suite of detectors with world-leading sensitivity. XENON1T was the first TPC with a target above the tonne-scale. The current generation of detectors, XENONnT~\cite{Aprile:2020vtw}, LUX-ZEPLIN (LZ)~\cite{Akerib:2018lyp}, and PandaX-4T~\cite{Zhang:2018xdp}, feature multi-tonne liquid xenon targets. Despite a lack of definitive signal so far, these detectors are clear leaders in sensitivity to WIMPs and other physics channels, and scale reliably in mass~\cite{Aprile:2018dbl}. It is for these reasons that a next-generation liquid xenon TPC is of such high interest to the international physics community.  

The Update of the European Strategy for Particle Physics (ESPP) from 2020 points out that the search for dark matter is a crucial part of the search for new physics and that experiments that offer ``potential high-impact" should be supported~\cite{ESPP:2020}. The APPEC Report on the Direct Detection of Dark Matter (2021) states that ``the search for dark matter with the aim of detecting a direct signal of dark matter particle interactions with a detector should be given top priority in astroparticle physics, and in all particle physics"~\cite{Billard:2021uyg}. Already in 2014, the U.S.~Particle Physics Project Prioritization Panel (P5) highlighted the identification of the new physics of dark matter as one of the five science drivers for all of particle physics and recommended that U.S.~funding agencies ``support one or more third-generation (G3) direct detection experiments...[with] a globally complementary program and increased international partnership in G3 experiments"~\cite{P5:2014}. Consolidation of the world-wide xenon community already took place when the members of the ZEPLIN collaboration joined LUX, and XMASS teamed up with XENON. Another important step towards the realization of this next-generation detector happened in 2021, when the scientists from the XENON/DARWIN and LUX-ZEPLIN collaborations agreed to join forces towards the realization of this observatory. 

The German~\cite{GerPS:2018}, Swiss~\cite{chipp2021} and Dutch~\cite{NL:2014} particle physics communities likewise identified the multi-tonne liquid xenon observatory DARWIN of particular interest for their national strategy roadmaps and support R\&D towards this goal via national funding programs. Other countries are strong members of the XENON experiment and it is expected that its follow-up project (e.g. DARWIN) will also be supported. The U.K.’s Particle Astrophysics roadmap also stresses the importance of a xenon-based next-generation (''G3'') observatory, explicitly recommending R\&D towards this detector as the highest priority in dark matter. Relevant R\&D is also supported through multiple European Research Council (ERC) grants.

\subsection{Neutrinoless Double Beta Decay}

Understanding the physics of neutrino mass is another important science driver for particle physics identified by the U.S. P5 and APPEC, which noted the importance of neutrinoless double beta decay searches in that context~\cite{P5:2014,APPEC:2017}. Such experiments are also a top priority in the 2015 US Long Range Plan for Nuclear Science~\cite{LRP:2015}, with one of the four main recommendations being the construction of a massive detector. The European APPEC double beta report (2019) states that "the search for neutrinoless double beta decay is a top priority in particle and astroparticle physics" and acknowledges that the $0\nu\beta\beta$ sensitivity of a next generation \emph{dark matter} detector opens up an exciting scenario~\cite{Giuliani:2019uno}. Similar statements of support for neutrinoless double beta decay detection, especially in dark matter detectors, can be found in UK~\cite{UK:2015}, Russian~\cite{Russia:2012}, CERN/European~\cite{CERN:2013}, and Chinese~\cite{China:2020} particle and nuclear physics priority planning documents. The APPEC Dark Matter Report states on this topic that ``the potential of dark matter detectors to search for rare nuclear decays has been demonstrated spectacularly when XENON1T observed for the first time double electron capture on $^{124}$Xe~\cite{XENON:2019dti}"~\cite{Billard:2021uyg}. 

\subsection{Neutrinos}

Recently, the observed phenomenon of coherent elastic neutrino-nucleus scattering~\cite{Akimov:2017ade} has made large dark matter detectors, such as the one discussed here, particularly desirable for studying neutrinos. Such a detector would be invaluable to the field of astrophysics for measuring Galactic supernovae neutrinos of all flavors. A next-generation liquid xenon detector would be able to probe multiple solar, atmospheric and supernova neutrino signals, which are invaluable measurements in their own right. The US Nuclear Physics community, in the 2015 Long Range Plan for Nuclear Science (LRP)~\cite{LRP:2015}, notes that measuring the CNO cycle and addressing the ``metallicity problem" in the Sun --- both accessible to this detector technology --- are the next big challenges in solar neutrino research.

Taken together, the experiment discussed here addresses a number of high-priority science issues. Spanning across (astro-)particle physics, astrophysics, and nuclear physics, such a detector will significantly advance fundamental science on a variety of fronts.

\section{Summary}\label{sec:summary}

The compelling and versatile science case for a next-generation liquid xenon experiment, combined with its mature technology and minimal technological risk, renders such a detector a paramount facility for the next decade of particle physics, nuclear physics, and astrophysics. This detector will be sensitive to many types of dark matter interactions. Probing the remaining, well-motivated parameter space for spin-independent WIMP scattering down to the neutrino fog will be a milestone in the quest to unravel the nature of dark matter. With its xenon target, this detector will have unprecedented sensitivity to a variety of dark matter models, including spin-dependent couplings, axion-like particles, dark photons, and sterile neutrinos. With the help of optimized analyses, it covers dark matter masses ranging from kilo-electronvolts all the way up to the Planck mass. This next-generation experiment will therefore have significant and lasting impact on dark matter physics.

Simultaneously, such a next-generation liquid xenon experiment will be a competitive experiment in the search for neutrinoless double-beta decay, using a very cost-effective natural xenon target. It will therefore directly address one of the most pressing problems of nuclear physics. Isotopic separation of the natural xenon target can be used to further this sensitivity, or to enable a direct measurement of solar CNO neutrinos. 

Furthermore, this next-generation experiment will be a true observatory for a number of relevant physics. Examples include a precision measurement of the Solar pp neutrino flux, a measurement of the Solar metallicity through boron-8 neutrinos, as well as a first measurement of atmospheric neutrinos in the mega-electronvolt energy range. This detector also has the chance to observe neutrinos from a Galactic supernova in a complementary, flavor-independent channel, if such an event were to occur in the lifetime of the experiment. 

Finally, this detector provides the opportunity to search for a host of signatures from physics beyond the standard model of particle physics. No other technology is capable of probing this many different signals, spanning areas from cosmology to nuclear physics, particle physics, and solar astrophysics.


\section{Acknowledgements}

This work has been supported 
by the US National Science Foundation (NSF, grants PHYS-1719271, PHYS-2112796, PHYS-2112801, PHYS-2112802, PHYS-2112803, PHYS-2112851, PHYS-2137911);
by the Department of Energy (DOE), Office of Science (grants DE-AC02-05CH11231, DE-AC02-07CH11359, DE-AC02-76SF00515, DE-AC52-07NA27344, DE-FG02-00ER41132, DE-FG02-10ER46709, DE-NA0003180, DE-SC0006605, DE-SC0008475, DE-SC0009999, DE-SC0010010, DE-SC0010072, DE-SC0010813, DE-SC0011640, DE-SC0011702, DE-SC0012161, DOE-SC0012447, DE-SC0012704, DE-SC0013542, DE-SC0014223, DE-SC0015535, DE-SC0015708, DE-SC0018982, DE-SC0019066, DE-SC0020216, UW PRJ82AJ);
by the U.K. Science \& Technology Facilities Council (grants ST/M003655/1, ST/M003981/1, ST/M003744/1, ST/M003639/1, ST/M003604/1, ST/R003181/1, ST/M003469/1); 
by the German Research Foundation (DFG) (grants KO 4820/4-1, EXC-2118, 279384907 (SFB 1245)); 
by the Max Planck Gesellschaft;
by the Dutch Research Council (NWO);
by the Swiss National Science Foundation (grants PCEFP2\_181117, 200020-188716);
by the European Research Council (ERC) under the European Union’s Horizon 2020 research and innovation programme (grants 742789, 101020842);
by the Portuguese Foundation for Science and Technology (FCT) (grants PTDC/FIS-PAR/28567/2017);
by the Institute for Basic Science, Korea (grant IBS-R016-D1);
by the Australian Research Council through the ARC Centre of Excellence for Dark Matter Particle Physics, CE200100008;
and by the Marie Skodowska-Curie grant agreement No 860881.
BvK acknowledges support from Emmy Noether Grant No. 420484612;
AM et al. acknowledge additional support from the STFC Boulby Underground Laboratory in the U.K., the GridPP \cite{GridPP:2006wnd,Britton:2009ser} and IRIS Collaborations, in particular at Imperial College London and additional support by the University College London (UCL) Cosmoparticle Initiative. This research used resources of the National Energy Research Scientific Computing Center, a DOE Office of Science User Facility supported by the Office of Science of the U.S. Department of Energy under Contract No. DE-AC02-05CH11231. The University of Edinburgh is a charitable body, registered in Scotland, with the registration number SC005336.
PK and SB from the National Research Foundation of Korea (NRF) grants NRF-2018R1A2A3075605 and NRF-2019R1A2C3005009 and by KIAS Individual Grant No. PG021403;
RC from an individual research grant from the Swedish Research Council (Dnr 2018-05029);
RC and TE from the Knut and Alice Wallenberg project Light Dark Matter (Dnr KAW 2019.0080 and 2019.0080);
PB and SS from the Dr.~Raja Ramanna Fellowship program of the Department of Atomic Energy (DAE), Government of India.
RB acknowledges ISF and the Pazy foundation.
Finally, RFL acknowledges support from the Purdue University Department of Physics and Astronomy and from the Purdue Research Foundation without which this paper could not have been realized.

\bibliographystyle{apsrev}
\bibliography{bibliography}

\begin{thebibliography}{1262}
\expandafter\ifx\csname natexlab\endcsname\relax\def\natexlab#1{#1}\fi
\expandafter\ifx\csname bibnamefont\endcsname\relax
  \def\bibnamefont#1{#1}\fi
\expandafter\ifx\csname bibfnamefont\endcsname\relax
  \def\bibfnamefont#1{#1}\fi
\expandafter\ifx\csname citenamefont\endcsname\relax
  \def\citenamefont#1{#1}\fi
\expandafter\ifx\csname url\endcsname\relax
  \def\url#1{\texttt{#1}}\fi
\expandafter\ifx\csname urlprefix\endcsname\relax\def\urlprefix{URL }\fi
\providecommand{\bibinfo}[2]{#2}
\providecommand{\eprint}[2][]{\url{#2}}

\bibitem[{\citenamefont{Aalbers et~al.}(2016)}]{Aalbers:2016jon}
\bibinfo{author}{\bibfnamefont{J.}~\bibnamefont{Aalbers}} \bibnamefont{et~al.}
  (\bibinfo{collaboration}{DARWIN}), \bibinfo{journal}{JCAP}
  \textbf{\bibinfo{volume}{11}}, \bibinfo{pages}{017} (\bibinfo{year}{2016}),
  \eprint{1606.07001}.

\bibitem[{\citenamefont{Bertone et~al.}(2005)\citenamefont{Bertone, Hooper, and
  Silk}}]{Bertone:2004pz}
\bibinfo{author}{\bibfnamefont{G.}~\bibnamefont{Bertone}},
  \bibinfo{author}{\bibfnamefont{D.}~\bibnamefont{Hooper}}, \bibnamefont{and}
  \bibinfo{author}{\bibfnamefont{J.}~\bibnamefont{Silk}},
  \bibinfo{journal}{Phys. Rept.} \textbf{\bibinfo{volume}{405}},
  \bibinfo{pages}{279} (\bibinfo{year}{2005}), \eprint{hep-ph/0404175}.

\bibitem[{\citenamefont{Bertone and Hooper}(2018)}]{Bertone:2016nfn}
\bibinfo{author}{\bibfnamefont{G.}~\bibnamefont{Bertone}} \bibnamefont{and}
  \bibinfo{author}{\bibfnamefont{D.}~\bibnamefont{Hooper}},
  \bibinfo{journal}{Rev. Mod. Phys.} \textbf{\bibinfo{volume}{90}},
  \bibinfo{pages}{045002} (\bibinfo{year}{2018}), \eprint{1605.04909}.

\bibitem[{\citenamefont{Kapteyn}(1922)}]{Kapteyn:1922zz}
\bibinfo{author}{\bibfnamefont{J.~C.} \bibnamefont{Kapteyn}},
  \bibinfo{journal}{Astrophys. J.} \textbf{\bibinfo{volume}{55}},
  \bibinfo{pages}{302} (\bibinfo{year}{1922}).

\bibitem[{\citenamefont{Jeans}(1922)}]{Jeans:1922}
\bibinfo{author}{\bibfnamefont{J.~H.} \bibnamefont{Jeans}},
  \bibinfo{journal}{Monthly Notices of the Royal Astronomical Society}
  \textbf{\bibinfo{volume}{82}}, \bibinfo{pages}{122} (\bibinfo{year}{1922}),
  ISSN \bibinfo{issn}{0035-8711},
  \urlprefix\url{https://doi.org/10.1093/mnras/82.3.122}.

\bibitem[{\citenamefont{Oort}(1932)}]{Oort:1932}
\bibinfo{author}{\bibfnamefont{J.~H.} \bibnamefont{Oort}},
  \bibinfo{journal}{Bulletin of the Astronomical Institutes of the Netherlands}
  \textbf{\bibinfo{volume}{6}}, \bibinfo{pages}{249} (\bibinfo{year}{1932}).

\bibitem[{\citenamefont{Zwicky}(1933)}]{Zwicky:1933gu}
\bibinfo{author}{\bibfnamefont{F.}~\bibnamefont{Zwicky}},
  \bibinfo{journal}{Helv. Phys. Acta} \textbf{\bibinfo{volume}{6}},
  \bibinfo{pages}{110} (\bibinfo{year}{1933}), \bibinfo{note}{[Gen. Rel. Grav.
  {\bf 41}, 207 (2009)]}.

\bibitem[{\citenamefont{Rubin and Ford}(1970)}]{Rubin:1970zza}
\bibinfo{author}{\bibfnamefont{V.~C.} \bibnamefont{Rubin}} \bibnamefont{and}
  \bibinfo{author}{\bibfnamefont{W.~K.} \bibnamefont{Ford},
  \bibfnamefont{Jr.}}, \bibinfo{journal}{Astrophys. J.}
  \textbf{\bibinfo{volume}{159}}, \bibinfo{pages}{379} (\bibinfo{year}{1970}).

\bibitem[{\citenamefont{Lisanti}(2016)}]{Lisanti:2016jxe}
\bibinfo{author}{\bibfnamefont{M.}~\bibnamefont{Lisanti}}, in
  \emph{\bibinfo{booktitle}{{Theoretical Advanced Study Institute in Elementary
  Particle Physics}: {New Frontiers in Fields and Strings}}}
  (\bibinfo{year}{2016}), \eprint{1603.03797}.

\bibitem[{\citenamefont{Kolb and Turner}(1990)}]{Kolb:1990vq}
\bibinfo{author}{\bibfnamefont{E.~W.} \bibnamefont{Kolb}} \bibnamefont{and}
  \bibinfo{author}{\bibfnamefont{M.~S.} \bibnamefont{Turner}},
  \bibinfo{journal}{Front. Phys.} \textbf{\bibinfo{volume}{69}},
  \bibinfo{pages}{1} (\bibinfo{year}{1990}).

\bibitem[{\citenamefont{Aghanim et~al.}(2020{\natexlab{a}})}]{Akrami:2018vks}
\bibinfo{author}{\bibfnamefont{N.}~\bibnamefont{Aghanim}} \bibnamefont{et~al.}
  (\bibinfo{collaboration}{Planck}), \bibinfo{journal}{Astron. Astrophys.}
  \textbf{\bibinfo{volume}{641}}, \bibinfo{pages}{A1}
  (\bibinfo{year}{2020}{\natexlab{a}}), \eprint{1807.06205}.

\bibitem[{\citenamefont{Aghanim et~al.}(2020{\natexlab{b}})}]{Aghanim:2018eyx}
\bibinfo{author}{\bibfnamefont{N.}~\bibnamefont{Aghanim}} \bibnamefont{et~al.}
  (\bibinfo{collaboration}{Planck}), \bibinfo{journal}{Astron. Astrophys.}
  \textbf{\bibinfo{volume}{641}}, \bibinfo{pages}{A6}
  (\bibinfo{year}{2020}{\natexlab{b}}), \eprint{1807.06209}.

\bibitem[{\citenamefont{Springel et~al.}(2006)\citenamefont{Springel, Frenk,
  and White}}]{Springel:2006vs}
\bibinfo{author}{\bibfnamefont{V.}~\bibnamefont{Springel}},
  \bibinfo{author}{\bibfnamefont{C.~S.} \bibnamefont{Frenk}}, \bibnamefont{and}
  \bibinfo{author}{\bibfnamefont{S.~D.~M.} \bibnamefont{White}},
  \bibinfo{journal}{Nature} \textbf{\bibinfo{volume}{440}},
  \bibinfo{pages}{1137} (\bibinfo{year}{2006}), \eprint{astro-ph/0604561}.

\bibitem[{\citenamefont{Knobel}(2012)}]{Knobel:2012wa}
\bibinfo{author}{\bibfnamefont{C.}~\bibnamefont{Knobel}}
  (\bibinfo{year}{2012}), \eprint{1208.5931}.

\bibitem[{\citenamefont{Coil}(2012)}]{Coil:2012vw}
\bibinfo{author}{\bibfnamefont{A.~L.} \bibnamefont{Coil}}
  (\bibinfo{year}{2012}), \eprint{1202.6633}.

\bibitem[{\citenamefont{Bartelmann and Schneider}(2001)}]{Bartelmann:1999yn}
\bibinfo{author}{\bibfnamefont{M.}~\bibnamefont{Bartelmann}} \bibnamefont{and}
  \bibinfo{author}{\bibfnamefont{P.}~\bibnamefont{Schneider}},
  \bibinfo{journal}{Phys. Rept.} \textbf{\bibinfo{volume}{340}},
  \bibinfo{pages}{291} (\bibinfo{year}{2001}), \eprint{astro-ph/9912508}.

\bibitem[{\citenamefont{Silk and Mamon}(2012)}]{Silk:2012ra}
\bibinfo{author}{\bibfnamefont{J.}~\bibnamefont{Silk}} \bibnamefont{and}
  \bibinfo{author}{\bibfnamefont{G.~A.} \bibnamefont{Mamon}},
  \bibinfo{journal}{Res. Astron. Astrophys.} \textbf{\bibinfo{volume}{12}},
  \bibinfo{pages}{917} (\bibinfo{year}{2012}), \eprint{1207.3080}.

\bibitem[{\citenamefont{Read}(2014)}]{Read:2014qva}
\bibinfo{author}{\bibfnamefont{J.~I.} \bibnamefont{Read}}, \bibinfo{journal}{J.
  Phys. G} \textbf{\bibinfo{volume}{41}}, \bibinfo{pages}{063101}
  (\bibinfo{year}{2014}), \eprint{1404.1938}.

\bibitem[{\citenamefont{Salucci and Persic}(1996)}]{Salucci:1996bf}
\bibinfo{author}{\bibfnamefont{P.}~\bibnamefont{Salucci}} \bibnamefont{and}
  \bibinfo{author}{\bibfnamefont{M.}~\bibnamefont{Persic}}
  (\bibinfo{year}{1996}), \bibinfo{note}{[ASP Conf. Ser. {\bf 117}, 1 (1997)]},
  \eprint{astro-ph/9703027}.

\bibitem[{\citenamefont{Richards et~al.}(2015)}]{Richards:2015gla}
\bibinfo{author}{\bibfnamefont{E.~E.} \bibnamefont{Richards}}
  \bibnamefont{et~al.}, \bibinfo{journal}{Mon. Not. Roy. Astron. Soc.}
  \textbf{\bibinfo{volume}{449}}, \bibinfo{pages}{3981} (\bibinfo{year}{2015}),
  \eprint{1503.05981}.

\bibitem[{\citenamefont{Sofue and Rubin}(2001)}]{Sofue:2000jx}
\bibinfo{author}{\bibfnamefont{Y.}~\bibnamefont{Sofue}} \bibnamefont{and}
  \bibinfo{author}{\bibfnamefont{V.}~\bibnamefont{Rubin}},
  \bibinfo{journal}{Ann. Rev. Astron. Astrophys.}
  \textbf{\bibinfo{volume}{39}}, \bibinfo{pages}{137} (\bibinfo{year}{2001}),
  \eprint{astro-ph/0010594}.

\bibitem[{\citenamefont{Foster and Cooper}(2011)}]{Foster:2010ri}
\bibinfo{author}{\bibfnamefont{T.}~\bibnamefont{Foster}} \bibnamefont{and}
  \bibinfo{author}{\bibfnamefont{B.}~\bibnamefont{Cooper}},
  \bibinfo{journal}{ASP Conf. Ser.} \textbf{\bibinfo{volume}{438}},
  \bibinfo{pages}{16} (\bibinfo{year}{2011}), \eprint{1009.3220}.

\bibitem[{\citenamefont{de~Salas}(2020)}]{deSalas:2019rdi}
\bibinfo{author}{\bibfnamefont{P.~F.} \bibnamefont{de~Salas}},
  \bibinfo{journal}{J. Phys. Conf. Ser.} \textbf{\bibinfo{volume}{1468}},
  \bibinfo{pages}{012020} (\bibinfo{year}{2020}), \eprint{1910.14366}.

\bibitem[{\citenamefont{Brown et~al.}(2018)}]{Brown:2018dum}
\bibinfo{author}{\bibfnamefont{A.}~\bibnamefont{Brown}} \bibnamefont{et~al.}
  (\bibinfo{collaboration}{Gaia}), \bibinfo{journal}{Astron. Astrophys.}
  \textbf{\bibinfo{volume}{616}}, \bibinfo{pages}{A1} (\bibinfo{year}{2018}),
  \eprint{1804.09365}.

\bibitem[{\citenamefont{de~Salas et~al.}(2019)\citenamefont{de~Salas, Malhan,
  Freese, Hattori, and Valluri}}]{deSalas:2019pee}
\bibinfo{author}{\bibfnamefont{P.~F.} \bibnamefont{de~Salas}},
  \bibinfo{author}{\bibfnamefont{K.}~\bibnamefont{Malhan}},
  \bibinfo{author}{\bibfnamefont{K.}~\bibnamefont{Freese}},
  \bibinfo{author}{\bibfnamefont{K.}~\bibnamefont{Hattori}}, \bibnamefont{and}
  \bibinfo{author}{\bibfnamefont{M.}~\bibnamefont{Valluri}},
  \bibinfo{journal}{JCAP} \textbf{\bibinfo{volume}{1910}}, \bibinfo{pages}{037}
  (\bibinfo{year}{2019}), \eprint{1906.06133}.

\bibitem[{\citenamefont{{Hagen} and {Helmi}}(2018)}]{Hagen18}
\bibinfo{author}{\bibfnamefont{J.~H.~J.} \bibnamefont{{Hagen}}}
  \bibnamefont{and} \bibinfo{author}{\bibfnamefont{A.}~\bibnamefont{{Helmi}}},
  \bibinfo{journal}{{Astron. Astrophys.}} \textbf{\bibinfo{volume}{615}},
  \bibinfo{eid}{A99} (\bibinfo{year}{2018}), \eprint{1802.09291}.

\bibitem[{\citenamefont{Buch et~al.}(2019)\citenamefont{Buch, Leung, and
  Fan}}]{Buch:2018qdr}
\bibinfo{author}{\bibfnamefont{J.}~\bibnamefont{Buch}},
  \bibinfo{author}{\bibfnamefont{S.~C.~J.} \bibnamefont{Leung}},
  \bibnamefont{and} \bibinfo{author}{\bibfnamefont{J.}~\bibnamefont{Fan}},
  \bibinfo{journal}{JCAP} \textbf{\bibinfo{volume}{1904}}, \bibinfo{pages}{026}
  (\bibinfo{year}{2019}), \eprint{1808.05603}.

\bibitem[{\citenamefont{Widmark}(2019)}]{Widmark:2018ylf}
\bibinfo{author}{\bibfnamefont{A.}~\bibnamefont{Widmark}},
  \bibinfo{journal}{Astron. Astrophys.} \textbf{\bibinfo{volume}{623}},
  \bibinfo{pages}{A30} (\bibinfo{year}{2019}), \eprint{1811.07911}.

\bibitem[{\citenamefont{Wu et~al.}(2019)\citenamefont{Wu, Freese, Kelso,
  Stengel, and Valluri}}]{Wu:2019nhd}
\bibinfo{author}{\bibfnamefont{Y.}~\bibnamefont{Wu}},
  \bibinfo{author}{\bibfnamefont{K.}~\bibnamefont{Freese}},
  \bibinfo{author}{\bibfnamefont{C.}~\bibnamefont{Kelso}},
  \bibinfo{author}{\bibfnamefont{P.}~\bibnamefont{Stengel}}, \bibnamefont{and}
  \bibinfo{author}{\bibfnamefont{M.}~\bibnamefont{Valluri}},
  \bibinfo{journal}{JCAP} \textbf{\bibinfo{volume}{10}}, \bibinfo{pages}{034}
  (\bibinfo{year}{2019}), \eprint{1904.04781}.

\bibitem[{\citenamefont{Baxter et~al.}(2021{\natexlab{a}})}]{Baxter:2021pqo}
\bibinfo{author}{\bibfnamefont{D.}~\bibnamefont{Baxter}} \bibnamefont{et~al.},
  \bibinfo{journal}{Eur. Phys. J. C} \textbf{\bibinfo{volume}{81}},
  \bibinfo{pages}{907} (\bibinfo{year}{2021}{\natexlab{a}}),
  \eprint{2105.00599}.

\bibitem[{\citenamefont{Green}(2012)}]{Green:2011bv}
\bibinfo{author}{\bibfnamefont{A.~M.} \bibnamefont{Green}},
  \bibinfo{journal}{Mod. Phys. Lett. A} \textbf{\bibinfo{volume}{27}},
  \bibinfo{pages}{1230004} (\bibinfo{year}{2012}), \eprint{1112.0524}.

\bibitem[{\citenamefont{Tanabashi et~al.}(2018)}]{Tanabashi:2018oca}
\bibinfo{author}{\bibfnamefont{M.}~\bibnamefont{Tanabashi}}
  \bibnamefont{et~al.} (\bibinfo{collaboration}{Particle Data Group}),
  \bibinfo{journal}{Phys. Rev. D} \textbf{\bibinfo{volume}{98}},
  \bibinfo{pages}{030001} (\bibinfo{year}{2018}).

\bibitem[{\citenamefont{Drukier and Stodolsky}(1984)}]{Drukier:1983gj}
\bibinfo{author}{\bibfnamefont{A.}~\bibnamefont{Drukier}} \bibnamefont{and}
  \bibinfo{author}{\bibfnamefont{L.}~\bibnamefont{Stodolsky}},
  \bibinfo{journal}{Phys. Rev. D} \textbf{\bibinfo{volume}{30}},
  \bibinfo{pages}{2295} (\bibinfo{year}{1984}).

\bibitem[{\citenamefont{Smith and Lewin}(1983)}]{Smith:1983jj}
\bibinfo{author}{\bibfnamefont{P.~F.} \bibnamefont{Smith}} \bibnamefont{and}
  \bibinfo{author}{\bibfnamefont{J.~D.} \bibnamefont{Lewin}},
  \bibinfo{journal}{Phys. Lett. B} \textbf{\bibinfo{volume}{127}},
  \bibinfo{pages}{185} (\bibinfo{year}{1983}).

\bibitem[{\citenamefont{Goodman and Witten}(1985)}]{Goodman:1984dc}
\bibinfo{author}{\bibfnamefont{M.~W.} \bibnamefont{Goodman}} \bibnamefont{and}
  \bibinfo{author}{\bibfnamefont{E.}~\bibnamefont{Witten}},
  \bibinfo{journal}{Phys. Rev. D} \textbf{\bibinfo{volume}{31}},
  \bibinfo{pages}{3059} (\bibinfo{year}{1985}).

\bibitem[{\citenamefont{Drukier et~al.}(1986)\citenamefont{Drukier, Freese, and
  Spergel}}]{Drukier:1986tm}
\bibinfo{author}{\bibfnamefont{A.~K.} \bibnamefont{Drukier}},
  \bibinfo{author}{\bibfnamefont{K.}~\bibnamefont{Freese}}, \bibnamefont{and}
  \bibinfo{author}{\bibfnamefont{D.~N.} \bibnamefont{Spergel}},
  \bibinfo{journal}{Phys. Rev. D} \textbf{\bibinfo{volume}{33}},
  \bibinfo{pages}{3495} (\bibinfo{year}{1986}).

\bibitem[{\citenamefont{Freese et~al.}(2013)\citenamefont{Freese, Lisanti, and
  Savage}}]{Freese:2012xd}
\bibinfo{author}{\bibfnamefont{K.}~\bibnamefont{Freese}},
  \bibinfo{author}{\bibfnamefont{M.}~\bibnamefont{Lisanti}}, \bibnamefont{and}
  \bibinfo{author}{\bibfnamefont{C.}~\bibnamefont{Savage}},
  \bibinfo{journal}{Rev. Mod. Phys.} \textbf{\bibinfo{volume}{85}},
  \bibinfo{pages}{1561} (\bibinfo{year}{2013}), \eprint{1209.3339}.

\bibitem[{\citenamefont{Copi and Krauss}(2003)}]{Copi:2002hm}
\bibinfo{author}{\bibfnamefont{C.~J.} \bibnamefont{Copi}} \bibnamefont{and}
  \bibinfo{author}{\bibfnamefont{L.~M.} \bibnamefont{Krauss}},
  \bibinfo{journal}{Phys. Rev. D} \textbf{\bibinfo{volume}{67}},
  \bibinfo{pages}{103507} (\bibinfo{year}{2003}), \eprint{astro-ph/0208010}.

\bibitem[{\citenamefont{Savage et~al.}(2006)\citenamefont{Savage, Freese, and
  Gondolo}}]{Savage:2006qr}
\bibinfo{author}{\bibfnamefont{C.}~\bibnamefont{Savage}},
  \bibinfo{author}{\bibfnamefont{K.}~\bibnamefont{Freese}}, \bibnamefont{and}
  \bibinfo{author}{\bibfnamefont{P.}~\bibnamefont{Gondolo}},
  \bibinfo{journal}{Phys. Rev. D} \textbf{\bibinfo{volume}{74}},
  \bibinfo{pages}{043531} (\bibinfo{year}{2006}), \eprint{astro-ph/0607121}.

\bibitem[{\citenamefont{Lang and Weiner}(2010)}]{Lang:2010cd}
\bibinfo{author}{\bibfnamefont{R.~F.} \bibnamefont{Lang}} \bibnamefont{and}
  \bibinfo{author}{\bibfnamefont{N.}~\bibnamefont{Weiner}},
  \bibinfo{journal}{JCAP} \textbf{\bibinfo{volume}{06}}, \bibinfo{pages}{032}
  (\bibinfo{year}{2010}), \eprint{1003.3664}.

\bibitem[{\citenamefont{Kuhlen et~al.}(2010)\citenamefont{Kuhlen, Weiner,
  Diemand, Madau, Moore, Potter, Stadel, and Zemp}}]{Kuhlen:2009vh}
\bibinfo{author}{\bibfnamefont{M.}~\bibnamefont{Kuhlen}},
  \bibinfo{author}{\bibfnamefont{N.}~\bibnamefont{Weiner}},
  \bibinfo{author}{\bibfnamefont{J.}~\bibnamefont{Diemand}},
  \bibinfo{author}{\bibfnamefont{P.}~\bibnamefont{Madau}},
  \bibinfo{author}{\bibfnamefont{B.}~\bibnamefont{Moore}},
  \bibinfo{author}{\bibfnamefont{D.}~\bibnamefont{Potter}},
  \bibinfo{author}{\bibfnamefont{J.}~\bibnamefont{Stadel}}, \bibnamefont{and}
  \bibinfo{author}{\bibfnamefont{M.}~\bibnamefont{Zemp}},
  \bibinfo{journal}{JCAP} \textbf{\bibinfo{volume}{02}}, \bibinfo{pages}{030}
  (\bibinfo{year}{2010}), \eprint{0912.2358}.

\bibitem[{\citenamefont{Schneider et~al.}(2010)\citenamefont{Schneider, Krauss,
  and Moore}}]{Schneider:2010jr}
\bibinfo{author}{\bibfnamefont{A.}~\bibnamefont{Schneider}},
  \bibinfo{author}{\bibfnamefont{L.}~\bibnamefont{Krauss}}, \bibnamefont{and}
  \bibinfo{author}{\bibfnamefont{B.}~\bibnamefont{Moore}},
  \bibinfo{journal}{Phys. Rev. D} \textbf{\bibinfo{volume}{82}},
  \bibinfo{pages}{063525} (\bibinfo{year}{2010}), \eprint{1004.5432}.

\bibitem[{\citenamefont{O'Hare et~al.}(2020)\citenamefont{O'Hare, Evans,
  McCabe, Myeong, and Belokurov}}]{OHare:2019qxc}
\bibinfo{author}{\bibfnamefont{C.~A.~J.} \bibnamefont{O'Hare}},
  \bibinfo{author}{\bibfnamefont{N.~W.} \bibnamefont{Evans}},
  \bibinfo{author}{\bibfnamefont{C.}~\bibnamefont{McCabe}},
  \bibinfo{author}{\bibfnamefont{G.}~\bibnamefont{Myeong}}, \bibnamefont{and}
  \bibinfo{author}{\bibfnamefont{V.}~\bibnamefont{Belokurov}},
  \bibinfo{journal}{Phys. Rev. D} \textbf{\bibinfo{volume}{101}},
  \bibinfo{pages}{023006} (\bibinfo{year}{2020}), \eprint{1909.04684}.

\bibitem[{\citenamefont{Ling et~al.}(2010)\citenamefont{Ling, Nezri,
  Athanassoula, and Teyssier}}]{Ling:2009eh}
\bibinfo{author}{\bibfnamefont{F.~S.} \bibnamefont{Ling}},
  \bibinfo{author}{\bibfnamefont{E.}~\bibnamefont{Nezri}},
  \bibinfo{author}{\bibfnamefont{E.}~\bibnamefont{Athanassoula}},
  \bibnamefont{and} \bibinfo{author}{\bibfnamefont{R.}~\bibnamefont{Teyssier}},
  \bibinfo{journal}{JCAP} \textbf{\bibinfo{volume}{02}}, \bibinfo{pages}{012}
  (\bibinfo{year}{2010}), \eprint{0909.2028}.

\bibitem[{\citenamefont{Damour and Krauss}(1999)}]{Damour:1998vg}
\bibinfo{author}{\bibfnamefont{T.}~\bibnamefont{Damour}} \bibnamefont{and}
  \bibinfo{author}{\bibfnamefont{L.~M.} \bibnamefont{Krauss}},
  \bibinfo{journal}{Phys. Rev. D} \textbf{\bibinfo{volume}{59}},
  \bibinfo{pages}{063509} (\bibinfo{year}{1999}), \eprint{astro-ph/9807099}.

\bibitem[{\citenamefont{Krauss and Romanelli}(1989)}]{Krauss:1988qm}
\bibinfo{author}{\bibfnamefont{L.~M.} \bibnamefont{Krauss}} \bibnamefont{and}
  \bibinfo{author}{\bibfnamefont{P.}~\bibnamefont{Romanelli}},
  \bibinfo{journal}{Phys. Rev. D} \textbf{\bibinfo{volume}{39}},
  \bibinfo{pages}{1225} (\bibinfo{year}{1989}).

\bibitem[{\citenamefont{Schumann}(2019)}]{Schumann:2019eaa}
\bibinfo{author}{\bibfnamefont{M.}~\bibnamefont{Schumann}},
  \bibinfo{journal}{J. Phys. G} \textbf{\bibinfo{volume}{46}},
  \bibinfo{pages}{103003} (\bibinfo{year}{2019}), \eprint{1903.03026}.

\bibitem[{\citenamefont{Aalseth et~al.}(2014)}]{Aalseth:2014eft}
\bibinfo{author}{\bibfnamefont{C.~E.} \bibnamefont{Aalseth}}
  \bibnamefont{et~al.} (\bibinfo{collaboration}{CoGeNT})
  (\bibinfo{year}{2014}), \eprint{1401.3295}.

\bibitem[{\citenamefont{Agnese et~al.}(2014)}]{Agnese:2014aze}
\bibinfo{author}{\bibfnamefont{R.}~\bibnamefont{Agnese}} \bibnamefont{et~al.}
  (\bibinfo{collaboration}{SuperCDMS}), \bibinfo{journal}{Phys. Rev. Lett.}
  \textbf{\bibinfo{volume}{112}}, \bibinfo{pages}{241302}
  (\bibinfo{year}{2014}), \eprint{1402.7137}.

\bibitem[{\citenamefont{Armengaud et~al.}(2016)}]{Armengaud:2016cvl}
\bibinfo{author}{\bibfnamefont{E.}~\bibnamefont{Armengaud}}
  \bibnamefont{et~al.} (\bibinfo{collaboration}{EDELWEISS}),
  \bibinfo{journal}{JCAP} \textbf{\bibinfo{volume}{1605}}, \bibinfo{pages}{019}
  (\bibinfo{year}{2016}), \eprint{1603.05120}.

\bibitem[{\citenamefont{Shields et~al.}(2015)\citenamefont{Shields, Xu, and
  Calaprice}}]{Shields:2015wka}
\bibinfo{author}{\bibfnamefont{E.}~\bibnamefont{Shields}},
  \bibinfo{author}{\bibfnamefont{J.}~\bibnamefont{Xu}}, \bibnamefont{and}
  \bibinfo{author}{\bibfnamefont{F.}~\bibnamefont{Calaprice}},
  \bibinfo{journal}{Phys. Procedia} \textbf{\bibinfo{volume}{61}},
  \bibinfo{pages}{169} (\bibinfo{year}{2015}).

\bibitem[{\citenamefont{Aguilar-Arevalo
  et~al.}(2016{\natexlab{a}})}]{Aguilar-Arevalo:2016ndq}
\bibinfo{author}{\bibfnamefont{A.}~\bibnamefont{Aguilar-Arevalo}}
  \bibnamefont{et~al.} (\bibinfo{collaboration}{DAMIC}),
  \bibinfo{journal}{Phys. Rev. D} \textbf{\bibinfo{volume}{94}},
  \bibinfo{pages}{082006} (\bibinfo{year}{2016}{\natexlab{a}}),
  \eprint{1607.07410}.

\bibitem[{\citenamefont{Abdelhameed et~al.}(2019)}]{CRESST:2019jnq}
\bibinfo{author}{\bibfnamefont{A.~H.} \bibnamefont{Abdelhameed}}
  \bibnamefont{et~al.} (\bibinfo{collaboration}{CRESST}),
  \bibinfo{journal}{Phys. Rev. D} \textbf{\bibinfo{volume}{100}},
  \bibinfo{pages}{102002} (\bibinfo{year}{2019}), \eprint{1904.00498}.

\bibitem[{\citenamefont{Crisler et~al.}(2018)\citenamefont{Crisler, Essig,
  Estrada, Fernandez, Tiffenberg, Sofo~haro, Volansky, and
  Yu}}]{Crisler:2018gci}
\bibinfo{author}{\bibfnamefont{M.}~\bibnamefont{Crisler}},
  \bibinfo{author}{\bibfnamefont{R.}~\bibnamefont{Essig}},
  \bibinfo{author}{\bibfnamefont{J.}~\bibnamefont{Estrada}},
  \bibinfo{author}{\bibfnamefont{G.}~\bibnamefont{Fernandez}},
  \bibinfo{author}{\bibfnamefont{J.}~\bibnamefont{Tiffenberg}},
  \bibinfo{author}{\bibfnamefont{M.}~\bibnamefont{Sofo~haro}},
  \bibinfo{author}{\bibfnamefont{T.}~\bibnamefont{Volansky}}, \bibnamefont{and}
  \bibinfo{author}{\bibfnamefont{T.-T.} \bibnamefont{Yu}}
  (\bibinfo{collaboration}{SENSEI}), \bibinfo{journal}{Phys. Rev. Lett.}
  \textbf{\bibinfo{volume}{121}}, \bibinfo{pages}{061803}
  (\bibinfo{year}{2018}), \eprint{1804.00088}.

\bibitem[{\citenamefont{Amaral et~al.}(2020{\natexlab{a}})}]{SuperCDMS:2020ymb}
\bibinfo{author}{\bibfnamefont{D.~W.} \bibnamefont{Amaral}}
  \bibnamefont{et~al.} (\bibinfo{collaboration}{SuperCDMS}),
  \bibinfo{journal}{Phys. Rev. D} \textbf{\bibinfo{volume}{102}},
  \bibinfo{pages}{091101} (\bibinfo{year}{2020}{\natexlab{a}}),
  \eprint{2005.14067}.

\bibitem[{\citenamefont{Alkhatib et~al.}(2021)}]{SuperCDMS:2020aus}
\bibinfo{author}{\bibfnamefont{I.}~\bibnamefont{Alkhatib}} \bibnamefont{et~al.}
  (\bibinfo{collaboration}{SuperCDMS}), \bibinfo{journal}{Phys. Rev. Lett.}
  \textbf{\bibinfo{volume}{127}}, \bibinfo{pages}{061801}
  (\bibinfo{year}{2021}), \eprint{2007.14289}.

\bibitem[{\citenamefont{Behnke et~al.}(2017)}]{Behnke:2016lsk}
\bibinfo{author}{\bibfnamefont{E.}~\bibnamefont{Behnke}} \bibnamefont{et~al.},
  \bibinfo{journal}{Astropart. Phys.} \textbf{\bibinfo{volume}{90}},
  \bibinfo{pages}{85} (\bibinfo{year}{2017}), \eprint{1611.01499}.

\bibitem[{\citenamefont{Szydagis et~al.}(2018)\citenamefont{Szydagis, Levy,
  Huang, Kamaha, Knight, Rischbieter, and Wilson}}]{Szydagis:2018wjp}
\bibinfo{author}{\bibfnamefont{M.}~\bibnamefont{Szydagis}},
  \bibinfo{author}{\bibfnamefont{C.}~\bibnamefont{Levy}},
  \bibinfo{author}{\bibfnamefont{Y.}~\bibnamefont{Huang}},
  \bibinfo{author}{\bibfnamefont{A.~C.} \bibnamefont{Kamaha}},
  \bibinfo{author}{\bibfnamefont{C.~C.} \bibnamefont{Knight}},
  \bibinfo{author}{\bibfnamefont{G.~R.~C.} \bibnamefont{Rischbieter}},
  \bibnamefont{and} \bibinfo{author}{\bibfnamefont{P.~W.} \bibnamefont{Wilson}}
  (\bibinfo{year}{2018}), \eprint{1807.09253}.

\bibitem[{\citenamefont{Aalseth et~al.}(2018)}]{Aalseth:2018gq}
\bibinfo{author}{\bibfnamefont{C.~E.} \bibnamefont{Aalseth}}
  \bibnamefont{et~al.}, \bibinfo{journal}{Eur. Phys. J. Plus}
  \textbf{\bibinfo{volume}{133}}, \bibinfo{pages}{131} (\bibinfo{year}{2018}),
  \eprint{1707.08145}.

\bibitem[{\citenamefont{Akerib et~al.}(2017{\natexlab{a}})}]{Akerib:2016vxi}
\bibinfo{author}{\bibfnamefont{D.~S.} \bibnamefont{Akerib}}
  \bibnamefont{et~al.} (\bibinfo{collaboration}{LUX}), \bibinfo{journal}{Phys.
  Rev. Lett.} \textbf{\bibinfo{volume}{118}}, \bibinfo{pages}{021303}
  (\bibinfo{year}{2017}{\natexlab{a}}), \eprint{1608.07648}.

\bibitem[{\citenamefont{Aprile et~al.}(2017{\natexlab{a}})}]{Aprile:2017aty}
\bibinfo{author}{\bibfnamefont{E.}~\bibnamefont{Aprile}} \bibnamefont{et~al.}
  (\bibinfo{collaboration}{XENON}), \bibinfo{journal}{Eur. Phys. J. C}
  \textbf{\bibinfo{volume}{77}}, \bibinfo{pages}{881}
  (\bibinfo{year}{2017}{\natexlab{a}}), \eprint{1708.07051}.

\bibitem[{\citenamefont{Cui et~al.}(2017{\natexlab{a}})}]{Cui:2017nnn}
\bibinfo{author}{\bibfnamefont{X.}~\bibnamefont{Cui}} \bibnamefont{et~al.}
  (\bibinfo{collaboration}{PandaX-II}), \bibinfo{journal}{Phys. Rev. Lett.}
  \textbf{\bibinfo{volume}{119}}, \bibinfo{pages}{181302}
  (\bibinfo{year}{2017}{\natexlab{a}}), \eprint{1708.06917}.

\bibitem[{\citenamefont{Kobayashi et~al.}(2019)}]{Kobayashi:2018jky}
\bibinfo{author}{\bibfnamefont{M.}~\bibnamefont{Kobayashi}}
  \bibnamefont{et~al.} (\bibinfo{collaboration}{XMASS}),
  \bibinfo{journal}{Phys. Lett. B} \textbf{\bibinfo{volume}{795}},
  \bibinfo{pages}{308} (\bibinfo{year}{2019}), \eprint{1808.06177}.

\bibitem[{\citenamefont{Amole et~al.}(2017)}]{Amole:2017dex}
\bibinfo{author}{\bibfnamefont{C.}~\bibnamefont{Amole}} \bibnamefont{et~al.}
  (\bibinfo{collaboration}{PICO}), \bibinfo{journal}{Phys. Rev. Lett.}
  \textbf{\bibinfo{volume}{118}}, \bibinfo{pages}{251301}
  (\bibinfo{year}{2017}), \eprint{1702.07666}.

\bibitem[{\citenamefont{Bolozdynya}(2010)}]{Bolozdynya:2010}
\bibinfo{author}{\bibfnamefont{A.~I.} \bibnamefont{Bolozdynya}},
  \emph{\bibinfo{title}{Emission Detectors}} (\bibinfo{publisher}{World
  Scientific Publishing Co Pte Ltd}, \bibinfo{year}{2010}).

\bibitem[{\citenamefont{Suzuki and Hitachi}(2010)}]{hatano_applications_2010}
\bibinfo{author}{\bibfnamefont{S.}~\bibnamefont{Suzuki}} \bibnamefont{and}
  \bibinfo{author}{\bibfnamefont{A.}~\bibnamefont{Hitachi}}, in
  \emph{\bibinfo{booktitle}{Charged {Particle} and {Photon} {Interactions} with
  {Matter}}}, edited by
  \bibinfo{editor}{\bibfnamefont{Y.}~\bibnamefont{Hatano}},
  \bibinfo{editor}{\bibfnamefont{Y.}~\bibnamefont{Katsumura}},
  \bibnamefont{and} \bibinfo{editor}{\bibfnamefont{A.}~\bibnamefont{Mozumder}}
  (\bibinfo{publisher}{CRC Press}, \bibinfo{year}{2010}), ISBN
  \bibinfo{isbn}{9780429134265}.

\bibitem[{\citenamefont{Akimov et~al.}(2021{\natexlab{a}})\citenamefont{Akimov,
  Bolozdynya, Buzulutskov, and Chepel}}]{Akimov:2021}
\bibinfo{author}{\bibfnamefont{D.~Y.} \bibnamefont{Akimov}},
  \bibinfo{author}{\bibfnamefont{A.~I.} \bibnamefont{Bolozdynya}},
  \bibinfo{author}{\bibfnamefont{A.~F.} \bibnamefont{Buzulutskov}},
  \bibnamefont{and} \bibinfo{author}{\bibfnamefont{V.}~\bibnamefont{Chepel}},
  \emph{\bibinfo{title}{Two-Phase Emission Detectors}}
  (\bibinfo{publisher}{World Scientific Publishing Co Pte Ltd},
  \bibinfo{year}{2021}{\natexlab{a}}).

\bibitem[{\citenamefont{Dolgoshein et~al.}(1970)\citenamefont{Dolgoshein,
  Lebedenko, and Rodionov}}]{Dolgoshein:1970}
\bibinfo{author}{\bibfnamefont{B.~A.} \bibnamefont{Dolgoshein}},
  \bibinfo{author}{\bibfnamefont{V.~N.} \bibnamefont{Lebedenko}},
  \bibnamefont{and} \bibinfo{author}{\bibfnamefont{B.~U.}
  \bibnamefont{Rodionov}}, \bibinfo{journal}{Sov. Phys. JETP Lett.}
  \textbf{\bibinfo{volume}{11}}, \bibinfo{pages}{351} (\bibinfo{year}{1970}).

\bibitem[{\citenamefont{Bolozdynya et~al.}(1995)\citenamefont{Bolozdynya,
  Egorov, Rodinov, and Miroshnichenko}}]{Bolozdynya:1995}
\bibinfo{author}{\bibfnamefont{A.~I.} \bibnamefont{Bolozdynya}},
  \bibinfo{author}{\bibfnamefont{O.~K.} \bibnamefont{Egorov}},
  \bibinfo{author}{\bibfnamefont{B.~U.} \bibnamefont{Rodinov}},
  \bibnamefont{and} \bibinfo{author}{\bibfnamefont{V.~P.}
  \bibnamefont{Miroshnichenko}}, \bibinfo{journal}{IEEE Trans. Nucl. Sci.}
  \textbf{\bibinfo{volume}{42}}, \bibinfo{pages}{565} (\bibinfo{year}{1995}).

\bibitem[{\citenamefont{Cline et~al.}(2000)}]{Cline:2000}
\bibinfo{author}{\bibfnamefont{D.}~\bibnamefont{Cline}} \bibnamefont{et~al.},
  \bibinfo{journal}{Astropart. Phys.} \textbf{\bibinfo{volume}{12}},
  \bibinfo{pages}{373} (\bibinfo{year}{2000}).

\bibitem[{\citenamefont{Aprile et~al.}(2002)}]{Aprile:2002ef}
\bibinfo{author}{\bibfnamefont{E.}~\bibnamefont{Aprile}} \bibnamefont{et~al.},
  in \emph{\bibinfo{booktitle}{{Technique and application of xenon detectors.
  Proceedings, International Workshop, Kashiwa, Japan, December 3-4, 2001}}}
  (\bibinfo{year}{2002}), pp. \bibinfo{pages}{165--178},
  \eprint{astro-ph/0207670}.

\bibitem[{\citenamefont{Alner et~al.}(2005)}]{Alner:2005pa}
\bibinfo{author}{\bibfnamefont{G.~J.} \bibnamefont{Alner}} \bibnamefont{et~al.}
  (\bibinfo{collaboration}{UK Dark Matter}), \bibinfo{journal}{Astropart.
  Phys.} \textbf{\bibinfo{volume}{23}}, \bibinfo{pages}{444}
  (\bibinfo{year}{2005}).

\bibitem[{\citenamefont{Alner et~al.}(2007)}]{Alner:2007ja}
\bibinfo{author}{\bibfnamefont{G.~J.} \bibnamefont{Alner}}
  \bibnamefont{et~al.}, \bibinfo{journal}{Astropart. Phys.}
  \textbf{\bibinfo{volume}{28}}, \bibinfo{pages}{287} (\bibinfo{year}{2007}),
  \eprint{astro-ph/0701858}.

\bibitem[{\citenamefont{Lebedenko et~al.}(2009)}]{Lebedenko:2008gb}
\bibinfo{author}{\bibfnamefont{V.~N.} \bibnamefont{Lebedenko}}
  \bibnamefont{et~al.}, \bibinfo{journal}{Phys. Rev. D}
  \textbf{\bibinfo{volume}{80}}, \bibinfo{pages}{052010}
  (\bibinfo{year}{2009}), \eprint{0812.1150}.

\bibitem[{\citenamefont{Araujo et~al.}(2012)}]{Araujo:2011as}
\bibinfo{author}{\bibfnamefont{H.~M.} \bibnamefont{Araujo}}
  \bibnamefont{et~al.}, \bibinfo{journal}{Astropart. Phys.}
  \textbf{\bibinfo{volume}{35}}, \bibinfo{pages}{495} (\bibinfo{year}{2012}),
  \eprint{1104.3538}.

\bibitem[{\citenamefont{Angle et~al.}(2008)}]{Angle:2007uj}
\bibinfo{author}{\bibfnamefont{J.}~\bibnamefont{Angle}} \bibnamefont{et~al.}
  (\bibinfo{collaboration}{XENON}), \bibinfo{journal}{Phys. Rev. Lett.}
  \textbf{\bibinfo{volume}{100}}, \bibinfo{pages}{021303}
  (\bibinfo{year}{2008}), \eprint{0706.0039}.

\bibitem[{\citenamefont{Abe et~al.}(2019{\natexlab{a}})}]{XMASS:2018bid}
\bibinfo{author}{\bibfnamefont{K.}~\bibnamefont{Abe}} \bibnamefont{et~al.}
  (\bibinfo{collaboration}{XMASS}), \bibinfo{journal}{Phys. Lett. B}
  \textbf{\bibinfo{volume}{789}}, \bibinfo{pages}{45}
  (\bibinfo{year}{2019}{\natexlab{a}}), \eprint{1804.02180}.

\bibitem[{\citenamefont{Aprile et~al.}(2012{\natexlab{a}})}]{Aprile:2011dd}
\bibinfo{author}{\bibfnamefont{E.}~\bibnamefont{Aprile}} \bibnamefont{et~al.}
  (\bibinfo{collaboration}{XENON100}), \bibinfo{journal}{Astropart. Phys.}
  \textbf{\bibinfo{volume}{35}}, \bibinfo{pages}{573}
  (\bibinfo{year}{2012}{\natexlab{a}}), \eprint{1107.2155}.

\bibitem[{\citenamefont{Aprile et~al.}(2012{\natexlab{b}})}]{Aprile:2012nq}
\bibinfo{author}{\bibfnamefont{E.}~\bibnamefont{Aprile}} \bibnamefont{et~al.}
  (\bibinfo{collaboration}{XENON100}), \bibinfo{journal}{Phys. Rev. Lett.}
  \textbf{\bibinfo{volume}{109}}, \bibinfo{pages}{181301}
  (\bibinfo{year}{2012}{\natexlab{b}}), \eprint{1207.5988}.

\bibitem[{\citenamefont{Xiao et~al.}(2014)}]{Xiao:2014xyn}
\bibinfo{author}{\bibfnamefont{M.}~\bibnamefont{Xiao}} \bibnamefont{et~al.}
  (\bibinfo{collaboration}{PandaX}), \bibinfo{journal}{Sci. China Phys. Mech.
  Astron.} \textbf{\bibinfo{volume}{57}}, \bibinfo{pages}{2024}
  (\bibinfo{year}{2014}), \eprint{1408.5114}.

\bibitem[{\citenamefont{Wang et~al.}(2020{\natexlab{a}})}]{Wang:2020coa}
\bibinfo{author}{\bibfnamefont{Q.}~\bibnamefont{Wang}} \bibnamefont{et~al.}
  (\bibinfo{collaboration}{PandaX-II}), \bibinfo{journal}{Chin. Phys. C}
  \textbf{\bibinfo{volume}{44}}, \bibinfo{pages}{125001}
  (\bibinfo{year}{2020}{\natexlab{a}}), \eprint{2007.15469}.

\bibitem[{\citenamefont{Aprile et~al.}(2018{\natexlab{a}})}]{Aprile:2018dbl}
\bibinfo{author}{\bibfnamefont{E.}~\bibnamefont{Aprile}} \bibnamefont{et~al.}
  (\bibinfo{collaboration}{XENON}), \bibinfo{journal}{Phys. Rev. Lett.}
  \textbf{\bibinfo{volume}{121}}, \bibinfo{pages}{111302}
  (\bibinfo{year}{2018}{\natexlab{a}}), \eprint{1805.12562}.

\bibitem[{\citenamefont{Zhang et~al.}(2019)}]{Zhang:2018xdp}
\bibinfo{author}{\bibfnamefont{H.}~\bibnamefont{Zhang}} \bibnamefont{et~al.}
  (\bibinfo{collaboration}{PandaX}), \bibinfo{journal}{Sci. China Phys. Mech.
  Astron.} \textbf{\bibinfo{volume}{62}}, \bibinfo{pages}{31011}
  (\bibinfo{year}{2019}), \eprint{1806.02229}.

\bibitem[{\citenamefont{Aprile et~al.}(2020{\natexlab{a}})}]{Aprile:2020vtw}
\bibinfo{author}{\bibfnamefont{E.}~\bibnamefont{Aprile}} \bibnamefont{et~al.}
  (\bibinfo{collaboration}{XENON}), \bibinfo{journal}{JCAP}
  \textbf{\bibinfo{volume}{11}}, \bibinfo{pages}{031}
  (\bibinfo{year}{2020}{\natexlab{a}}), \eprint{2007.08796}.

\bibitem[{\citenamefont{Akerib et~al.}(2020{\natexlab{a}})}]{Akerib:2018lyp}
\bibinfo{author}{\bibfnamefont{D.~S.} \bibnamefont{Akerib}}
  \bibnamefont{et~al.} (\bibinfo{collaboration}{LUX-ZEPLIN}),
  \bibinfo{journal}{Phys. Rev. D} \textbf{\bibinfo{volume}{101}},
  \bibinfo{pages}{052002} (\bibinfo{year}{2020}{\natexlab{a}}),
  \eprint{1802.06039}.

\bibitem[{\citenamefont{Schumann et~al.}(2015)\citenamefont{Schumann, Baudis,
  B\"utikofer, Kish, and Selvi}}]{Schumann:2015cpa}
\bibinfo{author}{\bibfnamefont{M.}~\bibnamefont{Schumann}},
  \bibinfo{author}{\bibfnamefont{L.}~\bibnamefont{Baudis}},
  \bibinfo{author}{\bibfnamefont{L.}~\bibnamefont{B\"utikofer}},
  \bibinfo{author}{\bibfnamefont{A.}~\bibnamefont{Kish}}, \bibnamefont{and}
  \bibinfo{author}{\bibfnamefont{M.}~\bibnamefont{Selvi}},
  \bibinfo{journal}{JCAP} \textbf{\bibinfo{volume}{10}}, \bibinfo{pages}{016}
  (\bibinfo{year}{2015}), \eprint{1506.08309}.

\bibitem[{\citenamefont{Aprile et~al.}(2005)}]{Aprile:2004ey}
\bibinfo{author}{\bibfnamefont{E.}~\bibnamefont{Aprile}} \bibnamefont{et~al.},
  \bibinfo{journal}{Nucl. Phys. Proc. Suppl.} \textbf{\bibinfo{volume}{138}},
  \bibinfo{pages}{156} (\bibinfo{year}{2005}), \eprint{astro-ph/0407575}.

\bibitem[{\citenamefont{Akerib et~al.}(2013)}]{Akerib:2012da}
\bibinfo{author}{\bibfnamefont{D.~S.} \bibnamefont{Akerib}}
  \bibnamefont{et~al.}, \bibinfo{journal}{Nucl. Instrum. Meth. A}
  \textbf{\bibinfo{volume}{703}}, \bibinfo{pages}{1} (\bibinfo{year}{2013}),
  \eprint{1205.2272}.

\bibitem[{\citenamefont{Aprile et~al.}(2015{\natexlab{a}})}]{XENON:2015ara}
\bibinfo{author}{\bibfnamefont{E.}~\bibnamefont{Aprile}} \bibnamefont{et~al.}
  (\bibinfo{collaboration}{XENON}), \bibinfo{journal}{Eur. Phys. J. C}
  \textbf{\bibinfo{volume}{75}}, \bibinfo{pages}{546}
  (\bibinfo{year}{2015}{\natexlab{a}}), \eprint{1503.07698}.

\bibitem[{\citenamefont{Aprile et~al.}(2006{\natexlab{a}})\citenamefont{Aprile,
  Cushman, Ni, and Shagin}}]{Aprile:2005qq}
\bibinfo{author}{\bibfnamefont{E.}~\bibnamefont{Aprile}},
  \bibinfo{author}{\bibfnamefont{P.}~\bibnamefont{Cushman}},
  \bibinfo{author}{\bibfnamefont{K.}~\bibnamefont{Ni}}, \bibnamefont{and}
  \bibinfo{author}{\bibfnamefont{P.}~\bibnamefont{Shagin}},
  \bibinfo{journal}{Nucl. Instrum. Meth. A} \textbf{\bibinfo{volume}{556}},
  \bibinfo{pages}{215} (\bibinfo{year}{2006}{\natexlab{a}}),
  \eprint{physics/0501002}.

\bibitem[{\citenamefont{Baudis et~al.}(2020)\citenamefont{Baudis, Biondi,
  Galloway, Girard, Hochrein, Reichard, Sanchez-Lucas, Thieme, and
  Wulf}}]{Baudis:2020nwe}
\bibinfo{author}{\bibfnamefont{L.}~\bibnamefont{Baudis}},
  \bibinfo{author}{\bibfnamefont{Y.}~\bibnamefont{Biondi}},
  \bibinfo{author}{\bibfnamefont{M.}~\bibnamefont{Galloway}},
  \bibinfo{author}{\bibfnamefont{F.}~\bibnamefont{Girard}},
  \bibinfo{author}{\bibfnamefont{S.}~\bibnamefont{Hochrein}},
  \bibinfo{author}{\bibfnamefont{S.}~\bibnamefont{Reichard}},
  \bibinfo{author}{\bibfnamefont{P.}~\bibnamefont{Sanchez-Lucas}},
  \bibinfo{author}{\bibfnamefont{K.}~\bibnamefont{Thieme}}, \bibnamefont{and}
  \bibinfo{author}{\bibfnamefont{J.}~\bibnamefont{Wulf}},
  \bibinfo{journal}{Eur. Phys. J. C} \textbf{\bibinfo{volume}{80}},
  \bibinfo{pages}{477} (\bibinfo{year}{2020}), \eprint{2003.01731}.

\bibitem[{\citenamefont{Lansiart et~al.}(1976)}]{Lansiart:1976}
\bibinfo{author}{\bibfnamefont{A.}~\bibnamefont{Lansiart}}
  \bibnamefont{et~al.}, \bibinfo{journal}{Nucl. Instrum. Meth.}
  \textbf{\bibinfo{volume}{135}}, \bibinfo{pages}{47} (\bibinfo{year}{1976}),
  ISSN \bibinfo{issn}{0029-554X}.

\bibitem[{\citenamefont{Yamashita et~al.}(2003)\citenamefont{Yamashita, Doke,
  Kikuchi, and Suzuki}}]{Yamashita:2003rc}
\bibinfo{author}{\bibfnamefont{M.}~\bibnamefont{Yamashita}},
  \bibinfo{author}{\bibfnamefont{T.}~\bibnamefont{Doke}},
  \bibinfo{author}{\bibfnamefont{J.}~\bibnamefont{Kikuchi}}, \bibnamefont{and}
  \bibinfo{author}{\bibfnamefont{S.}~\bibnamefont{Suzuki}},
  \bibinfo{journal}{Astropart. Phys.} \textbf{\bibinfo{volume}{20}},
  \bibinfo{pages}{79} (\bibinfo{year}{2003}).

\bibitem[{\citenamefont{Mount et~al.}(2017)}]{Mount:2017qzi}
\bibinfo{author}{\bibfnamefont{B.~J.} \bibnamefont{Mount}} \bibnamefont{et~al.}
  (\bibinfo{year}{2017}), \eprint{1703.09144}.

\bibitem[{\citenamefont{Zhang et~al.}(2021)}]{PANDA-X:2021jua}
\bibinfo{author}{\bibfnamefont{D.}~\bibnamefont{Zhang}} \bibnamefont{et~al.}
  (\bibinfo{collaboration}{PandaX}), \bibinfo{journal}{JINST}
  \textbf{\bibinfo{volume}{16}}, \bibinfo{pages}{P11040}
  (\bibinfo{year}{2021}), \eprint{2106.08380}.

\bibitem[{\citenamefont{Liang et~al.}(2021{\natexlab{a}})\citenamefont{Liang,
  Higuera, Peters, Roy, Bajwa, Shatkay, and Tunnell}}]{Liang:2021nsz}
\bibinfo{author}{\bibfnamefont{S.}~\bibnamefont{Liang}},
  \bibinfo{author}{\bibfnamefont{A.}~\bibnamefont{Higuera}},
  \bibinfo{author}{\bibfnamefont{C.}~\bibnamefont{Peters}},
  \bibinfo{author}{\bibfnamefont{V.}~\bibnamefont{Roy}},
  \bibinfo{author}{\bibfnamefont{W.~U.} \bibnamefont{Bajwa}},
  \bibinfo{author}{\bibfnamefont{H.}~\bibnamefont{Shatkay}}, \bibnamefont{and}
  \bibinfo{author}{\bibfnamefont{C.~D.} \bibnamefont{Tunnell}}
  (\bibinfo{year}{2021}{\natexlab{a}}), \eprint{2112.07995}.

\bibitem[{\citenamefont{Akerib et~al.}(2018{\natexlab{a}})}]{LUX:2017lif}
\bibinfo{author}{\bibfnamefont{D.~S.} \bibnamefont{Akerib}}
  \bibnamefont{et~al.} (\bibinfo{collaboration}{LUX}), \bibinfo{journal}{JINST}
  \textbf{\bibinfo{volume}{13}}, \bibinfo{pages}{P02001}
  (\bibinfo{year}{2018}{\natexlab{a}}), \eprint{1710.02752}.

\bibitem[{\citenamefont{Angle et~al.}(2007)}]{Angle:2006rj}
\bibinfo{author}{\bibfnamefont{J.}~\bibnamefont{Angle}} \bibnamefont{et~al.}
  (\bibinfo{collaboration}{XENON}), \bibinfo{journal}{Nucl. Phys. Proc. Suppl.}
  \textbf{\bibinfo{volume}{173}}, \bibinfo{pages}{117} (\bibinfo{year}{2007}),
  \bibinfo{note}{[Erratum: Nucl. Phys. Proc. Suppl. {\bf 175}, E3 (2008)]},
  \eprint{astro-ph/0609714}.

\bibitem[{\citenamefont{Aprile}(2005)}]{Aprile:2005mz}
\bibinfo{author}{\bibfnamefont{E.}~\bibnamefont{Aprile}}
  (\bibinfo{collaboration}{XENON}) (\bibinfo{year}{2005}),
  \eprint{astro-ph/0502279}.

\bibitem[{\citenamefont{Akerib et~al.}(2020{\natexlab{b}})}]{Akerib:2020lkv}
\bibinfo{author}{\bibfnamefont{D.~S.} \bibnamefont{Akerib}}
  \bibnamefont{et~al.} (\bibinfo{collaboration}{LUX}), \bibinfo{journal}{Phys.
  Rev. D} \textbf{\bibinfo{volume}{102}}, \bibinfo{pages}{112002}
  (\bibinfo{year}{2020}{\natexlab{b}}), \eprint{2004.06304}.

\bibitem[{\citenamefont{Aprile et~al.}(2006{\natexlab{b}})\citenamefont{Aprile,
  Dahl, DeViveiros, Gaitskell, Giboni, Kwong, Majewski, Ni, Shutt, and
  Yamashita}}]{Aprile:2006kx}
\bibinfo{author}{\bibfnamefont{E.}~\bibnamefont{Aprile}},
  \bibinfo{author}{\bibfnamefont{C.}~\bibnamefont{Dahl}},
  \bibinfo{author}{\bibfnamefont{L.}~\bibnamefont{DeViveiros}},
  \bibinfo{author}{\bibfnamefont{R.}~\bibnamefont{Gaitskell}},
  \bibinfo{author}{\bibfnamefont{K.}~\bibnamefont{Giboni}},
  \bibinfo{author}{\bibfnamefont{J.}~\bibnamefont{Kwong}},
  \bibinfo{author}{\bibfnamefont{P.}~\bibnamefont{Majewski}},
  \bibinfo{author}{\bibfnamefont{K.}~\bibnamefont{Ni}},
  \bibinfo{author}{\bibfnamefont{T.}~\bibnamefont{Shutt}}, \bibnamefont{and}
  \bibinfo{author}{\bibfnamefont{M.}~\bibnamefont{Yamashita}},
  \bibinfo{journal}{Phys. Rev. Lett.} \textbf{\bibinfo{volume}{97}},
  \bibinfo{pages}{081302} (\bibinfo{year}{2006}{\natexlab{b}}),
  \eprint{astro-ph/0601552}.

\bibitem[{\citenamefont{Aprile et~al.}(2020{\natexlab{b}})}]{XENON:2020iwh}
\bibinfo{author}{\bibfnamefont{E.}~\bibnamefont{Aprile}} \bibnamefont{et~al.}
  (\bibinfo{collaboration}{XENON}), \bibinfo{journal}{Eur. Phys. J. C}
  \textbf{\bibinfo{volume}{80}}, \bibinfo{pages}{785}
  (\bibinfo{year}{2020}{\natexlab{b}}), \eprint{2003.03825}.

\bibitem[{\citenamefont{Doke et~al.}(2002)\citenamefont{Doke, Hitachi, Kikuchi,
  Masuda, Okada, and Shibamura}}]{Doke:2002oab}
\bibinfo{author}{\bibfnamefont{T.}~\bibnamefont{Doke}},
  \bibinfo{author}{\bibfnamefont{A.}~\bibnamefont{Hitachi}},
  \bibinfo{author}{\bibfnamefont{J.}~\bibnamefont{Kikuchi}},
  \bibinfo{author}{\bibfnamefont{K.}~\bibnamefont{Masuda}},
  \bibinfo{author}{\bibfnamefont{H.}~\bibnamefont{Okada}}, \bibnamefont{and}
  \bibinfo{author}{\bibfnamefont{E.}~\bibnamefont{Shibamura}},
  \bibinfo{journal}{Jap. J. Appl. Phys.} \textbf{\bibinfo{volume}{41}},
  \bibinfo{pages}{1538} (\bibinfo{year}{2002}).

\bibitem[{\citenamefont{Aprile et~al.}(2008)\citenamefont{Aprile, Bolotnikov,
  Bolozdynya, and Doke}}]{Aprile:2008bga}
\bibinfo{author}{\bibfnamefont{E.}~\bibnamefont{Aprile}},
  \bibinfo{author}{\bibfnamefont{A.~E.} \bibnamefont{Bolotnikov}},
  \bibinfo{author}{\bibfnamefont{A.~L.} \bibnamefont{Bolozdynya}},
  \bibnamefont{and} \bibinfo{author}{\bibfnamefont{T.}~\bibnamefont{Doke}},
  \emph{\bibinfo{title}{{Noble Gas Detectors}}} (\bibinfo{publisher}{Wiley},
  \bibinfo{year}{2008}), ISBN \bibinfo{isbn}{9783527405978, 9783527610020}.

\bibitem[{\citenamefont{Baudis et~al.}(2021)\citenamefont{Baudis,
  Sanchez-Lucas, and Thieme}}]{Baudis:2021dsq}
\bibinfo{author}{\bibfnamefont{L.}~\bibnamefont{Baudis}},
  \bibinfo{author}{\bibfnamefont{P.}~\bibnamefont{Sanchez-Lucas}},
  \bibnamefont{and} \bibinfo{author}{\bibfnamefont{K.}~\bibnamefont{Thieme}},
  \bibinfo{journal}{Eur. Phys. J. C} \textbf{\bibinfo{volume}{81}},
  \bibinfo{pages}{1060} (\bibinfo{year}{2021}), \eprint{2109.07151}.

\bibitem[{\citenamefont{Chepel and Araujo}(2013)}]{Chepel:2012sj}
\bibinfo{author}{\bibfnamefont{V.}~\bibnamefont{Chepel}} \bibnamefont{and}
  \bibinfo{author}{\bibfnamefont{H.}~\bibnamefont{Araujo}},
  \bibinfo{journal}{JINST} \textbf{\bibinfo{volume}{8}},
  \bibinfo{pages}{R04001} (\bibinfo{year}{2013}), \eprint{1207.2292}.

\bibitem[{\citenamefont{Lenardo et~al.}(2015)\citenamefont{Lenardo, Kazkaz,
  Manalaysay, Mock, Szydagis, and Tripathi}}]{Lenardo:2014cva}
\bibinfo{author}{\bibfnamefont{B.}~\bibnamefont{Lenardo}},
  \bibinfo{author}{\bibfnamefont{K.}~\bibnamefont{Kazkaz}},
  \bibinfo{author}{\bibfnamefont{A.}~\bibnamefont{Manalaysay}},
  \bibinfo{author}{\bibfnamefont{J.}~\bibnamefont{Mock}},
  \bibinfo{author}{\bibfnamefont{M.}~\bibnamefont{Szydagis}}, \bibnamefont{and}
  \bibinfo{author}{\bibfnamefont{M.}~\bibnamefont{Tripathi}},
  \bibinfo{journal}{IEEE Trans. Nucl. Sci.} \textbf{\bibinfo{volume}{62}},
  \bibinfo{pages}{3387} (\bibinfo{year}{2015}), \eprint{1412.4417}.

\bibitem[{\citenamefont{Aprile et~al.}(2018{\natexlab{b}})\citenamefont{Aprile,
  Anthony, Lin, Greene, De~Perio, Gao, Howlett, Plante, Zhang, and
  Zhu}}]{Aprile:2018jvg}
\bibinfo{author}{\bibfnamefont{E.}~\bibnamefont{Aprile}},
  \bibinfo{author}{\bibfnamefont{M.}~\bibnamefont{Anthony}},
  \bibinfo{author}{\bibfnamefont{Q.}~\bibnamefont{Lin}},
  \bibinfo{author}{\bibfnamefont{Z.}~\bibnamefont{Greene}},
  \bibinfo{author}{\bibfnamefont{P.}~\bibnamefont{De~Perio}},
  \bibinfo{author}{\bibfnamefont{F.}~\bibnamefont{Gao}},
  \bibinfo{author}{\bibfnamefont{J.}~\bibnamefont{Howlett}},
  \bibinfo{author}{\bibfnamefont{G.}~\bibnamefont{Plante}},
  \bibinfo{author}{\bibfnamefont{Y.}~\bibnamefont{Zhang}}, \bibnamefont{and}
  \bibinfo{author}{\bibfnamefont{T.}~\bibnamefont{Zhu}},
  \bibinfo{journal}{Phys. Rev. D} \textbf{\bibinfo{volume}{98}},
  \bibinfo{pages}{112003} (\bibinfo{year}{2018}{\natexlab{b}}),
  \eprint{1809.02072}.

\bibitem[{\citenamefont{Yan et~al.}(2021)}]{PandaX-II:2021jmq}
\bibinfo{author}{\bibfnamefont{B.}~\bibnamefont{Yan}} \bibnamefont{et~al.}
  (\bibinfo{collaboration}{PandaX-II}), \bibinfo{journal}{Chin. Phys. C}
  \textbf{\bibinfo{volume}{45}}, \bibinfo{pages}{075001}
  (\bibinfo{year}{2021}), \eprint{2102.09158}.

\bibitem[{\citenamefont{Solovov et~al.}(2012)}]{Solovov:2011aa}
\bibinfo{author}{\bibfnamefont{V.~N.} \bibnamefont{Solovov}}
  \bibnamefont{et~al.}, \bibinfo{journal}{IEEE Trans. Nucl. Sci.}
  \textbf{\bibinfo{volume}{59}}, \bibinfo{pages}{3286} (\bibinfo{year}{2012}),
  \eprint{1112.1481}.

\bibitem[{\citenamefont{Aprile et~al.}(2014{\natexlab{a}})}]{Aprile:2012vw}
\bibinfo{author}{\bibfnamefont{E.}~\bibnamefont{Aprile}} \bibnamefont{et~al.}
  (\bibinfo{collaboration}{XENON100}), \bibinfo{journal}{Astropart. Phys.}
  \textbf{\bibinfo{volume}{54}}, \bibinfo{pages}{11}
  (\bibinfo{year}{2014}{\natexlab{a}}), \eprint{1207.3458}.

\bibitem[{\citenamefont{Szydagis et~al.}(2011)\citenamefont{Szydagis, Barry,
  Kazkaz, Mock, Stolp, Sweany, Tripathi, Uvarov, Walsh, and
  Woods}}]{Szydagis:2011tk}
\bibinfo{author}{\bibfnamefont{M.}~\bibnamefont{Szydagis}},
  \bibinfo{author}{\bibfnamefont{N.}~\bibnamefont{Barry}},
  \bibinfo{author}{\bibfnamefont{K.}~\bibnamefont{Kazkaz}},
  \bibinfo{author}{\bibfnamefont{J.}~\bibnamefont{Mock}},
  \bibinfo{author}{\bibfnamefont{D.}~\bibnamefont{Stolp}},
  \bibinfo{author}{\bibfnamefont{M.}~\bibnamefont{Sweany}},
  \bibinfo{author}{\bibfnamefont{M.}~\bibnamefont{Tripathi}},
  \bibinfo{author}{\bibfnamefont{S.}~\bibnamefont{Uvarov}},
  \bibinfo{author}{\bibfnamefont{N.}~\bibnamefont{Walsh}}, \bibnamefont{and}
  \bibinfo{author}{\bibfnamefont{M.}~\bibnamefont{Woods}},
  \bibinfo{journal}{JINST} \textbf{\bibinfo{volume}{6}},
  \bibinfo{pages}{P10002} (\bibinfo{year}{2011}), \eprint{1106.1613}.

\bibitem[{\citenamefont{Szydagis et~al.}(2013)\citenamefont{Szydagis, Fyhrie,
  Thorngren, and Tripathi}}]{Szydagis:2013sih}
\bibinfo{author}{\bibfnamefont{M.}~\bibnamefont{Szydagis}},
  \bibinfo{author}{\bibfnamefont{A.}~\bibnamefont{Fyhrie}},
  \bibinfo{author}{\bibfnamefont{D.}~\bibnamefont{Thorngren}},
  \bibnamefont{and} \bibinfo{author}{\bibfnamefont{M.}~\bibnamefont{Tripathi}},
  \bibinfo{journal}{JINST} \textbf{\bibinfo{volume}{8}},
  \bibinfo{pages}{C10003} (\bibinfo{year}{2013}), \eprint{1307.6601}.

\bibitem[{\citenamefont{Szydagis et~al.}(2022)\citenamefont{Szydagis, Balajthy,
  Brodsky, Cutter, Farrell, Huang, Kozlova, Lenardo, Manalaysay, McKinsey
  et~al.}}]{szydagis_m_2022_6028483}
\bibinfo{author}{\bibfnamefont{M.}~\bibnamefont{Szydagis}},
  \bibinfo{author}{\bibfnamefont{J.}~\bibnamefont{Balajthy}},
  \bibinfo{author}{\bibfnamefont{J.}~\bibnamefont{Brodsky}},
  \bibinfo{author}{\bibfnamefont{J.}~\bibnamefont{Cutter}},
  \bibinfo{author}{\bibfnamefont{S.}~\bibnamefont{Farrell}},
  \bibinfo{author}{\bibfnamefont{J.}~\bibnamefont{Huang}},
  \bibinfo{author}{\bibfnamefont{E.}~\bibnamefont{Kozlova}},
  \bibinfo{author}{\bibfnamefont{B.}~\bibnamefont{Lenardo}},
  \bibinfo{author}{\bibfnamefont{A.}~\bibnamefont{Manalaysay}},
  \bibinfo{author}{\bibfnamefont{D.}~\bibnamefont{McKinsey}},
  \bibnamefont{et~al.} (\bibinfo{collaboration}{NEST Collaboration}),
  \emph{\bibinfo{title}{Noble element simulation technique v2.3.5}}
  (\bibinfo{year}{2022}), \bibinfo{note}{zenodo:6028483},
  \urlprefix\url{https://zenodo.org/record/6028483}.

\bibitem[{\citenamefont{Sorensen et~al.}(2009)}]{Sorensen:2008ec}
\bibinfo{author}{\bibfnamefont{P.}~\bibnamefont{Sorensen}} \bibnamefont{et~al.}
  (\bibinfo{collaboration}{XENON10}), \bibinfo{journal}{Nucl. Instrum. Meth. A}
  \textbf{\bibinfo{volume}{601}}, \bibinfo{pages}{339} (\bibinfo{year}{2009}),
  \eprint{0807.0459}.

\bibitem[{\citenamefont{Dahl}(2009)}]{Dahl:2009nta}
\bibinfo{author}{\bibfnamefont{C.~E.} \bibnamefont{Dahl}}, Ph.D. thesis,
  \bibinfo{school}{Princeton U.} (\bibinfo{year}{2009}),
  \urlprefix\url{https://www.princeton.edu/physics/graduate-program/theses/theses-from-2009/E.Dahlthesis.pdf}.

\bibitem[{\citenamefont{Bezrukov et~al.}(2011)\citenamefont{Bezrukov,
  Kahlhoefer, Lindner, Kahlhoefer, and Lindner}}]{Bezrukov:2010qa}
\bibinfo{author}{\bibfnamefont{F.}~\bibnamefont{Bezrukov}},
  \bibinfo{author}{\bibfnamefont{F.}~\bibnamefont{Kahlhoefer}},
  \bibinfo{author}{\bibfnamefont{M.}~\bibnamefont{Lindner}},
  \bibinfo{author}{\bibfnamefont{F.}~\bibnamefont{Kahlhoefer}},
  \bibnamefont{and} \bibinfo{author}{\bibfnamefont{M.}~\bibnamefont{Lindner}},
  \bibinfo{journal}{Astropart. Phys.} \textbf{\bibinfo{volume}{35}},
  \bibinfo{pages}{119} (\bibinfo{year}{2011}), \eprint{1011.3990}.

\bibitem[{\citenamefont{Sorensen}(2010)}]{Sorensen:2010hq}
\bibinfo{author}{\bibfnamefont{P.}~\bibnamefont{Sorensen}},
  \bibinfo{journal}{JCAP} \textbf{\bibinfo{volume}{1009}}, \bibinfo{pages}{033}
  (\bibinfo{year}{2010}), \eprint{1007.3549}.

\bibitem[{\citenamefont{Mu et~al.}(2015)\citenamefont{Mu, Xiong, and
  Ji}}]{Mu:2013dga}
\bibinfo{author}{\bibfnamefont{W.}~\bibnamefont{Mu}},
  \bibinfo{author}{\bibfnamefont{X.}~\bibnamefont{Xiong}}, \bibnamefont{and}
  \bibinfo{author}{\bibfnamefont{X.}~\bibnamefont{Ji}},
  \bibinfo{journal}{Astropart. Phys.} \textbf{\bibinfo{volume}{61}},
  \bibinfo{pages}{56} (\bibinfo{year}{2015}), \bibinfo{note}{[Erratum:
  Astropart. Phys. {\bf 72}, 109 (2016)]}, \eprint{1306.0170}.

\bibitem[{\citenamefont{Mu and Ji}(2015)}]{Mu:2013pja}
\bibinfo{author}{\bibfnamefont{W.}~\bibnamefont{Mu}} \bibnamefont{and}
  \bibinfo{author}{\bibfnamefont{X.}~\bibnamefont{Ji}},
  \bibinfo{journal}{Astropart. Phys.} \textbf{\bibinfo{volume}{62}},
  \bibinfo{pages}{108} (\bibinfo{year}{2015}), \eprint{1310.2094}.

\bibitem[{\citenamefont{Wang and Mei}(2017)}]{Wang:2016obw}
\bibinfo{author}{\bibfnamefont{L.}~\bibnamefont{Wang}} \bibnamefont{and}
  \bibinfo{author}{\bibfnamefont{D.~M.} \bibnamefont{Mei}},
  \bibinfo{journal}{J. Phys. G} \textbf{\bibinfo{volume}{44}},
  \bibinfo{pages}{055001} (\bibinfo{year}{2017}), \eprint{1604.01083}.

\bibitem[{\citenamefont{Abe et~al.}(2009)}]{Abe:2008py}
\bibinfo{author}{\bibfnamefont{K.}~\bibnamefont{Abe}} \bibnamefont{et~al.}
  (\bibinfo{collaboration}{XMASS}), \bibinfo{journal}{Astropart. Phys.}
  \textbf{\bibinfo{volume}{31}}, \bibinfo{pages}{290} (\bibinfo{year}{2009}),
  \eprint{0809.4413}.

\bibitem[{\citenamefont{Aprile et~al.}(2016{\natexlab{a}})}]{Aprile:2016swn}
\bibinfo{author}{\bibfnamefont{E.}~\bibnamefont{Aprile}} \bibnamefont{et~al.}
  (\bibinfo{collaboration}{XENON100}), \bibinfo{journal}{Phys. Rev. D}
  \textbf{\bibinfo{volume}{94}}, \bibinfo{pages}{122001}
  (\bibinfo{year}{2016}{\natexlab{a}}), \eprint{1609.06154}.

\bibitem[{\citenamefont{Albert et~al.}(2014)}]{Albert:2014awa}
\bibinfo{author}{\bibfnamefont{J.~B.} \bibnamefont{Albert}}
  \bibnamefont{et~al.} (\bibinfo{collaboration}{EXO-200}),
  \bibinfo{journal}{Nature} \textbf{\bibinfo{volume}{510}},
  \bibinfo{pages}{229} (\bibinfo{year}{2014}), \eprint{1402.6956}.

\bibitem[{\citenamefont{Aalbers et~al.}(2022)}]{LUX-ZEPLIN:2022sad}
\bibinfo{author}{\bibfnamefont{J.}~\bibnamefont{Aalbers}} \bibnamefont{et~al.}
  (\bibinfo{collaboration}{LUX-ZEPLIN}) (\bibinfo{year}{2022}),
  \eprint{2201.02858}.

\bibitem[{\citenamefont{Jungman et~al.}(1996)\citenamefont{Jungman,
  Kamionkowski, and Griest}}]{Jungman:1995df}
\bibinfo{author}{\bibfnamefont{G.}~\bibnamefont{Jungman}},
  \bibinfo{author}{\bibfnamefont{M.}~\bibnamefont{Kamionkowski}},
  \bibnamefont{and} \bibinfo{author}{\bibfnamefont{K.}~\bibnamefont{Griest}},
  \bibinfo{journal}{Phys. Rept.} \textbf{\bibinfo{volume}{267}},
  \bibinfo{pages}{195} (\bibinfo{year}{1996}), \eprint{hep-ph/9506380}.

\bibitem[{\citenamefont{Gelmini and Gondolo}(2010)}]{Gelmini:2010zh}
\bibinfo{author}{\bibfnamefont{G.}~\bibnamefont{Gelmini}} \bibnamefont{and}
  \bibinfo{author}{\bibfnamefont{P.}~\bibnamefont{Gondolo}}, pp.
  \bibinfo{pages}{121--141} (\bibinfo{year}{2010}), \eprint{1009.3690}.

\bibitem[{\citenamefont{Arcadi et~al.}(2018{\natexlab{a}})\citenamefont{Arcadi,
  Dutra, Ghosh, Lindner, Mambrini, Pierre, Profumo, and
  Queiroz}}]{Arcadi:2017kky}
\bibinfo{author}{\bibfnamefont{G.}~\bibnamefont{Arcadi}},
  \bibinfo{author}{\bibfnamefont{M.}~\bibnamefont{Dutra}},
  \bibinfo{author}{\bibfnamefont{P.}~\bibnamefont{Ghosh}},
  \bibinfo{author}{\bibfnamefont{M.}~\bibnamefont{Lindner}},
  \bibinfo{author}{\bibfnamefont{Y.}~\bibnamefont{Mambrini}},
  \bibinfo{author}{\bibfnamefont{M.}~\bibnamefont{Pierre}},
  \bibinfo{author}{\bibfnamefont{S.}~\bibnamefont{Profumo}}, \bibnamefont{and}
  \bibinfo{author}{\bibfnamefont{F.~S.} \bibnamefont{Queiroz}},
  \bibinfo{journal}{Eur. Phys. J. C} \textbf{\bibinfo{volume}{78}},
  \bibinfo{pages}{203} (\bibinfo{year}{2018}{\natexlab{a}}),
  \eprint{1703.07364}.

\bibitem[{\citenamefont{Witten}(1981)}]{Witten:1981nf}
\bibinfo{author}{\bibfnamefont{E.}~\bibnamefont{Witten}},
  \bibinfo{journal}{Nucl. Phys. B} \textbf{\bibinfo{volume}{188}},
  \bibinfo{pages}{513} (\bibinfo{year}{1981}).

\bibitem[{\citenamefont{Bertone}(2010)}]{Bertone:2010zza}
\bibinfo{editor}{\bibfnamefont{G.}~\bibnamefont{Bertone}}, ed.,
  \emph{\bibinfo{title}{{Particle Dark Matter: Observations, Models and
  Searches}}} (\bibinfo{publisher}{Cambridge Univ. Press},
  \bibinfo{address}{Cambridge}, \bibinfo{year}{2010}), ISBN
  \bibinfo{isbn}{9781107653924},
  \urlprefix\url{http://www.cambridge.org/uk/catalogue/catalogue.asp?isbn=9780521763684}.

\bibitem[{\citenamefont{Steigman et~al.}(2012)\citenamefont{Steigman, Dasgupta,
  and Beacom}}]{Steigman:2012nb}
\bibinfo{author}{\bibfnamefont{G.}~\bibnamefont{Steigman}},
  \bibinfo{author}{\bibfnamefont{B.}~\bibnamefont{Dasgupta}}, \bibnamefont{and}
  \bibinfo{author}{\bibfnamefont{J.~F.} \bibnamefont{Beacom}},
  \bibinfo{journal}{Phys. Rev. D} \textbf{\bibinfo{volume}{86}},
  \bibinfo{pages}{023506} (\bibinfo{year}{2012}), \eprint{1204.3622}.

\bibitem[{\citenamefont{Kolb and Olive}(1986)}]{Kolb:1985nn}
\bibinfo{author}{\bibfnamefont{E.~W.} \bibnamefont{Kolb}} \bibnamefont{and}
  \bibinfo{author}{\bibfnamefont{K.~A.} \bibnamefont{Olive}},
  \bibinfo{journal}{Phys. Rev. D} \textbf{\bibinfo{volume}{33}},
  \bibinfo{pages}{1202} (\bibinfo{year}{1986}), \bibinfo{note}{[Erratum: Phys.
  Rev. D {\bf 34}, 2531 (1986)]}.

\bibitem[{\citenamefont{Bal\'azs et~al.}(2014)\citenamefont{Bal\'azs, Li, and
  Newstead}}]{Balazs:2014rsa}
\bibinfo{author}{\bibfnamefont{C.}~\bibnamefont{Bal\'azs}},
  \bibinfo{author}{\bibfnamefont{T.}~\bibnamefont{Li}}, \bibnamefont{and}
  \bibinfo{author}{\bibfnamefont{J.~L.} \bibnamefont{Newstead}},
  \bibinfo{journal}{JHEP} \textbf{\bibinfo{volume}{08}}, \bibinfo{pages}{061}
  (\bibinfo{year}{2014}), \eprint{1403.5829}.

\bibitem[{\citenamefont{Roszkowski et~al.}(2018)\citenamefont{Roszkowski,
  Sessolo, and Trojanowski}}]{Roszkowski:2017nbc}
\bibinfo{author}{\bibfnamefont{L.}~\bibnamefont{Roszkowski}},
  \bibinfo{author}{\bibfnamefont{E.~M.} \bibnamefont{Sessolo}},
  \bibnamefont{and}
  \bibinfo{author}{\bibfnamefont{S.}~\bibnamefont{Trojanowski}},
  \bibinfo{journal}{Rept. Prog. Phys.} \textbf{\bibinfo{volume}{81}},
  \bibinfo{pages}{066201} (\bibinfo{year}{2018}), \eprint{1707.06277}.

\bibitem[{\citenamefont{Kahlhoefer}(2017)}]{Kahlhoefer:2017dnp}
\bibinfo{author}{\bibfnamefont{F.}~\bibnamefont{Kahlhoefer}},
  \bibinfo{journal}{Int. J. Mod. Phys. A} \textbf{\bibinfo{volume}{32}},
  \bibinfo{pages}{1730006} (\bibinfo{year}{2017}), \eprint{1702.02430}.

\bibitem[{\citenamefont{Gaskins}(2016)}]{Gaskins:2016cha}
\bibinfo{author}{\bibfnamefont{J.~M.} \bibnamefont{Gaskins}},
  \bibinfo{journal}{Contemp. Phys.} \textbf{\bibinfo{volume}{57}},
  \bibinfo{pages}{496} (\bibinfo{year}{2016}), \eprint{1604.00014}.

\bibitem[{\citenamefont{Marrodán~Undagoitia and
  Rauch}(2016)}]{Undagoitia:2015gya}
\bibinfo{author}{\bibfnamefont{T.}~\bibnamefont{Marrodán~Undagoitia}}
  \bibnamefont{and} \bibinfo{author}{\bibfnamefont{L.}~\bibnamefont{Rauch}},
  \bibinfo{journal}{J. Phys. G} \textbf{\bibinfo{volume}{43}},
  \bibinfo{pages}{013001} (\bibinfo{year}{2016}), \eprint{1509.08767}.

\bibitem[{\citenamefont{Ong}(2018)}]{Ong:2017ihp}
\bibinfo{author}{\bibfnamefont{R.~A.} \bibnamefont{Ong}}
  (\bibinfo{collaboration}{CTA}), \bibinfo{journal}{PoS}
  \textbf{\bibinfo{volume}{ICRC2017}}, \bibinfo{pages}{1071}
  (\bibinfo{year}{2018}), \eprint{1709.05434}.

\bibitem[{\citenamefont{Arduini et~al.}(2016)}]{Arduini:2016xsb}
\bibinfo{author}{\bibfnamefont{G.}~\bibnamefont{Arduini}} \bibnamefont{et~al.},
  \bibinfo{journal}{JINST} \textbf{\bibinfo{volume}{11}},
  \bibinfo{pages}{C12081} (\bibinfo{year}{2016}).

\bibitem[{\citenamefont{Smith and Lewin}(1985)}]{Smith:1986ms}
\bibinfo{author}{\bibfnamefont{P.~F.} \bibnamefont{Smith}} \bibnamefont{and}
  \bibinfo{author}{\bibfnamefont{J.~D.} \bibnamefont{Lewin}},
  \bibinfo{journal}{Acta Phys. Polon. B} \textbf{\bibinfo{volume}{16}},
  \bibinfo{pages}{837} (\bibinfo{year}{1985}).

\bibitem[{\citenamefont{Bramante et~al.}(2018)\citenamefont{Bramante, Broerman,
  Lang, and Raj}}]{Bramante:2018qbc}
\bibinfo{author}{\bibfnamefont{J.}~\bibnamefont{Bramante}},
  \bibinfo{author}{\bibfnamefont{B.}~\bibnamefont{Broerman}},
  \bibinfo{author}{\bibfnamefont{R.~F.} \bibnamefont{Lang}}, \bibnamefont{and}
  \bibinfo{author}{\bibfnamefont{N.}~\bibnamefont{Raj}},
  \bibinfo{journal}{Phys. Rev. D} \textbf{\bibinfo{volume}{98}},
  \bibinfo{pages}{083516} (\bibinfo{year}{2018}), \eprint{1803.08044}.

\bibitem[{\citenamefont{Newstead et~al.}(2013)\citenamefont{Newstead, Jacques,
  Krauss, Dent, and Ferrer}}]{Newstead:2013pea}
\bibinfo{author}{\bibfnamefont{J.~L.} \bibnamefont{Newstead}},
  \bibinfo{author}{\bibfnamefont{T.~D.} \bibnamefont{Jacques}},
  \bibinfo{author}{\bibfnamefont{L.~M.} \bibnamefont{Krauss}},
  \bibinfo{author}{\bibfnamefont{J.~B.} \bibnamefont{Dent}}, \bibnamefont{and}
  \bibinfo{author}{\bibfnamefont{F.}~\bibnamefont{Ferrer}},
  \bibinfo{journal}{Phys. Rev. D} \textbf{\bibinfo{volume}{88}},
  \bibinfo{pages}{076011} (\bibinfo{year}{2013}), \eprint{1306.3244}.

\bibitem[{\citenamefont{Akerib et~al.}(2018{\natexlab{b}})}]{Akerib:2017vbi}
\bibinfo{author}{\bibfnamefont{D.~S.} \bibnamefont{Akerib}}
  \bibnamefont{et~al.} (\bibinfo{collaboration}{LUX}), \bibinfo{journal}{Phys.
  Rev. D} \textbf{\bibinfo{volume}{97}}, \bibinfo{pages}{102008}
  (\bibinfo{year}{2018}{\natexlab{b}}), \eprint{1712.05696}.

\bibitem[{\citenamefont{Aprile et~al.}(2019{\natexlab{a}})}]{Aprile:2019bbb}
\bibinfo{author}{\bibfnamefont{E.}~\bibnamefont{Aprile}} \bibnamefont{et~al.}
  (\bibinfo{collaboration}{XENON}), \bibinfo{journal}{Phys. Rev. D}
  \textbf{\bibinfo{volume}{100}}, \bibinfo{pages}{052014}
  (\bibinfo{year}{2019}{\natexlab{a}}), \eprint{1906.04717}.

\bibitem[{\citenamefont{Wilks}(1938)}]{Wilks:1938dza}
\bibinfo{author}{\bibfnamefont{S.}~\bibnamefont{Wilks}},
  \bibinfo{journal}{Annals Math. Statist.} \textbf{\bibinfo{volume}{9}},
  \bibinfo{pages}{60} (\bibinfo{year}{1938}).

\bibitem[{\citenamefont{Akerib et~al.}(2020{\natexlab{c}})}]{LUX:2019ius}
\bibinfo{author}{\bibfnamefont{D.~S.} \bibnamefont{Akerib}}
  \bibnamefont{et~al.} (\bibinfo{collaboration}{LUX}), \bibinfo{journal}{JINST}
  \textbf{\bibinfo{volume}{15}}, \bibinfo{pages}{T02007}
  (\bibinfo{year}{2020}{\natexlab{c}}), \eprint{1910.04211}.

\bibitem[{Kot()}]{KotilaIachelloWebSite}
\emph{\bibinfo{title}{Double beta decay phase-space factors, low-energy region
  files}},
  \bibinfo{howpublished}{\url{https://nucleartheory.yale.edu/double-beta-decay-phase-space-factors}},
  \bibinfo{note}{\url{https://nucleartheory.yale.edu/sites/default/files/files/136Xe_2vbb.zip}}.

\bibitem[{\citenamefont{Chen et~al.}(2017{\natexlab{a}})\citenamefont{Chen,
  Chi, Liu, and Wu}}]{Chen:2016eab}
\bibinfo{author}{\bibfnamefont{J.-W.} \bibnamefont{Chen}},
  \bibinfo{author}{\bibfnamefont{H.-C.} \bibnamefont{Chi}},
  \bibinfo{author}{\bibfnamefont{C.~P.} \bibnamefont{Liu}}, \bibnamefont{and}
  \bibinfo{author}{\bibfnamefont{C.-P.} \bibnamefont{Wu}},
  \bibinfo{journal}{Phys. Lett. B} \textbf{\bibinfo{volume}{774}},
  \bibinfo{pages}{656} (\bibinfo{year}{2017}{\natexlab{a}}),
  \eprint{1610.04177}.

\bibitem[{\citenamefont{Billard et~al.}(2014)\citenamefont{Billard, Strigari,
  and Figueroa-Feliciano}}]{Billard:2013qya}
\bibinfo{author}{\bibfnamefont{J.}~\bibnamefont{Billard}},
  \bibinfo{author}{\bibfnamefont{L.}~\bibnamefont{Strigari}}, \bibnamefont{and}
  \bibinfo{author}{\bibfnamefont{E.}~\bibnamefont{Figueroa-Feliciano}},
  \bibinfo{journal}{Phys. Rev. D} \textbf{\bibinfo{volume}{89}},
  \bibinfo{pages}{023524} (\bibinfo{year}{2014}), \eprint{1307.5458}.

\bibitem[{\citenamefont{Newstead et~al.}(2021)\citenamefont{Newstead, Lang, and
  Strigari}}]{Newstead:2020fie}
\bibinfo{author}{\bibfnamefont{J.~L.} \bibnamefont{Newstead}},
  \bibinfo{author}{\bibfnamefont{R.~F.} \bibnamefont{Lang}}, \bibnamefont{and}
  \bibinfo{author}{\bibfnamefont{L.~E.} \bibnamefont{Strigari}},
  \bibinfo{journal}{Phys. Rev. D} \textbf{\bibinfo{volume}{104}},
  \bibinfo{pages}{115022} (\bibinfo{year}{2021}), \eprint{2002.08566}.

\bibitem[{\citenamefont{Aalbers et~al.}(2019)\citenamefont{Aalbers, Pelssers,
  and Morå}}]{wimprates}
\bibinfo{author}{\bibfnamefont{J.}~\bibnamefont{Aalbers}},
  \bibinfo{author}{\bibfnamefont{B.}~\bibnamefont{Pelssers}}, \bibnamefont{and}
  \bibinfo{author}{\bibfnamefont{K.~D.} \bibnamefont{Morå}},
  \emph{\bibinfo{title}{wimprates: v0.3.0}} (\bibinfo{year}{2019}),
  \urlprefix\url{https://doi.org/10.5281/zenodo.3345959}.

\bibitem[{\citenamefont{Lewin and Smith}(1996)}]{Lewin:1995rx}
\bibinfo{author}{\bibfnamefont{J.~D.} \bibnamefont{Lewin}} \bibnamefont{and}
  \bibinfo{author}{\bibfnamefont{P.~F.} \bibnamefont{Smith}},
  \bibinfo{journal}{Astropart. Phys.} \textbf{\bibinfo{volume}{6}},
  \bibinfo{pages}{87} (\bibinfo{year}{1996}).

\bibitem[{\citenamefont{Klos et~al.}(2013)\citenamefont{Klos, Men{\'e}ndez,
  Gazit, and Schwenk}}]{Klos:2013rwa}
\bibinfo{author}{\bibfnamefont{P.}~\bibnamefont{Klos}},
  \bibinfo{author}{\bibfnamefont{J.}~\bibnamefont{Men{\'e}ndez}},
  \bibinfo{author}{\bibfnamefont{D.}~\bibnamefont{Gazit}}, \bibnamefont{and}
  \bibinfo{author}{\bibfnamefont{A.}~\bibnamefont{Schwenk}},
  \bibinfo{journal}{Phys. Rev. D} \textbf{\bibinfo{volume}{88}},
  \bibinfo{pages}{083516} (\bibinfo{year}{2013}), \bibinfo{note}{[Erratum:
  Phys. Rev. D {\bf 89}, 029901 (2014)]}, \eprint{1304.7684}.

\bibitem[{\citenamefont{Aprile et~al.}(2019{\natexlab{b}})}]{Aprile:2018cxk}
\bibinfo{author}{\bibfnamefont{E.}~\bibnamefont{Aprile}} \bibnamefont{et~al.}
  (\bibinfo{collaboration}{XENON}), \bibinfo{journal}{Phys. Rev. Lett.}
  \textbf{\bibinfo{volume}{122}}, \bibinfo{pages}{071301}
  (\bibinfo{year}{2019}{\natexlab{b}}), \eprint{1811.12482}.

\bibitem[{\citenamefont{Hoferichter et~al.}(2019)\citenamefont{Hoferichter,
  Klos, Men\'endez, and Schwenk}}]{Hoferichter:2018acd}
\bibinfo{author}{\bibfnamefont{M.}~\bibnamefont{Hoferichter}},
  \bibinfo{author}{\bibfnamefont{P.}~\bibnamefont{Klos}},
  \bibinfo{author}{\bibfnamefont{J.}~\bibnamefont{Men\'endez}},
  \bibnamefont{and} \bibinfo{author}{\bibfnamefont{A.}~\bibnamefont{Schwenk}},
  \bibinfo{journal}{Phys. Rev. D} \textbf{\bibinfo{volume}{99}},
  \bibinfo{pages}{055031} (\bibinfo{year}{2019}), \eprint{1812.05617}.

\bibitem[{\citenamefont{Billard et~al.}(2015)\citenamefont{Billard, Strigari,
  and Figueroa-Feliciano}}]{Billard:2014yka}
\bibinfo{author}{\bibfnamefont{J.}~\bibnamefont{Billard}},
  \bibinfo{author}{\bibfnamefont{L.~E.} \bibnamefont{Strigari}},
  \bibnamefont{and}
  \bibinfo{author}{\bibfnamefont{E.}~\bibnamefont{Figueroa-Feliciano}},
  \bibinfo{journal}{Phys. Rev. D} \textbf{\bibinfo{volume}{91}},
  \bibinfo{pages}{095023} (\bibinfo{year}{2015}), \eprint{1409.0050}.

\bibitem[{\citenamefont{O'Hare}(2016)}]{OHare:2016pjy}
\bibinfo{author}{\bibfnamefont{C.~A.~J.} \bibnamefont{O'Hare}},
  \bibinfo{journal}{Phys. Rev. D} \textbf{\bibinfo{volume}{94}},
  \bibinfo{pages}{063527} (\bibinfo{year}{2016}), \eprint{1604.03858}.

\bibitem[{\citenamefont{Meng et~al.}(2021)}]{PandaX-4T:2021bab}
\bibinfo{author}{\bibfnamefont{Y.}~\bibnamefont{Meng}} \bibnamefont{et~al.}
  (\bibinfo{collaboration}{PandaX-4T}), \bibinfo{journal}{Phys. Rev. Lett.}
  \textbf{\bibinfo{volume}{127}}, \bibinfo{pages}{261802}
  (\bibinfo{year}{2021}), \eprint{2107.13438}.

\bibitem[{\citenamefont{Monroe and Fisher}(2007)}]{Monroe:2007xp}
\bibinfo{author}{\bibfnamefont{J.}~\bibnamefont{Monroe}} \bibnamefont{and}
  \bibinfo{author}{\bibfnamefont{P.}~\bibnamefont{Fisher}},
  \bibinfo{journal}{Phys. Rev. D} \textbf{\bibinfo{volume}{76}},
  \bibinfo{pages}{033007} (\bibinfo{year}{2007}), \eprint{0706.3019}.

\bibitem[{\citenamefont{Feng et~al.}(2015)\citenamefont{Feng, Profumo, and
  Ubaldi}}]{Feng:2014vea}
\bibinfo{author}{\bibfnamefont{L.}~\bibnamefont{Feng}},
  \bibinfo{author}{\bibfnamefont{S.}~\bibnamefont{Profumo}}, \bibnamefont{and}
  \bibinfo{author}{\bibfnamefont{L.}~\bibnamefont{Ubaldi}},
  \bibinfo{journal}{JHEP} \textbf{\bibinfo{volume}{03}}, \bibinfo{pages}{045}
  (\bibinfo{year}{2015}), \eprint{1412.1105}.

\bibitem[{\citenamefont{Pato et~al.}(2011)\citenamefont{Pato, Baudis, Bertone,
  Ruiz~de Austri, Strigari, and Trotta}}]{Pato:2010zk}
\bibinfo{author}{\bibfnamefont{M.}~\bibnamefont{Pato}},
  \bibinfo{author}{\bibfnamefont{L.}~\bibnamefont{Baudis}},
  \bibinfo{author}{\bibfnamefont{G.}~\bibnamefont{Bertone}},
  \bibinfo{author}{\bibfnamefont{R.}~\bibnamefont{Ruiz~de Austri}},
  \bibinfo{author}{\bibfnamefont{L.~E.} \bibnamefont{Strigari}},
  \bibnamefont{and} \bibinfo{author}{\bibfnamefont{R.}~\bibnamefont{Trotta}},
  \bibinfo{journal}{Phys. Rev. D} \textbf{\bibinfo{volume}{83}},
  \bibinfo{pages}{083505} (\bibinfo{year}{2011}), \eprint{1012.3458}.

\bibitem[{\citenamefont{Kamionkowski et~al.}(1994)\citenamefont{Kamionkowski,
  Krauss, and Ressell}}]{Kamionkowski:1994rm}
\bibinfo{author}{\bibfnamefont{M.}~\bibnamefont{Kamionkowski}},
  \bibinfo{author}{\bibfnamefont{L.~M.} \bibnamefont{Krauss}},
  \bibnamefont{and} \bibinfo{author}{\bibfnamefont{M.~T.}
  \bibnamefont{Ressell}} (\bibinfo{year}{1994}), \eprint{hep-ph/9402353}.

\bibitem[{\citenamefont{Chang et~al.}(2010{\natexlab{a}})\citenamefont{Chang,
  Liu, Pierce, Weiner, and Yavin}}]{Chang:2010yk}
\bibinfo{author}{\bibfnamefont{S.}~\bibnamefont{Chang}},
  \bibinfo{author}{\bibfnamefont{J.}~\bibnamefont{Liu}},
  \bibinfo{author}{\bibfnamefont{A.}~\bibnamefont{Pierce}},
  \bibinfo{author}{\bibfnamefont{N.}~\bibnamefont{Weiner}}, \bibnamefont{and}
  \bibinfo{author}{\bibfnamefont{I.}~\bibnamefont{Yavin}},
  \bibinfo{journal}{JCAP} \textbf{\bibinfo{volume}{08}}, \bibinfo{pages}{018}
  (\bibinfo{year}{2010}{\natexlab{a}}), \eprint{1004.0697}.

\bibitem[{\citenamefont{Feng et~al.}(2011)\citenamefont{Feng, Kumar, Marfatia,
  and Sanford}}]{Feng:2011vu}
\bibinfo{author}{\bibfnamefont{J.~L.} \bibnamefont{Feng}},
  \bibinfo{author}{\bibfnamefont{J.}~\bibnamefont{Kumar}},
  \bibinfo{author}{\bibfnamefont{D.}~\bibnamefont{Marfatia}}, \bibnamefont{and}
  \bibinfo{author}{\bibfnamefont{D.}~\bibnamefont{Sanford}},
  \bibinfo{journal}{Phys. Lett. B} \textbf{\bibinfo{volume}{703}},
  \bibinfo{pages}{124} (\bibinfo{year}{2011}), \eprint{1102.4331}.

\bibitem[{\citenamefont{Frandsen
  et~al.}(2011{\natexlab{a}})\citenamefont{Frandsen, Kahlhoefer, March-Russell,
  McCabe, McCullough, and Schmidt-Hoberg}}]{Frandsen:2011ts}
\bibinfo{author}{\bibfnamefont{M.~T.} \bibnamefont{Frandsen}},
  \bibinfo{author}{\bibfnamefont{F.}~\bibnamefont{Kahlhoefer}},
  \bibinfo{author}{\bibfnamefont{J.}~\bibnamefont{March-Russell}},
  \bibinfo{author}{\bibfnamefont{C.}~\bibnamefont{McCabe}},
  \bibinfo{author}{\bibfnamefont{M.}~\bibnamefont{McCullough}},
  \bibnamefont{and}
  \bibinfo{author}{\bibfnamefont{K.}~\bibnamefont{Schmidt-Hoberg}},
  \bibinfo{journal}{Phys. Rev. D} \textbf{\bibinfo{volume}{84}},
  \bibinfo{pages}{041301} (\bibinfo{year}{2011}{\natexlab{a}}),
  \eprint{1105.3734}.

\bibitem[{\citenamefont{Aprile et~al.}(2019{\natexlab{c}})}]{Aprile:2019xxb}
\bibinfo{author}{\bibfnamefont{E.}~\bibnamefont{Aprile}} \bibnamefont{et~al.}
  (\bibinfo{collaboration}{XENON}), \bibinfo{journal}{Phys. Rev. Lett.}
  \textbf{\bibinfo{volume}{123}}, \bibinfo{pages}{251801}
  (\bibinfo{year}{2019}{\natexlab{c}}), \eprint{1907.11485}.

\bibitem[{\citenamefont{Iachello et~al.}(1991)\citenamefont{Iachello, Krauss,
  and Maino}}]{Iachello:1990ut}
\bibinfo{author}{\bibfnamefont{F.}~\bibnamefont{Iachello}},
  \bibinfo{author}{\bibfnamefont{L.~M.} \bibnamefont{Krauss}},
  \bibnamefont{and} \bibinfo{author}{\bibfnamefont{G.}~\bibnamefont{Maino}},
  \bibinfo{journal}{Phys. Lett. B} \textbf{\bibinfo{volume}{254}},
  \bibinfo{pages}{220} (\bibinfo{year}{1991}).

\bibitem[{\citenamefont{Engel et~al.}(1992)\citenamefont{Engel, Pittel, and
  Vogel}}]{Engel:1992bf}
\bibinfo{author}{\bibfnamefont{J.}~\bibnamefont{Engel}},
  \bibinfo{author}{\bibfnamefont{S.}~\bibnamefont{Pittel}}, \bibnamefont{and}
  \bibinfo{author}{\bibfnamefont{P.}~\bibnamefont{Vogel}},
  \bibinfo{journal}{Int. J. Mod. Phys. E} \textbf{\bibinfo{volume}{1}},
  \bibinfo{pages}{1} (\bibinfo{year}{1992}).

\bibitem[{\citenamefont{Tovey et~al.}(2000)\citenamefont{Tovey, Gaitskell,
  Gondolo, Ramachers, and Roszkowski}}]{Tovey:2000mm}
\bibinfo{author}{\bibfnamefont{D.~R.} \bibnamefont{Tovey}},
  \bibinfo{author}{\bibfnamefont{R.~J.} \bibnamefont{Gaitskell}},
  \bibinfo{author}{\bibfnamefont{P.}~\bibnamefont{Gondolo}},
  \bibinfo{author}{\bibfnamefont{Y.~A.} \bibnamefont{Ramachers}},
  \bibnamefont{and}
  \bibinfo{author}{\bibfnamefont{L.}~\bibnamefont{Roszkowski}},
  \bibinfo{journal}{Phys. Lett. B} \textbf{\bibinfo{volume}{488}},
  \bibinfo{pages}{17} (\bibinfo{year}{2000}), \eprint{hep-ph/0005041}.

\bibitem[{\citenamefont{Fan et~al.}(2010)\citenamefont{Fan, Reece, and
  Wang}}]{Fan:2010gt}
\bibinfo{author}{\bibfnamefont{J.}~\bibnamefont{Fan}},
  \bibinfo{author}{\bibfnamefont{M.}~\bibnamefont{Reece}}, \bibnamefont{and}
  \bibinfo{author}{\bibfnamefont{L.-T.} \bibnamefont{Wang}},
  \bibinfo{journal}{JCAP} \textbf{\bibinfo{volume}{1011}}, \bibinfo{pages}{042}
  (\bibinfo{year}{2010}), \eprint{1008.1591}.

\bibitem[{\citenamefont{Fitzpatrick et~al.}(2013)\citenamefont{Fitzpatrick,
  Haxton, Katz, Lubbers, and Xu}}]{Fitzpatrick:2012ix}
\bibinfo{author}{\bibfnamefont{A.~L.} \bibnamefont{Fitzpatrick}},
  \bibinfo{author}{\bibfnamefont{W.}~\bibnamefont{Haxton}},
  \bibinfo{author}{\bibfnamefont{E.}~\bibnamefont{Katz}},
  \bibinfo{author}{\bibfnamefont{N.}~\bibnamefont{Lubbers}}, \bibnamefont{and}
  \bibinfo{author}{\bibfnamefont{Y.}~\bibnamefont{Xu}}, \bibinfo{journal}{JCAP}
  \textbf{\bibinfo{volume}{1302}}, \bibinfo{pages}{004} (\bibinfo{year}{2013}),
  \eprint{1203.3542}.

\bibitem[{\citenamefont{Anand et~al.}(2014)\citenamefont{Anand, Fitzpatrick,
  and Haxton}}]{Anand:2013yka}
\bibinfo{author}{\bibfnamefont{N.}~\bibnamefont{Anand}},
  \bibinfo{author}{\bibfnamefont{A.~L.} \bibnamefont{Fitzpatrick}},
  \bibnamefont{and} \bibinfo{author}{\bibfnamefont{W.~C.}
  \bibnamefont{Haxton}}, \bibinfo{journal}{Phys. Rev. C}
  \textbf{\bibinfo{volume}{89}}, \bibinfo{pages}{065501}
  (\bibinfo{year}{2014}), \eprint{1308.6288}.

\bibitem[{\citenamefont{Schneck et~al.}(2015)}]{Schneck:2015eqa}
\bibinfo{author}{\bibfnamefont{K.}~\bibnamefont{Schneck}} \bibnamefont{et~al.}
  (\bibinfo{collaboration}{SuperCDMS}), \bibinfo{journal}{Phys. Rev. D}
  \textbf{\bibinfo{volume}{91}}, \bibinfo{pages}{092004}
  (\bibinfo{year}{2015}), \eprint{1503.03379}.

\bibitem[{\citenamefont{Aprile et~al.}(2017{\natexlab{b}})}]{Aprile:2017aas}
\bibinfo{author}{\bibfnamefont{E.}~\bibnamefont{Aprile}} \bibnamefont{et~al.}
  (\bibinfo{collaboration}{XENON}), \bibinfo{journal}{Phys. Rev. D}
  \textbf{\bibinfo{volume}{96}}, \bibinfo{pages}{042004}
  (\bibinfo{year}{2017}{\natexlab{b}}), \eprint{1705.02614}.

\bibitem[{\citenamefont{Xia et~al.}(2019)}]{Xia:2018qgs}
\bibinfo{author}{\bibfnamefont{J.}~\bibnamefont{Xia}} \bibnamefont{et~al.}
  (\bibinfo{collaboration}{PandaX-II}), \bibinfo{journal}{Phys. Lett. B}
  \textbf{\bibinfo{volume}{792}}, \bibinfo{pages}{193} (\bibinfo{year}{2019}),
  \eprint{1807.01936}.

\bibitem[{\citenamefont{Angloher et~al.}(2019)}]{Angloher:2018fcs}
\bibinfo{author}{\bibfnamefont{G.}~\bibnamefont{Angloher}} \bibnamefont{et~al.}
  (\bibinfo{collaboration}{CRESST}), \bibinfo{journal}{Eur. Phys. J. C}
  \textbf{\bibinfo{volume}{79}}, \bibinfo{pages}{43} (\bibinfo{year}{2019}),
  \eprint{1809.03753}.

\bibitem[{\citenamefont{Dent et~al.}(2015)\citenamefont{Dent, Krauss, Newstead,
  and Sabharwal}}]{Dent:2015zpa}
\bibinfo{author}{\bibfnamefont{J.~B.} \bibnamefont{Dent}},
  \bibinfo{author}{\bibfnamefont{L.~M.} \bibnamefont{Krauss}},
  \bibinfo{author}{\bibfnamefont{J.~L.} \bibnamefont{Newstead}},
  \bibnamefont{and}
  \bibinfo{author}{\bibfnamefont{S.}~\bibnamefont{Sabharwal}},
  \bibinfo{journal}{Phys. Rev. D} \textbf{\bibinfo{volume}{92}},
  \bibinfo{pages}{063515} (\bibinfo{year}{2015}), \eprint{1505.03117}.

\bibitem[{\citenamefont{Catena et~al.}(2019)\citenamefont{Catena, Fridell, and
  Krauss}}]{Catena:2019hzw}
\bibinfo{author}{\bibfnamefont{R.}~\bibnamefont{Catena}},
  \bibinfo{author}{\bibfnamefont{K.}~\bibnamefont{Fridell}}, \bibnamefont{and}
  \bibinfo{author}{\bibfnamefont{M.~B.} \bibnamefont{Krauss}},
  \bibinfo{journal}{JHEP} \textbf{\bibinfo{volume}{08}}, \bibinfo{pages}{030}
  (\bibinfo{year}{2019}), \eprint{1907.02910}.

\bibitem[{\citenamefont{Gondolo et~al.}(2021)\citenamefont{Gondolo, Kang,
  Scopel, and Tomar}}]{Gondolo:2020wge}
\bibinfo{author}{\bibfnamefont{P.}~\bibnamefont{Gondolo}},
  \bibinfo{author}{\bibfnamefont{S.}~\bibnamefont{Kang}},
  \bibinfo{author}{\bibfnamefont{S.}~\bibnamefont{Scopel}}, \bibnamefont{and}
  \bibinfo{author}{\bibfnamefont{G.}~\bibnamefont{Tomar}},
  \bibinfo{journal}{Phys. Rev. D} \textbf{\bibinfo{volume}{104}},
  \bibinfo{pages}{063017} (\bibinfo{year}{2021}), \eprint{2008.05120}.

\bibitem[{\citenamefont{Krauss and Newstead}(2018)}]{Krauss:2018pvg}
\bibinfo{author}{\bibfnamefont{L.~M.} \bibnamefont{Krauss}} \bibnamefont{and}
  \bibinfo{author}{\bibfnamefont{J.~L.} \bibnamefont{Newstead}}
  (\bibinfo{year}{2018}), \eprint{1801.08523}.

\bibitem[{\citenamefont{Epelbaum et~al.}(2009)\citenamefont{Epelbaum, Hammer,
  and Mei{\ss}ner}}]{Epelbaum:2008ga}
\bibinfo{author}{\bibfnamefont{E.}~\bibnamefont{Epelbaum}},
  \bibinfo{author}{\bibfnamefont{H.-W.} \bibnamefont{Hammer}},
  \bibnamefont{and} \bibinfo{author}{\bibfnamefont{U.-G.}
  \bibnamefont{Mei{\ss}ner}}, \bibinfo{journal}{Rev. Mod. Phys.}
  \textbf{\bibinfo{volume}{81}}, \bibinfo{pages}{1773} (\bibinfo{year}{2009}),
  \eprint{0811.1338}.

\bibitem[{\citenamefont{Machleidt and Entem}(2011)}]{Machleidt:2011zz}
\bibinfo{author}{\bibfnamefont{R.}~\bibnamefont{Machleidt}} \bibnamefont{and}
  \bibinfo{author}{\bibfnamefont{D.~R.} \bibnamefont{Entem}},
  \bibinfo{journal}{Phys. Rept.} \textbf{\bibinfo{volume}{503}},
  \bibinfo{pages}{1} (\bibinfo{year}{2011}), \eprint{1105.2919}.

\bibitem[{\citenamefont{Hammer et~al.}(2013)\citenamefont{Hammer, Nogga, and
  Schwenk}}]{Hammer:2012id}
\bibinfo{author}{\bibfnamefont{H.-W.} \bibnamefont{Hammer}},
  \bibinfo{author}{\bibfnamefont{A.}~\bibnamefont{Nogga}}, \bibnamefont{and}
  \bibinfo{author}{\bibfnamefont{A.}~\bibnamefont{Schwenk}},
  \bibinfo{journal}{Rev. Mod. Phys.} \textbf{\bibinfo{volume}{85}},
  \bibinfo{pages}{197} (\bibinfo{year}{2013}), \eprint{1210.4273}.

\bibitem[{\citenamefont{Hoferichter
  et~al.}(2015{\natexlab{a}})\citenamefont{Hoferichter, Klos, and
  Schwenk}}]{Hoferichter:2015ipa}
\bibinfo{author}{\bibfnamefont{M.}~\bibnamefont{Hoferichter}},
  \bibinfo{author}{\bibfnamefont{P.}~\bibnamefont{Klos}}, \bibnamefont{and}
  \bibinfo{author}{\bibfnamefont{A.}~\bibnamefont{Schwenk}},
  \bibinfo{journal}{Phys. Lett. B} \textbf{\bibinfo{volume}{746}},
  \bibinfo{pages}{410} (\bibinfo{year}{2015}{\natexlab{a}}),
  \eprint{1503.04811}.

\bibitem[{\citenamefont{Bishara
  et~al.}(2017{\natexlab{a}})\citenamefont{Bishara, Brod, Grinstein, and
  Zupan}}]{Bishara:2016hek}
\bibinfo{author}{\bibfnamefont{F.}~\bibnamefont{Bishara}},
  \bibinfo{author}{\bibfnamefont{J.}~\bibnamefont{Brod}},
  \bibinfo{author}{\bibfnamefont{B.}~\bibnamefont{Grinstein}},
  \bibnamefont{and} \bibinfo{author}{\bibfnamefont{J.}~\bibnamefont{Zupan}},
  \bibinfo{journal}{JCAP} \textbf{\bibinfo{volume}{1702}}, \bibinfo{pages}{009}
  (\bibinfo{year}{2017}{\natexlab{a}}), \eprint{1611.00368}.

\bibitem[{\citenamefont{Bishara
  et~al.}(2017{\natexlab{b}})\citenamefont{Bishara, Brod, Grinstein, and
  Zupan}}]{Bishara:2017pfq}
\bibinfo{author}{\bibfnamefont{F.}~\bibnamefont{Bishara}},
  \bibinfo{author}{\bibfnamefont{J.}~\bibnamefont{Brod}},
  \bibinfo{author}{\bibfnamefont{B.}~\bibnamefont{Grinstein}},
  \bibnamefont{and} \bibinfo{author}{\bibfnamefont{J.}~\bibnamefont{Zupan}},
  \bibinfo{journal}{JHEP} \textbf{\bibinfo{volume}{11}}, \bibinfo{pages}{059}
  (\bibinfo{year}{2017}{\natexlab{b}}), \eprint{1707.06998}.

\bibitem[{\citenamefont{Pr{\'e}zeau
  et~al.}(2003{\natexlab{a}})\citenamefont{Pr{\'e}zeau, Kurylov, Kamionkowski,
  and Vogel}}]{Prezeau:2003sv}
\bibinfo{author}{\bibfnamefont{G.}~\bibnamefont{Pr{\'e}zeau}},
  \bibinfo{author}{\bibfnamefont{A.}~\bibnamefont{Kurylov}},
  \bibinfo{author}{\bibfnamefont{M.}~\bibnamefont{Kamionkowski}},
  \bibnamefont{and} \bibinfo{author}{\bibfnamefont{P.}~\bibnamefont{Vogel}},
  \bibinfo{journal}{Phys. Rev. Lett.} \textbf{\bibinfo{volume}{91}},
  \bibinfo{pages}{231301} (\bibinfo{year}{2003}{\natexlab{a}}),
  \eprint{astro-ph/0309115}.

\bibitem[{\citenamefont{Cirigliano et~al.}(2012)\citenamefont{Cirigliano,
  Graesser, and Ovanesyan}}]{Cirigliano:2012pq}
\bibinfo{author}{\bibfnamefont{V.}~\bibnamefont{Cirigliano}},
  \bibinfo{author}{\bibfnamefont{M.~L.} \bibnamefont{Graesser}},
  \bibnamefont{and}
  \bibinfo{author}{\bibfnamefont{G.}~\bibnamefont{Ovanesyan}},
  \bibinfo{journal}{JHEP} \textbf{\bibinfo{volume}{10}}, \bibinfo{pages}{025}
  (\bibinfo{year}{2012}), \eprint{1205.2695}.

\bibitem[{\citenamefont{Men{\'e}ndez et~al.}(2012)\citenamefont{Men{\'e}ndez,
  Gazit, and Schwenk}}]{Menendez:2012tm}
\bibinfo{author}{\bibfnamefont{J.}~\bibnamefont{Men{\'e}ndez}},
  \bibinfo{author}{\bibfnamefont{D.}~\bibnamefont{Gazit}}, \bibnamefont{and}
  \bibinfo{author}{\bibfnamefont{A.}~\bibnamefont{Schwenk}},
  \bibinfo{journal}{Phys. Rev. D} \textbf{\bibinfo{volume}{86}},
  \bibinfo{pages}{103511} (\bibinfo{year}{2012}), \eprint{1208.1094}.

\bibitem[{\citenamefont{Cirigliano et~al.}(2014)\citenamefont{Cirigliano,
  Graesser, Ovanesyan, and Shoemaker}}]{Cirigliano:2013zta}
\bibinfo{author}{\bibfnamefont{V.}~\bibnamefont{Cirigliano}},
  \bibinfo{author}{\bibfnamefont{M.~L.} \bibnamefont{Graesser}},
  \bibinfo{author}{\bibfnamefont{G.}~\bibnamefont{Ovanesyan}},
  \bibnamefont{and} \bibinfo{author}{\bibfnamefont{I.~M.}
  \bibnamefont{Shoemaker}}, \bibinfo{journal}{Phys. Lett. B}
  \textbf{\bibinfo{volume}{739}}, \bibinfo{pages}{293} (\bibinfo{year}{2014}),
  \eprint{1311.5886}.

\bibitem[{\citenamefont{Hoferichter et~al.}(2016)\citenamefont{Hoferichter,
  Klos, Men{\'e}ndez, and Schwenk}}]{Hoferichter:2016nvd}
\bibinfo{author}{\bibfnamefont{M.}~\bibnamefont{Hoferichter}},
  \bibinfo{author}{\bibfnamefont{P.}~\bibnamefont{Klos}},
  \bibinfo{author}{\bibfnamefont{J.}~\bibnamefont{Men{\'e}ndez}},
  \bibnamefont{and} \bibinfo{author}{\bibfnamefont{A.}~\bibnamefont{Schwenk}},
  \bibinfo{journal}{Phys. Rev. D} \textbf{\bibinfo{volume}{94}},
  \bibinfo{pages}{063505} (\bibinfo{year}{2016}), \eprint{1605.08043}.

\bibitem[{\citenamefont{K{\"o}rber et~al.}(2017)\citenamefont{K{\"o}rber,
  Nogga, and de~Vries}}]{Korber:2017ery}
\bibinfo{author}{\bibfnamefont{C.}~\bibnamefont{K{\"o}rber}},
  \bibinfo{author}{\bibfnamefont{A.}~\bibnamefont{Nogga}}, \bibnamefont{and}
  \bibinfo{author}{\bibfnamefont{J.}~\bibnamefont{de~Vries}},
  \bibinfo{journal}{Phys. Rev. C} \textbf{\bibinfo{volume}{96}},
  \bibinfo{pages}{035805} (\bibinfo{year}{2017}), \eprint{1704.01150}.

\bibitem[{\citenamefont{Hoferichter et~al.}(2017)\citenamefont{Hoferichter,
  Klos, Men{\'e}ndez, and Schwenk}}]{Hoferichter:2017olk}
\bibinfo{author}{\bibfnamefont{M.}~\bibnamefont{Hoferichter}},
  \bibinfo{author}{\bibfnamefont{P.}~\bibnamefont{Klos}},
  \bibinfo{author}{\bibfnamefont{J.}~\bibnamefont{Men{\'e}ndez}},
  \bibnamefont{and} \bibinfo{author}{\bibfnamefont{A.}~\bibnamefont{Schwenk}},
  \bibinfo{journal}{Phys. Rev. Lett.} \textbf{\bibinfo{volume}{119}},
  \bibinfo{pages}{181803} (\bibinfo{year}{2017}), \eprint{1708.02245}.

\bibitem[{\citenamefont{Andreoli et~al.}(2019)\citenamefont{Andreoli,
  Cirigliano, Gandolfi, and Pederiva}}]{Andreoli:2018etf}
\bibinfo{author}{\bibfnamefont{L.}~\bibnamefont{Andreoli}},
  \bibinfo{author}{\bibfnamefont{V.}~\bibnamefont{Cirigliano}},
  \bibinfo{author}{\bibfnamefont{S.}~\bibnamefont{Gandolfi}}, \bibnamefont{and}
  \bibinfo{author}{\bibfnamefont{F.}~\bibnamefont{Pederiva}},
  \bibinfo{journal}{Phys. Rev. C} \textbf{\bibinfo{volume}{99}},
  \bibinfo{pages}{025501} (\bibinfo{year}{2019}), \eprint{1811.01843}.

\bibitem[{\citenamefont{Hu et~al.}(2022)\citenamefont{Hu, Padua-Arg\"uelles,
  Leutheusser, Miyagi, Stroberg, and Holt}}]{Hu:2021awl}
\bibinfo{author}{\bibfnamefont{B.~S.} \bibnamefont{Hu}},
  \bibinfo{author}{\bibfnamefont{J.}~\bibnamefont{Padua-Arg\"uelles}},
  \bibinfo{author}{\bibfnamefont{S.}~\bibnamefont{Leutheusser}},
  \bibinfo{author}{\bibfnamefont{T.}~\bibnamefont{Miyagi}},
  \bibinfo{author}{\bibfnamefont{S.~R.} \bibnamefont{Stroberg}},
  \bibnamefont{and} \bibinfo{author}{\bibfnamefont{J.~D.} \bibnamefont{Holt}},
  \bibinfo{journal}{Phys. Rev. Lett.} \textbf{\bibinfo{volume}{128}},
  \bibinfo{pages}{072502} (\bibinfo{year}{2022}), \eprint{2109.00193}.

\bibitem[{\citenamefont{Goodman et~al.}(2011)\citenamefont{Goodman, Ibe,
  Rajaraman, Shepherd, Tait, and Yu}}]{Goodman:2010yf}
\bibinfo{author}{\bibfnamefont{J.}~\bibnamefont{Goodman}},
  \bibinfo{author}{\bibfnamefont{M.}~\bibnamefont{Ibe}},
  \bibinfo{author}{\bibfnamefont{A.}~\bibnamefont{Rajaraman}},
  \bibinfo{author}{\bibfnamefont{W.}~\bibnamefont{Shepherd}},
  \bibinfo{author}{\bibfnamefont{T.~M.} \bibnamefont{Tait}}, \bibnamefont{and}
  \bibinfo{author}{\bibfnamefont{H.-B.} \bibnamefont{Yu}},
  \bibinfo{journal}{Phys. Lett. B} \textbf{\bibinfo{volume}{695}},
  \bibinfo{pages}{185} (\bibinfo{year}{2011}), \eprint{1005.1286}.

\bibitem[{\citenamefont{Goodman et~al.}(2010)\citenamefont{Goodman, Ibe,
  Rajaraman, Shepherd, Tait, and Yu}}]{Goodman:2010ku}
\bibinfo{author}{\bibfnamefont{J.}~\bibnamefont{Goodman}},
  \bibinfo{author}{\bibfnamefont{M.}~\bibnamefont{Ibe}},
  \bibinfo{author}{\bibfnamefont{A.}~\bibnamefont{Rajaraman}},
  \bibinfo{author}{\bibfnamefont{W.}~\bibnamefont{Shepherd}},
  \bibinfo{author}{\bibfnamefont{T.~M.} \bibnamefont{Tait}}, \bibnamefont{and}
  \bibinfo{author}{\bibfnamefont{H.-B.} \bibnamefont{Yu}},
  \bibinfo{journal}{Phys. Rev. D} \textbf{\bibinfo{volume}{82}},
  \bibinfo{pages}{116010} (\bibinfo{year}{2010}), \eprint{1008.1783}.

\bibitem[{\citenamefont{Fox et~al.}(2012)\citenamefont{Fox, Harnik, Kopp, and
  Tsai}}]{Fox:2011pm}
\bibinfo{author}{\bibfnamefont{P.~J.} \bibnamefont{Fox}},
  \bibinfo{author}{\bibfnamefont{R.}~\bibnamefont{Harnik}},
  \bibinfo{author}{\bibfnamefont{J.}~\bibnamefont{Kopp}}, \bibnamefont{and}
  \bibinfo{author}{\bibfnamefont{Y.}~\bibnamefont{Tsai}},
  \bibinfo{journal}{Phys. Rev. D} \textbf{\bibinfo{volume}{85}},
  \bibinfo{pages}{056011} (\bibinfo{year}{2012}), \eprint{1109.4398}.

\bibitem[{\citenamefont{Crivellin
  et~al.}(2014{\natexlab{a}})\citenamefont{Crivellin, D'Eramo, and
  Procura}}]{Crivellin:2014qxa}
\bibinfo{author}{\bibfnamefont{A.}~\bibnamefont{Crivellin}},
  \bibinfo{author}{\bibfnamefont{F.}~\bibnamefont{D'Eramo}}, \bibnamefont{and}
  \bibinfo{author}{\bibfnamefont{M.}~\bibnamefont{Procura}},
  \bibinfo{journal}{Phys. Rev. Lett.} \textbf{\bibinfo{volume}{112}},
  \bibinfo{pages}{191304} (\bibinfo{year}{2014}{\natexlab{a}}),
  \eprint{1402.1173}.

\bibitem[{\citenamefont{D'Eramo and Procura}(2015)}]{DEramo:2014nmf}
\bibinfo{author}{\bibfnamefont{F.}~\bibnamefont{D'Eramo}} \bibnamefont{and}
  \bibinfo{author}{\bibfnamefont{M.}~\bibnamefont{Procura}},
  \bibinfo{journal}{JHEP} \textbf{\bibinfo{volume}{04}}, \bibinfo{pages}{054}
  (\bibinfo{year}{2015}), \eprint{1411.3342}.

\bibitem[{\citenamefont{Hill and Solon}(2015{\natexlab{a}})}]{Hill:2014yxa}
\bibinfo{author}{\bibfnamefont{R.~J.} \bibnamefont{Hill}} \bibnamefont{and}
  \bibinfo{author}{\bibfnamefont{M.~P.} \bibnamefont{Solon}},
  \bibinfo{journal}{Phys. Rev. D} \textbf{\bibinfo{volume}{91}},
  \bibinfo{pages}{043505} (\bibinfo{year}{2015}{\natexlab{a}}),
  \eprint{1409.8290}.

\bibitem[{\citenamefont{Hill and Solon}(2015{\natexlab{b}})}]{Hill:2014yka}
\bibinfo{author}{\bibfnamefont{R.~J.} \bibnamefont{Hill}} \bibnamefont{and}
  \bibinfo{author}{\bibfnamefont{M.~P.} \bibnamefont{Solon}},
  \bibinfo{journal}{Phys. Rev. D} \textbf{\bibinfo{volume}{91}},
  \bibinfo{pages}{043504} (\bibinfo{year}{2015}{\natexlab{b}}),
  \eprint{1401.3339}.

\bibitem[{\citenamefont{Bishara
  et~al.}(2017{\natexlab{c}})\citenamefont{Bishara, Brod, Grinstein, and
  Zupan}}]{Bishara:2017nnn}
\bibinfo{author}{\bibfnamefont{F.}~\bibnamefont{Bishara}},
  \bibinfo{author}{\bibfnamefont{J.}~\bibnamefont{Brod}},
  \bibinfo{author}{\bibfnamefont{B.}~\bibnamefont{Grinstein}},
  \bibnamefont{and} \bibinfo{author}{\bibfnamefont{J.}~\bibnamefont{Zupan}}
  (\bibinfo{year}{2017}{\natexlab{c}}), \eprint{1708.02678}.

\bibitem[{\citenamefont{Brod et~al.}(2018{\natexlab{a}})\citenamefont{Brod,
  Gootjes-Dreesbach, Tammaro, and Zupan}}]{Brod:2017bsw}
\bibinfo{author}{\bibfnamefont{J.}~\bibnamefont{Brod}},
  \bibinfo{author}{\bibfnamefont{A.}~\bibnamefont{Gootjes-Dreesbach}},
  \bibinfo{author}{\bibfnamefont{M.}~\bibnamefont{Tammaro}}, \bibnamefont{and}
  \bibinfo{author}{\bibfnamefont{J.}~\bibnamefont{Zupan}},
  \bibinfo{journal}{JHEP} \textbf{\bibinfo{volume}{10}}, \bibinfo{pages}{065}
  (\bibinfo{year}{2018}{\natexlab{a}}), \eprint{1710.10218}.

\bibitem[{\citenamefont{Hill and Solon}(2012)}]{Hill:2011be}
\bibinfo{author}{\bibfnamefont{R.~J.} \bibnamefont{Hill}} \bibnamefont{and}
  \bibinfo{author}{\bibfnamefont{M.~P.} \bibnamefont{Solon}},
  \bibinfo{journal}{Phys. Lett. B} \textbf{\bibinfo{volume}{707}},
  \bibinfo{pages}{539} (\bibinfo{year}{2012}), \eprint{1111.0016}.

\bibitem[{\citenamefont{Chen and Hill}(2020)}]{Chen:2019gtm}
\bibinfo{author}{\bibfnamefont{Q.}~\bibnamefont{Chen}} \bibnamefont{and}
  \bibinfo{author}{\bibfnamefont{R.~J.} \bibnamefont{Hill}},
  \bibinfo{journal}{Phys. Lett. B} \textbf{\bibinfo{volume}{804}},
  \bibinfo{pages}{135364} (\bibinfo{year}{2020}), \eprint{1912.07795}.

\bibitem[{\citenamefont{Brod et~al.}(2018{\natexlab{b}})\citenamefont{Brod,
  Grinstein, Stamou, and Zupan}}]{Brod:2018ust}
\bibinfo{author}{\bibfnamefont{J.}~\bibnamefont{Brod}},
  \bibinfo{author}{\bibfnamefont{B.}~\bibnamefont{Grinstein}},
  \bibinfo{author}{\bibfnamefont{E.}~\bibnamefont{Stamou}}, \bibnamefont{and}
  \bibinfo{author}{\bibfnamefont{J.}~\bibnamefont{Zupan}},
  \bibinfo{journal}{JHEP} \textbf{\bibinfo{volume}{02}}, \bibinfo{pages}{174}
  (\bibinfo{year}{2018}{\natexlab{b}}), \eprint{1801.04240}.

\bibitem[{\citenamefont{Bishara et~al.}(2020)\citenamefont{Bishara, Brod,
  Grinstein, and Zupan}}]{Bishara:2018vix}
\bibinfo{author}{\bibfnamefont{F.}~\bibnamefont{Bishara}},
  \bibinfo{author}{\bibfnamefont{J.}~\bibnamefont{Brod}},
  \bibinfo{author}{\bibfnamefont{B.}~\bibnamefont{Grinstein}},
  \bibnamefont{and} \bibinfo{author}{\bibfnamefont{J.}~\bibnamefont{Zupan}},
  \bibinfo{journal}{JHEP} \textbf{\bibinfo{volume}{03}}, \bibinfo{pages}{089}
  (\bibinfo{year}{2020}), \eprint{1809.03506}.

\bibitem[{\citenamefont{Crivellin and Haisch}(2014)}]{Crivellin:2014gpa}
\bibinfo{author}{\bibfnamefont{A.}~\bibnamefont{Crivellin}} \bibnamefont{and}
  \bibinfo{author}{\bibfnamefont{U.}~\bibnamefont{Haisch}},
  \bibinfo{journal}{Phys. Rev. D} \textbf{\bibinfo{volume}{90}},
  \bibinfo{pages}{115011} (\bibinfo{year}{2014}), \eprint{1408.5046}.

\bibitem[{\citenamefont{Serot}(1978)}]{Serot:1978vj}
\bibinfo{author}{\bibfnamefont{B.~D.} \bibnamefont{Serot}},
  \bibinfo{journal}{Nucl. Phys. A} \textbf{\bibinfo{volume}{308}},
  \bibinfo{pages}{457} (\bibinfo{year}{1978}).

\bibitem[{\citenamefont{Donnelly and Haxton}(1979)}]{Donnelly:1979ezn}
\bibinfo{author}{\bibfnamefont{T.~W.} \bibnamefont{Donnelly}} \bibnamefont{and}
  \bibinfo{author}{\bibfnamefont{W.~C.} \bibnamefont{Haxton}},
  \bibinfo{journal}{Atom. Data Nucl. Data Tabl.} \textbf{\bibinfo{volume}{23}},
  \bibinfo{pages}{103} (\bibinfo{year}{1979}).

\bibitem[{\citenamefont{Serot}(1979)}]{Serot:1979yk}
\bibinfo{author}{\bibfnamefont{B.~D.} \bibnamefont{Serot}},
  \bibinfo{journal}{Nucl. Phys. A} \textbf{\bibinfo{volume}{322}},
  \bibinfo{pages}{408} (\bibinfo{year}{1979}).

\bibitem[{\citenamefont{Bottino et~al.}(2000)\citenamefont{Bottino, Donato,
  Fornengo, and Scopel}}]{Bottino:1999ei}
\bibinfo{author}{\bibfnamefont{A.}~\bibnamefont{Bottino}},
  \bibinfo{author}{\bibfnamefont{F.}~\bibnamefont{Donato}},
  \bibinfo{author}{\bibfnamefont{N.}~\bibnamefont{Fornengo}}, \bibnamefont{and}
  \bibinfo{author}{\bibfnamefont{S.}~\bibnamefont{Scopel}},
  \bibinfo{journal}{Astropart. Phys.} \textbf{\bibinfo{volume}{13}},
  \bibinfo{pages}{215} (\bibinfo{year}{2000}), \eprint{hep-ph/9909228}.

\bibitem[{\citenamefont{Bottino et~al.}(2002)\citenamefont{Bottino, Donato,
  Fornengo, and Scopel}}]{Bottino:2001dj}
\bibinfo{author}{\bibfnamefont{A.}~\bibnamefont{Bottino}},
  \bibinfo{author}{\bibfnamefont{F.}~\bibnamefont{Donato}},
  \bibinfo{author}{\bibfnamefont{N.}~\bibnamefont{Fornengo}}, \bibnamefont{and}
  \bibinfo{author}{\bibfnamefont{S.}~\bibnamefont{Scopel}},
  \bibinfo{journal}{Astropart. Phys.} \textbf{\bibinfo{volume}{18}},
  \bibinfo{pages}{205} (\bibinfo{year}{2002}), \eprint{hep-ph/0111229}.

\bibitem[{\citenamefont{Ellis et~al.}(2008)\citenamefont{Ellis, Olive, and
  Savage}}]{Ellis:2008hf}
\bibinfo{author}{\bibfnamefont{J.~R.} \bibnamefont{Ellis}},
  \bibinfo{author}{\bibfnamefont{K.~A.} \bibnamefont{Olive}}, \bibnamefont{and}
  \bibinfo{author}{\bibfnamefont{C.}~\bibnamefont{Savage}},
  \bibinfo{journal}{Phys. Rev. D} \textbf{\bibinfo{volume}{77}},
  \bibinfo{pages}{065026} (\bibinfo{year}{2008}), \eprint{0801.3656}.

\bibitem[{\citenamefont{Crivellin
  et~al.}(2014{\natexlab{b}})\citenamefont{Crivellin, Hoferichter, and
  Procura}}]{Crivellin:2013ipa}
\bibinfo{author}{\bibfnamefont{A.}~\bibnamefont{Crivellin}},
  \bibinfo{author}{\bibfnamefont{M.}~\bibnamefont{Hoferichter}},
  \bibnamefont{and} \bibinfo{author}{\bibfnamefont{M.}~\bibnamefont{Procura}},
  \bibinfo{journal}{Phys. Rev. D} \textbf{\bibinfo{volume}{89}},
  \bibinfo{pages}{054021} (\bibinfo{year}{2014}{\natexlab{b}}),
  \eprint{1312.4951}.

\bibitem[{\citenamefont{Hoferichter
  et~al.}(2015{\natexlab{b}})\citenamefont{Hoferichter, Ruiz~de Elvira, Kubis,
  and Mei\ss{}ner}}]{Hoferichter:2015dsa}
\bibinfo{author}{\bibfnamefont{M.}~\bibnamefont{Hoferichter}},
  \bibinfo{author}{\bibfnamefont{J.}~\bibnamefont{Ruiz~de Elvira}},
  \bibinfo{author}{\bibfnamefont{B.}~\bibnamefont{Kubis}}, \bibnamefont{and}
  \bibinfo{author}{\bibfnamefont{U.-G.} \bibnamefont{Mei\ss{}ner}},
  \bibinfo{journal}{Phys. Rev. Lett.} \textbf{\bibinfo{volume}{115}},
  \bibinfo{pages}{092301} (\bibinfo{year}{2015}{\natexlab{b}}),
  \eprint{1506.04142}.

\bibitem[{\citenamefont{Gupta et~al.}(2021)\citenamefont{Gupta, Park,
  Hoferichter, Mereghetti, Yoon, and Bhattacharya}}]{Gupta:2021ahb}
\bibinfo{author}{\bibfnamefont{R.}~\bibnamefont{Gupta}},
  \bibinfo{author}{\bibfnamefont{S.}~\bibnamefont{Park}},
  \bibinfo{author}{\bibfnamefont{M.}~\bibnamefont{Hoferichter}},
  \bibinfo{author}{\bibfnamefont{E.}~\bibnamefont{Mereghetti}},
  \bibinfo{author}{\bibfnamefont{B.}~\bibnamefont{Yoon}}, \bibnamefont{and}
  \bibinfo{author}{\bibfnamefont{T.}~\bibnamefont{Bhattacharya}},
  \bibinfo{journal}{Phys. Rev. Lett.} \textbf{\bibinfo{volume}{127}},
  \bibinfo{pages}{242002} (\bibinfo{year}{2021}), \eprint{2105.12095}.

\bibitem[{\citenamefont{Helm}(1956)}]{Helm:1956zz}
\bibinfo{author}{\bibfnamefont{R.~H.} \bibnamefont{Helm}},
  \bibinfo{journal}{Phys. Rev.} \textbf{\bibinfo{volume}{104}},
  \bibinfo{pages}{1466} (\bibinfo{year}{1956}).

\bibitem[{\citenamefont{Vietze et~al.}(2015)\citenamefont{Vietze, Klos,
  Men{\'e}ndez, Haxton, and Schwenk}}]{Vietze:2014vsa}
\bibinfo{author}{\bibfnamefont{L.}~\bibnamefont{Vietze}},
  \bibinfo{author}{\bibfnamefont{P.}~\bibnamefont{Klos}},
  \bibinfo{author}{\bibfnamefont{J.}~\bibnamefont{Men{\'e}ndez}},
  \bibinfo{author}{\bibfnamefont{W.~C.} \bibnamefont{Haxton}},
  \bibnamefont{and} \bibinfo{author}{\bibfnamefont{A.}~\bibnamefont{Schwenk}},
  \bibinfo{journal}{Phys. Rev. D} \textbf{\bibinfo{volume}{91}},
  \bibinfo{pages}{043520} (\bibinfo{year}{2015}), \eprint{1412.6091}.

\bibitem[{\citenamefont{Caurier et~al.}(2005)\citenamefont{Caurier,
  Mart{\'i}nez-Pinedo, Nowacki, Poves, and Zuker}}]{Caurier:2004gf}
\bibinfo{author}{\bibfnamefont{E.}~\bibnamefont{Caurier}},
  \bibinfo{author}{\bibfnamefont{G.}~\bibnamefont{Mart{\'i}nez-Pinedo}},
  \bibinfo{author}{\bibfnamefont{F.}~\bibnamefont{Nowacki}},
  \bibinfo{author}{\bibfnamefont{A.}~\bibnamefont{Poves}}, \bibnamefont{and}
  \bibinfo{author}{\bibfnamefont{A.~P.} \bibnamefont{Zuker}},
  \bibinfo{journal}{Rev. Mod. Phys.} \textbf{\bibinfo{volume}{77}},
  \bibinfo{pages}{427} (\bibinfo{year}{2005}), \eprint{nucl-th/0402046}.

\bibitem[{\citenamefont{Kamada et~al.}(2001)}]{Kamada:2001tv}
\bibinfo{author}{\bibfnamefont{H.}~\bibnamefont{Kamada}} \bibnamefont{et~al.},
  \bibinfo{journal}{Phys. Rev. C} \textbf{\bibinfo{volume}{64}},
  \bibinfo{pages}{044001} (\bibinfo{year}{2001}), \eprint{nucl-th/0104057}.

\bibitem[{\citenamefont{Carlson et~al.}(2015)\citenamefont{Carlson, Gandolfi,
  Pederiva, Pieper, Schiavilla, Schmidt, and Wiringa}}]{Carlson:2014vla}
\bibinfo{author}{\bibfnamefont{J.}~\bibnamefont{Carlson}},
  \bibinfo{author}{\bibfnamefont{S.}~\bibnamefont{Gandolfi}},
  \bibinfo{author}{\bibfnamefont{F.}~\bibnamefont{Pederiva}},
  \bibinfo{author}{\bibfnamefont{S.~C.} \bibnamefont{Pieper}},
  \bibinfo{author}{\bibfnamefont{R.}~\bibnamefont{Schiavilla}},
  \bibinfo{author}{\bibfnamefont{K.~E.} \bibnamefont{Schmidt}},
  \bibnamefont{and} \bibinfo{author}{\bibfnamefont{R.~B.}
  \bibnamefont{Wiringa}}, \bibinfo{journal}{Rev. Mod. Phys.}
  \textbf{\bibinfo{volume}{87}}, \bibinfo{pages}{1067} (\bibinfo{year}{2015}),
  \eprint{1412.3081}.

\bibitem[{\citenamefont{Navr\'atil et~al.}(2016)\citenamefont{Navr\'atil,
  Quaglioni, Hupin, Romero-Redondo, and Calci}}]{Navratil:2016ycn}
\bibinfo{author}{\bibfnamefont{P.}~\bibnamefont{Navr\'atil}},
  \bibinfo{author}{\bibfnamefont{S.}~\bibnamefont{Quaglioni}},
  \bibinfo{author}{\bibfnamefont{G.}~\bibnamefont{Hupin}},
  \bibinfo{author}{\bibfnamefont{C.}~\bibnamefont{Romero-Redondo}},
  \bibnamefont{and} \bibinfo{author}{\bibfnamefont{A.}~\bibnamefont{Calci}},
  \bibinfo{journal}{Phys. Scripta} \textbf{\bibinfo{volume}{91}},
  \bibinfo{pages}{053002} (\bibinfo{year}{2016}), \eprint{1601.03765}.

\bibitem[{\citenamefont{Hagen et~al.}(2014)\citenamefont{Hagen, Papenbrock,
  Hjorth-Jensen, and Dean}}]{Hagen:2013nca}
\bibinfo{author}{\bibfnamefont{G.}~\bibnamefont{Hagen}},
  \bibinfo{author}{\bibfnamefont{T.}~\bibnamefont{Papenbrock}},
  \bibinfo{author}{\bibfnamefont{M.}~\bibnamefont{Hjorth-Jensen}},
  \bibnamefont{and} \bibinfo{author}{\bibfnamefont{D.~J.} \bibnamefont{Dean}},
  \bibinfo{journal}{Rept. Prog. Phys.} \textbf{\bibinfo{volume}{77}},
  \bibinfo{pages}{096302} (\bibinfo{year}{2014}), \eprint{1312.7872}.

\bibitem[{\citenamefont{Hergert et~al.}(2016)\citenamefont{Hergert, Bogner,
  Morris, Schwenk, and Tsukiyama}}]{Hergert:2015awm}
\bibinfo{author}{\bibfnamefont{H.}~\bibnamefont{Hergert}},
  \bibinfo{author}{\bibfnamefont{S.~K.} \bibnamefont{Bogner}},
  \bibinfo{author}{\bibfnamefont{T.~D.} \bibnamefont{Morris}},
  \bibinfo{author}{\bibfnamefont{A.}~\bibnamefont{Schwenk}}, \bibnamefont{and}
  \bibinfo{author}{\bibfnamefont{K.}~\bibnamefont{Tsukiyama}},
  \bibinfo{journal}{Phys. Rept.} \textbf{\bibinfo{volume}{621}},
  \bibinfo{pages}{165} (\bibinfo{year}{2016}), \eprint{1512.06956}.

\bibitem[{\citenamefont{Stroberg et~al.}(2019)\citenamefont{Stroberg, Bogner,
  Hergert, and Holt}}]{Stroberg:2019mxo}
\bibinfo{author}{\bibfnamefont{S.~R.} \bibnamefont{Stroberg}},
  \bibinfo{author}{\bibfnamefont{S.~K.} \bibnamefont{Bogner}},
  \bibinfo{author}{\bibfnamefont{H.}~\bibnamefont{Hergert}}, \bibnamefont{and}
  \bibinfo{author}{\bibfnamefont{J.~D.} \bibnamefont{Holt}},
  \bibinfo{journal}{Ann. Rev. Nucl. Part. Sci.} \textbf{\bibinfo{volume}{69}},
  \bibinfo{pages}{307} (\bibinfo{year}{2019}), \eprint{1902.06154}.

\bibitem[{\citenamefont{Gazda et~al.}(2017)\citenamefont{Gazda, Catena, and
  Forssén}}]{Gazda:2016mrp}
\bibinfo{author}{\bibfnamefont{D.}~\bibnamefont{Gazda}},
  \bibinfo{author}{\bibfnamefont{R.}~\bibnamefont{Catena}}, \bibnamefont{and}
  \bibinfo{author}{\bibfnamefont{C.}~\bibnamefont{Forssén}},
  \bibinfo{journal}{Phys. Rev. D} \textbf{\bibinfo{volume}{95}},
  \bibinfo{pages}{103011} (\bibinfo{year}{2017}), \eprint{1612.09165}.

\bibitem[{\citenamefont{Ellis et~al.}(1988{\natexlab{a}})\citenamefont{Ellis,
  Flores, and Lewin}}]{Ellis:1988nb}
\bibinfo{author}{\bibfnamefont{J.~R.} \bibnamefont{Ellis}},
  \bibinfo{author}{\bibfnamefont{R.~A.} \bibnamefont{Flores}},
  \bibnamefont{and} \bibinfo{author}{\bibfnamefont{J.~D.} \bibnamefont{Lewin}},
  \bibinfo{journal}{Phys. Lett. B} \textbf{\bibinfo{volume}{212}},
  \bibinfo{pages}{375} (\bibinfo{year}{1988}{\natexlab{a}}).

\bibitem[{\citenamefont{Baudis et~al.}(2013)\citenamefont{Baudis, Kessler,
  Klos, Lang, Men{\'e}ndez, Reichard, and Schwenk}}]{Baudis:2013bba}
\bibinfo{author}{\bibfnamefont{L.}~\bibnamefont{Baudis}},
  \bibinfo{author}{\bibfnamefont{G.}~\bibnamefont{Kessler}},
  \bibinfo{author}{\bibfnamefont{P.}~\bibnamefont{Klos}},
  \bibinfo{author}{\bibfnamefont{R.~F.} \bibnamefont{Lang}},
  \bibinfo{author}{\bibfnamefont{J.}~\bibnamefont{Men{\'e}ndez}},
  \bibinfo{author}{\bibfnamefont{S.}~\bibnamefont{Reichard}}, \bibnamefont{and}
  \bibinfo{author}{\bibfnamefont{A.}~\bibnamefont{Schwenk}},
  \bibinfo{journal}{Phys. Rev. D} \textbf{\bibinfo{volume}{88}},
  \bibinfo{pages}{115014} (\bibinfo{year}{2013}), \eprint{1309.0825}.

\bibitem[{\citenamefont{Aprile et~al.}(2017{\natexlab{c}})}]{Aprile:2017ngb}
\bibinfo{author}{\bibfnamefont{E.}~\bibnamefont{Aprile}} \bibnamefont{et~al.}
  (\bibinfo{collaboration}{XENON}), \bibinfo{journal}{Phys. Rev. D}
  \textbf{\bibinfo{volume}{96}}, \bibinfo{pages}{022008}
  (\bibinfo{year}{2017}{\natexlab{c}}), \eprint{1705.05830}.

\bibitem[{\citenamefont{Aprile et~al.}(2021{\natexlab{a}})}]{Aprile:2020sfu}
\bibinfo{author}{\bibfnamefont{E.}~\bibnamefont{Aprile}} \bibnamefont{et~al.}
  (\bibinfo{collaboration}{XENON}), \bibinfo{journal}{Phys. Rev. D}
  \textbf{\bibinfo{volume}{103}}, \bibinfo{pages}{063028}
  (\bibinfo{year}{2021}{\natexlab{a}}), \eprint{2011.10431}.

\bibitem[{\citenamefont{Suzuki et~al.}(2019)}]{Suzuki:2019ine}
\bibinfo{author}{\bibfnamefont{T.}~\bibnamefont{Suzuki}} \bibnamefont{et~al.}
  (\bibinfo{collaboration}{XMASS}), \bibinfo{journal}{Astropart. Phys.}
  \textbf{\bibinfo{volume}{110}}, \bibinfo{pages}{1} (\bibinfo{year}{2019}),
  \eprint{1809.05358}.

\bibitem[{\citenamefont{McCabe}(2016)}]{McCabe:2015eia}
\bibinfo{author}{\bibfnamefont{C.}~\bibnamefont{McCabe}},
  \bibinfo{journal}{JCAP} \textbf{\bibinfo{volume}{1605}}, \bibinfo{pages}{033}
  (\bibinfo{year}{2016}), \eprint{1512.00460}.

\bibitem[{\citenamefont{Fieguth et~al.}(2018)\citenamefont{Fieguth,
  Hoferichter, Klos, Men{\'e}ndez, Schwenk, and Weinheimer}}]{Fieguth:2018vob}
\bibinfo{author}{\bibfnamefont{A.}~\bibnamefont{Fieguth}},
  \bibinfo{author}{\bibfnamefont{M.}~\bibnamefont{Hoferichter}},
  \bibinfo{author}{\bibfnamefont{P.}~\bibnamefont{Klos}},
  \bibinfo{author}{\bibfnamefont{J.}~\bibnamefont{Men{\'e}ndez}},
  \bibinfo{author}{\bibfnamefont{A.}~\bibnamefont{Schwenk}}, \bibnamefont{and}
  \bibinfo{author}{\bibfnamefont{C.}~\bibnamefont{Weinheimer}},
  \bibinfo{journal}{Phys. Rev. D} \textbf{\bibinfo{volume}{97}},
  \bibinfo{pages}{103532} (\bibinfo{year}{2018}), \eprint{1802.04294}.

\bibitem[{\citenamefont{Rogers et~al.}(2017)\citenamefont{Rogers, Cerde{\~n}o,
  Cushman, Livet, and Mandic}}]{Rogers:2016jrx}
\bibinfo{author}{\bibfnamefont{H.}~\bibnamefont{Rogers}},
  \bibinfo{author}{\bibfnamefont{D.~G.} \bibnamefont{Cerde{\~n}o}},
  \bibinfo{author}{\bibfnamefont{P.}~\bibnamefont{Cushman}},
  \bibinfo{author}{\bibfnamefont{F.}~\bibnamefont{Livet}}, \bibnamefont{and}
  \bibinfo{author}{\bibfnamefont{V.}~\bibnamefont{Mandic}},
  \bibinfo{journal}{Phys. Rev. D} \textbf{\bibinfo{volume}{95}},
  \bibinfo{pages}{082003} (\bibinfo{year}{2017}), \eprint{1612.09038}.

\bibitem[{\citenamefont{Tucker-Smith and Weiner}(2001)}]{TuckerSmith:2001hy}
\bibinfo{author}{\bibfnamefont{D.}~\bibnamefont{Tucker-Smith}}
  \bibnamefont{and} \bibinfo{author}{\bibfnamefont{N.}~\bibnamefont{Weiner}},
  \bibinfo{journal}{Phys. Rev. D} \textbf{\bibinfo{volume}{64}},
  \bibinfo{pages}{043502} (\bibinfo{year}{2001}), \eprint{hep-ph/0101138}.

\bibitem[{\citenamefont{Graham et~al.}(2010)\citenamefont{Graham, Harnik,
  Rajendran, and Saraswat}}]{Graham:2010ca}
\bibinfo{author}{\bibfnamefont{P.~W.} \bibnamefont{Graham}},
  \bibinfo{author}{\bibfnamefont{R.}~\bibnamefont{Harnik}},
  \bibinfo{author}{\bibfnamefont{S.}~\bibnamefont{Rajendran}},
  \bibnamefont{and} \bibinfo{author}{\bibfnamefont{P.}~\bibnamefont{Saraswat}},
  \bibinfo{journal}{Phys. Rev. D} \textbf{\bibinfo{volume}{82}},
  \bibinfo{pages}{063512} (\bibinfo{year}{2010}), \eprint{1004.0937}.

\bibitem[{\citenamefont{Dienes et~al.}(2015)\citenamefont{Dienes, Kumar,
  Thomas, and Yaylali}}]{Dienes:2014via}
\bibinfo{author}{\bibfnamefont{K.~R.} \bibnamefont{Dienes}},
  \bibinfo{author}{\bibfnamefont{J.}~\bibnamefont{Kumar}},
  \bibinfo{author}{\bibfnamefont{B.}~\bibnamefont{Thomas}}, \bibnamefont{and}
  \bibinfo{author}{\bibfnamefont{D.}~\bibnamefont{Yaylali}},
  \bibinfo{journal}{Phys. Rev. Lett.} \textbf{\bibinfo{volume}{114}},
  \bibinfo{pages}{051301} (\bibinfo{year}{2015}), \eprint{1406.4868}.

\bibitem[{\citenamefont{Hardy et~al.}(2015{\natexlab{a}})\citenamefont{Hardy,
  Lasenby, March-Russell, and West}}]{Hardy:2015boa}
\bibinfo{author}{\bibfnamefont{E.}~\bibnamefont{Hardy}},
  \bibinfo{author}{\bibfnamefont{R.}~\bibnamefont{Lasenby}},
  \bibinfo{author}{\bibfnamefont{J.}~\bibnamefont{March-Russell}},
  \bibnamefont{and} \bibinfo{author}{\bibfnamefont{S.~M.} \bibnamefont{West}},
  \bibinfo{journal}{JHEP} \textbf{\bibinfo{volume}{07}}, \bibinfo{pages}{133}
  (\bibinfo{year}{2015}{\natexlab{a}}), \eprint{1504.05419}.

\bibitem[{\citenamefont{Bramante et~al.}(2016)\citenamefont{Bramante, Fox,
  Kribs, and Martin}}]{Bramante:2016rdh}
\bibinfo{author}{\bibfnamefont{J.}~\bibnamefont{Bramante}},
  \bibinfo{author}{\bibfnamefont{P.~J.} \bibnamefont{Fox}},
  \bibinfo{author}{\bibfnamefont{G.~D.} \bibnamefont{Kribs}}, \bibnamefont{and}
  \bibinfo{author}{\bibfnamefont{A.}~\bibnamefont{Martin}},
  \bibinfo{journal}{Phys. Rev. D} \textbf{\bibinfo{volume}{94}},
  \bibinfo{pages}{115026} (\bibinfo{year}{2016}), \eprint{1608.02662}.

\bibitem[{\citenamefont{Gluscevic et~al.}(2015)\citenamefont{Gluscevic,
  Gresham, McDermott, Peter, and Zurek}}]{Gluscevic:2015sqa}
\bibinfo{author}{\bibfnamefont{V.}~\bibnamefont{Gluscevic}},
  \bibinfo{author}{\bibfnamefont{M.~I.} \bibnamefont{Gresham}},
  \bibinfo{author}{\bibfnamefont{S.~D.} \bibnamefont{McDermott}},
  \bibinfo{author}{\bibfnamefont{A.~H.~G.} \bibnamefont{Peter}},
  \bibnamefont{and} \bibinfo{author}{\bibfnamefont{K.~M.} \bibnamefont{Zurek}},
  \bibinfo{journal}{JCAP} \textbf{\bibinfo{volume}{12}}, \bibinfo{pages}{057}
  (\bibinfo{year}{2015}), \eprint{1506.04454}.

\bibitem[{\citenamefont{Gelmini et~al.}(2018)\citenamefont{Gelmini, Takhistov,
  and Witte}}]{Gelmini:2018ogy}
\bibinfo{author}{\bibfnamefont{G.~B.} \bibnamefont{Gelmini}},
  \bibinfo{author}{\bibfnamefont{V.}~\bibnamefont{Takhistov}},
  \bibnamefont{and} \bibinfo{author}{\bibfnamefont{S.~J.} \bibnamefont{Witte}},
  \bibinfo{journal}{JCAP} \textbf{\bibinfo{volume}{1807}}, \bibinfo{pages}{009}
  (\bibinfo{year}{2018}), \eprint{1804.01638}.

\bibitem[{\citenamefont{Bozorgnia et~al.}(2018)\citenamefont{Bozorgnia,
  Cerdeño, Cheek, and Penning}}]{Bozorgnia:2018jep}
\bibinfo{author}{\bibfnamefont{N.}~\bibnamefont{Bozorgnia}},
  \bibinfo{author}{\bibfnamefont{D.~G.} \bibnamefont{Cerdeño}},
  \bibinfo{author}{\bibfnamefont{A.}~\bibnamefont{Cheek}}, \bibnamefont{and}
  \bibinfo{author}{\bibfnamefont{B.}~\bibnamefont{Penning}},
  \bibinfo{journal}{JCAP} \textbf{\bibinfo{volume}{1812}}, \bibinfo{pages}{013}
  (\bibinfo{year}{2018}), \eprint{1810.05576}.

\bibitem[{\citenamefont{Freese et~al.}(2005)\citenamefont{Freese, Gondolo, and
  Newberg}}]{Freese:2003tt}
\bibinfo{author}{\bibfnamefont{K.}~\bibnamefont{Freese}},
  \bibinfo{author}{\bibfnamefont{P.}~\bibnamefont{Gondolo}}, \bibnamefont{and}
  \bibinfo{author}{\bibfnamefont{H.~J.} \bibnamefont{Newberg}},
  \bibinfo{journal}{Phys. Rev. D} \textbf{\bibinfo{volume}{71}},
  \bibinfo{pages}{043516} (\bibinfo{year}{2005}), \eprint{astro-ph/0309279}.

\bibitem[{\citenamefont{Helmi}(2004)}]{Helmi:2004id}
\bibinfo{author}{\bibfnamefont{A.}~\bibnamefont{Helmi}},
  \bibinfo{journal}{Astrophys. J. Lett.} \textbf{\bibinfo{volume}{610}},
  \bibinfo{pages}{L97} (\bibinfo{year}{2004}), \eprint{astro-ph/0406396}.

\bibitem[{\citenamefont{Vogelsberger et~al.}(2008)\citenamefont{Vogelsberger,
  White, Helmi, and Springel}}]{Vogelsberger:2007ny}
\bibinfo{author}{\bibfnamefont{M.}~\bibnamefont{Vogelsberger}},
  \bibinfo{author}{\bibfnamefont{S.~D.~M.} \bibnamefont{White}},
  \bibinfo{author}{\bibfnamefont{A.}~\bibnamefont{Helmi}}, \bibnamefont{and}
  \bibinfo{author}{\bibfnamefont{V.}~\bibnamefont{Springel}},
  \bibinfo{journal}{Mon. Not. Roy. Astron. Soc.}
  \textbf{\bibinfo{volume}{385}}, \bibinfo{pages}{236} (\bibinfo{year}{2008}),
  \eprint{0711.1105}.

\bibitem[{\citenamefont{Purcell et~al.}(2012)\citenamefont{Purcell, Zentner,
  and Wang}}]{Purcell:2012sh}
\bibinfo{author}{\bibfnamefont{C.~W.} \bibnamefont{Purcell}},
  \bibinfo{author}{\bibfnamefont{A.~R.} \bibnamefont{Zentner}},
  \bibnamefont{and} \bibinfo{author}{\bibfnamefont{M.-Y.} \bibnamefont{Wang}},
  \bibinfo{journal}{JCAP} \textbf{\bibinfo{volume}{08}}, \bibinfo{pages}{027}
  (\bibinfo{year}{2012}), \eprint{1203.6617}.

\bibitem[{\citenamefont{Buckley et~al.}(2019)\citenamefont{Buckley, Mohlabeng,
  and Murphy}}]{Buckley:2019skk}
\bibinfo{author}{\bibfnamefont{M.~R.} \bibnamefont{Buckley}},
  \bibinfo{author}{\bibfnamefont{G.}~\bibnamefont{Mohlabeng}},
  \bibnamefont{and} \bibinfo{author}{\bibfnamefont{C.~W.}
  \bibnamefont{Murphy}}, \bibinfo{journal}{Phys. Rev. D}
  \textbf{\bibinfo{volume}{100}}, \bibinfo{pages}{055039}
  (\bibinfo{year}{2019}), \eprint{1905.05189}.

\bibitem[{\citenamefont{McCabe}(2010)}]{McCabe:2010zh}
\bibinfo{author}{\bibfnamefont{C.}~\bibnamefont{McCabe}},
  \bibinfo{journal}{Phys. Rev. D} \textbf{\bibinfo{volume}{82}},
  \bibinfo{pages}{023530} (\bibinfo{year}{2010}), \eprint{1005.0579}.

\bibitem[{\citenamefont{O'Hare et~al.}(2018)\citenamefont{O'Hare, McCabe,
  Evans, Myeong, and Belokurov}}]{OHare:2018trr}
\bibinfo{author}{\bibfnamefont{C.~A.~J.} \bibnamefont{O'Hare}},
  \bibinfo{author}{\bibfnamefont{C.}~\bibnamefont{McCabe}},
  \bibinfo{author}{\bibfnamefont{N.~W.} \bibnamefont{Evans}},
  \bibinfo{author}{\bibfnamefont{G.}~\bibnamefont{Myeong}}, \bibnamefont{and}
  \bibinfo{author}{\bibfnamefont{V.}~\bibnamefont{Belokurov}},
  \bibinfo{journal}{Phys. Rev. D} \textbf{\bibinfo{volume}{98}},
  \bibinfo{pages}{103006} (\bibinfo{year}{2018}), \eprint{1807.09004}.

\bibitem[{\citenamefont{Adhikari et~al.}(2020)}]{Adhikari:2020gxw}
\bibinfo{author}{\bibfnamefont{P.}~\bibnamefont{Adhikari}} \bibnamefont{et~al.}
  (\bibinfo{collaboration}{DEAP}), \bibinfo{journal}{Phys. Rev. D}
  \textbf{\bibinfo{volume}{102}}, \bibinfo{pages}{082001}
  (\bibinfo{year}{2020}), \eprint{2005.14667}.

\bibitem[{\citenamefont{Bertone et~al.}(2007)\citenamefont{Bertone, Cerdeno,
  Collar, and Odom}}]{Bertone:2007xj}
\bibinfo{author}{\bibfnamefont{G.}~\bibnamefont{Bertone}},
  \bibinfo{author}{\bibfnamefont{D.~G.} \bibnamefont{Cerdeno}},
  \bibinfo{author}{\bibfnamefont{J.}~\bibnamefont{Collar}}, \bibnamefont{and}
  \bibinfo{author}{\bibfnamefont{B.~C.} \bibnamefont{Odom}},
  \bibinfo{journal}{Phys. Rev. Lett.} \textbf{\bibinfo{volume}{99}},
  \bibinfo{pages}{151301} (\bibinfo{year}{2007}), \eprint{0705.2502}.

\bibitem[{\citenamefont{Cerdeño et~al.}(2013)}]{Cerdeno:2013gqa}
\bibinfo{author}{\bibfnamefont{D.}~\bibnamefont{Cerdeño}}
  \bibnamefont{et~al.}, \bibinfo{journal}{JCAP} \textbf{\bibinfo{volume}{07}},
  \bibinfo{pages}{028} (\bibinfo{year}{2013}), \bibinfo{note}{[Erratum: JCAP
  {\bf 09}, E01 (2013)]}, \eprint{1304.1758}.

\bibitem[{\citenamefont{Peter et~al.}(2014)\citenamefont{Peter, Gluscevic,
  Green, Kavanagh, and Lee}}]{Peter:2013aha}
\bibinfo{author}{\bibfnamefont{A.~H.} \bibnamefont{Peter}},
  \bibinfo{author}{\bibfnamefont{V.}~\bibnamefont{Gluscevic}},
  \bibinfo{author}{\bibfnamefont{A.~M.} \bibnamefont{Green}},
  \bibinfo{author}{\bibfnamefont{B.~J.} \bibnamefont{Kavanagh}},
  \bibnamefont{and} \bibinfo{author}{\bibfnamefont{S.~K.} \bibnamefont{Lee}},
  \bibinfo{journal}{Phys. Dark Univ.} \textbf{\bibinfo{volume}{5-6}},
  \bibinfo{pages}{45} (\bibinfo{year}{2014}), \eprint{1310.7039}.

\bibitem[{\citenamefont{Edwards
  et~al.}(2018{\natexlab{a}})\citenamefont{Edwards, Kavanagh, and
  Weniger}}]{Edwards:2018lsl}
\bibinfo{author}{\bibfnamefont{T.~D.} \bibnamefont{Edwards}},
  \bibinfo{author}{\bibfnamefont{B.~J.} \bibnamefont{Kavanagh}},
  \bibnamefont{and} \bibinfo{author}{\bibfnamefont{C.}~\bibnamefont{Weniger}},
  \bibinfo{journal}{Phys. Rev. Lett.} \textbf{\bibinfo{volume}{121}},
  \bibinfo{pages}{181101} (\bibinfo{year}{2018}{\natexlab{a}}),
  \eprint{1805.04117}.

\bibitem[{\citenamefont{Hisano et~al.}(2015{\natexlab{a}})\citenamefont{Hisano,
  Nagai, and Nagata}}]{Hisano:2015bma}
\bibinfo{author}{\bibfnamefont{J.}~\bibnamefont{Hisano}},
  \bibinfo{author}{\bibfnamefont{R.}~\bibnamefont{Nagai}}, \bibnamefont{and}
  \bibinfo{author}{\bibfnamefont{N.}~\bibnamefont{Nagata}},
  \bibinfo{journal}{JHEP} \textbf{\bibinfo{volume}{05}}, \bibinfo{pages}{037}
  (\bibinfo{year}{2015}{\natexlab{a}}), \eprint{1502.02244}.

\bibitem[{\citenamefont{De~Simone and Jacques}(2016)}]{DeSimone:2016fbz}
\bibinfo{author}{\bibfnamefont{A.}~\bibnamefont{De~Simone}} \bibnamefont{and}
  \bibinfo{author}{\bibfnamefont{T.}~\bibnamefont{Jacques}},
  \bibinfo{journal}{Eur. Phys. J. C} \textbf{\bibinfo{volume}{76}},
  \bibinfo{pages}{367} (\bibinfo{year}{2016}), \eprint{1603.08002}.

\bibitem[{\citenamefont{Abdallah et~al.}(2015)}]{Abdallah:2015ter}
\bibinfo{author}{\bibfnamefont{J.}~\bibnamefont{Abdallah}}
  \bibnamefont{et~al.}, \bibinfo{journal}{Phys. Dark Univ.}
  \textbf{\bibinfo{volume}{9-10}}, \bibinfo{pages}{8} (\bibinfo{year}{2015}),
  \eprint{1506.03116}.

\bibitem[{\citenamefont{DiFranzo et~al.}(2013)\citenamefont{DiFranzo, Nagao,
  Rajaraman, and Tait}}]{DiFranzo:2013vra}
\bibinfo{author}{\bibfnamefont{A.}~\bibnamefont{DiFranzo}},
  \bibinfo{author}{\bibfnamefont{K.~I.} \bibnamefont{Nagao}},
  \bibinfo{author}{\bibfnamefont{A.}~\bibnamefont{Rajaraman}},
  \bibnamefont{and} \bibinfo{author}{\bibfnamefont{T.~M.} \bibnamefont{Tait}},
  \bibinfo{journal}{JHEP} \textbf{\bibinfo{volume}{11}}, \bibinfo{pages}{014}
  (\bibinfo{year}{2013}), \bibinfo{note}{[Erratum: JHEP {\bf 01}, 162 (2014)]},
  \eprint{1308.2679}.

\bibitem[{\citenamefont{Abercrombie et~al.}(2020)}]{Abercrombie:2015wmb}
\bibinfo{author}{\bibfnamefont{D.}~\bibnamefont{Abercrombie}}
  \bibnamefont{et~al.}, \bibinfo{journal}{Phys. Dark Univ.}
  \textbf{\bibinfo{volume}{27}}, \bibinfo{pages}{100371}
  (\bibinfo{year}{2020}), \eprint{1507.00966}.

\bibitem[{\citenamefont{Arina et~al.}(2015)\citenamefont{Arina, Del~Nobile, and
  Panci}}]{Arina:2014yna}
\bibinfo{author}{\bibfnamefont{C.}~\bibnamefont{Arina}},
  \bibinfo{author}{\bibfnamefont{E.}~\bibnamefont{Del~Nobile}},
  \bibnamefont{and} \bibinfo{author}{\bibfnamefont{P.}~\bibnamefont{Panci}},
  \bibinfo{journal}{Phys. Rev. Lett.} \textbf{\bibinfo{volume}{114}},
  \bibinfo{pages}{011301} (\bibinfo{year}{2015}), \eprint{1406.5542}.

\bibitem[{\citenamefont{Hisano et~al.}(2018)\citenamefont{Hisano, Nagai, and
  Nagata}}]{Hisano:2018bpz}
\bibinfo{author}{\bibfnamefont{J.}~\bibnamefont{Hisano}},
  \bibinfo{author}{\bibfnamefont{R.}~\bibnamefont{Nagai}}, \bibnamefont{and}
  \bibinfo{author}{\bibfnamefont{N.}~\bibnamefont{Nagata}},
  \bibinfo{journal}{JHEP} \textbf{\bibinfo{volume}{12}}, \bibinfo{pages}{059}
  (\bibinfo{year}{2018}), \eprint{1808.06301}.

\bibitem[{\citenamefont{Balázs et~al.}(2017)\citenamefont{Balázs, Conrad,
  Farmer, Jacques, Li, Meyer, Queiroz, and Sánchez-Conde}}]{Balazs:2017hxh}
\bibinfo{author}{\bibfnamefont{C.}~\bibnamefont{Balázs}},
  \bibinfo{author}{\bibfnamefont{J.}~\bibnamefont{Conrad}},
  \bibinfo{author}{\bibfnamefont{B.}~\bibnamefont{Farmer}},
  \bibinfo{author}{\bibfnamefont{T.}~\bibnamefont{Jacques}},
  \bibinfo{author}{\bibfnamefont{T.}~\bibnamefont{Li}},
  \bibinfo{author}{\bibfnamefont{M.}~\bibnamefont{Meyer}},
  \bibinfo{author}{\bibfnamefont{F.~S.} \bibnamefont{Queiroz}},
  \bibnamefont{and} \bibinfo{author}{\bibfnamefont{M.~A.}
  \bibnamefont{Sánchez-Conde}}, \bibinfo{journal}{Phys. Rev. D}
  \textbf{\bibinfo{volume}{96}}, \bibinfo{pages}{083002}
  (\bibinfo{year}{2017}), \eprint{1706.01505}.

\bibitem[{\citenamefont{Jacques et~al.}(2016)\citenamefont{Jacques, Katz,
  Morgante, Racco, Rameez, and Riotto}}]{Jacques:2016dqz}
\bibinfo{author}{\bibfnamefont{T.}~\bibnamefont{Jacques}},
  \bibinfo{author}{\bibfnamefont{A.}~\bibnamefont{Katz}},
  \bibinfo{author}{\bibfnamefont{E.}~\bibnamefont{Morgante}},
  \bibinfo{author}{\bibfnamefont{D.}~\bibnamefont{Racco}},
  \bibinfo{author}{\bibfnamefont{M.}~\bibnamefont{Rameez}}, \bibnamefont{and}
  \bibinfo{author}{\bibfnamefont{A.}~\bibnamefont{Riotto}},
  \bibinfo{journal}{JHEP} \textbf{\bibinfo{volume}{10}}, \bibinfo{pages}{071}
  (\bibinfo{year}{2016}), \bibinfo{note}{[Erratum: JHEP {\bf 01}, 127 (2019)]},
  \eprint{1605.06513}.

\bibitem[{\citenamefont{Buckley et~al.}(2015)\citenamefont{Buckley, Feld, and
  Goncalves}}]{Buckley:2014fba}
\bibinfo{author}{\bibfnamefont{M.~R.} \bibnamefont{Buckley}},
  \bibinfo{author}{\bibfnamefont{D.}~\bibnamefont{Feld}}, \bibnamefont{and}
  \bibinfo{author}{\bibfnamefont{D.}~\bibnamefont{Goncalves}},
  \bibinfo{journal}{Phys. Rev. D} \textbf{\bibinfo{volume}{91}},
  \bibinfo{pages}{015017} (\bibinfo{year}{2015}), \eprint{1410.6497}.

\bibitem[{\citenamefont{Berlin et~al.}(2014)\citenamefont{Berlin, Hooper, and
  McDermott}}]{Berlin:2014tja}
\bibinfo{author}{\bibfnamefont{A.}~\bibnamefont{Berlin}},
  \bibinfo{author}{\bibfnamefont{D.}~\bibnamefont{Hooper}}, \bibnamefont{and}
  \bibinfo{author}{\bibfnamefont{S.~D.} \bibnamefont{McDermott}},
  \bibinfo{journal}{Phys. Rev. D} \textbf{\bibinfo{volume}{89}},
  \bibinfo{pages}{115022} (\bibinfo{year}{2014}), \eprint{1404.0022}.

\bibitem[{\citenamefont{Blanco et~al.}(2019{\natexlab{a}})\citenamefont{Blanco,
  Escudero, Hooper, and Witte}}]{Blanco:2019hah}
\bibinfo{author}{\bibfnamefont{C.}~\bibnamefont{Blanco}},
  \bibinfo{author}{\bibfnamefont{M.}~\bibnamefont{Escudero}},
  \bibinfo{author}{\bibfnamefont{D.}~\bibnamefont{Hooper}}, \bibnamefont{and}
  \bibinfo{author}{\bibfnamefont{S.~J.} \bibnamefont{Witte}},
  \bibinfo{journal}{JCAP} \textbf{\bibinfo{volume}{11}}, \bibinfo{pages}{024}
  (\bibinfo{year}{2019}{\natexlab{a}}), \eprint{1907.05893}.

\bibitem[{\citenamefont{Drees and Nojiri}(1993)}]{Drees:1993bu}
\bibinfo{author}{\bibfnamefont{M.}~\bibnamefont{Drees}} \bibnamefont{and}
  \bibinfo{author}{\bibfnamefont{M.}~\bibnamefont{Nojiri}},
  \bibinfo{journal}{Phys. Rev. D} \textbf{\bibinfo{volume}{48}},
  \bibinfo{pages}{3483} (\bibinfo{year}{1993}), \eprint{hep-ph/9307208}.

\bibitem[{\citenamefont{Hisano et~al.}(2010)\citenamefont{Hisano, Ishiwata, and
  Nagata}}]{Hisano:2010ct}
\bibinfo{author}{\bibfnamefont{J.}~\bibnamefont{Hisano}},
  \bibinfo{author}{\bibfnamefont{K.}~\bibnamefont{Ishiwata}}, \bibnamefont{and}
  \bibinfo{author}{\bibfnamefont{N.}~\bibnamefont{Nagata}},
  \bibinfo{journal}{Phys. Rev. D} \textbf{\bibinfo{volume}{82}},
  \bibinfo{pages}{115007} (\bibinfo{year}{2010}), \eprint{1007.2601}.

\bibitem[{\citenamefont{Baek et~al.}(2016)\citenamefont{Baek, Ko, and
  Wu}}]{Baek:2016lnv}
\bibinfo{author}{\bibfnamefont{S.}~\bibnamefont{Baek}},
  \bibinfo{author}{\bibfnamefont{P.}~\bibnamefont{Ko}}, \bibnamefont{and}
  \bibinfo{author}{\bibfnamefont{P.}~\bibnamefont{Wu}}, \bibinfo{journal}{JHEP}
  \textbf{\bibinfo{volume}{10}}, \bibinfo{pages}{117} (\bibinfo{year}{2016}),
  \eprint{1606.00072}.

\bibitem[{\citenamefont{Baek et~al.}(2018)\citenamefont{Baek, Ko, and
  Wu}}]{Baek:2017ykw}
\bibinfo{author}{\bibfnamefont{S.}~\bibnamefont{Baek}},
  \bibinfo{author}{\bibfnamefont{P.}~\bibnamefont{Ko}}, \bibnamefont{and}
  \bibinfo{author}{\bibfnamefont{P.}~\bibnamefont{Wu}}, \bibinfo{journal}{JCAP}
  \textbf{\bibinfo{volume}{07}}, \bibinfo{pages}{008} (\bibinfo{year}{2018}),
  \eprint{1709.00697}.

\bibitem[{\citenamefont{Arcadi et~al.}(2018{\natexlab{b}})\citenamefont{Arcadi,
  Lindner, Queiroz, Rodejohann, and Vogl}}]{Arcadi:2017wqi}
\bibinfo{author}{\bibfnamefont{G.}~\bibnamefont{Arcadi}},
  \bibinfo{author}{\bibfnamefont{M.}~\bibnamefont{Lindner}},
  \bibinfo{author}{\bibfnamefont{F.~S.} \bibnamefont{Queiroz}},
  \bibinfo{author}{\bibfnamefont{W.}~\bibnamefont{Rodejohann}},
  \bibnamefont{and} \bibinfo{author}{\bibfnamefont{S.}~\bibnamefont{Vogl}},
  \bibinfo{journal}{JCAP} \textbf{\bibinfo{volume}{03}}, \bibinfo{pages}{042}
  (\bibinfo{year}{2018}{\natexlab{b}}), \eprint{1711.02110}.

\bibitem[{\citenamefont{Li}(2018)}]{Li:2018qip}
\bibinfo{author}{\bibfnamefont{T.}~\bibnamefont{Li}}, \bibinfo{journal}{Phys.
  Lett. B} \textbf{\bibinfo{volume}{782}}, \bibinfo{pages}{497}
  (\bibinfo{year}{2018}), \eprint{1804.02120}.

\bibitem[{\citenamefont{Abe et~al.}(2019{\natexlab{b}})\citenamefont{Abe,
  Fujiwara, and Hisano}}]{Abe:2018emu}
\bibinfo{author}{\bibfnamefont{T.}~\bibnamefont{Abe}},
  \bibinfo{author}{\bibfnamefont{M.}~\bibnamefont{Fujiwara}}, \bibnamefont{and}
  \bibinfo{author}{\bibfnamefont{J.}~\bibnamefont{Hisano}},
  \bibinfo{journal}{JHEP} \textbf{\bibinfo{volume}{02}}, \bibinfo{pages}{028}
  (\bibinfo{year}{2019}{\natexlab{b}}), \eprint{1810.01039}.

\bibitem[{\citenamefont{Li and Wu}(2019)}]{Li:2019fnn}
\bibinfo{author}{\bibfnamefont{T.}~\bibnamefont{Li}} \bibnamefont{and}
  \bibinfo{author}{\bibfnamefont{P.}~\bibnamefont{Wu}}, \bibinfo{journal}{Chin.
  Phys. C} \textbf{\bibinfo{volume}{43}}, \bibinfo{pages}{113102}
  (\bibinfo{year}{2019}), \eprint{1904.03407}.

\bibitem[{\citenamefont{Mohan et~al.}(2019)\citenamefont{Mohan, Sengupta, Tait,
  Yan, and Yuan}}]{Mohan:2019zrk}
\bibinfo{author}{\bibfnamefont{K.~A.} \bibnamefont{Mohan}},
  \bibinfo{author}{\bibfnamefont{D.}~\bibnamefont{Sengupta}},
  \bibinfo{author}{\bibfnamefont{T.~M.} \bibnamefont{Tait}},
  \bibinfo{author}{\bibfnamefont{B.}~\bibnamefont{Yan}}, \bibnamefont{and}
  \bibinfo{author}{\bibfnamefont{C.-P.} \bibnamefont{Yuan}},
  \bibinfo{journal}{JHEP} \textbf{\bibinfo{volume}{05}}, \bibinfo{pages}{115}
  (\bibinfo{year}{2019}), \eprint{1903.05650}.

\bibitem[{\citenamefont{Ertas and Kahlhoefer}(2019)}]{Ertas:2019dew}
\bibinfo{author}{\bibfnamefont{F.}~\bibnamefont{Ertas}} \bibnamefont{and}
  \bibinfo{author}{\bibfnamefont{F.}~\bibnamefont{Kahlhoefer}},
  \bibinfo{journal}{JHEP} \textbf{\bibinfo{volume}{06}}, \bibinfo{pages}{052}
  (\bibinfo{year}{2019}), \eprint{1902.11070}.

\bibitem[{\citenamefont{Giacchino et~al.}(2016)\citenamefont{Giacchino, Ibarra,
  Lopez~Honorez, Tytgat, and Wild}}]{Giacchino:2015hvk}
\bibinfo{author}{\bibfnamefont{F.}~\bibnamefont{Giacchino}},
  \bibinfo{author}{\bibfnamefont{A.}~\bibnamefont{Ibarra}},
  \bibinfo{author}{\bibfnamefont{L.}~\bibnamefont{Lopez~Honorez}},
  \bibinfo{author}{\bibfnamefont{M.~H.~G.} \bibnamefont{Tytgat}},
  \bibnamefont{and} \bibinfo{author}{\bibfnamefont{S.}~\bibnamefont{Wild}},
  \bibinfo{journal}{JCAP} \textbf{\bibinfo{volume}{02}}, \bibinfo{pages}{002}
  (\bibinfo{year}{2016}), \eprint{1511.04452}.

\bibitem[{\citenamefont{Giacchino et~al.}(2014)\citenamefont{Giacchino,
  Lopez-Honorez, and Tytgat}}]{Giacchino:2014moa}
\bibinfo{author}{\bibfnamefont{F.}~\bibnamefont{Giacchino}},
  \bibinfo{author}{\bibfnamefont{L.}~\bibnamefont{Lopez-Honorez}},
  \bibnamefont{and} \bibinfo{author}{\bibfnamefont{M.~H.}
  \bibnamefont{Tytgat}}, \bibinfo{journal}{JCAP} \textbf{\bibinfo{volume}{08}},
  \bibinfo{pages}{046} (\bibinfo{year}{2014}), \eprint{1405.6921}.

\bibitem[{\citenamefont{Ibarra et~al.}(2014)\citenamefont{Ibarra, Toma,
  Totzauer, and Wild}}]{Ibarra:2014qma}
\bibinfo{author}{\bibfnamefont{A.}~\bibnamefont{Ibarra}},
  \bibinfo{author}{\bibfnamefont{T.}~\bibnamefont{Toma}},
  \bibinfo{author}{\bibfnamefont{M.}~\bibnamefont{Totzauer}}, \bibnamefont{and}
  \bibinfo{author}{\bibfnamefont{S.}~\bibnamefont{Wild}},
  \bibinfo{journal}{Phys. Rev. D} \textbf{\bibinfo{volume}{90}},
  \bibinfo{pages}{043526} (\bibinfo{year}{2014}), \eprint{1405.6917}.

\bibitem[{\citenamefont{Colucci
  et~al.}(2018{\natexlab{a}})\citenamefont{Colucci, Fuks, Giacchino,
  Lopez~Honorez, Tytgat, and Vandecasteele}}]{Colucci:2018vxz}
\bibinfo{author}{\bibfnamefont{S.}~\bibnamefont{Colucci}},
  \bibinfo{author}{\bibfnamefont{B.}~\bibnamefont{Fuks}},
  \bibinfo{author}{\bibfnamefont{F.}~\bibnamefont{Giacchino}},
  \bibinfo{author}{\bibfnamefont{L.}~\bibnamefont{Lopez~Honorez}},
  \bibinfo{author}{\bibfnamefont{M.~H.} \bibnamefont{Tytgat}},
  \bibnamefont{and}
  \bibinfo{author}{\bibfnamefont{J.}~\bibnamefont{Vandecasteele}},
  \bibinfo{journal}{Phys. Rev. D} \textbf{\bibinfo{volume}{98}},
  \bibinfo{pages}{035002} (\bibinfo{year}{2018}{\natexlab{a}}),
  \eprint{1804.05068}.

\bibitem[{\citenamefont{Colucci
  et~al.}(2018{\natexlab{b}})\citenamefont{Colucci, Giacchino, Tytgat, and
  Vandecasteele}}]{Colucci:2018qml}
\bibinfo{author}{\bibfnamefont{S.}~\bibnamefont{Colucci}},
  \bibinfo{author}{\bibfnamefont{F.}~\bibnamefont{Giacchino}},
  \bibinfo{author}{\bibfnamefont{M.~H.} \bibnamefont{Tytgat}},
  \bibnamefont{and}
  \bibinfo{author}{\bibfnamefont{J.}~\bibnamefont{Vandecasteele}},
  \bibinfo{journal}{Phys. Rev. D} \textbf{\bibinfo{volume}{98}},
  \bibinfo{pages}{115029} (\bibinfo{year}{2018}{\natexlab{b}}),
  \eprint{1805.10173}.

\bibitem[{\citenamefont{Chao}(2019)}]{Chao:2019lhb}
\bibinfo{author}{\bibfnamefont{W.}~\bibnamefont{Chao}}, \bibinfo{journal}{JHEP}
  \textbf{\bibinfo{volume}{11}}, \bibinfo{pages}{013} (\bibinfo{year}{2019}),
  \eprint{1904.09785}.

\bibitem[{\citenamefont{LaFontaine et~al.}(2021)\citenamefont{LaFontaine,
  Tallman, Ellis, Croteau, Torres, Hernandez, Guerrero, Jaksik, Lubanski, and
  Allen}}]{LaFontaine:2021cin}
\bibinfo{author}{\bibfnamefont{C.}~\bibnamefont{LaFontaine}},
  \bibinfo{author}{\bibfnamefont{B.}~\bibnamefont{Tallman}},
  \bibinfo{author}{\bibfnamefont{S.}~\bibnamefont{Ellis}},
  \bibinfo{author}{\bibfnamefont{T.}~\bibnamefont{Croteau}},
  \bibinfo{author}{\bibfnamefont{B.}~\bibnamefont{Torres}},
  \bibinfo{author}{\bibfnamefont{S.}~\bibnamefont{Hernandez}},
  \bibinfo{author}{\bibfnamefont{D.~C.} \bibnamefont{Guerrero}},
  \bibinfo{author}{\bibfnamefont{J.}~\bibnamefont{Jaksik}},
  \bibinfo{author}{\bibfnamefont{D.}~\bibnamefont{Lubanski}}, \bibnamefont{and}
  \bibinfo{author}{\bibfnamefont{R.~E.} \bibnamefont{Allen}},
  \bibinfo{journal}{Universe} \textbf{\bibinfo{volume}{7}},
  \bibinfo{pages}{270} (\bibinfo{year}{2021}), \eprint{2107.14390}.

\bibitem[{\citenamefont{Cirelli et~al.}(2006)\citenamefont{Cirelli, Fornengo,
  and Strumia}}]{Cirelli:2005uq}
\bibinfo{author}{\bibfnamefont{M.}~\bibnamefont{Cirelli}},
  \bibinfo{author}{\bibfnamefont{N.}~\bibnamefont{Fornengo}}, \bibnamefont{and}
  \bibinfo{author}{\bibfnamefont{A.}~\bibnamefont{Strumia}},
  \bibinfo{journal}{Nucl. Phys. B} \textbf{\bibinfo{volume}{753}},
  \bibinfo{pages}{178} (\bibinfo{year}{2006}), \eprint{hep-ph/0512090}.

\bibitem[{\citenamefont{Cirelli and Strumia}(2009)}]{Cirelli:2009uv}
\bibinfo{author}{\bibfnamefont{M.}~\bibnamefont{Cirelli}} \bibnamefont{and}
  \bibinfo{author}{\bibfnamefont{A.}~\bibnamefont{Strumia}},
  \bibinfo{journal}{New J. Phys.} \textbf{\bibinfo{volume}{11}},
  \bibinfo{pages}{105005} (\bibinfo{year}{2009}), \eprint{0903.3381}.

\bibitem[{\citenamefont{Di~Luzio et~al.}(2019)\citenamefont{Di~Luzio, Gr\"ober,
  and Panico}}]{DiLuzio:2018jwd}
\bibinfo{author}{\bibfnamefont{L.}~\bibnamefont{Di~Luzio}},
  \bibinfo{author}{\bibfnamefont{R.}~\bibnamefont{Gr\"ober}}, \bibnamefont{and}
  \bibinfo{author}{\bibfnamefont{G.}~\bibnamefont{Panico}},
  \bibinfo{journal}{JHEP} \textbf{\bibinfo{volume}{01}}, \bibinfo{pages}{011}
  (\bibinfo{year}{2019}), \eprint{1810.10993}.

\bibitem[{\citenamefont{Thomas and Wells}(1998)}]{Thomas:1998wy}
\bibinfo{author}{\bibfnamefont{S.~D.} \bibnamefont{Thomas}} \bibnamefont{and}
  \bibinfo{author}{\bibfnamefont{J.~D.} \bibnamefont{Wells}},
  \bibinfo{journal}{Phys. Rev. Lett.} \textbf{\bibinfo{volume}{81}},
  \bibinfo{pages}{34} (\bibinfo{year}{1998}), \eprint{hep-ph/9804359}.

\bibitem[{\citenamefont{Buckley et~al.}(2011)\citenamefont{Buckley, Randall,
  and Shuve}}]{Buckley:2009kv}
\bibinfo{author}{\bibfnamefont{M.~R.} \bibnamefont{Buckley}},
  \bibinfo{author}{\bibfnamefont{L.}~\bibnamefont{Randall}}, \bibnamefont{and}
  \bibinfo{author}{\bibfnamefont{B.}~\bibnamefont{Shuve}},
  \bibinfo{journal}{JHEP} \textbf{\bibinfo{volume}{05}}, \bibinfo{pages}{097}
  (\bibinfo{year}{2011}), \eprint{0909.4549}.

\bibitem[{\citenamefont{Ibe et~al.}(2013)\citenamefont{Ibe, Matsumoto, and
  Sato}}]{Ibe:2012sx}
\bibinfo{author}{\bibfnamefont{M.}~\bibnamefont{Ibe}},
  \bibinfo{author}{\bibfnamefont{S.}~\bibnamefont{Matsumoto}},
  \bibnamefont{and} \bibinfo{author}{\bibfnamefont{R.}~\bibnamefont{Sato}},
  \bibinfo{journal}{Phys. Lett. B} \textbf{\bibinfo{volume}{721}},
  \bibinfo{pages}{252} (\bibinfo{year}{2013}), \eprint{1212.5989}.

\bibitem[{\citenamefont{Bottaro et~al.}(2022)\citenamefont{Bottaro, Buttazzo,
  Costa, Franceschini, Panci, Redigolo, and Vittorio}}]{Bottaro:2021snn}
\bibinfo{author}{\bibfnamefont{S.}~\bibnamefont{Bottaro}},
  \bibinfo{author}{\bibfnamefont{D.}~\bibnamefont{Buttazzo}},
  \bibinfo{author}{\bibfnamefont{M.}~\bibnamefont{Costa}},
  \bibinfo{author}{\bibfnamefont{R.}~\bibnamefont{Franceschini}},
  \bibinfo{author}{\bibfnamefont{P.}~\bibnamefont{Panci}},
  \bibinfo{author}{\bibfnamefont{D.}~\bibnamefont{Redigolo}}, \bibnamefont{and}
  \bibinfo{author}{\bibfnamefont{L.}~\bibnamefont{Vittorio}},
  \bibinfo{journal}{Eur. Phys. J. C} \textbf{\bibinfo{volume}{82}},
  \bibinfo{pages}{31} (\bibinfo{year}{2022}), \eprint{2107.09688}.

\bibitem[{\citenamefont{Di~Luzio et~al.}(2015)\citenamefont{Di~Luzio, Gr\"ober,
  Kamenik, and Nardecchia}}]{DiLuzio:2015oha}
\bibinfo{author}{\bibfnamefont{L.}~\bibnamefont{Di~Luzio}},
  \bibinfo{author}{\bibfnamefont{R.}~\bibnamefont{Gr\"ober}},
  \bibinfo{author}{\bibfnamefont{J.~F.} \bibnamefont{Kamenik}},
  \bibnamefont{and}
  \bibinfo{author}{\bibfnamefont{M.}~\bibnamefont{Nardecchia}},
  \bibinfo{journal}{JHEP} \textbf{\bibinfo{volume}{07}}, \bibinfo{pages}{074}
  (\bibinfo{year}{2015}), \eprint{1504.00359}.

\bibitem[{\citenamefont{Pelaggi et~al.}(2018)\citenamefont{Pelaggi, Plascencia,
  Salvio, Sannino, Smirnov, and Strumia}}]{Pelaggi:2017abg}
\bibinfo{author}{\bibfnamefont{G.~M.} \bibnamefont{Pelaggi}},
  \bibinfo{author}{\bibfnamefont{A.~D.} \bibnamefont{Plascencia}},
  \bibinfo{author}{\bibfnamefont{A.}~\bibnamefont{Salvio}},
  \bibinfo{author}{\bibfnamefont{F.}~\bibnamefont{Sannino}},
  \bibinfo{author}{\bibfnamefont{J.}~\bibnamefont{Smirnov}}, \bibnamefont{and}
  \bibinfo{author}{\bibfnamefont{A.}~\bibnamefont{Strumia}},
  \bibinfo{journal}{Phys. Rev. D} \textbf{\bibinfo{volume}{97}},
  \bibinfo{pages}{095013} (\bibinfo{year}{2018}), \eprint{1708.00437}.

\bibitem[{\citenamefont{Smirnov and Beacom}(2019)}]{Smirnov:2019ngs}
\bibinfo{author}{\bibfnamefont{J.}~\bibnamefont{Smirnov}} \bibnamefont{and}
  \bibinfo{author}{\bibfnamefont{J.~F.} \bibnamefont{Beacom}},
  \bibinfo{journal}{Phys. Rev. D} \textbf{\bibinfo{volume}{100}},
  \bibinfo{pages}{043029} (\bibinfo{year}{2019}), \eprint{1904.11503}.

\bibitem[{\citenamefont{Belotsky et~al.}(2005)\citenamefont{Belotsky, Khlopov,
  Legonkov, and Shibaev}}]{Belotsky:2005dk}
\bibinfo{author}{\bibfnamefont{K.~M.} \bibnamefont{Belotsky}},
  \bibinfo{author}{\bibfnamefont{M.}~\bibnamefont{Khlopov}},
  \bibinfo{author}{\bibfnamefont{S.}~\bibnamefont{Legonkov}}, \bibnamefont{and}
  \bibinfo{author}{\bibfnamefont{K.}~\bibnamefont{Shibaev}},
  \bibinfo{journal}{Grav. Cosmol.} \textbf{\bibinfo{volume}{11}},
  \bibinfo{pages}{27} (\bibinfo{year}{2005}), \eprint{astro-ph/0504621}.

\bibitem[{\citenamefont{Hisano et~al.}(2007)\citenamefont{Hisano, Matsumoto,
  Nagai, Saito, and Senami}}]{Hisano:2006nn}
\bibinfo{author}{\bibfnamefont{J.}~\bibnamefont{Hisano}},
  \bibinfo{author}{\bibfnamefont{S.}~\bibnamefont{Matsumoto}},
  \bibinfo{author}{\bibfnamefont{M.}~\bibnamefont{Nagai}},
  \bibinfo{author}{\bibfnamefont{O.}~\bibnamefont{Saito}}, \bibnamefont{and}
  \bibinfo{author}{\bibfnamefont{M.}~\bibnamefont{Senami}},
  \bibinfo{journal}{Phys. Lett. B} \textbf{\bibinfo{volume}{646}},
  \bibinfo{pages}{34} (\bibinfo{year}{2007}), \eprint{hep-ph/0610249}.

\bibitem[{\citenamefont{Cirelli et~al.}(2007)\citenamefont{Cirelli, Strumia,
  and Tamburini}}]{Cirelli:2007xd}
\bibinfo{author}{\bibfnamefont{M.}~\bibnamefont{Cirelli}},
  \bibinfo{author}{\bibfnamefont{A.}~\bibnamefont{Strumia}}, \bibnamefont{and}
  \bibinfo{author}{\bibfnamefont{M.}~\bibnamefont{Tamburini}},
  \bibinfo{journal}{Nucl. Phys. B} \textbf{\bibinfo{volume}{787}},
  \bibinfo{pages}{152} (\bibinfo{year}{2007}), \eprint{0706.4071}.

\bibitem[{\citenamefont{March-Russell et~al.}(2008)\citenamefont{March-Russell,
  West, Cumberbatch, and Hooper}}]{March-Russell:2008lng}
\bibinfo{author}{\bibfnamefont{J.}~\bibnamefont{March-Russell}},
  \bibinfo{author}{\bibfnamefont{S.~M.} \bibnamefont{West}},
  \bibinfo{author}{\bibfnamefont{D.}~\bibnamefont{Cumberbatch}},
  \bibnamefont{and} \bibinfo{author}{\bibfnamefont{D.}~\bibnamefont{Hooper}},
  \bibinfo{journal}{JHEP} \textbf{\bibinfo{volume}{07}}, \bibinfo{pages}{058}
  (\bibinfo{year}{2008}), \eprint{0801.3440}.

\bibitem[{\citenamefont{Arkani-Hamed et~al.}(2009)\citenamefont{Arkani-Hamed,
  Finkbeiner, Slatyer, and Weiner}}]{ArkaniHamed:2008qn}
\bibinfo{author}{\bibfnamefont{N.}~\bibnamefont{Arkani-Hamed}},
  \bibinfo{author}{\bibfnamefont{D.~P.} \bibnamefont{Finkbeiner}},
  \bibinfo{author}{\bibfnamefont{T.~R.} \bibnamefont{Slatyer}},
  \bibnamefont{and} \bibinfo{author}{\bibfnamefont{N.}~\bibnamefont{Weiner}},
  \bibinfo{journal}{Phys. Rev. D} \textbf{\bibinfo{volume}{79}},
  \bibinfo{pages}{015014} (\bibinfo{year}{2009}), \eprint{0810.0713}.

\bibitem[{\citenamefont{Cassel}(2010)}]{Cassel:2009wt}
\bibinfo{author}{\bibfnamefont{S.}~\bibnamefont{Cassel}}, \bibinfo{journal}{J.
  Phys. G} \textbf{\bibinfo{volume}{37}}, \bibinfo{pages}{105009}
  (\bibinfo{year}{2010}), \eprint{0903.5307}.

\bibitem[{\citenamefont{March-Russell and West}(2009)}]{March-Russell:2008klu}
\bibinfo{author}{\bibfnamefont{J.~D.} \bibnamefont{March-Russell}}
  \bibnamefont{and} \bibinfo{author}{\bibfnamefont{S.~M.} \bibnamefont{West}},
  \bibinfo{journal}{Phys. Lett. B} \textbf{\bibinfo{volume}{676}},
  \bibinfo{pages}{133} (\bibinfo{year}{2009}), \eprint{0812.0559}.

\bibitem[{\citenamefont{von Harling and Petraki}(2014)}]{vonHarling:2014kha}
\bibinfo{author}{\bibfnamefont{B.}~\bibnamefont{von Harling}} \bibnamefont{and}
  \bibinfo{author}{\bibfnamefont{K.}~\bibnamefont{Petraki}},
  \bibinfo{journal}{JCAP} \textbf{\bibinfo{volume}{12}}, \bibinfo{pages}{033}
  (\bibinfo{year}{2014}), \eprint{1407.7874}.

\bibitem[{\citenamefont{An et~al.}(2016)\citenamefont{An, Wise, and
  Zhang}}]{An:2016gad}
\bibinfo{author}{\bibfnamefont{H.}~\bibnamefont{An}},
  \bibinfo{author}{\bibfnamefont{M.~B.} \bibnamefont{Wise}}, \bibnamefont{and}
  \bibinfo{author}{\bibfnamefont{Y.}~\bibnamefont{Zhang}},
  \bibinfo{journal}{Phys. Rev. D} \textbf{\bibinfo{volume}{93}},
  \bibinfo{pages}{115020} (\bibinfo{year}{2016}), \eprint{1604.01776}.

\bibitem[{\citenamefont{Mitridate et~al.}(2017)\citenamefont{Mitridate, Redi,
  Smirnov, and Strumia}}]{Mitridate:2017izz}
\bibinfo{author}{\bibfnamefont{A.}~\bibnamefont{Mitridate}},
  \bibinfo{author}{\bibfnamefont{M.}~\bibnamefont{Redi}},
  \bibinfo{author}{\bibfnamefont{J.}~\bibnamefont{Smirnov}}, \bibnamefont{and}
  \bibinfo{author}{\bibfnamefont{A.}~\bibnamefont{Strumia}},
  \bibinfo{journal}{JCAP} \textbf{\bibinfo{volume}{05}}, \bibinfo{pages}{006}
  (\bibinfo{year}{2017}), \eprint{1702.01141}.

\bibitem[{\citenamefont{Binder et~al.}(2021)\citenamefont{Binder, Filimonova,
  Petraki, and White}}]{Binder:2021vfo}
\bibinfo{author}{\bibfnamefont{T.}~\bibnamefont{Binder}},
  \bibinfo{author}{\bibfnamefont{A.}~\bibnamefont{Filimonova}},
  \bibinfo{author}{\bibfnamefont{K.}~\bibnamefont{Petraki}}, \bibnamefont{and}
  \bibinfo{author}{\bibfnamefont{G.}~\bibnamefont{White}}
  (\bibinfo{year}{2021}), \eprint{2112.00042}.

\bibitem[{\citenamefont{Del~Nobile et~al.}(2016)\citenamefont{Del~Nobile,
  Nardecchia, and Panci}}]{DelNobile:2015bqo}
\bibinfo{author}{\bibfnamefont{E.}~\bibnamefont{Del~Nobile}},
  \bibinfo{author}{\bibfnamefont{M.}~\bibnamefont{Nardecchia}},
  \bibnamefont{and} \bibinfo{author}{\bibfnamefont{P.}~\bibnamefont{Panci}},
  \bibinfo{journal}{JCAP} \textbf{\bibinfo{volume}{04}}, \bibinfo{pages}{048}
  (\bibinfo{year}{2016}), \eprint{1512.05353}.

\bibitem[{\citenamefont{Low and Wang}(2014)}]{Low:2014cba}
\bibinfo{author}{\bibfnamefont{M.}~\bibnamefont{Low}} \bibnamefont{and}
  \bibinfo{author}{\bibfnamefont{L.-T.} \bibnamefont{Wang}},
  \bibinfo{journal}{JHEP} \textbf{\bibinfo{volume}{08}}, \bibinfo{pages}{161}
  (\bibinfo{year}{2014}), \eprint{1404.0682}.

\bibitem[{\citenamefont{Cirelli et~al.}(2014)\citenamefont{Cirelli, Sala, and
  Taoso}}]{Cirelli:2014dsa}
\bibinfo{author}{\bibfnamefont{M.}~\bibnamefont{Cirelli}},
  \bibinfo{author}{\bibfnamefont{F.}~\bibnamefont{Sala}}, \bibnamefont{and}
  \bibinfo{author}{\bibfnamefont{M.}~\bibnamefont{Taoso}},
  \bibinfo{journal}{JHEP} \textbf{\bibinfo{volume}{10}}, \bibinfo{pages}{033}
  (\bibinfo{year}{2014}), \bibinfo{note}{[Erratum: JHEP 01, 041 (2015)]},
  \eprint{1407.7058}.

\bibitem[{\citenamefont{Han et~al.}(2018)\citenamefont{Han, Mukhopadhyay, and
  Wang}}]{Han:2018wus}
\bibinfo{author}{\bibfnamefont{T.}~\bibnamefont{Han}},
  \bibinfo{author}{\bibfnamefont{S.}~\bibnamefont{Mukhopadhyay}},
  \bibnamefont{and} \bibinfo{author}{\bibfnamefont{X.}~\bibnamefont{Wang}},
  \bibinfo{journal}{Phys. Rev. D} \textbf{\bibinfo{volume}{98}},
  \bibinfo{pages}{035026} (\bibinfo{year}{2018}), \eprint{1805.00015}.

\bibitem[{\citenamefont{Cid~Vidal et~al.}(2019)}]{CidVidal:2018eel}
\bibinfo{author}{\bibfnamefont{X.}~\bibnamefont{Cid~Vidal}}
  \bibnamefont{et~al.}, \bibinfo{journal}{CERN Yellow Rep. Monogr.}
  \textbf{\bibinfo{volume}{7}}, \bibinfo{pages}{585} (\bibinfo{year}{2019}),
  \eprint{1812.07831}.

\bibitem[{\citenamefont{Ellis et~al.}(2019)}]{Strategy:2019vxc}
\bibinfo{author}{\bibfnamefont{R.~K.} \bibnamefont{Ellis}} \bibnamefont{et~al.}
  (\bibinfo{year}{2019}), \eprint{1910.11775}.

\bibitem[{\citenamefont{Han et~al.}(2021{\natexlab{a}})\citenamefont{Han, Liu,
  Wang, and Wang}}]{Han:2020uak}
\bibinfo{author}{\bibfnamefont{T.}~\bibnamefont{Han}},
  \bibinfo{author}{\bibfnamefont{Z.}~\bibnamefont{Liu}},
  \bibinfo{author}{\bibfnamefont{L.-T.} \bibnamefont{Wang}}, \bibnamefont{and}
  \bibinfo{author}{\bibfnamefont{X.}~\bibnamefont{Wang}},
  \bibinfo{journal}{Phys. Rev. D} \textbf{\bibinfo{volume}{103}},
  \bibinfo{pages}{075004} (\bibinfo{year}{2021}{\natexlab{a}}),
  \eprint{2009.11287}.

\bibitem[{\citenamefont{Capdevilla et~al.}(2021)\citenamefont{Capdevilla,
  Meloni, Simoniello, and Zurita}}]{Capdevilla:2021fmj}
\bibinfo{author}{\bibfnamefont{R.}~\bibnamefont{Capdevilla}},
  \bibinfo{author}{\bibfnamefont{F.}~\bibnamefont{Meloni}},
  \bibinfo{author}{\bibfnamefont{R.}~\bibnamefont{Simoniello}},
  \bibnamefont{and} \bibinfo{author}{\bibfnamefont{J.}~\bibnamefont{Zurita}},
  \bibinfo{journal}{JHEP} \textbf{\bibinfo{volume}{06}}, \bibinfo{pages}{133}
  (\bibinfo{year}{2021}), \eprint{2102.11292}.

\bibitem[{\citenamefont{Hisano et~al.}(2015{\natexlab{b}})\citenamefont{Hisano,
  Ishiwata, and Nagata}}]{Hisano:2015rsa}
\bibinfo{author}{\bibfnamefont{J.}~\bibnamefont{Hisano}},
  \bibinfo{author}{\bibfnamefont{K.}~\bibnamefont{Ishiwata}}, \bibnamefont{and}
  \bibinfo{author}{\bibfnamefont{N.}~\bibnamefont{Nagata}},
  \bibinfo{journal}{JHEP} \textbf{\bibinfo{volume}{06}}, \bibinfo{pages}{097}
  (\bibinfo{year}{2015}{\natexlab{b}}), \eprint{1504.00915}.

\bibitem[{\citenamefont{Hill and Solon}(2014)}]{Hill:2013hoa}
\bibinfo{author}{\bibfnamefont{R.~J.} \bibnamefont{Hill}} \bibnamefont{and}
  \bibinfo{author}{\bibfnamefont{M.~P.} \bibnamefont{Solon}},
  \bibinfo{journal}{Phys. Rev. Lett.} \textbf{\bibinfo{volume}{112}},
  \bibinfo{pages}{211602} (\bibinfo{year}{2014}), \eprint{1309.4092}.

\bibitem[{\citenamefont{He and Tandean}(2016)}]{He:2016mls}
\bibinfo{author}{\bibfnamefont{X.-G.} \bibnamefont{He}} \bibnamefont{and}
  \bibinfo{author}{\bibfnamefont{J.}~\bibnamefont{Tandean}},
  \bibinfo{journal}{JHEP} \textbf{\bibinfo{volume}{12}}, \bibinfo{pages}{074}
  (\bibinfo{year}{2016}), \eprint{1609.03551}.

\bibitem[{\citenamefont{Thornberry et~al.}(2021)\citenamefont{Thornberry,
  Throm, Frohaug, Killough, Blend, Erickson, Sun, Bays, and
  Allen}}]{Thornberry:2021ych}
\bibinfo{author}{\bibfnamefont{R.}~\bibnamefont{Thornberry}},
  \bibinfo{author}{\bibfnamefont{M.}~\bibnamefont{Throm}},
  \bibinfo{author}{\bibfnamefont{G.}~\bibnamefont{Frohaug}},
  \bibinfo{author}{\bibfnamefont{J.}~\bibnamefont{Killough}},
  \bibinfo{author}{\bibfnamefont{D.}~\bibnamefont{Blend}},
  \bibinfo{author}{\bibfnamefont{M.}~\bibnamefont{Erickson}},
  \bibinfo{author}{\bibfnamefont{B.}~\bibnamefont{Sun}},
  \bibinfo{author}{\bibfnamefont{B.}~\bibnamefont{Bays}}, \bibnamefont{and}
  \bibinfo{author}{\bibfnamefont{R.~E.} \bibnamefont{Allen}},
  \bibinfo{journal}{EPL} \textbf{\bibinfo{volume}{134}}, \bibinfo{pages}{49001}
  (\bibinfo{year}{2021}), \eprint{2104.11715}.

\bibitem[{\citenamefont{Akerib et~al.}(2015)}]{LZ:2015kxe}
\bibinfo{author}{\bibfnamefont{D.~S.} \bibnamefont{Akerib}}
  \bibnamefont{et~al.} (\bibinfo{collaboration}{LZ}) (\bibinfo{year}{2015}),
  \eprint{1509.02910}.

\bibitem[{\citenamefont{Martin}(1998)}]{Martin:1997ns}
\bibinfo{author}{\bibfnamefont{S.~P.} \bibnamefont{Martin}},
  \bibinfo{journal}{Adv. Ser. Direct. High Energy Phys.}
  \textbf{\bibinfo{volume}{18}}, \bibinfo{pages}{1} (\bibinfo{year}{1998}),
  \eprint{hep-ph/9709356}.

\bibitem[{\citenamefont{Arkani-Hamed et~al.}(2006)\citenamefont{Arkani-Hamed,
  Delgado, and Giudice}}]{ArkaniHamed:2006mb}
\bibinfo{author}{\bibfnamefont{N.}~\bibnamefont{Arkani-Hamed}},
  \bibinfo{author}{\bibfnamefont{A.}~\bibnamefont{Delgado}}, \bibnamefont{and}
  \bibinfo{author}{\bibfnamefont{G.~F.} \bibnamefont{Giudice}},
  \bibinfo{journal}{Nucl. Phys. B} \textbf{\bibinfo{volume}{741}},
  \bibinfo{pages}{108} (\bibinfo{year}{2006}), \eprint{hep-ph/0601041}.

\bibitem[{\citenamefont{Cheung et~al.}(2013)\citenamefont{Cheung, Hall, Pinner,
  and Ruderman}}]{Cheung:2012qy}
\bibinfo{author}{\bibfnamefont{C.}~\bibnamefont{Cheung}},
  \bibinfo{author}{\bibfnamefont{L.~J.} \bibnamefont{Hall}},
  \bibinfo{author}{\bibfnamefont{D.}~\bibnamefont{Pinner}}, \bibnamefont{and}
  \bibinfo{author}{\bibfnamefont{J.~T.} \bibnamefont{Ruderman}},
  \bibinfo{journal}{JHEP} \textbf{\bibinfo{volume}{05}}, \bibinfo{pages}{100}
  (\bibinfo{year}{2013}), \eprint{1211.4873}.

\bibitem[{\citenamefont{Huang and Wagner}(2014)}]{Huang:2014xua}
\bibinfo{author}{\bibfnamefont{P.}~\bibnamefont{Huang}} \bibnamefont{and}
  \bibinfo{author}{\bibfnamefont{C.~E.~M.} \bibnamefont{Wagner}},
  \bibinfo{journal}{Phys. Rev. D} \textbf{\bibinfo{volume}{90}},
  \bibinfo{pages}{015018} (\bibinfo{year}{2014}), \eprint{1404.0392}.

\bibitem[{\citenamefont{Huang et~al.}(2017)\citenamefont{Huang, Roglans,
  Spiegel, Sun, and Wagner}}]{Huang:2017kdh}
\bibinfo{author}{\bibfnamefont{P.}~\bibnamefont{Huang}},
  \bibinfo{author}{\bibfnamefont{R.~A.} \bibnamefont{Roglans}},
  \bibinfo{author}{\bibfnamefont{D.~D.} \bibnamefont{Spiegel}},
  \bibinfo{author}{\bibfnamefont{Y.}~\bibnamefont{Sun}}, \bibnamefont{and}
  \bibinfo{author}{\bibfnamefont{C.~E.~M.} \bibnamefont{Wagner}},
  \bibinfo{journal}{Phys. Rev. D} \textbf{\bibinfo{volume}{95}},
  \bibinfo{pages}{095021} (\bibinfo{year}{2017}), \eprint{1701.02737}.

\bibitem[{\citenamefont{Baum et~al.}(2018)\citenamefont{Baum, Carena, Shah, and
  Wagner}}]{Baum:2017enm}
\bibinfo{author}{\bibfnamefont{S.}~\bibnamefont{Baum}},
  \bibinfo{author}{\bibfnamefont{M.}~\bibnamefont{Carena}},
  \bibinfo{author}{\bibfnamefont{N.~R.} \bibnamefont{Shah}}, \bibnamefont{and}
  \bibinfo{author}{\bibfnamefont{C.~E.~M.} \bibnamefont{Wagner}},
  \bibinfo{journal}{JHEP} \textbf{\bibinfo{volume}{04}}, \bibinfo{pages}{069}
  (\bibinfo{year}{2018}), \eprint{1712.09873}.

\bibitem[{\citenamefont{Cabrera et~al.}(2020)\citenamefont{Cabrera, Casas,
  Delgado, and Robles}}]{Cabrera:2019gaq}
\bibinfo{author}{\bibfnamefont{M.~E.} \bibnamefont{Cabrera}},
  \bibinfo{author}{\bibfnamefont{J.~A.} \bibnamefont{Casas}},
  \bibinfo{author}{\bibfnamefont{A.}~\bibnamefont{Delgado}}, \bibnamefont{and}
  \bibinfo{author}{\bibfnamefont{S.}~\bibnamefont{Robles}},
  \bibinfo{journal}{JHEP} \textbf{\bibinfo{volume}{02}}, \bibinfo{pages}{166}
  (\bibinfo{year}{2020}), \eprint{1912.01758}.

\bibitem[{\citenamefont{Han et~al.}(2019)\citenamefont{Han, Liu, Mukhopadhyay,
  and Wang}}]{Han:2018gej}
\bibinfo{author}{\bibfnamefont{T.}~\bibnamefont{Han}},
  \bibinfo{author}{\bibfnamefont{H.}~\bibnamefont{Liu}},
  \bibinfo{author}{\bibfnamefont{S.}~\bibnamefont{Mukhopadhyay}},
  \bibnamefont{and} \bibinfo{author}{\bibfnamefont{X.}~\bibnamefont{Wang}},
  \bibinfo{journal}{JHEP} \textbf{\bibinfo{volume}{03}}, \bibinfo{pages}{080}
  (\bibinfo{year}{2019}), \eprint{1810.04679}.

\bibitem[{\citenamefont{Carena et~al.}(2018)\citenamefont{Carena, Osborne,
  Shah, and Wagner}}]{Carena:2018nlf}
\bibinfo{author}{\bibfnamefont{M.}~\bibnamefont{Carena}},
  \bibinfo{author}{\bibfnamefont{J.}~\bibnamefont{Osborne}},
  \bibinfo{author}{\bibfnamefont{N.~R.} \bibnamefont{Shah}}, \bibnamefont{and}
  \bibinfo{author}{\bibfnamefont{C.~E.~M.} \bibnamefont{Wagner}},
  \bibinfo{journal}{Phys. Rev. D} \textbf{\bibinfo{volume}{98}},
  \bibinfo{pages}{115010} (\bibinfo{year}{2018}), \eprint{1809.11082}.

\bibitem[{\citenamefont{Cao et~al.}(2020{\natexlab{a}})\citenamefont{Cao, Meng,
  Yue, Zhou, and Zhu}}]{Cao:2019qng}
\bibinfo{author}{\bibfnamefont{J.}~\bibnamefont{Cao}},
  \bibinfo{author}{\bibfnamefont{L.}~\bibnamefont{Meng}},
  \bibinfo{author}{\bibfnamefont{Y.}~\bibnamefont{Yue}},
  \bibinfo{author}{\bibfnamefont{H.}~\bibnamefont{Zhou}}, \bibnamefont{and}
  \bibinfo{author}{\bibfnamefont{P.}~\bibnamefont{Zhu}},
  \bibinfo{journal}{Phys. Rev. D} \textbf{\bibinfo{volume}{101}},
  \bibinfo{pages}{075003} (\bibinfo{year}{2020}{\natexlab{a}}),
  \eprint{1910.14317}.

\bibitem[{\citenamefont{Wang et~al.}(2021{\natexlab{a}})\citenamefont{Wang,
  Zhu, and Jie}}]{Wang:2020xta}
\bibinfo{author}{\bibfnamefont{K.}~\bibnamefont{Wang}},
  \bibinfo{author}{\bibfnamefont{J.}~\bibnamefont{Zhu}}, \bibnamefont{and}
  \bibinfo{author}{\bibfnamefont{Q.}~\bibnamefont{Jie}},
  \bibinfo{journal}{Chin. Phys. C} \textbf{\bibinfo{volume}{45}},
  \bibinfo{pages}{041003} (\bibinfo{year}{2021}{\natexlab{a}}),
  \eprint{2011.12848}.

\bibitem[{\citenamefont{Hisano et~al.}(2011)\citenamefont{Hisano, Ishiwata,
  Nagata, and Takesako}}]{Hisano:2011cs}
\bibinfo{author}{\bibfnamefont{J.}~\bibnamefont{Hisano}},
  \bibinfo{author}{\bibfnamefont{K.}~\bibnamefont{Ishiwata}},
  \bibinfo{author}{\bibfnamefont{N.}~\bibnamefont{Nagata}}, \bibnamefont{and}
  \bibinfo{author}{\bibfnamefont{T.}~\bibnamefont{Takesako}},
  \bibinfo{journal}{JHEP} \textbf{\bibinfo{volume}{07}}, \bibinfo{pages}{005}
  (\bibinfo{year}{2011}), \eprint{1104.0228}.

\bibitem[{\citenamefont{Cohen et~al.}(2013)\citenamefont{Cohen, Lisanti,
  Pierce, and Slatyer}}]{Cohen:2013ama}
\bibinfo{author}{\bibfnamefont{T.}~\bibnamefont{Cohen}},
  \bibinfo{author}{\bibfnamefont{M.}~\bibnamefont{Lisanti}},
  \bibinfo{author}{\bibfnamefont{A.}~\bibnamefont{Pierce}}, \bibnamefont{and}
  \bibinfo{author}{\bibfnamefont{T.~R.} \bibnamefont{Slatyer}},
  \bibinfo{journal}{JCAP} \textbf{\bibinfo{volume}{1310}}, \bibinfo{pages}{061}
  (\bibinfo{year}{2013}), \eprint{1307.4082}.

\bibitem[{\citenamefont{Fan and Reece}(2013)}]{Fan:2013faa}
\bibinfo{author}{\bibfnamefont{J.}~\bibnamefont{Fan}} \bibnamefont{and}
  \bibinfo{author}{\bibfnamefont{M.}~\bibnamefont{Reece}},
  \bibinfo{journal}{JHEP} \textbf{\bibinfo{volume}{10}}, \bibinfo{pages}{124}
  (\bibinfo{year}{2013}), \eprint{1307.4400}.

\bibitem[{\citenamefont{Ackermann et~al.}(2014)}]{Ackermann:2013yva}
\bibinfo{author}{\bibfnamefont{M.}~\bibnamefont{Ackermann}}
  \bibnamefont{et~al.} (\bibinfo{collaboration}{Fermi-LAT}),
  \bibinfo{journal}{Phys. Rev. D} \textbf{\bibinfo{volume}{89}},
  \bibinfo{pages}{042001} (\bibinfo{year}{2014}), \eprint{1310.0828}.

\bibitem[{\citenamefont{Bhattacherjee et~al.}(2014)\citenamefont{Bhattacherjee,
  Ibe, Ichikawa, Matsumoto, and Nishiyama}}]{Bhattacherjee:2014dya}
\bibinfo{author}{\bibfnamefont{B.}~\bibnamefont{Bhattacherjee}},
  \bibinfo{author}{\bibfnamefont{M.}~\bibnamefont{Ibe}},
  \bibinfo{author}{\bibfnamefont{K.}~\bibnamefont{Ichikawa}},
  \bibinfo{author}{\bibfnamefont{S.}~\bibnamefont{Matsumoto}},
  \bibnamefont{and}
  \bibinfo{author}{\bibfnamefont{K.}~\bibnamefont{Nishiyama}},
  \bibinfo{journal}{JHEP} \textbf{\bibinfo{volume}{07}}, \bibinfo{pages}{080}
  (\bibinfo{year}{2014}), \eprint{1405.4914}.

\bibitem[{\citenamefont{Baer et~al.}(2011)\citenamefont{Baer, Lessa,
  Rajagopalan, and Sreethawong}}]{Baer:2011hx}
\bibinfo{author}{\bibfnamefont{H.}~\bibnamefont{Baer}},
  \bibinfo{author}{\bibfnamefont{A.}~\bibnamefont{Lessa}},
  \bibinfo{author}{\bibfnamefont{S.}~\bibnamefont{Rajagopalan}},
  \bibnamefont{and}
  \bibinfo{author}{\bibfnamefont{W.}~\bibnamefont{Sreethawong}},
  \bibinfo{journal}{JCAP} \textbf{\bibinfo{volume}{1106}}, \bibinfo{pages}{031}
  (\bibinfo{year}{2011}), \eprint{1103.5413}.

\bibitem[{\citenamefont{Zurek}(2009)}]{Zurek:2008qg}
\bibinfo{author}{\bibfnamefont{K.~M.} \bibnamefont{Zurek}},
  \bibinfo{journal}{Phys. Rev. D} \textbf{\bibinfo{volume}{79}},
  \bibinfo{pages}{115002} (\bibinfo{year}{2009}), \eprint{0811.4429}.

\bibitem[{\citenamefont{Profumo et~al.}(2009)\citenamefont{Profumo, Sigurdson,
  and Ubaldi}}]{Profumo:2009tb}
\bibinfo{author}{\bibfnamefont{S.}~\bibnamefont{Profumo}},
  \bibinfo{author}{\bibfnamefont{K.}~\bibnamefont{Sigurdson}},
  \bibnamefont{and} \bibinfo{author}{\bibfnamefont{L.}~\bibnamefont{Ubaldi}},
  \bibinfo{journal}{JCAP} \textbf{\bibinfo{volume}{12}}, \bibinfo{pages}{016}
  (\bibinfo{year}{2009}), \eprint{0907.4374}.

\bibitem[{\citenamefont{Kajiyama et~al.}(2013)\citenamefont{Kajiyama, Okada,
  and Toma}}]{Kajiyama:2013rla}
\bibinfo{author}{\bibfnamefont{Y.}~\bibnamefont{Kajiyama}},
  \bibinfo{author}{\bibfnamefont{H.}~\bibnamefont{Okada}}, \bibnamefont{and}
  \bibinfo{author}{\bibfnamefont{T.}~\bibnamefont{Toma}},
  \bibinfo{journal}{Phys. Rev. D} \textbf{\bibinfo{volume}{88}},
  \bibinfo{pages}{015029} (\bibinfo{year}{2013}), \eprint{1303.7356}.

\bibitem[{\citenamefont{Herrero-Garcia
  et~al.}(2017)\citenamefont{Herrero-Garcia, Scaffidi, White, and
  Williams}}]{Herrero-Garcia:2017vrl}
\bibinfo{author}{\bibfnamefont{J.}~\bibnamefont{Herrero-Garcia}},
  \bibinfo{author}{\bibfnamefont{A.}~\bibnamefont{Scaffidi}},
  \bibinfo{author}{\bibfnamefont{M.}~\bibnamefont{White}}, \bibnamefont{and}
  \bibinfo{author}{\bibfnamefont{A.~G.} \bibnamefont{Williams}},
  \bibinfo{journal}{JCAP} \textbf{\bibinfo{volume}{11}}, \bibinfo{pages}{021}
  (\bibinfo{year}{2017}), \eprint{1709.01945}.

\bibitem[{\citenamefont{Herrero-Garcia
  et~al.}(2019)\citenamefont{Herrero-Garcia, Scaffidi, White, and
  Williams}}]{Herrero-Garcia:2018qnz}
\bibinfo{author}{\bibfnamefont{J.}~\bibnamefont{Herrero-Garcia}},
  \bibinfo{author}{\bibfnamefont{A.}~\bibnamefont{Scaffidi}},
  \bibinfo{author}{\bibfnamefont{M.}~\bibnamefont{White}}, \bibnamefont{and}
  \bibinfo{author}{\bibfnamefont{A.~G.} \bibnamefont{Williams}},
  \bibinfo{journal}{JCAP} \textbf{\bibinfo{volume}{01}}, \bibinfo{pages}{008}
  (\bibinfo{year}{2019}), \eprint{1809.06881}.

\bibitem[{\citenamefont{Scaffidi}(2020)}]{Scaffidi:2020wpa}
\bibinfo{author}{\bibfnamefont{A.}~\bibnamefont{Scaffidi}},
  \bibinfo{journal}{EPJ Web Conf.} \textbf{\bibinfo{volume}{245}},
  \bibinfo{pages}{06036} (\bibinfo{year}{2020}).

\bibitem[{\citenamefont{Bernabei et~al.}(2008)}]{Bernabei:2008yh}
\bibinfo{author}{\bibfnamefont{R.}~\bibnamefont{Bernabei}} \bibnamefont{et~al.}
  (\bibinfo{collaboration}{DAMA}), \bibinfo{journal}{Nucl. Instrum. Meth. A}
  \textbf{\bibinfo{volume}{592}}, \bibinfo{pages}{297} (\bibinfo{year}{2008}),
  \eprint{0804.2738}.

\bibitem[{\citenamefont{Bernabei et~al.}(2010)}]{Bernabei:2010mq}
\bibinfo{author}{\bibfnamefont{R.}~\bibnamefont{Bernabei}} \bibnamefont{et~al.}
  (\bibinfo{collaboration}{DAMA, LIBRA}), \bibinfo{journal}{Eur. Phys. J. C}
  \textbf{\bibinfo{volume}{67}}, \bibinfo{pages}{39} (\bibinfo{year}{2010}),
  \eprint{1002.1028}.

\bibitem[{\citenamefont{Bernabei et~al.}(2018)}]{Bernabei:2018yyw}
\bibinfo{author}{\bibfnamefont{R.}~\bibnamefont{Bernabei}}
  \bibnamefont{et~al.}, \bibinfo{journal}{Nucl. Phys. Atom. Energy}
  \textbf{\bibinfo{volume}{19}}, \bibinfo{pages}{307} (\bibinfo{year}{2018}),
  \eprint{1805.10486}.

\bibitem[{\citenamefont{Ullio et~al.}(2001)\citenamefont{Ullio, Kamionkowski,
  and Vogel}}]{Ullio:2000bv}
\bibinfo{author}{\bibfnamefont{P.}~\bibnamefont{Ullio}},
  \bibinfo{author}{\bibfnamefont{M.}~\bibnamefont{Kamionkowski}},
  \bibnamefont{and} \bibinfo{author}{\bibfnamefont{P.}~\bibnamefont{Vogel}},
  \bibinfo{journal}{JHEP} \textbf{\bibinfo{volume}{07}}, \bibinfo{pages}{044}
  (\bibinfo{year}{2001}), \eprint{hep-ph/0010036}.

\bibitem[{\citenamefont{Tucker-Smith and Weiner}(2005)}]{TuckerSmith:2004jv}
\bibinfo{author}{\bibfnamefont{D.}~\bibnamefont{Tucker-Smith}}
  \bibnamefont{and} \bibinfo{author}{\bibfnamefont{N.}~\bibnamefont{Weiner}},
  \bibinfo{journal}{Phys. Rev. D} \textbf{\bibinfo{volume}{72}},
  \bibinfo{pages}{063509} (\bibinfo{year}{2005}), \eprint{hep-ph/0402065}.

\bibitem[{\citenamefont{Chang et~al.}(2009)\citenamefont{Chang, Kribs,
  Tucker-Smith, and Weiner}}]{Chang:2008gd}
\bibinfo{author}{\bibfnamefont{S.}~\bibnamefont{Chang}},
  \bibinfo{author}{\bibfnamefont{G.~D.} \bibnamefont{Kribs}},
  \bibinfo{author}{\bibfnamefont{D.}~\bibnamefont{Tucker-Smith}},
  \bibnamefont{and} \bibinfo{author}{\bibfnamefont{N.}~\bibnamefont{Weiner}},
  \bibinfo{journal}{Phys. Rev. D} \textbf{\bibinfo{volume}{79}},
  \bibinfo{pages}{043513} (\bibinfo{year}{2009}), \eprint{0807.2250}.

\bibitem[{\citenamefont{Cui et~al.}(2009)\citenamefont{Cui, Morrissey, Poland,
  and Randall}}]{Cui:2009xq}
\bibinfo{author}{\bibfnamefont{Y.}~\bibnamefont{Cui}},
  \bibinfo{author}{\bibfnamefont{D.~E.} \bibnamefont{Morrissey}},
  \bibinfo{author}{\bibfnamefont{D.}~\bibnamefont{Poland}}, \bibnamefont{and}
  \bibinfo{author}{\bibfnamefont{L.}~\bibnamefont{Randall}},
  \bibinfo{journal}{JHEP} \textbf{\bibinfo{volume}{05}}, \bibinfo{pages}{076}
  (\bibinfo{year}{2009}), \eprint{0901.0557}.

\bibitem[{\citenamefont{Aprile et~al.}(2011{\natexlab{a}})}]{Aprile:2011ts}
\bibinfo{author}{\bibfnamefont{E.}~\bibnamefont{Aprile}} \bibnamefont{et~al.}
  (\bibinfo{collaboration}{XENON100}), \bibinfo{journal}{Phys. Rev. D}
  \textbf{\bibinfo{volume}{84}}, \bibinfo{pages}{061101}
  (\bibinfo{year}{2011}{\natexlab{a}}), \eprint{1104.3121}.

\bibitem[{\citenamefont{Chang et~al.}(2011)\citenamefont{Chang, Lang, and
  Weiner}}]{Chang:2010pr}
\bibinfo{author}{\bibfnamefont{S.}~\bibnamefont{Chang}},
  \bibinfo{author}{\bibfnamefont{R.~F.} \bibnamefont{Lang}}, \bibnamefont{and}
  \bibinfo{author}{\bibfnamefont{N.}~\bibnamefont{Weiner}},
  \bibinfo{journal}{Phys. Rev. Lett.} \textbf{\bibinfo{volume}{106}},
  \bibinfo{pages}{011301} (\bibinfo{year}{2011}), \eprint{1007.2688}.

\bibitem[{\citenamefont{March-Russell et~al.}(2009)\citenamefont{March-Russell,
  McCabe, and McCullough}}]{March-Russell:2008rkh}
\bibinfo{author}{\bibfnamefont{J.}~\bibnamefont{March-Russell}},
  \bibinfo{author}{\bibfnamefont{C.}~\bibnamefont{McCabe}}, \bibnamefont{and}
  \bibinfo{author}{\bibfnamefont{M.}~\bibnamefont{McCullough}},
  \bibinfo{journal}{JHEP} \textbf{\bibinfo{volume}{05}}, \bibinfo{pages}{071}
  (\bibinfo{year}{2009}), \eprint{0812.1931}.

\bibitem[{\citenamefont{Smirnov and Beacom}(2020)}]{Smirnov:2020zwf}
\bibinfo{author}{\bibfnamefont{J.}~\bibnamefont{Smirnov}} \bibnamefont{and}
  \bibinfo{author}{\bibfnamefont{J.~F.} \bibnamefont{Beacom}},
  \bibinfo{journal}{Phys. Rev. Lett.} \textbf{\bibinfo{volume}{125}},
  \bibinfo{pages}{131301} (\bibinfo{year}{2020}), \eprint{2002.04038}.

\bibitem[{\citenamefont{Barello et~al.}(2014)\citenamefont{Barello, Chang, and
  Newby}}]{Barello:2014uda}
\bibinfo{author}{\bibfnamefont{G.}~\bibnamefont{Barello}},
  \bibinfo{author}{\bibfnamefont{S.}~\bibnamefont{Chang}}, \bibnamefont{and}
  \bibinfo{author}{\bibfnamefont{C.~A.} \bibnamefont{Newby}},
  \bibinfo{journal}{Phys. Rev. D} \textbf{\bibinfo{volume}{90}},
  \bibinfo{pages}{094027} (\bibinfo{year}{2014}), \eprint{1409.0536}.

\bibitem[{\citenamefont{Chen et~al.}(2017{\natexlab{b}})}]{Chen:2017prd}
\bibinfo{author}{\bibfnamefont{X.}~\bibnamefont{Chen}} \bibnamefont{et~al.}
  (\bibinfo{collaboration}{PandaX-II}), \bibinfo{journal}{Phys. Rev. D}
  \textbf{\bibinfo{volume}{96}}, \bibinfo{pages}{102007}
  (\bibinfo{year}{2017}{\natexlab{b}}), \eprint{1708.05825}.

\bibitem[{\citenamefont{Akerib et~al.}(2021{\natexlab{a}})}]{LUX:2021ksq}
\bibinfo{author}{\bibfnamefont{D.~S.} \bibnamefont{Akerib}}
  \bibnamefont{et~al.} (\bibinfo{collaboration}{LUX}), \bibinfo{journal}{Phys.
  Rev. D} \textbf{\bibinfo{volume}{104}}, \bibinfo{pages}{062005}
  (\bibinfo{year}{2021}{\natexlab{a}}), \eprint{2102.06998}.

\bibitem[{\citenamefont{Del~Nobile et~al.}(2015)\citenamefont{Del~Nobile,
  Kaplinghat, and Yu}}]{DelNobile:2015uua}
\bibinfo{author}{\bibfnamefont{E.}~\bibnamefont{Del~Nobile}},
  \bibinfo{author}{\bibfnamefont{M.}~\bibnamefont{Kaplinghat}},
  \bibnamefont{and} \bibinfo{author}{\bibfnamefont{H.-B.} \bibnamefont{Yu}},
  \bibinfo{journal}{JCAP} \textbf{\bibinfo{volume}{1510}}, \bibinfo{pages}{055}
  (\bibinfo{year}{2015}), \eprint{1507.04007}.

\bibitem[{\citenamefont{Spergel and Steinhardt}(2000)}]{Spergel:1999mh}
\bibinfo{author}{\bibfnamefont{D.~N.} \bibnamefont{Spergel}} \bibnamefont{and}
  \bibinfo{author}{\bibfnamefont{P.~J.} \bibnamefont{Steinhardt}},
  \bibinfo{journal}{Phys. Rev. Lett.} \textbf{\bibinfo{volume}{84}},
  \bibinfo{pages}{3760} (\bibinfo{year}{2000}), \eprint{astro-ph/9909386}.

\bibitem[{\citenamefont{Kaplinghat et~al.}(2016)\citenamefont{Kaplinghat,
  Tulin, and Yu}}]{Kaplinghat:2015aga}
\bibinfo{author}{\bibfnamefont{M.}~\bibnamefont{Kaplinghat}},
  \bibinfo{author}{\bibfnamefont{S.}~\bibnamefont{Tulin}}, \bibnamefont{and}
  \bibinfo{author}{\bibfnamefont{H.-B.} \bibnamefont{Yu}},
  \bibinfo{journal}{Phys. Rev. Lett.} \textbf{\bibinfo{volume}{116}},
  \bibinfo{pages}{041302} (\bibinfo{year}{2016}), \eprint{1508.03339}.

\bibitem[{\citenamefont{Tulin and Yu}(2018)}]{Tulin:2017ara}
\bibinfo{author}{\bibfnamefont{S.}~\bibnamefont{Tulin}} \bibnamefont{and}
  \bibinfo{author}{\bibfnamefont{H.-B.} \bibnamefont{Yu}},
  \bibinfo{journal}{Phys. Rept.} \textbf{\bibinfo{volume}{730}},
  \bibinfo{pages}{1} (\bibinfo{year}{2018}), \eprint{1705.02358}.

\bibitem[{\citenamefont{Kaplinghat
  et~al.}(2014{\natexlab{a}})\citenamefont{Kaplinghat, Keeley, Linden, and
  Yu}}]{Kaplinghat:2013xca}
\bibinfo{author}{\bibfnamefont{M.}~\bibnamefont{Kaplinghat}},
  \bibinfo{author}{\bibfnamefont{R.~E.} \bibnamefont{Keeley}},
  \bibinfo{author}{\bibfnamefont{T.}~\bibnamefont{Linden}}, \bibnamefont{and}
  \bibinfo{author}{\bibfnamefont{H.-B.} \bibnamefont{Yu}},
  \bibinfo{journal}{Phys. Rev. Lett.} \textbf{\bibinfo{volume}{113}},
  \bibinfo{pages}{021302} (\bibinfo{year}{2014}{\natexlab{a}}),
  \eprint{1311.6524}.

\bibitem[{\citenamefont{Kamada et~al.}(2017)\citenamefont{Kamada, Kaplinghat,
  Pace, and Yu}}]{Kamada:2016euw}
\bibinfo{author}{\bibfnamefont{A.}~\bibnamefont{Kamada}},
  \bibinfo{author}{\bibfnamefont{M.}~\bibnamefont{Kaplinghat}},
  \bibinfo{author}{\bibfnamefont{A.~B.} \bibnamefont{Pace}}, \bibnamefont{and}
  \bibinfo{author}{\bibfnamefont{H.-B.} \bibnamefont{Yu}},
  \bibinfo{journal}{Phys. Rev. Lett.} \textbf{\bibinfo{volume}{119}},
  \bibinfo{pages}{111102} (\bibinfo{year}{2017}), \eprint{1611.02716}.

\bibitem[{\citenamefont{Creasey et~al.}(2017)\citenamefont{Creasey, Sameie,
  Sales, Yu, Vogelsberger, and Zavala}}]{Creasey:2016jaq}
\bibinfo{author}{\bibfnamefont{P.}~\bibnamefont{Creasey}},
  \bibinfo{author}{\bibfnamefont{O.}~\bibnamefont{Sameie}},
  \bibinfo{author}{\bibfnamefont{L.~V.} \bibnamefont{Sales}},
  \bibinfo{author}{\bibfnamefont{H.-B.} \bibnamefont{Yu}},
  \bibinfo{author}{\bibfnamefont{M.}~\bibnamefont{Vogelsberger}},
  \bibnamefont{and} \bibinfo{author}{\bibfnamefont{J.}~\bibnamefont{Zavala}},
  \bibinfo{journal}{Mon. Not. Roy. Astron. Soc.}
  \textbf{\bibinfo{volume}{468}}, \bibinfo{pages}{2283} (\bibinfo{year}{2017}),
  \eprint{1612.03903}.

\bibitem[{\citenamefont{Ren et~al.}(2019)\citenamefont{Ren, Kwa, Kaplinghat,
  and Yu}}]{Ren:2018jpt}
\bibinfo{author}{\bibfnamefont{T.}~\bibnamefont{Ren}},
  \bibinfo{author}{\bibfnamefont{A.}~\bibnamefont{Kwa}},
  \bibinfo{author}{\bibfnamefont{M.}~\bibnamefont{Kaplinghat}},
  \bibnamefont{and} \bibinfo{author}{\bibfnamefont{H.-B.} \bibnamefont{Yu}},
  \bibinfo{journal}{Phys. Rev. X} \textbf{\bibinfo{volume}{9}},
  \bibinfo{pages}{031020} (\bibinfo{year}{2019}), \eprint{1808.05695}.

\bibitem[{\citenamefont{Feng et~al.}(2010)\citenamefont{Feng, Kaplinghat, and
  Yu}}]{Feng:2009hw}
\bibinfo{author}{\bibfnamefont{J.~L.} \bibnamefont{Feng}},
  \bibinfo{author}{\bibfnamefont{M.}~\bibnamefont{Kaplinghat}},
  \bibnamefont{and} \bibinfo{author}{\bibfnamefont{H.-B.} \bibnamefont{Yu}},
  \bibinfo{journal}{Phys. Rev. Lett.} \textbf{\bibinfo{volume}{104}},
  \bibinfo{pages}{151301} (\bibinfo{year}{2010}), \eprint{0911.0422}.

\bibitem[{\citenamefont{Buckley and Fox}(2010)}]{Buckley:2009in}
\bibinfo{author}{\bibfnamefont{M.~R.} \bibnamefont{Buckley}} \bibnamefont{and}
  \bibinfo{author}{\bibfnamefont{P.~J.} \bibnamefont{Fox}},
  \bibinfo{journal}{Phys. Rev. D} \textbf{\bibinfo{volume}{81}},
  \bibinfo{pages}{083522} (\bibinfo{year}{2010}), \eprint{0911.3898}.

\bibitem[{\citenamefont{Loeb and Weiner}(2011)}]{Loeb:2010gj}
\bibinfo{author}{\bibfnamefont{A.}~\bibnamefont{Loeb}} \bibnamefont{and}
  \bibinfo{author}{\bibfnamefont{N.}~\bibnamefont{Weiner}},
  \bibinfo{journal}{Phys. Rev. Lett.} \textbf{\bibinfo{volume}{106}},
  \bibinfo{pages}{171302} (\bibinfo{year}{2011}), \eprint{1011.6374}.

\bibitem[{\citenamefont{March-Russell and
  McCullough}(2012)}]{March-Russell:2011ang}
\bibinfo{author}{\bibfnamefont{J.}~\bibnamefont{March-Russell}}
  \bibnamefont{and}
  \bibinfo{author}{\bibfnamefont{M.}~\bibnamefont{McCullough}},
  \bibinfo{journal}{JCAP} \textbf{\bibinfo{volume}{03}}, \bibinfo{pages}{019}
  (\bibinfo{year}{2012}), \eprint{1106.4319}.

\bibitem[{\citenamefont{van~den Aarssen et~al.}(2012)\citenamefont{van~den
  Aarssen, Bringmann, and Pfrommer}}]{Aarssen:2012fx}
\bibinfo{author}{\bibfnamefont{L.~G.} \bibnamefont{van~den Aarssen}},
  \bibinfo{author}{\bibfnamefont{T.}~\bibnamefont{Bringmann}},
  \bibnamefont{and} \bibinfo{author}{\bibfnamefont{C.}~\bibnamefont{Pfrommer}},
  \bibinfo{journal}{Phys. Rev. Lett.} \textbf{\bibinfo{volume}{109}},
  \bibinfo{pages}{231301} (\bibinfo{year}{2012}), \eprint{1205.5809}.

\bibitem[{\citenamefont{Tulin et~al.}(2013)\citenamefont{Tulin, Yu, and
  Zurek}}]{Tulin:2013teo}
\bibinfo{author}{\bibfnamefont{S.}~\bibnamefont{Tulin}},
  \bibinfo{author}{\bibfnamefont{H.-B.} \bibnamefont{Yu}}, \bibnamefont{and}
  \bibinfo{author}{\bibfnamefont{K.~M.} \bibnamefont{Zurek}},
  \bibinfo{journal}{Phys. Rev. D} \textbf{\bibinfo{volume}{87}},
  \bibinfo{pages}{115007} (\bibinfo{year}{2013}), \eprint{1302.3898}.

\bibitem[{\citenamefont{Garcia~Garcia
  et~al.}(2015{\natexlab{a}})\citenamefont{Garcia~Garcia, Lasenby, and
  March-Russell}}]{GarciaGarcia:2015pnn}
\bibinfo{author}{\bibfnamefont{I.}~\bibnamefont{Garcia~Garcia}},
  \bibinfo{author}{\bibfnamefont{R.}~\bibnamefont{Lasenby}}, \bibnamefont{and}
  \bibinfo{author}{\bibfnamefont{J.}~\bibnamefont{March-Russell}},
  \bibinfo{journal}{Phys. Rev. Lett.} \textbf{\bibinfo{volume}{115}},
  \bibinfo{pages}{121801} (\bibinfo{year}{2015}{\natexlab{a}}),
  \eprint{1505.07410}.

\bibitem[{\citenamefont{Kahlhoefer
  et~al.}(2017{\natexlab{a}})\citenamefont{Kahlhoefer, Schmidt-Hoberg, and
  Wild}}]{Kahlhoefer:2017umn}
\bibinfo{author}{\bibfnamefont{F.}~\bibnamefont{Kahlhoefer}},
  \bibinfo{author}{\bibfnamefont{K.}~\bibnamefont{Schmidt-Hoberg}},
  \bibnamefont{and} \bibinfo{author}{\bibfnamefont{S.}~\bibnamefont{Wild}},
  \bibinfo{journal}{JCAP} \textbf{\bibinfo{volume}{1708}}, \bibinfo{pages}{003}
  (\bibinfo{year}{2017}{\natexlab{a}}), \eprint{1704.02149}.

\bibitem[{\citenamefont{Chu et~al.}(2020)\citenamefont{Chu, Garcia-Cely, and
  Murayama}}]{Chu:2018faw}
\bibinfo{author}{\bibfnamefont{X.}~\bibnamefont{Chu}},
  \bibinfo{author}{\bibfnamefont{C.}~\bibnamefont{Garcia-Cely}},
  \bibnamefont{and} \bibinfo{author}{\bibfnamefont{H.}~\bibnamefont{Murayama}},
  \bibinfo{journal}{Phys. Rev. Lett.} \textbf{\bibinfo{volume}{124}},
  \bibinfo{pages}{041101} (\bibinfo{year}{2020}), \eprint{1901.00075}.

\bibitem[{\citenamefont{Kaplinghat
  et~al.}(2014{\natexlab{b}})\citenamefont{Kaplinghat, Tulin, and
  Yu}}]{Kaplinghat:2013yxa}
\bibinfo{author}{\bibfnamefont{M.}~\bibnamefont{Kaplinghat}},
  \bibinfo{author}{\bibfnamefont{S.}~\bibnamefont{Tulin}}, \bibnamefont{and}
  \bibinfo{author}{\bibfnamefont{H.-B.} \bibnamefont{Yu}},
  \bibinfo{journal}{Phys. Rev. D} \textbf{\bibinfo{volume}{89}},
  \bibinfo{pages}{035009} (\bibinfo{year}{2014}{\natexlab{b}}),
  \eprint{1310.7945}.

\bibitem[{\citenamefont{Kahlhoefer
  et~al.}(2017{\natexlab{b}})\citenamefont{Kahlhoefer, Kulkarni, and
  Wild}}]{Kahlhoefer:2017ddj}
\bibinfo{author}{\bibfnamefont{F.}~\bibnamefont{Kahlhoefer}},
  \bibinfo{author}{\bibfnamefont{S.}~\bibnamefont{Kulkarni}}, \bibnamefont{and}
  \bibinfo{author}{\bibfnamefont{S.}~\bibnamefont{Wild}},
  \bibinfo{journal}{JCAP} \textbf{\bibinfo{volume}{1711}}, \bibinfo{pages}{016}
  (\bibinfo{year}{2017}{\natexlab{b}}), \eprint{1707.08571}.

\bibitem[{\citenamefont{Ren et~al.}(2018)}]{Ren:2018gyx}
\bibinfo{author}{\bibfnamefont{X.}~\bibnamefont{Ren}} \bibnamefont{et~al.}
  (\bibinfo{collaboration}{PandaX-II}), \bibinfo{journal}{Phys. Rev. Lett.}
  \textbf{\bibinfo{volume}{121}}, \bibinfo{pages}{021304}
  (\bibinfo{year}{2018}), \eprint{1802.06912}.

\bibitem[{\citenamefont{Yang et~al.}(2021)}]{PandaX-II:2021lap}
\bibinfo{author}{\bibfnamefont{J.}~\bibnamefont{Yang}} \bibnamefont{et~al.}
  (\bibinfo{collaboration}{PandaX-II}), \bibinfo{journal}{Sci. China Phys.
  Mech. Astron.} \textbf{\bibinfo{volume}{64}}, \bibinfo{pages}{111062}
  (\bibinfo{year}{2021}), \eprint{2104.14724}.

\bibitem[{\citenamefont{Tsai et~al.}(2020)\citenamefont{Tsai, McGehee, and
  Murayama}}]{Tsai:2020vpi}
\bibinfo{author}{\bibfnamefont{Y.-D.} \bibnamefont{Tsai}},
  \bibinfo{author}{\bibfnamefont{R.}~\bibnamefont{McGehee}}, \bibnamefont{and}
  \bibinfo{author}{\bibfnamefont{H.}~\bibnamefont{Murayama}}
  (\bibinfo{year}{2020}), \eprint{2008.08608}.

\bibitem[{\citenamefont{Adriani et~al.}(2013)}]{Adriani:2013uda}
\bibinfo{author}{\bibfnamefont{O.}~\bibnamefont{Adriani}} \bibnamefont{et~al.}
  (\bibinfo{collaboration}{PAMELA}), \bibinfo{journal}{Phys. Rev. Lett.}
  \textbf{\bibinfo{volume}{111}}, \bibinfo{pages}{081102}
  (\bibinfo{year}{2013}), \eprint{1308.0133}.

\bibitem[{\citenamefont{Aguilar et~al.}(2019)}]{Aguilar:2019owu}
\bibinfo{author}{\bibfnamefont{M.}~\bibnamefont{Aguilar}} \bibnamefont{et~al.}
  (\bibinfo{collaboration}{AMS}), \bibinfo{journal}{Phys. Rev. Lett.}
  \textbf{\bibinfo{volume}{122}}, \bibinfo{pages}{041102}
  (\bibinfo{year}{2019}).

\bibitem[{\citenamefont{Knodlseder et~al.}(2005)}]{Knodlseder:2005yq}
\bibinfo{author}{\bibfnamefont{J.}~\bibnamefont{Knodlseder}}
  \bibnamefont{et~al.}, \bibinfo{journal}{Astron. Astrophys.}
  \textbf{\bibinfo{volume}{441}}, \bibinfo{pages}{513} (\bibinfo{year}{2005}),
  \eprint{astro-ph/0506026}.

\bibitem[{\citenamefont{Ambrosi et~al.}(2017)}]{Ambrosi:2017wek}
\bibinfo{author}{\bibfnamefont{G.}~\bibnamefont{Ambrosi}} \bibnamefont{et~al.}
  (\bibinfo{collaboration}{DAMPE}), \bibinfo{journal}{Nature}
  \textbf{\bibinfo{volume}{552}}, \bibinfo{pages}{63} (\bibinfo{year}{2017}),
  \eprint{1711.10981}.

\bibitem[{\citenamefont{Fox and Poppitz}(2009)}]{Fox:2008kb}
\bibinfo{author}{\bibfnamefont{P.~J.} \bibnamefont{Fox}} \bibnamefont{and}
  \bibinfo{author}{\bibfnamefont{E.}~\bibnamefont{Poppitz}},
  \bibinfo{journal}{Phys. Rev. D} \textbf{\bibinfo{volume}{79}},
  \bibinfo{pages}{083528} (\bibinfo{year}{2009}), \eprint{0811.0399}.

\bibitem[{\citenamefont{Bell et~al.}(2014)\citenamefont{Bell, Cai, Leane, and
  Medina}}]{Bell:2014tta}
\bibinfo{author}{\bibfnamefont{N.~F.} \bibnamefont{Bell}},
  \bibinfo{author}{\bibfnamefont{Y.}~\bibnamefont{Cai}},
  \bibinfo{author}{\bibfnamefont{R.~K.} \bibnamefont{Leane}}, \bibnamefont{and}
  \bibinfo{author}{\bibfnamefont{A.~D.} \bibnamefont{Medina}},
  \bibinfo{journal}{Phys. Rev. D} \textbf{\bibinfo{volume}{90}},
  \bibinfo{pages}{035027} (\bibinfo{year}{2014}), \eprint{1407.3001}.

\bibitem[{\citenamefont{Chun et~al.}(2010)\citenamefont{Chun, Park, and
  Scopel}}]{Chun:2009zx}
\bibinfo{author}{\bibfnamefont{E.~J.} \bibnamefont{Chun}},
  \bibinfo{author}{\bibfnamefont{J.-C.} \bibnamefont{Park}}, \bibnamefont{and}
  \bibinfo{author}{\bibfnamefont{S.}~\bibnamefont{Scopel}},
  \bibinfo{journal}{JCAP} \textbf{\bibinfo{volume}{1002}}, \bibinfo{pages}{015}
  (\bibinfo{year}{2010}), \eprint{arXiv:0911.5273}.

\bibitem[{\citenamefont{Bringmann et~al.}(2012)\citenamefont{Bringmann, Huang,
  Ibarra, Vogl, and Weniger}}]{Bringmann:2012vr}
\bibinfo{author}{\bibfnamefont{T.}~\bibnamefont{Bringmann}},
  \bibinfo{author}{\bibfnamefont{X.}~\bibnamefont{Huang}},
  \bibinfo{author}{\bibfnamefont{A.}~\bibnamefont{Ibarra}},
  \bibinfo{author}{\bibfnamefont{S.}~\bibnamefont{Vogl}}, \bibnamefont{and}
  \bibinfo{author}{\bibfnamefont{C.}~\bibnamefont{Weniger}},
  \bibinfo{journal}{JCAP} \textbf{\bibinfo{volume}{07}}, \bibinfo{pages}{054}
  (\bibinfo{year}{2012}), \eprint{1203.1312}.

\bibitem[{\citenamefont{Agrawal et~al.}(2014)\citenamefont{Agrawal, Chacko, and
  Verhaaren}}]{Agrawal:2014ufa}
\bibinfo{author}{\bibfnamefont{P.}~\bibnamefont{Agrawal}},
  \bibinfo{author}{\bibfnamefont{Z.}~\bibnamefont{Chacko}}, \bibnamefont{and}
  \bibinfo{author}{\bibfnamefont{C.~B.} \bibnamefont{Verhaaren}},
  \bibinfo{journal}{JHEP} \textbf{\bibinfo{volume}{08}}, \bibinfo{pages}{147}
  (\bibinfo{year}{2014}), \eprint{1402.7369}.

\bibitem[{\citenamefont{Kopp et~al.}(2014)\citenamefont{Kopp, Michaels, and
  Smirnov}}]{Kopp:2014tsa}
\bibinfo{author}{\bibfnamefont{J.}~\bibnamefont{Kopp}},
  \bibinfo{author}{\bibfnamefont{L.}~\bibnamefont{Michaels}}, \bibnamefont{and}
  \bibinfo{author}{\bibfnamefont{J.}~\bibnamefont{Smirnov}},
  \bibinfo{journal}{JCAP} \textbf{\bibinfo{volume}{1404}}, \bibinfo{pages}{022}
  (\bibinfo{year}{2014}), \eprint{1401.6457}.

\bibitem[{\citenamefont{Fukushima et~al.}(2014)\citenamefont{Fukushima, Kelso,
  Kumar, Sandick, and Yamamoto}}]{Fukushima:2014yia}
\bibinfo{author}{\bibfnamefont{K.}~\bibnamefont{Fukushima}},
  \bibinfo{author}{\bibfnamefont{C.}~\bibnamefont{Kelso}},
  \bibinfo{author}{\bibfnamefont{J.}~\bibnamefont{Kumar}},
  \bibinfo{author}{\bibfnamefont{P.}~\bibnamefont{Sandick}}, \bibnamefont{and}
  \bibinfo{author}{\bibfnamefont{T.}~\bibnamefont{Yamamoto}},
  \bibinfo{journal}{Phys. Rev. D} \textbf{\bibinfo{volume}{90}},
  \bibinfo{pages}{095007} (\bibinfo{year}{2014}), \eprint{1406.4903}.

\bibitem[{\citenamefont{Kopp et~al.}(2009)\citenamefont{Kopp, Niro, Schwetz,
  and Zupan}}]{Kopp:2009et}
\bibinfo{author}{\bibfnamefont{J.}~\bibnamefont{Kopp}},
  \bibinfo{author}{\bibfnamefont{V.}~\bibnamefont{Niro}},
  \bibinfo{author}{\bibfnamefont{T.}~\bibnamefont{Schwetz}}, \bibnamefont{and}
  \bibinfo{author}{\bibfnamefont{J.}~\bibnamefont{Zupan}},
  \bibinfo{journal}{Phys. Rev.} \textbf{\bibinfo{volume}{D80}},
  \bibinfo{pages}{083502} (\bibinfo{year}{2009}), \eprint{0907.3159}.

\bibitem[{\citenamefont{Schmidt et~al.}(2012)\citenamefont{Schmidt, Schwetz,
  and Toma}}]{Schmidt:2012yg}
\bibinfo{author}{\bibfnamefont{D.}~\bibnamefont{Schmidt}},
  \bibinfo{author}{\bibfnamefont{T.}~\bibnamefont{Schwetz}}, \bibnamefont{and}
  \bibinfo{author}{\bibfnamefont{T.}~\bibnamefont{Toma}},
  \bibinfo{journal}{Phys. Rev. D} \textbf{\bibinfo{volume}{85}},
  \bibinfo{pages}{073009} (\bibinfo{year}{2012}), \eprint{1201.0906}.

\bibitem[{\citenamefont{Chang et~al.}(2014)\citenamefont{Chang, Edezhath,
  Hutchinson, and Luty}}]{Chang:2014tea}
\bibinfo{author}{\bibfnamefont{S.}~\bibnamefont{Chang}},
  \bibinfo{author}{\bibfnamefont{R.}~\bibnamefont{Edezhath}},
  \bibinfo{author}{\bibfnamefont{J.}~\bibnamefont{Hutchinson}},
  \bibnamefont{and} \bibinfo{author}{\bibfnamefont{M.}~\bibnamefont{Luty}},
  \bibinfo{journal}{Phys. Rev. D} \textbf{\bibinfo{volume}{90}},
  \bibinfo{pages}{015011} (\bibinfo{year}{2014}), \eprint{1402.7358}.

\bibitem[{\citenamefont{Bai and Berger}(2014)}]{Bai:2014osa}
\bibinfo{author}{\bibfnamefont{Y.}~\bibnamefont{Bai}} \bibnamefont{and}
  \bibinfo{author}{\bibfnamefont{J.}~\bibnamefont{Berger}},
  \bibinfo{journal}{JHEP} \textbf{\bibinfo{volume}{08}}, \bibinfo{pages}{153}
  (\bibinfo{year}{2014}), \eprint{1402.6696}.

\bibitem[{\citenamefont{Kile et~al.}(2015)\citenamefont{Kile, Kobach, and
  Soni}}]{Kile:2014jea}
\bibinfo{author}{\bibfnamefont{J.}~\bibnamefont{Kile}},
  \bibinfo{author}{\bibfnamefont{A.}~\bibnamefont{Kobach}}, \bibnamefont{and}
  \bibinfo{author}{\bibfnamefont{A.}~\bibnamefont{Soni}},
  \bibinfo{journal}{Phys. Lett. B} \textbf{\bibinfo{volume}{744}},
  \bibinfo{pages}{330} (\bibinfo{year}{2015}), \eprint{1411.1407}.

\bibitem[{\citenamefont{Roberts
  et~al.}(2016{\natexlab{a}})\citenamefont{Roberts, Dzuba, Flambaum, Pospelov,
  and Stadnik}}]{Roberts:2016xfw}
\bibinfo{author}{\bibfnamefont{B.~M.} \bibnamefont{Roberts}},
  \bibinfo{author}{\bibfnamefont{V.~A.} \bibnamefont{Dzuba}},
  \bibinfo{author}{\bibfnamefont{V.~V.} \bibnamefont{Flambaum}},
  \bibinfo{author}{\bibfnamefont{M.}~\bibnamefont{Pospelov}}, \bibnamefont{and}
  \bibinfo{author}{\bibfnamefont{Y.~V.} \bibnamefont{Stadnik}},
  \bibinfo{journal}{Phys. Rev. D} \textbf{\bibinfo{volume}{93}},
  \bibinfo{pages}{115037} (\bibinfo{year}{2016}{\natexlab{a}}),
  \eprint{1604.04559}.

\bibitem[{\citenamefont{D'Eramo et~al.}(2017)\citenamefont{D'Eramo, Kavanagh,
  and Panci}}]{DEramo:2017zqw}
\bibinfo{author}{\bibfnamefont{F.}~\bibnamefont{D'Eramo}},
  \bibinfo{author}{\bibfnamefont{B.~J.} \bibnamefont{Kavanagh}},
  \bibnamefont{and} \bibinfo{author}{\bibfnamefont{P.}~\bibnamefont{Panci}},
  \bibinfo{journal}{Phys. Lett. B} \textbf{\bibinfo{volume}{771}},
  \bibinfo{pages}{339} (\bibinfo{year}{2017}), \eprint{1702.00016}.

\bibitem[{\citenamefont{Roberts
  et~al.}(2016{\natexlab{b}})\citenamefont{Roberts, Flambaum, and
  Gribakin}}]{Roberts:2015lga}
\bibinfo{author}{\bibfnamefont{B.~M.} \bibnamefont{Roberts}},
  \bibinfo{author}{\bibfnamefont{V.~V.} \bibnamefont{Flambaum}},
  \bibnamefont{and} \bibinfo{author}{\bibfnamefont{G.~F.}
  \bibnamefont{Gribakin}}, \bibinfo{journal}{Phys. Rev. Lett.}
  \textbf{\bibinfo{volume}{116}}, \bibinfo{pages}{023201}
  (\bibinfo{year}{2016}{\natexlab{b}}), \eprint{1509.09044}.

\bibitem[{\citenamefont{Aprile et~al.}(2015{\natexlab{b}})}]{Aprile:2015ade}
\bibinfo{author}{\bibfnamefont{E.}~\bibnamefont{Aprile}} \bibnamefont{et~al.}
  (\bibinfo{collaboration}{XENON100}), \bibinfo{journal}{Science}
  \textbf{\bibinfo{volume}{349}}, \bibinfo{pages}{851}
  (\bibinfo{year}{2015}{\natexlab{b}}), \eprint{1507.07747}.

\bibitem[{\citenamefont{Aprile et~al.}(2015{\natexlab{c}})}]{Aprile:2015ibr}
\bibinfo{author}{\bibfnamefont{E.}~\bibnamefont{Aprile}} \bibnamefont{et~al.}
  (\bibinfo{collaboration}{XENON100}), \bibinfo{journal}{Phys. Rev. Lett.}
  \textbf{\bibinfo{volume}{115}}, \bibinfo{pages}{091302}
  (\bibinfo{year}{2015}{\natexlab{c}}), \eprint{1507.07748}.

\bibitem[{\citenamefont{Aprile et~al.}(2017{\natexlab{d}})}]{Aprile:2017yea}
\bibinfo{author}{\bibfnamefont{E.}~\bibnamefont{Aprile}} \bibnamefont{et~al.}
  (\bibinfo{collaboration}{XENON}), \bibinfo{journal}{Phys. Rev. Lett.}
  \textbf{\bibinfo{volume}{118}}, \bibinfo{pages}{101101}
  (\bibinfo{year}{2017}{\natexlab{d}}), \eprint{1701.00769}.

\bibitem[{\citenamefont{Akerib et~al.}(2018{\natexlab{c}})}]{Akerib:2018zoq}
\bibinfo{author}{\bibfnamefont{D.~S.} \bibnamefont{Akerib}}
  \bibnamefont{et~al.} (\bibinfo{collaboration}{LUX}), \bibinfo{journal}{Phys.
  Rev. D} \textbf{\bibinfo{volume}{98}}, \bibinfo{pages}{062005}
  (\bibinfo{year}{2018}{\natexlab{c}}), \eprint{1807.07113}.

\bibitem[{\citenamefont{Essig et~al.}(2012{\natexlab{a}})\citenamefont{Essig,
  Mardon, and Volansky}}]{Essig:2011nj}
\bibinfo{author}{\bibfnamefont{R.}~\bibnamefont{Essig}},
  \bibinfo{author}{\bibfnamefont{J.}~\bibnamefont{Mardon}}, \bibnamefont{and}
  \bibinfo{author}{\bibfnamefont{T.}~\bibnamefont{Volansky}},
  \bibinfo{journal}{Phys. Rev. D} \textbf{\bibinfo{volume}{85}},
  \bibinfo{pages}{076007} (\bibinfo{year}{2012}{\natexlab{a}}),
  \eprint{1108.5383}.

\bibitem[{\citenamefont{Garani and Palomares-Ruiz}(2017)}]{Garani:2017jcj}
\bibinfo{author}{\bibfnamefont{R.}~\bibnamefont{Garani}} \bibnamefont{and}
  \bibinfo{author}{\bibfnamefont{S.}~\bibnamefont{Palomares-Ruiz}},
  \bibinfo{journal}{JCAP} \textbf{\bibinfo{volume}{05}}, \bibinfo{pages}{007}
  (\bibinfo{year}{2017}), \eprint{1702.02768}.

\bibitem[{\citenamefont{Liang et~al.}(2018)\citenamefont{Liang, Tang, and
  Yang}}]{Liang:2018cjn}
\bibinfo{author}{\bibfnamefont{Z.-L.} \bibnamefont{Liang}},
  \bibinfo{author}{\bibfnamefont{Y.-L.} \bibnamefont{Tang}}, \bibnamefont{and}
  \bibinfo{author}{\bibfnamefont{Z.-Q.} \bibnamefont{Yang}},
  \bibinfo{journal}{JCAP} \textbf{\bibinfo{volume}{1810}}, \bibinfo{pages}{035}
  (\bibinfo{year}{2018}), \eprint{1802.01005}.

\bibitem[{\citenamefont{Fox et~al.}(2011)\citenamefont{Fox, Harnik, Kopp, and
  Tsai}}]{Fox:2011fx}
\bibinfo{author}{\bibfnamefont{P.~J.} \bibnamefont{Fox}},
  \bibinfo{author}{\bibfnamefont{R.}~\bibnamefont{Harnik}},
  \bibinfo{author}{\bibfnamefont{J.}~\bibnamefont{Kopp}}, \bibnamefont{and}
  \bibinfo{author}{\bibfnamefont{Y.}~\bibnamefont{Tsai}},
  \bibinfo{journal}{Phys. Rev. D} \textbf{\bibinfo{volume}{84}},
  \bibinfo{pages}{014028} (\bibinfo{year}{2011}), \eprint{1103.0240}.

\bibitem[{\citenamefont{Dreiner et~al.}(2013)\citenamefont{Dreiner, Huck,
  Kr\"amer, Schmeier, and Tattersall}}]{Dreiner:2012xm}
\bibinfo{author}{\bibfnamefont{H.}~\bibnamefont{Dreiner}},
  \bibinfo{author}{\bibfnamefont{M.}~\bibnamefont{Huck}},
  \bibinfo{author}{\bibfnamefont{M.}~\bibnamefont{Kr\"amer}},
  \bibinfo{author}{\bibfnamefont{D.}~\bibnamefont{Schmeier}}, \bibnamefont{and}
  \bibinfo{author}{\bibfnamefont{J.}~\bibnamefont{Tattersall}},
  \bibinfo{journal}{Phys. Rev. D} \textbf{\bibinfo{volume}{87}},
  \bibinfo{pages}{075015} (\bibinfo{year}{2013}), \eprint{1211.2254}.

\bibitem[{\citenamefont{Freitas and Westhoff}(2014)}]{Freitas:2014jla}
\bibinfo{author}{\bibfnamefont{A.}~\bibnamefont{Freitas}} \bibnamefont{and}
  \bibinfo{author}{\bibfnamefont{S.}~\bibnamefont{Westhoff}},
  \bibinfo{journal}{JHEP} \textbf{\bibinfo{volume}{10}}, \bibinfo{pages}{116}
  (\bibinfo{year}{2014}), \eprint{1408.1959}.

\bibitem[{\citenamefont{Dutta et~al.}(2017)\citenamefont{Dutta, Sachdeva, and
  Rawat}}]{Rawat:2017fak}
\bibinfo{author}{\bibfnamefont{S.}~\bibnamefont{Dutta}},
  \bibinfo{author}{\bibfnamefont{D.}~\bibnamefont{Sachdeva}}, \bibnamefont{and}
  \bibinfo{author}{\bibfnamefont{B.}~\bibnamefont{Rawat}},
  \bibinfo{journal}{Eur. Phys. J. C} \textbf{\bibinfo{volume}{77}},
  \bibinfo{pages}{639} (\bibinfo{year}{2017}), \eprint{1704.03994}.

\bibitem[{\citenamefont{Freese et~al.}(1988)\citenamefont{Freese, Frieman, and
  Gould}}]{Freese:1987wu}
\bibinfo{author}{\bibfnamefont{K.}~\bibnamefont{Freese}},
  \bibinfo{author}{\bibfnamefont{J.~A.} \bibnamefont{Frieman}},
  \bibnamefont{and} \bibinfo{author}{\bibfnamefont{A.}~\bibnamefont{Gould}},
  \bibinfo{journal}{Phys. Rev. D} \textbf{\bibinfo{volume}{37}},
  \bibinfo{pages}{3388} (\bibinfo{year}{1988}).

\bibitem[{\citenamefont{Copi et~al.}(1999)\citenamefont{Copi, Heo, and
  Krauss}}]{Copi:1999pw}
\bibinfo{author}{\bibfnamefont{C.~J.} \bibnamefont{Copi}},
  \bibinfo{author}{\bibfnamefont{J.}~\bibnamefont{Heo}}, \bibnamefont{and}
  \bibinfo{author}{\bibfnamefont{L.~M.} \bibnamefont{Krauss}},
  \bibinfo{journal}{Phys. Lett. B} \textbf{\bibinfo{volume}{461}},
  \bibinfo{pages}{43} (\bibinfo{year}{1999}), \eprint{hep-ph/9904499}.

\bibitem[{\citenamefont{Copi and Krauss}(2001)}]{Copi:2000tv}
\bibinfo{author}{\bibfnamefont{C.~J.} \bibnamefont{Copi}} \bibnamefont{and}
  \bibinfo{author}{\bibfnamefont{L.~M.} \bibnamefont{Krauss}},
  \bibinfo{journal}{Phys. Rev. D} \textbf{\bibinfo{volume}{63}},
  \bibinfo{pages}{043507} (\bibinfo{year}{2001}), \eprint{astro-ph/0009467}.

\bibitem[{\citenamefont{Baum et~al.}(2019)\citenamefont{Baum, Freese, and
  Kelso}}]{Baum:2018ekm}
\bibinfo{author}{\bibfnamefont{S.}~\bibnamefont{Baum}},
  \bibinfo{author}{\bibfnamefont{K.}~\bibnamefont{Freese}}, \bibnamefont{and}
  \bibinfo{author}{\bibfnamefont{C.}~\bibnamefont{Kelso}},
  \bibinfo{journal}{Phys. Lett. B} \textbf{\bibinfo{volume}{789}},
  \bibinfo{pages}{262} (\bibinfo{year}{2019}), \eprint{1804.01231}.

\bibitem[{\citenamefont{Amaré et~al.}(2019)}]{Amare:2019jul}
\bibinfo{author}{\bibfnamefont{J.}~\bibnamefont{Amaré}} \bibnamefont{et~al.},
  \bibinfo{journal}{Phys. Rev. Lett.} \textbf{\bibinfo{volume}{123}},
  \bibinfo{pages}{031301} (\bibinfo{year}{2019}), \eprint{1903.03973}.

\bibitem[{\citenamefont{Amare et~al.}(2021)}]{Amare:2021yyu}
\bibinfo{author}{\bibfnamefont{J.}~\bibnamefont{Amare}} \bibnamefont{et~al.},
  \bibinfo{journal}{Phys. Rev. D} \textbf{\bibinfo{volume}{103}},
  \bibinfo{pages}{102005} (\bibinfo{year}{2021}), \eprint{2103.01175}.

\bibitem[{\citenamefont{Adhikari et~al.}(2019)}]{Adhikari:2019off}
\bibinfo{author}{\bibfnamefont{G.}~\bibnamefont{Adhikari}} \bibnamefont{et~al.}
  (\bibinfo{collaboration}{COSINE-100}), \bibinfo{journal}{Phys. Rev. Lett.}
  \textbf{\bibinfo{volume}{123}}, \bibinfo{pages}{031302}
  (\bibinfo{year}{2019}), \eprint{1903.10098}.

\bibitem[{\citenamefont{Adhikari et~al.}(2021)}]{Adhikari:2021szr}
\bibinfo{author}{\bibfnamefont{G.}~\bibnamefont{Adhikari}} \bibnamefont{et~al.}
  (\bibinfo{collaboration}{COSINE-100}), \bibinfo{journal}{Sci. Adv.}
  \textbf{\bibinfo{volume}{7}}, \bibinfo{pages}{abk2699}
  (\bibinfo{year}{2021}), \eprint{2104.03537}.

\bibitem[{\citenamefont{Antonello et~al.}(2019)}]{SABRE:2018lfp}
\bibinfo{author}{\bibfnamefont{M.}~\bibnamefont{Antonello}}
  \bibnamefont{et~al.} (\bibinfo{collaboration}{SABRE}), \bibinfo{journal}{Eur.
  Phys. J. C} \textbf{\bibinfo{volume}{79}}, \bibinfo{pages}{363}
  (\bibinfo{year}{2019}), \eprint{1806.09340}.

\bibitem[{\citenamefont{Collar and Avignone}(1993)}]{Collar:1993ss}
\bibinfo{author}{\bibfnamefont{J.}~\bibnamefont{Collar}} \bibnamefont{and}
  \bibinfo{author}{\bibfnamefont{I.}~\bibnamefont{Avignone},
  \bibfnamefont{F.T.}}, \bibinfo{journal}{Phys. Rev. D}
  \textbf{\bibinfo{volume}{47}}, \bibinfo{pages}{5238} (\bibinfo{year}{1993}).

\bibitem[{\citenamefont{Hasenbalg et~al.}(1997)\citenamefont{Hasenbalg,
  Abriola, Avignone, Collar, Di~Gregorio, Gattone, Huck, Tomasi, and
  Urteaga}}]{Hasenbalg:1997hs}
\bibinfo{author}{\bibfnamefont{F.}~\bibnamefont{Hasenbalg}},
  \bibinfo{author}{\bibfnamefont{D.}~\bibnamefont{Abriola}},
  \bibinfo{author}{\bibfnamefont{F.}~\bibnamefont{Avignone}},
  \bibinfo{author}{\bibfnamefont{J.}~\bibnamefont{Collar}},
  \bibinfo{author}{\bibfnamefont{D.}~\bibnamefont{Di~Gregorio}},
  \bibinfo{author}{\bibfnamefont{A.}~\bibnamefont{Gattone}},
  \bibinfo{author}{\bibfnamefont{H.}~\bibnamefont{Huck}},
  \bibinfo{author}{\bibfnamefont{D.}~\bibnamefont{Tomasi}}, \bibnamefont{and}
  \bibinfo{author}{\bibfnamefont{I.}~\bibnamefont{Urteaga}},
  \bibinfo{journal}{Phys. Rev. D} \textbf{\bibinfo{volume}{55}},
  \bibinfo{pages}{7350} (\bibinfo{year}{1997}), \eprint{astro-ph/9702165}.

\bibitem[{\citenamefont{Kavanagh et~al.}(2017)\citenamefont{Kavanagh, Catena,
  and Kouvaris}}]{Kavanagh:2016pyr}
\bibinfo{author}{\bibfnamefont{B.~J.} \bibnamefont{Kavanagh}},
  \bibinfo{author}{\bibfnamefont{R.}~\bibnamefont{Catena}}, \bibnamefont{and}
  \bibinfo{author}{\bibfnamefont{C.}~\bibnamefont{Kouvaris}},
  \bibinfo{journal}{JCAP} \textbf{\bibinfo{volume}{01}}, \bibinfo{pages}{012}
  (\bibinfo{year}{2017}), \eprint{1611.05453}.

\bibitem[{\citenamefont{Emken and Kouvaris}(2017)}]{Emken:2017qmp}
\bibinfo{author}{\bibfnamefont{T.}~\bibnamefont{Emken}} \bibnamefont{and}
  \bibinfo{author}{\bibfnamefont{C.}~\bibnamefont{Kouvaris}},
  \bibinfo{journal}{JCAP} \textbf{\bibinfo{volume}{10}}, \bibinfo{pages}{031}
  (\bibinfo{year}{2017}), \eprint{1706.02249}.

\bibitem[{\citenamefont{Andriamirado et~al.}(2021)}]{PROSPECT:2021awi}
\bibinfo{author}{\bibfnamefont{M.}~\bibnamefont{Andriamirado}}
  \bibnamefont{et~al.} (\bibinfo{collaboration}{PROSPECT, (PROSPECT
  Collaboration)*}), \bibinfo{journal}{Phys. Rev. D}
  \textbf{\bibinfo{volume}{104}}, \bibinfo{pages}{012009}
  (\bibinfo{year}{2021}), \eprint{2104.11219}.

\bibitem[{\citenamefont{Kastens et~al.}(2009)\citenamefont{Kastens, Cahn,
  Manzur, and McKinsey}}]{Kastens:2009pa}
\bibinfo{author}{\bibfnamefont{L.}~\bibnamefont{Kastens}},
  \bibinfo{author}{\bibfnamefont{S.}~\bibnamefont{Cahn}},
  \bibinfo{author}{\bibfnamefont{A.}~\bibnamefont{Manzur}}, \bibnamefont{and}
  \bibinfo{author}{\bibfnamefont{D.}~\bibnamefont{McKinsey}},
  \bibinfo{journal}{Phys. Rev. C} \textbf{\bibinfo{volume}{80}},
  \bibinfo{pages}{045809} (\bibinfo{year}{2009}), \eprint{0905.1766}.

\bibitem[{\citenamefont{Manalaysay et~al.}(2010)}]{Manalaysay:2009yq}
\bibinfo{author}{\bibfnamefont{A.}~\bibnamefont{Manalaysay}}
  \bibnamefont{et~al.}, \bibinfo{journal}{Rev. Sci. Instrum.}
  \textbf{\bibinfo{volume}{81}}, \bibinfo{pages}{073303}
  (\bibinfo{year}{2010}), \eprint{0908.0616}.

\bibitem[{\citenamefont{Strigari}(2009)}]{Strigari:2009bq}
\bibinfo{author}{\bibfnamefont{L.~E.} \bibnamefont{Strigari}},
  \bibinfo{journal}{New J. Phys.} \textbf{\bibinfo{volume}{11}},
  \bibinfo{pages}{105011} (\bibinfo{year}{2009}), \eprint{0903.3630}.

\bibitem[{\citenamefont{Wyenberg and Shoemaker}(2018)}]{Wyenberg:2018eyv}
\bibinfo{author}{\bibfnamefont{J.}~\bibnamefont{Wyenberg}} \bibnamefont{and}
  \bibinfo{author}{\bibfnamefont{I.~M.} \bibnamefont{Shoemaker}},
  \bibinfo{journal}{Phys. Rev. D} \textbf{\bibinfo{volume}{97}},
  \bibinfo{pages}{115026} (\bibinfo{year}{2018}), \eprint{1803.08146}.

\bibitem[{\citenamefont{Essig et~al.}(2018)\citenamefont{Essig, Sholapurkar,
  and Yu}}]{Essig:2018tss}
\bibinfo{author}{\bibfnamefont{R.}~\bibnamefont{Essig}},
  \bibinfo{author}{\bibfnamefont{M.}~\bibnamefont{Sholapurkar}},
  \bibnamefont{and} \bibinfo{author}{\bibfnamefont{T.-T.} \bibnamefont{Yu}},
  \bibinfo{journal}{Phys. Rev. D} \textbf{\bibinfo{volume}{97}},
  \bibinfo{pages}{095029} (\bibinfo{year}{2018}), \eprint{1801.10159}.

\bibitem[{\citenamefont{Baudis et~al.}(2014)\citenamefont{Baudis, Ferella,
  Kish, Manalaysay, Marrodan~Undagoitia, and Schumann}}]{Baudis:2013qla}
\bibinfo{author}{\bibfnamefont{L.}~\bibnamefont{Baudis}},
  \bibinfo{author}{\bibfnamefont{A.}~\bibnamefont{Ferella}},
  \bibinfo{author}{\bibfnamefont{A.}~\bibnamefont{Kish}},
  \bibinfo{author}{\bibfnamefont{A.}~\bibnamefont{Manalaysay}},
  \bibinfo{author}{\bibfnamefont{T.}~\bibnamefont{Marrodan~Undagoitia}},
  \bibnamefont{and} \bibinfo{author}{\bibfnamefont{M.}~\bibnamefont{Schumann}},
  \bibinfo{journal}{JCAP} \textbf{\bibinfo{volume}{1401}}, \bibinfo{pages}{044}
  (\bibinfo{year}{2014}), \eprint{1309.7024}.

\bibitem[{\citenamefont{Dent et~al.}(2016)\citenamefont{Dent, Dutta, Newstead,
  and Strigari}}]{Dent:2016iht}
\bibinfo{author}{\bibfnamefont{J.~B.} \bibnamefont{Dent}},
  \bibinfo{author}{\bibfnamefont{B.}~\bibnamefont{Dutta}},
  \bibinfo{author}{\bibfnamefont{J.~L.} \bibnamefont{Newstead}},
  \bibnamefont{and} \bibinfo{author}{\bibfnamefont{L.~E.}
  \bibnamefont{Strigari}}, \bibinfo{journal}{Phys. Rev. D}
  \textbf{\bibinfo{volume}{93}}, \bibinfo{pages}{075018}
  (\bibinfo{year}{2016}), \eprint{1602.05300}.

\bibitem[{\citenamefont{Dent et~al.}(2017)\citenamefont{Dent, Dutta, Newstead,
  and Strigari}}]{Dent:2016wor}
\bibinfo{author}{\bibfnamefont{J.~B.} \bibnamefont{Dent}},
  \bibinfo{author}{\bibfnamefont{B.}~\bibnamefont{Dutta}},
  \bibinfo{author}{\bibfnamefont{J.~L.} \bibnamefont{Newstead}},
  \bibnamefont{and} \bibinfo{author}{\bibfnamefont{L.~E.}
  \bibnamefont{Strigari}}, \bibinfo{journal}{Phys. Rev. D}
  \textbf{\bibinfo{volume}{95}}, \bibinfo{pages}{051701}
  (\bibinfo{year}{2017}), \eprint{1607.01468}.

\bibitem[{\citenamefont{Dent et~al.}(2020{\natexlab{a}})\citenamefont{Dent,
  Dutta, Newstead, and Shoemaker}}]{Dent:2019krz}
\bibinfo{author}{\bibfnamefont{J.~B.} \bibnamefont{Dent}},
  \bibinfo{author}{\bibfnamefont{B.}~\bibnamefont{Dutta}},
  \bibinfo{author}{\bibfnamefont{J.~L.} \bibnamefont{Newstead}},
  \bibnamefont{and} \bibinfo{author}{\bibfnamefont{I.~M.}
  \bibnamefont{Shoemaker}}, \bibinfo{journal}{Phys. Rev. D}
  \textbf{\bibinfo{volume}{101}}, \bibinfo{pages}{116007}
  (\bibinfo{year}{2020}{\natexlab{a}}), \eprint{1907.03782}.

\bibitem[{\citenamefont{O'Hare}(2021)}]{OHare:2021utq}
\bibinfo{author}{\bibfnamefont{C.~A.~J.} \bibnamefont{O'Hare}}
  (\bibinfo{year}{2021}), \eprint{2109.03116}.

\bibitem[{\citenamefont{Papoulias et~al.}(2018)\citenamefont{Papoulias, Sahu,
  Kosmas, Kota, and Nayak}}]{Papoulias:2018uzy}
\bibinfo{author}{\bibfnamefont{D.~K.} \bibnamefont{Papoulias}},
  \bibinfo{author}{\bibfnamefont{R.}~\bibnamefont{Sahu}},
  \bibinfo{author}{\bibfnamefont{T.~S.} \bibnamefont{Kosmas}},
  \bibinfo{author}{\bibfnamefont{V.~K.~B.} \bibnamefont{Kota}},
  \bibnamefont{and} \bibinfo{author}{\bibfnamefont{B.}~\bibnamefont{Nayak}},
  \bibinfo{journal}{Adv. High Energy Phys.} \textbf{\bibinfo{volume}{2018}},
  \bibinfo{pages}{6031362} (\bibinfo{year}{2018}), \eprint{1804.11319}.

\bibitem[{\citenamefont{Aristizabal~Sierra
  et~al.}(2018)\citenamefont{Aristizabal~Sierra, Rojas, and
  Tytgat}}]{AristizabalSierra:2017joc}
\bibinfo{author}{\bibfnamefont{D.}~\bibnamefont{Aristizabal~Sierra}},
  \bibinfo{author}{\bibfnamefont{N.}~\bibnamefont{Rojas}}, \bibnamefont{and}
  \bibinfo{author}{\bibfnamefont{M.~H.~G.} \bibnamefont{Tytgat}},
  \bibinfo{journal}{JHEP} \textbf{\bibinfo{volume}{03}}, \bibinfo{pages}{197}
  (\bibinfo{year}{2018}), \eprint{1712.09667}.

\bibitem[{\citenamefont{Gonzalez-Garcia
  et~al.}(2018)\citenamefont{Gonzalez-Garcia, Maltoni, Perez-Gonzalez, and
  Zukanovich~Funchal}}]{Gonzalez-Garcia:2018dep}
\bibinfo{author}{\bibfnamefont{M.~C.} \bibnamefont{Gonzalez-Garcia}},
  \bibinfo{author}{\bibfnamefont{M.}~\bibnamefont{Maltoni}},
  \bibinfo{author}{\bibfnamefont{Y.~F.} \bibnamefont{Perez-Gonzalez}},
  \bibnamefont{and}
  \bibinfo{author}{\bibfnamefont{R.}~\bibnamefont{Zukanovich~Funchal}},
  \bibinfo{journal}{JHEP} \textbf{\bibinfo{volume}{07}}, \bibinfo{pages}{019}
  (\bibinfo{year}{2018}), \eprint{1803.03650}.

\bibitem[{\citenamefont{Bœhm et~al.}(2019)\citenamefont{Bœhm, Cerdeño,
  Machado, Olivares-Del~Campo, Perdomo, and Reid}}]{Boehm:2018sux}
\bibinfo{author}{\bibfnamefont{C.}~\bibnamefont{Bœhm}},
  \bibinfo{author}{\bibfnamefont{D.~G.} \bibnamefont{Cerdeño}},
  \bibinfo{author}{\bibfnamefont{P.~A.~N.} \bibnamefont{Machado}},
  \bibinfo{author}{\bibfnamefont{A.}~\bibnamefont{Olivares-Del~Campo}},
  \bibinfo{author}{\bibfnamefont{E.}~\bibnamefont{Perdomo}}, \bibnamefont{and}
  \bibinfo{author}{\bibfnamefont{E.}~\bibnamefont{Reid}},
  \bibinfo{journal}{JCAP} \textbf{\bibinfo{volume}{1901}}, \bibinfo{pages}{043}
  (\bibinfo{year}{2019}), \eprint{1809.06385}.

\bibitem[{\citenamefont{Davis}(2015)}]{Davis:2014ama}
\bibinfo{author}{\bibfnamefont{J.~H.} \bibnamefont{Davis}},
  \bibinfo{journal}{JCAP} \textbf{\bibinfo{volume}{1503}}, \bibinfo{pages}{012}
  (\bibinfo{year}{2015}), \eprint{1412.1475}.

\bibitem[{\citenamefont{Ruppin et~al.}(2014)\citenamefont{Ruppin, Billard,
  Figueroa-Feliciano, and Strigari}}]{Ruppin:2014bra}
\bibinfo{author}{\bibfnamefont{F.}~\bibnamefont{Ruppin}},
  \bibinfo{author}{\bibfnamefont{J.}~\bibnamefont{Billard}},
  \bibinfo{author}{\bibfnamefont{E.}~\bibnamefont{Figueroa-Feliciano}},
  \bibnamefont{and} \bibinfo{author}{\bibfnamefont{L.}~\bibnamefont{Strigari}},
  \bibinfo{journal}{Phys. Rev. D} \textbf{\bibinfo{volume}{90}},
  \bibinfo{pages}{083510} (\bibinfo{year}{2014}), \eprint{1408.3581}.

\bibitem[{\citenamefont{Grothaus et~al.}(2014)\citenamefont{Grothaus,
  Fairbairn, and Monroe}}]{Grothaus:2014hja}
\bibinfo{author}{\bibfnamefont{P.}~\bibnamefont{Grothaus}},
  \bibinfo{author}{\bibfnamefont{M.}~\bibnamefont{Fairbairn}},
  \bibnamefont{and} \bibinfo{author}{\bibfnamefont{J.}~\bibnamefont{Monroe}},
  \bibinfo{journal}{Phys. Rev. D} \textbf{\bibinfo{volume}{90}},
  \bibinfo{pages}{055018} (\bibinfo{year}{2014}), \eprint{1406.5047}.

\bibitem[{\citenamefont{O'Hare et~al.}(2015)\citenamefont{O'Hare, Green,
  Billard, Figueroa-Feliciano, and Strigari}}]{O'Hare:2015mda}
\bibinfo{author}{\bibfnamefont{C.~A.~J.} \bibnamefont{O'Hare}},
  \bibinfo{author}{\bibfnamefont{A.~M.} \bibnamefont{Green}},
  \bibinfo{author}{\bibfnamefont{J.}~\bibnamefont{Billard}},
  \bibinfo{author}{\bibfnamefont{E.}~\bibnamefont{Figueroa-Feliciano}},
  \bibnamefont{and} \bibinfo{author}{\bibfnamefont{L.~E.}
  \bibnamefont{Strigari}}, \bibinfo{journal}{Phys. Rev. D}
  \textbf{\bibinfo{volume}{92}}, \bibinfo{pages}{063518}
  (\bibinfo{year}{2015}), \eprint{1505.08061}.

\bibitem[{\citenamefont{Mayet et~al.}(2016)}]{Mayet:2016zxu}
\bibinfo{author}{\bibfnamefont{F.}~\bibnamefont{Mayet}} \bibnamefont{et~al.},
  \bibinfo{journal}{Phys. Rept.} \textbf{\bibinfo{volume}{627}},
  \bibinfo{pages}{1} (\bibinfo{year}{2016}), \eprint{1602.03781}.

\bibitem[{\citenamefont{Franarin and Fairbairn}(2016)}]{Franarin:2016ppr}
\bibinfo{author}{\bibfnamefont{T.}~\bibnamefont{Franarin}} \bibnamefont{and}
  \bibinfo{author}{\bibfnamefont{M.}~\bibnamefont{Fairbairn}},
  \bibinfo{journal}{Phys. Rev. D} \textbf{\bibinfo{volume}{94}},
  \bibinfo{pages}{053004} (\bibinfo{year}{2016}), \eprint{1605.08727}.

\bibitem[{\citenamefont{O'Hare et~al.}(2017)\citenamefont{O'Hare, Kavanagh, and
  Green}}]{OHare:2017rag}
\bibinfo{author}{\bibfnamefont{C.~A.~J.} \bibnamefont{O'Hare}},
  \bibinfo{author}{\bibfnamefont{B.~J.} \bibnamefont{Kavanagh}},
  \bibnamefont{and} \bibinfo{author}{\bibfnamefont{A.~M.} \bibnamefont{Green}},
  \bibinfo{journal}{Phys. Rev. D} \textbf{\bibinfo{volume}{96}},
  \bibinfo{pages}{083011} (\bibinfo{year}{2017}), \eprint{1708.02959}.

\bibitem[{\citenamefont{Bergstrom et~al.}(2016)\citenamefont{Bergstrom,
  Gonzalez-Garcia, Maltoni, Pena-Garay, Serenelli, and
  Song}}]{Bergstrom:2016cbh}
\bibinfo{author}{\bibfnamefont{J.}~\bibnamefont{Bergstrom}},
  \bibinfo{author}{\bibfnamefont{M.}~\bibnamefont{Gonzalez-Garcia}},
  \bibinfo{author}{\bibfnamefont{M.}~\bibnamefont{Maltoni}},
  \bibinfo{author}{\bibfnamefont{C.}~\bibnamefont{Pena-Garay}},
  \bibinfo{author}{\bibfnamefont{A.~M.} \bibnamefont{Serenelli}},
  \bibnamefont{and} \bibinfo{author}{\bibfnamefont{N.}~\bibnamefont{Song}},
  \bibinfo{journal}{JHEP} \textbf{\bibinfo{volume}{03}}, \bibinfo{pages}{132}
  (\bibinfo{year}{2016}), \eprint{1601.00972}.

\bibitem[{\citenamefont{Caden}(2020)}]{Caden:2017htb}
\bibinfo{author}{\bibfnamefont{E.}~\bibnamefont{Caden}}
  (\bibinfo{collaboration}{SNO+}), \bibinfo{journal}{J. Phys. Conf. Ser.}
  \textbf{\bibinfo{volume}{1342}}, \bibinfo{pages}{012022}
  (\bibinfo{year}{2020}), \eprint{1711.11094}.

\bibitem[{\citenamefont{Beacom et~al.}(2017)}]{Jinping:2016iiq}
\bibinfo{author}{\bibfnamefont{J.~F.} \bibnamefont{Beacom}}
  \bibnamefont{et~al.} (\bibinfo{collaboration}{Jinping}),
  \bibinfo{journal}{Chin. Phys. C} \textbf{\bibinfo{volume}{41}},
  \bibinfo{pages}{023002} (\bibinfo{year}{2017}), \eprint{1602.01733}.

\bibitem[{\citenamefont{Abi et~al.}(2018)}]{Abi:2018dnh}
\bibinfo{author}{\bibfnamefont{B.}~\bibnamefont{Abi}} \bibnamefont{et~al.}
  (\bibinfo{collaboration}{DUNE}) (\bibinfo{year}{2018}), \eprint{1807.10334}.

\bibitem[{\citenamefont{Kelly et~al.}(2019)\citenamefont{Kelly, Machado,
  Martinez~Soler, Parke, and Perez~Gonzalez}}]{Kelly:2019itm}
\bibinfo{author}{\bibfnamefont{K.~J.} \bibnamefont{Kelly}},
  \bibinfo{author}{\bibfnamefont{P.~A.} \bibnamefont{Machado}},
  \bibinfo{author}{\bibfnamefont{I.}~\bibnamefont{Martinez~Soler}},
  \bibinfo{author}{\bibfnamefont{S.~J.} \bibnamefont{Parke}}, \bibnamefont{and}
  \bibinfo{author}{\bibfnamefont{Y.~F.} \bibnamefont{Perez~Gonzalez}},
  \bibinfo{journal}{Phys. Rev. Lett.} \textbf{\bibinfo{volume}{123}},
  \bibinfo{pages}{081801} (\bibinfo{year}{2019}), \eprint{1904.02751}.

\bibitem[{\citenamefont{O'Hare}(2020)}]{OHare:2020lva}
\bibinfo{author}{\bibfnamefont{C.~A.~J.} \bibnamefont{O'Hare}},
  \bibinfo{journal}{Phys. Rev. D} \textbf{\bibinfo{volume}{102}},
  \bibinfo{pages}{063024} (\bibinfo{year}{2020}), \eprint{2002.07499}.

\bibitem[{\citenamefont{Roszkowski et~al.}(2015)\citenamefont{Roszkowski,
  Sessolo, and Williams}}]{Roszkowski:2014iqa}
\bibinfo{author}{\bibfnamefont{L.}~\bibnamefont{Roszkowski}},
  \bibinfo{author}{\bibfnamefont{E.~M.} \bibnamefont{Sessolo}},
  \bibnamefont{and} \bibinfo{author}{\bibfnamefont{A.~J.}
  \bibnamefont{Williams}}, \bibinfo{journal}{JHEP}
  \textbf{\bibinfo{volume}{02}}, \bibinfo{pages}{014} (\bibinfo{year}{2015}),
  \eprint{1411.5214}.

\bibitem[{\citenamefont{Athron et~al.}(2017)}]{Athron:2017qdc}
\bibinfo{author}{\bibfnamefont{P.}~\bibnamefont{Athron}} \bibnamefont{et~al.}
  (\bibinfo{collaboration}{GAMBIT}), \bibinfo{journal}{Eur. Phys. J. C}
  \textbf{\bibinfo{volume}{77}}, \bibinfo{pages}{824} (\bibinfo{year}{2017}),
  \eprint{1705.07935}.

\bibitem[{\citenamefont{Kobakhidze and Talia}(2019)}]{Kobakhidze:2018vuy}
\bibinfo{author}{\bibfnamefont{A.}~\bibnamefont{Kobakhidze}} \bibnamefont{and}
  \bibinfo{author}{\bibfnamefont{M.}~\bibnamefont{Talia}},
  \bibinfo{journal}{JHEP} \textbf{\bibinfo{volume}{08}}, \bibinfo{pages}{105}
  (\bibinfo{year}{2019}), \eprint{1806.08502}.

\bibitem[{\citenamefont{Baker et~al.}(2020)\citenamefont{Baker, Kopp, and
  Long}}]{Baker:2019ndr}
\bibinfo{author}{\bibfnamefont{M.~J.} \bibnamefont{Baker}},
  \bibinfo{author}{\bibfnamefont{J.}~\bibnamefont{Kopp}}, \bibnamefont{and}
  \bibinfo{author}{\bibfnamefont{A.~J.} \bibnamefont{Long}},
  \bibinfo{journal}{Phys. Rev. Lett.} \textbf{\bibinfo{volume}{125}},
  \bibinfo{pages}{151102} (\bibinfo{year}{2020}), \eprint{1912.02830}.

\bibitem[{\citenamefont{Arina et~al.}(2020)\citenamefont{Arina, Beniwal,
  Degrande, Heisig, and Scaffidi}}]{Arina:2019tib}
\bibinfo{author}{\bibfnamefont{C.}~\bibnamefont{Arina}},
  \bibinfo{author}{\bibfnamefont{A.}~\bibnamefont{Beniwal}},
  \bibinfo{author}{\bibfnamefont{C.}~\bibnamefont{Degrande}},
  \bibinfo{author}{\bibfnamefont{J.}~\bibnamefont{Heisig}}, \bibnamefont{and}
  \bibinfo{author}{\bibfnamefont{A.}~\bibnamefont{Scaffidi}},
  \bibinfo{journal}{JHEP} \textbf{\bibinfo{volume}{04}}, \bibinfo{pages}{015}
  (\bibinfo{year}{2020}), \eprint{1912.04008}.

\bibitem[{\citenamefont{Battaglieri et~al.}(2017)}]{Battaglieri:2017aum}
\bibinfo{author}{\bibfnamefont{M.}~\bibnamefont{Battaglieri}}
  \bibnamefont{et~al.} (\bibinfo{year}{2017}), \eprint{1707.04591}.

\bibitem[{\citenamefont{Essig et~al.}(2013)}]{Essig:2013lka}
\bibinfo{author}{\bibfnamefont{R.}~\bibnamefont{Essig}} \bibnamefont{et~al.},
  in \emph{\bibinfo{booktitle}{{Community Summer Study 2013: Snowmass on the
  Mississippi (CSS2013) Minneapolis, MN, USA, July 29-August 6, 2013}}}
  (\bibinfo{year}{2013}), \eprint{1311.0029},
  \urlprefix\url{http://inspirehep.net/record/1263039/files/arXiv:1311.0029.pdf}.

\bibitem[{\citenamefont{Hochberg et~al.}(2014)\citenamefont{Hochberg, Kuflik,
  Volansky, and Wacker}}]{Hochberg:2014dra}
\bibinfo{author}{\bibfnamefont{Y.}~\bibnamefont{Hochberg}},
  \bibinfo{author}{\bibfnamefont{E.}~\bibnamefont{Kuflik}},
  \bibinfo{author}{\bibfnamefont{T.}~\bibnamefont{Volansky}}, \bibnamefont{and}
  \bibinfo{author}{\bibfnamefont{J.~G.} \bibnamefont{Wacker}},
  \bibinfo{journal}{Phys. Rev. Lett.} \textbf{\bibinfo{volume}{113}},
  \bibinfo{pages}{171301} (\bibinfo{year}{2014}),
  \urlprefix\url{https://link.aps.org/doi/10.1103/PhysRevLett.113.171301}.

\bibitem[{\citenamefont{Hochberg et~al.}(2015)\citenamefont{Hochberg, Kuflik,
  Murayama, Volansky, and Wacker}}]{Kuflik:2015isi}
\bibinfo{author}{\bibfnamefont{Y.}~\bibnamefont{Hochberg}},
  \bibinfo{author}{\bibfnamefont{E.}~\bibnamefont{Kuflik}},
  \bibinfo{author}{\bibfnamefont{H.}~\bibnamefont{Murayama}},
  \bibinfo{author}{\bibfnamefont{T.}~\bibnamefont{Volansky}}, \bibnamefont{and}
  \bibinfo{author}{\bibfnamefont{J.~G.} \bibnamefont{Wacker}},
  \bibinfo{journal}{Phys. Rev. Lett.} \textbf{\bibinfo{volume}{115}},
  \bibinfo{pages}{021301} (\bibinfo{year}{2015}),
  \urlprefix\url{https://link.aps.org/doi/10.1103/PhysRevLett.115.021301}.

\bibitem[{\citenamefont{Alexander et~al.}(2016)}]{Alexander:2016aln}
\bibinfo{author}{\bibfnamefont{J.}~\bibnamefont{Alexander}}
  \bibnamefont{et~al.}, \emph{\bibinfo{title}{{Dark Sectors 2016 Workshop:
  Community Report}}} (\bibinfo{year}{2016}), \eprint{1608.08632}.

\bibitem[{\citenamefont{Essig et~al.}(2012{\natexlab{b}})\citenamefont{Essig,
  Manalaysay, Mardon, Sorensen, and Volansky}}]{Essig:2012yx}
\bibinfo{author}{\bibfnamefont{R.}~\bibnamefont{Essig}},
  \bibinfo{author}{\bibfnamefont{A.}~\bibnamefont{Manalaysay}},
  \bibinfo{author}{\bibfnamefont{J.}~\bibnamefont{Mardon}},
  \bibinfo{author}{\bibfnamefont{P.}~\bibnamefont{Sorensen}}, \bibnamefont{and}
  \bibinfo{author}{\bibfnamefont{T.}~\bibnamefont{Volansky}},
  \bibinfo{journal}{Phys. Rev. Lett.} \textbf{\bibinfo{volume}{109}},
  \bibinfo{pages}{021301} (\bibinfo{year}{2012}{\natexlab{b}}),
  \eprint{1206.2644}.

\bibitem[{\citenamefont{Agnese et~al.}(2016)}]{Agnese:2015nto}
\bibinfo{author}{\bibfnamefont{R.}~\bibnamefont{Agnese}} \bibnamefont{et~al.}
  (\bibinfo{collaboration}{SuperCDMS}), \bibinfo{journal}{Phys. Rev. Lett.}
  \textbf{\bibinfo{volume}{116}}, \bibinfo{pages}{071301}
  (\bibinfo{year}{2016}), \eprint{1509.02448}.

\bibitem[{\citenamefont{Angloher
  et~al.}(2016{\natexlab{a}})}]{Angloher:2015ewa}
\bibinfo{author}{\bibfnamefont{G.}~\bibnamefont{Angloher}} \bibnamefont{et~al.}
  (\bibinfo{collaboration}{CRESST}), \bibinfo{journal}{Eur. Phys. J. C}
  \textbf{\bibinfo{volume}{76}}, \bibinfo{pages}{25}
  (\bibinfo{year}{2016}{\natexlab{a}}), \eprint{1509.01515}.

\bibitem[{\citenamefont{Petricca et~al.}(2017)}]{Petricca:2017zdp}
\bibinfo{author}{\bibfnamefont{F.}~\bibnamefont{Petricca}} \bibnamefont{et~al.}
  (\bibinfo{collaboration}{CRESST}), in \emph{\bibinfo{booktitle}{{15th
  International Conference on Topics in Astroparticle and Underground Physics
  (TAUP 2017) Sudbury, Ontario, Canada, July 24-28, 2017}}}
  (\bibinfo{year}{2017}), \eprint{1711.07692}.

\bibitem[{\citenamefont{Angloher et~al.}(2017)}]{Angloher:2017sxg}
\bibinfo{author}{\bibfnamefont{G.}~\bibnamefont{Angloher}} \bibnamefont{et~al.}
  (\bibinfo{collaboration}{CRESST}), \bibinfo{journal}{Eur. Phys. J. C}
  \textbf{\bibinfo{volume}{77}}, \bibinfo{pages}{637} (\bibinfo{year}{2017}),
  \eprint{1707.06749}.

\bibitem[{\citenamefont{Arnaud et~al.}(2018)}]{NEWS-G:2017pxg}
\bibinfo{author}{\bibfnamefont{Q.}~\bibnamefont{Arnaud}} \bibnamefont{et~al.}
  (\bibinfo{collaboration}{NEWS-G}), \bibinfo{journal}{Astropart. Phys.}
  \textbf{\bibinfo{volume}{97}}, \bibinfo{pages}{54} (\bibinfo{year}{2018}),
  \eprint{1706.04934}.

\bibitem[{\citenamefont{Agnes et~al.}(2018{\natexlab{a}})}]{Agnes:2018ves}
\bibinfo{author}{\bibfnamefont{P.}~\bibnamefont{Agnes}} \bibnamefont{et~al.}
  (\bibinfo{collaboration}{DarkSide}), \bibinfo{journal}{Phys. Rev. Lett.}
  \textbf{\bibinfo{volume}{121}}, \bibinfo{pages}{081307}
  (\bibinfo{year}{2018}{\natexlab{a}}), \eprint{1802.06994}.

\bibitem[{\citenamefont{Agnes et~al.}(2018{\natexlab{b}})}]{Agnes:2018oej}
\bibinfo{author}{\bibfnamefont{P.}~\bibnamefont{Agnes}} \bibnamefont{et~al.}
  (\bibinfo{collaboration}{DarkSide}), \bibinfo{journal}{Phys. Rev. Lett.}
  \textbf{\bibinfo{volume}{121}}, \bibinfo{pages}{111303}
  (\bibinfo{year}{2018}{\natexlab{b}}), \eprint{1802.06998}.

\bibitem[{\citenamefont{Essig et~al.}(2017)\citenamefont{Essig, Volansky, and
  Yu}}]{Essig:2017kqs}
\bibinfo{author}{\bibfnamefont{R.}~\bibnamefont{Essig}},
  \bibinfo{author}{\bibfnamefont{T.}~\bibnamefont{Volansky}}, \bibnamefont{and}
  \bibinfo{author}{\bibfnamefont{T.-T.} \bibnamefont{Yu}},
  \bibinfo{journal}{Phys. Rev. D} \textbf{\bibinfo{volume}{96}},
  \bibinfo{pages}{043017} (\bibinfo{year}{2017}), \eprint{1703.00910}.

\bibitem[{\citenamefont{Akerib et~al.}(2019{\natexlab{a}})}]{Akerib:2018hck}
\bibinfo{author}{\bibfnamefont{D.~S.} \bibnamefont{Akerib}}
  \bibnamefont{et~al.} (\bibinfo{collaboration}{LUX}), \bibinfo{journal}{Phys.
  Rev. Lett.} \textbf{\bibinfo{volume}{122}}, \bibinfo{pages}{131301}
  (\bibinfo{year}{2019}{\natexlab{a}}), \eprint{1811.11241}.

\bibitem[{\citenamefont{Aprile et~al.}(2020{\natexlab{c}})}]{Aprile:2020tmw}
\bibinfo{author}{\bibfnamefont{E.}~\bibnamefont{Aprile}} \bibnamefont{et~al.}
  (\bibinfo{collaboration}{XENON}), \bibinfo{journal}{Phys. Rev. D}
  \textbf{\bibinfo{volume}{102}}, \bibinfo{pages}{072004}
  (\bibinfo{year}{2020}{\natexlab{c}}), \eprint{2006.09721}.

\bibitem[{\citenamefont{Lenardo et~al.}(2019)}]{Lenardo:2019fcn}
\bibinfo{author}{\bibfnamefont{B.}~\bibnamefont{Lenardo}} \bibnamefont{et~al.}
  (\bibinfo{year}{2019}), \eprint{1908.00518}.

\bibitem[{\citenamefont{Kouvaris and Pradler}(2017)}]{Kouvaris:2016afs}
\bibinfo{author}{\bibfnamefont{C.}~\bibnamefont{Kouvaris}} \bibnamefont{and}
  \bibinfo{author}{\bibfnamefont{J.}~\bibnamefont{Pradler}},
  \bibinfo{journal}{Phys. Rev. Lett.} \textbf{\bibinfo{volume}{118}},
  \bibinfo{pages}{031803} (\bibinfo{year}{2017}), \eprint{1607.01789}.

\bibitem[{\citenamefont{Aprile et~al.}(2021{\natexlab{b}})}]{Aprile:2020thb}
\bibinfo{author}{\bibfnamefont{E.}~\bibnamefont{Aprile}} \bibnamefont{et~al.}
  (\bibinfo{collaboration}{XENON}), \bibinfo{journal}{Phys. Rev. Lett.}
  \textbf{\bibinfo{volume}{126}}, \bibinfo{pages}{091301}
  (\bibinfo{year}{2021}{\natexlab{b}}), \eprint{2012.02846}.

\bibitem[{\citenamefont{Fujii et~al.}(2015)\citenamefont{Fujii, Endo, Torigoe,
  Nakamura, Haruyama, Kasami, Mihara, Saito, Sasaki, and Tawara}}]{Fuji:2015}
\bibinfo{author}{\bibfnamefont{K.}~\bibnamefont{Fujii}},
  \bibinfo{author}{\bibfnamefont{Y.}~\bibnamefont{Endo}},
  \bibinfo{author}{\bibfnamefont{Y.}~\bibnamefont{Torigoe}},
  \bibinfo{author}{\bibfnamefont{S.}~\bibnamefont{Nakamura}},
  \bibinfo{author}{\bibfnamefont{T.}~\bibnamefont{Haruyama}},
  \bibinfo{author}{\bibfnamefont{K.}~\bibnamefont{Kasami}},
  \bibinfo{author}{\bibfnamefont{S.}~\bibnamefont{Mihara}},
  \bibinfo{author}{\bibfnamefont{K.}~\bibnamefont{Saito}},
  \bibinfo{author}{\bibfnamefont{S.}~\bibnamefont{Sasaki}}, \bibnamefont{and}
  \bibinfo{author}{\bibfnamefont{H.}~\bibnamefont{Tawara}},
  \bibinfo{journal}{Nuclear Instruments and Methods in Physics Research Section
  A: Accelerators, Spectrometers, Detectors and Associated Equipment}
  \textbf{\bibinfo{volume}{795}}, \bibinfo{pages}{293} (\bibinfo{year}{2015}),
  ISSN \bibinfo{issn}{0168-9002},
  \urlprefix\url{https://www.sciencedirect.com/science/article/pii/S016890021500724X}.

\bibitem[{\citenamefont{Faham et~al.}(2015)\citenamefont{Faham, Gehman, Currie,
  Dobi, Sorensen, and Gaitskell}}]{Faham:2015kqa}
\bibinfo{author}{\bibfnamefont{C.}~\bibnamefont{Faham}},
  \bibinfo{author}{\bibfnamefont{V.}~\bibnamefont{Gehman}},
  \bibinfo{author}{\bibfnamefont{A.}~\bibnamefont{Currie}},
  \bibinfo{author}{\bibfnamefont{A.}~\bibnamefont{Dobi}},
  \bibinfo{author}{\bibfnamefont{P.}~\bibnamefont{Sorensen}}, \bibnamefont{and}
  \bibinfo{author}{\bibfnamefont{R.}~\bibnamefont{Gaitskell}},
  \bibinfo{journal}{JINST} \textbf{\bibinfo{volume}{10}},
  \bibinfo{pages}{P09010} (\bibinfo{year}{2015}), \eprint{1506.08748}.

\bibitem[{\citenamefont{L\'opez~Paredes
  et~al.}(2018)\citenamefont{L\'opez~Paredes, Ara\'ujo, Froborg, Marangou,
  Olcina, Sumner, Taylor, Tom\'as, and Vacheret}}]{Paredes:2018hxp}
\bibinfo{author}{\bibfnamefont{B.}~\bibnamefont{L\'opez~Paredes}},
  \bibinfo{author}{\bibfnamefont{H.~M.} \bibnamefont{Ara\'ujo}},
  \bibinfo{author}{\bibfnamefont{F.}~\bibnamefont{Froborg}},
  \bibinfo{author}{\bibfnamefont{N.}~\bibnamefont{Marangou}},
  \bibinfo{author}{\bibfnamefont{I.}~\bibnamefont{Olcina}},
  \bibinfo{author}{\bibfnamefont{T.~J.} \bibnamefont{Sumner}},
  \bibinfo{author}{\bibfnamefont{R.}~\bibnamefont{Taylor}},
  \bibinfo{author}{\bibfnamefont{A.}~\bibnamefont{Tom\'as}}, \bibnamefont{and}
  \bibinfo{author}{\bibfnamefont{A.}~\bibnamefont{Vacheret}},
  \bibinfo{journal}{Astropart. Phys.} \textbf{\bibinfo{volume}{102}},
  \bibinfo{pages}{56} (\bibinfo{year}{2018}), \eprint{1801.01597}.

\bibitem[{\citenamefont{Ara\'ujo}(2020)}]{Araujo:2020rwg}
\bibinfo{author}{\bibfnamefont{H.}~\bibnamefont{Ara\'ujo}}
  (\bibinfo{year}{2020}), \eprint{2007.01683}.

\bibitem[{\citenamefont{Akerib et~al.}(2020{\natexlab{d}})}]{Akerib:2019zrt}
\bibinfo{author}{\bibfnamefont{D.}~\bibnamefont{Akerib}} \bibnamefont{et~al.}
  (\bibinfo{collaboration}{LUX}), \bibinfo{journal}{Phys. Rev. D}
  \textbf{\bibinfo{volume}{101}}, \bibinfo{pages}{042001}
  (\bibinfo{year}{2020}{\natexlab{d}}), \eprint{1907.06272}.

\bibitem[{\citenamefont{Akerib et~al.}(2016{\natexlab{a}})}]{Akerib:2015rjg}
\bibinfo{author}{\bibfnamefont{D.}~\bibnamefont{Akerib}} \bibnamefont{et~al.}
  (\bibinfo{collaboration}{LUX}), \bibinfo{journal}{Phys. Rev. Lett.}
  \textbf{\bibinfo{volume}{116}}, \bibinfo{pages}{161301}
  (\bibinfo{year}{2016}{\natexlab{a}}), \eprint{1512.03506}.

\bibitem[{\citenamefont{Akerib et~al.}(2021{\natexlab{b}})}]{Akerib:2021pfd}
\bibinfo{author}{\bibfnamefont{D.~S.} \bibnamefont{Akerib}}
  \bibnamefont{et~al.} (\bibinfo{year}{2021}{\natexlab{b}}),
  \eprint{2101.08753}.

\bibitem[{\citenamefont{Aprile et~al.}(2016{\natexlab{b}})}]{Aprile:2016wwo}
\bibinfo{author}{\bibfnamefont{E.}~\bibnamefont{Aprile}} \bibnamefont{et~al.}
  (\bibinfo{collaboration}{XENON}), \bibinfo{journal}{Phys. Rev. D}
  \textbf{\bibinfo{volume}{94}}, \bibinfo{pages}{092001}
  (\bibinfo{year}{2016}{\natexlab{b}}), \bibinfo{note}{[Erratum: Phys. Rev. D
  {\bf 95}, 059901 (2017)]}, \eprint{1605.06262}.

\bibitem[{\citenamefont{Aprile et~al.}(2021{\natexlab{c}})}]{XENON:2021myl}
\bibinfo{author}{\bibfnamefont{E.}~\bibnamefont{Aprile}} \bibnamefont{et~al.}
  (\bibinfo{collaboration}{XENON}) (\bibinfo{year}{2021}{\natexlab{c}}),
  \eprint{2112.12116}.

\bibitem[{\citenamefont{Cheng et~al.}(2021)}]{PandaX-II:2021nsg}
\bibinfo{author}{\bibfnamefont{C.}~\bibnamefont{Cheng}} \bibnamefont{et~al.}
  (\bibinfo{collaboration}{PandaX-II}), \bibinfo{journal}{Phys. Rev. Lett.}
  \textbf{\bibinfo{volume}{126}}, \bibinfo{pages}{211803}
  (\bibinfo{year}{2021}), \eprint{2101.07479}.

\bibitem[{\citenamefont{Aprile et~al.}(2014{\natexlab{b}})}]{Aprile:2013blg}
\bibinfo{author}{\bibfnamefont{E.}~\bibnamefont{Aprile}} \bibnamefont{et~al.}
  (\bibinfo{collaboration}{XENON100}), \bibinfo{journal}{J. Phys. G}
  \textbf{\bibinfo{volume}{41}}, \bibinfo{pages}{035201}
  (\bibinfo{year}{2014}{\natexlab{b}}), \eprint{1311.1088}.

\bibitem[{\citenamefont{Akimov et~al.}(2016)}]{Akimov:2016rbs}
\bibinfo{author}{\bibfnamefont{D.}~\bibnamefont{Akimov}} \bibnamefont{et~al.},
  \bibinfo{journal}{JINST} \textbf{\bibinfo{volume}{11}},
  \bibinfo{pages}{C03007} (\bibinfo{year}{2016}).

\bibitem[{\citenamefont{Sorensen and Kamdin}(2018)}]{Sorensen:2017kpl}
\bibinfo{author}{\bibfnamefont{P.}~\bibnamefont{Sorensen}} \bibnamefont{and}
  \bibinfo{author}{\bibfnamefont{K.}~\bibnamefont{Kamdin}},
  \bibinfo{journal}{JINST} \textbf{\bibinfo{volume}{13}},
  \bibinfo{pages}{P02032} (\bibinfo{year}{2018}), \eprint{1711.07025}.

\bibitem[{\citenamefont{Sorensen}(2017)}]{Sorensen:2017ymt}
\bibinfo{author}{\bibfnamefont{P.}~\bibnamefont{Sorensen}}
  (\bibinfo{year}{2017}), \eprint{1702.04805}.

\bibitem[{\citenamefont{Akerib et~al.}(2020{\natexlab{e}})}]{Akerib:2020jud}
\bibinfo{author}{\bibfnamefont{D.}~\bibnamefont{Akerib}} \bibnamefont{et~al.}
  (\bibinfo{collaboration}{LUX}), \bibinfo{journal}{Phys. Rev. D}
  \textbf{\bibinfo{volume}{102}}, \bibinfo{pages}{092004}
  (\bibinfo{year}{2020}{\natexlab{e}}), \eprint{2004.07791}.

\bibitem[{\citenamefont{Kopec et~al.}(2021)\citenamefont{Kopec, Baxter, Clark,
  Lang, Li, Qin, and Singh}}]{Kopec:2021ccm}
\bibinfo{author}{\bibfnamefont{A.}~\bibnamefont{Kopec}},
  \bibinfo{author}{\bibfnamefont{A.~L.} \bibnamefont{Baxter}},
  \bibinfo{author}{\bibfnamefont{M.}~\bibnamefont{Clark}},
  \bibinfo{author}{\bibfnamefont{R.~F.} \bibnamefont{Lang}},
  \bibinfo{author}{\bibfnamefont{S.}~\bibnamefont{Li}},
  \bibinfo{author}{\bibfnamefont{J.}~\bibnamefont{Qin}}, \bibnamefont{and}
  \bibinfo{author}{\bibfnamefont{R.}~\bibnamefont{Singh}},
  \bibinfo{journal}{JINST} \textbf{\bibinfo{volume}{16}},
  \bibinfo{pages}{P07014} (\bibinfo{year}{2021}), \eprint{2103.05077}.

\bibitem[{\citenamefont{Bodnia et~al.}(2021)}]{Bodnia:2021flk}
\bibinfo{author}{\bibfnamefont{E.}~\bibnamefont{Bodnia}} \bibnamefont{et~al.},
  \bibinfo{journal}{JINST} \textbf{\bibinfo{volume}{16}},
  \bibinfo{pages}{P12015} (\bibinfo{year}{2021}), \eprint{2101.03686}.

\bibitem[{\citenamefont{Szydagis et~al.}(2019)\citenamefont{Szydagis, Balajthy,
  Brodsky, Cutter, Huang, Kozlova, Lenardo, Manalaysay, McKinsey, Mooney
  et~al.}}]{szydagis_m_2019_3357973}
\bibinfo{author}{\bibfnamefont{M.}~\bibnamefont{Szydagis}},
  \bibinfo{author}{\bibfnamefont{J.}~\bibnamefont{Balajthy}},
  \bibinfo{author}{\bibfnamefont{J.}~\bibnamefont{Brodsky}},
  \bibinfo{author}{\bibfnamefont{J.}~\bibnamefont{Cutter}},
  \bibinfo{author}{\bibfnamefont{J.}~\bibnamefont{Huang}},
  \bibinfo{author}{\bibfnamefont{E.}~\bibnamefont{Kozlova}},
  \bibinfo{author}{\bibfnamefont{B.}~\bibnamefont{Lenardo}},
  \bibinfo{author}{\bibfnamefont{A.}~\bibnamefont{Manalaysay}},
  \bibinfo{author}{\bibfnamefont{D.}~\bibnamefont{McKinsey}},
  \bibinfo{author}{\bibfnamefont{M.}~\bibnamefont{Mooney}},
  \bibnamefont{et~al.}, \emph{\bibinfo{title}{Nestcollaboration/nest: New,
  flexible lxe nr yields and resolution model + g4 improvements + linear noise
  + much more}} (\bibinfo{year}{2019}),
  \urlprefix\url{https://doi.org/10.5281/zenodo.3357973}.

\bibitem[{\citenamefont{Catena et~al.}(2020{\natexlab{a}})\citenamefont{Catena,
  Emken, Spaldin, and Tarantino}}]{Catena:2019gfa}
\bibinfo{author}{\bibfnamefont{R.}~\bibnamefont{Catena}},
  \bibinfo{author}{\bibfnamefont{T.}~\bibnamefont{Emken}},
  \bibinfo{author}{\bibfnamefont{N.~A.} \bibnamefont{Spaldin}},
  \bibnamefont{and}
  \bibinfo{author}{\bibfnamefont{W.}~\bibnamefont{Tarantino}},
  \bibinfo{journal}{Phys. Rev. Res.} \textbf{\bibinfo{volume}{2}},
  \bibinfo{pages}{033195} (\bibinfo{year}{2020}{\natexlab{a}}),
  \eprint{1912.08204}.

\bibitem[{\citenamefont{Catena et~al.}(2020{\natexlab{b}})\citenamefont{Catena,
  Emken, and Ravanis}}]{Catena:2020tbv}
\bibinfo{author}{\bibfnamefont{R.}~\bibnamefont{Catena}},
  \bibinfo{author}{\bibfnamefont{T.}~\bibnamefont{Emken}}, \bibnamefont{and}
  \bibinfo{author}{\bibfnamefont{J.}~\bibnamefont{Ravanis}},
  \bibinfo{journal}{JCAP} \textbf{\bibinfo{volume}{06}}, \bibinfo{pages}{056}
  (\bibinfo{year}{2020}{\natexlab{b}}), \eprint{2003.04039}.

\bibitem[{\citenamefont{Aprile et~al.}(2019{\natexlab{d}})}]{Aprile:2019jmx}
\bibinfo{author}{\bibfnamefont{E.}~\bibnamefont{Aprile}} \bibnamefont{et~al.}
  (\bibinfo{collaboration}{XENON}), \bibinfo{journal}{Phys. Rev. Lett.}
  \textbf{\bibinfo{volume}{123}}, \bibinfo{pages}{241803}
  (\bibinfo{year}{2019}{\natexlab{d}}), \eprint{1907.12771}.

\bibitem[{\citenamefont{Liu et~al.}(2019)}]{Liu:2019kzq}
\bibinfo{author}{\bibfnamefont{Z.}~\bibnamefont{Liu}} \bibnamefont{et~al.}
  (\bibinfo{collaboration}{CDEX}), \bibinfo{journal}{Phys. Rev. Lett.}
  \textbf{\bibinfo{volume}{123}}, \bibinfo{pages}{161301}
  (\bibinfo{year}{2019}), \eprint{1905.00354}.

\bibitem[{\citenamefont{Migdal}(1941)}]{Migdal:1941}
\bibinfo{author}{\bibfnamefont{A.}~\bibnamefont{Migdal}}, \bibinfo{journal}{J.
  Phys. USSR} \textbf{\bibinfo{volume}{4}}, \bibinfo{pages}{449}
  (\bibinfo{year}{1941}).

\bibitem[{\citenamefont{Ibe et~al.}(2018)\citenamefont{Ibe, Nakano, Shoji, and
  Suzuki}}]{Ibe:2017yqa}
\bibinfo{author}{\bibfnamefont{M.}~\bibnamefont{Ibe}},
  \bibinfo{author}{\bibfnamefont{W.}~\bibnamefont{Nakano}},
  \bibinfo{author}{\bibfnamefont{Y.}~\bibnamefont{Shoji}}, \bibnamefont{and}
  \bibinfo{author}{\bibfnamefont{K.}~\bibnamefont{Suzuki}},
  \bibinfo{journal}{JHEP} \textbf{\bibinfo{volume}{03}}, \bibinfo{pages}{194}
  (\bibinfo{year}{2018}), \eprint{1707.07258}.

\bibitem[{\citenamefont{Dolan et~al.}(2018)\citenamefont{Dolan, Kahlhoefer, and
  McCabe}}]{Dolan:2017xbu}
\bibinfo{author}{\bibfnamefont{M.~J.} \bibnamefont{Dolan}},
  \bibinfo{author}{\bibfnamefont{F.}~\bibnamefont{Kahlhoefer}},
  \bibnamefont{and} \bibinfo{author}{\bibfnamefont{C.}~\bibnamefont{McCabe}},
  \bibinfo{journal}{Phys. Rev. Lett.} \textbf{\bibinfo{volume}{121}},
  \bibinfo{pages}{101801} (\bibinfo{year}{2018}), \eprint{1711.09906}.

\bibitem[{\citenamefont{Wang et~al.}(2021{\natexlab{b}})\citenamefont{Wang, Wu,
  Wu, and Zhu}}]{Wang:2021oha}
\bibinfo{author}{\bibfnamefont{W.}~\bibnamefont{Wang}},
  \bibinfo{author}{\bibfnamefont{K.-Y.} \bibnamefont{Wu}},
  \bibinfo{author}{\bibfnamefont{L.}~\bibnamefont{Wu}}, \bibnamefont{and}
  \bibinfo{author}{\bibfnamefont{B.}~\bibnamefont{Zhu}}
  (\bibinfo{year}{2021}{\natexlab{b}}), \eprint{2112.06492}.

\bibitem[{\citenamefont{Bell et~al.}(2021{\natexlab{a}})\citenamefont{Bell,
  Dent, Lang, Newstead, and Ritter}}]{Bell:2021ihi}
\bibinfo{author}{\bibfnamefont{N.~F.} \bibnamefont{Bell}},
  \bibinfo{author}{\bibfnamefont{J.~B.} \bibnamefont{Dent}},
  \bibinfo{author}{\bibfnamefont{R.~F.} \bibnamefont{Lang}},
  \bibinfo{author}{\bibfnamefont{J.~L.} \bibnamefont{Newstead}},
  \bibnamefont{and} \bibinfo{author}{\bibfnamefont{A.~C.} \bibnamefont{Ritter}}
  (\bibinfo{year}{2021}{\natexlab{a}}), \eprint{2112.08514}.

\bibitem[{\citenamefont{Bell et~al.}(2020{\natexlab{a}})\citenamefont{Bell,
  Dent, Newstead, Sabharwal, and Weiler}}]{Bell:2019egg}
\bibinfo{author}{\bibfnamefont{N.~F.} \bibnamefont{Bell}},
  \bibinfo{author}{\bibfnamefont{J.~B.} \bibnamefont{Dent}},
  \bibinfo{author}{\bibfnamefont{J.~L.} \bibnamefont{Newstead}},
  \bibinfo{author}{\bibfnamefont{S.}~\bibnamefont{Sabharwal}},
  \bibnamefont{and} \bibinfo{author}{\bibfnamefont{T.~J.}
  \bibnamefont{Weiler}}, \bibinfo{journal}{Phys. Rev. D}
  \textbf{\bibinfo{volume}{101}}, \bibinfo{pages}{015012}
  (\bibinfo{year}{2020}{\natexlab{a}}), \eprint{1905.00046}.

\bibitem[{\citenamefont{McCabe}(2017)}]{McCabe:2017rln}
\bibinfo{author}{\bibfnamefont{C.}~\bibnamefont{McCabe}},
  \bibinfo{journal}{Phys. Rev. D} \textbf{\bibinfo{volume}{96}},
  \bibinfo{pages}{043010} (\bibinfo{year}{2017}), \eprint{1702.04730}.

\bibitem[{\citenamefont{Akerib et~al.}(2020{\natexlab{f}})}]{Akerib:2021hydrox}
\bibinfo{author}{\bibfnamefont{D.}~\bibnamefont{Akerib}} \bibnamefont{et~al.},
  \emph{\bibinfo{title}{{HydroX - Using hydrogen doped in liquid xenon to
  search for dark matter}}} (\bibinfo{year}{2020}{\natexlab{f}}),
  \urlprefix\url{https://www.snowmass21.org/docs/files/summaries/CF/SNOWMASS21-CF1_CF0_Hugh_Lippincott-106.pdf}.

\bibitem[{\citenamefont{Beacom et~al.}(2002)\citenamefont{Beacom, Farr, and
  Vogel}}]{Beacom:2002hs}
\bibinfo{author}{\bibfnamefont{J.~F.} \bibnamefont{Beacom}},
  \bibinfo{author}{\bibfnamefont{W.~M.} \bibnamefont{Farr}}, \bibnamefont{and}
  \bibinfo{author}{\bibfnamefont{P.}~\bibnamefont{Vogel}},
  \bibinfo{journal}{Phys. Rev. D} \textbf{\bibinfo{volume}{66}},
  \bibinfo{pages}{033001} (\bibinfo{year}{2002}), \eprint{hep-ph/0205220}.

\bibitem[{\citenamefont{Bolozdynya}(1991)}]{Bolozdynya:1991}
\bibinfo{author}{\bibfnamefont{A.~I.} \bibnamefont{Bolozdynya}}, in
  \emph{\bibinfo{booktitle}{{Proceedings of the 3rd International Conference on
  Properties and Applications of Dielectric Materials: Tokyo, Japan, July 8-12,
  1991}}} (\bibinfo{year}{1991}), \bibinfo{number}{2}, pp.
  \bibinfo{pages}{841--844}.

\bibitem[{\citenamefont{Tezuka et~al.}(2004)}]{Tezuka:2004gza}
\bibinfo{author}{\bibfnamefont{C.}~\bibnamefont{Tezuka}} \bibnamefont{et~al.},
  in \emph{\bibinfo{booktitle}{{Proceedings, 2004 IEEE Nuclear Science
  Symposium and Medical Imaging Conference (NSS/MIC 2004): Rome, Italy, October
  16-22, 2004}}} (\bibinfo{year}{2004}), \bibinfo{number}{2}, pp.
  \bibinfo{pages}{1157--1159}.

\bibitem[{\citenamefont{Bringmann and Pospelov}(2019)}]{Bringmann:2018cvk}
\bibinfo{author}{\bibfnamefont{T.}~\bibnamefont{Bringmann}} \bibnamefont{and}
  \bibinfo{author}{\bibfnamefont{M.}~\bibnamefont{Pospelov}},
  \bibinfo{journal}{Phys. Rev. Lett.} \textbf{\bibinfo{volume}{122}},
  \bibinfo{pages}{171801} (\bibinfo{year}{2019}), \eprint{1810.10543}.

\bibitem[{\citenamefont{Alvey et~al.}(2019)\citenamefont{Alvey, Campos,
  Fairbairn, and You}}]{Alvey:2019zaa}
\bibinfo{author}{\bibfnamefont{J.}~\bibnamefont{Alvey}},
  \bibinfo{author}{\bibfnamefont{M.}~\bibnamefont{Campos}},
  \bibinfo{author}{\bibfnamefont{M.}~\bibnamefont{Fairbairn}},
  \bibnamefont{and} \bibinfo{author}{\bibfnamefont{T.}~\bibnamefont{You}},
  \bibinfo{journal}{Phys. Rev. Lett.} \textbf{\bibinfo{volume}{123}},
  \bibinfo{pages}{261802} (\bibinfo{year}{2019}), \eprint{1905.05776}.

\bibitem[{\citenamefont{Cappiello and Beacom}(2019)}]{Cappiello:2019qsw}
\bibinfo{author}{\bibfnamefont{C.}~\bibnamefont{Cappiello}} \bibnamefont{and}
  \bibinfo{author}{\bibfnamefont{J.~F.} \bibnamefont{Beacom}},
  \bibinfo{journal}{Phys. Rev. D} \textbf{\bibinfo{volume}{100}},
  \bibinfo{pages}{103011} (\bibinfo{year}{2019}), \eprint{1906.11283}.

\bibitem[{\citenamefont{Bondarenko et~al.}(2020)\citenamefont{Bondarenko,
  Boyarsky, Bringmann, Hufnagel, Schmidt-Hoberg, and
  Sokolenko}}]{Bondarenko:2019vrb}
\bibinfo{author}{\bibfnamefont{K.}~\bibnamefont{Bondarenko}},
  \bibinfo{author}{\bibfnamefont{A.}~\bibnamefont{Boyarsky}},
  \bibinfo{author}{\bibfnamefont{T.}~\bibnamefont{Bringmann}},
  \bibinfo{author}{\bibfnamefont{M.}~\bibnamefont{Hufnagel}},
  \bibinfo{author}{\bibfnamefont{K.}~\bibnamefont{Schmidt-Hoberg}},
  \bibnamefont{and}
  \bibinfo{author}{\bibfnamefont{A.}~\bibnamefont{Sokolenko}},
  \bibinfo{journal}{JHEP} \textbf{\bibinfo{volume}{03}}, \bibinfo{pages}{118}
  (\bibinfo{year}{2020}), \eprint{1909.08632}.

\bibitem[{\citenamefont{Wang et~al.}(2020{\natexlab{b}})\citenamefont{Wang, Wu,
  Yang, Zhou, and Zhu}}]{Wang:2019jtk}
\bibinfo{author}{\bibfnamefont{W.}~\bibnamefont{Wang}},
  \bibinfo{author}{\bibfnamefont{L.}~\bibnamefont{Wu}},
  \bibinfo{author}{\bibfnamefont{J.~M.} \bibnamefont{Yang}},
  \bibinfo{author}{\bibfnamefont{H.}~\bibnamefont{Zhou}}, \bibnamefont{and}
  \bibinfo{author}{\bibfnamefont{B.}~\bibnamefont{Zhu}},
  \bibinfo{journal}{JHEP} \textbf{\bibinfo{volume}{12}}, \bibinfo{pages}{072}
  (\bibinfo{year}{2020}{\natexlab{b}}), \bibinfo{note}{[Erratum: JHEP {\bf 02},
  052 (2021)]}, \eprint{1912.09904}.

\bibitem[{\citenamefont{Dent et~al.}(2021)\citenamefont{Dent, Dutta, Newstead,
  Shoemaker, and Arellano}}]{Dent:2020syp}
\bibinfo{author}{\bibfnamefont{J.~B.} \bibnamefont{Dent}},
  \bibinfo{author}{\bibfnamefont{B.}~\bibnamefont{Dutta}},
  \bibinfo{author}{\bibfnamefont{J.~L.} \bibnamefont{Newstead}},
  \bibinfo{author}{\bibfnamefont{I.~M.} \bibnamefont{Shoemaker}},
  \bibnamefont{and} \bibinfo{author}{\bibfnamefont{N.~T.}
  \bibnamefont{Arellano}}, \bibinfo{journal}{Phys. Rev. D}
  \textbf{\bibinfo{volume}{103}}, \bibinfo{pages}{095015}
  (\bibinfo{year}{2021}), \eprint{2010.09749}.

\bibitem[{\citenamefont{Bell et~al.}(2021{\natexlab{b}})\citenamefont{Bell,
  Dent, Dutta, Ghosh, Kumar, Newstead, and Shoemaker}}]{Bell:2021xff}
\bibinfo{author}{\bibfnamefont{N.~F.} \bibnamefont{Bell}},
  \bibinfo{author}{\bibfnamefont{J.~B.} \bibnamefont{Dent}},
  \bibinfo{author}{\bibfnamefont{B.}~\bibnamefont{Dutta}},
  \bibinfo{author}{\bibfnamefont{S.}~\bibnamefont{Ghosh}},
  \bibinfo{author}{\bibfnamefont{J.}~\bibnamefont{Kumar}},
  \bibinfo{author}{\bibfnamefont{J.~L.} \bibnamefont{Newstead}},
  \bibnamefont{and} \bibinfo{author}{\bibfnamefont{I.~M.}
  \bibnamefont{Shoemaker}}, \bibinfo{journal}{Phys. Rev. D}
  \textbf{\bibinfo{volume}{104}}, \bibinfo{pages}{076020}
  (\bibinfo{year}{2021}{\natexlab{b}}), \eprint{2108.00583}.

\bibitem[{\citenamefont{An et~al.}(2018)\citenamefont{An, Pospelov, Pradler,
  and Ritz}}]{An:2017ojc}
\bibinfo{author}{\bibfnamefont{H.}~\bibnamefont{An}},
  \bibinfo{author}{\bibfnamefont{M.}~\bibnamefont{Pospelov}},
  \bibinfo{author}{\bibfnamefont{J.}~\bibnamefont{Pradler}}, \bibnamefont{and}
  \bibinfo{author}{\bibfnamefont{A.}~\bibnamefont{Ritz}},
  \bibinfo{journal}{Phys. Rev. Lett.} \textbf{\bibinfo{volume}{120}},
  \bibinfo{pages}{141801} (\bibinfo{year}{2018}), \bibinfo{note}{[Erratum:
  Phys. Rev. Lett. {\bf 121}, 259903 (2018)]}, \eprint{1708.03642}.

\bibitem[{\citenamefont{Emken et~al.}(2018)\citenamefont{Emken, Kouvaris, and
  Nielsen}}]{Emken:2017hnp}
\bibinfo{author}{\bibfnamefont{T.}~\bibnamefont{Emken}},
  \bibinfo{author}{\bibfnamefont{C.}~\bibnamefont{Kouvaris}}, \bibnamefont{and}
  \bibinfo{author}{\bibfnamefont{N.~G.} \bibnamefont{Nielsen}},
  \bibinfo{journal}{Phys. Rev. D} \textbf{\bibinfo{volume}{97}},
  \bibinfo{pages}{063007} (\bibinfo{year}{2018}), \eprint{1709.06573}.

\bibitem[{\citenamefont{Liang et~al.}(2021{\natexlab{b}})\citenamefont{Liang,
  Mo, and Zhang}}]{Liang:2021zkg}
\bibinfo{author}{\bibfnamefont{Z.-L.} \bibnamefont{Liang}},
  \bibinfo{author}{\bibfnamefont{C.}~\bibnamefont{Mo}}, \bibnamefont{and}
  \bibinfo{author}{\bibfnamefont{P.}~\bibnamefont{Zhang}},
  \bibinfo{journal}{Phys. Rev. D} \textbf{\bibinfo{volume}{104}},
  \bibinfo{pages}{096001} (\bibinfo{year}{2021}{\natexlab{b}}),
  \eprint{2107.01209}.

\bibitem[{\citenamefont{An et~al.}(2021)\citenamefont{An, Nie, Pospelov,
  Pradler, and Ritz}}]{An:2021qdl}
\bibinfo{author}{\bibfnamefont{H.}~\bibnamefont{An}},
  \bibinfo{author}{\bibfnamefont{H.}~\bibnamefont{Nie}},
  \bibinfo{author}{\bibfnamefont{M.}~\bibnamefont{Pospelov}},
  \bibinfo{author}{\bibfnamefont{J.}~\bibnamefont{Pradler}}, \bibnamefont{and}
  \bibinfo{author}{\bibfnamefont{A.}~\bibnamefont{Ritz}},
  \bibinfo{journal}{Phys. Rev. D} \textbf{\bibinfo{volume}{104}},
  \bibinfo{pages}{103026} (\bibinfo{year}{2021}), \eprint{2108.10332}.

\bibitem[{\citenamefont{Emken}(2021)}]{Emken:2021lgc}
\bibinfo{author}{\bibfnamefont{T.}~\bibnamefont{Emken}} (\bibinfo{year}{2021}),
  \eprint{2102.12483}.

\bibitem[{\citenamefont{Cherry et~al.}(2014)\citenamefont{Cherry, Friedland,
  and Shoemaker}}]{Cherry:2014xra}
\bibinfo{author}{\bibfnamefont{J.~F.} \bibnamefont{Cherry}},
  \bibinfo{author}{\bibfnamefont{A.}~\bibnamefont{Friedland}},
  \bibnamefont{and} \bibinfo{author}{\bibfnamefont{I.~M.}
  \bibnamefont{Shoemaker}} (\bibinfo{year}{2014}), \eprint{1411.1071}.

\bibitem[{\citenamefont{Gonz\'alez~Mac\'\i{}as
  et~al.}(2015)\citenamefont{Gonz\'alez~Mac\'\i{}as, Wudka, and
  Illana}}]{GonzalezMacias:2015xbi}
\bibinfo{author}{\bibfnamefont{V.}~\bibnamefont{Gonz\'alez~Mac\'\i{}as}},
  \bibinfo{author}{\bibfnamefont{J.}~\bibnamefont{Wudka}}, \bibnamefont{and}
  \bibinfo{author}{\bibfnamefont{J.~I.} \bibnamefont{Illana}},
  \bibinfo{journal}{Acta Phys. Polon. B} \textbf{\bibinfo{volume}{46}},
  \bibinfo{pages}{2173} (\bibinfo{year}{2015}).

\bibitem[{\citenamefont{Becker}(2019)}]{Becker:2018rve}
\bibinfo{author}{\bibfnamefont{M.}~\bibnamefont{Becker}},
  \bibinfo{journal}{Eur. Phys. J. C} \textbf{\bibinfo{volume}{79}},
  \bibinfo{pages}{611} (\bibinfo{year}{2019}), \eprint{1806.08579}.

\bibitem[{\citenamefont{Lamprea et~al.}(2021)\citenamefont{Lamprea, Peinado,
  Smolenski, and Wudka}}]{Lamprea:2019qet}
\bibinfo{author}{\bibfnamefont{J.~M.} \bibnamefont{Lamprea}},
  \bibinfo{author}{\bibfnamefont{E.}~\bibnamefont{Peinado}},
  \bibinfo{author}{\bibfnamefont{S.}~\bibnamefont{Smolenski}},
  \bibnamefont{and} \bibinfo{author}{\bibfnamefont{J.}~\bibnamefont{Wudka}},
  \bibinfo{journal}{Phys. Rev. D} \textbf{\bibinfo{volume}{103}},
  \bibinfo{pages}{015017} (\bibinfo{year}{2021}), \eprint{1906.02340}.

\bibitem[{\citenamefont{Patel et~al.}(2020)\citenamefont{Patel, Profumo, and
  Shakya}}]{Patel:2019zky}
\bibinfo{author}{\bibfnamefont{H.~H.} \bibnamefont{Patel}},
  \bibinfo{author}{\bibfnamefont{S.}~\bibnamefont{Profumo}}, \bibnamefont{and}
  \bibinfo{author}{\bibfnamefont{B.}~\bibnamefont{Shakya}},
  \bibinfo{journal}{Phys. Rev. D} \textbf{\bibinfo{volume}{101}},
  \bibinfo{pages}{095001} (\bibinfo{year}{2020}), \eprint{1912.05581}.

\bibitem[{\citenamefont{McKeen and Raj}(2019)}]{McKeen:2018pbb}
\bibinfo{author}{\bibfnamefont{D.}~\bibnamefont{McKeen}} \bibnamefont{and}
  \bibinfo{author}{\bibfnamefont{N.}~\bibnamefont{Raj}},
  \bibinfo{journal}{Phys. Rev. D} \textbf{\bibinfo{volume}{99}},
  \bibinfo{pages}{103003} (\bibinfo{year}{2019}), \eprint{1812.05102}.

\bibitem[{\citenamefont{Marrod\'an~Undagoitia
  et~al.}(2021)\citenamefont{Marrod\'an~Undagoitia, Rodejohann, Wolf, and
  Yaguna}}]{Undagoitia:2021tza}
\bibinfo{author}{\bibfnamefont{T.}~\bibnamefont{Marrod\'an~Undagoitia}},
  \bibinfo{author}{\bibfnamefont{W.}~\bibnamefont{Rodejohann}},
  \bibinfo{author}{\bibfnamefont{T.}~\bibnamefont{Wolf}}, \bibnamefont{and}
  \bibinfo{author}{\bibfnamefont{C.~E.} \bibnamefont{Yaguna}}
  (\bibinfo{year}{2021}), \eprint{2107.05685}.

\bibitem[{\citenamefont{Slatyer}(2016)}]{Slatyer:2015jla}
\bibinfo{author}{\bibfnamefont{T.~R.} \bibnamefont{Slatyer}},
  \bibinfo{journal}{Phys. Rev. D} \textbf{\bibinfo{volume}{93}},
  \bibinfo{pages}{023527} (\bibinfo{year}{2016}), \eprint{1506.03811}.

\bibitem[{\citenamefont{Ng et~al.}(2019)\citenamefont{Ng, Roach, Perez, Beacom,
  Horiuchi, Krivonos, and Wik}}]{Ng:2019gch}
\bibinfo{author}{\bibfnamefont{K.~C.} \bibnamefont{Ng}},
  \bibinfo{author}{\bibfnamefont{B.~M.} \bibnamefont{Roach}},
  \bibinfo{author}{\bibfnamefont{K.}~\bibnamefont{Perez}},
  \bibinfo{author}{\bibfnamefont{J.~F.} \bibnamefont{Beacom}},
  \bibinfo{author}{\bibfnamefont{S.}~\bibnamefont{Horiuchi}},
  \bibinfo{author}{\bibfnamefont{R.}~\bibnamefont{Krivonos}}, \bibnamefont{and}
  \bibinfo{author}{\bibfnamefont{D.~R.} \bibnamefont{Wik}},
  \bibinfo{journal}{Phys. Rev. D} \textbf{\bibinfo{volume}{99}},
  \bibinfo{pages}{083005} (\bibinfo{year}{2019}), \eprint{1901.01262}.

\bibitem[{\citenamefont{McDonald}(2002)}]{McDonald:2001vt}
\bibinfo{author}{\bibfnamefont{J.}~\bibnamefont{McDonald}},
  \bibinfo{journal}{Phys. Rev. Lett.} \textbf{\bibinfo{volume}{88}},
  \bibinfo{pages}{091304} (\bibinfo{year}{2002}), \eprint{hep-ph/0106249}.

\bibitem[{\citenamefont{Hall et~al.}(2010{\natexlab{a}})\citenamefont{Hall,
  Jedamzik, March-Russell, and West}}]{Hall:2009bx}
\bibinfo{author}{\bibfnamefont{L.~J.} \bibnamefont{Hall}},
  \bibinfo{author}{\bibfnamefont{K.}~\bibnamefont{Jedamzik}},
  \bibinfo{author}{\bibfnamefont{J.}~\bibnamefont{March-Russell}},
  \bibnamefont{and} \bibinfo{author}{\bibfnamefont{S.~M.} \bibnamefont{West}},
  \bibinfo{journal}{JHEP} \textbf{\bibinfo{volume}{03}}, \bibinfo{pages}{080}
  (\bibinfo{year}{2010}{\natexlab{a}}), \eprint{0911.1120}.

\bibitem[{\citenamefont{Feng et~al.}(2003)\citenamefont{Feng, Rajaraman, and
  Takayama}}]{Feng:2003xh}
\bibinfo{author}{\bibfnamefont{J.~L.} \bibnamefont{Feng}},
  \bibinfo{author}{\bibfnamefont{A.}~\bibnamefont{Rajaraman}},
  \bibnamefont{and} \bibinfo{author}{\bibfnamefont{F.}~\bibnamefont{Takayama}},
  \bibinfo{journal}{Phys. Rev. Lett.} \textbf{\bibinfo{volume}{91}},
  \bibinfo{pages}{011302} (\bibinfo{year}{2003}), \eprint{hep-ph/0302215}.

\bibitem[{\citenamefont{Dodelson and Widrow}(1994)}]{Dodelson:1993je}
\bibinfo{author}{\bibfnamefont{S.}~\bibnamefont{Dodelson}} \bibnamefont{and}
  \bibinfo{author}{\bibfnamefont{L.~M.} \bibnamefont{Widrow}},
  \bibinfo{journal}{Phys. Rev. Lett.} \textbf{\bibinfo{volume}{72}},
  \bibinfo{pages}{17} (\bibinfo{year}{1994}), \eprint{hep-ph/9303287}.

\bibitem[{\citenamefont{Pospelov et~al.}(2008)\citenamefont{Pospelov, Ritz, and
  Voloshin}}]{Pospelov:2008jk}
\bibinfo{author}{\bibfnamefont{M.}~\bibnamefont{Pospelov}},
  \bibinfo{author}{\bibfnamefont{A.}~\bibnamefont{Ritz}}, \bibnamefont{and}
  \bibinfo{author}{\bibfnamefont{M.~B.} \bibnamefont{Voloshin}},
  \bibinfo{journal}{Phys. Rev. D} \textbf{\bibinfo{volume}{78}},
  \bibinfo{pages}{115012} (\bibinfo{year}{2008}), \eprint{0807.3279}.

\bibitem[{\citenamefont{Hambye et~al.}(2018)\citenamefont{Hambye, Tytgat,
  Vandecasteele, and Vanderheyden}}]{Hambye:2018dpi}
\bibinfo{author}{\bibfnamefont{T.}~\bibnamefont{Hambye}},
  \bibinfo{author}{\bibfnamefont{M.~H.~G.} \bibnamefont{Tytgat}},
  \bibinfo{author}{\bibfnamefont{J.}~\bibnamefont{Vandecasteele}},
  \bibnamefont{and}
  \bibinfo{author}{\bibfnamefont{L.}~\bibnamefont{Vanderheyden}},
  \bibinfo{journal}{Phys. Rev. D} \textbf{\bibinfo{volume}{98}},
  \bibinfo{pages}{075017} (\bibinfo{year}{2018}), \eprint{1807.05022}.

\bibitem[{\citenamefont{B\'elanger et~al.}(2020)\citenamefont{B\'elanger,
  Delaunay, Pukhov, and Zaldivar}}]{Belanger:2020npe}
\bibinfo{author}{\bibfnamefont{G.}~\bibnamefont{B\'elanger}},
  \bibinfo{author}{\bibfnamefont{C.}~\bibnamefont{Delaunay}},
  \bibinfo{author}{\bibfnamefont{A.}~\bibnamefont{Pukhov}}, \bibnamefont{and}
  \bibinfo{author}{\bibfnamefont{B.}~\bibnamefont{Zaldivar}},
  \bibinfo{journal}{Phys. Rev. D} \textbf{\bibinfo{volume}{102}},
  \bibinfo{pages}{035017} (\bibinfo{year}{2020}), \eprint{2005.06294}.

\bibitem[{\citenamefont{Elor et~al.}(2021)\citenamefont{Elor, McGehee, and
  Pierce}}]{Elor:2021swj}
\bibinfo{author}{\bibfnamefont{G.}~\bibnamefont{Elor}},
  \bibinfo{author}{\bibfnamefont{R.}~\bibnamefont{McGehee}}, \bibnamefont{and}
  \bibinfo{author}{\bibfnamefont{A.}~\bibnamefont{Pierce}}
  (\bibinfo{year}{2021}), \eprint{2112.03920}.

\bibitem[{\citenamefont{Graham et~al.}(2016)\citenamefont{Graham, Mardon, and
  Rajendran}}]{Graham:2015rva}
\bibinfo{author}{\bibfnamefont{P.~W.} \bibnamefont{Graham}},
  \bibinfo{author}{\bibfnamefont{J.}~\bibnamefont{Mardon}}, \bibnamefont{and}
  \bibinfo{author}{\bibfnamefont{S.}~\bibnamefont{Rajendran}},
  \bibinfo{journal}{Phys. Rev. D} \textbf{\bibinfo{volume}{93}},
  \bibinfo{pages}{103520} (\bibinfo{year}{2016}), \eprint{1504.02102}.

\bibitem[{\citenamefont{Holdom}(1986)}]{Holdom:1985ag}
\bibinfo{author}{\bibfnamefont{B.}~\bibnamefont{Holdom}},
  \bibinfo{journal}{Phys. Lett.} \textbf{\bibinfo{volume}{166B}},
  \bibinfo{pages}{196} (\bibinfo{year}{1986}).

\bibitem[{\citenamefont{Babu et~al.}(1996)\citenamefont{Babu, Kolda, and
  March-Russell}}]{Babu:1996vt}
\bibinfo{author}{\bibfnamefont{K.~S.} \bibnamefont{Babu}},
  \bibinfo{author}{\bibfnamefont{C.~F.} \bibnamefont{Kolda}}, \bibnamefont{and}
  \bibinfo{author}{\bibfnamefont{J.}~\bibnamefont{March-Russell}},
  \bibinfo{journal}{Phys. Rev. D} \textbf{\bibinfo{volume}{54}},
  \bibinfo{pages}{4635} (\bibinfo{year}{1996}), \eprint{hep-ph/9603212}.

\bibitem[{\citenamefont{Babu et~al.}(1998)\citenamefont{Babu, Kolda, and
  March-Russell}}]{Babu:1997st}
\bibinfo{author}{\bibfnamefont{K.~S.} \bibnamefont{Babu}},
  \bibinfo{author}{\bibfnamefont{C.~F.} \bibnamefont{Kolda}}, \bibnamefont{and}
  \bibinfo{author}{\bibfnamefont{J.}~\bibnamefont{March-Russell}},
  \bibinfo{journal}{Phys. Rev. D} \textbf{\bibinfo{volume}{57}},
  \bibinfo{pages}{6788} (\bibinfo{year}{1998}), \eprint{hep-ph/9710441}.

\bibitem[{\citenamefont{An et~al.}(2015)\citenamefont{An, Pospelov, Pradler,
  and Ritz}}]{An:2014twa}
\bibinfo{author}{\bibfnamefont{H.}~\bibnamefont{An}},
  \bibinfo{author}{\bibfnamefont{M.}~\bibnamefont{Pospelov}},
  \bibinfo{author}{\bibfnamefont{J.}~\bibnamefont{Pradler}}, \bibnamefont{and}
  \bibinfo{author}{\bibfnamefont{A.}~\bibnamefont{Ritz}},
  \bibinfo{journal}{Phys. Lett. B} \textbf{\bibinfo{volume}{747}},
  \bibinfo{pages}{331} (\bibinfo{year}{2015}), \eprint{1412.8378}.

\bibitem[{\citenamefont{Akerib et~al.}(2021{\natexlab{c}})}]{LZ:2021xov}
\bibinfo{author}{\bibfnamefont{D.~S.} \bibnamefont{Akerib}}
  \bibnamefont{et~al.} (\bibinfo{collaboration}{LZ}), \bibinfo{journal}{Phys.
  Rev. D} \textbf{\bibinfo{volume}{104}}, \bibinfo{pages}{092009}
  (\bibinfo{year}{2021}{\natexlab{c}}), \eprint{2102.11740}.

\bibitem[{\citenamefont{An et~al.}(2020)\citenamefont{An, Pospelov, Pradler,
  and Ritz}}]{An:2020bxd}
\bibinfo{author}{\bibfnamefont{H.}~\bibnamefont{An}},
  \bibinfo{author}{\bibfnamefont{M.}~\bibnamefont{Pospelov}},
  \bibinfo{author}{\bibfnamefont{J.}~\bibnamefont{Pradler}}, \bibnamefont{and}
  \bibinfo{author}{\bibfnamefont{A.}~\bibnamefont{Ritz}},
  \bibinfo{journal}{Phys. Rev. D} \textbf{\bibinfo{volume}{102}},
  \bibinfo{pages}{115022} (\bibinfo{year}{2020}), \eprint{2006.13929}.

\bibitem[{\citenamefont{Peccei and Quinn}(1977)}]{PecceiQuinn_1977}
\bibinfo{author}{\bibfnamefont{R.~D.} \bibnamefont{Peccei}} \bibnamefont{and}
  \bibinfo{author}{\bibfnamefont{H.~R.} \bibnamefont{Quinn}},
  \bibinfo{journal}{Phys. Rev. Lett.} \textbf{\bibinfo{volume}{38}},
  \bibinfo{pages}{1440} (\bibinfo{year}{1977}),
  \urlprefix\url{https://link.aps.org/doi/10.1103/PhysRevLett.38.1440}.

\bibitem[{\citenamefont{Weinberg}(1978)}]{Weinberg:1977ma}
\bibinfo{author}{\bibfnamefont{S.}~\bibnamefont{Weinberg}},
  \bibinfo{journal}{Phys. Rev. Lett.} \textbf{\bibinfo{volume}{40}},
  \bibinfo{pages}{223} (\bibinfo{year}{1978}).

\bibitem[{\citenamefont{Wilczek}(1978)}]{Wilczek:1977pj}
\bibinfo{author}{\bibfnamefont{F.}~\bibnamefont{Wilczek}},
  \bibinfo{journal}{Phys. Rev. Lett.} \textbf{\bibinfo{volume}{40}},
  \bibinfo{pages}{279} (\bibinfo{year}{1978}).

\bibitem[{\citenamefont{Preskill et~al.}(1983)\citenamefont{Preskill, Wise, and
  Wilczek}}]{Preskill:1982cy}
\bibinfo{author}{\bibfnamefont{J.}~\bibnamefont{Preskill}},
  \bibinfo{author}{\bibfnamefont{M.~B.} \bibnamefont{Wise}}, \bibnamefont{and}
  \bibinfo{author}{\bibfnamefont{F.}~\bibnamefont{Wilczek}},
  \bibinfo{journal}{Phys. Lett. B} \textbf{\bibinfo{volume}{120}},
  \bibinfo{pages}{127} (\bibinfo{year}{1983}).

\bibitem[{\citenamefont{Abbott and Sikivie}(1983)}]{Abbott:1982af}
\bibinfo{author}{\bibfnamefont{L.}~\bibnamefont{Abbott}} \bibnamefont{and}
  \bibinfo{author}{\bibfnamefont{P.}~\bibnamefont{Sikivie}},
  \bibinfo{journal}{Phys. Lett. B} \textbf{\bibinfo{volume}{120}},
  \bibinfo{pages}{133} (\bibinfo{year}{1983}).

\bibitem[{\citenamefont{Dine and Fischler}(1983)}]{Dine:1982ah}
\bibinfo{author}{\bibfnamefont{M.}~\bibnamefont{Dine}} \bibnamefont{and}
  \bibinfo{author}{\bibfnamefont{W.}~\bibnamefont{Fischler}},
  \bibinfo{journal}{Phys. Lett. B} \textbf{\bibinfo{volume}{120}},
  \bibinfo{pages}{137} (\bibinfo{year}{1983}).

\bibitem[{\citenamefont{Krauss}(1985)}]{Krauss:1985ww}
\bibinfo{author}{\bibfnamefont{L.~M.} \bibnamefont{Krauss}}, in
  \emph{\bibinfo{booktitle}{{Theoretical Advanced Study Institute in Elementary
  Particle Physics}}} (\bibinfo{year}{1985}).

\bibitem[{\citenamefont{Irastorza}(2021)}]{Irastorza:2021tdu}
\bibinfo{author}{\bibfnamefont{I.~G.} \bibnamefont{Irastorza}}, in
  \emph{\bibinfo{booktitle}{{Les Houches summer school on Dark Matter}}}
  (\bibinfo{year}{2021}), \eprint{2109.07376}.

\bibitem[{\citenamefont{Grilli~di Cortona et~al.}(2016)\citenamefont{Grilli~di
  Cortona, Hardy, Pardo~Vega, and Villadoro}}]{GrillidiCortona:2015jxo}
\bibinfo{author}{\bibfnamefont{G.}~\bibnamefont{Grilli~di Cortona}},
  \bibinfo{author}{\bibfnamefont{E.}~\bibnamefont{Hardy}},
  \bibinfo{author}{\bibfnamefont{J.}~\bibnamefont{Pardo~Vega}},
  \bibnamefont{and}
  \bibinfo{author}{\bibfnamefont{G.}~\bibnamefont{Villadoro}},
  \bibinfo{journal}{JHEP} \textbf{\bibinfo{volume}{01}}, \bibinfo{pages}{034}
  (\bibinfo{year}{2016}), \eprint{1511.02867}.

\bibitem[{\citenamefont{Zyla et~al.}(2020)}]{Zyla:2020zbs}
\bibinfo{author}{\bibfnamefont{P.}~\bibnamefont{Zyla}} \bibnamefont{et~al.}
  (\bibinfo{collaboration}{Particle Data Group}), \bibinfo{journal}{PTEP}
  \textbf{\bibinfo{volume}{2020}}, \bibinfo{pages}{083C01}
  (\bibinfo{year}{2020}).

\bibitem[{\citenamefont{Raffelt}(2008)}]{Raffelt:2006cw}
\bibinfo{author}{\bibfnamefont{G.~G.} \bibnamefont{Raffelt}},
  \bibinfo{journal}{Lect. Notes Phys.} \textbf{\bibinfo{volume}{741}},
  \bibinfo{pages}{51} (\bibinfo{year}{2008}), \eprint{hep-ph/0611350}.

\bibitem[{\citenamefont{Sikivie}(2008)}]{Sikivie:2006ni}
\bibinfo{author}{\bibfnamefont{P.}~\bibnamefont{Sikivie}},
  \bibinfo{journal}{Lect. Notes Phys.} \textbf{\bibinfo{volume}{741}},
  \bibinfo{pages}{19} (\bibinfo{year}{2008}), \eprint{astro-ph/0610440}.

\bibitem[{\citenamefont{Ayala et~al.}(2014)\citenamefont{Ayala, Domínguez,
  Giannotti, Mirizzi, and Straniero}}]{Ayala:2014pea}
\bibinfo{author}{\bibfnamefont{A.}~\bibnamefont{Ayala}},
  \bibinfo{author}{\bibfnamefont{I.}~\bibnamefont{Domínguez}},
  \bibinfo{author}{\bibfnamefont{M.}~\bibnamefont{Giannotti}},
  \bibinfo{author}{\bibfnamefont{A.}~\bibnamefont{Mirizzi}}, \bibnamefont{and}
  \bibinfo{author}{\bibfnamefont{O.}~\bibnamefont{Straniero}},
  \bibinfo{journal}{Phys. Rev. Lett.} \textbf{\bibinfo{volume}{113}},
  \bibinfo{pages}{191302} (\bibinfo{year}{2014}), \eprint{1406.6053}.

\bibitem[{\citenamefont{Chang et~al.}(2018)\citenamefont{Chang, Essig, and
  McDermott}}]{Chang:2018rso}
\bibinfo{author}{\bibfnamefont{J.~H.} \bibnamefont{Chang}},
  \bibinfo{author}{\bibfnamefont{R.}~\bibnamefont{Essig}}, \bibnamefont{and}
  \bibinfo{author}{\bibfnamefont{S.~D.} \bibnamefont{McDermott}},
  \bibinfo{journal}{JHEP} \textbf{\bibinfo{volume}{09}}, \bibinfo{pages}{051}
  (\bibinfo{year}{2018}), \eprint{1803.00993}.

\bibitem[{\citenamefont{Capozzi and Raffelt}(2020)}]{Capozzi:2020cbu}
\bibinfo{author}{\bibfnamefont{F.}~\bibnamefont{Capozzi}} \bibnamefont{and}
  \bibinfo{author}{\bibfnamefont{G.}~\bibnamefont{Raffelt}},
  \bibinfo{journal}{Phys. Rev. D} \textbf{\bibinfo{volume}{102}},
  \bibinfo{pages}{083007} (\bibinfo{year}{2020}), \eprint{2007.03694}.

\bibitem[{\citenamefont{Anastassopoulos
  et~al.}(2017)}]{Anastassopoulos:2017ftl}
\bibinfo{author}{\bibfnamefont{V.}~\bibnamefont{Anastassopoulos}}
  \bibnamefont{et~al.} (\bibinfo{collaboration}{CAST}),
  \bibinfo{journal}{Nature Phys.} \textbf{\bibinfo{volume}{13}},
  \bibinfo{pages}{584} (\bibinfo{year}{2017}), \eprint{1705.02290}.

\bibitem[{\citenamefont{Witten}(1984)}]{Witten:1983ar}
\bibinfo{author}{\bibfnamefont{E.}~\bibnamefont{Witten}},
  \bibinfo{journal}{Commun. Math. Phys.} \textbf{\bibinfo{volume}{92}},
  \bibinfo{pages}{455} (\bibinfo{year}{1984}).

\bibitem[{\citenamefont{Svrcek and Witten}(2006)}]{Svrcek:2006yi}
\bibinfo{author}{\bibfnamefont{P.}~\bibnamefont{Svrcek}} \bibnamefont{and}
  \bibinfo{author}{\bibfnamefont{E.}~\bibnamefont{Witten}},
  \bibinfo{journal}{JHEP} \textbf{\bibinfo{volume}{06}}, \bibinfo{pages}{051}
  (\bibinfo{year}{2006}), \eprint{hep-th/0605206}.

\bibitem[{\citenamefont{Conlon}(2006)}]{Conlon:2006tq}
\bibinfo{author}{\bibfnamefont{J.~P.} \bibnamefont{Conlon}},
  \bibinfo{journal}{JHEP} \textbf{\bibinfo{volume}{05}}, \bibinfo{pages}{078}
  (\bibinfo{year}{2006}), \eprint{hep-th/0602233}.

\bibitem[{\citenamefont{Arvanitaki et~al.}(2010)\citenamefont{Arvanitaki,
  Dimopoulos, Dubovsky, Kaloper, and March-Russell}}]{Arvanitaki:2009fg}
\bibinfo{author}{\bibfnamefont{A.}~\bibnamefont{Arvanitaki}},
  \bibinfo{author}{\bibfnamefont{S.}~\bibnamefont{Dimopoulos}},
  \bibinfo{author}{\bibfnamefont{S.}~\bibnamefont{Dubovsky}},
  \bibinfo{author}{\bibfnamefont{N.}~\bibnamefont{Kaloper}}, \bibnamefont{and}
  \bibinfo{author}{\bibfnamefont{J.}~\bibnamefont{March-Russell}},
  \bibinfo{journal}{Phys. Rev. D} \textbf{\bibinfo{volume}{81}},
  \bibinfo{pages}{123530} (\bibinfo{year}{2010}), \eprint{0905.4720}.

\bibitem[{\citenamefont{Cicoli et~al.}(2012)\citenamefont{Cicoli, Goodsell, and
  Ringwald}}]{Cicoli:2012sz}
\bibinfo{author}{\bibfnamefont{M.}~\bibnamefont{Cicoli}},
  \bibinfo{author}{\bibfnamefont{M.}~\bibnamefont{Goodsell}}, \bibnamefont{and}
  \bibinfo{author}{\bibfnamefont{A.}~\bibnamefont{Ringwald}},
  \bibinfo{journal}{JHEP} \textbf{\bibinfo{volume}{10}}, \bibinfo{pages}{146}
  (\bibinfo{year}{2012}), \eprint{1206.0819}.

\bibitem[{\citenamefont{Derevianko et~al.}(2010)\citenamefont{Derevianko,
  Dzuba, Flambaum, and Pospelov}}]{Derevianko:2010kz}
\bibinfo{author}{\bibfnamefont{A.}~\bibnamefont{Derevianko}},
  \bibinfo{author}{\bibfnamefont{V.~A.} \bibnamefont{Dzuba}},
  \bibinfo{author}{\bibfnamefont{V.~V.} \bibnamefont{Flambaum}},
  \bibnamefont{and} \bibinfo{author}{\bibfnamefont{M.}~\bibnamefont{Pospelov}},
  \bibinfo{journal}{Phys. Rev. D} \textbf{\bibinfo{volume}{82}},
  \bibinfo{pages}{065006} (\bibinfo{year}{2010}), \eprint{1007.1833}.

\bibitem[{\citenamefont{Aprile et~al.}(2014{\natexlab{c}})}]{Aprile:2014eoa}
\bibinfo{author}{\bibfnamefont{E.}~\bibnamefont{Aprile}} \bibnamefont{et~al.}
  (\bibinfo{collaboration}{XENON100}), \bibinfo{journal}{Phys. Rev. D}
  \textbf{\bibinfo{volume}{90}}, \bibinfo{pages}{062009}
  (\bibinfo{year}{2014}{\natexlab{c}}), \bibinfo{note}{[Erratum: Phys. Rev. D
  {\bf 95}, 029904 (2017)]}, \eprint{1404.1455}.

\bibitem[{\citenamefont{Abe et~al.}(2014)}]{Abe:2014zcd}
\bibinfo{author}{\bibfnamefont{K.}~\bibnamefont{Abe}} \bibnamefont{et~al.}
  (\bibinfo{collaboration}{XMASS}), \bibinfo{journal}{Phys. Rev. Lett.}
  \textbf{\bibinfo{volume}{113}}, \bibinfo{pages}{121301}
  (\bibinfo{year}{2014}), \eprint{1406.0502}.

\bibitem[{\citenamefont{Akerib et~al.}(2017{\natexlab{b}})}]{Akerib:2017uem}
\bibinfo{author}{\bibfnamefont{D.~S.} \bibnamefont{Akerib}}
  \bibnamefont{et~al.} (\bibinfo{collaboration}{LUX}), \bibinfo{journal}{Phys.
  Rev. Lett.} \textbf{\bibinfo{volume}{118}}, \bibinfo{pages}{261301}
  (\bibinfo{year}{2017}{\natexlab{b}}), \eprint{1704.02297}.

\bibitem[{\citenamefont{Fu et~al.}(2017)}]{Fu:2017lfc}
\bibinfo{author}{\bibfnamefont{C.}~\bibnamefont{Fu}} \bibnamefont{et~al.}
  (\bibinfo{collaboration}{PandaX}), \bibinfo{journal}{Phys. Rev. Lett.}
  \textbf{\bibinfo{volume}{119}}, \bibinfo{pages}{181806}
  (\bibinfo{year}{2017}), \eprint{1707.07921}.

\bibitem[{\citenamefont{Abe et~al.}(2018{\natexlab{a}})}]{Abe:2018owy}
\bibinfo{author}{\bibfnamefont{K.}~\bibnamefont{Abe}} \bibnamefont{et~al.}
  (\bibinfo{collaboration}{XMASS}), \bibinfo{journal}{Phys. Lett. B}
  \textbf{\bibinfo{volume}{787}}, \bibinfo{pages}{153}
  (\bibinfo{year}{2018}{\natexlab{a}}), \eprint{1807.08516}.

\bibitem[{\citenamefont{Zhou et~al.}(2020)}]{Zhou:2020bvf}
\bibinfo{author}{\bibfnamefont{X.}~\bibnamefont{Zhou}} \bibnamefont{et~al.}
  (\bibinfo{collaboration}{PandaX-II}) (\bibinfo{year}{2020}),
  \eprint{2008.06485}.

\bibitem[{\citenamefont{Davis}(1986)}]{Davis:1986xc}
\bibinfo{author}{\bibfnamefont{R.~L.} \bibnamefont{Davis}},
  \bibinfo{journal}{Phys. Lett. B} \textbf{\bibinfo{volume}{180}},
  \bibinfo{pages}{225} (\bibinfo{year}{1986}).

\bibitem[{\citenamefont{Sikivie}(1982)}]{Sikivie:1982qv}
\bibinfo{author}{\bibfnamefont{P.}~\bibnamefont{Sikivie}},
  \bibinfo{journal}{Phys. Rev. Lett.} \textbf{\bibinfo{volume}{48}},
  \bibinfo{pages}{1156} (\bibinfo{year}{1982}).

\bibitem[{\citenamefont{Visinelli and Gondolo}(2010)}]{Visinelli:2009kt}
\bibinfo{author}{\bibfnamefont{L.}~\bibnamefont{Visinelli}} \bibnamefont{and}
  \bibinfo{author}{\bibfnamefont{P.}~\bibnamefont{Gondolo}},
  \bibinfo{journal}{Phys. Rev. D} \textbf{\bibinfo{volume}{81}},
  \bibinfo{pages}{063508} (\bibinfo{year}{2010}), \eprint{0912.0015}.

\bibitem[{\citenamefont{Hiramatsu et~al.}(2011)\citenamefont{Hiramatsu,
  Kawasaki, and Saikawa}}]{Hiramatsu:2010yn}
\bibinfo{author}{\bibfnamefont{T.}~\bibnamefont{Hiramatsu}},
  \bibinfo{author}{\bibfnamefont{M.}~\bibnamefont{Kawasaki}}, \bibnamefont{and}
  \bibinfo{author}{\bibfnamefont{K.}~\bibnamefont{Saikawa}},
  \bibinfo{journal}{JCAP} \textbf{\bibinfo{volume}{08}}, \bibinfo{pages}{030}
  (\bibinfo{year}{2011}), \eprint{1012.4558}.

\bibitem[{\citenamefont{Co et~al.}(2018)\citenamefont{Co, Hall, and
  Harigaya}}]{Co:2017mop}
\bibinfo{author}{\bibfnamefont{R.~T.} \bibnamefont{Co}},
  \bibinfo{author}{\bibfnamefont{L.~J.} \bibnamefont{Hall}}, \bibnamefont{and}
  \bibinfo{author}{\bibfnamefont{K.}~\bibnamefont{Harigaya}},
  \bibinfo{journal}{Phys. Rev. Lett.} \textbf{\bibinfo{volume}{120}},
  \bibinfo{pages}{211602} (\bibinfo{year}{2018}), \eprint{1711.10486}.

\bibitem[{\citenamefont{Co et~al.}(2019)\citenamefont{Co, Gonzalez, and
  Harigaya}}]{Co:2018mho}
\bibinfo{author}{\bibfnamefont{R.~T.} \bibnamefont{Co}},
  \bibinfo{author}{\bibfnamefont{E.}~\bibnamefont{Gonzalez}}, \bibnamefont{and}
  \bibinfo{author}{\bibfnamefont{K.}~\bibnamefont{Harigaya}},
  \bibinfo{journal}{JHEP} \textbf{\bibinfo{volume}{05}}, \bibinfo{pages}{163}
  (\bibinfo{year}{2019}), \eprint{1812.11192}.

\bibitem[{\citenamefont{Co et~al.}(2020)\citenamefont{Co, Hall, and
  Harigaya}}]{Co:2019jts}
\bibinfo{author}{\bibfnamefont{R.~T.} \bibnamefont{Co}},
  \bibinfo{author}{\bibfnamefont{L.~J.} \bibnamefont{Hall}}, \bibnamefont{and}
  \bibinfo{author}{\bibfnamefont{K.}~\bibnamefont{Harigaya}},
  \bibinfo{journal}{Phys. Rev. Lett.} \textbf{\bibinfo{volume}{124}},
  \bibinfo{pages}{251802} (\bibinfo{year}{2020}), \eprint{1910.14152}.

\bibitem[{\citenamefont{Hook et~al.}(2020)\citenamefont{Hook, Marques-Tavares,
  and Tsai}}]{Hook:2019hdk}
\bibinfo{author}{\bibfnamefont{A.}~\bibnamefont{Hook}},
  \bibinfo{author}{\bibfnamefont{G.}~\bibnamefont{Marques-Tavares}},
  \bibnamefont{and} \bibinfo{author}{\bibfnamefont{Y.}~\bibnamefont{Tsai}},
  \bibinfo{journal}{Phys. Rev. Lett.} \textbf{\bibinfo{volume}{124}},
  \bibinfo{pages}{211801} (\bibinfo{year}{2020}), \eprint{1912.08817}.

\bibitem[{\citenamefont{Co et~al.}(2021)\citenamefont{Co, Hall, and
  Harigaya}}]{Co:2020xlh}
\bibinfo{author}{\bibfnamefont{R.~T.} \bibnamefont{Co}},
  \bibinfo{author}{\bibfnamefont{L.~J.} \bibnamefont{Hall}}, \bibnamefont{and}
  \bibinfo{author}{\bibfnamefont{K.}~\bibnamefont{Harigaya}},
  \bibinfo{journal}{JHEP} \textbf{\bibinfo{volume}{01}}, \bibinfo{pages}{172}
  (\bibinfo{year}{2021}), \eprint{2006.04809}.

\bibitem[{\citenamefont{Dent et~al.}(2020{\natexlab{b}})\citenamefont{Dent,
  Dutta, Newstead, and Thompson}}]{Dent:2020jhf}
\bibinfo{author}{\bibfnamefont{J.~B.} \bibnamefont{Dent}},
  \bibinfo{author}{\bibfnamefont{B.}~\bibnamefont{Dutta}},
  \bibinfo{author}{\bibfnamefont{J.~L.} \bibnamefont{Newstead}},
  \bibnamefont{and} \bibinfo{author}{\bibfnamefont{A.}~\bibnamefont{Thompson}},
  \bibinfo{journal}{Phys. Rev. Lett.} \textbf{\bibinfo{volume}{125}},
  \bibinfo{pages}{131805} (\bibinfo{year}{2020}{\natexlab{b}}),
  \eprint{2006.15118}.

\bibitem[{\citenamefont{Nussinov}(1985)}]{Nussinov:1985xr}
\bibinfo{author}{\bibfnamefont{S.}~\bibnamefont{Nussinov}},
  \bibinfo{journal}{Phys. Lett. B} \textbf{\bibinfo{volume}{165}},
  \bibinfo{pages}{55} (\bibinfo{year}{1985}).

\bibitem[{\citenamefont{Gelmini et~al.}(1987)\citenamefont{Gelmini, Hall, and
  Lin}}]{Gelmini:1986zz}
\bibinfo{author}{\bibfnamefont{G.~B.} \bibnamefont{Gelmini}},
  \bibinfo{author}{\bibfnamefont{L.~J.} \bibnamefont{Hall}}, \bibnamefont{and}
  \bibinfo{author}{\bibfnamefont{M.~J.} \bibnamefont{Lin}},
  \bibinfo{journal}{Nucl. Phys. B} \textbf{\bibinfo{volume}{281}},
  \bibinfo{pages}{726} (\bibinfo{year}{1987}).

\bibitem[{\citenamefont{Kaplan}(1992)}]{Kaplan:1991ah}
\bibinfo{author}{\bibfnamefont{D.~B.} \bibnamefont{Kaplan}},
  \bibinfo{journal}{Phys. Rev. Lett.} \textbf{\bibinfo{volume}{68}},
  \bibinfo{pages}{741} (\bibinfo{year}{1992}).

\bibitem[{\citenamefont{Hooper et~al.}(2005)\citenamefont{Hooper,
  March-Russell, and West}}]{Hooper:2004dc}
\bibinfo{author}{\bibfnamefont{D.}~\bibnamefont{Hooper}},
  \bibinfo{author}{\bibfnamefont{J.}~\bibnamefont{March-Russell}},
  \bibnamefont{and} \bibinfo{author}{\bibfnamefont{S.~M.} \bibnamefont{West}},
  \bibinfo{journal}{Phys. Lett. B} \textbf{\bibinfo{volume}{605}},
  \bibinfo{pages}{228} (\bibinfo{year}{2005}), \eprint{hep-ph/0410114}.

\bibitem[{\citenamefont{Kitano and Low}(2005)}]{Kitano:2004sv}
\bibinfo{author}{\bibfnamefont{R.}~\bibnamefont{Kitano}} \bibnamefont{and}
  \bibinfo{author}{\bibfnamefont{I.}~\bibnamefont{Low}},
  \bibinfo{journal}{Phys. Rev. D} \textbf{\bibinfo{volume}{71}},
  \bibinfo{pages}{023510} (\bibinfo{year}{2005}), \eprint{hep-ph/0411133}.

\bibitem[{\citenamefont{Cosme et~al.}(2005)\citenamefont{Cosme, Lopez~Honorez,
  and Tytgat}}]{Cosme:2005sb}
\bibinfo{author}{\bibfnamefont{N.}~\bibnamefont{Cosme}},
  \bibinfo{author}{\bibfnamefont{L.}~\bibnamefont{Lopez~Honorez}},
  \bibnamefont{and} \bibinfo{author}{\bibfnamefont{M.~H.~G.}
  \bibnamefont{Tytgat}}, \bibinfo{journal}{Phys. Rev. D}
  \textbf{\bibinfo{volume}{72}}, \bibinfo{pages}{043505}
  (\bibinfo{year}{2005}), \eprint{hep-ph/0506320}.

\bibitem[{\citenamefont{Kaplan et~al.}(2009)\citenamefont{Kaplan, Luty, and
  Zurek}}]{Kaplan:2009ag}
\bibinfo{author}{\bibfnamefont{D.~E.} \bibnamefont{Kaplan}},
  \bibinfo{author}{\bibfnamefont{M.~A.} \bibnamefont{Luty}}, \bibnamefont{and}
  \bibinfo{author}{\bibfnamefont{K.~M.} \bibnamefont{Zurek}},
  \bibinfo{journal}{Phys. Rev. D} \textbf{\bibinfo{volume}{79}},
  \bibinfo{pages}{115016} (\bibinfo{year}{2009}), \eprint{0901.4117}.

\bibitem[{\citenamefont{March-Russell et~al.}(2010)\citenamefont{March-Russell,
  McCabe, and McCullough}}]{March-Russell:2009vla}
\bibinfo{author}{\bibfnamefont{J.}~\bibnamefont{March-Russell}},
  \bibinfo{author}{\bibfnamefont{C.}~\bibnamefont{McCabe}}, \bibnamefont{and}
  \bibinfo{author}{\bibfnamefont{M.}~\bibnamefont{McCullough}},
  \bibinfo{journal}{JHEP} \textbf{\bibinfo{volume}{03}}, \bibinfo{pages}{108}
  (\bibinfo{year}{2010}), \eprint{0911.4489}.

\bibitem[{\citenamefont{Frandsen and Sarkar}(2010)}]{Frandsen:2010yj}
\bibinfo{author}{\bibfnamefont{M.~T.} \bibnamefont{Frandsen}} \bibnamefont{and}
  \bibinfo{author}{\bibfnamefont{S.}~\bibnamefont{Sarkar}},
  \bibinfo{journal}{Phys. Rev. Lett.} \textbf{\bibinfo{volume}{105}},
  \bibinfo{pages}{011301} (\bibinfo{year}{2010}), \eprint{1003.4505}.

\bibitem[{\citenamefont{Frandsen
  et~al.}(2011{\natexlab{b}})\citenamefont{Frandsen, Sarkar, and
  Schmidt-Hoberg}}]{Frandsen:2011kt}
\bibinfo{author}{\bibfnamefont{M.~T.} \bibnamefont{Frandsen}},
  \bibinfo{author}{\bibfnamefont{S.}~\bibnamefont{Sarkar}}, \bibnamefont{and}
  \bibinfo{author}{\bibfnamefont{K.}~\bibnamefont{Schmidt-Hoberg}},
  \bibinfo{journal}{Phys. Rev. D} \textbf{\bibinfo{volume}{84}},
  \bibinfo{pages}{051703} (\bibinfo{year}{2011}{\natexlab{b}}),
  \eprint{1103.4350}.

\bibitem[{\citenamefont{Petraki and Volkas}(2013)}]{Petraki:2013wwa}
\bibinfo{author}{\bibfnamefont{K.}~\bibnamefont{Petraki}} \bibnamefont{and}
  \bibinfo{author}{\bibfnamefont{R.~R.} \bibnamefont{Volkas}},
  \bibinfo{journal}{Int. J. Mod. Phys. A} \textbf{\bibinfo{volume}{28}},
  \bibinfo{pages}{1330028} (\bibinfo{year}{2013}), \eprint{1305.4939}.

\bibitem[{\citenamefont{Zurek}(2014)}]{Zurek:2013wia}
\bibinfo{author}{\bibfnamefont{K.~M.} \bibnamefont{Zurek}},
  \bibinfo{journal}{Phys. Rept.} \textbf{\bibinfo{volume}{537}},
  \bibinfo{pages}{91} (\bibinfo{year}{2014}), \eprint{1308.0338}.

\bibitem[{\citenamefont{Chacko et~al.}(2006)\citenamefont{Chacko, Goh, and
  Harnik}}]{Chacko:2005pe}
\bibinfo{author}{\bibfnamefont{Z.}~\bibnamefont{Chacko}},
  \bibinfo{author}{\bibfnamefont{H.-S.} \bibnamefont{Goh}}, \bibnamefont{and}
  \bibinfo{author}{\bibfnamefont{R.}~\bibnamefont{Harnik}},
  \bibinfo{journal}{Phys. Rev. Lett.} \textbf{\bibinfo{volume}{96}},
  \bibinfo{pages}{231802} (\bibinfo{year}{2006}), \eprint{hep-ph/0506256}.

\bibitem[{\citenamefont{Garcia~Garcia
  et~al.}(2015{\natexlab{b}})\citenamefont{Garcia~Garcia, Lasenby, and
  March-Russell}}]{GarciaGarcia:2015fol}
\bibinfo{author}{\bibfnamefont{I.}~\bibnamefont{Garcia~Garcia}},
  \bibinfo{author}{\bibfnamefont{R.}~\bibnamefont{Lasenby}}, \bibnamefont{and}
  \bibinfo{author}{\bibfnamefont{J.}~\bibnamefont{March-Russell}},
  \bibinfo{journal}{Phys. Rev. D} \textbf{\bibinfo{volume}{92}},
  \bibinfo{pages}{055034} (\bibinfo{year}{2015}{\natexlab{b}}),
  \eprint{1505.07109}.

\bibitem[{\citenamefont{Craig and Katz}(2015)}]{Craig:2015xla}
\bibinfo{author}{\bibfnamefont{N.}~\bibnamefont{Craig}} \bibnamefont{and}
  \bibinfo{author}{\bibfnamefont{A.}~\bibnamefont{Katz}},
  \bibinfo{journal}{JCAP} \textbf{\bibinfo{volume}{10}}, \bibinfo{pages}{054}
  (\bibinfo{year}{2015}), \eprint{1505.07113}.

\bibitem[{\citenamefont{Cacciapaglia et~al.}(2021)\citenamefont{Cacciapaglia,
  Frandsen, Huang, Rosenlyst, and S\o{}rensen}}]{Cacciapaglia:2021aex}
\bibinfo{author}{\bibfnamefont{G.}~\bibnamefont{Cacciapaglia}},
  \bibinfo{author}{\bibfnamefont{M.~T.} \bibnamefont{Frandsen}},
  \bibinfo{author}{\bibfnamefont{W.-C.} \bibnamefont{Huang}},
  \bibinfo{author}{\bibfnamefont{M.}~\bibnamefont{Rosenlyst}},
  \bibnamefont{and}
  \bibinfo{author}{\bibfnamefont{P.}~\bibnamefont{S\o{}rensen}}
  (\bibinfo{year}{2021}), \eprint{2111.09319}.

\bibitem[{\citenamefont{Hall et~al.}(2010{\natexlab{b}})\citenamefont{Hall,
  March-Russell, and West}}]{Hall:2010jx}
\bibinfo{author}{\bibfnamefont{L.~J.} \bibnamefont{Hall}},
  \bibinfo{author}{\bibfnamefont{J.}~\bibnamefont{March-Russell}},
  \bibnamefont{and} \bibinfo{author}{\bibfnamefont{S.~M.} \bibnamefont{West}}
  (\bibinfo{year}{2010}{\natexlab{b}}), \eprint{1010.0245}.

\bibitem[{\citenamefont{Buckley and Randall}(2011)}]{Buckley:2010ui}
\bibinfo{author}{\bibfnamefont{M.~R.} \bibnamefont{Buckley}} \bibnamefont{and}
  \bibinfo{author}{\bibfnamefont{L.}~\bibnamefont{Randall}},
  \bibinfo{journal}{JHEP} \textbf{\bibinfo{volume}{09}}, \bibinfo{pages}{009}
  (\bibinfo{year}{2011}), \eprint{1009.0270}.

\bibitem[{\citenamefont{Buckley}(2011)}]{Buckley:2011kk}
\bibinfo{author}{\bibfnamefont{M.~R.} \bibnamefont{Buckley}},
  \bibinfo{journal}{Phys. Rev. D} \textbf{\bibinfo{volume}{84}},
  \bibinfo{pages}{043510} (\bibinfo{year}{2011}), \eprint{1104.1429}.

\bibitem[{\citenamefont{March-Russell et~al.}(2012)\citenamefont{March-Russell,
  Unwin, and West}}]{March-Russell:2012elz}
\bibinfo{author}{\bibfnamefont{J.}~\bibnamefont{March-Russell}},
  \bibinfo{author}{\bibfnamefont{J.}~\bibnamefont{Unwin}}, \bibnamefont{and}
  \bibinfo{author}{\bibfnamefont{S.~M.} \bibnamefont{West}},
  \bibinfo{journal}{JHEP} \textbf{\bibinfo{volume}{08}}, \bibinfo{pages}{029}
  (\bibinfo{year}{2012}), \eprint{1203.4854}.

\bibitem[{\citenamefont{Cui and Shamma}(2020)}]{Cui:2020dly}
\bibinfo{author}{\bibfnamefont{Y.}~\bibnamefont{Cui}} \bibnamefont{and}
  \bibinfo{author}{\bibfnamefont{M.}~\bibnamefont{Shamma}},
  \bibinfo{journal}{JHEP} \textbf{\bibinfo{volume}{12}}, \bibinfo{pages}{046}
  (\bibinfo{year}{2020}), \eprint{2002.05170}.

\bibitem[{\citenamefont{Detmold
  et~al.}(2014{\natexlab{a}})\citenamefont{Detmold, McCullough, and
  Pochinsky}}]{Detmold:2014qqa}
\bibinfo{author}{\bibfnamefont{W.}~\bibnamefont{Detmold}},
  \bibinfo{author}{\bibfnamefont{M.}~\bibnamefont{McCullough}},
  \bibnamefont{and}
  \bibinfo{author}{\bibfnamefont{A.}~\bibnamefont{Pochinsky}},
  \bibinfo{journal}{Phys. Rev. D} \textbf{\bibinfo{volume}{90}},
  \bibinfo{pages}{115013} (\bibinfo{year}{2014}{\natexlab{a}}),
  \eprint{1406.2276}.

\bibitem[{\citenamefont{Hardy et~al.}(2015{\natexlab{b}})\citenamefont{Hardy,
  Lasenby, March-Russell, and West}}]{Hardy:2014mqa}
\bibinfo{author}{\bibfnamefont{E.}~\bibnamefont{Hardy}},
  \bibinfo{author}{\bibfnamefont{R.}~\bibnamefont{Lasenby}},
  \bibinfo{author}{\bibfnamefont{J.}~\bibnamefont{March-Russell}},
  \bibnamefont{and} \bibinfo{author}{\bibfnamefont{S.~M.} \bibnamefont{West}},
  \bibinfo{journal}{JHEP} \textbf{\bibinfo{volume}{06}}, \bibinfo{pages}{011}
  (\bibinfo{year}{2015}{\natexlab{b}}), \eprint{1411.3739}.

\bibitem[{\citenamefont{Detmold
  et~al.}(2014{\natexlab{b}})\citenamefont{Detmold, McCullough, and
  Pochinsky}}]{Detmold:2014kba}
\bibinfo{author}{\bibfnamefont{W.}~\bibnamefont{Detmold}},
  \bibinfo{author}{\bibfnamefont{M.}~\bibnamefont{McCullough}},
  \bibnamefont{and}
  \bibinfo{author}{\bibfnamefont{A.}~\bibnamefont{Pochinsky}},
  \bibinfo{journal}{Phys. Rev. D} \textbf{\bibinfo{volume}{90}},
  \bibinfo{pages}{114506} (\bibinfo{year}{2014}{\natexlab{b}}),
  \eprint{1406.4116}.

\bibitem[{\citenamefont{Wise and Zhang}(2015)}]{Wise:2014ola}
\bibinfo{author}{\bibfnamefont{M.~B.} \bibnamefont{Wise}} \bibnamefont{and}
  \bibinfo{author}{\bibfnamefont{Y.}~\bibnamefont{Zhang}},
  \bibinfo{journal}{JHEP} \textbf{\bibinfo{volume}{02}}, \bibinfo{pages}{023}
  (\bibinfo{year}{2015}), \bibinfo{note}{[Erratum: JHEP 10, 165 (2015)]},
  \eprint{1411.1772}.

\bibitem[{\citenamefont{Gresham et~al.}(2017)\citenamefont{Gresham, Lou, and
  Zurek}}]{Gresham:2017zqi}
\bibinfo{author}{\bibfnamefont{M.~I.} \bibnamefont{Gresham}},
  \bibinfo{author}{\bibfnamefont{H.~K.} \bibnamefont{Lou}}, \bibnamefont{and}
  \bibinfo{author}{\bibfnamefont{K.~M.} \bibnamefont{Zurek}},
  \bibinfo{journal}{Phys. Rev. D} \textbf{\bibinfo{volume}{96}},
  \bibinfo{pages}{096012} (\bibinfo{year}{2017}), \eprint{1707.02313}.

\bibitem[{\citenamefont{Bai et~al.}(2019)\citenamefont{Bai, Long, and
  Lu}}]{Bai:2018dxf}
\bibinfo{author}{\bibfnamefont{Y.}~\bibnamefont{Bai}},
  \bibinfo{author}{\bibfnamefont{A.~J.} \bibnamefont{Long}}, \bibnamefont{and}
  \bibinfo{author}{\bibfnamefont{S.}~\bibnamefont{Lu}}, \bibinfo{journal}{Phys.
  Rev. D} \textbf{\bibinfo{volume}{99}}, \bibinfo{pages}{055047}
  (\bibinfo{year}{2019}), \eprint{1810.04360}.

\bibitem[{\citenamefont{Kaplan et~al.}(2011)\citenamefont{Kaplan, Krnjaic,
  Rehermann, and Wells}}]{Kaplan:2011yj}
\bibinfo{author}{\bibfnamefont{D.~E.} \bibnamefont{Kaplan}},
  \bibinfo{author}{\bibfnamefont{G.~Z.} \bibnamefont{Krnjaic}},
  \bibinfo{author}{\bibfnamefont{K.~R.} \bibnamefont{Rehermann}},
  \bibnamefont{and} \bibinfo{author}{\bibfnamefont{C.~M.} \bibnamefont{Wells}},
  \bibinfo{journal}{JCAP} \textbf{\bibinfo{volume}{10}}, \bibinfo{pages}{011}
  (\bibinfo{year}{2011}), \eprint{1105.2073}.

\bibitem[{\citenamefont{Petraki et~al.}(2014)\citenamefont{Petraki, Pearce, and
  Kusenko}}]{Petraki:2014uza}
\bibinfo{author}{\bibfnamefont{K.}~\bibnamefont{Petraki}},
  \bibinfo{author}{\bibfnamefont{L.}~\bibnamefont{Pearce}}, \bibnamefont{and}
  \bibinfo{author}{\bibfnamefont{A.}~\bibnamefont{Kusenko}},
  \bibinfo{journal}{JCAP} \textbf{\bibinfo{volume}{07}}, \bibinfo{pages}{039}
  (\bibinfo{year}{2014}), \eprint{1403.1077}.

\bibitem[{\citenamefont{Wise and Zhang}(2014)}]{Wise:2014jva}
\bibinfo{author}{\bibfnamefont{M.~B.} \bibnamefont{Wise}} \bibnamefont{and}
  \bibinfo{author}{\bibfnamefont{Y.}~\bibnamefont{Zhang}},
  \bibinfo{journal}{Phys. Rev. D} \textbf{\bibinfo{volume}{90}},
  \bibinfo{pages}{055030} (\bibinfo{year}{2014}), \bibinfo{note}{[Erratum:
  Phys.Rev.D 91, 039907 (2015)]}, \eprint{1407.4121}.

\bibitem[{\citenamefont{Petraki et~al.}(2015)\citenamefont{Petraki, Postma, and
  Wiechers}}]{Petraki:2015hla}
\bibinfo{author}{\bibfnamefont{K.}~\bibnamefont{Petraki}},
  \bibinfo{author}{\bibfnamefont{M.}~\bibnamefont{Postma}}, \bibnamefont{and}
  \bibinfo{author}{\bibfnamefont{M.}~\bibnamefont{Wiechers}},
  \bibinfo{journal}{JHEP} \textbf{\bibinfo{volume}{06}}, \bibinfo{pages}{128}
  (\bibinfo{year}{2015}), \eprint{1505.00109}.

\bibitem[{\citenamefont{Laha}(2015)}]{Laha:2015yoa}
\bibinfo{author}{\bibfnamefont{R.}~\bibnamefont{Laha}}, \bibinfo{journal}{Phys.
  Rev. D} \textbf{\bibinfo{volume}{92}}, \bibinfo{pages}{083509}
  (\bibinfo{year}{2015}), \eprint{1505.02772}.

\bibitem[{\citenamefont{Coleman}(1985)}]{Coleman:1985ki}
\bibinfo{author}{\bibfnamefont{S.~R.} \bibnamefont{Coleman}},
  \bibinfo{journal}{Nucl. Phys. B} \textbf{\bibinfo{volume}{262}},
  \bibinfo{pages}{263} (\bibinfo{year}{1985}), \bibinfo{note}{[Addendum:
  Nucl.Phys.B 269, 744 (1986)]}.

\bibitem[{\citenamefont{Butcher et~al.}(2017)\citenamefont{Butcher, Kirk,
  Monroe, and West}}]{Butcher:2016hic}
\bibinfo{author}{\bibfnamefont{A.}~\bibnamefont{Butcher}},
  \bibinfo{author}{\bibfnamefont{R.}~\bibnamefont{Kirk}},
  \bibinfo{author}{\bibfnamefont{J.}~\bibnamefont{Monroe}}, \bibnamefont{and}
  \bibinfo{author}{\bibfnamefont{S.}~\bibnamefont{West}},
  \bibinfo{journal}{JCAP} \textbf{\bibinfo{volume}{10}}, \bibinfo{pages}{035}
  (\bibinfo{year}{2017}), \eprint{1610.01840}.

\bibitem[{\citenamefont{Grabowska et~al.}(2018)\citenamefont{Grabowska, Melia,
  and Rajendran}}]{Grabowska:2018lnd}
\bibinfo{author}{\bibfnamefont{D.~M.} \bibnamefont{Grabowska}},
  \bibinfo{author}{\bibfnamefont{T.}~\bibnamefont{Melia}}, \bibnamefont{and}
  \bibinfo{author}{\bibfnamefont{S.}~\bibnamefont{Rajendran}},
  \bibinfo{journal}{Phys. Rev. D} \textbf{\bibinfo{volume}{98}},
  \bibinfo{pages}{115020} (\bibinfo{year}{2018}), \eprint{1807.03788}.

\bibitem[{\citenamefont{Coskuner et~al.}(2019)\citenamefont{Coskuner,
  Grabowska, Knapen, and Zurek}}]{Coskuner:2018are}
\bibinfo{author}{\bibfnamefont{A.}~\bibnamefont{Coskuner}},
  \bibinfo{author}{\bibfnamefont{D.~M.} \bibnamefont{Grabowska}},
  \bibinfo{author}{\bibfnamefont{S.}~\bibnamefont{Knapen}}, \bibnamefont{and}
  \bibinfo{author}{\bibfnamefont{K.~M.} \bibnamefont{Zurek}},
  \bibinfo{journal}{Phys. Rev. D} \textbf{\bibinfo{volume}{100}},
  \bibinfo{pages}{035025} (\bibinfo{year}{2019}), \eprint{1812.07573}.

\bibitem[{\citenamefont{Foot and Vagnozzi}(2015)}]{Foot:2014osa}
\bibinfo{author}{\bibfnamefont{R.}~\bibnamefont{Foot}} \bibnamefont{and}
  \bibinfo{author}{\bibfnamefont{S.}~\bibnamefont{Vagnozzi}},
  \bibinfo{journal}{Phys. Lett. B} \textbf{\bibinfo{volume}{748}},
  \bibinfo{pages}{61} (\bibinfo{year}{2015}), \eprint{1412.0762}.

\bibitem[{\citenamefont{Hodges}(1993)}]{Hodges:1993yb}
\bibinfo{author}{\bibfnamefont{H.~M.} \bibnamefont{Hodges}},
  \bibinfo{journal}{Phys. Rev. D} \textbf{\bibinfo{volume}{47}},
  \bibinfo{pages}{456} (\bibinfo{year}{1993}).

\bibitem[{\citenamefont{Berezhiani}(2004)}]{Berezhiani:2003xm}
\bibinfo{author}{\bibfnamefont{Z.}~\bibnamefont{Berezhiani}},
  \bibinfo{journal}{Int. J. Mod. Phys. A} \textbf{\bibinfo{volume}{19}},
  \bibinfo{pages}{3775} (\bibinfo{year}{2004}), \eprint{hep-ph/0312335}.

\bibitem[{\citenamefont{Okun}(2007)}]{Okun:2006eb}
\bibinfo{author}{\bibfnamefont{L.~B.} \bibnamefont{Okun}},
  \bibinfo{journal}{Phys. Usp.} \textbf{\bibinfo{volume}{50}},
  \bibinfo{pages}{380} (\bibinfo{year}{2007}), \eprint{hep-ph/0606202}.

\bibitem[{\citenamefont{Foot}(2007)}]{Foot:2007nn}
\bibinfo{author}{\bibfnamefont{R.}~\bibnamefont{Foot}}, \bibinfo{journal}{Int.
  J. Mod. Phys. A} \textbf{\bibinfo{volume}{22}}, \bibinfo{pages}{4951}
  (\bibinfo{year}{2007}), \eprint{0706.2694}.

\bibitem[{\citenamefont{Foot}(2010)}]{Foot:2010hu}
\bibinfo{author}{\bibfnamefont{R.}~\bibnamefont{Foot}}, \bibinfo{journal}{Phys.
  Rev. D} \textbf{\bibinfo{volume}{82}}, \bibinfo{pages}{095001}
  (\bibinfo{year}{2010}), \eprint{1008.0685}.

\bibitem[{\citenamefont{Foot}(2014)}]{Foot:2014mia}
\bibinfo{author}{\bibfnamefont{R.}~\bibnamefont{Foot}}, \bibinfo{journal}{Int.
  J. Mod. Phys. A} \textbf{\bibinfo{volume}{29}}, \bibinfo{pages}{1430013}
  (\bibinfo{year}{2014}), \eprint{1401.3965}.

\bibitem[{\citenamefont{Foot}(2019)}]{Foot:2018jpo}
\bibinfo{author}{\bibfnamefont{R.}~\bibnamefont{Foot}}, \bibinfo{journal}{Phys.
  Lett. B} \textbf{\bibinfo{volume}{789}}, \bibinfo{pages}{592}
  (\bibinfo{year}{2019}), \eprint{1806.04293}.

\bibitem[{\citenamefont{Akerib et~al.}(2020{\natexlab{g}})}]{LUX:2019gwa}
\bibinfo{author}{\bibfnamefont{D.~S.} \bibnamefont{Akerib}}
  \bibnamefont{et~al.} (\bibinfo{collaboration}{LUX}), \bibinfo{journal}{Phys.
  Rev. D} \textbf{\bibinfo{volume}{101}}, \bibinfo{pages}{012003}
  (\bibinfo{year}{2020}{\natexlab{g}}), \eprint{1908.03479}.

\bibitem[{\citenamefont{Clarke and Foot}(2017)}]{Clarke:2016eac}
\bibinfo{author}{\bibfnamefont{J.~D.} \bibnamefont{Clarke}} \bibnamefont{and}
  \bibinfo{author}{\bibfnamefont{R.}~\bibnamefont{Foot}},
  \bibinfo{journal}{Phys. Lett. B} \textbf{\bibinfo{volume}{766}},
  \bibinfo{pages}{29} (\bibinfo{year}{2017}), \eprint{1606.09063}.

\bibitem[{\citenamefont{Feldstein et~al.}(2010)\citenamefont{Feldstein, Graham,
  and Rajendran}}]{Feldstein:2010su}
\bibinfo{author}{\bibfnamefont{B.}~\bibnamefont{Feldstein}},
  \bibinfo{author}{\bibfnamefont{P.~W.} \bibnamefont{Graham}},
  \bibnamefont{and}
  \bibinfo{author}{\bibfnamefont{S.}~\bibnamefont{Rajendran}},
  \bibinfo{journal}{Phys. Rev. D} \textbf{\bibinfo{volume}{82}},
  \bibinfo{pages}{075019} (\bibinfo{year}{2010}), \eprint{1008.1988}.

\bibitem[{\citenamefont{Pospelov et~al.}(2014)\citenamefont{Pospelov, Weiner,
  and Yavin}}]{Pospelov:2013nea}
\bibinfo{author}{\bibfnamefont{M.}~\bibnamefont{Pospelov}},
  \bibinfo{author}{\bibfnamefont{N.}~\bibnamefont{Weiner}}, \bibnamefont{and}
  \bibinfo{author}{\bibfnamefont{I.}~\bibnamefont{Yavin}},
  \bibinfo{journal}{Phys. Rev. D} \textbf{\bibinfo{volume}{89}},
  \bibinfo{pages}{055008} (\bibinfo{year}{2014}), \eprint{1312.1363}.

\bibitem[{\citenamefont{Bell et~al.}(2020{\natexlab{b}})\citenamefont{Bell,
  Dent, Dutta, Ghosh, Kumar, and Newstead}}]{Bell:2020bes}
\bibinfo{author}{\bibfnamefont{N.~F.} \bibnamefont{Bell}},
  \bibinfo{author}{\bibfnamefont{J.~B.} \bibnamefont{Dent}},
  \bibinfo{author}{\bibfnamefont{B.}~\bibnamefont{Dutta}},
  \bibinfo{author}{\bibfnamefont{S.}~\bibnamefont{Ghosh}},
  \bibinfo{author}{\bibfnamefont{J.}~\bibnamefont{Kumar}}, \bibnamefont{and}
  \bibinfo{author}{\bibfnamefont{J.~L.} \bibnamefont{Newstead}},
  \bibinfo{journal}{Phys. Rev. Lett.} \textbf{\bibinfo{volume}{125}},
  \bibinfo{pages}{161803} (\bibinfo{year}{2020}{\natexlab{b}}),
  \eprint{2006.12461}.

\bibitem[{\citenamefont{Chang et~al.}(2010{\natexlab{b}})\citenamefont{Chang,
  Weiner, and Yavin}}]{Chang:2010en}
\bibinfo{author}{\bibfnamefont{S.}~\bibnamefont{Chang}},
  \bibinfo{author}{\bibfnamefont{N.}~\bibnamefont{Weiner}}, \bibnamefont{and}
  \bibinfo{author}{\bibfnamefont{I.}~\bibnamefont{Yavin}},
  \bibinfo{journal}{Phys. Rev. D} \textbf{\bibinfo{volume}{82}},
  \bibinfo{pages}{125011} (\bibinfo{year}{2010}{\natexlab{b}}),
  \eprint{1007.4200}.

\bibitem[{\citenamefont{Kumar et~al.}(2012)\citenamefont{Kumar, Menon, and
  Tait}}]{Kumar:2011iy}
\bibinfo{author}{\bibfnamefont{K.}~\bibnamefont{Kumar}},
  \bibinfo{author}{\bibfnamefont{A.}~\bibnamefont{Menon}}, \bibnamefont{and}
  \bibinfo{author}{\bibfnamefont{T.~M.~P.} \bibnamefont{Tait}},
  \bibinfo{journal}{JHEP} \textbf{\bibinfo{volume}{02}}, \bibinfo{pages}{131}
  (\bibinfo{year}{2012}), \eprint{1111.2336}.

\bibitem[{\citenamefont{Patra and Rao}(2011)}]{Patra:2011aa}
\bibinfo{author}{\bibfnamefont{S.}~\bibnamefont{Patra}} \bibnamefont{and}
  \bibinfo{author}{\bibfnamefont{S.}~\bibnamefont{Rao}} (\bibinfo{year}{2011}),
  \eprint{1112.3454}.

\bibitem[{\citenamefont{Weiner and Yavin}(2013)}]{Weiner:2012gm}
\bibinfo{author}{\bibfnamefont{N.}~\bibnamefont{Weiner}} \bibnamefont{and}
  \bibinfo{author}{\bibfnamefont{I.}~\bibnamefont{Yavin}},
  \bibinfo{journal}{Phys. Rev. D} \textbf{\bibinfo{volume}{87}},
  \bibinfo{pages}{023523} (\bibinfo{year}{2013}), \eprint{1209.1093}.

\bibitem[{\citenamefont{Pierce and Zhang}(2014)}]{Pierce:2014spa}
\bibinfo{author}{\bibfnamefont{A.}~\bibnamefont{Pierce}} \bibnamefont{and}
  \bibinfo{author}{\bibfnamefont{Z.}~\bibnamefont{Zhang}},
  \bibinfo{journal}{Phys. Rev. D} \textbf{\bibinfo{volume}{90}},
  \bibinfo{pages}{015026} (\bibinfo{year}{2014}), \eprint{1405.1937}.

\bibitem[{\citenamefont{Lin and Finkbeiner}(2011)}]{Lin:2010sb}
\bibinfo{author}{\bibfnamefont{T.}~\bibnamefont{Lin}} \bibnamefont{and}
  \bibinfo{author}{\bibfnamefont{D.~P.} \bibnamefont{Finkbeiner}},
  \bibinfo{journal}{Phys. Rev. D} \textbf{\bibinfo{volume}{83}},
  \bibinfo{pages}{083510} (\bibinfo{year}{2011}), \eprint{1011.3052}.

\bibitem[{\citenamefont{Aprile et~al.}(2017{\natexlab{e}})}]{Aprile:2017kek}
\bibinfo{author}{\bibfnamefont{E.}~\bibnamefont{Aprile}} \bibnamefont{et~al.}
  (\bibinfo{collaboration}{XENON}), \bibinfo{journal}{JCAP}
  \textbf{\bibinfo{volume}{1710}}, \bibinfo{pages}{039}
  (\bibinfo{year}{2017}{\natexlab{e}}), \eprint{1704.05804}.

\bibitem[{\citenamefont{Kolb and Long}(2017)}]{Kolb:2017jvz}
\bibinfo{author}{\bibfnamefont{E.~W.} \bibnamefont{Kolb}} \bibnamefont{and}
  \bibinfo{author}{\bibfnamefont{A.~J.} \bibnamefont{Long}},
  \bibinfo{journal}{Phys. Rev. D} \textbf{\bibinfo{volume}{96}},
  \bibinfo{pages}{103540} (\bibinfo{year}{2017}), \eprint{1708.04293}.

\bibitem[{\citenamefont{Fujita et~al.}(2014)\citenamefont{Fujita, Kawasaki,
  Harigaya, and Matsuda}}]{Fujita:2014hha}
\bibinfo{author}{\bibfnamefont{T.}~\bibnamefont{Fujita}},
  \bibinfo{author}{\bibfnamefont{M.}~\bibnamefont{Kawasaki}},
  \bibinfo{author}{\bibfnamefont{K.}~\bibnamefont{Harigaya}}, \bibnamefont{and}
  \bibinfo{author}{\bibfnamefont{R.}~\bibnamefont{Matsuda}},
  \bibinfo{journal}{Phys. Rev. D} \textbf{\bibinfo{volume}{89}},
  \bibinfo{pages}{103501} (\bibinfo{year}{2014}), \eprint{1401.1909}.

\bibitem[{\citenamefont{Lennon et~al.}(2018)\citenamefont{Lennon,
  March-Russell, Petrossian-Byrne, and Tillim}}]{Lennon:2017tqq}
\bibinfo{author}{\bibfnamefont{O.}~\bibnamefont{Lennon}},
  \bibinfo{author}{\bibfnamefont{J.}~\bibnamefont{March-Russell}},
  \bibinfo{author}{\bibfnamefont{R.}~\bibnamefont{Petrossian-Byrne}},
  \bibnamefont{and} \bibinfo{author}{\bibfnamefont{H.}~\bibnamefont{Tillim}},
  \bibinfo{journal}{JCAP} \textbf{\bibinfo{volume}{04}}, \bibinfo{pages}{009}
  (\bibinfo{year}{2018}), \eprint{1712.07664}.

\bibitem[{\citenamefont{Asadi et~al.}(2021)\citenamefont{Asadi, Slatyer, and
  Smirnov}}]{Asadi:2021bxp}
\bibinfo{author}{\bibfnamefont{P.}~\bibnamefont{Asadi}},
  \bibinfo{author}{\bibfnamefont{T.~R.} \bibnamefont{Slatyer}},
  \bibnamefont{and} \bibinfo{author}{\bibfnamefont{J.}~\bibnamefont{Smirnov}}
  (\bibinfo{year}{2021}), \eprint{2111.11444}.

\bibitem[{\citenamefont{Chung et~al.}(1999)\citenamefont{Chung, Kolb, and
  Riotto}}]{Chung:1998zb}
\bibinfo{author}{\bibfnamefont{D.~J.~H.} \bibnamefont{Chung}},
  \bibinfo{author}{\bibfnamefont{E.~W.} \bibnamefont{Kolb}}, \bibnamefont{and}
  \bibinfo{author}{\bibfnamefont{A.}~\bibnamefont{Riotto}},
  \bibinfo{journal}{Phys. Rev. D} \textbf{\bibinfo{volume}{59}},
  \bibinfo{pages}{023501} (\bibinfo{year}{1999}), \eprint{hep-ph/9802238}.

\bibitem[{\citenamefont{Hamdan and Unwin}(2018)}]{Hamdan:2017psw}
\bibinfo{author}{\bibfnamefont{S.}~\bibnamefont{Hamdan}} \bibnamefont{and}
  \bibinfo{author}{\bibfnamefont{J.}~\bibnamefont{Unwin}},
  \bibinfo{journal}{Mod. Phys. Lett. A} \textbf{\bibinfo{volume}{33}},
  \bibinfo{pages}{1850181} (\bibinfo{year}{2018}), \eprint{1710.03758}.

\bibitem[{\citenamefont{Davoudiasl and Mohlabeng}(2018)}]{Davoudiasl:2018wxz}
\bibinfo{author}{\bibfnamefont{H.}~\bibnamefont{Davoudiasl}} \bibnamefont{and}
  \bibinfo{author}{\bibfnamefont{G.}~\bibnamefont{Mohlabeng}},
  \bibinfo{journal}{Phys. Rev. D} \textbf{\bibinfo{volume}{98}},
  \bibinfo{pages}{115035} (\bibinfo{year}{2018}), \eprint{1809.07768}.

\bibitem[{\citenamefont{Bramante
  et~al.}(2019{\natexlab{a}})\citenamefont{Bramante, Broerman, Kumar, Lang,
  Pospelov, and Raj}}]{Bramante:2018tos}
\bibinfo{author}{\bibfnamefont{J.}~\bibnamefont{Bramante}},
  \bibinfo{author}{\bibfnamefont{B.}~\bibnamefont{Broerman}},
  \bibinfo{author}{\bibfnamefont{J.}~\bibnamefont{Kumar}},
  \bibinfo{author}{\bibfnamefont{R.~F.} \bibnamefont{Lang}},
  \bibinfo{author}{\bibfnamefont{M.}~\bibnamefont{Pospelov}}, \bibnamefont{and}
  \bibinfo{author}{\bibfnamefont{N.}~\bibnamefont{Raj}},
  \bibinfo{journal}{Phys. Rev. D} \textbf{\bibinfo{volume}{99}},
  \bibinfo{pages}{083010} (\bibinfo{year}{2019}{\natexlab{a}}),
  \eprint{1812.09325}.

\bibitem[{\citenamefont{Bramante
  et~al.}(2019{\natexlab{b}})\citenamefont{Bramante, Kumar, and
  Raj}}]{Bramante:2019yss}
\bibinfo{author}{\bibfnamefont{J.}~\bibnamefont{Bramante}},
  \bibinfo{author}{\bibfnamefont{J.}~\bibnamefont{Kumar}}, \bibnamefont{and}
  \bibinfo{author}{\bibfnamefont{N.}~\bibnamefont{Raj}},
  \bibinfo{journal}{Phys. Rev. D} \textbf{\bibinfo{volume}{100}},
  \bibinfo{pages}{123016} (\bibinfo{year}{2019}{\natexlab{b}}),
  \eprint{1910.05380}.

\bibitem[{\citenamefont{Clark et~al.}(2020)\citenamefont{Clark, Depoian,
  Elshimy, Kopec, Lang, and Qin}}]{Clark:2020mna}
\bibinfo{author}{\bibfnamefont{M.}~\bibnamefont{Clark}},
  \bibinfo{author}{\bibfnamefont{A.}~\bibnamefont{Depoian}},
  \bibinfo{author}{\bibfnamefont{B.}~\bibnamefont{Elshimy}},
  \bibinfo{author}{\bibfnamefont{A.}~\bibnamefont{Kopec}},
  \bibinfo{author}{\bibfnamefont{R.~F.} \bibnamefont{Lang}}, \bibnamefont{and}
  \bibinfo{author}{\bibfnamefont{J.}~\bibnamefont{Qin}},
  \bibinfo{journal}{Phys. Rev. D} \textbf{\bibinfo{volume}{102}},
  \bibinfo{pages}{123026} (\bibinfo{year}{2020}), \eprint{2009.07909}.

\bibitem[{\citenamefont{Adhikari et~al.}(2022)}]{Adhikari:2021fum}
\bibinfo{author}{\bibfnamefont{P.}~\bibnamefont{Adhikari}} \bibnamefont{et~al.}
  (\bibinfo{collaboration}{(DEAP Collaboration)\textdaggerdbl{}, DEAP}),
  \bibinfo{journal}{Phys. Rev. Lett.} \textbf{\bibinfo{volume}{128}},
  \bibinfo{pages}{011801} (\bibinfo{year}{2022}), \eprint{2108.09405}.

\bibitem[{\citenamefont{Acevedo et~al.}(2022)\citenamefont{Acevedo, Bramante,
  and Goodman}}]{Acevedo:2021kly}
\bibinfo{author}{\bibfnamefont{J.~F.} \bibnamefont{Acevedo}},
  \bibinfo{author}{\bibfnamefont{J.}~\bibnamefont{Bramante}}, \bibnamefont{and}
  \bibinfo{author}{\bibfnamefont{A.}~\bibnamefont{Goodman}},
  \bibinfo{journal}{Phys. Rev. D} \textbf{\bibinfo{volume}{105}},
  \bibinfo{pages}{023012} (\bibinfo{year}{2022}), \eprint{2108.10889}.

\bibitem[{\citenamefont{Doi and Kotani}(1993)}]{Doi:1992dm}
\bibinfo{author}{\bibfnamefont{M.}~\bibnamefont{Doi}} \bibnamefont{and}
  \bibinfo{author}{\bibfnamefont{T.}~\bibnamefont{Kotani}},
  \bibinfo{journal}{Prog. Theor. Phys.} \textbf{\bibinfo{volume}{89}},
  \bibinfo{pages}{139} (\bibinfo{year}{1993}).

\bibitem[{\citenamefont{Avignone et~al.}(2008)\citenamefont{Avignone, Elliott,
  and Engel}}]{Avignone:2007fu}
\bibinfo{author}{\bibfnamefont{I.}~\bibnamefont{Avignone},
  \bibfnamefont{Frank~T.}}, \bibinfo{author}{\bibfnamefont{S.~R.}
  \bibnamefont{Elliott}}, \bibnamefont{and}
  \bibinfo{author}{\bibfnamefont{J.}~\bibnamefont{Engel}},
  \bibinfo{journal}{Rev. Mod. Phys.} \textbf{\bibinfo{volume}{80}},
  \bibinfo{pages}{481} (\bibinfo{year}{2008}), \eprint{0708.1033}.

\bibitem[{\citenamefont{Dolinski et~al.}(2019)\citenamefont{Dolinski, Poon, and
  Rodejohann}}]{Dolinski:2019nrj}
\bibinfo{author}{\bibfnamefont{M.~J.} \bibnamefont{Dolinski}},
  \bibinfo{author}{\bibfnamefont{A.~W.} \bibnamefont{Poon}}, \bibnamefont{and}
  \bibinfo{author}{\bibfnamefont{W.}~\bibnamefont{Rodejohann}},
  \bibinfo{journal}{Ann. Rev. Nucl. Part. Sci.} \textbf{\bibinfo{volume}{69}},
  \bibinfo{pages}{219} (\bibinfo{year}{2019}), \eprint{1902.04097}.

\bibitem[{\citenamefont{Agostini et~al.}(2022)\citenamefont{Agostini, Benato,
  Detwiler, Men\'endez, and Vissani}}]{Agostini:2022zub}
\bibinfo{author}{\bibfnamefont{M.}~\bibnamefont{Agostini}},
  \bibinfo{author}{\bibfnamefont{G.}~\bibnamefont{Benato}},
  \bibinfo{author}{\bibfnamefont{J.~A.} \bibnamefont{Detwiler}},
  \bibinfo{author}{\bibfnamefont{J.}~\bibnamefont{Men\'endez}},
  \bibnamefont{and} \bibinfo{author}{\bibfnamefont{F.}~\bibnamefont{Vissani}}
  (\bibinfo{year}{2022}), \eprint{2202.01787}.

\bibitem[{\citenamefont{Akerib et~al.}(2021{\natexlab{d}})}]{LZ:2021rff}
\bibinfo{author}{\bibfnamefont{D.~S.} \bibnamefont{Akerib}}
  \bibnamefont{et~al.} (\bibinfo{collaboration}{LUX-ZEPLIN, LZ}),
  \bibinfo{journal}{Phys. Rev. C} \textbf{\bibinfo{volume}{104}},
  \bibinfo{pages}{065501} (\bibinfo{year}{2021}{\natexlab{d}}),
  \eprint{2104.13374}.

\bibitem[{\citenamefont{Gando et~al.}(2016)}]{KamLAND-Zen:2016pfg}
\bibinfo{author}{\bibfnamefont{A.}~\bibnamefont{Gando}} \bibnamefont{et~al.}
  (\bibinfo{collaboration}{KamLAND-Zen}), \bibinfo{journal}{Phys. Rev. Lett.}
  \textbf{\bibinfo{volume}{117}}, \bibinfo{pages}{082503}
  (\bibinfo{year}{2016}), \bibinfo{note}{[Addendum: Phys. Rev. Lett. {\bf 117},
  109903 (2016)]}, \eprint{1605.02889}.

\bibitem[{\citenamefont{Adams et~al.}(2021{\natexlab{a}})}]{CUORE:2021gpk}
\bibinfo{author}{\bibfnamefont{D.~Q.} \bibnamefont{Adams}} \bibnamefont{et~al.}
  (\bibinfo{collaboration}{CUORE}) (\bibinfo{year}{2021}{\natexlab{a}}),
  \eprint{2104.06906}.

\bibitem[{\citenamefont{Agostini et~al.}(2020{\natexlab{a}})}]{GERDA:2020xhi}
\bibinfo{author}{\bibfnamefont{M.}~\bibnamefont{Agostini}} \bibnamefont{et~al.}
  (\bibinfo{collaboration}{GERDA}), \bibinfo{journal}{Phys. Rev. Lett.}
  \textbf{\bibinfo{volume}{125}}, \bibinfo{pages}{252502}
  (\bibinfo{year}{2020}{\natexlab{a}}), \eprint{2009.06079}.

\bibitem[{\citenamefont{Armengaud et~al.}(2021)}]{CUPID:2020aow}
\bibinfo{author}{\bibfnamefont{E.}~\bibnamefont{Armengaud}}
  \bibnamefont{et~al.} (\bibinfo{collaboration}{CUPID}),
  \bibinfo{journal}{Phys. Rev. Lett.} \textbf{\bibinfo{volume}{126}},
  \bibinfo{pages}{181802} (\bibinfo{year}{2021}), \eprint{2011.13243}.

\bibitem[{\citenamefont{Alvis et~al.}(2019)}]{Majorana:2019nbd}
\bibinfo{author}{\bibfnamefont{S.~I.} \bibnamefont{Alvis}} \bibnamefont{et~al.}
  (\bibinfo{collaboration}{Majorana}), \bibinfo{journal}{Phys. Rev. C}
  \textbf{\bibinfo{volume}{100}}, \bibinfo{pages}{025501}
  (\bibinfo{year}{2019}), \eprint{1902.02299}.

\bibitem[{\citenamefont{Anton et~al.}(2019)}]{EXO-200:2019rkq}
\bibinfo{author}{\bibfnamefont{G.}~\bibnamefont{Anton}} \bibnamefont{et~al.}
  (\bibinfo{collaboration}{EXO-200}), \bibinfo{journal}{Phys. Rev. Lett.}
  \textbf{\bibinfo{volume}{123}}, \bibinfo{pages}{161802}
  (\bibinfo{year}{2019}), \eprint{1906.02723}.

\bibitem[{\citenamefont{Azzolini et~al.}(2019)}]{CUPID:2019gpc}
\bibinfo{author}{\bibfnamefont{O.}~\bibnamefont{Azzolini}} \bibnamefont{et~al.}
  (\bibinfo{collaboration}{CUPID}), \bibinfo{journal}{Phys. Rev. Lett.}
  \textbf{\bibinfo{volume}{123}}, \bibinfo{pages}{032501}
  (\bibinfo{year}{2019}), \eprint{1906.05001}.

\bibitem[{\citenamefont{Albert et~al.}(2018{\natexlab{a}})}]{Albert:2017owj}
\bibinfo{author}{\bibfnamefont{J.}~\bibnamefont{Albert}} \bibnamefont{et~al.}
  (\bibinfo{collaboration}{EXO}), \bibinfo{journal}{Phys. Rev. Lett.}
  \textbf{\bibinfo{volume}{120}}, \bibinfo{pages}{072701}
  (\bibinfo{year}{2018}{\natexlab{a}}), \eprint{1707.08707}.

\bibitem[{\citenamefont{Ni et~al.}(2019)}]{Ni:2019kms}
\bibinfo{author}{\bibfnamefont{K.}~\bibnamefont{Ni}} \bibnamefont{et~al.}
  (\bibinfo{collaboration}{PandaX-II}), \bibinfo{journal}{Chin. Phys. C}
  \textbf{\bibinfo{volume}{43}}, \bibinfo{pages}{113001}
  (\bibinfo{year}{2019}), \eprint{1906.11457}.

\bibitem[{\citenamefont{Agostini
  et~al.}(2020{\natexlab{b}})}]{Agostini:2020adk}
\bibinfo{author}{\bibfnamefont{F.}~\bibnamefont{Agostini}} \bibnamefont{et~al.}
  (\bibinfo{collaboration}{DARWIN}), \bibinfo{journal}{Eur. Phys. J. C}
  \textbf{\bibinfo{volume}{80}}, \bibinfo{pages}{808}
  (\bibinfo{year}{2020}{\natexlab{b}}), \eprint{2003.13407}.

\bibitem[{\citenamefont{Gomez-Cadenas}(2019)}]{Gomez_NEXT:2019}
\bibinfo{author}{\bibfnamefont{J.~J.} \bibnamefont{Gomez-Cadenas}}, pp.
  \bibinfo{pages}{201--206} (\bibinfo{year}{2019}), \eprint{1906.01743}.

\bibitem[{\citenamefont{Chen et~al.}(2017{\natexlab{c}})}]{Chen:2016qcd}
\bibinfo{author}{\bibfnamefont{X.}~\bibnamefont{Chen}} \bibnamefont{et~al.},
  \bibinfo{journal}{Sci. China Phys. Mech. Astron.}
  \textbf{\bibinfo{volume}{60}}, \bibinfo{pages}{061011}
  (\bibinfo{year}{2017}{\natexlab{c}}), \eprint{1610.08883}.

\bibitem[{\citenamefont{Albert et~al.}(2018{\natexlab{b}})}]{Albert:2017hjq}
\bibinfo{author}{\bibfnamefont{J.~B.} \bibnamefont{Albert}}
  \bibnamefont{et~al.} (\bibinfo{collaboration}{nEXO}), \bibinfo{journal}{Phys.
  Rev. C} \textbf{\bibinfo{volume}{97}}, \bibinfo{pages}{065503}
  (\bibinfo{year}{2018}{\natexlab{b}}), \eprint{1710.05075}.

\bibitem[{\citenamefont{Barabash}(2015)}]{Barabash:2015eza}
\bibinfo{author}{\bibfnamefont{A.~S.} \bibnamefont{Barabash}},
  \bibinfo{journal}{Nucl. Phys. A} \textbf{\bibinfo{volume}{935}},
  \bibinfo{pages}{52} (\bibinfo{year}{2015}), \eprint{1501.05133}.

\bibitem[{\citenamefont{Cebri\'an}(2020)}]{Cebrian:2020bwn}
\bibinfo{author}{\bibfnamefont{S.}~\bibnamefont{Cebri\'an}},
  \bibinfo{journal}{Universe} \textbf{\bibinfo{volume}{6}},
  \bibinfo{pages}{162} (\bibinfo{year}{2020}), \eprint{2010.02381}.

\bibitem[{\citenamefont{Rogers et~al.}(2020)}]{Rogers:2020npx}
\bibinfo{author}{\bibfnamefont{L.}~\bibnamefont{Rogers}} \bibnamefont{et~al.}
  (\bibinfo{collaboration}{NEXT}), \bibinfo{journal}{J. Phys. G}
  \textbf{\bibinfo{volume}{47}}, \bibinfo{pages}{075001}
  (\bibinfo{year}{2020}), \eprint{2001.11147}.

\bibitem[{\citenamefont{Pas et~al.}(1999)\citenamefont{Pas, Hirsch,
  Klapdor-Kleingrothaus, and Kovalenko}}]{Pas:1999fc}
\bibinfo{author}{\bibfnamefont{H.}~\bibnamefont{Pas}},
  \bibinfo{author}{\bibfnamefont{M.}~\bibnamefont{Hirsch}},
  \bibinfo{author}{\bibfnamefont{H.~V.} \bibnamefont{Klapdor-Kleingrothaus}},
  \bibnamefont{and} \bibinfo{author}{\bibfnamefont{S.~G.}
  \bibnamefont{Kovalenko}}, \bibinfo{journal}{Phys. Lett. B}
  \textbf{\bibinfo{volume}{453}}, \bibinfo{pages}{194} (\bibinfo{year}{1999}).

\bibitem[{\citenamefont{Pas et~al.}(2001)\citenamefont{Pas, Hirsch,
  Klapdor-Kleingrothaus, and Kovalenko}}]{Pas:2000vn}
\bibinfo{author}{\bibfnamefont{H.}~\bibnamefont{Pas}},
  \bibinfo{author}{\bibfnamefont{M.}~\bibnamefont{Hirsch}},
  \bibinfo{author}{\bibfnamefont{H.~V.} \bibnamefont{Klapdor-Kleingrothaus}},
  \bibnamefont{and} \bibinfo{author}{\bibfnamefont{S.~G.}
  \bibnamefont{Kovalenko}}, \bibinfo{journal}{Phys. Lett. B}
  \textbf{\bibinfo{volume}{498}}, \bibinfo{pages}{35} (\bibinfo{year}{2001}),
  \eprint{hep-ph/0008182}.

\bibitem[{\citenamefont{Pr{\'e}zeau
  et~al.}(2003{\natexlab{b}})\citenamefont{Pr{\'e}zeau, Ramsey-Musolf, and
  Vogel}}]{Prezeau:2003xn}
\bibinfo{author}{\bibfnamefont{G.}~\bibnamefont{Pr{\'e}zeau}},
  \bibinfo{author}{\bibfnamefont{M.}~\bibnamefont{Ramsey-Musolf}},
  \bibnamefont{and} \bibinfo{author}{\bibfnamefont{P.}~\bibnamefont{Vogel}},
  \bibinfo{journal}{Phys. Rev. D} \textbf{\bibinfo{volume}{68}},
  \bibinfo{pages}{034016} (\bibinfo{year}{2003}{\natexlab{b}}),
  \eprint{hep-ph/0303205}.

\bibitem[{\citenamefont{de~Gouvea and Jenkins}(2008)}]{deGouvea:2007qla}
\bibinfo{author}{\bibfnamefont{A.}~\bibnamefont{de~Gouvea}} \bibnamefont{and}
  \bibinfo{author}{\bibfnamefont{J.}~\bibnamefont{Jenkins}},
  \bibinfo{journal}{Phys. Rev. D} \textbf{\bibinfo{volume}{77}},
  \bibinfo{pages}{013008} (\bibinfo{year}{2008}), \eprint{0708.1344}.

\bibitem[{\citenamefont{Cirigliano et~al.}(2017)\citenamefont{Cirigliano,
  Dekens, de~Vries, Graesser, and Mereghetti}}]{Cirigliano:2017djv}
\bibinfo{author}{\bibfnamefont{V.}~\bibnamefont{Cirigliano}},
  \bibinfo{author}{\bibfnamefont{W.}~\bibnamefont{Dekens}},
  \bibinfo{author}{\bibfnamefont{J.}~\bibnamefont{de~Vries}},
  \bibinfo{author}{\bibfnamefont{M.}~\bibnamefont{Graesser}}, \bibnamefont{and}
  \bibinfo{author}{\bibfnamefont{E.}~\bibnamefont{Mereghetti}},
  \bibinfo{journal}{JHEP} \textbf{\bibinfo{volume}{12}}, \bibinfo{pages}{082}
  (\bibinfo{year}{2017}), \eprint{1708.09390}.

\bibitem[{\citenamefont{Graf et~al.}(2018)\citenamefont{Graf, Deppisch,
  Iachello, and Kotila}}]{Graf:2018ozy}
\bibinfo{author}{\bibfnamefont{L.}~\bibnamefont{Graf}},
  \bibinfo{author}{\bibfnamefont{F.~F.} \bibnamefont{Deppisch}},
  \bibinfo{author}{\bibfnamefont{F.}~\bibnamefont{Iachello}}, \bibnamefont{and}
  \bibinfo{author}{\bibfnamefont{J.}~\bibnamefont{Kotila}},
  \bibinfo{journal}{Phys. Rev. D} \textbf{\bibinfo{volume}{98}},
  \bibinfo{pages}{095023} (\bibinfo{year}{2018}), \eprint{1806.06058}.

\bibitem[{\citenamefont{Cirigliano
  et~al.}(2018{\natexlab{a}})\citenamefont{Cirigliano, Dekens, de~Vries,
  Graesser, and Mereghetti}}]{Cirigliano:2018yza}
\bibinfo{author}{\bibfnamefont{V.}~\bibnamefont{Cirigliano}},
  \bibinfo{author}{\bibfnamefont{W.}~\bibnamefont{Dekens}},
  \bibinfo{author}{\bibfnamefont{J.}~\bibnamefont{de~Vries}},
  \bibinfo{author}{\bibfnamefont{M.}~\bibnamefont{Graesser}}, \bibnamefont{and}
  \bibinfo{author}{\bibfnamefont{E.}~\bibnamefont{Mereghetti}},
  \bibinfo{journal}{JHEP} \textbf{\bibinfo{volume}{12}}, \bibinfo{pages}{097}
  (\bibinfo{year}{2018}{\natexlab{a}}), \eprint{1806.02780}.

\bibitem[{\citenamefont{Dekens et~al.}(2020)\citenamefont{Dekens, de~Vries,
  Fuyuto, Mereghetti, and Zhou}}]{Dekens:2020ttz}
\bibinfo{author}{\bibfnamefont{W.}~\bibnamefont{Dekens}},
  \bibinfo{author}{\bibfnamefont{J.}~\bibnamefont{de~Vries}},
  \bibinfo{author}{\bibfnamefont{K.}~\bibnamefont{Fuyuto}},
  \bibinfo{author}{\bibfnamefont{E.}~\bibnamefont{Mereghetti}},
  \bibnamefont{and} \bibinfo{author}{\bibfnamefont{G.}~\bibnamefont{Zhou}},
  \bibinfo{journal}{JHEP} \textbf{\bibinfo{volume}{06}}, \bibinfo{pages}{097}
  (\bibinfo{year}{2020}), \eprint{2002.07182}.

\bibitem[{\citenamefont{Deppisch
  et~al.}(2020{\natexlab{a}})\citenamefont{Deppisch, Graf, Iachello, and
  Kotila}}]{Deppisch:2020ztt}
\bibinfo{author}{\bibfnamefont{F.~F.} \bibnamefont{Deppisch}},
  \bibinfo{author}{\bibfnamefont{L.}~\bibnamefont{Graf}},
  \bibinfo{author}{\bibfnamefont{F.}~\bibnamefont{Iachello}}, \bibnamefont{and}
  \bibinfo{author}{\bibfnamefont{J.}~\bibnamefont{Kotila}},
  \bibinfo{journal}{Phys. Rev. D} \textbf{\bibinfo{volume}{102}},
  \bibinfo{pages}{095016} (\bibinfo{year}{2020}{\natexlab{a}}),
  \eprint{2009.10119}.

\bibitem[{\citenamefont{Kotila et~al.}(2021)\citenamefont{Kotila, Ferretti, and
  Iachello}}]{Kotila:2021xgw}
\bibinfo{author}{\bibfnamefont{J.}~\bibnamefont{Kotila}},
  \bibinfo{author}{\bibfnamefont{J.}~\bibnamefont{Ferretti}}, \bibnamefont{and}
  \bibinfo{author}{\bibfnamefont{F.}~\bibnamefont{Iachello}}
  (\bibinfo{year}{2021}), \eprint{2110.09141}.

\bibitem[{\citenamefont{Men{\'e}ndez et~al.}(2011)\citenamefont{Men{\'e}ndez,
  Gazit, and Schwenk}}]{Menendez:2011qq}
\bibinfo{author}{\bibfnamefont{J.}~\bibnamefont{Men{\'e}ndez}},
  \bibinfo{author}{\bibfnamefont{D.}~\bibnamefont{Gazit}}, \bibnamefont{and}
  \bibinfo{author}{\bibfnamefont{A.}~\bibnamefont{Schwenk}},
  \bibinfo{journal}{Phys. Rev. Lett.} \textbf{\bibinfo{volume}{107}},
  \bibinfo{pages}{062501} (\bibinfo{year}{2011}), \eprint{1103.3622}.

\bibitem[{\citenamefont{Wang et~al.}(2018)\citenamefont{Wang, Engel, and
  Yao}}]{Wang:2018htk}
\bibinfo{author}{\bibfnamefont{L.-J.} \bibnamefont{Wang}},
  \bibinfo{author}{\bibfnamefont{J.}~\bibnamefont{Engel}}, \bibnamefont{and}
  \bibinfo{author}{\bibfnamefont{J.~M.} \bibnamefont{Yao}},
  \bibinfo{journal}{Phys. Rev. C} \textbf{\bibinfo{volume}{98}},
  \bibinfo{pages}{031301} (\bibinfo{year}{2018}), \eprint{1805.10276}.

\bibitem[{\citenamefont{Cirigliano
  et~al.}(2018{\natexlab{b}})\citenamefont{Cirigliano, Dekens, Mereghetti, and
  Walker-Loud}}]{Cirigliano:2017tvr}
\bibinfo{author}{\bibfnamefont{V.}~\bibnamefont{Cirigliano}},
  \bibinfo{author}{\bibfnamefont{W.}~\bibnamefont{Dekens}},
  \bibinfo{author}{\bibfnamefont{E.}~\bibnamefont{Mereghetti}},
  \bibnamefont{and}
  \bibinfo{author}{\bibfnamefont{A.}~\bibnamefont{Walker-Loud}},
  \bibinfo{journal}{Phys. Rev. C} \textbf{\bibinfo{volume}{97}},
  \bibinfo{pages}{065501} (\bibinfo{year}{2018}{\natexlab{b}}),
  \bibinfo{note}{[Erratum: Phys. Rev. C {\bf 100}, 019903 (2019)]},
  \eprint{1710.01729}.

\bibitem[{\citenamefont{Pastore et~al.}(2018)\citenamefont{Pastore, Carlson,
  Cirigliano, Dekens, Mereghetti, and Wiringa}}]{Pastore:2017ofx}
\bibinfo{author}{\bibfnamefont{S.}~\bibnamefont{Pastore}},
  \bibinfo{author}{\bibfnamefont{J.}~\bibnamefont{Carlson}},
  \bibinfo{author}{\bibfnamefont{V.}~\bibnamefont{Cirigliano}},
  \bibinfo{author}{\bibfnamefont{W.}~\bibnamefont{Dekens}},
  \bibinfo{author}{\bibfnamefont{E.}~\bibnamefont{Mereghetti}},
  \bibnamefont{and} \bibinfo{author}{\bibfnamefont{R.}~\bibnamefont{Wiringa}},
  \bibinfo{journal}{Phys. Rev. C} \textbf{\bibinfo{volume}{97}},
  \bibinfo{pages}{014606} (\bibinfo{year}{2018}), \eprint{1710.05026}.

\bibitem[{\citenamefont{Cirigliano
  et~al.}(2018{\natexlab{c}})\citenamefont{Cirigliano, Dekens, de~Vries,
  Graesser, Mereghetti, Pastore, and Van~Kolck}}]{Cirigliano:2018hja}
\bibinfo{author}{\bibfnamefont{V.}~\bibnamefont{Cirigliano}},
  \bibinfo{author}{\bibfnamefont{W.}~\bibnamefont{Dekens}},
  \bibinfo{author}{\bibfnamefont{J.}~\bibnamefont{de~Vries}},
  \bibinfo{author}{\bibfnamefont{M.~L.} \bibnamefont{Graesser}},
  \bibinfo{author}{\bibfnamefont{E.}~\bibnamefont{Mereghetti}},
  \bibinfo{author}{\bibfnamefont{S.}~\bibnamefont{Pastore}}, \bibnamefont{and}
  \bibinfo{author}{\bibfnamefont{U.}~\bibnamefont{Van~Kolck}},
  \bibinfo{journal}{Phys. Rev. Lett.} \textbf{\bibinfo{volume}{120}},
  \bibinfo{pages}{202001} (\bibinfo{year}{2018}{\natexlab{c}}),
  \eprint{1802.10097}.

\bibitem[{\citenamefont{Cirigliano et~al.}(2019)\citenamefont{Cirigliano,
  Dekens, de~Vries, Graesser, Mereghetti, Pastore, Piarulli, Van~Kolck, and
  Wiringa}}]{Cirigliano:2019vdj}
\bibinfo{author}{\bibfnamefont{V.}~\bibnamefont{Cirigliano}},
  \bibinfo{author}{\bibfnamefont{W.}~\bibnamefont{Dekens}},
  \bibinfo{author}{\bibfnamefont{J.}~\bibnamefont{de~Vries}},
  \bibinfo{author}{\bibfnamefont{M.}~\bibnamefont{Graesser}},
  \bibinfo{author}{\bibfnamefont{E.}~\bibnamefont{Mereghetti}},
  \bibinfo{author}{\bibfnamefont{S.}~\bibnamefont{Pastore}},
  \bibinfo{author}{\bibfnamefont{M.}~\bibnamefont{Piarulli}},
  \bibinfo{author}{\bibfnamefont{U.}~\bibnamefont{Van~Kolck}},
  \bibnamefont{and} \bibinfo{author}{\bibfnamefont{R.}~\bibnamefont{Wiringa}},
  \bibinfo{journal}{Phys. Rev. C} \textbf{\bibinfo{volume}{100}},
  \bibinfo{pages}{055504} (\bibinfo{year}{2019}), \eprint{1907.11254}.

\bibitem[{\citenamefont{Cirigliano
  et~al.}(2021{\natexlab{a}})\citenamefont{Cirigliano, Dekens, de~Vries,
  Hoferichter, and Mereghetti}}]{Cirigliano:2020dmx}
\bibinfo{author}{\bibfnamefont{V.}~\bibnamefont{Cirigliano}},
  \bibinfo{author}{\bibfnamefont{W.}~\bibnamefont{Dekens}},
  \bibinfo{author}{\bibfnamefont{J.}~\bibnamefont{de~Vries}},
  \bibinfo{author}{\bibfnamefont{M.}~\bibnamefont{Hoferichter}},
  \bibnamefont{and}
  \bibinfo{author}{\bibfnamefont{E.}~\bibnamefont{Mereghetti}},
  \bibinfo{journal}{Phys. Rev. Lett.} \textbf{\bibinfo{volume}{126}},
  \bibinfo{pages}{172002} (\bibinfo{year}{2021}{\natexlab{a}}),
  \eprint{2012.11602}.

\bibitem[{\citenamefont{Cirigliano
  et~al.}(2021{\natexlab{b}})\citenamefont{Cirigliano, Dekens, de~Vries,
  Hoferichter, and Mereghetti}}]{Cirigliano:2021qko}
\bibinfo{author}{\bibfnamefont{V.}~\bibnamefont{Cirigliano}},
  \bibinfo{author}{\bibfnamefont{W.}~\bibnamefont{Dekens}},
  \bibinfo{author}{\bibfnamefont{J.}~\bibnamefont{de~Vries}},
  \bibinfo{author}{\bibfnamefont{M.}~\bibnamefont{Hoferichter}},
  \bibnamefont{and}
  \bibinfo{author}{\bibfnamefont{E.}~\bibnamefont{Mereghetti}},
  \bibinfo{journal}{JHEP} \textbf{\bibinfo{volume}{05}}, \bibinfo{pages}{289}
  (\bibinfo{year}{2021}{\natexlab{b}}), \eprint{2102.03371}.

\bibitem[{\citenamefont{Jokiniemi et~al.}(2021)\citenamefont{Jokiniemi,
  Soriano, and Men\'endez}}]{Jokiniemi:2021qqv}
\bibinfo{author}{\bibfnamefont{L.}~\bibnamefont{Jokiniemi}},
  \bibinfo{author}{\bibfnamefont{P.}~\bibnamefont{Soriano}}, \bibnamefont{and}
  \bibinfo{author}{\bibfnamefont{J.}~\bibnamefont{Men\'endez}},
  \bibinfo{journal}{Phys. Lett. B} \textbf{\bibinfo{volume}{823}},
  \bibinfo{pages}{136720} (\bibinfo{year}{2021}), \eprint{2107.13354}.

\bibitem[{\citenamefont{Wirth et~al.}(2021)\citenamefont{Wirth, Yao, and
  Hergert}}]{Wirth:2021pij}
\bibinfo{author}{\bibfnamefont{R.}~\bibnamefont{Wirth}},
  \bibinfo{author}{\bibfnamefont{J.~M.} \bibnamefont{Yao}}, \bibnamefont{and}
  \bibinfo{author}{\bibfnamefont{H.}~\bibnamefont{Hergert}},
  \bibinfo{journal}{Phys. Rev. Lett.} \textbf{\bibinfo{volume}{127}},
  \bibinfo{pages}{242502} (\bibinfo{year}{2021}), \eprint{2105.05415}.

\bibitem[{\citenamefont{Barabash}(2020)}]{Barabash:2020nck}
\bibinfo{author}{\bibfnamefont{A.}~\bibnamefont{Barabash}},
  \bibinfo{journal}{Universe} \textbf{\bibinfo{volume}{6}},
  \bibinfo{pages}{159} (\bibinfo{year}{2020}), \eprint{2009.14451}.

\bibitem[{\citenamefont{Saakyan}(2013)}]{Saakyan:2013yna}
\bibinfo{author}{\bibfnamefont{R.}~\bibnamefont{Saakyan}},
  \bibinfo{journal}{Ann. Rev. Nucl. Part. Sci.} \textbf{\bibinfo{volume}{63}},
  \bibinfo{pages}{503} (\bibinfo{year}{2013}).

\bibitem[{\citenamefont{Engel and Men\'endez}(2017)}]{Engel:2016xgb}
\bibinfo{author}{\bibfnamefont{J.}~\bibnamefont{Engel}} \bibnamefont{and}
  \bibinfo{author}{\bibfnamefont{J.}~\bibnamefont{Men\'endez}},
  \bibinfo{journal}{Rept. Prog. Phys.} \textbf{\bibinfo{volume}{80}},
  \bibinfo{pages}{046301} (\bibinfo{year}{2017}), \eprint{1610.06548}.

\bibitem[{\citenamefont{Rodriguez and
  Martinez-Pinedo}(2010)}]{Rodriguez:2010mn}
\bibinfo{author}{\bibfnamefont{T.~R.} \bibnamefont{Rodriguez}}
  \bibnamefont{and}
  \bibinfo{author}{\bibfnamefont{G.}~\bibnamefont{Martinez-Pinedo}},
  \bibinfo{journal}{Phys. Rev. Lett.} \textbf{\bibinfo{volume}{105}},
  \bibinfo{pages}{252503} (\bibinfo{year}{2010}), \eprint{1008.5260}.

\bibitem[{\citenamefont{Mustonen and Engel}(2013)}]{Mustonen:2013zu}
\bibinfo{author}{\bibfnamefont{M.~T.} \bibnamefont{Mustonen}} \bibnamefont{and}
  \bibinfo{author}{\bibfnamefont{J.}~\bibnamefont{Engel}},
  \bibinfo{journal}{Phys. Rev. C} \textbf{\bibinfo{volume}{87}},
  \bibinfo{pages}{064302} (\bibinfo{year}{2013}), \eprint{1301.6997}.

\bibitem[{\citenamefont{Hyv\"arinen and Suhonen}(2015)}]{Hyvarinen:2015bda}
\bibinfo{author}{\bibfnamefont{J.}~\bibnamefont{Hyv\"arinen}} \bibnamefont{and}
  \bibinfo{author}{\bibfnamefont{J.}~\bibnamefont{Suhonen}},
  \bibinfo{journal}{Phys. Rev. C} \textbf{\bibinfo{volume}{91}},
  \bibinfo{pages}{024613} (\bibinfo{year}{2015}).

\bibitem[{\citenamefont{Horoi and Neacsu}(2016)}]{Horoi:2015tkc}
\bibinfo{author}{\bibfnamefont{M.}~\bibnamefont{Horoi}} \bibnamefont{and}
  \bibinfo{author}{\bibfnamefont{A.}~\bibnamefont{Neacsu}},
  \bibinfo{journal}{Phys. Rev. C} \textbf{\bibinfo{volume}{93}},
  \bibinfo{pages}{024308} (\bibinfo{year}{2016}), \eprint{1511.03711}.

\bibitem[{\citenamefont{Men\'endez}(2018)}]{Menendez:2017fdf}
\bibinfo{author}{\bibfnamefont{J.}~\bibnamefont{Men\'endez}},
  \bibinfo{journal}{J. Phys. G} \textbf{\bibinfo{volume}{45}},
  \bibinfo{pages}{014003} (\bibinfo{year}{2018}), \eprint{1804.02105}.

\bibitem[{\citenamefont{Song et~al.}(2017)\citenamefont{Song, Yao, Ring, and
  Meng}}]{Song:2017ktj}
\bibinfo{author}{\bibfnamefont{L.~S.} \bibnamefont{Song}},
  \bibinfo{author}{\bibfnamefont{J.~M.} \bibnamefont{Yao}},
  \bibinfo{author}{\bibfnamefont{P.}~\bibnamefont{Ring}}, \bibnamefont{and}
  \bibinfo{author}{\bibfnamefont{J.}~\bibnamefont{Meng}},
  \bibinfo{journal}{Phys. Rev. C} \textbf{\bibinfo{volume}{95}},
  \bibinfo{pages}{024305} (\bibinfo{year}{2017}), \eprint{1702.02448}.

\bibitem[{\citenamefont{\v{S}imkovic et~al.}(2018)\citenamefont{\v{S}imkovic,
  Smetana, and Vogel}}]{Simkovic:2018hiq}
\bibinfo{author}{\bibfnamefont{F.}~\bibnamefont{\v{S}imkovic}},
  \bibinfo{author}{\bibfnamefont{A.}~\bibnamefont{Smetana}}, \bibnamefont{and}
  \bibinfo{author}{\bibfnamefont{P.}~\bibnamefont{Vogel}},
  \bibinfo{journal}{Phys. Rev. C} \textbf{\bibinfo{volume}{98}},
  \bibinfo{pages}{064325} (\bibinfo{year}{2018}), \eprint{1808.05016}.

\bibitem[{\citenamefont{Fang et~al.}(2018)\citenamefont{Fang, Faessler, and
  Simkovic}}]{Fang:2018tui}
\bibinfo{author}{\bibfnamefont{D.-L.} \bibnamefont{Fang}},
  \bibinfo{author}{\bibfnamefont{A.}~\bibnamefont{Faessler}}, \bibnamefont{and}
  \bibinfo{author}{\bibfnamefont{F.}~\bibnamefont{Simkovic}},
  \bibinfo{journal}{Phys. Rev. C} \textbf{\bibinfo{volume}{97}},
  \bibinfo{pages}{045503} (\bibinfo{year}{2018}), \eprint{1803.09195}.

\bibitem[{\citenamefont{Coraggio et~al.}(2020)\citenamefont{Coraggio, Gargano,
  Itaco, Mancino, and Nowacki}}]{Coraggio:2020hwx}
\bibinfo{author}{\bibfnamefont{L.}~\bibnamefont{Coraggio}},
  \bibinfo{author}{\bibfnamefont{A.}~\bibnamefont{Gargano}},
  \bibinfo{author}{\bibfnamefont{N.}~\bibnamefont{Itaco}},
  \bibinfo{author}{\bibfnamefont{R.}~\bibnamefont{Mancino}}, \bibnamefont{and}
  \bibinfo{author}{\bibfnamefont{F.}~\bibnamefont{Nowacki}},
  \bibinfo{journal}{Phys. Rev. C} \textbf{\bibinfo{volume}{101}},
  \bibinfo{pages}{044315} (\bibinfo{year}{2020}), \eprint{2001.00890}.

\bibitem[{\citenamefont{Yao et~al.}(2020)\citenamefont{Yao, Bally, Engel,
  Wirth, Rodr\'\i{}guez, and Hergert}}]{Yao:2019rck}
\bibinfo{author}{\bibfnamefont{J.~M.} \bibnamefont{Yao}},
  \bibinfo{author}{\bibfnamefont{B.}~\bibnamefont{Bally}},
  \bibinfo{author}{\bibfnamefont{J.}~\bibnamefont{Engel}},
  \bibinfo{author}{\bibfnamefont{R.}~\bibnamefont{Wirth}},
  \bibinfo{author}{\bibfnamefont{T.~R.} \bibnamefont{Rodr\'\i{}guez}},
  \bibnamefont{and} \bibinfo{author}{\bibfnamefont{H.}~\bibnamefont{Hergert}},
  \bibinfo{journal}{Phys. Rev. Lett.} \textbf{\bibinfo{volume}{124}},
  \bibinfo{pages}{232501} (\bibinfo{year}{2020}), \eprint{1908.05424}.

\bibitem[{\citenamefont{Novario et~al.}(2021)\citenamefont{Novario, Gysbers,
  Engel, Hagen, Jansen, Morris, Navr\'atil, Papenbrock, and
  Quaglioni}}]{Novario:2020dmr}
\bibinfo{author}{\bibfnamefont{S.}~\bibnamefont{Novario}},
  \bibinfo{author}{\bibfnamefont{P.}~\bibnamefont{Gysbers}},
  \bibinfo{author}{\bibfnamefont{J.}~\bibnamefont{Engel}},
  \bibinfo{author}{\bibfnamefont{G.}~\bibnamefont{Hagen}},
  \bibinfo{author}{\bibfnamefont{G.~R.} \bibnamefont{Jansen}},
  \bibinfo{author}{\bibfnamefont{T.~D.} \bibnamefont{Morris}},
  \bibinfo{author}{\bibfnamefont{P.}~\bibnamefont{Navr\'atil}},
  \bibinfo{author}{\bibfnamefont{T.}~\bibnamefont{Papenbrock}},
  \bibnamefont{and}
  \bibinfo{author}{\bibfnamefont{S.}~\bibnamefont{Quaglioni}},
  \bibinfo{journal}{Phys. Rev. Lett.} \textbf{\bibinfo{volume}{126}},
  \bibinfo{pages}{182502} (\bibinfo{year}{2021}), \eprint{2008.09696}.

\bibitem[{\citenamefont{Belley et~al.}(2021)\citenamefont{Belley, Payne,
  Stroberg, Miyagi, and Holt}}]{Belley:2020ejd}
\bibinfo{author}{\bibfnamefont{A.}~\bibnamefont{Belley}},
  \bibinfo{author}{\bibfnamefont{C.~G.} \bibnamefont{Payne}},
  \bibinfo{author}{\bibfnamefont{S.~R.} \bibnamefont{Stroberg}},
  \bibinfo{author}{\bibfnamefont{T.}~\bibnamefont{Miyagi}}, \bibnamefont{and}
  \bibinfo{author}{\bibfnamefont{J.~D.} \bibnamefont{Holt}},
  \bibinfo{journal}{Phys. Rev. Lett.} \textbf{\bibinfo{volume}{126}},
  \bibinfo{pages}{042502} (\bibinfo{year}{2021}), \eprint{2008.06588}.

\bibitem[{\citenamefont{Gando et~al.}(2019)}]{KamLAND-Zen:2019imh}
\bibinfo{author}{\bibfnamefont{A.}~\bibnamefont{Gando}} \bibnamefont{et~al.}
  (\bibinfo{collaboration}{KamLAND-Zen}), \bibinfo{journal}{Phys. Rev. Lett.}
  \textbf{\bibinfo{volume}{122}}, \bibinfo{pages}{192501}
  (\bibinfo{year}{2019}), \eprint{1901.03871}.

\bibitem[{\citenamefont{Deppisch
  et~al.}(2020{\natexlab{b}})\citenamefont{Deppisch, Graf, and
  \v{S}imkovic}}]{Deppisch:2020mxv}
\bibinfo{author}{\bibfnamefont{F.~F.} \bibnamefont{Deppisch}},
  \bibinfo{author}{\bibfnamefont{L.}~\bibnamefont{Graf}}, \bibnamefont{and}
  \bibinfo{author}{\bibfnamefont{F.}~\bibnamefont{\v{S}imkovic}},
  \bibinfo{journal}{Phys. Rev. Lett.} \textbf{\bibinfo{volume}{125}},
  \bibinfo{pages}{171801} (\bibinfo{year}{2020}{\natexlab{b}}),
  \eprint{2003.11836}.

\bibitem[{\citenamefont{Bolton et~al.}(2021)\citenamefont{Bolton, Deppisch,
  Gr\'af, and \v{S}imkovic}}]{Bolton:2020ncv}
\bibinfo{author}{\bibfnamefont{P.~D.} \bibnamefont{Bolton}},
  \bibinfo{author}{\bibfnamefont{F.~F.} \bibnamefont{Deppisch}},
  \bibinfo{author}{\bibfnamefont{L.}~\bibnamefont{Gr\'af}}, \bibnamefont{and}
  \bibinfo{author}{\bibfnamefont{F.}~\bibnamefont{\v{S}imkovic}},
  \bibinfo{journal}{Phys. Rev. D} \textbf{\bibinfo{volume}{103}},
  \bibinfo{pages}{055019} (\bibinfo{year}{2021}), \eprint{2011.13387}.

\bibitem[{\citenamefont{Agostini et~al.}(2021)\citenamefont{Agostini, Bossio,
  Ibarra, and Marcano}}]{Agostini:2020cpz}
\bibinfo{author}{\bibfnamefont{M.}~\bibnamefont{Agostini}},
  \bibinfo{author}{\bibfnamefont{E.}~\bibnamefont{Bossio}},
  \bibinfo{author}{\bibfnamefont{A.}~\bibnamefont{Ibarra}}, \bibnamefont{and}
  \bibinfo{author}{\bibfnamefont{X.}~\bibnamefont{Marcano}},
  \bibinfo{journal}{Phys. Lett. B} \textbf{\bibinfo{volume}{815}},
  \bibinfo{pages}{136127} (\bibinfo{year}{2021}), \eprint{2012.09281}.

\bibitem[{\citenamefont{Deppisch
  et~al.}(2020{\natexlab{c}})\citenamefont{Deppisch, Graf, Rodejohann, and
  Xu}}]{Deppisch:2020sqh}
\bibinfo{author}{\bibfnamefont{F.~F.} \bibnamefont{Deppisch}},
  \bibinfo{author}{\bibfnamefont{L.}~\bibnamefont{Graf}},
  \bibinfo{author}{\bibfnamefont{W.}~\bibnamefont{Rodejohann}},
  \bibnamefont{and} \bibinfo{author}{\bibfnamefont{X.-J.} \bibnamefont{Xu}},
  \bibinfo{journal}{Phys. Rev. D} \textbf{\bibinfo{volume}{102}},
  \bibinfo{pages}{051701} (\bibinfo{year}{2020}{\natexlab{c}}),
  \eprint{2004.11919}.

\bibitem[{\citenamefont{Winter}(1955)}]{Winter:1955zz}
\bibinfo{author}{\bibfnamefont{R.~G.} \bibnamefont{Winter}},
  \bibinfo{journal}{Phys. Rev.} \textbf{\bibinfo{volume}{100}},
  \bibinfo{pages}{142} (\bibinfo{year}{1955}).

\bibitem[{\citenamefont{Blaum et~al.}(2020)\citenamefont{Blaum, Eliseev,
  Danevich, Tretyak, Kovalenko, Krivoruchenko, Novikov, and
  Suhonen}}]{Blaum:2020ogl}
\bibinfo{author}{\bibfnamefont{K.}~\bibnamefont{Blaum}},
  \bibinfo{author}{\bibfnamefont{S.}~\bibnamefont{Eliseev}},
  \bibinfo{author}{\bibfnamefont{F.~A.} \bibnamefont{Danevich}},
  \bibinfo{author}{\bibfnamefont{V.~I.} \bibnamefont{Tretyak}},
  \bibinfo{author}{\bibfnamefont{S.}~\bibnamefont{Kovalenko}},
  \bibinfo{author}{\bibfnamefont{M.~I.} \bibnamefont{Krivoruchenko}},
  \bibinfo{author}{\bibfnamefont{Y.~N.} \bibnamefont{Novikov}},
  \bibnamefont{and} \bibinfo{author}{\bibfnamefont{J.}~\bibnamefont{Suhonen}},
  \bibinfo{journal}{Rev. Mod. Phys.} \textbf{\bibinfo{volume}{92}},
  \bibinfo{pages}{045007} (\bibinfo{year}{2020}), \eprint{2007.14908}.

\bibitem[{\citenamefont{Nesterenko et~al.}(2012)}]{Nesterenko:2012xp}
\bibinfo{author}{\bibfnamefont{D.}~\bibnamefont{Nesterenko}}
  \bibnamefont{et~al.}, \bibinfo{journal}{Phys. Rev. C}
  \textbf{\bibinfo{volume}{86}}, \bibinfo{pages}{044313}
  (\bibinfo{year}{2012}).

\bibitem[{\citenamefont{Doi and Kotani}(1992)}]{Doi:1991xf}
\bibinfo{author}{\bibfnamefont{M.}~\bibnamefont{Doi}} \bibnamefont{and}
  \bibinfo{author}{\bibfnamefont{T.}~\bibnamefont{Kotani}},
  \bibinfo{journal}{Prog. Theor. Phys.} \textbf{\bibinfo{volume}{87}},
  \bibinfo{pages}{1207} (\bibinfo{year}{1992}).

\bibitem[{\citenamefont{Suhonen}(2013)}]{Suhonen:2013rca}
\bibinfo{author}{\bibfnamefont{J.}~\bibnamefont{Suhonen}}, \bibinfo{journal}{J.
  Phys. G} \textbf{\bibinfo{volume}{40}}, \bibinfo{pages}{075102}
  (\bibinfo{year}{2013}).

\bibitem[{\citenamefont{Pirinen and Suhonen}(2015)}]{Pirinen:2015sma}
\bibinfo{author}{\bibfnamefont{P.}~\bibnamefont{Pirinen}} \bibnamefont{and}
  \bibinfo{author}{\bibfnamefont{J.}~\bibnamefont{Suhonen}},
  \bibinfo{journal}{Phys. Rev. C} \textbf{\bibinfo{volume}{91}},
  \bibinfo{pages}{054309} (\bibinfo{year}{2015}).

\bibitem[{\citenamefont{Coello~Pérez et~al.}(2019)\citenamefont{Coello~Pérez,
  Menéndez, and Schwenk}}]{Perez:2018cly}
\bibinfo{author}{\bibfnamefont{E.}~\bibnamefont{Coello~Pérez}},
  \bibinfo{author}{\bibfnamefont{J.}~\bibnamefont{Menéndez}},
  \bibnamefont{and} \bibinfo{author}{\bibfnamefont{A.}~\bibnamefont{Schwenk}},
  \bibinfo{journal}{Phys. Lett. B} \textbf{\bibinfo{volume}{797}},
  \bibinfo{pages}{134885} (\bibinfo{year}{2019}), \eprint{1809.04443}.

\bibitem[{\citenamefont{Abe et~al.}(2018{\natexlab{b}})}]{Abe:2018gyq}
\bibinfo{author}{\bibfnamefont{K.}~\bibnamefont{Abe}} \bibnamefont{et~al.}
  (\bibinfo{collaboration}{XMASS}), \bibinfo{journal}{PTEP}
  \textbf{\bibinfo{volume}{2018}}, \bibinfo{pages}{053D03}
  (\bibinfo{year}{2018}{\natexlab{b}}), \eprint{1801.03251}.

\bibitem[{\citenamefont{Aprile et~al.}(2019{\natexlab{e}})}]{XENON:2019dti}
\bibinfo{author}{\bibfnamefont{E.}~\bibnamefont{Aprile}} \bibnamefont{et~al.}
  (\bibinfo{collaboration}{XENON}), \bibinfo{journal}{Nature}
  \textbf{\bibinfo{volume}{568}}, \bibinfo{pages}{532}
  (\bibinfo{year}{2019}{\natexlab{e}}), \eprint{1904.11002}.

\bibitem[{\citenamefont{Bernabeu et~al.}(1983)\citenamefont{Bernabeu,
  De~Rujula, and Jarlskog}}]{Bernabeu:1983yb}
\bibinfo{author}{\bibfnamefont{J.}~\bibnamefont{Bernabeu}},
  \bibinfo{author}{\bibfnamefont{A.}~\bibnamefont{De~Rujula}},
  \bibnamefont{and} \bibinfo{author}{\bibfnamefont{C.}~\bibnamefont{Jarlskog}},
  \bibinfo{journal}{Nucl. Phys. B} \textbf{\bibinfo{volume}{223}},
  \bibinfo{pages}{15} (\bibinfo{year}{1983}).

\bibitem[{\citenamefont{Sujkowski and Wycech}(2004)}]{Sujkowski:2003mb}
\bibinfo{author}{\bibfnamefont{Z.}~\bibnamefont{Sujkowski}} \bibnamefont{and}
  \bibinfo{author}{\bibfnamefont{S.}~\bibnamefont{Wycech}},
  \bibinfo{journal}{Phys. Rev. C} \textbf{\bibinfo{volume}{70}},
  \bibinfo{pages}{052501} (\bibinfo{year}{2004}), \eprint{hep-ph/0312040}.

\bibitem[{\citenamefont{Kotila et~al.}(2014)\citenamefont{Kotila, Barea, and
  Iachello}}]{Kotila:2014zya}
\bibinfo{author}{\bibfnamefont{J.}~\bibnamefont{Kotila}},
  \bibinfo{author}{\bibfnamefont{J.}~\bibnamefont{Barea}}, \bibnamefont{and}
  \bibinfo{author}{\bibfnamefont{F.}~\bibnamefont{Iachello}},
  \bibinfo{journal}{Phys. Rev. C} \textbf{\bibinfo{volume}{89}},
  \bibinfo{pages}{064319} (\bibinfo{year}{2014}), \eprint{1509.01927}.

\bibitem[{\citenamefont{Wittweg et~al.}(2020)\citenamefont{Wittweg, Lenardo,
  Fieguth, and Weinheimer}}]{Wittweg:2020fak}
\bibinfo{author}{\bibfnamefont{C.}~\bibnamefont{Wittweg}},
  \bibinfo{author}{\bibfnamefont{B.}~\bibnamefont{Lenardo}},
  \bibinfo{author}{\bibfnamefont{A.}~\bibnamefont{Fieguth}}, \bibnamefont{and}
  \bibinfo{author}{\bibfnamefont{C.}~\bibnamefont{Weinheimer}},
  \bibinfo{journal}{Eur. Phys. J. C} \textbf{\bibinfo{volume}{80}},
  \bibinfo{pages}{1161} (\bibinfo{year}{2020}), \eprint{2002.04239}.

\bibitem[{\citenamefont{Kim and Kubodera}(1983)}]{Kim:1982vi}
\bibinfo{author}{\bibfnamefont{C.}~\bibnamefont{Kim}} \bibnamefont{and}
  \bibinfo{author}{\bibfnamefont{K.}~\bibnamefont{Kubodera}},
  \bibinfo{journal}{Phys. Rev. D} \textbf{\bibinfo{volume}{27}},
  \bibinfo{pages}{2765} (\bibinfo{year}{1983}).

\bibitem[{\citenamefont{Kotila and Iachello}(2013)}]{Kotila:2013gea}
\bibinfo{author}{\bibfnamefont{J.}~\bibnamefont{Kotila}} \bibnamefont{and}
  \bibinfo{author}{\bibfnamefont{F.}~\bibnamefont{Iachello}},
  \bibinfo{journal}{Phys. Rev. C} \textbf{\bibinfo{volume}{87}},
  \bibinfo{pages}{024313} (\bibinfo{year}{2013}), \eprint{1303.4124}.

\bibitem[{\citenamefont{Barros et~al.}(2014)\citenamefont{Barros, Thurn, and
  Zuber}}]{Barros:2014exa}
\bibinfo{author}{\bibfnamefont{N.}~\bibnamefont{Barros}},
  \bibinfo{author}{\bibfnamefont{J.}~\bibnamefont{Thurn}}, \bibnamefont{and}
  \bibinfo{author}{\bibfnamefont{K.}~\bibnamefont{Zuber}}, \bibinfo{journal}{J.
  Phys. G} \textbf{\bibinfo{volume}{41}}, \bibinfo{pages}{115105}
  (\bibinfo{year}{2014}), \eprint{1409.8308}.

\bibitem[{\citenamefont{Bolozdynya
  et~al.}(1997{\natexlab{a}})\citenamefont{Bolozdynya, Egorov, Koutchenkov
  et~al.}}]{Bolozdynya:1997pdbd}
\bibinfo{author}{\bibfnamefont{A.}~\bibnamefont{Bolozdynya}},
  \bibinfo{author}{\bibfnamefont{V.}~\bibnamefont{Egorov}},
  \bibinfo{author}{\bibfnamefont{A.}~\bibnamefont{Koutchenkov}},
  \bibnamefont{et~al.}, \bibinfo{journal}{IEEE Trans. Nucl. Sci.}
  \textbf{\bibinfo{volume}{44}}, \bibinfo{pages}{1046}
  (\bibinfo{year}{1997}{\natexlab{a}}).

\bibitem[{\citenamefont{Rath et~al.}(2009)\citenamefont{Rath, Chandra,
  Chaturvedi, Raina, and Hirsch}}]{Rath:2009dr}
\bibinfo{author}{\bibfnamefont{P.}~\bibnamefont{Rath}},
  \bibinfo{author}{\bibfnamefont{R.}~\bibnamefont{Chandra}},
  \bibinfo{author}{\bibfnamefont{K.}~\bibnamefont{Chaturvedi}},
  \bibinfo{author}{\bibfnamefont{P.}~\bibnamefont{Raina}}, \bibnamefont{and}
  \bibinfo{author}{\bibfnamefont{J.}~\bibnamefont{Hirsch}},
  \bibinfo{journal}{Phys. Rev. C} \textbf{\bibinfo{volume}{80}},
  \bibinfo{pages}{044303} (\bibinfo{year}{2009}), \eprint{0906.4476}.

\bibitem[{\citenamefont{Barea et~al.}(2013)\citenamefont{Barea, Kotila, and
  Iachello}}]{Barea:2013wga}
\bibinfo{author}{\bibfnamefont{J.}~\bibnamefont{Barea}},
  \bibinfo{author}{\bibfnamefont{J.}~\bibnamefont{Kotila}}, \bibnamefont{and}
  \bibinfo{author}{\bibfnamefont{F.}~\bibnamefont{Iachello}},
  \bibinfo{journal}{Phys. Rev. C} \textbf{\bibinfo{volume}{87}},
  \bibinfo{pages}{057301} (\bibinfo{year}{2013}), \eprint{1509.05157}.

\bibitem[{\citenamefont{Kotila and Iachello}(2012)}]{Kotila:2012zza}
\bibinfo{author}{\bibfnamefont{J.}~\bibnamefont{Kotila}} \bibnamefont{and}
  \bibinfo{author}{\bibfnamefont{F.}~\bibnamefont{Iachello}},
  \bibinfo{journal}{Phys. Rev. C} \textbf{\bibinfo{volume}{85}},
  \bibinfo{pages}{034316} (\bibinfo{year}{2012}), \eprint{1209.5722}.

\bibitem[{\citenamefont{Angloher
  et~al.}(2016{\natexlab{b}})}]{Angloher:2016ktr}
\bibinfo{author}{\bibfnamefont{G.}~\bibnamefont{Angloher}}
  \bibnamefont{et~al.}, \bibinfo{journal}{J. Phys. G}
  \textbf{\bibinfo{volume}{43}}, \bibinfo{pages}{095202}
  (\bibinfo{year}{2016}{\natexlab{b}}), \eprint{1604.08493}.

\bibitem[{\citenamefont{Lehnert et~al.}(2016)\citenamefont{Lehnert, Degering,
  Frotscher, Michel, and Zuber}}]{Lehnert:2016gra}
\bibinfo{author}{\bibfnamefont{B.}~\bibnamefont{Lehnert}},
  \bibinfo{author}{\bibfnamefont{D.}~\bibnamefont{Degering}},
  \bibinfo{author}{\bibfnamefont{A.}~\bibnamefont{Frotscher}},
  \bibinfo{author}{\bibfnamefont{T.}~\bibnamefont{Michel}}, \bibnamefont{and}
  \bibinfo{author}{\bibfnamefont{K.}~\bibnamefont{Zuber}}, \bibinfo{journal}{J.
  Phys. G} \textbf{\bibinfo{volume}{43}}, \bibinfo{pages}{065201}
  (\bibinfo{year}{2016}), \eprint{1605.03287}.

\bibitem[{\citenamefont{Alduino et~al.}(2018)}]{CUORE:2017dbf}
\bibinfo{author}{\bibfnamefont{C.}~\bibnamefont{Alduino}} \bibnamefont{et~al.}
  (\bibinfo{collaboration}{CUORE}), \bibinfo{journal}{Phys. Rev. C}
  \textbf{\bibinfo{volume}{97}}, \bibinfo{pages}{055502}
  (\bibinfo{year}{2018}), \eprint{1710.07459}.

\bibitem[{\citenamefont{Hirsch et~al.}(1994)\citenamefont{Hirsch, Muto, Oda,
  and Klapdor-Kleingrothaus}}]{Hirsch:1994es}
\bibinfo{author}{\bibfnamefont{M.}~\bibnamefont{Hirsch}},
  \bibinfo{author}{\bibfnamefont{K.}~\bibnamefont{Muto}},
  \bibinfo{author}{\bibfnamefont{T.}~\bibnamefont{Oda}}, \bibnamefont{and}
  \bibinfo{author}{\bibfnamefont{H.}~\bibnamefont{Klapdor-Kleingrothaus}},
  \bibinfo{journal}{Z. Phys. A} \textbf{\bibinfo{volume}{347}},
  \bibinfo{pages}{151} (\bibinfo{year}{1994}).

\bibitem[{\citenamefont{Deppisch and Pas}(2007)}]{Deppisch:2006hb}
\bibinfo{author}{\bibfnamefont{F.}~\bibnamefont{Deppisch}} \bibnamefont{and}
  \bibinfo{author}{\bibfnamefont{H.}~\bibnamefont{Pas}},
  \bibinfo{journal}{Phys. Rev. Lett.} \textbf{\bibinfo{volume}{98}},
  \bibinfo{pages}{232501} (\bibinfo{year}{2007}), \eprint{hep-ph/0612165}.

\bibitem[{\citenamefont{Gehman and Elliott}(2007)}]{Gehman:2007qg}
\bibinfo{author}{\bibfnamefont{V.~M.} \bibnamefont{Gehman}} \bibnamefont{and}
  \bibinfo{author}{\bibfnamefont{S.~R.} \bibnamefont{Elliott}},
  \bibinfo{journal}{J. Phys. G} \textbf{\bibinfo{volume}{34}},
  \bibinfo{pages}{667} (\bibinfo{year}{2007}), \bibinfo{note}{[Erratum:
  J.Phys.G 35, 029701 (2008)]}, \eprint{hep-ph/0701099}.

\bibitem[{\citenamefont{Vitagliano et~al.}(2020)\citenamefont{Vitagliano,
  Tamborra, and Raffelt}}]{Vitagliano:2019yzm}
\bibinfo{author}{\bibfnamefont{E.}~\bibnamefont{Vitagliano}},
  \bibinfo{author}{\bibfnamefont{I.}~\bibnamefont{Tamborra}}, \bibnamefont{and}
  \bibinfo{author}{\bibfnamefont{G.}~\bibnamefont{Raffelt}},
  \bibinfo{journal}{Rev. Mod. Phys.} \textbf{\bibinfo{volume}{92}},
  \bibinfo{pages}{45006} (\bibinfo{year}{2020}), \eprint{1910.11878}.

\bibitem[{\citenamefont{Gann et~al.}(2021)\citenamefont{Gann, Zuber, Bemmerer,
  and Serenelli}}]{Gann:2021ndb}
\bibinfo{author}{\bibfnamefont{G.~D.~O.} \bibnamefont{Gann}},
  \bibinfo{author}{\bibfnamefont{K.}~\bibnamefont{Zuber}},
  \bibinfo{author}{\bibfnamefont{D.}~\bibnamefont{Bemmerer}}, \bibnamefont{and}
  \bibinfo{author}{\bibfnamefont{A.}~\bibnamefont{Serenelli}},
  \bibinfo{journal}{Ann. Rev. Nucl. Part. Sci.} \textbf{\bibinfo{volume}{71}},
  \bibinfo{pages}{491} (\bibinfo{year}{2021}), \eprint{2107.08613}.

\bibitem[{\citenamefont{Dutta and Strigari}(2019)}]{Dutta:2019oaj}
\bibinfo{author}{\bibfnamefont{B.}~\bibnamefont{Dutta}} \bibnamefont{and}
  \bibinfo{author}{\bibfnamefont{L.~E.} \bibnamefont{Strigari}},
  \bibinfo{journal}{Ann. Rev. Nucl. Part. Sci.} \textbf{\bibinfo{volume}{69}},
  \bibinfo{pages}{137} (\bibinfo{year}{2019}), \eprint{1901.08876}.

\bibitem[{\citenamefont{Abe et~al.}(2018{\natexlab{c}})}]{Abe:2018uyc}
\bibinfo{author}{\bibfnamefont{K.}~\bibnamefont{Abe}} \bibnamefont{et~al.}
  (\bibinfo{collaboration}{Hyper-Kamiokande})
  (\bibinfo{year}{2018}{\natexlab{c}}), \eprint{1805.04163}.

\bibitem[{\citenamefont{Acciarri et~al.}(2016)}]{Acciarri:2016crz}
\bibinfo{author}{\bibfnamefont{R.}~\bibnamefont{Acciarri}} \bibnamefont{et~al.}
  (\bibinfo{collaboration}{DUNE}) (\bibinfo{year}{2016}), \eprint{1601.05471}.

\bibitem[{\citenamefont{Engel}(1991)}]{Engel:1991wq}
\bibinfo{author}{\bibfnamefont{J.}~\bibnamefont{Engel}},
  \bibinfo{journal}{Phys. Lett. B} \textbf{\bibinfo{volume}{264}},
  \bibinfo{pages}{114} (\bibinfo{year}{1991}).

\bibitem[{\citenamefont{Hoferichter et~al.}(2020)\citenamefont{Hoferichter,
  Men\'endez, and Schwenk}}]{Hoferichter:2020osn}
\bibinfo{author}{\bibfnamefont{M.}~\bibnamefont{Hoferichter}},
  \bibinfo{author}{\bibfnamefont{J.}~\bibnamefont{Men\'endez}},
  \bibnamefont{and} \bibinfo{author}{\bibfnamefont{A.}~\bibnamefont{Schwenk}},
  \bibinfo{journal}{Phys. Rev. D} \textbf{\bibinfo{volume}{102}},
  \bibinfo{pages}{074018} (\bibinfo{year}{2020}), \eprint{2007.08529}.

\bibitem[{\citenamefont{Akimov et~al.}(2017{\natexlab{a}})}]{Akimov:2017ade}
\bibinfo{author}{\bibfnamefont{D.}~\bibnamefont{Akimov}} \bibnamefont{et~al.}
  (\bibinfo{collaboration}{COHERENT}), \bibinfo{journal}{Science}
  \textbf{\bibinfo{volume}{357}}, \bibinfo{pages}{1123}
  (\bibinfo{year}{2017}{\natexlab{a}}), \eprint{1708.01294}.

\bibitem[{\citenamefont{Cabrera et~al.}(1985)\citenamefont{Cabrera, Krauss, and
  Wilczek}}]{Cabrera:1984rr}
\bibinfo{author}{\bibfnamefont{B.}~\bibnamefont{Cabrera}},
  \bibinfo{author}{\bibfnamefont{L.~M.} \bibnamefont{Krauss}},
  \bibnamefont{and} \bibinfo{author}{\bibfnamefont{F.}~\bibnamefont{Wilczek}},
  \bibinfo{journal}{Phys. Rev. Lett.} \textbf{\bibinfo{volume}{55}},
  \bibinfo{pages}{25} (\bibinfo{year}{1985}).

\bibitem[{\citenamefont{Krauss and Wilczek}(1985)}]{Krauss:1985pf}
\bibinfo{author}{\bibfnamefont{L.}~\bibnamefont{Krauss}} \bibnamefont{and}
  \bibinfo{author}{\bibfnamefont{F.}~\bibnamefont{Wilczek}},
  \bibinfo{journal}{Phys. Rev. Lett.} \textbf{\bibinfo{volume}{55}},
  \bibinfo{pages}{122} (\bibinfo{year}{1985}).

\bibitem[{\citenamefont{Bahcall and Davis}(1976)}]{Bahcall:1976zz}
\bibinfo{author}{\bibfnamefont{J.~N.} \bibnamefont{Bahcall}} \bibnamefont{and}
  \bibinfo{author}{\bibfnamefont{R.}~\bibnamefont{Davis}},
  \bibinfo{journal}{Science} \textbf{\bibinfo{volume}{191}},
  \bibinfo{pages}{264} (\bibinfo{year}{1976}).

\bibitem[{\citenamefont{Haxton et~al.}(2013)\citenamefont{Haxton,
  Hamish~Robertson, and Serenelli}}]{Robertson:2012ib}
\bibinfo{author}{\bibfnamefont{W.}~\bibnamefont{Haxton}},
  \bibinfo{author}{\bibfnamefont{R.}~\bibnamefont{Hamish~Robertson}},
  \bibnamefont{and} \bibinfo{author}{\bibfnamefont{A.~M.}
  \bibnamefont{Serenelli}}, \bibinfo{journal}{Ann. Rev. Astron. Astrophys.}
  \textbf{\bibinfo{volume}{51}}, \bibinfo{pages}{21} (\bibinfo{year}{2013}),
  \eprint{1208.5723}.

\bibitem[{\citenamefont{de~Holanda and Smirnov}(2003)}]{deHolanda:2002dko}
\bibinfo{author}{\bibfnamefont{P.~C.} \bibnamefont{de~Holanda}}
  \bibnamefont{and} \bibinfo{author}{\bibfnamefont{A.}~\bibnamefont{Smirnov}},
  \bibinfo{journal}{JCAP} \textbf{\bibinfo{volume}{02}}, \bibinfo{pages}{001}
  (\bibinfo{year}{2003}), \eprint{hep-ph/0212270}.

\bibitem[{\citenamefont{Abe et~al.}(2016)}]{Abe:2016nxk}
\bibinfo{author}{\bibfnamefont{K.}~\bibnamefont{Abe}} \bibnamefont{et~al.}
  (\bibinfo{collaboration}{Super-Kamiokande}), \bibinfo{journal}{Phys. Rev. D}
  \textbf{\bibinfo{volume}{94}}, \bibinfo{pages}{052010}
  (\bibinfo{year}{2016}), \eprint{1606.07538}.

\bibitem[{\citenamefont{Aharmim et~al.}(2013)}]{Aharmim:2011vm}
\bibinfo{author}{\bibfnamefont{B.}~\bibnamefont{Aharmim}} \bibnamefont{et~al.}
  (\bibinfo{collaboration}{SNO}), \bibinfo{journal}{Phys. Rev. C}
  \textbf{\bibinfo{volume}{88}}, \bibinfo{pages}{025501}
  (\bibinfo{year}{2013}), \eprint{1109.0763}.

\bibitem[{\citenamefont{Agostini et~al.}(2018)}]{Agostini:2018uly}
\bibinfo{author}{\bibfnamefont{M.}~\bibnamefont{Agostini}} \bibnamefont{et~al.}
  (\bibinfo{collaboration}{BOREXINO}), \bibinfo{journal}{Nature}
  \textbf{\bibinfo{volume}{562}}, \bibinfo{pages}{505} (\bibinfo{year}{2018}).

\bibitem[{\citenamefont{Gando et~al.}(2011)}]{Gando:2010aa}
\bibinfo{author}{\bibfnamefont{A.}~\bibnamefont{Gando}} \bibnamefont{et~al.}
  (\bibinfo{collaboration}{KamLAND}), \bibinfo{journal}{Phys. Rev. D}
  \textbf{\bibinfo{volume}{83}}, \bibinfo{pages}{052002}
  (\bibinfo{year}{2011}), \eprint{1009.4771}.

\bibitem[{\citenamefont{Liao et~al.}(2017)\citenamefont{Liao, Marfatia, and
  Whisnant}}]{Liao:2017awz}
\bibinfo{author}{\bibfnamefont{J.}~\bibnamefont{Liao}},
  \bibinfo{author}{\bibfnamefont{D.}~\bibnamefont{Marfatia}}, \bibnamefont{and}
  \bibinfo{author}{\bibfnamefont{K.}~\bibnamefont{Whisnant}},
  \bibinfo{journal}{Phys. Lett. B} \textbf{\bibinfo{volume}{771}},
  \bibinfo{pages}{247} (\bibinfo{year}{2017}), \eprint{1704.04711}.

\bibitem[{\citenamefont{Newstead et~al.}(2019)\citenamefont{Newstead, Strigari,
  and Lang}}]{Newstead:2018muu}
\bibinfo{author}{\bibfnamefont{J.~L.} \bibnamefont{Newstead}},
  \bibinfo{author}{\bibfnamefont{L.~E.} \bibnamefont{Strigari}},
  \bibnamefont{and} \bibinfo{author}{\bibfnamefont{R.~F.} \bibnamefont{Lang}},
  \bibinfo{journal}{Phys. Rev. D} \textbf{\bibinfo{volume}{99}},
  \bibinfo{pages}{043006} (\bibinfo{year}{2019}), \eprint{1807.07169}.

\bibitem[{\citenamefont{Grevesse and Sauval}(1998)}]{Grevesse:1998bj}
\bibinfo{author}{\bibfnamefont{N.}~\bibnamefont{Grevesse}} \bibnamefont{and}
  \bibinfo{author}{\bibfnamefont{A.~J.} \bibnamefont{Sauval}},
  \bibinfo{journal}{Space Sci. Rev.} \textbf{\bibinfo{volume}{85}},
  \bibinfo{pages}{161} (\bibinfo{year}{1998}).

\bibitem[{\citenamefont{Asplund et~al.}(2009)\citenamefont{Asplund, Grevesse,
  Sauval, and Scott}}]{Asplund:2009fu}
\bibinfo{author}{\bibfnamefont{M.}~\bibnamefont{Asplund}},
  \bibinfo{author}{\bibfnamefont{N.}~\bibnamefont{Grevesse}},
  \bibinfo{author}{\bibfnamefont{A.~J.} \bibnamefont{Sauval}},
  \bibnamefont{and} \bibinfo{author}{\bibfnamefont{P.}~\bibnamefont{Scott}},
  \bibinfo{journal}{Annual Review of Astronomy and Astrophysics}
  \textbf{\bibinfo{volume}{47}}, \bibinfo{pages}{481–522}
  (\bibinfo{year}{2009}), ISSN \bibinfo{issn}{1545-4282},
  \urlprefix\url{http://dx.doi.org/10.1146/annurev.astro.46.060407.145222}.

\bibitem[{\citenamefont{Scott et~al.}(2015{\natexlab{a}})\citenamefont{Scott,
  Grevesse, Asplund, Sauval, Lind, Takeda, Collet, Trampedach, and
  Hayek}}]{Scott:2014lka}
\bibinfo{author}{\bibfnamefont{P.}~\bibnamefont{Scott}},
  \bibinfo{author}{\bibfnamefont{N.}~\bibnamefont{Grevesse}},
  \bibinfo{author}{\bibfnamefont{M.}~\bibnamefont{Asplund}},
  \bibinfo{author}{\bibfnamefont{A.~J.} \bibnamefont{Sauval}},
  \bibinfo{author}{\bibfnamefont{K.}~\bibnamefont{Lind}},
  \bibinfo{author}{\bibfnamefont{Y.}~\bibnamefont{Takeda}},
  \bibinfo{author}{\bibfnamefont{R.}~\bibnamefont{Collet}},
  \bibinfo{author}{\bibfnamefont{R.}~\bibnamefont{Trampedach}},
  \bibnamefont{and} \bibinfo{author}{\bibfnamefont{W.}~\bibnamefont{Hayek}},
  \bibinfo{journal}{Astron. Astrophys.} \textbf{\bibinfo{volume}{573}},
  \bibinfo{pages}{A25} (\bibinfo{year}{2015}{\natexlab{a}}),
  \eprint{1405.0279}.

\bibitem[{\citenamefont{Scott et~al.}(2015{\natexlab{b}})\citenamefont{Scott,
  Asplund, Grevesse, Bergemann, and Sauval}}]{Scott:2014mka}
\bibinfo{author}{\bibfnamefont{P.}~\bibnamefont{Scott}},
  \bibinfo{author}{\bibfnamefont{M.}~\bibnamefont{Asplund}},
  \bibinfo{author}{\bibfnamefont{N.}~\bibnamefont{Grevesse}},
  \bibinfo{author}{\bibfnamefont{M.}~\bibnamefont{Bergemann}},
  \bibnamefont{and} \bibinfo{author}{\bibfnamefont{A.~J.}
  \bibnamefont{Sauval}}, \bibinfo{journal}{Astron. Astrophys.}
  \textbf{\bibinfo{volume}{573}}, \bibinfo{pages}{A26}
  (\bibinfo{year}{2015}{\natexlab{b}}), \eprint{1405.0287}.

\bibitem[{\citenamefont{Grevesse et~al.}(2015)\citenamefont{Grevesse, Scott,
  Asplund, and Sauval}}]{Grevesse:2014nka}
\bibinfo{author}{\bibfnamefont{N.}~\bibnamefont{Grevesse}},
  \bibinfo{author}{\bibfnamefont{P.}~\bibnamefont{Scott}},
  \bibinfo{author}{\bibfnamefont{M.}~\bibnamefont{Asplund}}, \bibnamefont{and}
  \bibinfo{author}{\bibfnamefont{A.~J.} \bibnamefont{Sauval}},
  \bibinfo{journal}{Astron. Astrophys.} \textbf{\bibinfo{volume}{573}},
  \bibinfo{pages}{A27} (\bibinfo{year}{2015}), \eprint{1405.0288}.

\bibitem[{\citenamefont{Serenelli et~al.}(2009)\citenamefont{Serenelli, Basu,
  Ferguson, and Asplund}}]{Serenelli:2009yc}
\bibinfo{author}{\bibfnamefont{A.}~\bibnamefont{Serenelli}},
  \bibinfo{author}{\bibfnamefont{S.}~\bibnamefont{Basu}},
  \bibinfo{author}{\bibfnamefont{J.~W.} \bibnamefont{Ferguson}},
  \bibnamefont{and} \bibinfo{author}{\bibfnamefont{M.}~\bibnamefont{Asplund}},
  \bibinfo{journal}{Astrophys. J. Lett.} \textbf{\bibinfo{volume}{705}},
  \bibinfo{pages}{L123} (\bibinfo{year}{2009}), \eprint{0909.2668}.

\bibitem[{\citenamefont{Agostini
  et~al.}(2020{\natexlab{c}})}]{Agostini:2020mfq}
\bibinfo{author}{\bibfnamefont{M.}~\bibnamefont{Agostini}} \bibnamefont{et~al.}
  (\bibinfo{collaboration}{BOREXINO}), \bibinfo{journal}{Nature}
  \textbf{\bibinfo{volume}{587}}, \bibinfo{pages}{577}
  (\bibinfo{year}{2020}{\natexlab{c}}), \eprint{2006.15115}.

\bibitem[{\citenamefont{Agostini
  et~al.}(2020{\natexlab{d}})}]{Agostini:2017cav}
\bibinfo{author}{\bibfnamefont{M.}~\bibnamefont{Agostini}} \bibnamefont{et~al.}
  (\bibinfo{collaboration}{Borexino}), \bibinfo{journal}{Phys. Rev. D}
  \textbf{\bibinfo{volume}{101}}, \bibinfo{pages}{062001}
  (\bibinfo{year}{2020}{\natexlab{d}}), \eprint{1709.00756}.

\bibitem[{\citenamefont{Aharmim et~al.}(2006)}]{Aharmim:2006wq}
\bibinfo{author}{\bibfnamefont{B.}~\bibnamefont{Aharmim}} \bibnamefont{et~al.}
  (\bibinfo{collaboration}{SNO}), \bibinfo{journal}{Astrophys. J.}
  \textbf{\bibinfo{volume}{653}}, \bibinfo{pages}{1545} (\bibinfo{year}{2006}),
  \eprint{hep-ex/0607010}.

\bibitem[{\citenamefont{Suzuki}(2000)}]{Suzuki:2000ch}
\bibinfo{author}{\bibfnamefont{Y.}~\bibnamefont{Suzuki}}
  (\bibinfo{collaboration}{XMASS}), in \emph{\bibinfo{booktitle}{{Workshop on
  Solar Neutrinos below 1-MeV: NuLow}}} (\bibinfo{year}{2000}),
  \eprint{hep-ph/0008296}.

\bibitem[{\citenamefont{Smirnov et~al.}(2016)}]{Smirnov:2015lxy}
\bibinfo{author}{\bibfnamefont{O.}~\bibnamefont{Smirnov}} \bibnamefont{et~al.}
  (\bibinfo{collaboration}{Borexino}), \bibinfo{journal}{Phys. Part. Nucl.}
  \textbf{\bibinfo{volume}{47}}, \bibinfo{pages}{995} (\bibinfo{year}{2016}),
  \eprint{1507.02432}.

\bibitem[{\citenamefont{Bahcall}(2002)}]{Bahcall:2001pf}
\bibinfo{author}{\bibfnamefont{J.~N.} \bibnamefont{Bahcall}},
  \bibinfo{journal}{Phys. Rev. C} \textbf{\bibinfo{volume}{65}},
  \bibinfo{pages}{025801} (\bibinfo{year}{2002}), \eprint{hep-ph/0108148}.

\bibitem[{\citenamefont{Aalbers et~al.}(2020)}]{Aalbers:2020gsn}
\bibinfo{author}{\bibfnamefont{J.}~\bibnamefont{Aalbers}} \bibnamefont{et~al.}
  (\bibinfo{collaboration}{DARWIN}), \bibinfo{journal}{Eur. Phys. J. C}
  \textbf{\bibinfo{volume}{80}}, \bibinfo{pages}{1133} (\bibinfo{year}{2020}),
  \eprint{2006.03114}.

\bibitem[{\citenamefont{Georgadze et~al.}(1997)\citenamefont{Georgadze,
  Klapdor-Kleingrothaus, Pas, and Zdesenko}}]{Georgadze:1997zv}
\bibinfo{author}{\bibfnamefont{A.}~\bibnamefont{Georgadze}},
  \bibinfo{author}{\bibfnamefont{H.}~\bibnamefont{Klapdor-Kleingrothaus}},
  \bibinfo{author}{\bibfnamefont{H.}~\bibnamefont{Pas}}, \bibnamefont{and}
  \bibinfo{author}{\bibfnamefont{Y.}~\bibnamefont{Zdesenko}},
  \bibinfo{journal}{Astropart. Phys.} \textbf{\bibinfo{volume}{7}},
  \bibinfo{pages}{173} (\bibinfo{year}{1997}), \eprint{nucl-ex/9707006}.

\bibitem[{\citenamefont{Haselschwardt
  et~al.}(2020{\natexlab{a}})\citenamefont{Haselschwardt, Lenardo, Pirinen, and
  Suhonen}}]{Haselschwardt:2020ffr}
\bibinfo{author}{\bibfnamefont{S.}~\bibnamefont{Haselschwardt}},
  \bibinfo{author}{\bibfnamefont{B.}~\bibnamefont{Lenardo}},
  \bibinfo{author}{\bibfnamefont{P.}~\bibnamefont{Pirinen}}, \bibnamefont{and}
  \bibinfo{author}{\bibfnamefont{J.}~\bibnamefont{Suhonen}},
  \bibinfo{journal}{Phys. Rev. D} \textbf{\bibinfo{volume}{102}},
  \bibinfo{pages}{072009} (\bibinfo{year}{2020}{\natexlab{a}}),
  \eprint{2009.00535}.

\bibitem[{\citenamefont{Battistoni et~al.}(2005)\citenamefont{Battistoni,
  Ferrari, Montaruli, and Sala}}]{Battistoni:2005pd}
\bibinfo{author}{\bibfnamefont{G.}~\bibnamefont{Battistoni}},
  \bibinfo{author}{\bibfnamefont{A.}~\bibnamefont{Ferrari}},
  \bibinfo{author}{\bibfnamefont{T.}~\bibnamefont{Montaruli}},
  \bibnamefont{and} \bibinfo{author}{\bibfnamefont{P.~R.} \bibnamefont{Sala}},
  \bibinfo{journal}{Astropart. Phys.} \textbf{\bibinfo{volume}{23}},
  \bibinfo{pages}{526} (\bibinfo{year}{2005}).

\bibitem[{\citenamefont{Zhuang et~al.}(2022)\citenamefont{Zhuang, Strigari, and
  Lang}}]{Zhuang:2021rsg}
\bibinfo{author}{\bibfnamefont{Y.}~\bibnamefont{Zhuang}},
  \bibinfo{author}{\bibfnamefont{L.~E.} \bibnamefont{Strigari}},
  \bibnamefont{and} \bibinfo{author}{\bibfnamefont{R.~F.} \bibnamefont{Lang}},
  \bibinfo{journal}{Phys. Rev. D} \textbf{\bibinfo{volume}{105}},
  \bibinfo{pages}{043001} (\bibinfo{year}{2022}), \eprint{2110.14723}.

\bibitem[{\citenamefont{Janka et~al.}(2007)\citenamefont{Janka, Langanke,
  Marek, Martinez-Pinedo, and Mueller}}]{Janka:2006fh}
\bibinfo{author}{\bibfnamefont{H.-T.} \bibnamefont{Janka}},
  \bibinfo{author}{\bibfnamefont{K.}~\bibnamefont{Langanke}},
  \bibinfo{author}{\bibfnamefont{A.}~\bibnamefont{Marek}},
  \bibinfo{author}{\bibfnamefont{G.}~\bibnamefont{Martinez-Pinedo}},
  \bibnamefont{and} \bibinfo{author}{\bibfnamefont{B.}~\bibnamefont{Mueller}},
  \bibinfo{journal}{Phys. Rept.} \textbf{\bibinfo{volume}{442}},
  \bibinfo{pages}{38} (\bibinfo{year}{2007}), \eprint{astro-ph/0612072}.

\bibitem[{\citenamefont{Janka}(2012)}]{Janka:2012wk}
\bibinfo{author}{\bibfnamefont{H.-T.} \bibnamefont{Janka}},
  \bibinfo{journal}{Ann. Rev. Nucl. Part. Sci.} \textbf{\bibinfo{volume}{62}},
  \bibinfo{pages}{407} (\bibinfo{year}{2012}), \eprint{1206.2503}.

\bibitem[{\citenamefont{Krauss et~al.}(1992)\citenamefont{Krauss, Romanelli,
  and Schramm}}]{Krauss:1991xv}
\bibinfo{author}{\bibfnamefont{L.~M.} \bibnamefont{Krauss}},
  \bibinfo{author}{\bibfnamefont{P.}~\bibnamefont{Romanelli}},
  \bibnamefont{and} \bibinfo{author}{\bibfnamefont{D.~N.}
  \bibnamefont{Schramm}}, \bibinfo{journal}{Nucl. Phys. B}
  \textbf{\bibinfo{volume}{380}}, \bibinfo{pages}{507} (\bibinfo{year}{1992}).

\bibitem[{\citenamefont{Scholberg}(2012)}]{Scholberg:2012id}
\bibinfo{author}{\bibfnamefont{K.}~\bibnamefont{Scholberg}},
  \bibinfo{journal}{Ann. Rev. Nucl. Part. Sci.} \textbf{\bibinfo{volume}{62}},
  \bibinfo{pages}{81} (\bibinfo{year}{2012}), \eprint{1205.6003}.

\bibitem[{\citenamefont{Baxter et~al.}(2021{\natexlab{b}})}]{Baxter:2021stl}
\bibinfo{author}{\bibfnamefont{A.~L.} \bibnamefont{Baxter}}
  \bibnamefont{et~al.} (\bibinfo{collaboration}{SNEWS}), \bibinfo{journal}{J.
  Open Source Softw.} \textbf{\bibinfo{volume}{6}}, \bibinfo{pages}{67}
  (\bibinfo{year}{2021}{\natexlab{b}}), \eprint{2109.08188}.

\bibitem[{\citenamefont{Freedman et~al.}(1977)\citenamefont{Freedman, Schramm,
  and Tubbs}}]{Freedman:1977xn}
\bibinfo{author}{\bibfnamefont{D.~Z.} \bibnamefont{Freedman}},
  \bibinfo{author}{\bibfnamefont{D.~N.} \bibnamefont{Schramm}},
  \bibnamefont{and} \bibinfo{author}{\bibfnamefont{D.~L.} \bibnamefont{Tubbs}},
  \bibinfo{journal}{Ann. Rev. Nucl. Part. Sci.} \textbf{\bibinfo{volume}{27}},
  \bibinfo{pages}{167} (\bibinfo{year}{1977}).

\bibitem[{\citenamefont{Munoz et~al.}(2021)\citenamefont{Munoz, Takhistov,
  Witte, and Fuller}}]{Munoz:2021sad}
\bibinfo{author}{\bibfnamefont{V.}~\bibnamefont{Munoz}},
  \bibinfo{author}{\bibfnamefont{V.}~\bibnamefont{Takhistov}},
  \bibinfo{author}{\bibfnamefont{S.~J.} \bibnamefont{Witte}}, \bibnamefont{and}
  \bibinfo{author}{\bibfnamefont{G.~M.} \bibnamefont{Fuller}},
  \bibinfo{journal}{JCAP} \textbf{\bibinfo{volume}{11}}, \bibinfo{pages}{020}
  (\bibinfo{year}{2021}), \eprint{2102.00885}.

\bibitem[{\citenamefont{Horowitz et~al.}(2003)\citenamefont{Horowitz, Coakley,
  and McKinsey}}]{Horowitz:2003cz}
\bibinfo{author}{\bibfnamefont{C.~J.} \bibnamefont{Horowitz}},
  \bibinfo{author}{\bibfnamefont{K.~J.} \bibnamefont{Coakley}},
  \bibnamefont{and} \bibinfo{author}{\bibfnamefont{D.~N.}
  \bibnamefont{McKinsey}}, \bibinfo{journal}{Phys. Rev. D}
  \textbf{\bibinfo{volume}{68}}, \bibinfo{pages}{023005}
  (\bibinfo{year}{2003}), \eprint{astro-ph/0302071}.

\bibitem[{\citenamefont{Abe et~al.}(2017)}]{XMASS:2016cmy}
\bibinfo{author}{\bibfnamefont{K.}~\bibnamefont{Abe}} \bibnamefont{et~al.}
  (\bibinfo{collaboration}{XMASS}), \bibinfo{journal}{Astropart. Phys.}
  \textbf{\bibinfo{volume}{89}}, \bibinfo{pages}{51} (\bibinfo{year}{2017}),
  \eprint{1604.01218}.

\bibitem[{\citenamefont{Chakraborty et~al.}(2014)\citenamefont{Chakraborty,
  Bhattacharjee, and Kar}}]{Chakraborty:2013zua}
\bibinfo{author}{\bibfnamefont{S.}~\bibnamefont{Chakraborty}},
  \bibinfo{author}{\bibfnamefont{P.}~\bibnamefont{Bhattacharjee}},
  \bibnamefont{and} \bibinfo{author}{\bibfnamefont{K.}~\bibnamefont{Kar}},
  \bibinfo{journal}{Phys. Rev. D} \textbf{\bibinfo{volume}{89}},
  \bibinfo{pages}{013011} (\bibinfo{year}{2014}), \eprint{1309.4492}.

\bibitem[{\citenamefont{Lang et~al.}(2016{\natexlab{a}})\citenamefont{Lang,
  McCabe, Reichard, Selvi, and Tamborra}}]{Lang:2016zhv}
\bibinfo{author}{\bibfnamefont{R.~F.} \bibnamefont{Lang}},
  \bibinfo{author}{\bibfnamefont{C.}~\bibnamefont{McCabe}},
  \bibinfo{author}{\bibfnamefont{S.}~\bibnamefont{Reichard}},
  \bibinfo{author}{\bibfnamefont{M.}~\bibnamefont{Selvi}}, \bibnamefont{and}
  \bibinfo{author}{\bibfnamefont{I.}~\bibnamefont{Tamborra}},
  \bibinfo{journal}{Phys. Rev. D} \textbf{\bibinfo{volume}{94}},
  \bibinfo{pages}{103009} (\bibinfo{year}{2016}{\natexlab{a}}),
  \eprint{1606.09243}.

\bibitem[{\citenamefont{Raj}(2020)}]{Raj:2019sci}
\bibinfo{author}{\bibfnamefont{N.}~\bibnamefont{Raj}}, \bibinfo{journal}{Phys.
  Rev. Lett.} \textbf{\bibinfo{volume}{124}}, \bibinfo{pages}{141802}
  (\bibinfo{year}{2020}), \eprint{1907.05533}.

\bibitem[{\citenamefont{Pirinen et~al.}(2018)\citenamefont{Pirinen, Suhonen,
  and Ydrefors}}]{Pirinen:2018gsd}
\bibinfo{author}{\bibfnamefont{P.}~\bibnamefont{Pirinen}},
  \bibinfo{author}{\bibfnamefont{J.}~\bibnamefont{Suhonen}}, \bibnamefont{and}
  \bibinfo{author}{\bibfnamefont{E.}~\bibnamefont{Ydrefors}},
  \bibinfo{journal}{Adv. High Energy Phys.} \textbf{\bibinfo{volume}{2018}},
  \bibinfo{pages}{9163586} (\bibinfo{year}{2018}), \eprint{1804.08995}.

\bibitem[{\citenamefont{Ydrefors and Suhonen}(2015)}]{Ydrefors2015}
\bibinfo{author}{\bibfnamefont{E.}~\bibnamefont{Ydrefors}} \bibnamefont{and}
  \bibinfo{author}{\bibfnamefont{J.}~\bibnamefont{Suhonen}}
  (\bibinfo{year}{2015}), \bibinfo{note}{presented at the 2015 Neutrinos and
  Dark Matter Conference in Jyv{\"a}skyl{\"a}, Finland},
  \urlprefix\url{https://indico.cern.ch/event/394248/contributions/1831651/attachments/789193/1081737/ndm15_e_ydrefors.pdf}.

\bibitem[{\citenamefont{Litvinovich et~al.}(2017)\citenamefont{Litvinovich,
  Machulin, Pugachev, and Skorokhvatov}}]{Litvinovich:2017smi}
\bibinfo{author}{\bibfnamefont{E.~A.} \bibnamefont{Litvinovich}},
  \bibinfo{author}{\bibfnamefont{I.~N.} \bibnamefont{Machulin}},
  \bibinfo{author}{\bibfnamefont{D.~A.} \bibnamefont{Pugachev}},
  \bibnamefont{and} \bibinfo{author}{\bibfnamefont{M.~D.}
  \bibnamefont{Skorokhvatov}}, \bibinfo{journal}{J. Phys. Conf. Ser.}
  \textbf{\bibinfo{volume}{798}}, \bibinfo{pages}{012117}
  (\bibinfo{year}{2017}).

\bibitem[{\citenamefont{Bhattacharjee and Kar}(2021)}]{Bhattacharjee:2020rhs}
\bibinfo{author}{\bibfnamefont{P.}~\bibnamefont{Bhattacharjee}}
  \bibnamefont{and} \bibinfo{author}{\bibfnamefont{K.}~\bibnamefont{Kar}},
  \bibinfo{journal}{Eur. Phys. J. ST} \textbf{\bibinfo{volume}{230}},
  \bibinfo{pages}{505} (\bibinfo{year}{2021}), \eprint{2012.14888}.

\bibitem[{\citenamefont{Bhattacharjee et~al.}(2020)\citenamefont{Bhattacharjee,
  Bandyopadhyay, Chakraborty, Ghosh, Kar, and Saha}}]{Bhattacharjee:2020qrj}
\bibinfo{author}{\bibfnamefont{P.}~\bibnamefont{Bhattacharjee}},
  \bibinfo{author}{\bibfnamefont{A.}~\bibnamefont{Bandyopadhyay}},
  \bibinfo{author}{\bibfnamefont{S.}~\bibnamefont{Chakraborty}},
  \bibinfo{author}{\bibfnamefont{S.}~\bibnamefont{Ghosh}},
  \bibinfo{author}{\bibfnamefont{K.}~\bibnamefont{Kar}}, \bibnamefont{and}
  \bibinfo{author}{\bibfnamefont{S.}~\bibnamefont{Saha}}
  (\bibinfo{year}{2020}), \eprint{2012.13986}.

\bibitem[{\citenamefont{Suliga and Tamborra}(2021)}]{Suliga:2020jfa}
\bibinfo{author}{\bibfnamefont{A.~M.} \bibnamefont{Suliga}} \bibnamefont{and}
  \bibinfo{author}{\bibfnamefont{I.}~\bibnamefont{Tamborra}},
  \bibinfo{journal}{Phys. Rev. D} \textbf{\bibinfo{volume}{103}},
  \bibinfo{pages}{083002} (\bibinfo{year}{2021}), \eprint{2010.14545}.

\bibitem[{\citenamefont{Raj et~al.}(2020)\citenamefont{Raj, Takhistov, and
  Witte}}]{Raj:2019wpy}
\bibinfo{author}{\bibfnamefont{N.}~\bibnamefont{Raj}},
  \bibinfo{author}{\bibfnamefont{V.}~\bibnamefont{Takhistov}},
  \bibnamefont{and} \bibinfo{author}{\bibfnamefont{S.~J.} \bibnamefont{Witte}},
  \bibinfo{journal}{Phys. Rev. D} \textbf{\bibinfo{volume}{101}},
  \bibinfo{pages}{043008} (\bibinfo{year}{2020}), \eprint{1905.09283}.

\bibitem[{\citenamefont{Odrzywolek
  et~al.}(2004{\natexlab{a}})\citenamefont{Odrzywolek, Misiaszek, and
  Kutschera}}]{Odrzywolek:2003vn}
\bibinfo{author}{\bibfnamefont{A.}~\bibnamefont{Odrzywolek}},
  \bibinfo{author}{\bibfnamefont{M.}~\bibnamefont{Misiaszek}},
  \bibnamefont{and}
  \bibinfo{author}{\bibfnamefont{M.}~\bibnamefont{Kutschera}},
  \bibinfo{journal}{Astropart. Phys.} \textbf{\bibinfo{volume}{21}},
  \bibinfo{pages}{303} (\bibinfo{year}{2004}{\natexlab{a}}),
  \eprint{astro-ph/0311012}.

\bibitem[{\citenamefont{Odrzywolek
  et~al.}(2004{\natexlab{b}})\citenamefont{Odrzywolek, Misiaszek, and
  Kutschera}}]{Odrzywolek:2004em}
\bibinfo{author}{\bibfnamefont{A.}~\bibnamefont{Odrzywolek}},
  \bibinfo{author}{\bibfnamefont{M.}~\bibnamefont{Misiaszek}},
  \bibnamefont{and}
  \bibinfo{author}{\bibfnamefont{M.}~\bibnamefont{Kutschera}},
  \bibinfo{journal}{Acta Phys. Polon. B} \textbf{\bibinfo{volume}{35}},
  \bibinfo{pages}{1981} (\bibinfo{year}{2004}{\natexlab{b}}),
  \eprint{astro-ph/0405006}.

\bibitem[{\citenamefont{Kato et~al.}(2020)\citenamefont{Kato, Ishidoshiro, and
  Yoshida}}]{Kato:2020hlc}
\bibinfo{author}{\bibfnamefont{C.}~\bibnamefont{Kato}},
  \bibinfo{author}{\bibfnamefont{K.}~\bibnamefont{Ishidoshiro}},
  \bibnamefont{and} \bibinfo{author}{\bibfnamefont{T.}~\bibnamefont{Yoshida}},
  \bibinfo{journal}{Ann. Rev. Nucl. Part. Sci.} \textbf{\bibinfo{volume}{70}},
  \bibinfo{pages}{121} (\bibinfo{year}{2020}), \eprint{2006.02519}.

\bibitem[{\citenamefont{Sieverding et~al.}(2021)\citenamefont{Sieverding,
  Rrapaj, Guo, and Qian}}]{Sieverding:2021jfa}
\bibinfo{author}{\bibfnamefont{A.}~\bibnamefont{Sieverding}},
  \bibinfo{author}{\bibfnamefont{E.}~\bibnamefont{Rrapaj}},
  \bibinfo{author}{\bibfnamefont{G.}~\bibnamefont{Guo}}, \bibnamefont{and}
  \bibinfo{author}{\bibfnamefont{Y.~Z.} \bibnamefont{Qian}},
  \bibinfo{journal}{Astrophys. J.} \textbf{\bibinfo{volume}{912}},
  \bibinfo{pages}{13} (\bibinfo{year}{2021}), \eprint{2101.08672}.

\bibitem[{\citenamefont{Mori et~al.}(2021)\citenamefont{Mori, Takiwaki, and
  Kotake}}]{Mori:2021muf}
\bibinfo{author}{\bibfnamefont{K.}~\bibnamefont{Mori}},
  \bibinfo{author}{\bibfnamefont{T.}~\bibnamefont{Takiwaki}}, \bibnamefont{and}
  \bibinfo{author}{\bibfnamefont{K.}~\bibnamefont{Kotake}}
  (\bibinfo{year}{2021}), \eprint{2107.12661}.

\bibitem[{\citenamefont{Ge et~al.}(2020{\natexlab{a}})\citenamefont{Ge,
  Hamaguchi, Ichimura, Ishidoshiro, Kanazawa, Kishimoto, Nagata, and
  Zheng}}]{Ge:2020zww}
\bibinfo{author}{\bibfnamefont{S.-F.} \bibnamefont{Ge}},
  \bibinfo{author}{\bibfnamefont{K.}~\bibnamefont{Hamaguchi}},
  \bibinfo{author}{\bibfnamefont{K.}~\bibnamefont{Ichimura}},
  \bibinfo{author}{\bibfnamefont{K.}~\bibnamefont{Ishidoshiro}},
  \bibinfo{author}{\bibfnamefont{Y.}~\bibnamefont{Kanazawa}},
  \bibinfo{author}{\bibfnamefont{Y.}~\bibnamefont{Kishimoto}},
  \bibinfo{author}{\bibfnamefont{N.}~\bibnamefont{Nagata}}, \bibnamefont{and}
  \bibinfo{author}{\bibfnamefont{J.}~\bibnamefont{Zheng}},
  \bibinfo{journal}{JCAP} \textbf{\bibinfo{volume}{11}}, \bibinfo{pages}{059}
  (\bibinfo{year}{2020}{\natexlab{a}}), \eprint{2008.03924}.

\bibitem[{sne()}]{snewsweb}
\bibinfo{howpublished}{\url{http://snews.bnl.gov/}}.

\bibitem[{\citenamefont{Antonioli et~al.}(2004)}]{Antonioli:2004zb}
\bibinfo{author}{\bibfnamefont{P.}~\bibnamefont{Antonioli}}
  \bibnamefont{et~al.}, \bibinfo{journal}{New J. Phys.}
  \textbf{\bibinfo{volume}{6}}, \bibinfo{pages}{114} (\bibinfo{year}{2004}),
  \eprint{astro-ph/0406214}.

\bibitem[{\citenamefont{Al~Kharusi et~al.}(2021)}]{Kharusi:2020ovw}
\bibinfo{author}{\bibfnamefont{S.}~\bibnamefont{Al~Kharusi}}
  \bibnamefont{et~al.} (\bibinfo{collaboration}{SNEWS}), \bibinfo{journal}{New
  J. Phys.} \textbf{\bibinfo{volume}{23}}, \bibinfo{pages}{031201}
  (\bibinfo{year}{2021}), \eprint{2011.00035}.

\bibitem[{\citenamefont{Baxter et~al.}(2021{\natexlab{c}})\citenamefont{Baxter,
  BenZvi, Bonivento, Brazier, Clark, Coleiro, Collom, Colomer-Molla, Cousins,
  Orellana et~al.}}]{Baxter:2021aaa}
\bibinfo{author}{\bibfnamefont{A.~L.} \bibnamefont{Baxter}},
  \bibinfo{author}{\bibfnamefont{S.~Y.} \bibnamefont{BenZvi}},
  \bibinfo{author}{\bibfnamefont{W.}~\bibnamefont{Bonivento}},
  \bibinfo{author}{\bibfnamefont{A.}~\bibnamefont{Brazier}},
  \bibinfo{author}{\bibfnamefont{M.}~\bibnamefont{Clark}},
  \bibinfo{author}{\bibfnamefont{A.}~\bibnamefont{Coleiro}},
  \bibinfo{author}{\bibfnamefont{D.}~\bibnamefont{Collom}},
  \bibinfo{author}{\bibfnamefont{M.}~\bibnamefont{Colomer-Molla}},
  \bibinfo{author}{\bibfnamefont{B.}~\bibnamefont{Cousins}},
  \bibinfo{author}{\bibfnamefont{A.~D.} \bibnamefont{Orellana}},
  \bibnamefont{et~al.}, \emph{\bibinfo{title}{Agile scrum development in an ad
  hoc software collaboration}} (\bibinfo{year}{2021}{\natexlab{c}}),
  \eprint{2101.07779}.

\bibitem[{\citenamefont{Krauss et~al.}(1984{\natexlab{a}})\citenamefont{Krauss,
  Glashow, and Schramm}}]{Krauss:1983zn}
\bibinfo{author}{\bibfnamefont{L.~M.} \bibnamefont{Krauss}},
  \bibinfo{author}{\bibfnamefont{S.~L.} \bibnamefont{Glashow}},
  \bibnamefont{and} \bibinfo{author}{\bibfnamefont{D.~N.}
  \bibnamefont{Schramm}}, \bibinfo{journal}{Nature}
  \textbf{\bibinfo{volume}{310}}, \bibinfo{pages}{191}
  (\bibinfo{year}{1984}{\natexlab{a}}).

\bibitem[{\citenamefont{Lunardini}(2016)}]{Lunardini:2010ab}
\bibinfo{author}{\bibfnamefont{C.}~\bibnamefont{Lunardini}},
  \bibinfo{journal}{Astropart. Phys.} \textbf{\bibinfo{volume}{79}},
  \bibinfo{pages}{49} (\bibinfo{year}{2016}), \eprint{1007.3252}.

\bibitem[{\citenamefont{Beacom}(2010)}]{Beacom:2010kk}
\bibinfo{author}{\bibfnamefont{J.~F.} \bibnamefont{Beacom}},
  \bibinfo{journal}{Ann. Rev. Nucl. Part. Sci.} \textbf{\bibinfo{volume}{60}},
  \bibinfo{pages}{439} (\bibinfo{year}{2010}), \eprint{1004.3311}.

\bibitem[{\citenamefont{Horiuchi et~al.}(2009)\citenamefont{Horiuchi, Beacom,
  and Dwek}}]{Horiuchi:2008jz}
\bibinfo{author}{\bibfnamefont{S.}~\bibnamefont{Horiuchi}},
  \bibinfo{author}{\bibfnamefont{J.~F.} \bibnamefont{Beacom}},
  \bibnamefont{and} \bibinfo{author}{\bibfnamefont{E.}~\bibnamefont{Dwek}},
  \bibinfo{journal}{Phys. Rev. D} \textbf{\bibinfo{volume}{79}},
  \bibinfo{pages}{083013} (\bibinfo{year}{2009}), \eprint{0812.3157}.

\bibitem[{\citenamefont{Hopkins and Beacom}(2006)}]{Hopkins:2006bw}
\bibinfo{author}{\bibfnamefont{A.~M.} \bibnamefont{Hopkins}} \bibnamefont{and}
  \bibinfo{author}{\bibfnamefont{J.~F.} \bibnamefont{Beacom}},
  \bibinfo{journal}{Astrophys. J.} \textbf{\bibinfo{volume}{651}},
  \bibinfo{pages}{142} (\bibinfo{year}{2006}), \eprint{astro-ph/0601463}.

\bibitem[{\citenamefont{Bays et~al.}(2012)}]{Bays:2011si}
\bibinfo{author}{\bibfnamefont{K.}~\bibnamefont{Bays}} \bibnamefont{et~al.}
  (\bibinfo{collaboration}{Super-Kamiokande}), \bibinfo{journal}{Phys. Rev. D}
  \textbf{\bibinfo{volume}{85}}, \bibinfo{pages}{052007}
  (\bibinfo{year}{2012}), \eprint{1111.5031}.

\bibitem[{\citenamefont{Suliga et~al.}(2022)\citenamefont{Suliga, Beacom, and
  Tamborra}}]{Suliga:2021hek}
\bibinfo{author}{\bibfnamefont{A.~M.} \bibnamefont{Suliga}},
  \bibinfo{author}{\bibfnamefont{J.~F.} \bibnamefont{Beacom}},
  \bibnamefont{and} \bibinfo{author}{\bibfnamefont{I.}~\bibnamefont{Tamborra}},
  \bibinfo{journal}{Phys. Rev. D} \textbf{\bibinfo{volume}{105}},
  \bibinfo{pages}{043008} (\bibinfo{year}{2022}), \eprint{2112.09168}.

\bibitem[{\citenamefont{Araki et~al.}(2005)}]{Araki:2005qa}
\bibinfo{author}{\bibfnamefont{T.}~\bibnamefont{Araki}} \bibnamefont{et~al.},
  \bibinfo{journal}{Nature} \textbf{\bibinfo{volume}{436}},
  \bibinfo{pages}{499} (\bibinfo{year}{2005}).

\bibitem[{\citenamefont{Bellini et~al.}(2010)}]{Borexino:2010dli}
\bibinfo{author}{\bibfnamefont{G.}~\bibnamefont{Bellini}} \bibnamefont{et~al.}
  (\bibinfo{collaboration}{Borexino}), \bibinfo{journal}{Phys. Lett. B}
  \textbf{\bibinfo{volume}{687}}, \bibinfo{pages}{299} (\bibinfo{year}{2010}),
  \eprint{1003.0284}.

\bibitem[{\citenamefont{Bouchiat and Piketty}(1983)}]{Bouchiat:1983uf}
\bibinfo{author}{\bibfnamefont{C.}~\bibnamefont{Bouchiat}} \bibnamefont{and}
  \bibinfo{author}{\bibfnamefont{C.}~\bibnamefont{Piketty}},
  \bibinfo{journal}{Phys. Lett. B} \textbf{\bibinfo{volume}{128}},
  \bibinfo{pages}{73} (\bibinfo{year}{1983}).

\bibitem[{\citenamefont{Cerdeño et~al.}(2016)\citenamefont{Cerdeño,
  Fairbairn, Jubb, Machado, Vincent, and B\oe~hm}}]{Cerdeno:2016sfi}
\bibinfo{author}{\bibfnamefont{D.~G.} \bibnamefont{Cerdeño}},
  \bibinfo{author}{\bibfnamefont{M.}~\bibnamefont{Fairbairn}},
  \bibinfo{author}{\bibfnamefont{T.}~\bibnamefont{Jubb}},
  \bibinfo{author}{\bibfnamefont{P.~A.~N.} \bibnamefont{Machado}},
  \bibinfo{author}{\bibfnamefont{A.~C.} \bibnamefont{Vincent}},
  \bibnamefont{and} \bibinfo{author}{\bibfnamefont{C.}~\bibnamefont{B\oe~hm}},
  \bibinfo{journal}{JHEP} \textbf{\bibinfo{volume}{05}}, \bibinfo{pages}{118}
  (\bibinfo{year}{2016}), \bibinfo{note}{[Erratum: JHEP {\bf 09}, 048 (2016)]},
  \eprint{1604.01025}.

\bibitem[{\citenamefont{Cadeddu et~al.}(2020)\citenamefont{Cadeddu, Dordei,
  Giunti, Li, and Zhang}}]{Cadeddu:2019eta}
\bibinfo{author}{\bibfnamefont{M.}~\bibnamefont{Cadeddu}},
  \bibinfo{author}{\bibfnamefont{F.}~\bibnamefont{Dordei}},
  \bibinfo{author}{\bibfnamefont{C.}~\bibnamefont{Giunti}},
  \bibinfo{author}{\bibfnamefont{Y.~F.} \bibnamefont{Li}}, \bibnamefont{and}
  \bibinfo{author}{\bibfnamefont{Y.~Y.} \bibnamefont{Zhang}},
  \bibinfo{journal}{Phys. Rev. D} \textbf{\bibinfo{volume}{101}},
  \bibinfo{pages}{033004} (\bibinfo{year}{2020}), \eprint{1908.06045}.

\bibitem[{\citenamefont{Davoudiasl et~al.}(2014)\citenamefont{Davoudiasl, Lee,
  and Marciano}}]{Davoudiasl:2014kua}
\bibinfo{author}{\bibfnamefont{H.}~\bibnamefont{Davoudiasl}},
  \bibinfo{author}{\bibfnamefont{H.-S.} \bibnamefont{Lee}}, \bibnamefont{and}
  \bibinfo{author}{\bibfnamefont{W.~J.} \bibnamefont{Marciano}},
  \bibinfo{journal}{Phys. Rev. D} \textbf{\bibinfo{volume}{89}},
  \bibinfo{pages}{095006} (\bibinfo{year}{2014}), \eprint{1402.3620}.

\bibitem[{\citenamefont{Aoyama et~al.}(2020)}]{Aoyama:2020ynm}
\bibinfo{author}{\bibfnamefont{T.}~\bibnamefont{Aoyama}} \bibnamefont{et~al.},
  \bibinfo{journal}{Phys. Rept.} \textbf{\bibinfo{volume}{887}},
  \bibinfo{pages}{1} (\bibinfo{year}{2020}), \eprint{2006.04822}.

\bibitem[{\citenamefont{Abi et~al.}(2021)}]{Muong-2:2021ojo}
\bibinfo{author}{\bibfnamefont{B.}~\bibnamefont{Abi}} \bibnamefont{et~al.}
  (\bibinfo{collaboration}{Muon $g-2$}), \bibinfo{journal}{Phys. Rev. Lett.}
  \textbf{\bibinfo{volume}{126}}, \bibinfo{pages}{141801}
  (\bibinfo{year}{2021}), \eprint{2104.03281}.

\bibitem[{\citenamefont{de~Gouv\^ea et~al.}(2021)\citenamefont{de~Gouv\^ea,
  McGinness, Martinez-Soler, and Perez-Gonzalez}}]{deGouvea:2021ymm}
\bibinfo{author}{\bibfnamefont{A.}~\bibnamefont{de~Gouv\^ea}},
  \bibinfo{author}{\bibfnamefont{E.}~\bibnamefont{McGinness}},
  \bibinfo{author}{\bibfnamefont{I.}~\bibnamefont{Martinez-Soler}},
  \bibnamefont{and} \bibinfo{author}{\bibfnamefont{Y.~F.}
  \bibnamefont{Perez-Gonzalez}} (\bibinfo{year}{2021}), \eprint{2111.02421}.

\bibitem[{Dev(2019)}]{Dev:2019anc}
\emph{\bibinfo{title}{{Neutrino Non-Standard Interactions: A Status Report}}},
  vol.~\bibinfo{volume}{2} (\bibinfo{year}{2019}), \eprint{1907.00991}.

\bibitem[{\citenamefont{de~Gouv\^ea and Kelly}(2016)}]{deGouvea:2015ndi}
\bibinfo{author}{\bibfnamefont{A.}~\bibnamefont{de~Gouv\^ea}} \bibnamefont{and}
  \bibinfo{author}{\bibfnamefont{K.~J.} \bibnamefont{Kelly}},
  \bibinfo{journal}{Nucl. Phys. B} \textbf{\bibinfo{volume}{908}},
  \bibinfo{pages}{318} (\bibinfo{year}{2016}), \eprint{1511.05562}.

\bibitem[{\citenamefont{Datta et~al.}(2018)\citenamefont{Datta, Kumar, Liao,
  and Marfatia}}]{Datta:2017ezo}
\bibinfo{author}{\bibfnamefont{A.}~\bibnamefont{Datta}},
  \bibinfo{author}{\bibfnamefont{J.}~\bibnamefont{Kumar}},
  \bibinfo{author}{\bibfnamefont{J.}~\bibnamefont{Liao}}, \bibnamefont{and}
  \bibinfo{author}{\bibfnamefont{D.}~\bibnamefont{Marfatia}},
  \bibinfo{journal}{Phys. Rev. D} \textbf{\bibinfo{volume}{97}},
  \bibinfo{pages}{115038} (\bibinfo{year}{2018}), \eprint{1705.08423}.

\bibitem[{\citenamefont{Aristizabal~Sierra
  et~al.}(2020{\natexlab{a}})\citenamefont{Aristizabal~Sierra, De~Romeri,
  Flores, and Papoulias}}]{AristizabalSierra:2020edu}
\bibinfo{author}{\bibfnamefont{D.}~\bibnamefont{Aristizabal~Sierra}},
  \bibinfo{author}{\bibfnamefont{V.}~\bibnamefont{De~Romeri}},
  \bibinfo{author}{\bibfnamefont{L.~J.} \bibnamefont{Flores}},
  \bibnamefont{and} \bibinfo{author}{\bibfnamefont{D.~K.}
  \bibnamefont{Papoulias}}, \bibinfo{journal}{Phys. Lett. B}
  \textbf{\bibinfo{volume}{809}}, \bibinfo{pages}{135681}
  (\bibinfo{year}{2020}{\natexlab{a}}), \eprint{2006.12457}.

\bibitem[{\citenamefont{Khan}(2020)}]{Khan:2020vaf}
\bibinfo{author}{\bibfnamefont{A.~N.} \bibnamefont{Khan}},
  \bibinfo{journal}{Phys. Lett. B} \textbf{\bibinfo{volume}{809}},
  \bibinfo{pages}{135782} (\bibinfo{year}{2020}), \eprint{2006.12887}.

\bibitem[{\citenamefont{Fernandez-Moroni
  et~al.}(2021)\citenamefont{Fernandez-Moroni, Harnik, Machado, Martinez-Soler,
  Perez-Gonzalez, Rodrigues, and Rosauro-Alcaraz}}]{Fernandez-Moroni:2021nap}
\bibinfo{author}{\bibfnamefont{G.}~\bibnamefont{Fernandez-Moroni}},
  \bibinfo{author}{\bibfnamefont{R.}~\bibnamefont{Harnik}},
  \bibinfo{author}{\bibfnamefont{P.~A.~N.} \bibnamefont{Machado}},
  \bibinfo{author}{\bibfnamefont{I.}~\bibnamefont{Martinez-Soler}},
  \bibinfo{author}{\bibfnamefont{Y.~F.} \bibnamefont{Perez-Gonzalez}},
  \bibinfo{author}{\bibfnamefont{D.}~\bibnamefont{Rodrigues}},
  \bibnamefont{and}
  \bibinfo{author}{\bibfnamefont{S.}~\bibnamefont{Rosauro-Alcaraz}}
  (\bibinfo{year}{2021}), \eprint{2108.07310}.

\bibitem[{\citenamefont{Goldhagen et~al.}(2021)\citenamefont{Goldhagen,
  Maltoni, Reichard, and Schwetz}}]{Goldhagen:2021kxe}
\bibinfo{author}{\bibfnamefont{K.}~\bibnamefont{Goldhagen}},
  \bibinfo{author}{\bibfnamefont{M.}~\bibnamefont{Maltoni}},
  \bibinfo{author}{\bibfnamefont{S.}~\bibnamefont{Reichard}}, \bibnamefont{and}
  \bibinfo{author}{\bibfnamefont{T.}~\bibnamefont{Schwetz}}
  (\bibinfo{year}{2021}), \eprint{2109.14898}.

\bibitem[{\citenamefont{Link and Xu}(2019)}]{Link:2019pbm}
\bibinfo{author}{\bibfnamefont{J.~M.} \bibnamefont{Link}} \bibnamefont{and}
  \bibinfo{author}{\bibfnamefont{X.-J.} \bibnamefont{Xu}},
  \bibinfo{journal}{JHEP} \textbf{\bibinfo{volume}{08}}, \bibinfo{pages}{004}
  (\bibinfo{year}{2019}), \eprint{1903.09891}.

\bibitem[{\citenamefont{Krauss}(1991)}]{Krauss:1991ba}
\bibinfo{author}{\bibfnamefont{L.~M.} \bibnamefont{Krauss}},
  \bibinfo{journal}{Phys. Lett. B} \textbf{\bibinfo{volume}{269}},
  \bibinfo{pages}{407} (\bibinfo{year}{1991}).

\bibitem[{\citenamefont{Khan et~al.}(2020)\citenamefont{Khan, Rodejohann, and
  Xu}}]{Khan:2019jvr}
\bibinfo{author}{\bibfnamefont{A.~N.} \bibnamefont{Khan}},
  \bibinfo{author}{\bibfnamefont{W.}~\bibnamefont{Rodejohann}},
  \bibnamefont{and} \bibinfo{author}{\bibfnamefont{X.-J.} \bibnamefont{Xu}},
  \bibinfo{journal}{Phys. Rev. D} \textbf{\bibinfo{volume}{101}},
  \bibinfo{pages}{055047} (\bibinfo{year}{2020}), \eprint{1906.12102}.

\bibitem[{\citenamefont{Kamada and Yu}(2015)}]{Kamada:2015era}
\bibinfo{author}{\bibfnamefont{A.}~\bibnamefont{Kamada}} \bibnamefont{and}
  \bibinfo{author}{\bibfnamefont{H.-B.} \bibnamefont{Yu}},
  \bibinfo{journal}{Phys. Rev. D} \textbf{\bibinfo{volume}{92}},
  \bibinfo{pages}{113004} (\bibinfo{year}{2015}), \eprint{1504.00711}.

\bibitem[{\citenamefont{Dutta et~al.}(2020)\citenamefont{Dutta, Lang, Liao,
  Sinha, Strigari, and Thompson}}]{Dutta:2020che}
\bibinfo{author}{\bibfnamefont{B.}~\bibnamefont{Dutta}},
  \bibinfo{author}{\bibfnamefont{R.~F.} \bibnamefont{Lang}},
  \bibinfo{author}{\bibfnamefont{S.}~\bibnamefont{Liao}},
  \bibinfo{author}{\bibfnamefont{S.}~\bibnamefont{Sinha}},
  \bibinfo{author}{\bibfnamefont{L.}~\bibnamefont{Strigari}}, \bibnamefont{and}
  \bibinfo{author}{\bibfnamefont{A.}~\bibnamefont{Thompson}},
  \bibinfo{journal}{JHEP} \textbf{\bibinfo{volume}{09}}, \bibinfo{pages}{106}
  (\bibinfo{year}{2020}), \eprint{2002.03066}.

\bibitem[{\citenamefont{Abe et~al.}(2021)\citenamefont{Abe, Hamaguchi, and
  Nagata}}]{Abe:2021ocf}
\bibinfo{author}{\bibfnamefont{T.}~\bibnamefont{Abe}},
  \bibinfo{author}{\bibfnamefont{K.}~\bibnamefont{Hamaguchi}},
  \bibnamefont{and} \bibinfo{author}{\bibfnamefont{N.}~\bibnamefont{Nagata}},
  \bibinfo{journal}{Phys. Lett. B} \textbf{\bibinfo{volume}{815}},
  \bibinfo{pages}{136174} (\bibinfo{year}{2021}), \eprint{2012.02508}.

\bibitem[{\citenamefont{Abellan et~al.}(2020)\citenamefont{Abellan, Murgia,
  Poulin, and Lavalle}}]{Abellan:2020pmw}
\bibinfo{author}{\bibfnamefont{G.~F.} \bibnamefont{Abellan}},
  \bibinfo{author}{\bibfnamefont{R.}~\bibnamefont{Murgia}},
  \bibinfo{author}{\bibfnamefont{V.}~\bibnamefont{Poulin}}, \bibnamefont{and}
  \bibinfo{author}{\bibfnamefont{J.}~\bibnamefont{Lavalle}}
  (\bibinfo{year}{2020}), \eprint{2008.09615}.

\bibitem[{\citenamefont{Aboubrahim et~al.}(2021)\citenamefont{Aboubrahim,
  Klasen, and Nath}}]{Aboubrahim:2020iwb}
\bibinfo{author}{\bibfnamefont{A.}~\bibnamefont{Aboubrahim}},
  \bibinfo{author}{\bibfnamefont{M.}~\bibnamefont{Klasen}}, \bibnamefont{and}
  \bibinfo{author}{\bibfnamefont{P.}~\bibnamefont{Nath}},
  \bibinfo{journal}{JHEP} \textbf{\bibinfo{volume}{02}}, \bibinfo{pages}{229}
  (\bibinfo{year}{2021}), \eprint{2011.08053}.

\bibitem[{\citenamefont{Alhazmi et~al.}(2021)\citenamefont{Alhazmi, Kim, Kong,
  Mohlabeng, Park, and Shin}}]{Alhazmi:2020fju}
\bibinfo{author}{\bibfnamefont{H.}~\bibnamefont{Alhazmi}},
  \bibinfo{author}{\bibfnamefont{D.}~\bibnamefont{Kim}},
  \bibinfo{author}{\bibfnamefont{K.}~\bibnamefont{Kong}},
  \bibinfo{author}{\bibfnamefont{G.}~\bibnamefont{Mohlabeng}},
  \bibinfo{author}{\bibfnamefont{J.-C.} \bibnamefont{Park}}, \bibnamefont{and}
  \bibinfo{author}{\bibfnamefont{S.}~\bibnamefont{Shin}},
  \bibinfo{journal}{JHEP} \textbf{\bibinfo{volume}{05}}, \bibinfo{pages}{055}
  (\bibinfo{year}{2021}), \eprint{2006.16252}.

\bibitem[{\citenamefont{Alonso-\'Alvarez
  et~al.}(2020)\citenamefont{Alonso-\'Alvarez, Ertas, Jaeckel, Kahlhoefer, and
  Thormaehlen}}]{Alonso-Alvarez:2020cdv}
\bibinfo{author}{\bibfnamefont{G.}~\bibnamefont{Alonso-\'Alvarez}},
  \bibinfo{author}{\bibfnamefont{F.}~\bibnamefont{Ertas}},
  \bibinfo{author}{\bibfnamefont{J.}~\bibnamefont{Jaeckel}},
  \bibinfo{author}{\bibfnamefont{F.}~\bibnamefont{Kahlhoefer}},
  \bibnamefont{and} \bibinfo{author}{\bibfnamefont{L.~J.}
  \bibnamefont{Thormaehlen}}, \bibinfo{journal}{JCAP}
  \textbf{\bibinfo{volume}{11}}, \bibinfo{pages}{029} (\bibinfo{year}{2020}),
  \eprint{2006.11243}.

\bibitem[{\citenamefont{Amaral et~al.}(2020{\natexlab{b}})\citenamefont{Amaral,
  Cerdeno, Foldenauer, and Reid}}]{Amaral:2020tga}
\bibinfo{author}{\bibfnamefont{D.~W. P.~d.} \bibnamefont{Amaral}},
  \bibinfo{author}{\bibfnamefont{D.~G.} \bibnamefont{Cerdeno}},
  \bibinfo{author}{\bibfnamefont{P.}~\bibnamefont{Foldenauer}},
  \bibnamefont{and} \bibinfo{author}{\bibfnamefont{E.}~\bibnamefont{Reid}},
  \bibinfo{journal}{JHEP} \textbf{\bibinfo{volume}{12}}, \bibinfo{pages}{155}
  (\bibinfo{year}{2020}{\natexlab{b}}), \eprint{2006.11225}.

\bibitem[{\citenamefont{An and Yang}(2021)}]{An:2020tcg}
\bibinfo{author}{\bibfnamefont{H.}~\bibnamefont{An}} \bibnamefont{and}
  \bibinfo{author}{\bibfnamefont{D.}~\bibnamefont{Yang}},
  \bibinfo{journal}{Phys. Lett. B} \textbf{\bibinfo{volume}{818}},
  \bibinfo{pages}{136408} (\bibinfo{year}{2021}), \eprint{2006.15672}.

\bibitem[{\citenamefont{Anchordoqui et~al.}(2020)\citenamefont{Anchordoqui,
  Antoniadis, Benakli, and Lust}}]{Anchordoqui:2020tlp}
\bibinfo{author}{\bibfnamefont{L.~A.} \bibnamefont{Anchordoqui}},
  \bibinfo{author}{\bibfnamefont{I.}~\bibnamefont{Antoniadis}},
  \bibinfo{author}{\bibfnamefont{K.}~\bibnamefont{Benakli}}, \bibnamefont{and}
  \bibinfo{author}{\bibfnamefont{D.}~\bibnamefont{Lust}},
  \bibinfo{journal}{Phys. Lett. B} \textbf{\bibinfo{volume}{810}},
  \bibinfo{pages}{135838} (\bibinfo{year}{2020}), \eprint{2007.11697}.

\bibitem[{\citenamefont{Arcadi et~al.}(2021)\citenamefont{Arcadi, Bally,
  Goertz, Tame-Narvaez, Tenorth, and Vogl}}]{Arcadi:2020zni}
\bibinfo{author}{\bibfnamefont{G.}~\bibnamefont{Arcadi}},
  \bibinfo{author}{\bibfnamefont{A.}~\bibnamefont{Bally}},
  \bibinfo{author}{\bibfnamefont{F.}~\bibnamefont{Goertz}},
  \bibinfo{author}{\bibfnamefont{K.}~\bibnamefont{Tame-Narvaez}},
  \bibinfo{author}{\bibfnamefont{V.}~\bibnamefont{Tenorth}}, \bibnamefont{and}
  \bibinfo{author}{\bibfnamefont{S.}~\bibnamefont{Vogl}},
  \bibinfo{journal}{Phys. Rev. D} \textbf{\bibinfo{volume}{103}},
  \bibinfo{pages}{023024} (\bibinfo{year}{2021}), \eprint{2007.08500}.

\bibitem[{\citenamefont{Arg\"uelles~Delgado
  et~al.}(2021)\citenamefont{Arg\"uelles~Delgado, Kelly, and Mu\~noz
  Albornoz}}]{ArguellesDelgado:2021lek}
\bibinfo{author}{\bibfnamefont{C.~A.} \bibnamefont{Arg\"uelles~Delgado}},
  \bibinfo{author}{\bibfnamefont{K.~J.} \bibnamefont{Kelly}}, \bibnamefont{and}
  \bibinfo{author}{\bibfnamefont{V.}~\bibnamefont{Mu\~noz Albornoz}},
  \bibinfo{journal}{JHEP} \textbf{\bibinfo{volume}{11}}, \bibinfo{pages}{099}
  (\bibinfo{year}{2021}), \eprint{2104.13924}.

\bibitem[{\citenamefont{Arias-Arag\'on
  et~al.}(2020)\citenamefont{Arias-Arag\'on, D'eramo, Ferreira, Merlo, and
  Notari}}]{Arias-Aragon:2020qtn}
\bibinfo{author}{\bibfnamefont{F.}~\bibnamefont{Arias-Arag\'on}},
  \bibinfo{author}{\bibfnamefont{F.}~\bibnamefont{D'eramo}},
  \bibinfo{author}{\bibfnamefont{R.~Z.} \bibnamefont{Ferreira}},
  \bibinfo{author}{\bibfnamefont{L.}~\bibnamefont{Merlo}}, \bibnamefont{and}
  \bibinfo{author}{\bibfnamefont{A.}~\bibnamefont{Notari}},
  \bibinfo{journal}{JCAP} \textbf{\bibinfo{volume}{11}}, \bibinfo{pages}{025}
  (\bibinfo{year}{2020}), \eprint{2007.06579}.

\bibitem[{\citenamefont{Arias et~al.}(2021)\citenamefont{Arias, Arza, Jaeckel,
  and Vargas-Arancibia}}]{Arias:2020tzl}
\bibinfo{author}{\bibfnamefont{P.}~\bibnamefont{Arias}},
  \bibinfo{author}{\bibfnamefont{A.}~\bibnamefont{Arza}},
  \bibinfo{author}{\bibfnamefont{J.}~\bibnamefont{Jaeckel}}, \bibnamefont{and}
  \bibinfo{author}{\bibfnamefont{D.}~\bibnamefont{Vargas-Arancibia}},
  \bibinfo{journal}{JCAP} \textbf{\bibinfo{volume}{05}}, \bibinfo{pages}{070}
  (\bibinfo{year}{2021}), \eprint{2007.12585}.

\bibitem[{\citenamefont{Aristizabal~Sierra
  et~al.}(2020{\natexlab{b}})\citenamefont{Aristizabal~Sierra, Branada,
  Miranda, and Sanchez~Garcia}}]{AristizabalSierra:2020zod}
\bibinfo{author}{\bibfnamefont{D.}~\bibnamefont{Aristizabal~Sierra}},
  \bibinfo{author}{\bibfnamefont{R.}~\bibnamefont{Branada}},
  \bibinfo{author}{\bibfnamefont{O.~G.} \bibnamefont{Miranda}},
  \bibnamefont{and}
  \bibinfo{author}{\bibfnamefont{G.}~\bibnamefont{Sanchez~Garcia}},
  \bibinfo{journal}{JHEP} \textbf{\bibinfo{volume}{12}}, \bibinfo{pages}{178}
  (\bibinfo{year}{2020}{\natexlab{b}}), \eprint{2008.05080}.

\bibitem[{\citenamefont{Athron et~al.}(2021)}]{Athron:2020maw}
\bibinfo{author}{\bibfnamefont{P.}~\bibnamefont{Athron}} \bibnamefont{et~al.},
  \bibinfo{journal}{JHEP} \textbf{\bibinfo{volume}{05}}, \bibinfo{pages}{159}
  (\bibinfo{year}{2021}), \eprint{2007.05517}.

\bibitem[{\citenamefont{Babu et~al.}(2020)\citenamefont{Babu, Jana, and
  Lindner}}]{Babu:2020ivd}
\bibinfo{author}{\bibfnamefont{K.~S.} \bibnamefont{Babu}},
  \bibinfo{author}{\bibfnamefont{S.}~\bibnamefont{Jana}}, \bibnamefont{and}
  \bibinfo{author}{\bibfnamefont{M.}~\bibnamefont{Lindner}},
  \bibinfo{journal}{JHEP} \textbf{\bibinfo{volume}{10}}, \bibinfo{pages}{040}
  (\bibinfo{year}{2020}), \eprint{2007.04291}.

\bibitem[{\citenamefont{Babu et~al.}(2021)\citenamefont{Babu, Jana, Lindner,
  and K}}]{Babu:2021jnu}
\bibinfo{author}{\bibfnamefont{K.~S.} \bibnamefont{Babu}},
  \bibinfo{author}{\bibfnamefont{S.}~\bibnamefont{Jana}},
  \bibinfo{author}{\bibfnamefont{M.}~\bibnamefont{Lindner}}, \bibnamefont{and}
  \bibinfo{author}{\bibfnamefont{V.~P.} \bibnamefont{K}},
  \bibinfo{journal}{JHEP} \textbf{\bibinfo{volume}{10}}, \bibinfo{pages}{240}
  (\bibinfo{year}{2021}), \eprint{2104.03291}.

\bibitem[{\citenamefont{Baek et~al.}(2020)\citenamefont{Baek, Kim, and
  Ko}}]{Baek:2020owl}
\bibinfo{author}{\bibfnamefont{S.}~\bibnamefont{Baek}},
  \bibinfo{author}{\bibfnamefont{J.}~\bibnamefont{Kim}}, \bibnamefont{and}
  \bibinfo{author}{\bibfnamefont{P.}~\bibnamefont{Ko}}, \bibinfo{journal}{Phys.
  Lett. B} \textbf{\bibinfo{volume}{810}}, \bibinfo{pages}{135848}
  (\bibinfo{year}{2020}), \eprint{2006.16876}.

\bibitem[{\citenamefont{Baek}(2021)}]{Baek:2021yos}
\bibinfo{author}{\bibfnamefont{S.}~\bibnamefont{Baek}}, \bibinfo{journal}{JHEP}
  \textbf{\bibinfo{volume}{10}}, \bibinfo{pages}{135} (\bibinfo{year}{2021}),
  \eprint{2105.00877}.

\bibitem[{\citenamefont{Bally et~al.}(2020)\citenamefont{Bally, Jana, and
  Trautner}}]{Bally:2020yid}
\bibinfo{author}{\bibfnamefont{A.}~\bibnamefont{Bally}},
  \bibinfo{author}{\bibfnamefont{S.}~\bibnamefont{Jana}}, \bibnamefont{and}
  \bibinfo{author}{\bibfnamefont{A.}~\bibnamefont{Trautner}},
  \bibinfo{journal}{Phys. Rev. Lett.} \textbf{\bibinfo{volume}{125}},
  \bibinfo{pages}{161802} (\bibinfo{year}{2020}), \eprint{2006.11919}.

\bibitem[{\citenamefont{Baryakhtar et~al.}(2020)\citenamefont{Baryakhtar,
  Berlin, Liu, and Weiner}}]{Baryakhtar:2020rwy}
\bibinfo{author}{\bibfnamefont{M.}~\bibnamefont{Baryakhtar}},
  \bibinfo{author}{\bibfnamefont{A.}~\bibnamefont{Berlin}},
  \bibinfo{author}{\bibfnamefont{H.}~\bibnamefont{Liu}}, \bibnamefont{and}
  \bibinfo{author}{\bibfnamefont{N.}~\bibnamefont{Weiner}}
  (\bibinfo{year}{2020}), \eprint{2006.13918}.

\bibitem[{\citenamefont{Baym and Peng}(2021)}]{Baym:2020riw}
\bibinfo{author}{\bibfnamefont{G.}~\bibnamefont{Baym}} \bibnamefont{and}
  \bibinfo{author}{\bibfnamefont{J.-C.} \bibnamefont{Peng}},
  \bibinfo{journal}{Phys. Rev. Lett.} \textbf{\bibinfo{volume}{126}},
  \bibinfo{pages}{191803} (\bibinfo{year}{2021}), \eprint{2012.12421}.

\bibitem[{\citenamefont{Benakli et~al.}(2020)\citenamefont{Benakli, Branchina,
  and Lafforgue-Marmet}}]{Benakli:2020vng}
\bibinfo{author}{\bibfnamefont{K.}~\bibnamefont{Benakli}},
  \bibinfo{author}{\bibfnamefont{C.}~\bibnamefont{Branchina}},
  \bibnamefont{and}
  \bibinfo{author}{\bibfnamefont{G.}~\bibnamefont{Lafforgue-Marmet}},
  \bibinfo{journal}{Eur. Phys. J. C} \textbf{\bibinfo{volume}{80}},
  \bibinfo{pages}{1118} (\bibinfo{year}{2020}), \eprint{2007.02655}.

\bibitem[{\citenamefont{Bhattacherjee and
  Sengupta}(2021)}]{Bhattacherjee:2020qmv}
\bibinfo{author}{\bibfnamefont{B.}~\bibnamefont{Bhattacherjee}}
  \bibnamefont{and} \bibinfo{author}{\bibfnamefont{R.}~\bibnamefont{Sengupta}},
  \bibinfo{journal}{Phys. Lett. B} \textbf{\bibinfo{volume}{817}},
  \bibinfo{pages}{136305} (\bibinfo{year}{2021}), \eprint{2006.16172}.

\bibitem[{\citenamefont{Bloch et~al.}(2021)\citenamefont{Bloch, Caputo, Essig,
  Redigolo, Sholapurkar, and Volansky}}]{Bloch:2020uzh}
\bibinfo{author}{\bibfnamefont{I.~M.} \bibnamefont{Bloch}},
  \bibinfo{author}{\bibfnamefont{A.}~\bibnamefont{Caputo}},
  \bibinfo{author}{\bibfnamefont{R.}~\bibnamefont{Essig}},
  \bibinfo{author}{\bibfnamefont{D.}~\bibnamefont{Redigolo}},
  \bibinfo{author}{\bibfnamefont{M.}~\bibnamefont{Sholapurkar}},
  \bibnamefont{and} \bibinfo{author}{\bibfnamefont{T.}~\bibnamefont{Volansky}},
  \bibinfo{journal}{JHEP} \textbf{\bibinfo{volume}{01}}, \bibinfo{pages}{178}
  (\bibinfo{year}{2021}), \eprint{2006.14521}.

\bibitem[{\citenamefont{Boehm et~al.}(2020)\citenamefont{Boehm, Cerdeno,
  Fairbairn, Machado, and Vincent}}]{Boehm:2020ltd}
\bibinfo{author}{\bibfnamefont{C.}~\bibnamefont{Boehm}},
  \bibinfo{author}{\bibfnamefont{D.~G.} \bibnamefont{Cerdeno}},
  \bibinfo{author}{\bibfnamefont{M.}~\bibnamefont{Fairbairn}},
  \bibinfo{author}{\bibfnamefont{P.~A.~N.} \bibnamefont{Machado}},
  \bibnamefont{and} \bibinfo{author}{\bibfnamefont{A.~C.}
  \bibnamefont{Vincent}}, \bibinfo{journal}{Phys. Rev. D}
  \textbf{\bibinfo{volume}{102}}, \bibinfo{pages}{115013}
  (\bibinfo{year}{2020}), \eprint{2006.11250}.

\bibitem[{\citenamefont{Borah et~al.}(2020)\citenamefont{Borah, Mahapatra,
  Nanda, and Sahu}}]{Borah:2020jzi}
\bibinfo{author}{\bibfnamefont{D.}~\bibnamefont{Borah}},
  \bibinfo{author}{\bibfnamefont{S.}~\bibnamefont{Mahapatra}},
  \bibinfo{author}{\bibfnamefont{D.}~\bibnamefont{Nanda}}, \bibnamefont{and}
  \bibinfo{author}{\bibfnamefont{N.}~\bibnamefont{Sahu}},
  \bibinfo{journal}{Phys. Lett. B} \textbf{\bibinfo{volume}{811}},
  \bibinfo{pages}{135933} (\bibinfo{year}{2020}), \eprint{2007.10754}.

\bibitem[{\citenamefont{Borah et~al.}(2021{\natexlab{a}})\citenamefont{Borah,
  Mahapatra, and Sahu}}]{Borah:2020smw}
\bibinfo{author}{\bibfnamefont{D.}~\bibnamefont{Borah}},
  \bibinfo{author}{\bibfnamefont{S.}~\bibnamefont{Mahapatra}},
  \bibnamefont{and} \bibinfo{author}{\bibfnamefont{N.}~\bibnamefont{Sahu}},
  \bibinfo{journal}{Nucl. Phys. B} \textbf{\bibinfo{volume}{968}},
  \bibinfo{pages}{115407} (\bibinfo{year}{2021}{\natexlab{a}}),
  \eprint{2009.06294}.

\bibitem[{\citenamefont{Borah et~al.}(2021{\natexlab{b}})\citenamefont{Borah,
  Dutta, Mahapatra, and Sahu}}]{Borah:2021jzu}
\bibinfo{author}{\bibfnamefont{D.}~\bibnamefont{Borah}},
  \bibinfo{author}{\bibfnamefont{M.}~\bibnamefont{Dutta}},
  \bibinfo{author}{\bibfnamefont{S.}~\bibnamefont{Mahapatra}},
  \bibnamefont{and} \bibinfo{author}{\bibfnamefont{N.}~\bibnamefont{Sahu}},
  \bibinfo{journal}{Phys. Lett. B} \textbf{\bibinfo{volume}{820}},
  \bibinfo{pages}{136577} (\bibinfo{year}{2021}{\natexlab{b}}),
  \eprint{2104.05656}.

\bibitem[{\citenamefont{Bramante and Song}(2020)}]{Bramante:2020zos}
\bibinfo{author}{\bibfnamefont{J.}~\bibnamefont{Bramante}} \bibnamefont{and}
  \bibinfo{author}{\bibfnamefont{N.}~\bibnamefont{Song}},
  \bibinfo{journal}{Phys. Rev. Lett.} \textbf{\bibinfo{volume}{125}},
  \bibinfo{pages}{161805} (\bibinfo{year}{2020}), \eprint{2006.14089}.

\bibitem[{\citenamefont{Brdar et~al.}(2021)\citenamefont{Brdar, Greljo, Kopp,
  and Opferkuch}}]{Brdar:2020quo}
\bibinfo{author}{\bibfnamefont{V.}~\bibnamefont{Brdar}},
  \bibinfo{author}{\bibfnamefont{A.}~\bibnamefont{Greljo}},
  \bibinfo{author}{\bibfnamefont{J.}~\bibnamefont{Kopp}}, \bibnamefont{and}
  \bibinfo{author}{\bibfnamefont{T.}~\bibnamefont{Opferkuch}},
  \bibinfo{journal}{JCAP} \textbf{\bibinfo{volume}{01}}, \bibinfo{pages}{039}
  (\bibinfo{year}{2021}), \eprint{2007.15563}.

\bibitem[{\citenamefont{Buch et~al.}(2020)\citenamefont{Buch, Buen-Abad, Fan,
  and Leung}}]{Buch:2020mrg}
\bibinfo{author}{\bibfnamefont{J.}~\bibnamefont{Buch}},
  \bibinfo{author}{\bibfnamefont{M.~A.} \bibnamefont{Buen-Abad}},
  \bibinfo{author}{\bibfnamefont{J.}~\bibnamefont{Fan}}, \bibnamefont{and}
  \bibinfo{author}{\bibfnamefont{J.~S.~C.} \bibnamefont{Leung}},
  \bibinfo{journal}{JCAP} \textbf{\bibinfo{volume}{10}}, \bibinfo{pages}{051}
  (\bibinfo{year}{2020}), \eprint{2006.12488}.

\bibitem[{\citenamefont{Budnik et~al.}(2020)\citenamefont{Budnik, Kim,
  Matsedonskyi, Perez, and Soreq}}]{Budnik:2020nwz}
\bibinfo{author}{\bibfnamefont{R.}~\bibnamefont{Budnik}},
  \bibinfo{author}{\bibfnamefont{H.}~\bibnamefont{Kim}},
  \bibinfo{author}{\bibfnamefont{O.}~\bibnamefont{Matsedonskyi}},
  \bibinfo{author}{\bibfnamefont{G.}~\bibnamefont{Perez}}, \bibnamefont{and}
  \bibinfo{author}{\bibfnamefont{Y.}~\bibnamefont{Soreq}}
  (\bibinfo{year}{2020}), \eprint{2006.14568}.

\bibitem[{\citenamefont{Buttazzo et~al.}(2021)\citenamefont{Buttazzo, Panci,
  Teresi, and Ziegler}}]{Buttazzo:2020vfs}
\bibinfo{author}{\bibfnamefont{D.}~\bibnamefont{Buttazzo}},
  \bibinfo{author}{\bibfnamefont{P.}~\bibnamefont{Panci}},
  \bibinfo{author}{\bibfnamefont{D.}~\bibnamefont{Teresi}}, \bibnamefont{and}
  \bibinfo{author}{\bibfnamefont{R.}~\bibnamefont{Ziegler}},
  \bibinfo{journal}{Phys. Lett. B} \textbf{\bibinfo{volume}{817}},
  \bibinfo{pages}{136310} (\bibinfo{year}{2021}), \eprint{2011.08919}.

\bibitem[{\citenamefont{Cai et~al.}(2020)\citenamefont{Cai, Sun, Zhang, and
  Zhang}}]{Cai:2020kfq}
\bibinfo{author}{\bibfnamefont{R.-G.} \bibnamefont{Cai}},
  \bibinfo{author}{\bibfnamefont{S.}~\bibnamefont{Sun}},
  \bibinfo{author}{\bibfnamefont{B.}~\bibnamefont{Zhang}}, \bibnamefont{and}
  \bibinfo{author}{\bibfnamefont{Y.-L.} \bibnamefont{Zhang}}
  (\bibinfo{year}{2020}), \eprint{2009.02315}.

\bibitem[{\citenamefont{Cao et~al.}(2021)\citenamefont{Cao, Ding, and
  Xiang}}]{Cao:2020bwd}
\bibinfo{author}{\bibfnamefont{Q.-H.} \bibnamefont{Cao}},
  \bibinfo{author}{\bibfnamefont{R.}~\bibnamefont{Ding}}, \bibnamefont{and}
  \bibinfo{author}{\bibfnamefont{Q.-F.} \bibnamefont{Xiang}},
  \bibinfo{journal}{Chin. Phys. C} \textbf{\bibinfo{volume}{45}},
  \bibinfo{pages}{045002} (\bibinfo{year}{2021}), \eprint{2006.12767}.

\bibitem[{\citenamefont{Cao et~al.}(2020{\natexlab{b}})\citenamefont{Cao, Du,
  Li, Wang, and Zhang}}]{Cao:2020oxq}
\bibinfo{author}{\bibfnamefont{J.}~\bibnamefont{Cao}},
  \bibinfo{author}{\bibfnamefont{X.}~\bibnamefont{Du}},
  \bibinfo{author}{\bibfnamefont{Z.}~\bibnamefont{Li}},
  \bibinfo{author}{\bibfnamefont{F.}~\bibnamefont{Wang}}, \bibnamefont{and}
  \bibinfo{author}{\bibfnamefont{Y.}~\bibnamefont{Zhang}}
  (\bibinfo{year}{2020}{\natexlab{b}}), \eprint{2007.09981}.

\bibitem[{\citenamefont{Chakraborty et~al.}(2020)\citenamefont{Chakraborty,
  Jung, Loladze, Okui, and Tobioka}}]{Chakraborty:2020vec}
\bibinfo{author}{\bibfnamefont{S.}~\bibnamefont{Chakraborty}},
  \bibinfo{author}{\bibfnamefont{T.~H.} \bibnamefont{Jung}},
  \bibinfo{author}{\bibfnamefont{V.}~\bibnamefont{Loladze}},
  \bibinfo{author}{\bibfnamefont{T.}~\bibnamefont{Okui}}, \bibnamefont{and}
  \bibinfo{author}{\bibfnamefont{K.}~\bibnamefont{Tobioka}},
  \bibinfo{journal}{Phys. Rev. D} \textbf{\bibinfo{volume}{102}},
  \bibinfo{pages}{095029} (\bibinfo{year}{2020}), \eprint{2008.10610}.

\bibitem[{\citenamefont{Chala and Titov}(2020)}]{Chala:2020pbn}
\bibinfo{author}{\bibfnamefont{M.}~\bibnamefont{Chala}} \bibnamefont{and}
  \bibinfo{author}{\bibfnamefont{A.}~\bibnamefont{Titov}},
  \bibinfo{journal}{JHEP} \textbf{\bibinfo{volume}{09}}, \bibinfo{pages}{188}
  (\bibinfo{year}{2020}), \eprint{2006.14596}.

\bibitem[{\citenamefont{Chao et~al.}(2020)\citenamefont{Chao, Gao, and
  Jin}}]{Chao:2020yro}
\bibinfo{author}{\bibfnamefont{W.}~\bibnamefont{Chao}},
  \bibinfo{author}{\bibfnamefont{Y.}~\bibnamefont{Gao}}, \bibnamefont{and}
  \bibinfo{author}{\bibfnamefont{M.~j.} \bibnamefont{Jin}}
  (\bibinfo{year}{2020}), \eprint{2006.16145}.

\bibitem[{\citenamefont{Chen et~al.}(2021{\natexlab{a}})\citenamefont{Chen,
  Cui, Shu, Xue, Yuan, and Yuan}}]{Chen:2020gcl}
\bibinfo{author}{\bibfnamefont{Y.}~\bibnamefont{Chen}},
  \bibinfo{author}{\bibfnamefont{M.-Y.} \bibnamefont{Cui}},
  \bibinfo{author}{\bibfnamefont{J.}~\bibnamefont{Shu}},
  \bibinfo{author}{\bibfnamefont{X.}~\bibnamefont{Xue}},
  \bibinfo{author}{\bibfnamefont{G.-W.} \bibnamefont{Yuan}}, \bibnamefont{and}
  \bibinfo{author}{\bibfnamefont{Q.}~\bibnamefont{Yuan}},
  \bibinfo{journal}{JHEP} \textbf{\bibinfo{volume}{04}}, \bibinfo{pages}{282}
  (\bibinfo{year}{2021}{\natexlab{a}}), \eprint{2006.12447}.

\bibitem[{\citenamefont{Chen et~al.}(2021{\natexlab{b}})\citenamefont{Chen,
  Gelmini, and Takhistov}}]{Chen:2021qao}
\bibinfo{author}{\bibfnamefont{M.}~\bibnamefont{Chen}},
  \bibinfo{author}{\bibfnamefont{G.~B.} \bibnamefont{Gelmini}},
  \bibnamefont{and}
  \bibinfo{author}{\bibfnamefont{V.}~\bibnamefont{Takhistov}},
  \bibinfo{journal}{JCAP} \textbf{\bibinfo{volume}{12}}, \bibinfo{pages}{048}
  (\bibinfo{year}{2021}{\natexlab{b}}), \eprint{2105.08101}.

\bibitem[{\citenamefont{Chen et~al.}(2021{\natexlab{c}})\citenamefont{Chen, Li,
  and Liao}}]{Chen:2021uuw}
\bibinfo{author}{\bibfnamefont{Z.}~\bibnamefont{Chen}},
  \bibinfo{author}{\bibfnamefont{T.}~\bibnamefont{Li}}, \bibnamefont{and}
  \bibinfo{author}{\bibfnamefont{J.}~\bibnamefont{Liao}},
  \bibinfo{journal}{JHEP} \textbf{\bibinfo{volume}{05}}, \bibinfo{pages}{131}
  (\bibinfo{year}{2021}{\natexlab{c}}), \eprint{2102.09784}.

\bibitem[{\citenamefont{Chiang and Lu}(2020)}]{Chiang:2020hgb}
\bibinfo{author}{\bibfnamefont{C.-W.} \bibnamefont{Chiang}} \bibnamefont{and}
  \bibinfo{author}{\bibfnamefont{B.-Q.} \bibnamefont{Lu}},
  \bibinfo{journal}{Phys. Rev. D} \textbf{\bibinfo{volume}{102}},
  \bibinfo{pages}{123006} (\bibinfo{year}{2020}), \eprint{2007.06401}.

\bibitem[{\citenamefont{Chigusa et~al.}(2020)\citenamefont{Chigusa, Endo, and
  Kohri}}]{Chigusa:2020bgq}
\bibinfo{author}{\bibfnamefont{S.}~\bibnamefont{Chigusa}},
  \bibinfo{author}{\bibfnamefont{M.}~\bibnamefont{Endo}}, \bibnamefont{and}
  \bibinfo{author}{\bibfnamefont{K.}~\bibnamefont{Kohri}},
  \bibinfo{journal}{JCAP} \textbf{\bibinfo{volume}{10}}, \bibinfo{pages}{035}
  (\bibinfo{year}{2020}), \eprint{2007.01663}.

\bibitem[{\citenamefont{Choi et~al.}(2020{\natexlab{a}})\citenamefont{Choi,
  Yanagida, and Yokozaki}}]{Choi:2020kch}
\bibinfo{author}{\bibfnamefont{G.}~\bibnamefont{Choi}},
  \bibinfo{author}{\bibfnamefont{T.~T.} \bibnamefont{Yanagida}},
  \bibnamefont{and} \bibinfo{author}{\bibfnamefont{N.}~\bibnamefont{Yokozaki}},
  \bibinfo{journal}{Phys. Lett. B} \textbf{\bibinfo{volume}{810}},
  \bibinfo{pages}{135836} (\bibinfo{year}{2020}{\natexlab{a}}),
  \eprint{2007.04278}.

\bibitem[{\citenamefont{Choi et~al.}(2020{\natexlab{b}})\citenamefont{Choi,
  Suzuki, and Yanagida}}]{Choi:2020udy}
\bibinfo{author}{\bibfnamefont{G.}~\bibnamefont{Choi}},
  \bibinfo{author}{\bibfnamefont{M.}~\bibnamefont{Suzuki}}, \bibnamefont{and}
  \bibinfo{author}{\bibfnamefont{T.~T.} \bibnamefont{Yanagida}},
  \bibinfo{journal}{Phys. Lett. B} \textbf{\bibinfo{volume}{811}},
  \bibinfo{pages}{135976} (\bibinfo{year}{2020}{\natexlab{b}}),
  \eprint{2006.12348}.

\bibitem[{\citenamefont{Choi et~al.}(2021)\citenamefont{Choi, Lee, and
  Zhu}}]{Choi:2020ysq}
\bibinfo{author}{\bibfnamefont{S.-M.} \bibnamefont{Choi}},
  \bibinfo{author}{\bibfnamefont{H.~M.} \bibnamefont{Lee}}, \bibnamefont{and}
  \bibinfo{author}{\bibfnamefont{B.}~\bibnamefont{Zhu}},
  \bibinfo{journal}{JHEP} \textbf{\bibinfo{volume}{04}}, \bibinfo{pages}{251}
  (\bibinfo{year}{2021}), \eprint{2012.03713}.

\bibitem[{\citenamefont{Choudhury et~al.}(2021)\citenamefont{Choudhury,
  Maharana, Sachdeva, and Sahdev}}]{Choudhury:2020xui}
\bibinfo{author}{\bibfnamefont{D.}~\bibnamefont{Choudhury}},
  \bibinfo{author}{\bibfnamefont{S.}~\bibnamefont{Maharana}},
  \bibinfo{author}{\bibfnamefont{D.}~\bibnamefont{Sachdeva}}, \bibnamefont{and}
  \bibinfo{author}{\bibfnamefont{V.}~\bibnamefont{Sahdev}},
  \bibinfo{journal}{Phys. Rev. D} \textbf{\bibinfo{volume}{103}},
  \bibinfo{pages}{015006} (\bibinfo{year}{2021}), \eprint{2007.08205}.

\bibitem[{\citenamefont{Coloma et~al.}(2020)\citenamefont{Coloma, Huber, and
  Link}}]{Coloma:2020voz}
\bibinfo{author}{\bibfnamefont{P.}~\bibnamefont{Coloma}},
  \bibinfo{author}{\bibfnamefont{P.}~\bibnamefont{Huber}}, \bibnamefont{and}
  \bibinfo{author}{\bibfnamefont{J.~M.} \bibnamefont{Link}}
  (\bibinfo{year}{2020}), \eprint{2006.15767}.

\bibitem[{\citenamefont{Croon et~al.}(2021)\citenamefont{Croon, McDermott, and
  Sakstein}}]{Croon:2020ehi}
\bibinfo{author}{\bibfnamefont{D.}~\bibnamefont{Croon}},
  \bibinfo{author}{\bibfnamefont{S.~D.} \bibnamefont{McDermott}},
  \bibnamefont{and} \bibinfo{author}{\bibfnamefont{J.}~\bibnamefont{Sakstein}},
  \bibinfo{journal}{Phys. Dark Univ.} \textbf{\bibinfo{volume}{32}},
  \bibinfo{pages}{100801} (\bibinfo{year}{2021}), \eprint{2007.00650}.

\bibitem[{\citenamefont{Davighi et~al.}(2020)\citenamefont{Davighi, McCullough,
  and Tooby-Smith}}]{Davighi:2020vap}
\bibinfo{author}{\bibfnamefont{J.}~\bibnamefont{Davighi}},
  \bibinfo{author}{\bibfnamefont{M.}~\bibnamefont{McCullough}},
  \bibnamefont{and}
  \bibinfo{author}{\bibfnamefont{J.}~\bibnamefont{Tooby-Smith}},
  \bibinfo{journal}{JHEP} \textbf{\bibinfo{volume}{11}}, \bibinfo{pages}{120}
  (\bibinfo{year}{2020}), \eprint{2007.03662}.

\bibitem[{\citenamefont{Davoudiasl et~al.}(2020)\citenamefont{Davoudiasl,
  Denton, and Gehrlein}}]{Davoudiasl:2020ypv}
\bibinfo{author}{\bibfnamefont{H.}~\bibnamefont{Davoudiasl}},
  \bibinfo{author}{\bibfnamefont{P.~B.} \bibnamefont{Denton}},
  \bibnamefont{and} \bibinfo{author}{\bibfnamefont{J.}~\bibnamefont{Gehrlein}},
  \bibinfo{journal}{Phys. Rev. D} \textbf{\bibinfo{volume}{102}},
  \bibinfo{pages}{091701} (\bibinfo{year}{2020}), \eprint{2007.04989}.

\bibitem[{\citenamefont{Delle~Rose et~al.}(2021)\citenamefont{Delle~Rose,
  H\"utsi, Marzo, and Marzola}}]{DelleRose:2020pbh}
\bibinfo{author}{\bibfnamefont{L.}~\bibnamefont{Delle~Rose}},
  \bibinfo{author}{\bibfnamefont{G.}~\bibnamefont{H\"utsi}},
  \bibinfo{author}{\bibfnamefont{C.}~\bibnamefont{Marzo}}, \bibnamefont{and}
  \bibinfo{author}{\bibfnamefont{L.}~\bibnamefont{Marzola}},
  \bibinfo{journal}{JCAP} \textbf{\bibinfo{volume}{02}}, \bibinfo{pages}{031}
  (\bibinfo{year}{2021}), \eprint{2006.16078}.

\bibitem[{\citenamefont{DeRocco et~al.}(2020)\citenamefont{DeRocco, Graham, and
  Rajendran}}]{DeRocco:2020xdt}
\bibinfo{author}{\bibfnamefont{W.}~\bibnamefont{DeRocco}},
  \bibinfo{author}{\bibfnamefont{P.~W.} \bibnamefont{Graham}},
  \bibnamefont{and}
  \bibinfo{author}{\bibfnamefont{S.}~\bibnamefont{Rajendran}},
  \bibinfo{journal}{Phys. Rev. D} \textbf{\bibinfo{volume}{102}},
  \bibinfo{pages}{075015} (\bibinfo{year}{2020}), \eprint{2006.15112}.

\bibitem[{\citenamefont{Dessert et~al.}(2021)\citenamefont{Dessert, Foster,
  Kahn, and Safdi}}]{Dessert:2020vxy}
\bibinfo{author}{\bibfnamefont{C.}~\bibnamefont{Dessert}},
  \bibinfo{author}{\bibfnamefont{J.~W.} \bibnamefont{Foster}},
  \bibinfo{author}{\bibfnamefont{Y.}~\bibnamefont{Kahn}}, \bibnamefont{and}
  \bibinfo{author}{\bibfnamefont{B.~R.} \bibnamefont{Safdi}},
  \bibinfo{journal}{Phys. Dark Univ.} \textbf{\bibinfo{volume}{34}},
  \bibinfo{pages}{100878} (\bibinfo{year}{2021}), \eprint{2006.16220}.

\bibitem[{\citenamefont{Dey et~al.}(2020)\citenamefont{Dey, Maity, and
  Ray}}]{Dey:2020sai}
\bibinfo{author}{\bibfnamefont{U.~K.} \bibnamefont{Dey}},
  \bibinfo{author}{\bibfnamefont{T.~N.} \bibnamefont{Maity}}, \bibnamefont{and}
  \bibinfo{author}{\bibfnamefont{T.~S.} \bibnamefont{Ray}},
  \bibinfo{journal}{Phys. Lett. B} \textbf{\bibinfo{volume}{811}},
  \bibinfo{pages}{135900} (\bibinfo{year}{2020}), \eprint{2006.12529}.

\bibitem[{\citenamefont{Di~Luzio et~al.}(2020)\citenamefont{Di~Luzio, Fedele,
  Giannotti, Mescia, and Nardi}}]{DiLuzio:2020jjp}
\bibinfo{author}{\bibfnamefont{L.}~\bibnamefont{Di~Luzio}},
  \bibinfo{author}{\bibfnamefont{M.}~\bibnamefont{Fedele}},
  \bibinfo{author}{\bibfnamefont{M.}~\bibnamefont{Giannotti}},
  \bibinfo{author}{\bibfnamefont{F.}~\bibnamefont{Mescia}}, \bibnamefont{and}
  \bibinfo{author}{\bibfnamefont{E.}~\bibnamefont{Nardi}},
  \bibinfo{journal}{Phys. Rev. Lett.} \textbf{\bibinfo{volume}{125}},
  \bibinfo{pages}{131804} (\bibinfo{year}{2020}), \eprint{2006.12487}.

\bibitem[{\citenamefont{Dror et~al.}(2021)\citenamefont{Dror, Elor, McGehee,
  and Yu}}]{Dror:2020czw}
\bibinfo{author}{\bibfnamefont{J.~A.} \bibnamefont{Dror}},
  \bibinfo{author}{\bibfnamefont{G.}~\bibnamefont{Elor}},
  \bibinfo{author}{\bibfnamefont{R.}~\bibnamefont{McGehee}}, \bibnamefont{and}
  \bibinfo{author}{\bibfnamefont{T.-T.} \bibnamefont{Yu}},
  \bibinfo{journal}{Phys. Rev. D} \textbf{\bibinfo{volume}{103}},
  \bibinfo{pages}{035001} (\bibinfo{year}{2021}), \eprint{2011.01940}.

\bibitem[{\citenamefont{Du et~al.}(2022)\citenamefont{Du, Egana-Ugrinovic,
  Essig, and Sholapurkar}}]{Du:2020ldo}
\bibinfo{author}{\bibfnamefont{P.}~\bibnamefont{Du}},
  \bibinfo{author}{\bibfnamefont{D.}~\bibnamefont{Egana-Ugrinovic}},
  \bibinfo{author}{\bibfnamefont{R.}~\bibnamefont{Essig}}, \bibnamefont{and}
  \bibinfo{author}{\bibfnamefont{M.}~\bibnamefont{Sholapurkar}},
  \bibinfo{journal}{Phys. Rev. X} \textbf{\bibinfo{volume}{12}},
  \bibinfo{pages}{011009} (\bibinfo{year}{2022}), \eprint{2011.13939}.

\bibitem[{\citenamefont{Du et~al.}(2021)\citenamefont{Du, Liang, Liu, Tran, and
  Xue}}]{Du:2020ybt}
\bibinfo{author}{\bibfnamefont{M.}~\bibnamefont{Du}},
  \bibinfo{author}{\bibfnamefont{J.}~\bibnamefont{Liang}},
  \bibinfo{author}{\bibfnamefont{Z.}~\bibnamefont{Liu}},
  \bibinfo{author}{\bibfnamefont{V.~Q.} \bibnamefont{Tran}}, \bibnamefont{and}
  \bibinfo{author}{\bibfnamefont{Y.}~\bibnamefont{Xue}},
  \bibinfo{journal}{Chin. Phys. C} \textbf{\bibinfo{volume}{45}},
  \bibinfo{pages}{013114} (\bibinfo{year}{2021}), \eprint{2006.11949}.

\bibitem[{\citenamefont{Dutta et~al.}(2021{\natexlab{a}})\citenamefont{Dutta,
  Ghosh, Kar, and Mukhopadhyaya}}]{Dutta:2021nsy}
\bibinfo{author}{\bibfnamefont{K.}~\bibnamefont{Dutta}},
  \bibinfo{author}{\bibfnamefont{A.}~\bibnamefont{Ghosh}},
  \bibinfo{author}{\bibfnamefont{A.}~\bibnamefont{Kar}}, \bibnamefont{and}
  \bibinfo{author}{\bibfnamefont{B.}~\bibnamefont{Mukhopadhyaya}}
  (\bibinfo{year}{2021}{\natexlab{a}}), \eprint{2103.14664}.

\bibitem[{\citenamefont{Dutta et~al.}(2021{\natexlab{b}})\citenamefont{Dutta,
  Mahapatra, Borah, and Sahu}}]{Dutta:2021wbn}
\bibinfo{author}{\bibfnamefont{M.}~\bibnamefont{Dutta}},
  \bibinfo{author}{\bibfnamefont{S.}~\bibnamefont{Mahapatra}},
  \bibinfo{author}{\bibfnamefont{D.}~\bibnamefont{Borah}}, \bibnamefont{and}
  \bibinfo{author}{\bibfnamefont{N.}~\bibnamefont{Sahu}},
  \bibinfo{journal}{Phys. Rev. D} \textbf{\bibinfo{volume}{103}},
  \bibinfo{pages}{095018} (\bibinfo{year}{2021}{\natexlab{b}}),
  \eprint{2101.06472}.

\bibitem[{\citenamefont{Ema et~al.}(2021)\citenamefont{Ema, Sala, and
  Sato}}]{Ema:2020fit}
\bibinfo{author}{\bibfnamefont{Y.}~\bibnamefont{Ema}},
  \bibinfo{author}{\bibfnamefont{F.}~\bibnamefont{Sala}}, \bibnamefont{and}
  \bibinfo{author}{\bibfnamefont{R.}~\bibnamefont{Sato}},
  \bibinfo{journal}{Eur. Phys. J. C} \textbf{\bibinfo{volume}{81}},
  \bibinfo{pages}{129} (\bibinfo{year}{2021}), \eprint{2007.09105}.

\bibitem[{\citenamefont{Escribano and Vicente}(2021)}]{Escribano:2020wua}
\bibinfo{author}{\bibfnamefont{P.}~\bibnamefont{Escribano}} \bibnamefont{and}
  \bibinfo{author}{\bibfnamefont{A.}~\bibnamefont{Vicente}},
  \bibinfo{journal}{JHEP} \textbf{\bibinfo{volume}{03}}, \bibinfo{pages}{240}
  (\bibinfo{year}{2021}), \eprint{2008.01099}.

\bibitem[{\citenamefont{Farzan and Rajaee}(2020)}]{Farzan:2020llg}
\bibinfo{author}{\bibfnamefont{Y.}~\bibnamefont{Farzan}} \bibnamefont{and}
  \bibinfo{author}{\bibfnamefont{M.}~\bibnamefont{Rajaee}},
  \bibinfo{journal}{Phys. Rev. D} \textbf{\bibinfo{volume}{102}},
  \bibinfo{pages}{103532} (\bibinfo{year}{2020}), \eprint{2007.14421}.

\bibitem[{\citenamefont{Fayet}(2021)}]{Fayet:2020bmb}
\bibinfo{author}{\bibfnamefont{P.}~\bibnamefont{Fayet}},
  \bibinfo{journal}{Phys. Rev. D} \textbf{\bibinfo{volume}{103}},
  \bibinfo{pages}{035034} (\bibinfo{year}{2021}), \eprint{2010.04673}.

\bibitem[{\citenamefont{Fonseca and Morgante}(2021)}]{Fonseca:2020pjs}
\bibinfo{author}{\bibfnamefont{N.}~\bibnamefont{Fonseca}} \bibnamefont{and}
  \bibinfo{author}{\bibfnamefont{E.}~\bibnamefont{Morgante}},
  \bibinfo{journal}{Phys. Rev. D} \textbf{\bibinfo{volume}{103}},
  \bibinfo{pages}{015011} (\bibinfo{year}{2021}), \eprint{2009.10974}.

\bibitem[{\citenamefont{Foot}(2020)}]{Foot:2020ehn}
\bibinfo{author}{\bibfnamefont{R.}~\bibnamefont{Foot}} (\bibinfo{year}{2020}),
  \eprint{2011.02590}.

\bibitem[{\citenamefont{Fornal et~al.}(2020)\citenamefont{Fornal, Sandick, Shu,
  Su, and Zhao}}]{Fornal:2020npv}
\bibinfo{author}{\bibfnamefont{B.}~\bibnamefont{Fornal}},
  \bibinfo{author}{\bibfnamefont{P.}~\bibnamefont{Sandick}},
  \bibinfo{author}{\bibfnamefont{J.}~\bibnamefont{Shu}},
  \bibinfo{author}{\bibfnamefont{M.}~\bibnamefont{Su}}, \bibnamefont{and}
  \bibinfo{author}{\bibfnamefont{Y.}~\bibnamefont{Zhao}},
  \bibinfo{journal}{Phys. Rev. Lett.} \textbf{\bibinfo{volume}{125}},
  \bibinfo{pages}{161804} (\bibinfo{year}{2020}), \eprint{2006.11264}.

\bibitem[{\citenamefont{Gao et~al.}(2020)\citenamefont{Gao, Liu, Wang, Wang,
  Xue, and Zhong}}]{Gao:2020wer}
\bibinfo{author}{\bibfnamefont{C.}~\bibnamefont{Gao}},
  \bibinfo{author}{\bibfnamefont{J.}~\bibnamefont{Liu}},
  \bibinfo{author}{\bibfnamefont{L.-T.} \bibnamefont{Wang}},
  \bibinfo{author}{\bibfnamefont{X.-P.} \bibnamefont{Wang}},
  \bibinfo{author}{\bibfnamefont{W.}~\bibnamefont{Xue}}, \bibnamefont{and}
  \bibinfo{author}{\bibfnamefont{Y.-M.} \bibnamefont{Zhong}},
  \bibinfo{journal}{Phys. Rev. Lett.} \textbf{\bibinfo{volume}{125}},
  \bibinfo{pages}{131806} (\bibinfo{year}{2020}), \eprint{2006.14598}.

\bibitem[{\citenamefont{Gao and Li}(2020)}]{Gao:2020wfr}
\bibinfo{author}{\bibfnamefont{Y.}~\bibnamefont{Gao}} \bibnamefont{and}
  \bibinfo{author}{\bibfnamefont{T.}~\bibnamefont{Li}} (\bibinfo{year}{2020}),
  \eprint{2006.16192}.

\bibitem[{\citenamefont{Ge et~al.}(2020{\natexlab{b}})\citenamefont{Ge,
  Pasquini, and Sheng}}]{Ge:2020jfn}
\bibinfo{author}{\bibfnamefont{S.-F.} \bibnamefont{Ge}},
  \bibinfo{author}{\bibfnamefont{P.}~\bibnamefont{Pasquini}}, \bibnamefont{and}
  \bibinfo{author}{\bibfnamefont{J.}~\bibnamefont{Sheng}},
  \bibinfo{journal}{Phys. Lett. B} \textbf{\bibinfo{volume}{810}},
  \bibinfo{pages}{135787} (\bibinfo{year}{2020}{\natexlab{b}}),
  \eprint{2006.16069}.

\bibitem[{\citenamefont{Guo et~al.}(2020)\citenamefont{Guo, Tsai, Wu, and
  Yuan}}]{Guo:2020oum}
\bibinfo{author}{\bibfnamefont{G.}~\bibnamefont{Guo}},
  \bibinfo{author}{\bibfnamefont{Y.-L.~S.} \bibnamefont{Tsai}},
  \bibinfo{author}{\bibfnamefont{M.-R.} \bibnamefont{Wu}}, \bibnamefont{and}
  \bibinfo{author}{\bibfnamefont{Q.}~\bibnamefont{Yuan}},
  \bibinfo{journal}{Phys. Rev. D} \textbf{\bibinfo{volume}{102}},
  \bibinfo{pages}{103004} (\bibinfo{year}{2020}), \eprint{2008.12137}.

\bibitem[{\citenamefont{Han et~al.}(2021{\natexlab{b}})\citenamefont{Han,
  L\'opez-Ib\'a\~nez, Melis, Vives, and Yang}}]{Han:2020dwo}
\bibinfo{author}{\bibfnamefont{C.}~\bibnamefont{Han}},
  \bibinfo{author}{\bibfnamefont{M.~L.} \bibnamefont{L\'opez-Ib\'a\~nez}},
  \bibinfo{author}{\bibfnamefont{A.}~\bibnamefont{Melis}},
  \bibinfo{author}{\bibfnamefont{O.}~\bibnamefont{Vives}}, \bibnamefont{and}
  \bibinfo{author}{\bibfnamefont{J.~M.} \bibnamefont{Yang}},
  \bibinfo{journal}{Phys. Rev. D} \textbf{\bibinfo{volume}{103}},
  \bibinfo{pages}{035028} (\bibinfo{year}{2021}{\natexlab{b}}),
  \eprint{2007.08834}.

\bibitem[{\citenamefont{Harigaya et~al.}(2020)\citenamefont{Harigaya, Nakai,
  and Suzuki}}]{Harigaya:2020ckz}
\bibinfo{author}{\bibfnamefont{K.}~\bibnamefont{Harigaya}},
  \bibinfo{author}{\bibfnamefont{Y.}~\bibnamefont{Nakai}}, \bibnamefont{and}
  \bibinfo{author}{\bibfnamefont{M.}~\bibnamefont{Suzuki}},
  \bibinfo{journal}{Phys. Lett. B} \textbf{\bibinfo{volume}{809}},
  \bibinfo{pages}{135729} (\bibinfo{year}{2020}), \eprint{2006.11938}.

\bibitem[{\citenamefont{Harnik et~al.}(2021)\citenamefont{Harnik, Plestid,
  Pospelov, and Ramani}}]{Harnik:2020ugb}
\bibinfo{author}{\bibfnamefont{R.}~\bibnamefont{Harnik}},
  \bibinfo{author}{\bibfnamefont{R.}~\bibnamefont{Plestid}},
  \bibinfo{author}{\bibfnamefont{M.}~\bibnamefont{Pospelov}}, \bibnamefont{and}
  \bibinfo{author}{\bibfnamefont{H.}~\bibnamefont{Ramani}},
  \bibinfo{journal}{Phys. Rev. D} \textbf{\bibinfo{volume}{103}},
  \bibinfo{pages}{075029} (\bibinfo{year}{2021}), \eprint{2010.11190}.

\bibitem[{\citenamefont{Haselschwardt
  et~al.}(2020{\natexlab{b}})\citenamefont{Haselschwardt, Kostensalo, Mougeot,
  and Suhonen}}]{Haselschwardt:2020iey}
\bibinfo{author}{\bibfnamefont{S.~J.} \bibnamefont{Haselschwardt}},
  \bibinfo{author}{\bibfnamefont{J.}~\bibnamefont{Kostensalo}},
  \bibinfo{author}{\bibfnamefont{X.}~\bibnamefont{Mougeot}}, \bibnamefont{and}
  \bibinfo{author}{\bibfnamefont{J.}~\bibnamefont{Suhonen}},
  \bibinfo{journal}{Phys. Rev. C} \textbf{\bibinfo{volume}{102}},
  \bibinfo{pages}{065501} (\bibinfo{year}{2020}{\natexlab{b}}),
  \eprint{2007.13686}.

\bibitem[{\citenamefont{Hayen et~al.}(2020)\citenamefont{Hayen, Simonucci, and
  Taioli}}]{Hayen:2020mod}
\bibinfo{author}{\bibfnamefont{L.}~\bibnamefont{Hayen}},
  \bibinfo{author}{\bibfnamefont{S.}~\bibnamefont{Simonucci}},
  \bibnamefont{and} \bibinfo{author}{\bibfnamefont{S.}~\bibnamefont{Taioli}}
  (\bibinfo{year}{2020}), \eprint{2009.08303}.

\bibitem[{\citenamefont{He et~al.}(2021{\natexlab{a}})\citenamefont{He, Wang,
  and Zheng}}]{He:2020sat}
\bibinfo{author}{\bibfnamefont{H.-J.} \bibnamefont{He}},
  \bibinfo{author}{\bibfnamefont{Y.-C.} \bibnamefont{Wang}}, \bibnamefont{and}
  \bibinfo{author}{\bibfnamefont{J.}~\bibnamefont{Zheng}},
  \bibinfo{journal}{Phys. Rev. D} \textbf{\bibinfo{volume}{104}},
  \bibinfo{pages}{115033} (\bibinfo{year}{2021}{\natexlab{a}}),
  \eprint{2012.05891}.

\bibitem[{\citenamefont{He et~al.}(2021{\natexlab{b}})\citenamefont{He, Wang,
  and Zheng}}]{He:2020wjs}
\bibinfo{author}{\bibfnamefont{H.-J.} \bibnamefont{He}},
  \bibinfo{author}{\bibfnamefont{Y.-C.} \bibnamefont{Wang}}, \bibnamefont{and}
  \bibinfo{author}{\bibfnamefont{J.}~\bibnamefont{Zheng}},
  \bibinfo{journal}{JCAP} \textbf{\bibinfo{volume}{01}}, \bibinfo{pages}{042}
  (\bibinfo{year}{2021}{\natexlab{b}}), \eprint{2007.04963}.

\bibitem[{\citenamefont{Hoof et~al.}(2021)\citenamefont{Hoof, Jaeckel, and
  Thormaehlen}}]{Hoof:2021mld}
\bibinfo{author}{\bibfnamefont{S.}~\bibnamefont{Hoof}},
  \bibinfo{author}{\bibfnamefont{J.}~\bibnamefont{Jaeckel}}, \bibnamefont{and}
  \bibinfo{author}{\bibfnamefont{L.~J.} \bibnamefont{Thormaehlen}},
  \bibinfo{journal}{JCAP} \textbf{\bibinfo{volume}{09}}, \bibinfo{pages}{006}
  (\bibinfo{year}{2021}), \eprint{2101.08789}.

\bibitem[{\citenamefont{Hryczuk and Jod\l{}owski}(2020)}]{Hryczuk:2020jhi}
\bibinfo{author}{\bibfnamefont{A.}~\bibnamefont{Hryczuk}} \bibnamefont{and}
  \bibinfo{author}{\bibfnamefont{K.}~\bibnamefont{Jod\l{}owski}},
  \bibinfo{journal}{Phys. Rev. D} \textbf{\bibinfo{volume}{102}},
  \bibinfo{pages}{043024} (\bibinfo{year}{2020}), \eprint{2006.16139}.

\bibitem[{\citenamefont{Ibe et~al.}(2020)\citenamefont{Ibe, Kobayashi,
  Nakayama, and Shirai}}]{Ibe:2020dly}
\bibinfo{author}{\bibfnamefont{M.}~\bibnamefont{Ibe}},
  \bibinfo{author}{\bibfnamefont{S.}~\bibnamefont{Kobayashi}},
  \bibinfo{author}{\bibfnamefont{Y.}~\bibnamefont{Nakayama}}, \bibnamefont{and}
  \bibinfo{author}{\bibfnamefont{S.}~\bibnamefont{Shirai}},
  \bibinfo{journal}{JHEP} \textbf{\bibinfo{volume}{12}}, \bibinfo{pages}{004}
  (\bibinfo{year}{2020}), \eprint{2007.16105}.

\bibitem[{\citenamefont{Ilie and Levy}(2021)}]{Ilie:2021iyh}
\bibinfo{author}{\bibfnamefont{C.}~\bibnamefont{Ilie}} \bibnamefont{and}
  \bibinfo{author}{\bibfnamefont{C.}~\bibnamefont{Levy}},
  \bibinfo{journal}{Phys. Rev. D} \textbf{\bibinfo{volume}{104}},
  \bibinfo{pages}{083033} (\bibinfo{year}{2021}), \eprint{2105.09765}.

\bibitem[{\citenamefont{\.Inan and Kisselev}(2021)}]{Inan:2020kif}
\bibinfo{author}{\bibfnamefont{S.~C.} \bibnamefont{\.Inan}} \bibnamefont{and}
  \bibinfo{author}{\bibfnamefont{A.~V.} \bibnamefont{Kisselev}},
  \bibinfo{journal}{Chin. Phys. C} \textbf{\bibinfo{volume}{45}},
  \bibinfo{pages}{043109} (\bibinfo{year}{2021}), \eprint{2007.01693}.

\bibitem[{\citenamefont{Jaeckel and Yin}(2021)}]{Jaeckel:2020oet}
\bibinfo{author}{\bibfnamefont{J.}~\bibnamefont{Jaeckel}} \bibnamefont{and}
  \bibinfo{author}{\bibfnamefont{W.}~\bibnamefont{Yin}},
  \bibinfo{journal}{JCAP} \textbf{\bibinfo{volume}{02}}, \bibinfo{pages}{044}
  (\bibinfo{year}{2021}), \eprint{2007.15006}.

\bibitem[{\citenamefont{Jeong et~al.}(2021)\citenamefont{Jeong, Kim, and
  Youn}}]{Jeong:2021ivd}
\bibinfo{author}{\bibfnamefont{J.}~\bibnamefont{Jeong}},
  \bibinfo{author}{\bibfnamefont{J.~E.} \bibnamefont{Kim}}, \bibnamefont{and}
  \bibinfo{author}{\bibfnamefont{S.}~\bibnamefont{Youn}},
  \bibinfo{journal}{Int. J. Mod. Phys. A} \textbf{\bibinfo{volume}{36}},
  \bibinfo{pages}{2150182} (\bibinfo{year}{2021}), \eprint{2105.01842}.

\bibitem[{\citenamefont{Jho et~al.}(2020)\citenamefont{Jho, Park, Park, and
  Tseng}}]{Jho:2020sku}
\bibinfo{author}{\bibfnamefont{Y.}~\bibnamefont{Jho}},
  \bibinfo{author}{\bibfnamefont{J.-C.} \bibnamefont{Park}},
  \bibinfo{author}{\bibfnamefont{S.~C.} \bibnamefont{Park}}, \bibnamefont{and}
  \bibinfo{author}{\bibfnamefont{P.-Y.} \bibnamefont{Tseng}},
  \bibinfo{journal}{Phys. Lett. B} \textbf{\bibinfo{volume}{811}},
  \bibinfo{pages}{135863} (\bibinfo{year}{2020}), \eprint{2006.13910}.

\bibitem[{\citenamefont{Jia and Li}(2020)}]{Jia:2020omh}
\bibinfo{author}{\bibfnamefont{L.-B.} \bibnamefont{Jia}} \bibnamefont{and}
  \bibinfo{author}{\bibfnamefont{T.}~\bibnamefont{Li}} (\bibinfo{year}{2020}),
  \eprint{2012.07209}.

\bibitem[{\citenamefont{Kahlhoefer}(2021)}]{Kahlhoefer:2020gkz}
\bibinfo{author}{\bibfnamefont{F.}~\bibnamefont{Kahlhoefer}},
  \bibinfo{journal}{PoS} \textbf{\bibinfo{volume}{ICHEP2020}},
  \bibinfo{pages}{635} (\bibinfo{year}{2021}).

\bibitem[{\citenamefont{Kannike et~al.}(2020)\citenamefont{Kannike, Raidal,
  Veerm\"ae, Strumia, and Teresi}}]{Kannike:2020agf}
\bibinfo{author}{\bibfnamefont{K.}~\bibnamefont{Kannike}},
  \bibinfo{author}{\bibfnamefont{M.}~\bibnamefont{Raidal}},
  \bibinfo{author}{\bibfnamefont{H.}~\bibnamefont{Veerm\"ae}},
  \bibinfo{author}{\bibfnamefont{A.}~\bibnamefont{Strumia}}, \bibnamefont{and}
  \bibinfo{author}{\bibfnamefont{D.}~\bibnamefont{Teresi}},
  \bibinfo{journal}{Phys. Rev. D} \textbf{\bibinfo{volume}{102}},
  \bibinfo{pages}{095002} (\bibinfo{year}{2020}), \eprint{2006.10735}.

\bibitem[{\citenamefont{Karmakar and Pandey}(2020)}]{Karmakar:2020rbi}
\bibinfo{author}{\bibfnamefont{S.}~\bibnamefont{Karmakar}} \bibnamefont{and}
  \bibinfo{author}{\bibfnamefont{S.}~\bibnamefont{Pandey}}
  (\bibinfo{year}{2020}), \eprint{2007.11892}.

\bibitem[{\citenamefont{Karozas et~al.}(2021)\citenamefont{Karozas, King,
  Leontaris, and Papoulias}}]{Karozas:2020pun}
\bibinfo{author}{\bibfnamefont{A.}~\bibnamefont{Karozas}},
  \bibinfo{author}{\bibfnamefont{S.~F.} \bibnamefont{King}},
  \bibinfo{author}{\bibfnamefont{G.~K.} \bibnamefont{Leontaris}},
  \bibnamefont{and} \bibinfo{author}{\bibfnamefont{D.~K.}
  \bibnamefont{Papoulias}}, \bibinfo{journal}{Phys. Rev. D}
  \textbf{\bibinfo{volume}{103}}, \bibinfo{pages}{035019}
  (\bibinfo{year}{2021}), \eprint{2008.03295}.

\bibitem[{\citenamefont{Keung et~al.}(2021)\citenamefont{Keung, Marfatia, and
  Tseng}}]{Keung:2020uew}
\bibinfo{author}{\bibfnamefont{W.-Y.} \bibnamefont{Keung}},
  \bibinfo{author}{\bibfnamefont{D.}~\bibnamefont{Marfatia}}, \bibnamefont{and}
  \bibinfo{author}{\bibfnamefont{P.-Y.} \bibnamefont{Tseng}},
  \bibinfo{journal}{JHEAp} \textbf{\bibinfo{volume}{30}}, \bibinfo{pages}{9}
  (\bibinfo{year}{2021}), \eprint{2009.04444}.

\bibitem[{\citenamefont{Khan}(2021{\natexlab{a}})}]{Khan:2020csx}
\bibinfo{author}{\bibfnamefont{A.~N.} \bibnamefont{Khan}},
  \bibinfo{journal}{Phys. Lett. B} \textbf{\bibinfo{volume}{819}},
  \bibinfo{pages}{136415} (\bibinfo{year}{2021}{\natexlab{a}}),
  \eprint{2008.10279}.

\bibitem[{\citenamefont{Khan}(2021{\natexlab{b}})}]{Khan:2020pso}
\bibinfo{author}{\bibfnamefont{S.}~\bibnamefont{Khan}}, \bibinfo{journal}{Eur.
  Phys. J. C} \textbf{\bibinfo{volume}{81}}, \bibinfo{pages}{598}
  (\bibinfo{year}{2021}{\natexlab{b}}), \eprint{2007.13008}.

\bibitem[{\citenamefont{Khruschov}(2020)}]{Khruschov:2020cnf}
\bibinfo{author}{\bibfnamefont{V.~V.} \bibnamefont{Khruschov}}
  (\bibinfo{year}{2020}), \eprint{2008.03150}.

\bibitem[{\citenamefont{Kim et~al.}(2020)\citenamefont{Kim, Nomura, and
  Okada}}]{Kim:2020aua}
\bibinfo{author}{\bibfnamefont{J.}~\bibnamefont{Kim}},
  \bibinfo{author}{\bibfnamefont{T.}~\bibnamefont{Nomura}}, \bibnamefont{and}
  \bibinfo{author}{\bibfnamefont{H.}~\bibnamefont{Okada}},
  \bibinfo{journal}{Phys. Lett. B} \textbf{\bibinfo{volume}{811}},
  \bibinfo{pages}{135862} (\bibinfo{year}{2020}), \eprint{2007.09894}.

\bibitem[{\citenamefont{Ko and Tang}(2021)}]{Ko:2020gdg}
\bibinfo{author}{\bibfnamefont{P.}~\bibnamefont{Ko}} \bibnamefont{and}
  \bibinfo{author}{\bibfnamefont{Y.}~\bibnamefont{Tang}},
  \bibinfo{journal}{Phys. Lett. B} \textbf{\bibinfo{volume}{815}},
  \bibinfo{pages}{136181} (\bibinfo{year}{2021}), \eprint{2006.15822}.

\bibitem[{\citenamefont{Lee}(2021)}]{Lee:2020wmh}
\bibinfo{author}{\bibfnamefont{H.~M.} \bibnamefont{Lee}},
  \bibinfo{journal}{JHEP} \textbf{\bibinfo{volume}{01}}, \bibinfo{pages}{019}
  (\bibinfo{year}{2021}), \eprint{2006.13183}.

\bibitem[{\citenamefont{Li}(2020)}]{Li:2020naa}
\bibinfo{author}{\bibfnamefont{T.}~\bibnamefont{Li}} (\bibinfo{year}{2020}),
  \eprint{2007.00874}.

\bibitem[{\citenamefont{Lin}(2020)}]{Lin:2020mhx}
\bibinfo{author}{\bibfnamefont{T.}~\bibnamefont{Lin}}, \bibinfo{journal}{APS
  Physics} \textbf{\bibinfo{volume}{13}}, \bibinfo{pages}{135}
  (\bibinfo{year}{2020}).

\bibitem[{\citenamefont{Lindner et~al.}(2020)\citenamefont{Lindner, Mambrini,
  de~Melo, and Queiroz}}]{Lindner:2020kko}
\bibinfo{author}{\bibfnamefont{M.}~\bibnamefont{Lindner}},
  \bibinfo{author}{\bibfnamefont{Y.}~\bibnamefont{Mambrini}},
  \bibinfo{author}{\bibfnamefont{T.~B.} \bibnamefont{de~Melo}},
  \bibnamefont{and} \bibinfo{author}{\bibfnamefont{F.~S.}
  \bibnamefont{Queiroz}}, \bibinfo{journal}{Phys. Lett. B}
  \textbf{\bibinfo{volume}{811}}, \bibinfo{pages}{135972}
  (\bibinfo{year}{2020}), \eprint{2006.14590}.

\bibitem[{\citenamefont{Long et~al.}(2020)\citenamefont{Long, Soa, Binh, and
  C\'arcamo~Hern\'andez}}]{Long:2020uyf}
\bibinfo{author}{\bibfnamefont{H.~N.} \bibnamefont{Long}},
  \bibinfo{author}{\bibfnamefont{D.~V.} \bibnamefont{Soa}},
  \bibinfo{author}{\bibfnamefont{V.~H.} \bibnamefont{Binh}}, \bibnamefont{and}
  \bibinfo{author}{\bibfnamefont{A.~E.} \bibnamefont{C\'arcamo~Hern\'andez}}
  (\bibinfo{year}{2020}), \eprint{2007.05004}.

\bibitem[{\citenamefont{McKeen et~al.}(2020)\citenamefont{McKeen, Pospelov, and
  Raj}}]{McKeen:2020vpf}
\bibinfo{author}{\bibfnamefont{D.}~\bibnamefont{McKeen}},
  \bibinfo{author}{\bibfnamefont{M.}~\bibnamefont{Pospelov}}, \bibnamefont{and}
  \bibinfo{author}{\bibfnamefont{N.}~\bibnamefont{Raj}},
  \bibinfo{journal}{Phys. Rev. Lett.} \textbf{\bibinfo{volume}{125}},
  \bibinfo{pages}{231803} (\bibinfo{year}{2020}), \eprint{2006.15140}.

\bibitem[{\citenamefont{Miranda et~al.}(2020)\citenamefont{Miranda, Papoulias,
  T\'ortola, and Valle}}]{Miranda:2020kwy}
\bibinfo{author}{\bibfnamefont{O.~G.} \bibnamefont{Miranda}},
  \bibinfo{author}{\bibfnamefont{D.~K.} \bibnamefont{Papoulias}},
  \bibinfo{author}{\bibfnamefont{M.}~\bibnamefont{T\'ortola}},
  \bibnamefont{and} \bibinfo{author}{\bibfnamefont{J.~W.~F.}
  \bibnamefont{Valle}}, \bibinfo{journal}{Phys. Lett. B}
  \textbf{\bibinfo{volume}{808}}, \bibinfo{pages}{135685}
  (\bibinfo{year}{2020}), \eprint{2007.01765}.

\bibitem[{\citenamefont{Nakayama and Tang}(2020)}]{Nakayama:2020ikz}
\bibinfo{author}{\bibfnamefont{K.}~\bibnamefont{Nakayama}} \bibnamefont{and}
  \bibinfo{author}{\bibfnamefont{Y.}~\bibnamefont{Tang}},
  \bibinfo{journal}{Phys. Lett. B} \textbf{\bibinfo{volume}{811}},
  \bibinfo{pages}{135977} (\bibinfo{year}{2020}), \eprint{2006.13159}.

\bibitem[{\citenamefont{Okada et~al.}(2020)\citenamefont{Okada, Okada, Raut,
  and Shafi}}]{Okada:2020evk}
\bibinfo{author}{\bibfnamefont{N.}~\bibnamefont{Okada}},
  \bibinfo{author}{\bibfnamefont{S.}~\bibnamefont{Okada}},
  \bibinfo{author}{\bibfnamefont{D.}~\bibnamefont{Raut}}, \bibnamefont{and}
  \bibinfo{author}{\bibfnamefont{Q.}~\bibnamefont{Shafi}},
  \bibinfo{journal}{Phys. Lett. B} \textbf{\bibinfo{volume}{810}},
  \bibinfo{pages}{135785} (\bibinfo{year}{2020}), \eprint{2007.02898}.

\bibitem[{\citenamefont{Paz et~al.}(2021)\citenamefont{Paz, Petrov, Tammaro,
  and Zupan}}]{Paz:2020pbc}
\bibinfo{author}{\bibfnamefont{G.}~\bibnamefont{Paz}},
  \bibinfo{author}{\bibfnamefont{A.~A.} \bibnamefont{Petrov}},
  \bibinfo{author}{\bibfnamefont{M.}~\bibnamefont{Tammaro}}, \bibnamefont{and}
  \bibinfo{author}{\bibfnamefont{J.}~\bibnamefont{Zupan}},
  \bibinfo{journal}{Phys. Rev. D} \textbf{\bibinfo{volume}{103}},
  \bibinfo{pages}{L051703} (\bibinfo{year}{2021}), \eprint{2006.12462}.

\bibitem[{\citenamefont{Robinson}(2020)}]{Robinson:2020gfu}
\bibinfo{author}{\bibfnamefont{A.~E.} \bibnamefont{Robinson}}
  (\bibinfo{year}{2020}), \eprint{2006.13278}.

\bibitem[{\citenamefont{Seymour and Yagi}(2020)}]{Seymour:2020yle}
\bibinfo{author}{\bibfnamefont{B.~C.} \bibnamefont{Seymour}} \bibnamefont{and}
  \bibinfo{author}{\bibfnamefont{K.}~\bibnamefont{Yagi}},
  \bibinfo{journal}{Phys. Rev. D} \textbf{\bibinfo{volume}{102}},
  \bibinfo{pages}{104003} (\bibinfo{year}{2020}), \eprint{2007.14881}.

\bibitem[{\citenamefont{Shakeri et~al.}(2020)\citenamefont{Shakeri, Hajkarim,
  and Xue}}]{Shakeri:2020wvk}
\bibinfo{author}{\bibfnamefont{S.}~\bibnamefont{Shakeri}},
  \bibinfo{author}{\bibfnamefont{F.}~\bibnamefont{Hajkarim}}, \bibnamefont{and}
  \bibinfo{author}{\bibfnamefont{S.-S.} \bibnamefont{Xue}},
  \bibinfo{journal}{JHEP} \textbf{\bibinfo{volume}{12}}, \bibinfo{pages}{194}
  (\bibinfo{year}{2020}), \eprint{2008.05029}.

\bibitem[{\citenamefont{Shoemaker et~al.}(2021)\citenamefont{Shoemaker, Tsai,
  and Wyenberg}}]{Shoemaker:2020kji}
\bibinfo{author}{\bibfnamefont{I.~M.} \bibnamefont{Shoemaker}},
  \bibinfo{author}{\bibfnamefont{Y.-D.} \bibnamefont{Tsai}}, \bibnamefont{and}
  \bibinfo{author}{\bibfnamefont{J.}~\bibnamefont{Wyenberg}},
  \bibinfo{journal}{Phys. Rev. D} \textbf{\bibinfo{volume}{104}},
  \bibinfo{pages}{115026} (\bibinfo{year}{2021}), \eprint{2007.05513}.

\bibitem[{\citenamefont{Straniero et~al.}(2020)\citenamefont{Straniero,
  Pallanca, Dalessandro, Dominguez, Ferraro, Giannotti, Mirizzi, and
  Piersanti}}]{Straniero:2020iyi}
\bibinfo{author}{\bibfnamefont{O.}~\bibnamefont{Straniero}},
  \bibinfo{author}{\bibfnamefont{C.}~\bibnamefont{Pallanca}},
  \bibinfo{author}{\bibfnamefont{E.}~\bibnamefont{Dalessandro}},
  \bibinfo{author}{\bibfnamefont{I.}~\bibnamefont{Dominguez}},
  \bibinfo{author}{\bibfnamefont{F.~R.} \bibnamefont{Ferraro}},
  \bibinfo{author}{\bibfnamefont{M.}~\bibnamefont{Giannotti}},
  \bibinfo{author}{\bibfnamefont{A.}~\bibnamefont{Mirizzi}}, \bibnamefont{and}
  \bibinfo{author}{\bibfnamefont{L.}~\bibnamefont{Piersanti}},
  \bibinfo{journal}{Astron. Astrophys.} \textbf{\bibinfo{volume}{644}},
  \bibinfo{pages}{A166} (\bibinfo{year}{2020}), \eprint{2010.03833}.

\bibitem[{\citenamefont{Studenikin}(2021)}]{Studenikin:2021fai}
\bibinfo{author}{\bibfnamefont{A.}~\bibnamefont{Studenikin}},
  \bibinfo{journal}{PoS} \textbf{\bibinfo{volume}{ICHEP2020}},
  \bibinfo{pages}{180} (\bibinfo{year}{2021}), \eprint{2102.05468}.

\bibitem[{\citenamefont{Su et~al.}(2020)\citenamefont{Su, Wang, Wu, Yang, and
  Zhu}}]{Su:2020zny}
\bibinfo{author}{\bibfnamefont{L.}~\bibnamefont{Su}},
  \bibinfo{author}{\bibfnamefont{W.}~\bibnamefont{Wang}},
  \bibinfo{author}{\bibfnamefont{L.}~\bibnamefont{Wu}},
  \bibinfo{author}{\bibfnamefont{J.~M.} \bibnamefont{Yang}}, \bibnamefont{and}
  \bibinfo{author}{\bibfnamefont{B.}~\bibnamefont{Zhu}},
  \bibinfo{journal}{Phys. Rev. D} \textbf{\bibinfo{volume}{102}},
  \bibinfo{pages}{115028} (\bibinfo{year}{2020}), \eprint{2006.11837}.

\bibitem[{\citenamefont{Sun and He}(2020)}]{Sun:2020iim}
\bibinfo{author}{\bibfnamefont{J.}~\bibnamefont{Sun}} \bibnamefont{and}
  \bibinfo{author}{\bibfnamefont{X.-G.} \bibnamefont{He}},
  \bibinfo{journal}{Phys. Lett. B} \textbf{\bibinfo{volume}{811}},
  \bibinfo{pages}{135881} (\bibinfo{year}{2020}), \eprint{2006.16931}.

\bibitem[{\citenamefont{Szydagis et~al.}(2021)\citenamefont{Szydagis, Levy,
  Blockinger, Kamaha, Parveen, and Rischbieter}}]{Szydagis:2020isq}
\bibinfo{author}{\bibfnamefont{M.}~\bibnamefont{Szydagis}},
  \bibinfo{author}{\bibfnamefont{C.}~\bibnamefont{Levy}},
  \bibinfo{author}{\bibfnamefont{G.~M.} \bibnamefont{Blockinger}},
  \bibinfo{author}{\bibfnamefont{A.}~\bibnamefont{Kamaha}},
  \bibinfo{author}{\bibfnamefont{N.}~\bibnamefont{Parveen}}, \bibnamefont{and}
  \bibinfo{author}{\bibfnamefont{G.~R.~C.} \bibnamefont{Rischbieter}},
  \bibinfo{journal}{Phys. Rev. D} \textbf{\bibinfo{volume}{103}},
  \bibinfo{pages}{012002} (\bibinfo{year}{2021}), \eprint{2007.00528}.

\bibitem[{\citenamefont{Takahashi et~al.}(2020)\citenamefont{Takahashi, Yamada,
  and Yin}}]{Takahashi:2020bpq}
\bibinfo{author}{\bibfnamefont{F.}~\bibnamefont{Takahashi}},
  \bibinfo{author}{\bibfnamefont{M.}~\bibnamefont{Yamada}}, \bibnamefont{and}
  \bibinfo{author}{\bibfnamefont{W.}~\bibnamefont{Yin}},
  \bibinfo{journal}{Phys. Rev. Lett.} \textbf{\bibinfo{volume}{125}},
  \bibinfo{pages}{161801} (\bibinfo{year}{2020}), \eprint{2006.10035}.

\bibitem[{\citenamefont{Takahashi et~al.}(2021)\citenamefont{Takahashi, Yamada,
  and Yin}}]{Takahashi:2020uio}
\bibinfo{author}{\bibfnamefont{F.}~\bibnamefont{Takahashi}},
  \bibinfo{author}{\bibfnamefont{M.}~\bibnamefont{Yamada}}, \bibnamefont{and}
  \bibinfo{author}{\bibfnamefont{W.}~\bibnamefont{Yin}},
  \bibinfo{journal}{JHEP} \textbf{\bibinfo{volume}{01}}, \bibinfo{pages}{152}
  (\bibinfo{year}{2021}), \eprint{2007.10311}.

\bibitem[{\citenamefont{Tan et~al.}(2021)\citenamefont{Tan, Derevianko, Dzuba,
  and Flambaum}}]{Tan:2021nif}
\bibinfo{author}{\bibfnamefont{H.~B.~T.} \bibnamefont{Tan}},
  \bibinfo{author}{\bibfnamefont{A.}~\bibnamefont{Derevianko}},
  \bibinfo{author}{\bibfnamefont{V.~A.} \bibnamefont{Dzuba}}, \bibnamefont{and}
  \bibinfo{author}{\bibfnamefont{V.~V.} \bibnamefont{Flambaum}},
  \bibinfo{journal}{Phys. Rev. Lett.} \textbf{\bibinfo{volume}{127}},
  \bibinfo{pages}{081301} (\bibinfo{year}{2021}), \eprint{2105.08296}.

\bibitem[{\citenamefont{Vagnozzi et~al.}(2021)\citenamefont{Vagnozzi,
  Visinelli, Brax, Davis, and Sakstein}}]{Vagnozzi:2021quy}
\bibinfo{author}{\bibfnamefont{S.}~\bibnamefont{Vagnozzi}},
  \bibinfo{author}{\bibfnamefont{L.}~\bibnamefont{Visinelli}},
  \bibinfo{author}{\bibfnamefont{P.}~\bibnamefont{Brax}},
  \bibinfo{author}{\bibfnamefont{A.-C.} \bibnamefont{Davis}}, \bibnamefont{and}
  \bibinfo{author}{\bibfnamefont{J.}~\bibnamefont{Sakstein}},
  \bibinfo{journal}{Phys. Rev. D} \textbf{\bibinfo{volume}{104}},
  \bibinfo{pages}{063023} (\bibinfo{year}{2021}), \eprint{2103.15834}.

\bibitem[{\citenamefont{Van~Dong et~al.}(2021)\citenamefont{Van~Dong, Nam, and
  Van~Loi}}]{VanDong:2020bkg}
\bibinfo{author}{\bibfnamefont{P.}~\bibnamefont{Van~Dong}},
  \bibinfo{author}{\bibfnamefont{C.~H.} \bibnamefont{Nam}}, \bibnamefont{and}
  \bibinfo{author}{\bibfnamefont{D.}~\bibnamefont{Van~Loi}},
  \bibinfo{journal}{Phys. Rev. D} \textbf{\bibinfo{volume}{103}},
  \bibinfo{pages}{095016} (\bibinfo{year}{2021}), \eprint{2007.08957}.

\bibitem[{\citenamefont{Xu and Zheng}(2021)}]{Xu:2020qsy}
\bibinfo{author}{\bibfnamefont{S.}~\bibnamefont{Xu}} \bibnamefont{and}
  \bibinfo{author}{\bibfnamefont{S.}~\bibnamefont{Zheng}},
  \bibinfo{journal}{Eur. Phys. J. C} \textbf{\bibinfo{volume}{81}},
  \bibinfo{pages}{446} (\bibinfo{year}{2021}), \eprint{2012.10827}.

\bibitem[{\citenamefont{Ye et~al.}(2021)\citenamefont{Ye, Zhang, Xu, and
  Liu}}]{Ye:2021zso}
\bibinfo{author}{\bibfnamefont{Z.}~\bibnamefont{Ye}},
  \bibinfo{author}{\bibfnamefont{F.}~\bibnamefont{Zhang}},
  \bibinfo{author}{\bibfnamefont{D.}~\bibnamefont{Xu}}, \bibnamefont{and}
  \bibinfo{author}{\bibfnamefont{J.}~\bibnamefont{Liu}},
  \bibinfo{journal}{Chin. Phys. Lett.} \textbf{\bibinfo{volume}{38}},
  \bibinfo{pages}{111401} (\bibinfo{year}{2021}), \eprint{2103.11771}.

\bibitem[{\citenamefont{Lei et~al.}(2020)\citenamefont{Lei, Tang, and
  Zhang}}]{Zhang:2020htl}
\bibinfo{author}{\bibfnamefont{Z.-H.} \bibnamefont{Lei}},
  \bibinfo{author}{\bibfnamefont{J.}~\bibnamefont{Tang}}, \bibnamefont{and}
  \bibinfo{author}{\bibfnamefont{B.-L.} \bibnamefont{Zhang}}
  (\bibinfo{year}{2020}), \eprint{2008.07116}.

\bibitem[{\citenamefont{Zioutas et~al.}(2020)\citenamefont{Zioutas, Cantatore,
  Karuza, Kryemadhi, Maroudas, and Semertzidis}}]{Zioutas:2020cul}
\bibinfo{author}{\bibfnamefont{K.}~\bibnamefont{Zioutas}},
  \bibinfo{author}{\bibfnamefont{G.}~\bibnamefont{Cantatore}},
  \bibinfo{author}{\bibfnamefont{M.}~\bibnamefont{Karuza}},
  \bibinfo{author}{\bibfnamefont{A.}~\bibnamefont{Kryemadhi}},
  \bibinfo{author}{\bibfnamefont{M.}~\bibnamefont{Maroudas}}, \bibnamefont{and}
  \bibinfo{author}{\bibfnamefont{Y.~K.} \bibnamefont{Semertzidis}}
  (\bibinfo{year}{2020}), \eprint{2006.16907}.

\bibitem[{\citenamefont{Zu et~al.}(2021{\natexlab{a}})\citenamefont{Zu, Foot,
  Fan, and Feng}}]{Zu:2020bsx}
\bibinfo{author}{\bibfnamefont{L.}~\bibnamefont{Zu}},
  \bibinfo{author}{\bibfnamefont{R.}~\bibnamefont{Foot}},
  \bibinfo{author}{\bibfnamefont{Y.-Z.} \bibnamefont{Fan}}, \bibnamefont{and}
  \bibinfo{author}{\bibfnamefont{L.}~\bibnamefont{Feng}},
  \bibinfo{journal}{JCAP} \textbf{\bibinfo{volume}{01}}, \bibinfo{pages}{070}
  (\bibinfo{year}{2021}{\natexlab{a}}), \eprint{2007.15191}.

\bibitem[{\citenamefont{Zu et~al.}(2021{\natexlab{b}})\citenamefont{Zu, Yuan,
  Feng, and Fan}}]{Zu:2020idx}
\bibinfo{author}{\bibfnamefont{L.}~\bibnamefont{Zu}},
  \bibinfo{author}{\bibfnamefont{G.-W.} \bibnamefont{Yuan}},
  \bibinfo{author}{\bibfnamefont{L.}~\bibnamefont{Feng}}, \bibnamefont{and}
  \bibinfo{author}{\bibfnamefont{Y.-Z.} \bibnamefont{Fan}},
  \bibinfo{journal}{Nucl. Phys. B} \textbf{\bibinfo{volume}{965}},
  \bibinfo{pages}{115369} (\bibinfo{year}{2021}{\natexlab{b}}),
  \eprint{2006.14577}.

\bibitem[{\citenamefont{Ipser and Sikivie}(1983)}]{Ipser:1983mw}
\bibinfo{author}{\bibfnamefont{J.}~\bibnamefont{Ipser}} \bibnamefont{and}
  \bibinfo{author}{\bibfnamefont{P.}~\bibnamefont{Sikivie}},
  \bibinfo{journal}{Phys. Rev. Lett.} \textbf{\bibinfo{volume}{50}},
  \bibinfo{pages}{925} (\bibinfo{year}{1983}).

\bibitem[{\citenamefont{Duffy and van Bibber}(2009)}]{Duffy:2009ig}
\bibinfo{author}{\bibfnamefont{L.~D.} \bibnamefont{Duffy}} \bibnamefont{and}
  \bibinfo{author}{\bibfnamefont{K.}~\bibnamefont{van Bibber}},
  \bibinfo{journal}{New J. Phys.} \textbf{\bibinfo{volume}{11}},
  \bibinfo{pages}{105008} (\bibinfo{year}{2009}), \eprint{0904.3346}.

\bibitem[{\citenamefont{Krauss et~al.}(1985)\citenamefont{Krauss, Moody,
  Wilczek, and Morris}}]{Krauss:1985ub}
\bibinfo{author}{\bibfnamefont{L.}~\bibnamefont{Krauss}},
  \bibinfo{author}{\bibfnamefont{J.}~\bibnamefont{Moody}},
  \bibinfo{author}{\bibfnamefont{F.}~\bibnamefont{Wilczek}}, \bibnamefont{and}
  \bibinfo{author}{\bibfnamefont{D.~E.} \bibnamefont{Morris}},
  \bibinfo{journal}{Phys. Rev. Lett.} \textbf{\bibinfo{volume}{55}},
  \bibinfo{pages}{1797} (\bibinfo{year}{1985}).

\bibitem[{\citenamefont{Hagmann et~al.}(1990)\citenamefont{Hagmann, Sikivie,
  Sullivan, and Tanner}}]{Hagmann:1990tj}
\bibinfo{author}{\bibfnamefont{C.}~\bibnamefont{Hagmann}},
  \bibinfo{author}{\bibfnamefont{P.}~\bibnamefont{Sikivie}},
  \bibinfo{author}{\bibfnamefont{N.}~\bibnamefont{Sullivan}}, \bibnamefont{and}
  \bibinfo{author}{\bibfnamefont{D.}~\bibnamefont{Tanner}},
  \bibinfo{journal}{Phys. Rev. D} \textbf{\bibinfo{volume}{42}},
  \bibinfo{pages}{1297} (\bibinfo{year}{1990}).

\bibitem[{\citenamefont{Sikivie}(2000)}]{Sikivie:1999sy}
\bibinfo{author}{\bibfnamefont{P.}~\bibnamefont{Sikivie}},
  \bibinfo{journal}{Nucl. Phys. B Proc. Suppl.} \textbf{\bibinfo{volume}{87}},
  \bibinfo{pages}{41} (\bibinfo{year}{2000}), \eprint{hep-ph/0002154}.

\bibitem[{\citenamefont{Raffelt}(2007)}]{Raffelt:2006rj}
\bibinfo{author}{\bibfnamefont{G.~G.} \bibnamefont{Raffelt}},
  \bibinfo{journal}{J. Phys. A} \textbf{\bibinfo{volume}{40}},
  \bibinfo{pages}{6607} (\bibinfo{year}{2007}), \eprint{hep-ph/0611118}.

\bibitem[{\citenamefont{Aune et~al.}(2011)}]{Arik:2011rx}
\bibinfo{author}{\bibfnamefont{S.}~\bibnamefont{Aune}} \bibnamefont{et~al.}
  (\bibinfo{collaboration}{CAST}), \bibinfo{journal}{Phys. Rev. Lett.}
  \textbf{\bibinfo{volume}{107}}, \bibinfo{pages}{261302}
  (\bibinfo{year}{2011}), \eprint{1106.3919}.

\bibitem[{\citenamefont{Du et~al.}(2018)}]{Du:2018uak}
\bibinfo{author}{\bibfnamefont{N.}~\bibnamefont{Du}} \bibnamefont{et~al.}
  (\bibinfo{collaboration}{ADMX}), \bibinfo{journal}{Phys. Rev. Lett.}
  \textbf{\bibinfo{volume}{120}}, \bibinfo{pages}{151301}
  (\bibinfo{year}{2018}), \eprint{1804.05750}.

\bibitem[{\citenamefont{Hagmann et~al.}(1998)}]{Hagmann:1998cb}
\bibinfo{author}{\bibfnamefont{C.}~\bibnamefont{Hagmann}} \bibnamefont{et~al.}
  (\bibinfo{collaboration}{ADMX}), \bibinfo{journal}{Phys. Rev. Lett.}
  \textbf{\bibinfo{volume}{80}}, \bibinfo{pages}{2043} (\bibinfo{year}{1998}),
  \eprint{astro-ph/9801286}.

\bibitem[{\citenamefont{Krauss et~al.}(1984{\natexlab{b}})\citenamefont{Krauss,
  Moody, and Wilczek}}]{Krauss:1984gm}
\bibinfo{author}{\bibfnamefont{L.~M.} \bibnamefont{Krauss}},
  \bibinfo{author}{\bibfnamefont{J.~E.} \bibnamefont{Moody}}, \bibnamefont{and}
  \bibinfo{author}{\bibfnamefont{F.}~\bibnamefont{Wilczek}},
  \bibinfo{journal}{Phys. Lett. B} \textbf{\bibinfo{volume}{144}},
  \bibinfo{pages}{391} (\bibinfo{year}{1984}{\natexlab{b}}).

\bibitem[{\citenamefont{Dimopoulos et~al.}(1986)\citenamefont{Dimopoulos,
  Frieman, Lynn, and Starkman}}]{Dimopoulos:1986kc}
\bibinfo{author}{\bibfnamefont{S.}~\bibnamefont{Dimopoulos}},
  \bibinfo{author}{\bibfnamefont{J.~A.} \bibnamefont{Frieman}},
  \bibinfo{author}{\bibfnamefont{B.}~\bibnamefont{Lynn}}, \bibnamefont{and}
  \bibinfo{author}{\bibfnamefont{G.}~\bibnamefont{Starkman}},
  \bibinfo{journal}{Phys. Lett. B} \textbf{\bibinfo{volume}{179}},
  \bibinfo{pages}{223} (\bibinfo{year}{1986}).

\bibitem[{\citenamefont{Kim}(1979)}]{Kim:1979if}
\bibinfo{author}{\bibfnamefont{J.~E.} \bibnamefont{Kim}},
  \bibinfo{journal}{Phys. Rev. Lett.} \textbf{\bibinfo{volume}{43}},
  \bibinfo{pages}{103} (\bibinfo{year}{1979}).

\bibitem[{\citenamefont{Shifman et~al.}(1980)\citenamefont{Shifman, Vainshtein,
  and Zakharov}}]{Shifman:1979if}
\bibinfo{author}{\bibfnamefont{M.~A.} \bibnamefont{Shifman}},
  \bibinfo{author}{\bibfnamefont{A.~I.} \bibnamefont{Vainshtein}},
  \bibnamefont{and} \bibinfo{author}{\bibfnamefont{V.~I.}
  \bibnamefont{Zakharov}}, \bibinfo{journal}{Nucl. Phys. B}
  \textbf{\bibinfo{volume}{166}}, \bibinfo{pages}{493} (\bibinfo{year}{1980}).

\bibitem[{\citenamefont{Dine et~al.}(1981)\citenamefont{Dine, Fischler, and
  Srednicki}}]{Dine:1981rt}
\bibinfo{author}{\bibfnamefont{M.}~\bibnamefont{Dine}},
  \bibinfo{author}{\bibfnamefont{W.}~\bibnamefont{Fischler}}, \bibnamefont{and}
  \bibinfo{author}{\bibfnamefont{M.}~\bibnamefont{Srednicki}},
  \bibinfo{journal}{Phys. Lett. B} \textbf{\bibinfo{volume}{104}},
  \bibinfo{pages}{199} (\bibinfo{year}{1981}).

\bibitem[{\citenamefont{Zhitnitsky}(1980)}]{Zhitnitsky:1980tq}
\bibinfo{author}{\bibfnamefont{A.~R.} \bibnamefont{Zhitnitsky}},
  \bibinfo{journal}{Sov. J. Nucl. Phys.} \textbf{\bibinfo{volume}{31}},
  \bibinfo{pages}{260} (\bibinfo{year}{1980}), \bibinfo{note}{[Yad. Fiz. {\bf
  31}, 497 (1980)]}.

\bibitem[{\citenamefont{Redondo}(2013)}]{Redondo:2013wwa}
\bibinfo{author}{\bibfnamefont{J.}~\bibnamefont{Redondo}},
  \bibinfo{journal}{JCAP} \textbf{\bibinfo{volume}{1312}}, \bibinfo{pages}{008}
  (\bibinfo{year}{2013}), \eprint{1310.0823}.

\bibitem[{\citenamefont{Primakoff}(1951)}]{Pirmakoff:1951pj}
\bibinfo{author}{\bibfnamefont{H.}~\bibnamefont{Primakoff}},
  \bibinfo{journal}{Phys. Rev.} \textbf{\bibinfo{volume}{81}},
  \bibinfo{pages}{899} (\bibinfo{year}{1951}).

\bibitem[{\citenamefont{Moriyama}(1995)}]{Moriyama:1995bz}
\bibinfo{author}{\bibfnamefont{S.}~\bibnamefont{Moriyama}},
  \bibinfo{journal}{Phys. Rev. Lett.} \textbf{\bibinfo{volume}{75}},
  \bibinfo{pages}{3222} (\bibinfo{year}{1995}), \eprint{hep-ph/9504318}.

\bibitem[{\citenamefont{Armengaud et~al.}(2019)}]{Armengaud:2019uso}
\bibinfo{author}{\bibfnamefont{E.}~\bibnamefont{Armengaud}}
  \bibnamefont{et~al.} (\bibinfo{collaboration}{IAXO}), \bibinfo{journal}{JCAP}
  \textbf{\bibinfo{volume}{06}}, \bibinfo{pages}{047} (\bibinfo{year}{2019}),
  \eprint{1904.09155}.

\bibitem[{\citenamefont{Ahmed et~al.}(2009)}]{Ahmed:2009ht}
\bibinfo{author}{\bibfnamefont{Z.}~\bibnamefont{Ahmed}} \bibnamefont{et~al.}
  (\bibinfo{collaboration}{CDMS}), \bibinfo{journal}{Phys. Rev. Lett.}
  \textbf{\bibinfo{volume}{103}}, \bibinfo{pages}{141802}
  (\bibinfo{year}{2009}), \eprint{0902.4693}.

\bibitem[{\citenamefont{Abe et~al.}(2013)}]{Abe:2012ut}
\bibinfo{author}{\bibfnamefont{K.}~\bibnamefont{Abe}} \bibnamefont{et~al.}
  (\bibinfo{collaboration}{XMASS}), \bibinfo{journal}{Phys. Lett. B}
  \textbf{\bibinfo{volume}{724}}, \bibinfo{pages}{46} (\bibinfo{year}{2013}),
  \eprint{1212.6153}.

\bibitem[{\citenamefont{Harnik et~al.}(2012)\citenamefont{Harnik, Kopp, and
  Machado}}]{Harnik:2012ni}
\bibinfo{author}{\bibfnamefont{R.}~\bibnamefont{Harnik}},
  \bibinfo{author}{\bibfnamefont{J.}~\bibnamefont{Kopp}}, \bibnamefont{and}
  \bibinfo{author}{\bibfnamefont{P.~A.~N.} \bibnamefont{Machado}},
  \bibinfo{journal}{JCAP} \textbf{\bibinfo{volume}{1207}}, \bibinfo{pages}{026}
  (\bibinfo{year}{2012}), \eprint{1202.6073}.

\bibitem[{\citenamefont{Schwemberger and Yu}(2022)}]{Schwemberger:2022det}
\bibinfo{author}{\bibfnamefont{T.}~\bibnamefont{Schwemberger}}
  \bibnamefont{and} \bibinfo{author}{\bibfnamefont{T.-T.} \bibnamefont{Yu}},
  \emph{\bibinfo{title}{Detecting beyond the standard model interactions of
  solar neutrinos in low-threshold dark matter detectors}}
  (\bibinfo{year}{2022}), \eprint{2202.01254}.

\bibitem[{\citenamefont{Chen et~al.}(2017{\natexlab{d}})\citenamefont{Chen,
  Chi, Liu, and Wu}}]{Chen:2017plb}
\bibinfo{author}{\bibfnamefont{J.-W.} \bibnamefont{Chen}},
  \bibinfo{author}{\bibfnamefont{H.-C.} \bibnamefont{Chi}},
  \bibinfo{author}{\bibfnamefont{C.-P.} \bibnamefont{Liu}}, \bibnamefont{and}
  \bibinfo{author}{\bibfnamefont{C.-P.} \bibnamefont{Wu}},
  \bibinfo{journal}{Physics Letters B} \textbf{\bibinfo{volume}{774}},
  \bibinfo{pages}{656} (\bibinfo{year}{2017}{\natexlab{d}}).

\bibitem[{\citenamefont{Agostini et~al.}(2017)}]{Borexino:2017fbd}
\bibinfo{author}{\bibfnamefont{M.}~\bibnamefont{Agostini}} \bibnamefont{et~al.}
  (\bibinfo{collaboration}{Borexino}), \bibinfo{journal}{Phys. Rev. D}
  \textbf{\bibinfo{volume}{96}}, \bibinfo{pages}{091103}
  (\bibinfo{year}{2017}), \eprint{1707.09355}.

\bibitem[{\citenamefont{Bellini et~al.}(2011)}]{Bellini:2011rx}
\bibinfo{author}{\bibfnamefont{G.}~\bibnamefont{Bellini}} \bibnamefont{et~al.},
  \bibinfo{journal}{Phys. Rev. Lett.} \textbf{\bibinfo{volume}{107}},
  \bibinfo{pages}{141302} (\bibinfo{year}{2011}), \eprint{1104.1816}.

\bibitem[{\citenamefont{Beda et~al.}(2010)\citenamefont{Beda, Demidova,
  Starostin, Brudanin, Egorov, Medvedev, Shirchenko, and Vylov}}]{Beda:2009kx}
\bibinfo{author}{\bibfnamefont{A.~G.} \bibnamefont{Beda}},
  \bibinfo{author}{\bibfnamefont{E.~V.} \bibnamefont{Demidova}},
  \bibinfo{author}{\bibfnamefont{A.~S.} \bibnamefont{Starostin}},
  \bibinfo{author}{\bibfnamefont{V.~B.} \bibnamefont{Brudanin}},
  \bibinfo{author}{\bibfnamefont{V.~G.} \bibnamefont{Egorov}},
  \bibinfo{author}{\bibfnamefont{D.~V.} \bibnamefont{Medvedev}},
  \bibinfo{author}{\bibfnamefont{M.~V.} \bibnamefont{Shirchenko}},
  \bibnamefont{and} \bibinfo{author}{\bibfnamefont{T.}~\bibnamefont{Vylov}},
  \bibinfo{journal}{Phys. Part. Nucl. Lett.} \textbf{\bibinfo{volume}{7}},
  \bibinfo{pages}{406} (\bibinfo{year}{2010}), \eprint{0906.1926}.

\bibitem[{\citenamefont{Abe et~al.}(2020)}]{Abe:2020nwr}
\bibinfo{author}{\bibfnamefont{K.}~\bibnamefont{Abe}} \bibnamefont{et~al.}
  (\bibinfo{collaboration}{XMASS}), \bibinfo{journal}{Phys. Lett. B}
  \textbf{\bibinfo{volume}{809}}, \bibinfo{pages}{135741}
  (\bibinfo{year}{2020}), \eprint{2005.11891}.

\bibitem[{\citenamefont{Aguilar-Arevalo
  et~al.}(2016{\natexlab{b}})}]{Aguilar-Arevalo:2016khx}
\bibinfo{author}{\bibfnamefont{A.}~\bibnamefont{Aguilar-Arevalo}}
  \bibnamefont{et~al.} (\bibinfo{collaboration}{CONNIE}), \bibinfo{journal}{J.
  Phys. Conf. Ser.} \textbf{\bibinfo{volume}{761}}, \bibinfo{pages}{012057}
  (\bibinfo{year}{2016}{\natexlab{b}}), \eprint{1608.01565}.

\bibitem[{\citenamefont{Coloma et~al.}(2014)\citenamefont{Coloma, Huber, and
  Link}}]{Coloma:2014hka}
\bibinfo{author}{\bibfnamefont{P.}~\bibnamefont{Coloma}},
  \bibinfo{author}{\bibfnamefont{P.}~\bibnamefont{Huber}}, \bibnamefont{and}
  \bibinfo{author}{\bibfnamefont{J.~M.} \bibnamefont{Link}},
  \bibinfo{journal}{JHEP} \textbf{\bibinfo{volume}{11}}, \bibinfo{pages}{042}
  (\bibinfo{year}{2014}), \eprint{1406.4914}.

\bibitem[{\citenamefont{Dirac}(1931)}]{Dirac:1931kp}
\bibinfo{author}{\bibfnamefont{P.~A.~M.} \bibnamefont{Dirac}},
  \bibinfo{journal}{Proc. Roy. Soc. Lond. A} \textbf{\bibinfo{volume}{133}},
  \bibinfo{pages}{60} (\bibinfo{year}{1931}).

\bibitem[{\citenamefont{Pati and Salam}(1973)}]{Pati:1973uk}
\bibinfo{author}{\bibfnamefont{J.~C.} \bibnamefont{Pati}} \bibnamefont{and}
  \bibinfo{author}{\bibfnamefont{A.}~\bibnamefont{Salam}},
  \bibinfo{journal}{Phys. Rev. D} \textbf{\bibinfo{volume}{8}},
  \bibinfo{pages}{1240} (\bibinfo{year}{1973}).

\bibitem[{\citenamefont{Georgi and Glashow}(1974)}]{Georgi:1974sy}
\bibinfo{author}{\bibfnamefont{H.}~\bibnamefont{Georgi}} \bibnamefont{and}
  \bibinfo{author}{\bibfnamefont{S.~L.} \bibnamefont{Glashow}},
  \bibinfo{journal}{Phys. Rev. Lett.} \textbf{\bibinfo{volume}{32}},
  \bibinfo{pages}{438} (\bibinfo{year}{1974}).

\bibitem[{\citenamefont{Dobroliubov and Ignatiev}(1990)}]{Dobroliubov:1989mr}
\bibinfo{author}{\bibfnamefont{M.~I.} \bibnamefont{Dobroliubov}}
  \bibnamefont{and} \bibinfo{author}{\bibfnamefont{A.~{\relax Yu}.}
  \bibnamefont{Ignatiev}}, \bibinfo{journal}{Phys. Rev. Lett.}
  \textbf{\bibinfo{volume}{65}}, \bibinfo{pages}{679} (\bibinfo{year}{1990}).

\bibitem[{\citenamefont{Prinz et~al.}(1998)}]{Prinz:1998ua}
\bibinfo{author}{\bibfnamefont{A.~A.} \bibnamefont{Prinz}}
  \bibnamefont{et~al.}, \bibinfo{journal}{Phys. Rev. Lett.}
  \textbf{\bibinfo{volume}{81}}, \bibinfo{pages}{1175} (\bibinfo{year}{1998}),
  \eprint{hep-ex/9804008}.

\bibitem[{\citenamefont{Davidson et~al.}(2000)\citenamefont{Davidson,
  Hannestad, and Raffelt}}]{Davidson:2000hf}
\bibinfo{author}{\bibfnamefont{S.}~\bibnamefont{Davidson}},
  \bibinfo{author}{\bibfnamefont{S.}~\bibnamefont{Hannestad}},
  \bibnamefont{and} \bibinfo{author}{\bibfnamefont{G.}~\bibnamefont{Raffelt}},
  \bibinfo{journal}{JHEP} \textbf{\bibinfo{volume}{05}}, \bibinfo{pages}{003}
  (\bibinfo{year}{2000}), \eprint{hep-ph/0001179}.

\bibitem[{\citenamefont{Prinz}(2001)}]{Prinz:2001qz}
\bibinfo{author}{\bibfnamefont{A.~A.} \bibnamefont{Prinz}}, Ph.D. thesis,
  \bibinfo{school}{Stanford U., Phys. Dept.} (\bibinfo{year}{2001}),
  \urlprefix\url{http://wwwlib.umi.com/dissertations/fullcit?p3002033}.

\bibitem[{\citenamefont{Golowich and Robinett}(1987)}]{Golowich:1986tj}
\bibinfo{author}{\bibfnamefont{E.}~\bibnamefont{Golowich}} \bibnamefont{and}
  \bibinfo{author}{\bibfnamefont{R.~W.} \bibnamefont{Robinett}},
  \bibinfo{journal}{Phys. Rev. D} \textbf{\bibinfo{volume}{35}},
  \bibinfo{pages}{391} (\bibinfo{year}{1987}).

\bibitem[{\citenamefont{Babu et~al.}(1994)\citenamefont{Babu, Gould, and
  Rothstein}}]{Babu:1993yh}
\bibinfo{author}{\bibfnamefont{K.~S.} \bibnamefont{Babu}},
  \bibinfo{author}{\bibfnamefont{T.~M.} \bibnamefont{Gould}}, \bibnamefont{and}
  \bibinfo{author}{\bibfnamefont{I.~Z.} \bibnamefont{Rothstein}},
  \bibinfo{journal}{Phys. Lett. B} \textbf{\bibinfo{volume}{321}},
  \bibinfo{pages}{140} (\bibinfo{year}{1994}), \eprint{hep-ph/9310349}.

\bibitem[{\citenamefont{Gninenko et~al.}(2007)\citenamefont{Gninenko,
  Krasnikov, and Rubbia}}]{Gninenko:2006fi}
\bibinfo{author}{\bibfnamefont{S.~N.} \bibnamefont{Gninenko}},
  \bibinfo{author}{\bibfnamefont{N.~V.} \bibnamefont{Krasnikov}},
  \bibnamefont{and} \bibinfo{author}{\bibfnamefont{A.}~\bibnamefont{Rubbia}},
  \bibinfo{journal}{Phys. Rev. D} \textbf{\bibinfo{volume}{75}},
  \bibinfo{pages}{075014} (\bibinfo{year}{2007}), \eprint{hep-ph/0612203}.

\bibitem[{\citenamefont{Chatrchyan et~al.}(2013)}]{CMS:2012xi}
\bibinfo{author}{\bibfnamefont{S.}~\bibnamefont{Chatrchyan}}
  \bibnamefont{et~al.} (\bibinfo{collaboration}{CMS}), \bibinfo{journal}{Phys.
  Rev. D} \textbf{\bibinfo{volume}{87}}, \bibinfo{pages}{092008}
  (\bibinfo{year}{2013}), \eprint{1210.2311}.

\bibitem[{\citenamefont{Agnese et~al.}(2015)}]{Agnese:2014vxh}
\bibinfo{author}{\bibfnamefont{R.}~\bibnamefont{Agnese}} \bibnamefont{et~al.}
  (\bibinfo{collaboration}{CDMS}), \bibinfo{journal}{Phys. Rev. Lett.}
  \textbf{\bibinfo{volume}{114}}, \bibinfo{pages}{111302}
  (\bibinfo{year}{2015}), \eprint{1409.3270}.

\bibitem[{\citenamefont{Haas et~al.}(2015)\citenamefont{Haas, Hill, Izaguirre,
  and Yavin}}]{Haas:2014dda}
\bibinfo{author}{\bibfnamefont{A.}~\bibnamefont{Haas}},
  \bibinfo{author}{\bibfnamefont{C.~S.} \bibnamefont{Hill}},
  \bibinfo{author}{\bibfnamefont{E.}~\bibnamefont{Izaguirre}},
  \bibnamefont{and} \bibinfo{author}{\bibfnamefont{I.}~\bibnamefont{Yavin}},
  \bibinfo{journal}{Phys. Lett. B} \textbf{\bibinfo{volume}{746}},
  \bibinfo{pages}{117} (\bibinfo{year}{2015}), \eprint{1410.6816}.

\bibitem[{\citenamefont{Ball et~al.}(2016)}]{Ball:2016zrp}
\bibinfo{author}{\bibfnamefont{A.}~\bibnamefont{Ball}} \bibnamefont{et~al.}
  (\bibinfo{year}{2016}), \eprint{1607.04669}.

\bibitem[{\citenamefont{Alvis et~al.}(2018)}]{Alvis:2018yte}
\bibinfo{author}{\bibfnamefont{S.~I.} \bibnamefont{Alvis}} \bibnamefont{et~al.}
  (\bibinfo{collaboration}{Majorana}), \bibinfo{journal}{Phys. Rev. Lett.}
  \textbf{\bibinfo{volume}{120}}, \bibinfo{pages}{211804}
  (\bibinfo{year}{2018}), \eprint{1801.10145}.

\bibitem[{\citenamefont{Magill et~al.}(2019)\citenamefont{Magill, Plestid,
  Pospelov, and Tsai}}]{Magill:2018tbb}
\bibinfo{author}{\bibfnamefont{G.}~\bibnamefont{Magill}},
  \bibinfo{author}{\bibfnamefont{R.}~\bibnamefont{Plestid}},
  \bibinfo{author}{\bibfnamefont{M.}~\bibnamefont{Pospelov}}, \bibnamefont{and}
  \bibinfo{author}{\bibfnamefont{Y.-D.} \bibnamefont{Tsai}},
  \bibinfo{journal}{Phys. Rev. Lett.} \textbf{\bibinfo{volume}{122}},
  \bibinfo{pages}{071801} (\bibinfo{year}{2019}), \eprint{1806.03310}.

\bibitem[{\citenamefont{Kelly and Tsai}(2019)}]{Kelly:2018brz}
\bibinfo{author}{\bibfnamefont{K.~J.} \bibnamefont{Kelly}} \bibnamefont{and}
  \bibinfo{author}{\bibfnamefont{Y.-D.} \bibnamefont{Tsai}},
  \bibinfo{journal}{Phys. Rev. D} \textbf{\bibinfo{volume}{100}},
  \bibinfo{pages}{015043} (\bibinfo{year}{2019}), \eprint{1812.03998}.

\bibitem[{\citenamefont{Berlin et~al.}(2019)\citenamefont{Berlin, Blinov,
  Krnjaic, Schuster, and Toro}}]{Berlin:2018bsc}
\bibinfo{author}{\bibfnamefont{A.}~\bibnamefont{Berlin}},
  \bibinfo{author}{\bibfnamefont{N.}~\bibnamefont{Blinov}},
  \bibinfo{author}{\bibfnamefont{G.}~\bibnamefont{Krnjaic}},
  \bibinfo{author}{\bibfnamefont{P.}~\bibnamefont{Schuster}}, \bibnamefont{and}
  \bibinfo{author}{\bibfnamefont{N.}~\bibnamefont{Toro}},
  \bibinfo{journal}{Phys. Rev. D} \textbf{\bibinfo{volume}{99}},
  \bibinfo{pages}{075001} (\bibinfo{year}{2019}), \eprint{1807.01730}.

\bibitem[{\citenamefont{Dienes et~al.}(1997)\citenamefont{Dienes, Kolda, and
  March-Russell}}]{Dienes:1996zr}
\bibinfo{author}{\bibfnamefont{K.~R.} \bibnamefont{Dienes}},
  \bibinfo{author}{\bibfnamefont{C.~F.} \bibnamefont{Kolda}}, \bibnamefont{and}
  \bibinfo{author}{\bibfnamefont{J.}~\bibnamefont{March-Russell}},
  \bibinfo{journal}{Nucl. Phys. B} \textbf{\bibinfo{volume}{492}},
  \bibinfo{pages}{104} (\bibinfo{year}{1997}), \eprint{hep-ph/9610479}.

\bibitem[{\citenamefont{Abel et~al.}(2008)\citenamefont{Abel, Goodsell,
  Jaeckel, Khoze, and Ringwald}}]{Abel:2008ai}
\bibinfo{author}{\bibfnamefont{S.~A.} \bibnamefont{Abel}},
  \bibinfo{author}{\bibfnamefont{M.~D.} \bibnamefont{Goodsell}},
  \bibinfo{author}{\bibfnamefont{J.}~\bibnamefont{Jaeckel}},
  \bibinfo{author}{\bibfnamefont{V.~V.} \bibnamefont{Khoze}}, \bibnamefont{and}
  \bibinfo{author}{\bibfnamefont{A.}~\bibnamefont{Ringwald}},
  \bibinfo{journal}{JHEP} \textbf{\bibinfo{volume}{07}}, \bibinfo{pages}{124}
  (\bibinfo{year}{2008}), \eprint{0803.1449}.

\bibitem[{\citenamefont{Dubovsky et~al.}(2004)\citenamefont{Dubovsky, Gorbunov,
  and Rubtsov}}]{Dubovsky:2003yn}
\bibinfo{author}{\bibfnamefont{S.~L.} \bibnamefont{Dubovsky}},
  \bibinfo{author}{\bibfnamefont{D.~S.} \bibnamefont{Gorbunov}},
  \bibnamefont{and} \bibinfo{author}{\bibfnamefont{G.~I.}
  \bibnamefont{Rubtsov}}, \bibinfo{journal}{JETP Lett.}
  \textbf{\bibinfo{volume}{79}}, \bibinfo{pages}{1} (\bibinfo{year}{2004}),
  \eprint{hep-ph/0311189}.

\bibitem[{\citenamefont{Pospelov and Ramani}(2021)}]{Pospelov:2020ktu}
\bibinfo{author}{\bibfnamefont{M.}~\bibnamefont{Pospelov}} \bibnamefont{and}
  \bibinfo{author}{\bibfnamefont{H.}~\bibnamefont{Ramani}},
  \bibinfo{journal}{Phys. Rev. D} \textbf{\bibinfo{volume}{103}},
  \bibinfo{pages}{115031} (\bibinfo{year}{2021}), \eprint{2012.03957}.

\bibitem[{\citenamefont{Shiu et~al.}(2013)\citenamefont{Shiu, Soler, and
  Ye}}]{Shiu:2013wxa}
\bibinfo{author}{\bibfnamefont{G.}~\bibnamefont{Shiu}},
  \bibinfo{author}{\bibfnamefont{P.}~\bibnamefont{Soler}}, \bibnamefont{and}
  \bibinfo{author}{\bibfnamefont{F.}~\bibnamefont{Ye}}, \bibinfo{journal}{Phys.
  Rev. Lett.} \textbf{\bibinfo{volume}{110}}, \bibinfo{pages}{241304}
  (\bibinfo{year}{2013}), \eprint{1302.5471}.

\bibitem[{\citenamefont{Babu et~al.}(2013)}]{Babu:2013jba}
\bibinfo{author}{\bibfnamefont{K.~S.} \bibnamefont{Babu}} \bibnamefont{et~al.},
  in \emph{\bibinfo{booktitle}{{Proceedings, 2013 Community Summer Study on the
  Future of U.S. Particle Physics: Snowmass on the Mississippi (CSS2013):
  Minneapolis, MN, USA, July 29-August 6, 2013}}} (\bibinfo{year}{2013}),
  \eprint{1311.5285},
  \urlprefix\url{http://www.slac.stanford.edu/econf/C1307292/docs/IntensityFrontier/BaryonNo-13.pdf}.

\bibitem[{\citenamefont{Bernabei et~al.}(2000)}]{Bernabei:2000xp}
\bibinfo{author}{\bibfnamefont{R.}~\bibnamefont{Bernabei}}
  \bibnamefont{et~al.}, \bibinfo{journal}{Phys. Lett. B}
  \textbf{\bibinfo{volume}{493}}, \bibinfo{pages}{12} (\bibinfo{year}{2000}).

\bibitem[{\citenamefont{Bernabei et~al.}(2006)}]{Bernabei:2006tw}
\bibinfo{author}{\bibfnamefont{R.}~\bibnamefont{Bernabei}}
  \bibnamefont{et~al.}, \bibinfo{journal}{Eur. Phys. J. A}
  \textbf{\bibinfo{volume}{27}}, \bibinfo{pages}{35} (\bibinfo{year}{2006}),
  \bibinfo{note}{[,35(2006)]}.

\bibitem[{\citenamefont{Albert et~al.}(2018{\natexlab{c}})}]{Albert:2017qto}
\bibinfo{author}{\bibfnamefont{J.~B.} \bibnamefont{Albert}}
  \bibnamefont{et~al.} (\bibinfo{collaboration}{EXO-200}),
  \bibinfo{journal}{Phys. Rev. D} \textbf{\bibinfo{volume}{97}},
  \bibinfo{pages}{072007} (\bibinfo{year}{2018}{\natexlab{c}}),
  \eprint{1710.07670}.

\bibitem[{\citenamefont{Mohapatra and
  Perez-Lorenzana}(2003)}]{Mohapatra:2002ug}
\bibinfo{author}{\bibfnamefont{R.~N.} \bibnamefont{Mohapatra}}
  \bibnamefont{and}
  \bibinfo{author}{\bibfnamefont{A.}~\bibnamefont{Perez-Lorenzana}},
  \bibinfo{journal}{Phys. Rev. D} \textbf{\bibinfo{volume}{67}},
  \bibinfo{pages}{075015} (\bibinfo{year}{2003}), \eprint{hep-ph/0212254}.

\bibitem[{\citenamefont{Abazajian}(2017)}]{Abazajian:2017tcc}
\bibinfo{author}{\bibfnamefont{K.~N.} \bibnamefont{Abazajian}},
  \bibinfo{journal}{Phys. Rept.} \textbf{\bibinfo{volume}{711-712}},
  \bibinfo{pages}{1} (\bibinfo{year}{2017}), \eprint{1705.01837}.

\bibitem[{\citenamefont{Diaz et~al.}(2019)\citenamefont{Diaz, Argüelles,
  Collin, Conrad, and Shaevitz}}]{Diaz:2019fwt}
\bibinfo{author}{\bibfnamefont{A.}~\bibnamefont{Diaz}},
  \bibinfo{author}{\bibfnamefont{C.~A.} \bibnamefont{Argüelles}},
  \bibinfo{author}{\bibfnamefont{G.~H.} \bibnamefont{Collin}},
  \bibinfo{author}{\bibfnamefont{J.~M.} \bibnamefont{Conrad}},
  \bibnamefont{and} \bibinfo{author}{\bibfnamefont{M.~H.}
  \bibnamefont{Shaevitz}} (\bibinfo{year}{2019}), \eprint{1906.00045}.

\bibitem[{\citenamefont{Mei and Hime}(2006)}]{Mei:2005gm}
\bibinfo{author}{\bibfnamefont{D.}~\bibnamefont{Mei}} \bibnamefont{and}
  \bibinfo{author}{\bibfnamefont{A.}~\bibnamefont{Hime}},
  \bibinfo{journal}{Phys. Rev. D} \textbf{\bibinfo{volume}{73}},
  \bibinfo{pages}{053004} (\bibinfo{year}{2006}), \eprint{astro-ph/0512125}.

\bibitem[{\citenamefont{Kudryavtsev et~al.}(2008)\citenamefont{Kudryavtsev,
  Pandola, and Tomasello}}]{Kudryavtsev:2008fi}
\bibinfo{author}{\bibfnamefont{V.~A.} \bibnamefont{Kudryavtsev}},
  \bibinfo{author}{\bibfnamefont{L.}~\bibnamefont{Pandola}}, \bibnamefont{and}
  \bibinfo{author}{\bibfnamefont{V.}~\bibnamefont{Tomasello}},
  \bibinfo{journal}{Eur. Phys. J. A} \textbf{\bibinfo{volume}{36}},
  \bibinfo{pages}{171} (\bibinfo{year}{2008}), \eprint{0802.3566}.

\bibitem[{\citenamefont{Bettini}(2014)}]{Bettini:2014tva}
\bibinfo{author}{\bibfnamefont{A.}~\bibnamefont{Bettini}},
  \bibinfo{journal}{Phys. Dark Univ.} \textbf{\bibinfo{volume}{4}},
  \bibinfo{pages}{36} (\bibinfo{year}{2014}).

\bibitem[{\citenamefont{Votano}(2012)}]{Votano:2012fr}
\bibinfo{author}{\bibfnamefont{L.}~\bibnamefont{Votano}},
  \bibinfo{journal}{Eur. Phys. J. Plus} \textbf{\bibinfo{volume}{127}},
  \bibinfo{pages}{109} (\bibinfo{year}{2012}).

\bibitem[{\citenamefont{Wu et~al.}(2013)}]{Yu-Cheng:2013iaa}
\bibinfo{author}{\bibfnamefont{Y.-C.} \bibnamefont{Wu}} \bibnamefont{et~al.},
  \bibinfo{journal}{Chin. Phys. C} \textbf{\bibinfo{volume}{37}},
  \bibinfo{pages}{086001} (\bibinfo{year}{2013}), \eprint{1305.0899}.

\bibitem[{\citenamefont{Cheng et~al.}(2017)}]{Cheng:2018lcf}
\bibinfo{author}{\bibfnamefont{J.-P.} \bibnamefont{Cheng}}
  \bibnamefont{et~al.}, \bibinfo{journal}{Ann. Rev. Nucl. Part. Sci.}
  \textbf{\bibinfo{volume}{67}}, \bibinfo{pages}{231} (\bibinfo{year}{2017}),
  \eprint{1801.00587}.

\bibitem[{\citenamefont{Polaczek-Grelik
  et~al.}(2020)}]{Polaczek-Grelik:2020brg}
\bibinfo{author}{\bibfnamefont{K.}~\bibnamefont{Polaczek-Grelik}}
  \bibnamefont{et~al.}, \bibinfo{journal}{Nucl. Instrum. Meth. A}
  \textbf{\bibinfo{volume}{969}}, \bibinfo{pages}{164015}
  (\bibinfo{year}{2020}).

\bibitem[{\citenamefont{Trzaska et~al.}(2019)}]{Trzaska:2019kuk}
\bibinfo{author}{\bibfnamefont{W.~H.} \bibnamefont{Trzaska}}
  \bibnamefont{et~al.}, \bibinfo{journal}{Eur. Phys. J. C}
  \textbf{\bibinfo{volume}{79}}, \bibinfo{pages}{721} (\bibinfo{year}{2019}),
  \eprint{1902.00868}.

\bibitem[{\citenamefont{MORALES et~al.}(2005)\citenamefont{MORALES, BELTRÁN,
  CARMONA, CEBRIÁN, GARCÍA, IRASTORZA, GÓMEZ, LUZÓN, MARTÍNEZ, MORALES
  et~al.}}]{Morales:2005}
\bibinfo{author}{\bibfnamefont{J.}~\bibnamefont{MORALES}},
  \bibinfo{author}{\bibfnamefont{B.}~\bibnamefont{BELTRÁN}},
  \bibinfo{author}{\bibfnamefont{J.}~\bibnamefont{CARMONA}},
  \bibinfo{author}{\bibfnamefont{S.}~\bibnamefont{CEBRIÁN}},
  \bibinfo{author}{\bibfnamefont{E.}~\bibnamefont{GARCÍA}},
  \bibinfo{author}{\bibfnamefont{I.}~\bibnamefont{IRASTORZA}},
  \bibinfo{author}{\bibfnamefont{H.}~\bibnamefont{GÓMEZ}},
  \bibinfo{author}{\bibfnamefont{G.}~\bibnamefont{LUZÓN}},
  \bibinfo{author}{\bibfnamefont{M.}~\bibnamefont{MARTÍNEZ}},
  \bibinfo{author}{\bibfnamefont{A.}~\bibnamefont{MORALES}},
  \bibnamefont{et~al.}, in \emph{\bibinfo{booktitle}{The Identification of Dark
  Matter}} (\bibinfo{year}{2005}), pp. \bibinfo{pages}{447--452},
  \urlprefix\url{https://doi.org/10.1142/9789812701848_0067}.

\bibitem[{\citenamefont{Zhang and Mei}(2014)}]{Zhang:2014jsq}
\bibinfo{author}{\bibfnamefont{C.}~\bibnamefont{Zhang}} \bibnamefont{and}
  \bibinfo{author}{\bibfnamefont{D.~M.} \bibnamefont{Mei}},
  \bibinfo{journal}{Phys. Rev. D} \textbf{\bibinfo{volume}{90}},
  \bibinfo{pages}{122003} (\bibinfo{year}{2014}), \eprint{1407.3246}.

\bibitem[{\citenamefont{Zhang et~al.}(2016)}]{Super-Kamiokande:2015xra}
\bibinfo{author}{\bibfnamefont{Y.}~\bibnamefont{Zhang}} \bibnamefont{et~al.}
  (\bibinfo{collaboration}{Super-Kamiokande}), \bibinfo{journal}{Phys. Rev. D}
  \textbf{\bibinfo{volume}{93}}, \bibinfo{pages}{012004}
  (\bibinfo{year}{2016}), \eprint{1509.08168}.

\bibitem[{\citenamefont{Tang et~al.}(2006)\citenamefont{Tang, Horton-Smith,
  Kudryavtsev, and Tonazzo}}]{Tang:2006uu}
\bibinfo{author}{\bibfnamefont{A.}~\bibnamefont{Tang}},
  \bibinfo{author}{\bibfnamefont{G.}~\bibnamefont{Horton-Smith}},
  \bibinfo{author}{\bibfnamefont{V.~A.} \bibnamefont{Kudryavtsev}},
  \bibnamefont{and} \bibinfo{author}{\bibfnamefont{A.}~\bibnamefont{Tonazzo}},
  \bibinfo{journal}{Phys. Rev. D} \textbf{\bibinfo{volume}{74}},
  \bibinfo{pages}{053007} (\bibinfo{year}{2006}), \eprint{hep-ph/0604078}.

\bibitem[{\citenamefont{Reichhart et~al.}(2013)}]{Reichhart:2013xkd}
\bibinfo{author}{\bibfnamefont{L.}~\bibnamefont{Reichhart}}
  \bibnamefont{et~al.}, \bibinfo{journal}{Astropart. Phys.}
  \textbf{\bibinfo{volume}{47}}, \bibinfo{pages}{67} (\bibinfo{year}{2013}),
  \eprint{1302.4275}.

\bibitem[{\citenamefont{Agostini
  et~al.}(2019{\natexlab{a}})}]{Borexino:2018pev}
\bibinfo{author}{\bibfnamefont{M.}~\bibnamefont{Agostini}} \bibnamefont{et~al.}
  (\bibinfo{collaboration}{Borexino}), \bibinfo{journal}{JCAP}
  \textbf{\bibinfo{volume}{02}}, \bibinfo{pages}{046}
  (\bibinfo{year}{2019}{\natexlab{a}}), \eprint{1808.04207}.

\bibitem[{\citenamefont{Abgrall et~al.}(2017{\natexlab{a}})}]{MAJORANA:2016ifg}
\bibinfo{author}{\bibfnamefont{N.}~\bibnamefont{Abgrall}} \bibnamefont{et~al.}
  (\bibinfo{collaboration}{MAJORANA}), \bibinfo{journal}{Astropart. Phys.}
  \textbf{\bibinfo{volume}{93}}, \bibinfo{pages}{70}
  (\bibinfo{year}{2017}{\natexlab{a}}), \eprint{1602.07742}.

\bibitem[{\citenamefont{Cherry et~al.}(1983)\citenamefont{Cherry, Deakyne,
  Lande, Lee, Steinberg, Cleveland, and Fenyves}}]{Cherry:1983dp}
\bibinfo{author}{\bibfnamefont{M.~L.} \bibnamefont{Cherry}},
  \bibinfo{author}{\bibfnamefont{M.}~\bibnamefont{Deakyne}},
  \bibinfo{author}{\bibfnamefont{K.}~\bibnamefont{Lande}},
  \bibinfo{author}{\bibfnamefont{C.~K.} \bibnamefont{Lee}},
  \bibinfo{author}{\bibfnamefont{R.~I.} \bibnamefont{Steinberg}},
  \bibinfo{author}{\bibfnamefont{B.~T.} \bibnamefont{Cleveland}},
  \bibnamefont{and} \bibinfo{author}{\bibfnamefont{E.~J.}
  \bibnamefont{Fenyves}}, \bibinfo{journal}{Phys. Rev. D}
  \textbf{\bibinfo{volume}{27}}, \bibinfo{pages}{1444} (\bibinfo{year}{1983}).

\bibitem[{\citenamefont{Berger et~al.}(1989)}]{FREJUS:1989lko}
\bibinfo{author}{\bibfnamefont{C.}~\bibnamefont{Berger}} \bibnamefont{et~al.}
  (\bibinfo{collaboration}{FREJUS}), \bibinfo{journal}{Phys. Rev. D}
  \textbf{\bibinfo{volume}{40}}, \bibinfo{pages}{2163} (\bibinfo{year}{1989}).

\bibitem[{\citenamefont{Aharmim et~al.}(2009)}]{SNO:2009oor}
\bibinfo{author}{\bibfnamefont{B.}~\bibnamefont{Aharmim}} \bibnamefont{et~al.}
  (\bibinfo{collaboration}{SNO}), \bibinfo{journal}{Phys. Rev. D}
  \textbf{\bibinfo{volume}{80}}, \bibinfo{pages}{012001}
  (\bibinfo{year}{2009}), \eprint{0902.2776}.

\bibitem[{\citenamefont{Guo et~al.}(2021)}]{JNE:2020bwn}
\bibinfo{author}{\bibfnamefont{Z.}~\bibnamefont{Guo}} \bibnamefont{et~al.}
  (\bibinfo{collaboration}{JNE}), \bibinfo{journal}{Chin. Phys. C}
  \textbf{\bibinfo{volume}{45}}, \bibinfo{pages}{025001}
  (\bibinfo{year}{2021}), \eprint{2007.15925}.

\bibitem[{\citenamefont{Bettini}(2007)}]{Bettini:2007xc}
\bibinfo{author}{\bibfnamefont{A.}~\bibnamefont{Bettini}}
  (\bibinfo{year}{2007}), \eprint{0712.1051}.

\bibitem[{\citenamefont{Li et~al.}(2015)\citenamefont{Li, Ji, Haxton, and
  Wang}}]{Li:2014rca}
\bibinfo{author}{\bibfnamefont{J.}~\bibnamefont{Li}},
  \bibinfo{author}{\bibfnamefont{X.}~\bibnamefont{Ji}},
  \bibinfo{author}{\bibfnamefont{W.}~\bibnamefont{Haxton}}, \bibnamefont{and}
  \bibinfo{author}{\bibfnamefont{J.~S.~Y.} \bibnamefont{Wang}},
  \bibinfo{journal}{Phys. Procedia} \textbf{\bibinfo{volume}{61}},
  \bibinfo{pages}{576} (\bibinfo{year}{2015}), \eprint{1404.2651}.

\bibitem[{\citenamefont{Heise}(2017)}]{Heise:2017rpu}
\bibinfo{author}{\bibfnamefont{J.}~\bibnamefont{Heise}}, in
  \emph{\bibinfo{booktitle}{{15th International Conference on Topics in
  Astroparticle and Underground Physics (TAUP 2017) Sudbury, Ontario, Canada,
  July 24-28, 2017}}} (\bibinfo{year}{2017}), \eprint{1710.11584}.

\bibitem[{\citenamefont{Paling and Sadler}(2015)}]{Paling:2015ss}
\bibinfo{author}{\bibfnamefont{S.}~\bibnamefont{Paling}} \bibnamefont{and}
  \bibinfo{author}{\bibfnamefont{S.}~\bibnamefont{Sadler}},
  \bibinfo{journal}{Physics World} \textbf{\bibinfo{volume}{28}},
  \bibinfo{pages}{23} (\bibinfo{year}{2015}),
  \urlprefix\url{https://doi.org/10.1088\%2F2058-7058\%2F28\%2F5\%2F36}.

\bibitem[{\citenamefont{Piquemal}(2012)}]{Piquemal:2012fs}
\bibinfo{author}{\bibfnamefont{F.}~\bibnamefont{Piquemal}},
  \bibinfo{journal}{Eur. Phys. J. Plus} \textbf{\bibinfo{volume}{127}},
  \bibinfo{pages}{110} (\bibinfo{year}{2012}).

\bibitem[{\citenamefont{Lawson et~al.}(2012)\citenamefont{Lawson, Smith, and
  Vazquez~Jauregui}}]{Lawson:2012sga}
\bibinfo{author}{\bibfnamefont{I.}~\bibnamefont{Lawson}},
  \bibinfo{author}{\bibfnamefont{N.}~\bibnamefont{Smith}}, \bibnamefont{and}
  \bibinfo{author}{\bibfnamefont{E.}~\bibnamefont{Vazquez~Jauregui}},
  \bibinfo{journal}{Nucl. Phys. News} \textbf{\bibinfo{volume}{23}},
  \bibinfo{pages}{5} (\bibinfo{year}{2012}).

\bibitem[{\citenamefont{Aprile et~al.}(2011{\natexlab{b}})}]{Aprile:2011vb}
\bibinfo{author}{\bibfnamefont{E.}~\bibnamefont{Aprile}} \bibnamefont{et~al.}
  (\bibinfo{collaboration}{XENON100}), \bibinfo{journal}{Phys. Rev. D}
  \textbf{\bibinfo{volume}{83}}, \bibinfo{pages}{082001}
  (\bibinfo{year}{2011}{\natexlab{b}}), \bibinfo{note}{[Erratum: Phys. Rev. D
  {\bf 85}, 029904 (2012)]}, \eprint{1101.3866}.

\bibitem[{\citenamefont{Aprile et~al.}(2017{\natexlab{f}})}]{Aprile:2017ilq}
\bibinfo{author}{\bibfnamefont{E.}~\bibnamefont{Aprile}} \bibnamefont{et~al.}
  (\bibinfo{collaboration}{XENON}), \bibinfo{journal}{Eur. Phys. J. C}
  \textbf{\bibinfo{volume}{77}}, \bibinfo{pages}{890}
  (\bibinfo{year}{2017}{\natexlab{f}}), \eprint{1705.01828}.

\bibitem[{\citenamefont{Akerib et~al.}(2020{\natexlab{h}})}]{Akerib:2020com}
\bibinfo{author}{\bibfnamefont{D.~S.} \bibnamefont{Akerib}}
  \bibnamefont{et~al.} (\bibinfo{collaboration}{LZ}), \bibinfo{journal}{Eur.
  Phys. J. C} \textbf{\bibinfo{volume}{80}}, \bibinfo{pages}{1044}
  (\bibinfo{year}{2020}{\natexlab{h}}), \eprint{2006.02506}.

\bibitem[{\citenamefont{Aprile et~al.}(2021{\natexlab{d}})}]{Aprile:2020vmn}
\bibinfo{author}{\bibfnamefont{E.}~\bibnamefont{Aprile}} \bibnamefont{et~al.}
  (\bibinfo{collaboration}{XENON}), \bibinfo{journal}{Eur. Phys. J. C}
  \textbf{\bibinfo{volume}{81}}, \bibinfo{pages}{337}
  (\bibinfo{year}{2021}{\natexlab{d}}), \eprint{2009.13981}.

\bibitem[{\citenamefont{Wang et~al.}(2014)\citenamefont{Wang, Bao, Hao, and
  Ju}}]{Wang:2014ehv}
\bibinfo{author}{\bibfnamefont{Z.}~\bibnamefont{Wang}},
  \bibinfo{author}{\bibfnamefont{L.}~\bibnamefont{Bao}},
  \bibinfo{author}{\bibfnamefont{X.}~\bibnamefont{Hao}}, \bibnamefont{and}
  \bibinfo{author}{\bibfnamefont{Y.}~\bibnamefont{Ju}}, \bibinfo{journal}{Rev.
  Sci. Instrum.} \textbf{\bibinfo{volume}{85}}, \bibinfo{pages}{015116}
  (\bibinfo{year}{2014}).

\bibitem[{\citenamefont{Aprile et~al.}(2017{\natexlab{g}})}]{Aprile:2016xhi}
\bibinfo{author}{\bibfnamefont{E.}~\bibnamefont{Aprile}} \bibnamefont{et~al.}
  (\bibinfo{collaboration}{XENON}), \bibinfo{journal}{Eur. Phys. J. C}
  \textbf{\bibinfo{volume}{77}}, \bibinfo{pages}{275}
  (\bibinfo{year}{2017}{\natexlab{g}}), \eprint{1612.04284}.

\bibitem[{\citenamefont{Bolozdynya et~al.}(2007)\citenamefont{Bolozdynya,
  Brusov, Shutt, Dahl, and Kwong}}]{Bolozdynya:1997}
\bibinfo{author}{\bibfnamefont{A.}~\bibnamefont{Bolozdynya}},
  \bibinfo{author}{\bibfnamefont{P.}~\bibnamefont{Brusov}},
  \bibinfo{author}{\bibfnamefont{T.}~\bibnamefont{Shutt}},
  \bibinfo{author}{\bibfnamefont{C.}~\bibnamefont{Dahl}}, \bibnamefont{and}
  \bibinfo{author}{\bibfnamefont{J.}~\bibnamefont{Kwong}},
  \bibinfo{journal}{Nucl. Instr. Meth. A} \textbf{\bibinfo{volume}{579}},
  \bibinfo{pages}{50} (\bibinfo{year}{2007}).

\bibitem[{\citenamefont{Akerib et~al.}(2018{\natexlab{d}})}]{Akerib:2016hcd}
\bibinfo{author}{\bibfnamefont{D.~S.} \bibnamefont{Akerib}}
  \bibnamefont{et~al.} (\bibinfo{collaboration}{LUX}),
  \bibinfo{journal}{Astropart. Phys.} \textbf{\bibinfo{volume}{97}},
  \bibinfo{pages}{80} (\bibinfo{year}{2018}{\natexlab{d}}),
  \eprint{1605.03844}.

\bibitem[{\citenamefont{Aprile et~al.}(2021{\natexlab{e}})}]{XENON:2021fkt}
\bibinfo{author}{\bibfnamefont{E.}~\bibnamefont{Aprile}} \bibnamefont{et~al.}
  (\bibinfo{collaboration}{XENON}) (\bibinfo{year}{2021}{\natexlab{e}}),
  \eprint{2112.12231}.

\bibitem[{\citenamefont{Lindemann and Simgen}(2014)}]{Lindemann:2013kna}
\bibinfo{author}{\bibfnamefont{S.}~\bibnamefont{Lindemann}} \bibnamefont{and}
  \bibinfo{author}{\bibfnamefont{H.}~\bibnamefont{Simgen}},
  \bibinfo{journal}{Eur. Phys. J. C} \textbf{\bibinfo{volume}{74}},
  \bibinfo{pages}{2746} (\bibinfo{year}{2014}), \eprint{1308.4806}.

\bibitem[{\citenamefont{Aprile et~al.}(2017{\natexlab{h}})}]{Aprile:2017kop}
\bibinfo{author}{\bibfnamefont{E.}~\bibnamefont{Aprile}} \bibnamefont{et~al.}
  (\bibinfo{collaboration}{XENON100}), \bibinfo{journal}{Eur. Phys. J. C}
  \textbf{\bibinfo{volume}{77}}, \bibinfo{pages}{358}
  (\bibinfo{year}{2017}{\natexlab{h}}), \eprint{1702.06942}.

\bibitem[{\citenamefont{Burenkov et~al.}(2009)\citenamefont{Burenkov, Akimov,
  Grishkin, Kovalenko, Lebedenko, Solovov, Stekhanov, Neves, and
  Sumner}}]{Burenkov:2009zz}
\bibinfo{author}{\bibfnamefont{A.}~\bibnamefont{Burenkov}},
  \bibinfo{author}{\bibfnamefont{D.}~\bibnamefont{Akimov}},
  \bibinfo{author}{\bibfnamefont{Y.}~\bibnamefont{Grishkin}},
  \bibinfo{author}{\bibfnamefont{A.}~\bibnamefont{Kovalenko}},
  \bibinfo{author}{\bibfnamefont{V.}~\bibnamefont{Lebedenko}},
  \bibinfo{author}{\bibfnamefont{V.}~\bibnamefont{Solovov}},
  \bibinfo{author}{\bibfnamefont{V.}~\bibnamefont{Stekhanov}},
  \bibinfo{author}{\bibfnamefont{F.}~\bibnamefont{Neves}}, \bibnamefont{and}
  \bibinfo{author}{\bibfnamefont{T.}~\bibnamefont{Sumner}},
  \bibinfo{journal}{Phys. Atom. Nucl.} \textbf{\bibinfo{volume}{72}},
  \bibinfo{pages}{653} (\bibinfo{year}{2009}).

\bibitem[{\citenamefont{Santos et~al.}(2011)}]{Santos:2011ju}
\bibinfo{author}{\bibfnamefont{E.}~\bibnamefont{Santos}} \bibnamefont{et~al.}
  (\bibinfo{collaboration}{ZEPLIN-III}), \bibinfo{journal}{JHEP}
  \textbf{\bibinfo{volume}{12}}, \bibinfo{pages}{115} (\bibinfo{year}{2011}),
  \eprint{1110.3056}.

\bibitem[{\citenamefont{Angle et~al.}(2011)}]{Angle:2011th}
\bibinfo{author}{\bibfnamefont{J.}~\bibnamefont{Angle}} \bibnamefont{et~al.}
  (\bibinfo{collaboration}{XENON10}), \bibinfo{journal}{Phys. Rev. Lett.}
  \textbf{\bibinfo{volume}{107}}, \bibinfo{pages}{051301}
  (\bibinfo{year}{2011}), \bibinfo{note}{[Erratum: Phys. Rev. Lett. {\bf 110},
  249901 (2013)]}, \eprint{1104.3088}.

\bibitem[{\citenamefont{Edwards et~al.}(2008)}]{Edwards:2007nj}
\bibinfo{author}{\bibfnamefont{B.}~\bibnamefont{Edwards}} \bibnamefont{et~al.},
  \bibinfo{journal}{Astropart. Phys.} \textbf{\bibinfo{volume}{30}},
  \bibinfo{pages}{54} (\bibinfo{year}{2008}), \eprint{0708.0768}.

\bibitem[{\citenamefont{Edwards et~al.}(2018{\natexlab{b}})}]{Edwards:2017emx}
\bibinfo{author}{\bibfnamefont{B.}~\bibnamefont{Edwards}} \bibnamefont{et~al.},
  \bibinfo{journal}{JINST} \textbf{\bibinfo{volume}{13}},
  \bibinfo{pages}{P01005} (\bibinfo{year}{2018}{\natexlab{b}}),
  \eprint{1710.11032}.

\bibitem[{\citenamefont{Xu et~al.}(2019)\citenamefont{Xu, Pereverzev, Lenardo,
  Kingston, Naim, Bernstein, Kazkaz, and Tripathi}}]{Xu:2019dqb}
\bibinfo{author}{\bibfnamefont{J.}~\bibnamefont{Xu}},
  \bibinfo{author}{\bibfnamefont{S.}~\bibnamefont{Pereverzev}},
  \bibinfo{author}{\bibfnamefont{B.}~\bibnamefont{Lenardo}},
  \bibinfo{author}{\bibfnamefont{J.}~\bibnamefont{Kingston}},
  \bibinfo{author}{\bibfnamefont{D.}~\bibnamefont{Naim}},
  \bibinfo{author}{\bibfnamefont{A.}~\bibnamefont{Bernstein}},
  \bibinfo{author}{\bibfnamefont{K.}~\bibnamefont{Kazkaz}}, \bibnamefont{and}
  \bibinfo{author}{\bibfnamefont{M.}~\bibnamefont{Tripathi}},
  \bibinfo{journal}{Phys. Rev. D} \textbf{\bibinfo{volume}{99}},
  \bibinfo{pages}{103024} (\bibinfo{year}{2019}), \eprint{1904.02885}.

\bibitem[{\citenamefont{Tom\'as et~al.}(2018)\citenamefont{Tom\'as, Ara\'ujo,
  Bailey, Bayer, Chen, L\'opez~Paredes, and Sumner}}]{Tomas:2018pny}
\bibinfo{author}{\bibfnamefont{A.}~\bibnamefont{Tom\'as}},
  \bibinfo{author}{\bibfnamefont{H.~M.} \bibnamefont{Ara\'ujo}},
  \bibinfo{author}{\bibfnamefont{A.~J.} \bibnamefont{Bailey}},
  \bibinfo{author}{\bibfnamefont{A.}~\bibnamefont{Bayer}},
  \bibinfo{author}{\bibfnamefont{E.}~\bibnamefont{Chen}},
  \bibinfo{author}{\bibfnamefont{B.}~\bibnamefont{L\'opez~Paredes}},
  \bibnamefont{and} \bibinfo{author}{\bibfnamefont{T.~J.}
  \bibnamefont{Sumner}}, \bibinfo{journal}{Astropart. Phys.}
  \textbf{\bibinfo{volume}{103}}, \bibinfo{pages}{49} (\bibinfo{year}{2018}),
  \eprint{1801.07231}.

\bibitem[{\citenamefont{Aprile et~al.}(2019{\natexlab{f}})}]{Aprile:2019dme}
\bibinfo{author}{\bibfnamefont{E.}~\bibnamefont{Aprile}} \bibnamefont{et~al.}
  (\bibinfo{collaboration}{XENON}), \bibinfo{journal}{Phys. Rev. D}
  \textbf{\bibinfo{volume}{99}}, \bibinfo{pages}{112009}
  (\bibinfo{year}{2019}{\natexlab{f}}), \eprint{1902.11297}.

\bibitem[{\citenamefont{Agostinelli et~al.}(2003)}]{Agostinelli:2002hh}
\bibinfo{author}{\bibfnamefont{S.}~\bibnamefont{Agostinelli}}
  \bibnamefont{et~al.} (\bibinfo{collaboration}{GEANT4}),
  \bibinfo{journal}{Nucl. Instrum. Meth. A} \textbf{\bibinfo{volume}{506}},
  \bibinfo{pages}{250} (\bibinfo{year}{2003}).

\bibitem[{\citenamefont{Hagiwara et~al.}(2019)}]{Hagiwara:2019ptep}
\bibinfo{author}{\bibfnamefont{K.}~\bibnamefont{Hagiwara}}
  \bibnamefont{et~al.}, \bibinfo{journal}{PTEP}
  \textbf{\bibinfo{volume}{2019}}, \bibinfo{pages}{023D01}
  (\bibinfo{year}{2019}), \eprint{1809.02664}.

\bibitem[{\citenamefont{Tanaka et~al.}(2020)}]{Tanaka:2020ptep}
\bibinfo{author}{\bibfnamefont{T.}~\bibnamefont{Tanaka}} \bibnamefont{et~al.},
  \bibinfo{journal}{PTEP} \textbf{\bibinfo{volume}{2020}},
  \bibinfo{pages}{043D02} (\bibinfo{year}{2020}), \eprint{1907.00788}.

\bibitem[{\citenamefont{Bečvář}(1998)}]{Becvar:1998sd}
\bibinfo{author}{\bibfnamefont{F.}~\bibnamefont{Bečvář}},
  \bibinfo{journal}{Nuclear Instruments and Methods in Physics Research Section
  A: Accelerators, Spectrometers, Detectors and Associated Equipment}
  \textbf{\bibinfo{volume}{417}}, \bibinfo{pages}{434 } (\bibinfo{year}{1998}),
  ISSN \bibinfo{issn}{0168-9002},
  \urlprefix\url{http://www.sciencedirect.com/science/article/pii/S0168900298007876}.

\bibitem[{\citenamefont{Akerib et~al.}(2021{\natexlab{e}})}]{Akerib:2021ap}
\bibinfo{author}{\bibfnamefont{D.~S.} \bibnamefont{Akerib}}
  \bibnamefont{et~al.} (\bibinfo{collaboration}{LUX-ZEPLIN}),
  \bibinfo{journal}{Astropart. Phys.} \textbf{\bibinfo{volume}{125}},
  \bibinfo{pages}{102480} (\bibinfo{year}{2021}{\natexlab{e}}),
  \eprint{2001.09363}.

\bibitem[{\citenamefont{Sorensen et~al.}(2011)}]{Sorensen:2010hv}
\bibinfo{author}{\bibfnamefont{P.}~\bibnamefont{Sorensen}}
  \bibnamefont{et~al.}, \bibinfo{journal}{PoS}
  \textbf{\bibinfo{volume}{IDM2010}}, \bibinfo{pages}{017}
  (\bibinfo{year}{2011}), \eprint{1011.6439}.

\bibitem[{\citenamefont{Akerib et~al.}(2016{\natexlab{b}})}]{Akerib:2016mzi}
\bibinfo{author}{\bibfnamefont{D.}~\bibnamefont{Akerib}} \bibnamefont{et~al.}
  (\bibinfo{collaboration}{LUX}) (\bibinfo{year}{2016}{\natexlab{b}}),
  \eprint{1608.05381}.

\bibitem[{\citenamefont{Akerib et~al.}(2019{\natexlab{b}})}]{Akerib:2019jtm}
\bibinfo{author}{\bibfnamefont{D.}~\bibnamefont{Akerib}} \bibnamefont{et~al.}
  (\bibinfo{collaboration}{LUX}), \bibinfo{journal}{Phys. Rev. D}
  \textbf{\bibinfo{volume}{100}}, \bibinfo{pages}{022002}
  (\bibinfo{year}{2019}{\natexlab{b}}), \eprint{1903.12372}.

\bibitem[{\citenamefont{Akerib et~al.}(2016{\natexlab{c}})}]{Akerib:2015wdi}
\bibinfo{author}{\bibfnamefont{D.}~\bibnamefont{Akerib}} \bibnamefont{et~al.}
  (\bibinfo{collaboration}{LUX}), \bibinfo{journal}{Phys. Rev. D}
  \textbf{\bibinfo{volume}{93}}, \bibinfo{pages}{072009}
  (\bibinfo{year}{2016}{\natexlab{c}}), \eprint{1512.03133}.

\bibitem[{\citenamefont{Akerib et~al.}(2017{\natexlab{c}})}]{Akerib:2017hph}
\bibinfo{author}{\bibfnamefont{D.}~\bibnamefont{Akerib}} \bibnamefont{et~al.}
  (\bibinfo{collaboration}{LUX}), \bibinfo{journal}{Phys. Rev. D}
  \textbf{\bibinfo{volume}{96}}, \bibinfo{pages}{112011}
  (\bibinfo{year}{2017}{\natexlab{c}}), \eprint{1709.00800}.

\bibitem[{\citenamefont{Akerib et~al.}(2017{\natexlab{d}})}]{Akerib:2016qlr}
\bibinfo{author}{\bibfnamefont{D.}~\bibnamefont{Akerib}} \bibnamefont{et~al.}
  (\bibinfo{collaboration}{LUX}), \bibinfo{journal}{Phys. Rev. D}
  \textbf{\bibinfo{volume}{95}}, \bibinfo{pages}{012008}
  (\bibinfo{year}{2017}{\natexlab{d}}), \eprint{1610.02076}.

\bibitem[{\citenamefont{Aprile et~al.}(2018{\natexlab{c}})}]{Aprile:2017xxh}
\bibinfo{author}{\bibfnamefont{E.}~\bibnamefont{Aprile}} \bibnamefont{et~al.}
  (\bibinfo{collaboration}{XENON}), \bibinfo{journal}{Phys. Rev. D}
  \textbf{\bibinfo{volume}{97}}, \bibinfo{pages}{092007}
  (\bibinfo{year}{2018}{\natexlab{c}}), \eprint{1709.10149}.

\bibitem[{\citenamefont{Boulton et~al.}(2017)}]{Boulton:2017hub}
\bibinfo{author}{\bibfnamefont{E.}~\bibnamefont{Boulton}} \bibnamefont{et~al.},
  \bibinfo{journal}{JINST} \textbf{\bibinfo{volume}{12}},
  \bibinfo{pages}{P08004} (\bibinfo{year}{2017}), \eprint{1705.08958}.

\bibitem[{\citenamefont{Aprile et~al.}(2011{\natexlab{c}})}]{Aprile:2011hx}
\bibinfo{author}{\bibfnamefont{E.}~\bibnamefont{Aprile}} \bibnamefont{et~al.}
  (\bibinfo{collaboration}{XENON100}), \bibinfo{journal}{Phys. Rev. D}
  \textbf{\bibinfo{volume}{84}}, \bibinfo{pages}{052003}
  (\bibinfo{year}{2011}{\natexlab{c}}), \eprint{1103.0303}.

\bibitem[{\citenamefont{Mock et~al.}(2014)\citenamefont{Mock, Barry, Kazkaz,
  Szydagis, Tripathi, Uvarov, Woods, and Walsh}}]{Mock:2013ila}
\bibinfo{author}{\bibfnamefont{J.}~\bibnamefont{Mock}},
  \bibinfo{author}{\bibfnamefont{N.}~\bibnamefont{Barry}},
  \bibinfo{author}{\bibfnamefont{K.}~\bibnamefont{Kazkaz}},
  \bibinfo{author}{\bibfnamefont{M.}~\bibnamefont{Szydagis}},
  \bibinfo{author}{\bibfnamefont{M.}~\bibnamefont{Tripathi}},
  \bibinfo{author}{\bibfnamefont{S.}~\bibnamefont{Uvarov}},
  \bibinfo{author}{\bibfnamefont{M.}~\bibnamefont{Woods}}, \bibnamefont{and}
  \bibinfo{author}{\bibfnamefont{N.}~\bibnamefont{Walsh}},
  \bibinfo{journal}{JINST} \textbf{\bibinfo{volume}{9}},
  \bibinfo{pages}{T04002} (\bibinfo{year}{2014}), \eprint{1310.1117}.

\bibitem[{\citenamefont{Singh et~al.}(2020)}]{Singh:2019nrd}
\bibinfo{author}{\bibfnamefont{A.~G.} \bibnamefont{Singh}}
  \bibnamefont{et~al.}, \bibinfo{journal}{JINST} \textbf{\bibinfo{volume}{15}},
  \bibinfo{pages}{P01023} (\bibinfo{year}{2020}), \eprint{1911.03999}.

\bibitem[{\citenamefont{Plante et~al.}(2011)\citenamefont{Plante, Aprile,
  Budnik, Choi, Giboni, Goetzke, Lang, Lim, and
  Melgarejo~Fernandez}}]{Plante:2011hw}
\bibinfo{author}{\bibfnamefont{G.}~\bibnamefont{Plante}},
  \bibinfo{author}{\bibfnamefont{E.}~\bibnamefont{Aprile}},
  \bibinfo{author}{\bibfnamefont{R.}~\bibnamefont{Budnik}},
  \bibinfo{author}{\bibfnamefont{B.}~\bibnamefont{Choi}},
  \bibinfo{author}{\bibfnamefont{K.~L.} \bibnamefont{Giboni}},
  \bibinfo{author}{\bibfnamefont{L.~W.} \bibnamefont{Goetzke}},
  \bibinfo{author}{\bibfnamefont{R.~F.} \bibnamefont{Lang}},
  \bibinfo{author}{\bibfnamefont{K.~E.} \bibnamefont{Lim}}, \bibnamefont{and}
  \bibinfo{author}{\bibfnamefont{A.~J.} \bibnamefont{Melgarejo~Fernandez}},
  \bibinfo{journal}{Phys. Rev. C} \textbf{\bibinfo{volume}{84}},
  \bibinfo{pages}{045805} (\bibinfo{year}{2011}), \eprint{1104.2587}.

\bibitem[{\citenamefont{Goetzke et~al.}(2017)\citenamefont{Goetzke, Aprile,
  Anthony, Plante, and Weber}}]{Goetzke:2016lfg}
\bibinfo{author}{\bibfnamefont{L.~W.} \bibnamefont{Goetzke}},
  \bibinfo{author}{\bibfnamefont{E.}~\bibnamefont{Aprile}},
  \bibinfo{author}{\bibfnamefont{M.}~\bibnamefont{Anthony}},
  \bibinfo{author}{\bibfnamefont{G.}~\bibnamefont{Plante}}, \bibnamefont{and}
  \bibinfo{author}{\bibfnamefont{M.}~\bibnamefont{Weber}},
  \bibinfo{journal}{Phys. Rev. D} \textbf{\bibinfo{volume}{96}},
  \bibinfo{pages}{103007} (\bibinfo{year}{2017}), \eprint{1611.10322}.

\bibitem[{\citenamefont{Baudis et~al.}(2018)\citenamefont{Baudis, Biondi,
  Capelli, Galloway, Kazama, Kish, Pakarha, Piastra, and
  Wulf}}]{Baudis:2017xov}
\bibinfo{author}{\bibfnamefont{L.}~\bibnamefont{Baudis}},
  \bibinfo{author}{\bibfnamefont{Y.}~\bibnamefont{Biondi}},
  \bibinfo{author}{\bibfnamefont{C.}~\bibnamefont{Capelli}},
  \bibinfo{author}{\bibfnamefont{M.}~\bibnamefont{Galloway}},
  \bibinfo{author}{\bibfnamefont{S.}~\bibnamefont{Kazama}},
  \bibinfo{author}{\bibfnamefont{A.}~\bibnamefont{Kish}},
  \bibinfo{author}{\bibfnamefont{P.}~\bibnamefont{Pakarha}},
  \bibinfo{author}{\bibfnamefont{F.}~\bibnamefont{Piastra}}, \bibnamefont{and}
  \bibinfo{author}{\bibfnamefont{J.}~\bibnamefont{Wulf}},
  \bibinfo{journal}{Eur. Phys. J. C} \textbf{\bibinfo{volume}{78}},
  \bibinfo{pages}{351} (\bibinfo{year}{2018}), \eprint{1712.08607}.

\bibitem[{\citenamefont{{Dylan J Temples}}(2019)}]{Temples:2019taup}
\bibinfo{author}{\bibnamefont{{Dylan J Temples}}},
  \emph{\bibinfo{title}{{Understanding neutrino background implications in
  LXe-TPC dark matter searches using {127}Xe electron captures (TAUP 2019}}},
  \bibinfo{howpublished}{{\url{http://www-kam2.icrr.u-tokyo.ac.jp/indico/event/3/session/10/contribution/414/material/slides/0.pdf}}}
  (\bibinfo{year}{2019}), \bibinfo{note}{(TAUP 2019)}.

\bibitem[{\citenamefont{Temples et~al.}(2021)\citenamefont{Temples, McLaughlin,
  Bargemann, Baxter, Cottle, Dahl, Lippincott, Monte, and
  Phelan}}]{Temples:2021prd}
\bibinfo{author}{\bibfnamefont{D.~J.} \bibnamefont{Temples}},
  \bibinfo{author}{\bibfnamefont{J.}~\bibnamefont{McLaughlin}},
  \bibinfo{author}{\bibfnamefont{J.}~\bibnamefont{Bargemann}},
  \bibinfo{author}{\bibfnamefont{D.}~\bibnamefont{Baxter}},
  \bibinfo{author}{\bibfnamefont{A.}~\bibnamefont{Cottle}},
  \bibinfo{author}{\bibfnamefont{C.~E.} \bibnamefont{Dahl}},
  \bibinfo{author}{\bibfnamefont{W.~H.} \bibnamefont{Lippincott}},
  \bibinfo{author}{\bibfnamefont{A.}~\bibnamefont{Monte}}, \bibnamefont{and}
  \bibinfo{author}{\bibfnamefont{J.}~\bibnamefont{Phelan}},
  \bibinfo{journal}{Phys. Rev. D} \textbf{\bibinfo{volume}{104}},
  \bibinfo{pages}{112001} (\bibinfo{year}{2021}), \eprint{2109.11487}.

\bibitem[{\citenamefont{Lang et~al.}(2016{\natexlab{b}})\citenamefont{Lang,
  Brown, Brown, Cervantes, Macmullin, Masson, Schreiner, and
  Simgen}}]{Lang:2016zde}
\bibinfo{author}{\bibfnamefont{R.~F.} \bibnamefont{Lang}},
  \bibinfo{author}{\bibfnamefont{A.}~\bibnamefont{Brown}},
  \bibinfo{author}{\bibfnamefont{E.}~\bibnamefont{Brown}},
  \bibinfo{author}{\bibfnamefont{M.}~\bibnamefont{Cervantes}},
  \bibinfo{author}{\bibfnamefont{S.}~\bibnamefont{Macmullin}},
  \bibinfo{author}{\bibfnamefont{D.}~\bibnamefont{Masson}},
  \bibinfo{author}{\bibfnamefont{J.}~\bibnamefont{Schreiner}},
  \bibnamefont{and} \bibinfo{author}{\bibfnamefont{H.}~\bibnamefont{Simgen}},
  \bibinfo{journal}{JINST} \textbf{\bibinfo{volume}{11}},
  \bibinfo{pages}{P04004} (\bibinfo{year}{2016}{\natexlab{b}}),
  \eprint{1602.01138}.

\bibitem[{\citenamefont{Aprile et~al.}(2017{\natexlab{i}})}]{Aprile:2016pmc}
\bibinfo{author}{\bibfnamefont{E.}~\bibnamefont{Aprile}} \bibnamefont{et~al.}
  (\bibinfo{collaboration}{XENON}), \bibinfo{journal}{Phys. Rev. D}
  \textbf{\bibinfo{volume}{95}}, \bibinfo{pages}{072008}
  (\bibinfo{year}{2017}{\natexlab{i}}), \eprint{1611.03585}.

\bibitem[{\citenamefont{Collar}(2013)}]{Collar:2013xva}
\bibinfo{author}{\bibfnamefont{J.~I.} \bibnamefont{Collar}},
  \bibinfo{journal}{Phys. Rev. Lett.} \textbf{\bibinfo{volume}{110}},
  \bibinfo{pages}{211101} (\bibinfo{year}{2013}), \eprint{1303.2686}.

\bibitem[{\citenamefont{Aprile et~al.}(2013)}]{Aprile:2013teh}
\bibinfo{author}{\bibfnamefont{E.}~\bibnamefont{Aprile}} \bibnamefont{et~al.}
  (\bibinfo{collaboration}{XENON100}), \bibinfo{journal}{Phys. Rev. D}
  \textbf{\bibinfo{volume}{88}}, \bibinfo{pages}{012006}
  (\bibinfo{year}{2013}), \eprint{1304.1427}.

\bibitem[{\citenamefont{Srednicki et~al.}(1988)\citenamefont{Srednicki,
  Watkins, and Olive}}]{Srednicki:1988ce}
\bibinfo{author}{\bibfnamefont{M.}~\bibnamefont{Srednicki}},
  \bibinfo{author}{\bibfnamefont{R.}~\bibnamefont{Watkins}}, \bibnamefont{and}
  \bibinfo{author}{\bibfnamefont{K.~A.} \bibnamefont{Olive}},
  \bibinfo{journal}{Nucl. Phys. B} \textbf{\bibinfo{volume}{310}},
  \bibinfo{pages}{693} (\bibinfo{year}{1988}), \bibinfo{note}{[,247(1988)]}.

\bibitem[{\citenamefont{Feng}(2010)}]{Feng:2010gw}
\bibinfo{author}{\bibfnamefont{J.}~\bibnamefont{Feng}}, \bibinfo{journal}{Ann.
  Rev. Astron. Astrophys.} \textbf{\bibinfo{volume}{48}}, \bibinfo{pages}{495}
  (\bibinfo{year}{2010}), \eprint{1003.0904}.

\bibitem[{\citenamefont{Profumo et~al.}(2013)\citenamefont{Profumo, Shepherd,
  and Tait}}]{Profumo:2013hqa}
\bibinfo{author}{\bibfnamefont{S.}~\bibnamefont{Profumo}},
  \bibinfo{author}{\bibfnamefont{W.}~\bibnamefont{Shepherd}}, \bibnamefont{and}
  \bibinfo{author}{\bibfnamefont{T.}~\bibnamefont{Tait}},
  \bibinfo{journal}{Phys. Rev. D} \textbf{\bibinfo{volume}{88}},
  \bibinfo{pages}{056018} (\bibinfo{year}{2013}), \eprint{1307.6277}.

\bibitem[{\citenamefont{Evans and Bryant}(2008)}]{Evans:2008zzb}
\bibinfo{author}{\bibfnamefont{L.}~\bibnamefont{Evans}} \bibnamefont{and}
  \bibinfo{author}{\bibfnamefont{P.}~\bibnamefont{Bryant}},
  \bibinfo{journal}{JINST} \textbf{\bibinfo{volume}{3}},
  \bibinfo{pages}{S08001} (\bibinfo{year}{2008}).

\bibitem[{\citenamefont{Golfand and Likhtman}(1971)}]{Golfand:1971iw}
\bibinfo{author}{\bibfnamefont{{\relax Yu}.~A.} \bibnamefont{Golfand}}
  \bibnamefont{and} \bibinfo{author}{\bibfnamefont{E.~P.}
  \bibnamefont{Likhtman}}, \bibinfo{journal}{JETP Lett.}
  \textbf{\bibinfo{volume}{13}}, \bibinfo{pages}{323} (\bibinfo{year}{1971}),
  \bibinfo{note}{[Pisma Zh. Eksp. Teor. Fiz. {\bf 13}, 452 (1971)]}.

\bibitem[{\citenamefont{Clavelli and Ramond}(1971)}]{Clavelli:1970qy}
\bibinfo{author}{\bibfnamefont{L.}~\bibnamefont{Clavelli}} \bibnamefont{and}
  \bibinfo{author}{\bibfnamefont{P.}~\bibnamefont{Ramond}},
  \bibinfo{journal}{Phys. Rev. D} \textbf{\bibinfo{volume}{3}},
  \bibinfo{pages}{988} (\bibinfo{year}{1971}).

\bibitem[{\citenamefont{Penning}(2018)}]{Penning:2017tmb}
\bibinfo{author}{\bibfnamefont{B.}~\bibnamefont{Penning}}, \bibinfo{journal}{J.
  Phys. G} \textbf{\bibinfo{volume}{45}}, \bibinfo{pages}{063001}
  (\bibinfo{year}{2018}), \eprint{1712.01391}.

\bibitem[{\citenamefont{Bonilla et~al.}(2022)\citenamefont{Bonilla, Brivio,
  Machado-Rodr\'\i{}guez, and de~Troc\'oniz}}]{Bonilla:2022pxu}
\bibinfo{author}{\bibfnamefont{J.}~\bibnamefont{Bonilla}},
  \bibinfo{author}{\bibfnamefont{I.}~\bibnamefont{Brivio}},
  \bibinfo{author}{\bibfnamefont{J.}~\bibnamefont{Machado-Rodr\'\i{}guez}},
  \bibnamefont{and} \bibinfo{author}{\bibfnamefont{J.~F.}
  \bibnamefont{de~Troc\'oniz}} (\bibinfo{year}{2022}), \eprint{2202.03450}.

\bibitem[{\citenamefont{Aaij et~al.}(2018)}]{LHCb:2017trq}
\bibinfo{author}{\bibfnamefont{R.}~\bibnamefont{Aaij}} \bibnamefont{et~al.}
  (\bibinfo{collaboration}{LHCb}), \bibinfo{journal}{Phys. Rev. Lett.}
  \textbf{\bibinfo{volume}{120}}, \bibinfo{pages}{061801}
  (\bibinfo{year}{2018}), \eprint{1710.02867}.

\bibitem[{\citenamefont{Finkbeiner et~al.}(2012)\citenamefont{Finkbeiner,
  Galli, Lin, and Slatyer}}]{Finkbeiner:2011dx}
\bibinfo{author}{\bibfnamefont{D.~P.} \bibnamefont{Finkbeiner}},
  \bibinfo{author}{\bibfnamefont{S.}~\bibnamefont{Galli}},
  \bibinfo{author}{\bibfnamefont{T.}~\bibnamefont{Lin}}, \bibnamefont{and}
  \bibinfo{author}{\bibfnamefont{T.~R.} \bibnamefont{Slatyer}},
  \bibinfo{journal}{Phys. Rev. D} \textbf{\bibinfo{volume}{85}},
  \bibinfo{pages}{043522} (\bibinfo{year}{2012}), \eprint{1109.6322}.

\bibitem[{\citenamefont{Galli et~al.}(2013)\citenamefont{Galli, Slatyer,
  Valdes, and Iocco}}]{Galli:2013dna}
\bibinfo{author}{\bibfnamefont{S.}~\bibnamefont{Galli}},
  \bibinfo{author}{\bibfnamefont{T.~R.} \bibnamefont{Slatyer}},
  \bibinfo{author}{\bibfnamefont{M.}~\bibnamefont{Valdes}}, \bibnamefont{and}
  \bibinfo{author}{\bibfnamefont{F.}~\bibnamefont{Iocco}},
  \bibinfo{journal}{Phys. Rev. D} \textbf{\bibinfo{volume}{88}},
  \bibinfo{pages}{063502} (\bibinfo{year}{2013}), \eprint{1306.0563}.

\bibitem[{\citenamefont{Boyarsky et~al.}(2008)\citenamefont{Boyarsky, Malyshev,
  Neronov, and Ruchayskiy}}]{Boyarsky:2007ge}
\bibinfo{author}{\bibfnamefont{A.}~\bibnamefont{Boyarsky}},
  \bibinfo{author}{\bibfnamefont{D.}~\bibnamefont{Malyshev}},
  \bibinfo{author}{\bibfnamefont{A.}~\bibnamefont{Neronov}}, \bibnamefont{and}
  \bibinfo{author}{\bibfnamefont{O.}~\bibnamefont{Ruchayskiy}},
  \bibinfo{journal}{Mon. Not. Roy. Astron. Soc.}
  \textbf{\bibinfo{volume}{387}}, \bibinfo{pages}{1345} (\bibinfo{year}{2008}),
  \eprint{0710.4922}.

\bibitem[{\citenamefont{Yuksel et~al.}(2008)\citenamefont{Yuksel, Beacom, and
  Watson}}]{Yuksel:2007xh}
\bibinfo{author}{\bibfnamefont{H.}~\bibnamefont{Yuksel}},
  \bibinfo{author}{\bibfnamefont{J.~F.} \bibnamefont{Beacom}},
  \bibnamefont{and} \bibinfo{author}{\bibfnamefont{C.~R.}
  \bibnamefont{Watson}}, \bibinfo{journal}{Phys. Rev. Lett.}
  \textbf{\bibinfo{volume}{101}}, \bibinfo{pages}{121301}
  (\bibinfo{year}{2008}), \eprint{0706.4084}.

\bibitem[{\citenamefont{Perez et~al.}(2017)\citenamefont{Perez, Ng, Beacom,
  Hersh, Horiuchi, and Krivonos}}]{Perez:2016tcq}
\bibinfo{author}{\bibfnamefont{K.}~\bibnamefont{Perez}},
  \bibinfo{author}{\bibfnamefont{K.~C.~Y.} \bibnamefont{Ng}},
  \bibinfo{author}{\bibfnamefont{J.~F.} \bibnamefont{Beacom}},
  \bibinfo{author}{\bibfnamefont{C.}~\bibnamefont{Hersh}},
  \bibinfo{author}{\bibfnamefont{S.}~\bibnamefont{Horiuchi}}, \bibnamefont{and}
  \bibinfo{author}{\bibfnamefont{R.}~\bibnamefont{Krivonos}},
  \bibinfo{journal}{Phys. Rev. D} \textbf{\bibinfo{volume}{95}},
  \bibinfo{pages}{123002} (\bibinfo{year}{2017}), \eprint{1609.00667}.

\bibitem[{\citenamefont{Gunn et~al.}(1978)\citenamefont{Gunn, Lee, Lerche,
  Schramm, and Steigman}}]{Gunn:1978gr}
\bibinfo{author}{\bibfnamefont{J.~E.} \bibnamefont{Gunn}},
  \bibinfo{author}{\bibfnamefont{B.~W.} \bibnamefont{Lee}},
  \bibinfo{author}{\bibfnamefont{I.}~\bibnamefont{Lerche}},
  \bibinfo{author}{\bibfnamefont{D.~N.} \bibnamefont{Schramm}},
  \bibnamefont{and} \bibinfo{author}{\bibfnamefont{G.}~\bibnamefont{Steigman}},
  \bibinfo{journal}{Astrophys. J.} \textbf{\bibinfo{volume}{223}},
  \bibinfo{pages}{1015} (\bibinfo{year}{1978}), \bibinfo{note}{[,190(1978)]}.

\bibitem[{\citenamefont{Stecker}(1978)}]{Stecker:1978du}
\bibinfo{author}{\bibfnamefont{F.}~\bibnamefont{Stecker}},
  \bibinfo{journal}{Astrophys. J.} \textbf{\bibinfo{volume}{223}},
  \bibinfo{pages}{1032} (\bibinfo{year}{1978}).

\bibitem[{\citenamefont{Berezinsky et~al.}(1994)\citenamefont{Berezinsky,
  Bottino, and Mignola}}]{Berezinsky:1994wva}
\bibinfo{author}{\bibfnamefont{V.}~\bibnamefont{Berezinsky}},
  \bibinfo{author}{\bibfnamefont{A.}~\bibnamefont{Bottino}}, \bibnamefont{and}
  \bibinfo{author}{\bibfnamefont{G.}~\bibnamefont{Mignola}},
  \bibinfo{journal}{Phys. Lett. B} \textbf{\bibinfo{volume}{325}},
  \bibinfo{pages}{136} (\bibinfo{year}{1994}), \eprint{hep-ph/9402215}.

\bibitem[{\citenamefont{Bergstrom et~al.}(1998)\citenamefont{Bergstrom, Ullio,
  and Buckley}}]{Bergstrom:1997fj}
\bibinfo{author}{\bibfnamefont{L.}~\bibnamefont{Bergstrom}},
  \bibinfo{author}{\bibfnamefont{P.}~\bibnamefont{Ullio}}, \bibnamefont{and}
  \bibinfo{author}{\bibfnamefont{J.~H.} \bibnamefont{Buckley}},
  \bibinfo{journal}{Astropart. Phys.} \textbf{\bibinfo{volume}{9}},
  \bibinfo{pages}{137} (\bibinfo{year}{1998}), \eprint{astro-ph/9712318}.

\bibitem[{\citenamefont{Gondolo and Silk}(1999)}]{Gondolo:1999ef}
\bibinfo{author}{\bibfnamefont{P.}~\bibnamefont{Gondolo}} \bibnamefont{and}
  \bibinfo{author}{\bibfnamefont{J.}~\bibnamefont{Silk}},
  \bibinfo{journal}{Phys. Rev. Lett.} \textbf{\bibinfo{volume}{83}},
  \bibinfo{pages}{1719} (\bibinfo{year}{1999}), \eprint{astro-ph/9906391}.

\bibitem[{\citenamefont{Gehrels and Michelson}(1999)}]{Gehrels:1999ri}
\bibinfo{author}{\bibfnamefont{N.}~\bibnamefont{Gehrels}} \bibnamefont{and}
  \bibinfo{author}{\bibfnamefont{P.}~\bibnamefont{Michelson}},
  \bibinfo{journal}{Astropart. Phys.} \textbf{\bibinfo{volume}{11}},
  \bibinfo{pages}{277} (\bibinfo{year}{1999}).

\bibitem[{\citenamefont{Cesarini et~al.}(2004)\citenamefont{Cesarini, Fucito,
  Lionetto, Morselli, and Ullio}}]{Cesarini:2003nr}
\bibinfo{author}{\bibfnamefont{A.}~\bibnamefont{Cesarini}},
  \bibinfo{author}{\bibfnamefont{F.}~\bibnamefont{Fucito}},
  \bibinfo{author}{\bibfnamefont{A.}~\bibnamefont{Lionetto}},
  \bibinfo{author}{\bibfnamefont{A.}~\bibnamefont{Morselli}}, \bibnamefont{and}
  \bibinfo{author}{\bibfnamefont{P.}~\bibnamefont{Ullio}},
  \bibinfo{journal}{Astropart. Phys.} \textbf{\bibinfo{volume}{21}},
  \bibinfo{pages}{267} (\bibinfo{year}{2004}), \eprint{astro-ph/0305075}.

\bibitem[{\citenamefont{Peirani et~al.}(2004)\citenamefont{Peirani, Mohayaee,
  and de~Freitas~Pacheco}}]{Peirani:2004wy}
\bibinfo{author}{\bibfnamefont{S.}~\bibnamefont{Peirani}},
  \bibinfo{author}{\bibfnamefont{R.}~\bibnamefont{Mohayaee}}, \bibnamefont{and}
  \bibinfo{author}{\bibfnamefont{J.~A.} \bibnamefont{de~Freitas~Pacheco}},
  \bibinfo{journal}{Phys. Rev. D} \textbf{\bibinfo{volume}{70}},
  \bibinfo{pages}{043503} (\bibinfo{year}{2004}), \eprint{astro-ph/0401378}.

\bibitem[{\citenamefont{Dodelson et~al.}(2008)\citenamefont{Dodelson, Hooper,
  and Serpico}}]{Dodelson:2007gd}
\bibinfo{author}{\bibfnamefont{S.}~\bibnamefont{Dodelson}},
  \bibinfo{author}{\bibfnamefont{D.}~\bibnamefont{Hooper}}, \bibnamefont{and}
  \bibinfo{author}{\bibfnamefont{P.~D.} \bibnamefont{Serpico}},
  \bibinfo{journal}{Phys. Rev. D} \textbf{\bibinfo{volume}{77}},
  \bibinfo{pages}{063512} (\bibinfo{year}{2008}), \eprint{0711.4621}.

\bibitem[{\citenamefont{Actis et~al.}(2011)}]{Consortium:2010bc}
\bibinfo{author}{\bibfnamefont{M.}~\bibnamefont{Actis}} \bibnamefont{et~al.}
  (\bibinfo{collaboration}{CTA Consortium}), \bibinfo{journal}{Exper. Astron.}
  \textbf{\bibinfo{volume}{32}}, \bibinfo{pages}{193} (\bibinfo{year}{2011}),
  \eprint{1008.3703}.

\bibitem[{\citenamefont{Cholis et~al.}(2014)\citenamefont{Cholis, Hooper, and
  McDermott}}]{Cholis:2013ena}
\bibinfo{author}{\bibfnamefont{I.}~\bibnamefont{Cholis}},
  \bibinfo{author}{\bibfnamefont{D.}~\bibnamefont{Hooper}}, \bibnamefont{and}
  \bibinfo{author}{\bibfnamefont{S.~D.} \bibnamefont{McDermott}},
  \bibinfo{journal}{JCAP} \textbf{\bibinfo{volume}{02}}, \bibinfo{pages}{014}
  (\bibinfo{year}{2014}), \eprint{1312.0608}.

\bibitem[{\citenamefont{Ando and Ishiwata}(2015)}]{Ando:2015qda}
\bibinfo{author}{\bibfnamefont{S.}~\bibnamefont{Ando}} \bibnamefont{and}
  \bibinfo{author}{\bibfnamefont{K.}~\bibnamefont{Ishiwata}},
  \bibinfo{journal}{JCAP} \textbf{\bibinfo{volume}{05}}, \bibinfo{pages}{024}
  (\bibinfo{year}{2015}), \eprint{1502.02007}.

\bibitem[{\citenamefont{Ackermann et~al.}(2015)}]{Ackermann:2015tah}
\bibinfo{author}{\bibfnamefont{M.}~\bibnamefont{Ackermann}}
  \bibnamefont{et~al.} (\bibinfo{collaboration}{Fermi-LAT}),
  \bibinfo{journal}{JCAP} \textbf{\bibinfo{volume}{09}}, \bibinfo{pages}{008}
  (\bibinfo{year}{2015}), \eprint{1501.05464}.

\bibitem[{\citenamefont{Di~Mauro and Donato}(2015)}]{DiMauro:2015tfa}
\bibinfo{author}{\bibfnamefont{M.}~\bibnamefont{Di~Mauro}} \bibnamefont{and}
  \bibinfo{author}{\bibfnamefont{F.}~\bibnamefont{Donato}},
  \bibinfo{journal}{Phys. Rev. D} \textbf{\bibinfo{volume}{91}},
  \bibinfo{pages}{123001} (\bibinfo{year}{2015}), \eprint{1501.05316}.

\bibitem[{\citenamefont{Ajello et~al.}(2015)}]{Ajello:2015mfa}
\bibinfo{author}{\bibfnamefont{M.}~\bibnamefont{Ajello}} \bibnamefont{et~al.},
  \bibinfo{journal}{Astrophys. J. Lett.} \textbf{\bibinfo{volume}{800}},
  \bibinfo{pages}{L27} (\bibinfo{year}{2015}), \eprint{1501.05301}.

\bibitem[{\citenamefont{Abdallah et~al.}(2016)}]{Abdallah:2016ygi}
\bibinfo{author}{\bibfnamefont{H.}~\bibnamefont{Abdallah}} \bibnamefont{et~al.}
  (\bibinfo{collaboration}{H.E.S.S.}), \bibinfo{journal}{Phys. Rev. Lett.}
  \textbf{\bibinfo{volume}{117}}, \bibinfo{pages}{111301}
  (\bibinfo{year}{2016}), \eprint{1607.08142}.

\bibitem[{\citenamefont{Ahnen et~al.}(2016)}]{Ahnen:2016qkx}
\bibinfo{author}{\bibfnamefont{M.}~\bibnamefont{Ahnen}} \bibnamefont{et~al.}
  (\bibinfo{collaboration}{MAGIC, Fermi-LAT}), \bibinfo{journal}{JCAP}
  \textbf{\bibinfo{volume}{02}}, \bibinfo{pages}{039} (\bibinfo{year}{2016}),
  \eprint{1601.06590}.

\bibitem[{\citenamefont{Zitzer}(2016)}]{Zitzer:2016fvx}
\bibinfo{author}{\bibfnamefont{B.}~\bibnamefont{Zitzer}}
  (\bibinfo{collaboration}{VERITAS}), \bibinfo{journal}{PoS}
  \textbf{\bibinfo{volume}{ICHEP2016}}, \bibinfo{pages}{446}
  (\bibinfo{year}{2016}).

\bibitem[{\citenamefont{Blanco et~al.}(2018)\citenamefont{Blanco, Harding, and
  Hooper}}]{Blanco:2017sbc}
\bibinfo{author}{\bibfnamefont{C.}~\bibnamefont{Blanco}},
  \bibinfo{author}{\bibfnamefont{J.~P.} \bibnamefont{Harding}},
  \bibnamefont{and} \bibinfo{author}{\bibfnamefont{D.}~\bibnamefont{Hooper}},
  \bibinfo{journal}{JCAP} \textbf{\bibinfo{volume}{04}}, \bibinfo{pages}{060}
  (\bibinfo{year}{2018}), \eprint{1712.02805}.

\bibitem[{\citenamefont{Archambault et~al.}(2017)}]{Archambault:2017wyh}
\bibinfo{author}{\bibfnamefont{S.}~\bibnamefont{Archambault}}
  \bibnamefont{et~al.} (\bibinfo{collaboration}{VERITAS}),
  \bibinfo{journal}{Phys. Rev. D} \textbf{\bibinfo{volume}{95}},
  \bibinfo{pages}{082001} (\bibinfo{year}{2017}), \eprint{1703.04937}.

\bibitem[{\citenamefont{Lisanti et~al.}(2018)\citenamefont{Lisanti,
  Mishra-Sharma, Rodd, and Safdi}}]{Lisanti:2017qlb}
\bibinfo{author}{\bibfnamefont{M.}~\bibnamefont{Lisanti}},
  \bibinfo{author}{\bibfnamefont{S.}~\bibnamefont{Mishra-Sharma}},
  \bibinfo{author}{\bibfnamefont{N.~L.} \bibnamefont{Rodd}}, \bibnamefont{and}
  \bibinfo{author}{\bibfnamefont{B.~R.} \bibnamefont{Safdi}},
  \bibinfo{journal}{Phys. Rev. Lett.} \textbf{\bibinfo{volume}{120}},
  \bibinfo{pages}{101101} (\bibinfo{year}{2018}), \eprint{1708.09385}.

\bibitem[{\citenamefont{Ahnen et~al.}(2018)}]{Ahnen:2017pqx}
\bibinfo{author}{\bibfnamefont{M.}~\bibnamefont{Ahnen}} \bibnamefont{et~al.}
  (\bibinfo{collaboration}{MAGIC}), \bibinfo{journal}{JCAP}
  \textbf{\bibinfo{volume}{03}}, \bibinfo{pages}{009} (\bibinfo{year}{2018}),
  \eprint{1712.03095}.

\bibitem[{\citenamefont{Abeysekara et~al.}(2018)}]{Abeysekara:2017jxs}
\bibinfo{author}{\bibfnamefont{A.}~\bibnamefont{Abeysekara}}
  \bibnamefont{et~al.} (\bibinfo{collaboration}{HAWC}), \bibinfo{journal}{JCAP}
  \textbf{\bibinfo{volume}{02}}, \bibinfo{pages}{049} (\bibinfo{year}{2018}),
  \eprint{1710.10288}.

\bibitem[{\citenamefont{Blanco and Hooper}(2019)}]{Blanco:2018esa}
\bibinfo{author}{\bibfnamefont{C.}~\bibnamefont{Blanco}} \bibnamefont{and}
  \bibinfo{author}{\bibfnamefont{D.}~\bibnamefont{Hooper}},
  \bibinfo{journal}{JCAP} \textbf{\bibinfo{volume}{03}}, \bibinfo{pages}{019}
  (\bibinfo{year}{2019}), \eprint{1811.05988}.

\bibitem[{\citenamefont{Abdalla et~al.}(2018)}]{Abdalla:2018mve}
\bibinfo{author}{\bibfnamefont{H.}~\bibnamefont{Abdalla}} \bibnamefont{et~al.}
  (\bibinfo{collaboration}{HESS}), \bibinfo{journal}{JCAP}
  \textbf{\bibinfo{volume}{1811}}, \bibinfo{pages}{037} (\bibinfo{year}{2018}),
  \eprint{1810.00995}.

\bibitem[{\citenamefont{Abdallah et~al.}(2018)}]{Abdallah:2018qtu}
\bibinfo{author}{\bibfnamefont{H.}~\bibnamefont{Abdallah}} \bibnamefont{et~al.}
  (\bibinfo{collaboration}{HESS}), \bibinfo{journal}{Phys. Rev. Lett.}
  \textbf{\bibinfo{volume}{120}}, \bibinfo{pages}{201101}
  (\bibinfo{year}{2018}), \eprint{1805.05741}.

\bibitem[{\citenamefont{Blanco et~al.}(2019{\natexlab{b}})\citenamefont{Blanco,
  Delos, Erickcek, and Hooper}}]{Blanco:2019eij}
\bibinfo{author}{\bibfnamefont{C.}~\bibnamefont{Blanco}},
  \bibinfo{author}{\bibfnamefont{M.~S.} \bibnamefont{Delos}},
  \bibinfo{author}{\bibfnamefont{A.~L.} \bibnamefont{Erickcek}},
  \bibnamefont{and} \bibinfo{author}{\bibfnamefont{D.}~\bibnamefont{Hooper}},
  \bibinfo{journal}{Phys. Rev. D} \textbf{\bibinfo{volume}{100}},
  \bibinfo{pages}{103010} (\bibinfo{year}{2019}{\natexlab{b}}),
  \eprint{1906.00010}.

\bibitem[{\citenamefont{Aguilar et~al.}(2016)}]{Aguilar:2016kjl}
\bibinfo{author}{\bibfnamefont{M.}~\bibnamefont{Aguilar}} \bibnamefont{et~al.}
  (\bibinfo{collaboration}{AMS}), \bibinfo{journal}{Phys. Rev. Lett.}
  \textbf{\bibinfo{volume}{117}}, \bibinfo{pages}{091103}
  (\bibinfo{year}{2016}).

\bibitem[{\citenamefont{Cuoco et~al.}(2017{\natexlab{a}})\citenamefont{Cuoco,
  Krämer, and Korsmeier}}]{Cuoco:2016eej}
\bibinfo{author}{\bibfnamefont{A.}~\bibnamefont{Cuoco}},
  \bibinfo{author}{\bibfnamefont{M.}~\bibnamefont{Krämer}}, \bibnamefont{and}
  \bibinfo{author}{\bibfnamefont{M.}~\bibnamefont{Korsmeier}},
  \bibinfo{journal}{Phys. Rev. Lett.} \textbf{\bibinfo{volume}{118}},
  \bibinfo{pages}{191102} (\bibinfo{year}{2017}{\natexlab{a}}),
  \eprint{1610.03071}.

\bibitem[{\citenamefont{Cui et~al.}(2017{\natexlab{b}})\citenamefont{Cui, Yuan,
  Tsai, and Fan}}]{Cui:2016ppb}
\bibinfo{author}{\bibfnamefont{M.-Y.} \bibnamefont{Cui}},
  \bibinfo{author}{\bibfnamefont{Q.}~\bibnamefont{Yuan}},
  \bibinfo{author}{\bibfnamefont{Y.-L.~S.} \bibnamefont{Tsai}},
  \bibnamefont{and} \bibinfo{author}{\bibfnamefont{Y.-Z.} \bibnamefont{Fan}},
  \bibinfo{journal}{Phys. Rev. Lett.} \textbf{\bibinfo{volume}{118}},
  \bibinfo{pages}{191101} (\bibinfo{year}{2017}{\natexlab{b}}),
  \eprint{1610.03840}.

\bibitem[{\citenamefont{Cuoco et~al.}(2017{\natexlab{b}})\citenamefont{Cuoco,
  Heisig, Korsmeier, and Krämer}}]{Cuoco:2017rxb}
\bibinfo{author}{\bibfnamefont{A.}~\bibnamefont{Cuoco}},
  \bibinfo{author}{\bibfnamefont{J.}~\bibnamefont{Heisig}},
  \bibinfo{author}{\bibfnamefont{M.}~\bibnamefont{Korsmeier}},
  \bibnamefont{and} \bibinfo{author}{\bibfnamefont{M.}~\bibnamefont{Krämer}},
  \bibinfo{journal}{JCAP} \textbf{\bibinfo{volume}{10}}, \bibinfo{pages}{053}
  (\bibinfo{year}{2017}{\natexlab{b}}), \eprint{1704.08258}.

\bibitem[{\citenamefont{Cuoco et~al.}(2018)\citenamefont{Cuoco, Heisig,
  Korsmeier, and Krämer}}]{Cuoco:2017iax}
\bibinfo{author}{\bibfnamefont{A.}~\bibnamefont{Cuoco}},
  \bibinfo{author}{\bibfnamefont{J.}~\bibnamefont{Heisig}},
  \bibinfo{author}{\bibfnamefont{M.}~\bibnamefont{Korsmeier}},
  \bibnamefont{and} \bibinfo{author}{\bibfnamefont{M.}~\bibnamefont{Krämer}},
  \bibinfo{journal}{JCAP} \textbf{\bibinfo{volume}{04}}, \bibinfo{pages}{004}
  (\bibinfo{year}{2018}), \eprint{1711.05274}.

\bibitem[{\citenamefont{Cui et~al.}(2018)\citenamefont{Cui, Pan, Yuan, Fan, and
  Zong}}]{Cui:2018klo}
\bibinfo{author}{\bibfnamefont{M.-Y.} \bibnamefont{Cui}},
  \bibinfo{author}{\bibfnamefont{X.}~\bibnamefont{Pan}},
  \bibinfo{author}{\bibfnamefont{Q.}~\bibnamefont{Yuan}},
  \bibinfo{author}{\bibfnamefont{Y.-Z.} \bibnamefont{Fan}}, \bibnamefont{and}
  \bibinfo{author}{\bibfnamefont{H.-S.} \bibnamefont{Zong}},
  \bibinfo{journal}{JCAP} \textbf{\bibinfo{volume}{06}}, \bibinfo{pages}{024}
  (\bibinfo{year}{2018}), \eprint{1803.02163}.

\bibitem[{\citenamefont{Cholis et~al.}(2009)\citenamefont{Cholis, Goodenough,
  Hooper, Simet, and Weiner}}]{Cholis:2008hb}
\bibinfo{author}{\bibfnamefont{I.}~\bibnamefont{Cholis}},
  \bibinfo{author}{\bibfnamefont{L.}~\bibnamefont{Goodenough}},
  \bibinfo{author}{\bibfnamefont{D.}~\bibnamefont{Hooper}},
  \bibinfo{author}{\bibfnamefont{M.}~\bibnamefont{Simet}}, \bibnamefont{and}
  \bibinfo{author}{\bibfnamefont{N.}~\bibnamefont{Weiner}},
  \bibinfo{journal}{Phys. Rev. D} \textbf{\bibinfo{volume}{80}},
  \bibinfo{pages}{123511} (\bibinfo{year}{2009}), \eprint{0809.1683}.

\bibitem[{\citenamefont{Bergstrom et~al.}(2008)\citenamefont{Bergstrom,
  Bringmann, and Edsjo}}]{Bergstrom:2008gr}
\bibinfo{author}{\bibfnamefont{L.}~\bibnamefont{Bergstrom}},
  \bibinfo{author}{\bibfnamefont{T.}~\bibnamefont{Bringmann}},
  \bibnamefont{and} \bibinfo{author}{\bibfnamefont{J.}~\bibnamefont{Edsjo}},
  \bibinfo{journal}{Phys. Rev. D} \textbf{\bibinfo{volume}{78}},
  \bibinfo{pages}{103520} (\bibinfo{year}{2008}), \eprint{0808.3725}.

\bibitem[{\citenamefont{Harnik and Kribs}(2009)}]{Harnik:2008uu}
\bibinfo{author}{\bibfnamefont{R.}~\bibnamefont{Harnik}} \bibnamefont{and}
  \bibinfo{author}{\bibfnamefont{G.~D.} \bibnamefont{Kribs}},
  \bibinfo{journal}{Phys. Rev. D} \textbf{\bibinfo{volume}{79}},
  \bibinfo{pages}{095007} (\bibinfo{year}{2009}), \eprint{0810.5557}.

\bibitem[{\citenamefont{Cirelli and Strumia}(2008)}]{Cirelli:2008jk}
\bibinfo{author}{\bibfnamefont{M.}~\bibnamefont{Cirelli}} \bibnamefont{and}
  \bibinfo{author}{\bibfnamefont{A.}~\bibnamefont{Strumia}},
  \bibinfo{journal}{PoS} \textbf{\bibinfo{volume}{IDM2008}},
  \bibinfo{pages}{089} (\bibinfo{year}{2008}), \eprint{0808.3867}.

\bibitem[{\citenamefont{Hooper et~al.}(2009)\citenamefont{Hooper, Blasi, and
  Serpico}}]{Hooper:2008kg}
\bibinfo{author}{\bibfnamefont{D.}~\bibnamefont{Hooper}},
  \bibinfo{author}{\bibfnamefont{P.}~\bibnamefont{Blasi}}, \bibnamefont{and}
  \bibinfo{author}{\bibfnamefont{P.~D.} \bibnamefont{Serpico}},
  \bibinfo{journal}{JCAP} \textbf{\bibinfo{volume}{01}}, \bibinfo{pages}{025}
  (\bibinfo{year}{2009}), \eprint{0810.1527}.

\bibitem[{\citenamefont{Silk et~al.}(1985)\citenamefont{Silk, Olive, and
  Srednicki}}]{Silk:1985ax}
\bibinfo{author}{\bibfnamefont{J.}~\bibnamefont{Silk}},
  \bibinfo{author}{\bibfnamefont{K.~A.} \bibnamefont{Olive}}, \bibnamefont{and}
  \bibinfo{author}{\bibfnamefont{M.}~\bibnamefont{Srednicki}},
  \bibinfo{journal}{Phys. Rev. Lett.} \textbf{\bibinfo{volume}{55}},
  \bibinfo{pages}{257} (\bibinfo{year}{1985}).

\bibitem[{\citenamefont{Hagelin et~al.}(1986)\citenamefont{Hagelin, Ng, and
  Olive}}]{Hagelin:1986gv}
\bibinfo{author}{\bibfnamefont{J.~S.} \bibnamefont{Hagelin}},
  \bibinfo{author}{\bibfnamefont{K.}~\bibnamefont{Ng}}, \bibnamefont{and}
  \bibinfo{author}{\bibfnamefont{K.~A.} \bibnamefont{Olive}},
  \bibinfo{journal}{Phys. Lett. B} \textbf{\bibinfo{volume}{180}},
  \bibinfo{pages}{375} (\bibinfo{year}{1986}).

\bibitem[{\citenamefont{Freese}(1986)}]{Freese:1985qw}
\bibinfo{author}{\bibfnamefont{K.}~\bibnamefont{Freese}},
  \bibinfo{journal}{Phys. Lett. B} \textbf{\bibinfo{volume}{167}},
  \bibinfo{pages}{295} (\bibinfo{year}{1986}).

\bibitem[{\citenamefont{Krauss et~al.}(1986)\citenamefont{Krauss, Srednicki,
  and Wilczek}}]{Krauss:1985aaa}
\bibinfo{author}{\bibfnamefont{L.~M.} \bibnamefont{Krauss}},
  \bibinfo{author}{\bibfnamefont{M.}~\bibnamefont{Srednicki}},
  \bibnamefont{and} \bibinfo{author}{\bibfnamefont{F.}~\bibnamefont{Wilczek}},
  \bibinfo{journal}{Phys. Rev. D} \textbf{\bibinfo{volume}{33}},
  \bibinfo{pages}{2079} (\bibinfo{year}{1986}).

\bibitem[{\citenamefont{Gaisser et~al.}(1986)\citenamefont{Gaisser, Steigman,
  and Tilav}}]{Gaisser:1986ha}
\bibinfo{author}{\bibfnamefont{T.}~\bibnamefont{Gaisser}},
  \bibinfo{author}{\bibfnamefont{G.}~\bibnamefont{Steigman}}, \bibnamefont{and}
  \bibinfo{author}{\bibfnamefont{S.}~\bibnamefont{Tilav}},
  \bibinfo{journal}{Phys. Rev. D} \textbf{\bibinfo{volume}{34}},
  \bibinfo{pages}{2206} (\bibinfo{year}{1986}).

\bibitem[{\citenamefont{Desai et~al.}(2004)}]{Desai:2004pq}
\bibinfo{author}{\bibfnamefont{S.}~\bibnamefont{Desai}} \bibnamefont{et~al.}
  (\bibinfo{collaboration}{Super-Kamiokande}), \bibinfo{journal}{Phys. Rev. D}
  \textbf{\bibinfo{volume}{70}}, \bibinfo{pages}{083523}
  (\bibinfo{year}{2004}), \bibinfo{note}{[Erratum: Phys. Rev. D {\bf 70},
  109901 (2004)]}, \eprint{hep-ex/0404025}.

\bibitem[{\citenamefont{Palomares-Ruiz}(2008)}]{PalomaresRuiz:2007ry}
\bibinfo{author}{\bibfnamefont{S.}~\bibnamefont{Palomares-Ruiz}},
  \bibinfo{journal}{Phys. Lett. B} \textbf{\bibinfo{volume}{665}},
  \bibinfo{pages}{50} (\bibinfo{year}{2008}), \eprint{0712.1937}.

\bibitem[{\citenamefont{Murase and Beacom}(2012)}]{Murase:2012xs}
\bibinfo{author}{\bibfnamefont{K.}~\bibnamefont{Murase}} \bibnamefont{and}
  \bibinfo{author}{\bibfnamefont{J.~F.} \bibnamefont{Beacom}},
  \bibinfo{journal}{JCAP} \textbf{\bibinfo{volume}{10}}, \bibinfo{pages}{043}
  (\bibinfo{year}{2012}), \eprint{1206.2595}.

\bibitem[{\citenamefont{Aartsen et~al.}(2017{\natexlab{a}})}]{Aartsen:2016zhm}
\bibinfo{author}{\bibfnamefont{M.}~\bibnamefont{Aartsen}} \bibnamefont{et~al.}
  (\bibinfo{collaboration}{IceCube}), \bibinfo{journal}{Eur. Phys. J. C}
  \textbf{\bibinfo{volume}{77}}, \bibinfo{pages}{146}
  (\bibinfo{year}{2017}{\natexlab{a}}), \bibinfo{note}{[Erratum: Eur. Phys. J.
  C {\bf 79}, 214 (2019)]}, \eprint{1612.05949}.

\bibitem[{\citenamefont{Aartsen et~al.}(2017{\natexlab{b}})}]{Aartsen:2016fep}
\bibinfo{author}{\bibfnamefont{M.}~\bibnamefont{Aartsen}} \bibnamefont{et~al.}
  (\bibinfo{collaboration}{IceCube}), \bibinfo{journal}{Eur. Phys. J. C}
  \textbf{\bibinfo{volume}{77}}, \bibinfo{pages}{82}
  (\bibinfo{year}{2017}{\natexlab{b}}), \eprint{1609.01492}.

\bibitem[{\citenamefont{Aartsen et~al.}(2016)}]{Aartsen:2016pfc}
\bibinfo{author}{\bibfnamefont{M.}~\bibnamefont{Aartsen}} \bibnamefont{et~al.}
  (\bibinfo{collaboration}{IceCube}), \bibinfo{journal}{Eur. Phys. J. C}
  \textbf{\bibinfo{volume}{76}}, \bibinfo{pages}{531} (\bibinfo{year}{2016}),
  \eprint{1606.00209}.

\bibitem[{\citenamefont{Aartsen et~al.}(2017{\natexlab{c}})}]{Aartsen:2017ulx}
\bibinfo{author}{\bibfnamefont{M.}~\bibnamefont{Aartsen}} \bibnamefont{et~al.}
  (\bibinfo{collaboration}{IceCube}), \bibinfo{journal}{Eur. Phys. J. C}
  \textbf{\bibinfo{volume}{77}}, \bibinfo{pages}{627}
  (\bibinfo{year}{2017}{\natexlab{c}}), \eprint{1705.08103}.

\bibitem[{\citenamefont{Ellis et~al.}(1988{\natexlab{b}})\citenamefont{Ellis,
  Flores, Freese, Ritz, Seckel, and Silk}}]{Ellis:1988qp}
\bibinfo{author}{\bibfnamefont{J.~R.} \bibnamefont{Ellis}},
  \bibinfo{author}{\bibfnamefont{R.}~\bibnamefont{Flores}},
  \bibinfo{author}{\bibfnamefont{K.}~\bibnamefont{Freese}},
  \bibinfo{author}{\bibfnamefont{S.}~\bibnamefont{Ritz}},
  \bibinfo{author}{\bibfnamefont{D.}~\bibnamefont{Seckel}}, \bibnamefont{and}
  \bibinfo{author}{\bibfnamefont{J.}~\bibnamefont{Silk}},
  \bibinfo{journal}{Phys. Lett. B} \textbf{\bibinfo{volume}{214}},
  \bibinfo{pages}{403} (\bibinfo{year}{1988}{\natexlab{b}}).

\bibitem[{\citenamefont{Donato et~al.}(2000)\citenamefont{Donato, Fornengo, and
  Salati}}]{Donato:1999gy}
\bibinfo{author}{\bibfnamefont{F.}~\bibnamefont{Donato}},
  \bibinfo{author}{\bibfnamefont{N.}~\bibnamefont{Fornengo}}, \bibnamefont{and}
  \bibinfo{author}{\bibfnamefont{P.}~\bibnamefont{Salati}},
  \bibinfo{journal}{Phys. Rev. D} \textbf{\bibinfo{volume}{62}},
  \bibinfo{pages}{043003} (\bibinfo{year}{2000}), \eprint{hep-ph/9904481}.

\bibitem[{\citenamefont{Fuke et~al.}(2005)}]{Fuke:2005it}
\bibinfo{author}{\bibfnamefont{H.}~\bibnamefont{Fuke}} \bibnamefont{et~al.},
  \bibinfo{journal}{Phys. Rev. Lett.} \textbf{\bibinfo{volume}{95}},
  \bibinfo{pages}{081101} (\bibinfo{year}{2005}), \eprint{astro-ph/0504361}.

\bibitem[{\citenamefont{Donato et~al.}(2008)\citenamefont{Donato, Fornengo, and
  Maurin}}]{Donato:2008yx}
\bibinfo{author}{\bibfnamefont{F.}~\bibnamefont{Donato}},
  \bibinfo{author}{\bibfnamefont{N.}~\bibnamefont{Fornengo}}, \bibnamefont{and}
  \bibinfo{author}{\bibfnamefont{D.}~\bibnamefont{Maurin}},
  \bibinfo{journal}{Phys. Rev. D} \textbf{\bibinfo{volume}{78}},
  \bibinfo{pages}{043506} (\bibinfo{year}{2008}), \eprint{0803.2640}.

\bibitem[{\citenamefont{Ibarra and Wild}(2013)}]{Ibarra:2013qt}
\bibinfo{author}{\bibfnamefont{A.}~\bibnamefont{Ibarra}} \bibnamefont{and}
  \bibinfo{author}{\bibfnamefont{S.}~\bibnamefont{Wild}},
  \bibinfo{journal}{Phys. Rev. D} \textbf{\bibinfo{volume}{88}},
  \bibinfo{pages}{023014} (\bibinfo{year}{2013}), \eprint{1301.3820}.

\bibitem[{\citenamefont{Hryczuk et~al.}(2014)\citenamefont{Hryczuk, Cholis,
  Iengo, Tavakoli, and Ullio}}]{Hryczuk:2014hpa}
\bibinfo{author}{\bibfnamefont{A.}~\bibnamefont{Hryczuk}},
  \bibinfo{author}{\bibfnamefont{I.}~\bibnamefont{Cholis}},
  \bibinfo{author}{\bibfnamefont{R.}~\bibnamefont{Iengo}},
  \bibinfo{author}{\bibfnamefont{M.}~\bibnamefont{Tavakoli}}, \bibnamefont{and}
  \bibinfo{author}{\bibfnamefont{P.}~\bibnamefont{Ullio}},
  \bibinfo{journal}{JCAP} \textbf{\bibinfo{volume}{07}}, \bibinfo{pages}{031}
  (\bibinfo{year}{2014}), \eprint{1401.6212}.

\bibitem[{\citenamefont{Carlson et~al.}(2014)\citenamefont{Carlson, Coogan,
  Linden, Profumo, Ibarra, and Wild}}]{Carlson:2014ssa}
\bibinfo{author}{\bibfnamefont{E.}~\bibnamefont{Carlson}},
  \bibinfo{author}{\bibfnamefont{A.}~\bibnamefont{Coogan}},
  \bibinfo{author}{\bibfnamefont{T.}~\bibnamefont{Linden}},
  \bibinfo{author}{\bibfnamefont{S.}~\bibnamefont{Profumo}},
  \bibinfo{author}{\bibfnamefont{A.}~\bibnamefont{Ibarra}}, \bibnamefont{and}
  \bibinfo{author}{\bibfnamefont{S.}~\bibnamefont{Wild}},
  \bibinfo{journal}{Phys. Rev. D} \textbf{\bibinfo{volume}{89}},
  \bibinfo{pages}{076005} (\bibinfo{year}{2014}), \eprint{1401.2461}.

\bibitem[{\citenamefont{Aramaki et~al.}(2016)\citenamefont{Aramaki, Hailey,
  Boggs, von Doetinchem, Fuke, Mognet, Ong, Perez, and
  Zweerink}}]{Aramaki:2015laa}
\bibinfo{author}{\bibfnamefont{T.}~\bibnamefont{Aramaki}},
  \bibinfo{author}{\bibfnamefont{C.}~\bibnamefont{Hailey}},
  \bibinfo{author}{\bibfnamefont{S.}~\bibnamefont{Boggs}},
  \bibinfo{author}{\bibfnamefont{P.}~\bibnamefont{von Doetinchem}},
  \bibinfo{author}{\bibfnamefont{H.}~\bibnamefont{Fuke}},
  \bibinfo{author}{\bibfnamefont{S.}~\bibnamefont{Mognet}},
  \bibinfo{author}{\bibfnamefont{R.}~\bibnamefont{Ong}},
  \bibinfo{author}{\bibfnamefont{K.}~\bibnamefont{Perez}}, \bibnamefont{and}
  \bibinfo{author}{\bibfnamefont{J.}~\bibnamefont{Zweerink}}
  (\bibinfo{collaboration}{GAPS}), \bibinfo{journal}{Astropart. Phys.}
  \textbf{\bibinfo{volume}{74}}, \bibinfo{pages}{6} (\bibinfo{year}{2016}),
  \eprint{1506.02513}.

\bibitem[{\citenamefont{Korsmeier et~al.}(2018)\citenamefont{Korsmeier, Donato,
  and Fornengo}}]{Korsmeier:2017xzj}
\bibinfo{author}{\bibfnamefont{M.}~\bibnamefont{Korsmeier}},
  \bibinfo{author}{\bibfnamefont{F.}~\bibnamefont{Donato}}, \bibnamefont{and}
  \bibinfo{author}{\bibfnamefont{N.}~\bibnamefont{Fornengo}},
  \bibinfo{journal}{Phys. Rev. D} \textbf{\bibinfo{volume}{97}},
  \bibinfo{pages}{103011} (\bibinfo{year}{2018}), \eprint{1711.08465}.

\bibitem[{\citenamefont{Reinert and Winkler}(2018)}]{Reinert:2017aga}
\bibinfo{author}{\bibfnamefont{A.}~\bibnamefont{Reinert}} \bibnamefont{and}
  \bibinfo{author}{\bibfnamefont{M.~W.} \bibnamefont{Winkler}},
  \bibinfo{journal}{JCAP} \textbf{\bibinfo{volume}{01}}, \bibinfo{pages}{055}
  (\bibinfo{year}{2018}), \eprint{1712.00002}.

\bibitem[{\citenamefont{Arg\"uelles et~al.}(2021)\citenamefont{Arg\"uelles,
  Diaz, Kheirandish, Olivares-Del-Campo, Safa, and
  Vincent}}]{Arguelles:2019ouk}
\bibinfo{author}{\bibfnamefont{C.~A.} \bibnamefont{Arg\"uelles}},
  \bibinfo{author}{\bibfnamefont{A.}~\bibnamefont{Diaz}},
  \bibinfo{author}{\bibfnamefont{A.}~\bibnamefont{Kheirandish}},
  \bibinfo{author}{\bibfnamefont{A.}~\bibnamefont{Olivares-Del-Campo}},
  \bibinfo{author}{\bibfnamefont{I.}~\bibnamefont{Safa}}, \bibnamefont{and}
  \bibinfo{author}{\bibfnamefont{A.~C.} \bibnamefont{Vincent}},
  \bibinfo{journal}{Rev. Mod. Phys.} \textbf{\bibinfo{volume}{93}},
  \bibinfo{pages}{035007} (\bibinfo{year}{2021}), \eprint{1912.09486}.

\bibitem[{\citenamefont{Hisano et~al.}(2005)\citenamefont{Hisano, Matsumoto,
  Nojiri, and Saito}}]{Hisano:2004ds}
\bibinfo{author}{\bibfnamefont{J.}~\bibnamefont{Hisano}},
  \bibinfo{author}{\bibfnamefont{S.}~\bibnamefont{Matsumoto}},
  \bibinfo{author}{\bibfnamefont{M.~M.} \bibnamefont{Nojiri}},
  \bibnamefont{and} \bibinfo{author}{\bibfnamefont{O.}~\bibnamefont{Saito}},
  \bibinfo{journal}{Phys. Rev. D} \textbf{\bibinfo{volume}{71}},
  \bibinfo{pages}{063528} (\bibinfo{year}{2005}), \eprint{hep-ph/0412403}.

\bibitem[{\citenamefont{Gelmini et~al.}(2006)\citenamefont{Gelmini, Gondolo,
  Soldatenko, and Yaguna}}]{Gelmini:2006pq}
\bibinfo{author}{\bibfnamefont{G.}~\bibnamefont{Gelmini}},
  \bibinfo{author}{\bibfnamefont{P.}~\bibnamefont{Gondolo}},
  \bibinfo{author}{\bibfnamefont{A.}~\bibnamefont{Soldatenko}},
  \bibnamefont{and} \bibinfo{author}{\bibfnamefont{C.~E.}
  \bibnamefont{Yaguna}}, \bibinfo{journal}{Phys. Rev. D}
  \textbf{\bibinfo{volume}{74}}, \bibinfo{pages}{083514}
  (\bibinfo{year}{2006}), \eprint{hep-ph/0605016}.

\bibitem[{\citenamefont{Gelmini and Gondolo}(2006)}]{Gelmini:2006pw}
\bibinfo{author}{\bibfnamefont{G.~B.} \bibnamefont{Gelmini}} \bibnamefont{and}
  \bibinfo{author}{\bibfnamefont{P.}~\bibnamefont{Gondolo}},
  \bibinfo{journal}{Phys. Rev. D} \textbf{\bibinfo{volume}{74}},
  \bibinfo{pages}{023510} (\bibinfo{year}{2006}), \eprint{hep-ph/0602230}.

\bibitem[{\citenamefont{Merle et~al.}(2014)\citenamefont{Merle, Niro, and
  Schmidt}}]{Merle:2013wta}
\bibinfo{author}{\bibfnamefont{A.}~\bibnamefont{Merle}},
  \bibinfo{author}{\bibfnamefont{V.}~\bibnamefont{Niro}}, \bibnamefont{and}
  \bibinfo{author}{\bibfnamefont{D.}~\bibnamefont{Schmidt}},
  \bibinfo{journal}{JCAP} \textbf{\bibinfo{volume}{03}}, \bibinfo{pages}{028}
  (\bibinfo{year}{2014}), \eprint{1306.3996}.

\bibitem[{\citenamefont{König et~al.}(2016)\citenamefont{König, Merle, and
  Totzauer}}]{Konig:2016dzg}
\bibinfo{author}{\bibfnamefont{J.}~\bibnamefont{König}},
  \bibinfo{author}{\bibfnamefont{A.}~\bibnamefont{Merle}}, \bibnamefont{and}
  \bibinfo{author}{\bibfnamefont{M.}~\bibnamefont{Totzauer}},
  \bibinfo{journal}{JCAP} \textbf{\bibinfo{volume}{11}}, \bibinfo{pages}{038}
  (\bibinfo{year}{2016}), \eprint{1609.01289}.

\bibitem[{\citenamefont{Graesser et~al.}(2011)\citenamefont{Graesser,
  Shoemaker, and Vecchi}}]{Graesser:2011wi}
\bibinfo{author}{\bibfnamefont{M.~L.} \bibnamefont{Graesser}},
  \bibinfo{author}{\bibfnamefont{I.~M.} \bibnamefont{Shoemaker}},
  \bibnamefont{and} \bibinfo{author}{\bibfnamefont{L.}~\bibnamefont{Vecchi}},
  \bibinfo{journal}{JHEP} \textbf{\bibinfo{volume}{10}}, \bibinfo{pages}{110}
  (\bibinfo{year}{2011}), \eprint{1103.2771}.

\bibitem[{\citenamefont{Lin et~al.}(2012)\citenamefont{Lin, Yu, and
  Zurek}}]{Lin:2011gj}
\bibinfo{author}{\bibfnamefont{T.}~\bibnamefont{Lin}},
  \bibinfo{author}{\bibfnamefont{H.-B.} \bibnamefont{Yu}}, \bibnamefont{and}
  \bibinfo{author}{\bibfnamefont{K.~M.} \bibnamefont{Zurek}},
  \bibinfo{journal}{Phys. Rev. D} \textbf{\bibinfo{volume}{85}},
  \bibinfo{pages}{063503} (\bibinfo{year}{2012}), \eprint{1111.0293}.

\bibitem[{\citenamefont{Iminniyaz et~al.}(2011)\citenamefont{Iminniyaz, Drees,
  and Chen}}]{Iminniyaz:2011yp}
\bibinfo{author}{\bibfnamefont{H.}~\bibnamefont{Iminniyaz}},
  \bibinfo{author}{\bibfnamefont{M.}~\bibnamefont{Drees}}, \bibnamefont{and}
  \bibinfo{author}{\bibfnamefont{X.}~\bibnamefont{Chen}},
  \bibinfo{journal}{JCAP} \textbf{\bibinfo{volume}{07}}, \bibinfo{pages}{003}
  (\bibinfo{year}{2011}), \eprint{1104.5548}.

\bibitem[{\citenamefont{Griest and Seckel}(1991)}]{Griest:1990kh}
\bibinfo{author}{\bibfnamefont{K.}~\bibnamefont{Griest}} \bibnamefont{and}
  \bibinfo{author}{\bibfnamefont{D.}~\bibnamefont{Seckel}},
  \bibinfo{journal}{Phys. Rev. D} \textbf{\bibinfo{volume}{43}},
  \bibinfo{pages}{3191} (\bibinfo{year}{1991}).

\bibitem[{\citenamefont{Edsjo and Gondolo}(1997)}]{Edsjo:1997bg}
\bibinfo{author}{\bibfnamefont{J.}~\bibnamefont{Edsjo}} \bibnamefont{and}
  \bibinfo{author}{\bibfnamefont{P.}~\bibnamefont{Gondolo}},
  \bibinfo{journal}{Phys. Rev. D} \textbf{\bibinfo{volume}{56}},
  \bibinfo{pages}{1879} (\bibinfo{year}{1997}), \eprint{hep-ph/9704361}.

\bibitem[{\citenamefont{Ellis et~al.}(1998)\citenamefont{Ellis, Falk, and
  Olive}}]{Ellis:1998kh}
\bibinfo{author}{\bibfnamefont{J.~R.} \bibnamefont{Ellis}},
  \bibinfo{author}{\bibfnamefont{T.}~\bibnamefont{Falk}}, \bibnamefont{and}
  \bibinfo{author}{\bibfnamefont{K.~A.} \bibnamefont{Olive}},
  \bibinfo{journal}{Phys. Lett. B} \textbf{\bibinfo{volume}{444}},
  \bibinfo{pages}{367} (\bibinfo{year}{1998}), \eprint{hep-ph/9810360}.

\bibitem[{\citenamefont{Fornengo et~al.}(2003)\citenamefont{Fornengo, Riotto,
  and Scopel}}]{Fornengo:2002db}
\bibinfo{author}{\bibfnamefont{N.}~\bibnamefont{Fornengo}},
  \bibinfo{author}{\bibfnamefont{A.}~\bibnamefont{Riotto}}, \bibnamefont{and}
  \bibinfo{author}{\bibfnamefont{S.}~\bibnamefont{Scopel}},
  \bibinfo{journal}{Phys. Rev. D} \textbf{\bibinfo{volume}{67}},
  \bibinfo{pages}{023514} (\bibinfo{year}{2003}), \eprint{hep-ph/0208072}.

\bibitem[{\citenamefont{Boeckel and Schaffner-Bielich}(2010)}]{Boeckel:2009ej}
\bibinfo{author}{\bibfnamefont{T.}~\bibnamefont{Boeckel}} \bibnamefont{and}
  \bibinfo{author}{\bibfnamefont{J.}~\bibnamefont{Schaffner-Bielich}},
  \bibinfo{journal}{Phys. Rev. Lett.} \textbf{\bibinfo{volume}{105}},
  \bibinfo{pages}{041301} (\bibinfo{year}{2010}), \bibinfo{note}{[Erratum:
  Phys. Rev. Lett. {\bf 106}, 069901 (2011)]}, \eprint{0906.4520}.

\bibitem[{\citenamefont{Boeckel and Schaffner-Bielich}(2012)}]{Boeckel:2011yj}
\bibinfo{author}{\bibfnamefont{T.}~\bibnamefont{Boeckel}} \bibnamefont{and}
  \bibinfo{author}{\bibfnamefont{J.}~\bibnamefont{Schaffner-Bielich}},
  \bibinfo{journal}{Phys. Rev. D} \textbf{\bibinfo{volume}{85}},
  \bibinfo{pages}{103506} (\bibinfo{year}{2012}), \eprint{1105.0832}.

\bibitem[{\citenamefont{Kane et~al.}(2015)\citenamefont{Kane, Sinha, and
  Watson}}]{Kane:2015jia}
\bibinfo{author}{\bibfnamefont{G.}~\bibnamefont{Kane}},
  \bibinfo{author}{\bibfnamefont{K.}~\bibnamefont{Sinha}}, \bibnamefont{and}
  \bibinfo{author}{\bibfnamefont{S.}~\bibnamefont{Watson}},
  \bibinfo{journal}{Int. J. Mod. Phys. D} \textbf{\bibinfo{volume}{24}},
  \bibinfo{pages}{1530022} (\bibinfo{year}{2015}), \eprint{1502.07746}.

\bibitem[{\citenamefont{Davoudiasl et~al.}(2016)\citenamefont{Davoudiasl,
  Hooper, and McDermott}}]{Davoudiasl:2015vba}
\bibinfo{author}{\bibfnamefont{H.}~\bibnamefont{Davoudiasl}},
  \bibinfo{author}{\bibfnamefont{D.}~\bibnamefont{Hooper}}, \bibnamefont{and}
  \bibinfo{author}{\bibfnamefont{S.~D.} \bibnamefont{McDermott}},
  \bibinfo{journal}{Phys. Rev. Lett.} \textbf{\bibinfo{volume}{116}},
  \bibinfo{pages}{031303} (\bibinfo{year}{2016}), \eprint{1507.08660}.

\bibitem[{\citenamefont{Berlin et~al.}(2016{\natexlab{a}})\citenamefont{Berlin,
  Hooper, and Krnjaic}}]{Berlin:2016gtr}
\bibinfo{author}{\bibfnamefont{A.}~\bibnamefont{Berlin}},
  \bibinfo{author}{\bibfnamefont{D.}~\bibnamefont{Hooper}}, \bibnamefont{and}
  \bibinfo{author}{\bibfnamefont{G.}~\bibnamefont{Krnjaic}},
  \bibinfo{journal}{Phys. Rev. D} \textbf{\bibinfo{volume}{94}},
  \bibinfo{pages}{095019} (\bibinfo{year}{2016}{\natexlab{a}}),
  \eprint{1609.02555}.

\bibitem[{\citenamefont{Berlin et~al.}(2016{\natexlab{b}})\citenamefont{Berlin,
  Hooper, and Krnjaic}}]{Berlin:2016vnh}
\bibinfo{author}{\bibfnamefont{A.}~\bibnamefont{Berlin}},
  \bibinfo{author}{\bibfnamefont{D.}~\bibnamefont{Hooper}}, \bibnamefont{and}
  \bibinfo{author}{\bibfnamefont{G.}~\bibnamefont{Krnjaic}},
  \bibinfo{journal}{Phys. Lett. B} \textbf{\bibinfo{volume}{760}},
  \bibinfo{pages}{106} (\bibinfo{year}{2016}{\natexlab{b}}),
  \eprint{1602.08490}.

\bibitem[{\citenamefont{Abdalla et~al.}(2016)}]{Abdalla:2016olq}
\bibinfo{author}{\bibfnamefont{H.}~\bibnamefont{Abdalla}} \bibnamefont{et~al.}
  (\bibinfo{collaboration}{H.E.S.S.}), \bibinfo{journal}{Phys. Rev. Lett.}
  \textbf{\bibinfo{volume}{117}}, \bibinfo{pages}{151302}
  (\bibinfo{year}{2016}), \eprint{1609.08091}.

\bibitem[{\citenamefont{Chiappo et~al.}(2019)\citenamefont{Chiappo,
  Cohen-Tanugi, Conrad, and Strigari}}]{Chiappo:2018mlt}
\bibinfo{author}{\bibfnamefont{A.}~\bibnamefont{Chiappo}},
  \bibinfo{author}{\bibfnamefont{J.}~\bibnamefont{Cohen-Tanugi}},
  \bibinfo{author}{\bibfnamefont{J.}~\bibnamefont{Conrad}}, \bibnamefont{and}
  \bibinfo{author}{\bibfnamefont{L.}~\bibnamefont{Strigari}},
  \bibinfo{journal}{Mon. Not. Roy. Astron. Soc.}
  \textbf{\bibinfo{volume}{488}}, \bibinfo{pages}{2616} (\bibinfo{year}{2019}),
  \eprint{1810.09917}.

\bibitem[{\citenamefont{Navarro et~al.}(1997)\citenamefont{Navarro, Frenk, and
  White}}]{Navarro:1996gj}
\bibinfo{author}{\bibfnamefont{J.~F.} \bibnamefont{Navarro}},
  \bibinfo{author}{\bibfnamefont{C.~S.} \bibnamefont{Frenk}}, \bibnamefont{and}
  \bibinfo{author}{\bibfnamefont{S.~D.~M.} \bibnamefont{White}},
  \bibinfo{journal}{Astrophys. J.} \textbf{\bibinfo{volume}{490}},
  \bibinfo{pages}{493} (\bibinfo{year}{1997}), \eprint{astro-ph/9611107}.

\bibitem[{\citenamefont{Navarro et~al.}(1996)\citenamefont{Navarro, Eke, and
  Frenk}}]{Navarro:1996bv}
\bibinfo{author}{\bibfnamefont{J.~F.} \bibnamefont{Navarro}},
  \bibinfo{author}{\bibfnamefont{V.~R.} \bibnamefont{Eke}}, \bibnamefont{and}
  \bibinfo{author}{\bibfnamefont{C.~S.} \bibnamefont{Frenk}},
  \bibinfo{journal}{Mon. Not. Roy. Astron. Soc.}
  \textbf{\bibinfo{volume}{283}}, \bibinfo{pages}{L72} (\bibinfo{year}{1996}),
  \eprint{astro-ph/9610187}.

\bibitem[{\citenamefont{Seidel et~al.}(2013)\citenamefont{Seidel, Simon, Tesar,
  and Poss}}]{Seidel:2013sqa}
\bibinfo{author}{\bibfnamefont{K.}~\bibnamefont{Seidel}},
  \bibinfo{author}{\bibfnamefont{F.}~\bibnamefont{Simon}},
  \bibinfo{author}{\bibfnamefont{M.}~\bibnamefont{Tesar}}, \bibnamefont{and}
  \bibinfo{author}{\bibfnamefont{S.}~\bibnamefont{Poss}},
  \bibinfo{journal}{Eur. Phys. J. C} \textbf{\bibinfo{volume}{73}},
  \bibinfo{pages}{2530} (\bibinfo{year}{2013}), \eprint{1303.3758}.

\bibitem[{\citenamefont{Horiguchi et~al.}(2013)\citenamefont{Horiguchi,
  Ishikawa, Suehara, Fujii, Sumino, Kiyo, and Yamamoto}}]{Horiguchi:2013wra}
\bibinfo{author}{\bibfnamefont{T.}~\bibnamefont{Horiguchi}},
  \bibinfo{author}{\bibfnamefont{A.}~\bibnamefont{Ishikawa}},
  \bibinfo{author}{\bibfnamefont{T.}~\bibnamefont{Suehara}},
  \bibinfo{author}{\bibfnamefont{K.}~\bibnamefont{Fujii}},
  \bibinfo{author}{\bibfnamefont{Y.}~\bibnamefont{Sumino}},
  \bibinfo{author}{\bibfnamefont{Y.}~\bibnamefont{Kiyo}}, \bibnamefont{and}
  \bibinfo{author}{\bibfnamefont{H.}~\bibnamefont{Yamamoto}}
  (\bibinfo{year}{2013}), \eprint{1310.0563}.

\bibitem[{\citenamefont{Kiyo et~al.}(2015)\citenamefont{Kiyo, Mishima, and
  Sumino}}]{Kiyo:2015ooa}
\bibinfo{author}{\bibfnamefont{Y.}~\bibnamefont{Kiyo}},
  \bibinfo{author}{\bibfnamefont{G.}~\bibnamefont{Mishima}}, \bibnamefont{and}
  \bibinfo{author}{\bibfnamefont{Y.}~\bibnamefont{Sumino}},
  \bibinfo{journal}{JHEP} \textbf{\bibinfo{volume}{11}}, \bibinfo{pages}{084}
  (\bibinfo{year}{2015}), \eprint{1506.06542}.

\bibitem[{\citenamefont{Beneke et~al.}(2015)\citenamefont{Beneke, Kiyo,
  Marquard, Penin, Piclum, and Steinhauser}}]{Beneke:2015kwa}
\bibinfo{author}{\bibfnamefont{M.}~\bibnamefont{Beneke}},
  \bibinfo{author}{\bibfnamefont{Y.}~\bibnamefont{Kiyo}},
  \bibinfo{author}{\bibfnamefont{P.}~\bibnamefont{Marquard}},
  \bibinfo{author}{\bibfnamefont{A.}~\bibnamefont{Penin}},
  \bibinfo{author}{\bibfnamefont{J.}~\bibnamefont{Piclum}}, \bibnamefont{and}
  \bibinfo{author}{\bibfnamefont{M.}~\bibnamefont{Steinhauser}},
  \bibinfo{journal}{Phys. Rev. Lett.} \textbf{\bibinfo{volume}{115}},
  \bibinfo{pages}{192001} (\bibinfo{year}{2015}), \eprint{1506.06864}.

\bibitem[{\citenamefont{Bicer et~al.}(2014)}]{Gomez-Ceballos:2013zzn}
\bibinfo{author}{\bibfnamefont{M.}~\bibnamefont{Bicer}} \bibnamefont{et~al.}
  (\bibinfo{collaboration}{TLEP Design Study Working Group}),
  \bibinfo{journal}{JHEP} \textbf{\bibinfo{volume}{01}}, \bibinfo{pages}{164}
  (\bibinfo{year}{2014}), \eprint{1308.6176}.

\bibitem[{\citenamefont{Lepage et~al.}(2014)\citenamefont{Lepage, Mackenzie,
  and Peskin}}]{Lepage:2014fla}
\bibinfo{author}{\bibfnamefont{G.~P.} \bibnamefont{Lepage}},
  \bibinfo{author}{\bibfnamefont{P.~B.} \bibnamefont{Mackenzie}},
  \bibnamefont{and} \bibinfo{author}{\bibfnamefont{M.~E.} \bibnamefont{Peskin}}
  (\bibinfo{year}{2014}), \eprint{1404.0319}.

\bibitem[{\citenamefont{Dunsky et~al.}(2019)\citenamefont{Dunsky, Hall, and
  Harigaya}}]{Dunsky:2019api}
\bibinfo{author}{\bibfnamefont{D.}~\bibnamefont{Dunsky}},
  \bibinfo{author}{\bibfnamefont{L.~J.} \bibnamefont{Hall}}, \bibnamefont{and}
  \bibinfo{author}{\bibfnamefont{K.}~\bibnamefont{Harigaya}},
  \bibinfo{journal}{JHEP} \textbf{\bibinfo{volume}{07}}, \bibinfo{pages}{016}
  (\bibinfo{year}{2019}), \eprint{1902.07726}.

\bibitem[{\citenamefont{Dunsky et~al.}(2021)\citenamefont{Dunsky, Hall, and
  Harigaya}}]{Dunsky:2020yhv}
\bibinfo{author}{\bibfnamefont{D.}~\bibnamefont{Dunsky}},
  \bibinfo{author}{\bibfnamefont{L.~J.} \bibnamefont{Hall}}, \bibnamefont{and}
  \bibinfo{author}{\bibfnamefont{K.}~\bibnamefont{Harigaya}},
  \bibinfo{journal}{JHEP} \textbf{\bibinfo{volume}{04}}, \bibinfo{pages}{052}
  (\bibinfo{year}{2021}), \eprint{2011.12302}.

\bibitem[{\citenamefont{Bernstein et~al.}(2020)}]{Bernstein:2020cpc}
\bibinfo{author}{\bibfnamefont{A.}~\bibnamefont{Bernstein}}
  \bibnamefont{et~al.}, \bibinfo{journal}{J. Phys. Conf. Ser.}
  \textbf{\bibinfo{volume}{1468}}, \bibinfo{pages}{012035}
  (\bibinfo{year}{2020}), \eprint{2001.09311}.

\bibitem[{\citenamefont{Dell'Oro et~al.}(2016)\citenamefont{Dell'Oro, Marcocci,
  Viel, and Vissani}}]{DellOro:2016tmg}
\bibinfo{author}{\bibfnamefont{S.}~\bibnamefont{Dell'Oro}},
  \bibinfo{author}{\bibfnamefont{S.}~\bibnamefont{Marcocci}},
  \bibinfo{author}{\bibfnamefont{M.}~\bibnamefont{Viel}}, \bibnamefont{and}
  \bibinfo{author}{\bibfnamefont{F.}~\bibnamefont{Vissani}},
  \bibinfo{journal}{Adv. High Energy Phys.} \textbf{\bibinfo{volume}{2016}},
  \bibinfo{pages}{2162659} (\bibinfo{year}{2016}), \eprint{1601.07512}.

\bibitem[{\citenamefont{Agostini
  et~al.}(2019{\natexlab{b}})}]{Agostini:2019hzm}
\bibinfo{author}{\bibfnamefont{M.}~\bibnamefont{Agostini}} \bibnamefont{et~al.}
  (\bibinfo{collaboration}{GERDA}), \bibinfo{journal}{Science}
  \textbf{\bibinfo{volume}{365}}, \bibinfo{pages}{1445}
  (\bibinfo{year}{2019}{\natexlab{b}}), \eprint{1909.02726}.

\bibitem[{\citenamefont{Adams et~al.}(2020)}]{Adams:2019jhp}
\bibinfo{author}{\bibfnamefont{D.}~\bibnamefont{Adams}} \bibnamefont{et~al.}
  (\bibinfo{collaboration}{CUORE}), \bibinfo{journal}{Phys. Rev. Lett.}
  \textbf{\bibinfo{volume}{124}}, \bibinfo{pages}{122501}
  (\bibinfo{year}{2020}), \eprint{1912.10966}.

\bibitem[{\citenamefont{Gando}(2020)}]{Gando:2020cxo}
\bibinfo{author}{\bibfnamefont{Y.}~\bibnamefont{Gando}}
  (\bibinfo{collaboration}{KamLAND-Zen}), \bibinfo{journal}{J. Phys. Conf.
  Ser.} \textbf{\bibinfo{volume}{1468}}, \bibinfo{pages}{012142}
  (\bibinfo{year}{2020}).

\bibitem[{\citenamefont{Abgrall et~al.}(2017{\natexlab{b}})}]{Abgrall:2017syy}
\bibinfo{author}{\bibfnamefont{N.}~\bibnamefont{Abgrall}} \bibnamefont{et~al.}
  (\bibinfo{collaboration}{LEGEND}), \bibinfo{journal}{AIP Conf. Proc.}
  \textbf{\bibinfo{volume}{1894}}, \bibinfo{pages}{020027}
  (\bibinfo{year}{2017}{\natexlab{b}}), \eprint{1709.01980}.

\bibitem[{\citenamefont{Adams et~al.}(2021{\natexlab{b}})}]{Adams:2020cye}
\bibinfo{author}{\bibfnamefont{C.}~\bibnamefont{Adams}} \bibnamefont{et~al.}
  (\bibinfo{collaboration}{NEXT}), \bibinfo{journal}{JHEP}
  \textbf{\bibinfo{volume}{2021}}, \bibinfo{pages}{164}
  (\bibinfo{year}{2021}{\natexlab{b}}), \eprint{2005.06467}.

\bibitem[{\citenamefont{Armstrong et~al.}(2019)}]{CUPIDInterestGroup:2019inu}
\bibinfo{author}{\bibfnamefont{W.}~\bibnamefont{Armstrong}}
  \bibnamefont{et~al.} (\bibinfo{collaboration}{CUPID}) (\bibinfo{year}{2019}),
  \eprint{1907.09376}.

\bibitem[{\citenamefont{Nakamura et~al.}(2020)\citenamefont{Nakamura,
  Sambonsugi, Shiraishi, and Wada}}]{Nakamura:2020szx}
\bibinfo{author}{\bibfnamefont{R.}~\bibnamefont{Nakamura}},
  \bibinfo{author}{\bibfnamefont{H.}~\bibnamefont{Sambonsugi}},
  \bibinfo{author}{\bibfnamefont{K.}~\bibnamefont{Shiraishi}},
  \bibnamefont{and} \bibinfo{author}{\bibfnamefont{Y.}~\bibnamefont{Wada}},
  \bibinfo{journal}{J. Phys. Conf. Ser.} \textbf{\bibinfo{volume}{1468}},
  \bibinfo{pages}{012256} (\bibinfo{year}{2020}).

\bibitem[{\citenamefont{Andringa et~al.}(2016)}]{Andringa:2015tza}
\bibinfo{author}{\bibfnamefont{S.}~\bibnamefont{Andringa}} \bibnamefont{et~al.}
  (\bibinfo{collaboration}{SNO+}), \bibinfo{journal}{Adv. High Energy Phys.}
  \textbf{\bibinfo{volume}{2016}}, \bibinfo{pages}{6194250}
  (\bibinfo{year}{2016}), \eprint{1508.05759}.

\bibitem[{\citenamefont{Akimov et~al.}(2018)}]{Akimov:2018ghi}
\bibinfo{author}{\bibfnamefont{D.}~\bibnamefont{Akimov}} \bibnamefont{et~al.}
  (\bibinfo{collaboration}{COHERENT}) (\bibinfo{year}{2018}),
  \eprint{1803.09183}.

\bibitem[{\citenamefont{de~la Vega et~al.}(2021)\citenamefont{de~la Vega,
  Flores, Nath, and Peinado}}]{delaVega:2021mhj}
\bibinfo{author}{\bibfnamefont{L.~M.~G.} \bibnamefont{de~la Vega}},
  \bibinfo{author}{\bibfnamefont{L.~J.} \bibnamefont{Flores}},
  \bibinfo{author}{\bibfnamefont{N.}~\bibnamefont{Nath}}, \bibnamefont{and}
  \bibinfo{author}{\bibfnamefont{E.}~\bibnamefont{Peinado}}
  (\bibinfo{year}{2021}), \eprint{2107.04037}.

\bibitem[{\citenamefont{Ciuffoli et~al.}(2018)\citenamefont{Ciuffoli, Evslin,
  Fu, and Tang}}]{Ciuffoli:2018qem}
\bibinfo{author}{\bibfnamefont{E.}~\bibnamefont{Ciuffoli}},
  \bibinfo{author}{\bibfnamefont{J.}~\bibnamefont{Evslin}},
  \bibinfo{author}{\bibfnamefont{Q.}~\bibnamefont{Fu}}, \bibnamefont{and}
  \bibinfo{author}{\bibfnamefont{J.}~\bibnamefont{Tang}},
  \bibinfo{journal}{Phys. Rev. D} \textbf{\bibinfo{volume}{97}},
  \bibinfo{pages}{113003} (\bibinfo{year}{2018}), \eprint{1801.02166}.

\bibitem[{\citenamefont{Coloma et~al.}(2017{\natexlab{a}})\citenamefont{Coloma,
  Denton, Gonzalez-Garcia, Maltoni, and Schwetz}}]{Coloma:2017egw}
\bibinfo{author}{\bibfnamefont{P.}~\bibnamefont{Coloma}},
  \bibinfo{author}{\bibfnamefont{P.~B.} \bibnamefont{Denton}},
  \bibinfo{author}{\bibfnamefont{M.~C.} \bibnamefont{Gonzalez-Garcia}},
  \bibinfo{author}{\bibfnamefont{M.}~\bibnamefont{Maltoni}}, \bibnamefont{and}
  \bibinfo{author}{\bibfnamefont{T.}~\bibnamefont{Schwetz}},
  \bibinfo{journal}{JHEP} \textbf{\bibinfo{volume}{04}}, \bibinfo{pages}{116}
  (\bibinfo{year}{2017}{\natexlab{a}}), \eprint{1701.04828}.

\bibitem[{\citenamefont{Coloma et~al.}(2017{\natexlab{b}})\citenamefont{Coloma,
  Gonzalez-Garcia, Maltoni, and Schwetz}}]{Coloma:2017ncl}
\bibinfo{author}{\bibfnamefont{P.}~\bibnamefont{Coloma}},
  \bibinfo{author}{\bibfnamefont{M.~C.} \bibnamefont{Gonzalez-Garcia}},
  \bibinfo{author}{\bibfnamefont{M.}~\bibnamefont{Maltoni}}, \bibnamefont{and}
  \bibinfo{author}{\bibfnamefont{T.}~\bibnamefont{Schwetz}},
  \bibinfo{journal}{Phys. Rev. D} \textbf{\bibinfo{volume}{96}},
  \bibinfo{pages}{115007} (\bibinfo{year}{2017}{\natexlab{b}}),
  \eprint{1708.02899}.

\bibitem[{\citenamefont{Liao and Marfatia}(2017)}]{Liao:2017uzy}
\bibinfo{author}{\bibfnamefont{J.}~\bibnamefont{Liao}} \bibnamefont{and}
  \bibinfo{author}{\bibfnamefont{D.}~\bibnamefont{Marfatia}},
  \bibinfo{journal}{Phys. Lett. B} \textbf{\bibinfo{volume}{775}},
  \bibinfo{pages}{54} (\bibinfo{year}{2017}), \eprint{1708.04255}.

\bibitem[{\citenamefont{Akimov et~al.}(2021{\natexlab{b}})}]{Akimov:2020pdx}
\bibinfo{author}{\bibfnamefont{D.}~\bibnamefont{Akimov}} \bibnamefont{et~al.}
  (\bibinfo{collaboration}{COHERENT}), \bibinfo{journal}{Phys. Rev. Lett.}
  \textbf{\bibinfo{volume}{126}}, \bibinfo{pages}{012002}
  (\bibinfo{year}{2021}{\natexlab{b}}), \eprint{2003.10630}.

\bibitem[{\citenamefont{Akimov et~al.}(2013)}]{RED:2012hpm}
\bibinfo{author}{\bibfnamefont{D.~Y.} \bibnamefont{Akimov}}
  \bibnamefont{et~al.} (\bibinfo{collaboration}{RED}), \bibinfo{journal}{JINST}
  \textbf{\bibinfo{volume}{8}}, \bibinfo{pages}{P10023} (\bibinfo{year}{2013}),
  \eprint{1212.1938}.

\bibitem[{\citenamefont{Soma et~al.}(2016)}]{Soma:2014zgm}
\bibinfo{author}{\bibfnamefont{A.~K.} \bibnamefont{Soma}} \bibnamefont{et~al.}
  (\bibinfo{collaboration}{TEXONO}), \bibinfo{journal}{Nucl. Instrum. Meth. A}
  \textbf{\bibinfo{volume}{836}}, \bibinfo{pages}{67} (\bibinfo{year}{2016}),
  \eprint{1411.4802}.

\bibitem[{\citenamefont{Agnolet et~al.}(2017)}]{Agnolet:2016zir}
\bibinfo{author}{\bibfnamefont{G.}~\bibnamefont{Agnolet}} \bibnamefont{et~al.}
  (\bibinfo{collaboration}{MINER}), \bibinfo{journal}{Nucl. Instrum. Meth. A}
  \textbf{\bibinfo{volume}{853}}, \bibinfo{pages}{53} (\bibinfo{year}{2017}),
  \eprint{1609.02066}.

\bibitem[{\citenamefont{Akimov et~al.}(2017{\natexlab{b}})}]{Akimov:2017hee}
\bibinfo{author}{\bibfnamefont{D.~Y.} \bibnamefont{Akimov}}
  \bibnamefont{et~al.}, \bibinfo{journal}{JINST} \textbf{\bibinfo{volume}{12}},
  \bibinfo{pages}{C06018} (\bibinfo{year}{2017}{\natexlab{b}}).

\bibitem[{\citenamefont{Strauss et~al.}(2017)}]{Strauss:2017cuu}
\bibinfo{author}{\bibfnamefont{R.}~\bibnamefont{Strauss}} \bibnamefont{et~al.},
  \bibinfo{journal}{Eur. Phys. J. C} \textbf{\bibinfo{volume}{77}},
  \bibinfo{pages}{506} (\bibinfo{year}{2017}), \eprint{1704.04320}.

\bibitem[{\citenamefont{Leder et~al.}(2018)}]{Leder:2017lva}
\bibinfo{author}{\bibfnamefont{A.}~\bibnamefont{Leder}} \bibnamefont{et~al.},
  \bibinfo{journal}{JINST} \textbf{\bibinfo{volume}{13}},
  \bibinfo{pages}{P02004} (\bibinfo{year}{2018}), \eprint{1710.00802}.

\bibitem[{\citenamefont{Akimov et~al.}(2020)}]{Akimov:2019ogx}
\bibinfo{author}{\bibfnamefont{D.~Y.} \bibnamefont{Akimov}}
  \bibnamefont{et~al.} (\bibinfo{collaboration}{RED-100}),
  \bibinfo{journal}{JINST} \textbf{\bibinfo{volume}{15}},
  \bibinfo{pages}{P02020} (\bibinfo{year}{2020}), \eprint{1910.06190}.

\bibitem[{\citenamefont{Ni et~al.}(2021)\citenamefont{Ni, Qi, Shockley, and
  Wei}}]{Ni:2021mwa}
\bibinfo{author}{\bibfnamefont{K.}~\bibnamefont{Ni}},
  \bibinfo{author}{\bibfnamefont{J.}~\bibnamefont{Qi}},
  \bibinfo{author}{\bibfnamefont{E.}~\bibnamefont{Shockley}}, \bibnamefont{and}
  \bibinfo{author}{\bibfnamefont{Y.}~\bibnamefont{Wei}},
  \bibinfo{journal}{Universe} \textbf{\bibinfo{volume}{7}}, \bibinfo{pages}{54}
  (\bibinfo{year}{2021}).

\bibitem[{\citenamefont{Arpesella et~al.}(2008)}]{Arpesella:2007xf}
\bibinfo{author}{\bibfnamefont{C.}~\bibnamefont{Arpesella}}
  \bibnamefont{et~al.} (\bibinfo{collaboration}{Borexino}),
  \bibinfo{journal}{Phys. Lett. B} \textbf{\bibinfo{volume}{658}},
  \bibinfo{pages}{101} (\bibinfo{year}{2008}), \eprint{0708.2251}.

\bibitem[{\citenamefont{Zaklad et~al.}(1973)\citenamefont{Zaklad, Derenzo,
  Muller, and Smits}}]{Zaklad4326941}
\bibinfo{author}{\bibfnamefont{H.}~\bibnamefont{Zaklad}},
  \bibinfo{author}{\bibfnamefont{S.~E.} \bibnamefont{Derenzo}},
  \bibinfo{author}{\bibfnamefont{R.~A.} \bibnamefont{Muller}},
  \bibnamefont{and} \bibinfo{author}{\bibfnamefont{R.~G.} \bibnamefont{Smits}},
  \bibinfo{journal}{IEEE Transactions on Nuclear Science}
  \textbf{\bibinfo{volume}{20}}, \bibinfo{pages}{429} (\bibinfo{year}{1973}).

\bibitem[{\citenamefont{Lavoie}(1976)}]{Lavoie594289}
\bibinfo{author}{\bibfnamefont{L.}~\bibnamefont{Lavoie}},
  \bibinfo{journal}{Medical Physics} \textbf{\bibinfo{volume}{3}},
  \bibinfo{pages}{283} (\bibinfo{year}{1976}).

\bibitem[{\citenamefont{Egorov et~al.}(1983)}]{Egorov:1983}
\bibinfo{author}{\bibfnamefont{V.}~\bibnamefont{Egorov}} \bibnamefont{et~al.},
  \bibinfo{journal}{Nucl. Instr. Meth.} \textbf{\bibinfo{volume}{205}},
  \bibinfo{pages}{373} (\bibinfo{year}{1983}).

\bibitem[{\citenamefont{Bolozdynya
  et~al.}(1997{\natexlab{b}})\citenamefont{Bolozdynya, Egorov, Koutchenkov
  et~al.}}]{Bolozdynya:1997ecc}
\bibinfo{author}{\bibfnamefont{A.}~\bibnamefont{Bolozdynya}},
  \bibinfo{author}{\bibfnamefont{V.}~\bibnamefont{Egorov}},
  \bibinfo{author}{\bibfnamefont{A.}~\bibnamefont{Koutchenkov}},
  \bibnamefont{et~al.}, \bibinfo{journal}{IEEE Trans. Nucl. Sci.}
  \textbf{\bibinfo{volume}{44}}, \bibinfo{pages}{2408}
  (\bibinfo{year}{1997}{\natexlab{b}}).

\bibitem[{\citenamefont{Bolozdynya
  et~al.}(1997{\natexlab{c}})\citenamefont{Bolozdynya, Ordonez, and
  Chang}}]{Bolozdynya:1997ccc}
\bibinfo{author}{\bibfnamefont{A.}~\bibnamefont{Bolozdynya}},
  \bibinfo{author}{\bibfnamefont{C.}~\bibnamefont{Ordonez}}, \bibnamefont{and}
  \bibinfo{author}{\bibfnamefont{W.}~\bibnamefont{Chang}},
  \bibinfo{journal}{1997 IEEE Nucl. Sci. Sym. Med. Imag. Conf. Rec.}
  \textbf{\bibinfo{volume}{2}}, \bibinfo{pages}{1047}
  (\bibinfo{year}{1997}{\natexlab{c}}).

\bibitem[{\citenamefont{Rogers et~al.}(2004)\citenamefont{Rogers, Clinthorne,
  and Bolozdynya}}]{Rogers:2004cc}
\bibinfo{author}{\bibfnamefont{W.}~\bibnamefont{Rogers}},
  \bibinfo{author}{\bibfnamefont{N.}~\bibnamefont{Clinthorne}},
  \bibnamefont{and}
  \bibinfo{author}{\bibfnamefont{A.}~\bibnamefont{Bolozdynya}}, in
  \emph{\bibinfo{booktitle}{{Emission Tomography: the fundamentals of PET and
  SPECT}}}, edited by \bibinfo{editor}{\bibfnamefont{M.}~\bibnamefont{Wernick}}
  \bibnamefont{and} \bibinfo{editor}{\bibfnamefont{J.}~\bibnamefont{Aarsvold}}
  (\bibinfo{publisher}{Elsevier ScienceDirect}, \bibinfo{year}{2004}), pp.
  \bibinfo{pages}{383--419},
  \urlprefix\url{https://www.sciencedirect.com/science/article/pii/B9780127444826500223}.

\bibitem[{\citenamefont{Chepel}(1993)}]{Chepel1993ANL}
\bibinfo{author}{\bibfnamefont{V.}~\bibnamefont{Chepel}},
  \bibinfo{journal}{Nuclear Tracks and Radiation Measurements}
  \textbf{\bibinfo{volume}{21}}, \bibinfo{pages}{47} (\bibinfo{year}{1993}).

\bibitem[{\citenamefont{Chepel et~al.}(1999)\citenamefont{Chepel, Solovov, Van
  Der~Marel, Lopes, Crespo, Janeiro, Santos, Marques, and
  Policarpo}}]{Chepel790822}
\bibinfo{author}{\bibfnamefont{V.}~\bibnamefont{Chepel}},
  \bibinfo{author}{\bibfnamefont{V.}~\bibnamefont{Solovov}},
  \bibinfo{author}{\bibfnamefont{J.}~\bibnamefont{Van Der~Marel}},
  \bibinfo{author}{\bibfnamefont{M.}~\bibnamefont{Lopes}},
  \bibinfo{author}{\bibfnamefont{P.}~\bibnamefont{Crespo}},
  \bibinfo{author}{\bibfnamefont{L.}~\bibnamefont{Janeiro}},
  \bibinfo{author}{\bibfnamefont{D.}~\bibnamefont{Santos}},
  \bibinfo{author}{\bibfnamefont{R.}~\bibnamefont{Marques}}, \bibnamefont{and}
  \bibinfo{author}{\bibfnamefont{A.}~\bibnamefont{Policarpo}},
  \bibinfo{journal}{IEEE Transactions on Nuclear Science}
  \textbf{\bibinfo{volume}{46}}, \bibinfo{pages}{1038} (\bibinfo{year}{1999}).

\bibitem[{\citenamefont{Chepel et~al.}(2002)\citenamefont{Chepel, Lopes,
  Solovov, Ferreira~Marques, and Policarpo}}]{Chepel:2002ucm}
\bibinfo{author}{\bibfnamefont{V.}~\bibnamefont{Chepel}},
  \bibinfo{author}{\bibfnamefont{M.~I.} \bibnamefont{Lopes}},
  \bibinfo{author}{\bibfnamefont{V.}~\bibnamefont{Solovov}},
  \bibinfo{author}{\bibfnamefont{R.}~\bibnamefont{Ferreira~Marques}},
  \bibnamefont{and} \bibinfo{author}{\bibfnamefont{A.~J. P.~L.}
  \bibnamefont{Policarpo}}, in \emph{\bibinfo{booktitle}{{Technique and
  application of xenon detectors. Proceedings, International Workshop, Kashiwa,
  Japan, December 3-4, 2001}}} (\bibinfo{year}{2002}), pp.
  \bibinfo{pages}{28--40}, \eprint{physics/0211117}.

\bibitem[{\citenamefont{Gallego~Manzano et~al.}(2015)}]{GallegoManzano:2015hkg}
\bibinfo{author}{\bibfnamefont{L.}~\bibnamefont{Gallego~Manzano}}
  \bibnamefont{et~al.}, \bibinfo{journal}{Nucl. Instrum. Meth. A}
  \textbf{\bibinfo{volume}{787}}, \bibinfo{pages}{89} (\bibinfo{year}{2015}).

\bibitem[{\citenamefont{Ferrario}(2018)}]{Ferrario:2017sgq}
\bibinfo{author}{\bibfnamefont{P.}~\bibnamefont{Ferrario}},
  \bibinfo{journal}{JINST} \textbf{\bibinfo{volume}{13}},
  \bibinfo{pages}{C01044} (\bibinfo{year}{2018}), \eprint{1712.05751}.

\bibitem[{\citenamefont{Zhu et~al.}(2019)}]{Zhu:2019wbt}
\bibinfo{author}{\bibfnamefont{Y.}~\bibnamefont{Zhu}} \bibnamefont{et~al.}, in
  \emph{\bibinfo{booktitle}{{IEEE International Conference on Dielectric
  Liquids}}} (\bibinfo{year}{2019}).

\bibitem[{\citenamefont{Giovagnoli et~al.}(2021)}]{Giovagnoli:2021bwc}
\bibinfo{author}{\bibfnamefont{D.}~\bibnamefont{Giovagnoli}}
  \bibnamefont{et~al.}, \bibinfo{journal}{IEEE Trans. Rad. Plasma Med. Sci.}
  \textbf{\bibinfo{volume}{5}}, \bibinfo{pages}{826} (\bibinfo{year}{2021}).

\bibitem[{\citenamefont{Kevles}(1998)}]{Kevles:1998}
\bibinfo{author}{\bibfnamefont{B.~H.} \bibnamefont{Kevles}},
  \emph{\bibinfo{title}{Naked to the Bone; Medical Imaging in the Twentieth
  Century}} (\bibinfo{publisher}{Basic Books}, \bibinfo{year}{1998}), ISBN
  \bibinfo{isbn}{9780201328332}.

\bibitem[{\citenamefont{Renner et~al.}(2021)}]{Renner:2021vyz}
\bibinfo{author}{\bibfnamefont{J.}~\bibnamefont{Renner}} \bibnamefont{et~al.}
  (\bibinfo{year}{2021}), \eprint{2109.12899}.

\bibitem[{\citenamefont{Lynch~III et~al.}(2000)\citenamefont{Lynch~III, Baum,
  and Tenbrinck}}]{Lynch:2000}
\bibinfo{author}{\bibfnamefont{C.}~\bibnamefont{Lynch~III}},
  \bibinfo{author}{\bibfnamefont{J.}~\bibnamefont{Baum}}, \bibnamefont{and}
  \bibinfo{author}{\bibfnamefont{R.}~\bibnamefont{Tenbrinck}},
  \bibinfo{journal}{Anesthesiology} \textbf{\bibinfo{volume}{92}},
  \bibinfo{pages}{865} (\bibinfo{year}{2000}).

\bibitem[{\citenamefont{Law et~al.}(2016)\citenamefont{Law, Lo, and
  Gan}}]{Law:2016}
\bibinfo{author}{\bibfnamefont{L.}~\bibnamefont{Law}},
  \bibinfo{author}{\bibfnamefont{E.}~\bibnamefont{Lo}}, \bibnamefont{and}
  \bibinfo{author}{\bibfnamefont{T.}~\bibnamefont{Gan}},
  \bibinfo{journal}{Anesth Analg} \textbf{\bibinfo{volume}{122(3)}},
  \bibinfo{pages}{678} (\bibinfo{year}{2016}).

\bibitem[{\citenamefont{et~al.}(2015)}]{Campos-Pires}
\bibinfo{author}{\bibfnamefont{R.~C.-P.} \bibnamefont{et~al.}},
  \bibinfo{journal}{Neurologic Critical Care} \textbf{\bibinfo{volume}{43(1)}}
  (\bibinfo{year}{2015}).

\bibitem[{\citenamefont{Dobrovolsky et~al.}(2018)\citenamefont{Dobrovolsky,
  Bogin, and Meloni}}]{Dobrovolsky}
\bibinfo{author}{\bibfnamefont{A.}~\bibnamefont{Dobrovolsky}},
  \bibinfo{author}{\bibfnamefont{V.}~\bibnamefont{Bogin}}, \bibnamefont{and}
  \bibinfo{author}{\bibfnamefont{E.}~\bibnamefont{Meloni}},
  \bibinfo{journal}{Ann. Psychiatry Ment. Health}
  \textbf{\bibinfo{volume}{6(3)}}, \bibinfo{pages}{1133}
  (\bibinfo{year}{2018}).

\bibitem[{\citenamefont{Nikkel et~al.}(2012)\citenamefont{Nikkel, Gozani,
  Brown, Kwong, McKinsey, Shin, Kane, Gary, and Firestone}}]{Nikkel:2012zz}
\bibinfo{author}{\bibfnamefont{J.~A.} \bibnamefont{Nikkel}},
  \bibinfo{author}{\bibfnamefont{T.}~\bibnamefont{Gozani}},
  \bibinfo{author}{\bibfnamefont{C.}~\bibnamefont{Brown}},
  \bibinfo{author}{\bibfnamefont{J.}~\bibnamefont{Kwong}},
  \bibinfo{author}{\bibfnamefont{D.~N.} \bibnamefont{McKinsey}},
  \bibinfo{author}{\bibfnamefont{Y.}~\bibnamefont{Shin}},
  \bibinfo{author}{\bibfnamefont{S.}~\bibnamefont{Kane}},
  \bibinfo{author}{\bibfnamefont{C.}~\bibnamefont{Gary}}, \bibnamefont{and}
  \bibinfo{author}{\bibfnamefont{M.}~\bibnamefont{Firestone}},
  \bibinfo{journal}{JINST} \textbf{\bibinfo{volume}{7}},
  \bibinfo{pages}{C03007} (\bibinfo{year}{2012}).

\bibitem[{\citenamefont{Boireau et~al.}(2016)}]{Boireau:2015dda}
\bibinfo{author}{\bibfnamefont{G.}~\bibnamefont{Boireau}} \bibnamefont{et~al.}
  (\bibinfo{collaboration}{NUCIFER}), \bibinfo{journal}{Phys. Rev. D}
  \textbf{\bibinfo{volume}{93}}, \bibinfo{pages}{112006}
  (\bibinfo{year}{2016}), \eprint{1509.05610}.

\bibitem[{\citenamefont{Khosa et~al.}(2020)\citenamefont{Khosa, Mars, Richards,
  and Sanz}}]{Khosa:2019qgp}
\bibinfo{author}{\bibfnamefont{C.~K.} \bibnamefont{Khosa}},
  \bibinfo{author}{\bibfnamefont{L.}~\bibnamefont{Mars}},
  \bibinfo{author}{\bibfnamefont{J.}~\bibnamefont{Richards}}, \bibnamefont{and}
  \bibinfo{author}{\bibfnamefont{V.}~\bibnamefont{Sanz}}, \bibinfo{journal}{J.
  Phys. G} \textbf{\bibinfo{volume}{47}}, \bibinfo{pages}{095201}
  (\bibinfo{year}{2020}), \eprint{1911.09210}.

\bibitem[{\citenamefont{Barrand et~al.}(2001)}]{Barrand:2001ny}
\bibinfo{author}{\bibfnamefont{G.}~\bibnamefont{Barrand}} \bibnamefont{et~al.},
  \bibinfo{journal}{Comput. Phys. Commun.} \textbf{\bibinfo{volume}{140}},
  \bibinfo{pages}{45} (\bibinfo{year}{2001}).

\bibitem[{\citenamefont{Jayatilaka et~al.}(2017)\citenamefont{Jayatilaka,
  Levshina, Sehgal, Gardner, Rynge, and W{\"u}rthwein}}]{Jayatilaka:2017twe}
\bibinfo{author}{\bibfnamefont{B.}~\bibnamefont{Jayatilaka}},
  \bibinfo{author}{\bibfnamefont{T.}~\bibnamefont{Levshina}},
  \bibinfo{author}{\bibfnamefont{C.}~\bibnamefont{Sehgal}},
  \bibinfo{author}{\bibfnamefont{R.}~\bibnamefont{Gardner}},
  \bibinfo{author}{\bibfnamefont{M.}~\bibnamefont{Rynge}}, \bibnamefont{and}
  \bibinfo{author}{\bibfnamefont{F.}~\bibnamefont{W{\"u}rthwein}},
  \bibinfo{journal}{J. Phys. Conf. Ser.} \textbf{\bibinfo{volume}{898}},
  \bibinfo{pages}{082048} (\bibinfo{year}{2017}).

\bibitem[{\citenamefont{Albrecht et~al.}(2019)}]{Alves:2017she}
\bibinfo{author}{\bibfnamefont{J.}~\bibnamefont{Albrecht}} \bibnamefont{et~al.}
  (\bibinfo{collaboration}{HEP Software Foundation}), \bibinfo{journal}{Comput.
  Softw. Big Sci.} \textbf{\bibinfo{volume}{3}}, \bibinfo{pages}{7}
  (\bibinfo{year}{2019}), \eprint{1712.06982}.

\bibitem[{\citenamefont{van~den Berg et~al.}(2011)}]{ASPERA2011}
\bibinfo{author}{\bibfnamefont{A.}~\bibnamefont{van~den Berg}}
  \bibnamefont{et~al.}, \emph{\bibinfo{title}{{Astroparticle physics - The
  European Roadmap}}} (\bibinfo{year}{2011}).

\bibitem[{\citenamefont{Cushman et~al.}(2013)}]{Snowmass:2013}
\bibinfo{author}{\bibfnamefont{P.}~\bibnamefont{Cushman}} \bibnamefont{et~al.},
  \emph{\bibinfo{title}{Snowmass \uppercase{CF1} summary: \uppercase{WIMP} dark
  matter direct detection}},
  \bibinfo{howpublished}{\url{https://arxiv.org/abs/1310.8327.pdf}}
  (\bibinfo{year}{2013}), \eprint{1310.8327}.

\bibitem[{\citenamefont{APPEC}(2017)}]{APPEC:2017}
\bibinfo{author}{\bibfnamefont{A.~P. E.~C.} \bibnamefont{APPEC}},
  \emph{\bibinfo{title}{{European Astroparticle Physics Strategy 2017-2026}}},
  \bibinfo{howpublished}{\url{http://www.appec.org/wp-content/uploads/2017/08/APPEC-Strategy-Book-Proof-23-Nov-2.pdf}}
  (\bibinfo{year}{2017}).

\bibitem[{\citenamefont{of~Particles and
  Fields}(2018{\natexlab{a}})}]{DPF:2018a}
\bibinfo{author}{\bibfnamefont{D.}~\bibnamefont{of~Particles}}
  \bibnamefont{and} \bibinfo{author}{\bibnamefont{Fields}},
  \emph{\bibinfo{title}{{APS Division of Particles and Fields Response to
  European Strategy Group Call for White Papers: Community Planning and Science
  Drivers}}},
  \bibinfo{howpublished}{\url{https://www.aps.org/units/dpf/upload/DPF-strategy.pdf}}
  (\bibinfo{year}{2018}{\natexlab{a}}).

\bibitem[{\citenamefont{of~Particles and
  Fields}(2018{\natexlab{b}})}]{DPF:2018b}
\bibinfo{author}{\bibfnamefont{D.}~\bibnamefont{of~Particles}}
  \bibnamefont{and} \bibinfo{author}{\bibnamefont{Fields}},
  \emph{\bibinfo{title}{{APS Division of Particles and Fields Response to
  European Strategy Group Call for White Papers: Tools for Particle Physics}}},
  \bibinfo{howpublished}{\url{https://www.aps.org/units/dpf/upload/DPF-tools.pdf}}
  (\bibinfo{year}{2018}{\natexlab{b}}).

\bibitem[{\citenamefont{Cao et~al.}(in preparation)}]{China:2021}
\bibinfo{author}{\bibfnamefont{J.}~\bibnamefont{Cao}} \bibnamefont{et~al.},
  \emph{\bibinfo{title}{{Non-accelerator Particle Physics in China: A Strategic
  Plan, in Chinese}}} (\bibinfo{year}{in preparation}).

\bibitem[{\citenamefont{Billard et~al.}(2021)}]{Billard:2021uyg}
\bibinfo{author}{\bibfnamefont{J.}~\bibnamefont{Billard}} \bibnamefont{et~al.}
  (\bibinfo{year}{2021}), \eprint{2104.07634}.

\bibitem[{\citenamefont{Group}(2020)}]{ESPP:2020}
\bibinfo{author}{\bibfnamefont{E.~S.} \bibnamefont{Group}},
  \emph{\bibinfo{title}{{The European Strategy for Particle Physics Update
  2020}}} (\bibinfo{year}{2020}).

\bibitem[{\citenamefont{Ritz et~al.}(2014)}]{P5:2014}
\bibinfo{author}{\bibfnamefont{S.}~\bibnamefont{Ritz}} \bibnamefont{et~al.},
  \emph{\bibinfo{title}{{Building for Discovery: Strategic Plan for U.S.
  Particle Physics in the Global Context}}},
  \bibinfo{howpublished}{\url{https://science.energy.gov/~/media/hep/hepap/pdf/May-2014/FINAL_P5_Report_Interactive_060214.pdf}}
  (\bibinfo{year}{2014}).

\bibitem[{\citenamefont{Zeitnitz}(2018)}]{GerPS:2018}
\bibinfo{author}{\bibfnamefont{C.}~\bibnamefont{Zeitnitz}},
  \emph{\bibinfo{title}{{Particle Physics Strategy in Germany}}},
  \bibinfo{howpublished}{\url{https://indico.desy.de/indico/event/20166/contribution/6/material/slides/0.pdf}}
  (\bibinfo{year}{2018}).

\bibitem[{\citenamefont{Wallny et~al.}(2021)}]{chipp2021}
\bibinfo{author}{\bibfnamefont{R.}~\bibnamefont{Wallny}} \bibnamefont{et~al.},
  \bibinfo{journal}{Swiss Academies Reports} \textbf{\bibinfo{volume}{16}},
  \bibinfo{pages}{6} (\bibinfo{year}{2021}),
  \urlprefix\url{https://api.swiss-academies.ch/site/assets/files/24379/chipp_roadmap_2021.pdf}.

\bibitem[{\citenamefont{van~den Berg et~al.}(2014)}]{NL:2014}
\bibinfo{author}{\bibfnamefont{A.}~\bibnamefont{van~den Berg}}
  \bibnamefont{et~al.}, \emph{\bibinfo{title}{{Strategic Plan for Astroparticle
  Physicsin the Netherlands2014–2024}}} (\bibinfo{year}{2014}).

\bibitem[{\citenamefont{Geesaman et~al.}(2015)}]{LRP:2015}
\bibinfo{author}{\bibfnamefont{D.}~\bibnamefont{Geesaman}}
  \bibnamefont{et~al.}, \emph{\bibinfo{title}{{Reaching for the Horizon: The
  2015 Long Range Plan for Nuclear Science}}},
  \bibinfo{howpublished}{\url{https://science.energy.gov/~/media/np/nsac/pdf/2015LRP/2015_LRPNS_091815.pdf}}
  (\bibinfo{year}{2015}).

\bibitem[{\citenamefont{Giuliani et~al.}(2019)\citenamefont{Giuliani,
  Gomez~Cadenas, Pascoli, Previtali, Saakyan, Sch\"affner, and
  Sch\"onert}}]{Giuliani:2019uno}
\bibinfo{author}{\bibfnamefont{A.}~\bibnamefont{Giuliani}},
  \bibinfo{author}{\bibfnamefont{J.~J.} \bibnamefont{Gomez~Cadenas}},
  \bibinfo{author}{\bibfnamefont{S.}~\bibnamefont{Pascoli}},
  \bibinfo{author}{\bibfnamefont{E.}~\bibnamefont{Previtali}},
  \bibinfo{author}{\bibfnamefont{R.}~\bibnamefont{Saakyan}},
  \bibinfo{author}{\bibfnamefont{K.}~\bibnamefont{Sch\"affner}},
  \bibnamefont{and}
  \bibinfo{author}{\bibfnamefont{S.}~\bibnamefont{Sch\"onert}}
  (\bibinfo{collaboration}{APPEC Committee}) (\bibinfo{year}{2019}),
  \eprint{1910.04688}.

\bibitem[{\citenamefont{Panesor}(2015)}]{UK:2015}
\bibinfo{author}{\bibfnamefont{T.}~\bibnamefont{Panesor}},
  \emph{\bibinfo{title}{{A Review of UK Astroparticle Physics Research}}}
  (\bibinfo{year}{2015}).

\bibitem[{\citenamefont{Tyurin}(2012)}]{Russia:2012}
\bibinfo{author}{\bibfnamefont{N.}~\bibnamefont{Tyurin}},
  \emph{\bibinfo{title}{{PARTICLE PHYSICS IN RUSSIA}}} (\bibinfo{year}{2012}).

\bibitem[{\citenamefont{Nakada et~al.}(2013)}]{CERN:2013}
\bibinfo{author}{\bibfnamefont{T.}~\bibnamefont{Nakada}} \bibnamefont{et~al.},
  \emph{\bibinfo{title}{{The European Strategy for Particle Physics Update
  2013}}} (\bibinfo{year}{2013}).

\bibitem[{\citenamefont{Bai et~al.}(2020)}]{China:2020}
\bibinfo{author}{\bibfnamefont{C.}~\bibnamefont{Bai}} \bibnamefont{et~al.},
  \emph{\bibinfo{title}{{Neutrinoless Double Beta Decay: A Study of Strategic
  Development by Chinese Academy of Sciences, in Chinese}}}
  (\bibinfo{year}{2020}).

\bibitem[{\citenamefont{Faulkner et~al.}(2006)}]{GridPP:2006wnd}
\bibinfo{author}{\bibfnamefont{P.~J.~W.} \bibnamefont{Faulkner}}
  \bibnamefont{et~al.} (\bibinfo{collaboration}{GridPP}), \bibinfo{journal}{J.
  Phys. G} \textbf{\bibinfo{volume}{32}}, \bibinfo{pages}{N1}
  (\bibinfo{year}{2006}).

\bibitem[{\citenamefont{Britton et~al.}(2009)}]{Britton:2009ser}
\bibinfo{author}{\bibfnamefont{D.}~\bibnamefont{Britton}} \bibnamefont{et~al.},
  \bibinfo{journal}{Phil. Trans. Roy. Soc. Lond. A}
  \textbf{\bibinfo{volume}{367}}, \bibinfo{pages}{2447} (\bibinfo{year}{2009}).

\end{thebibliography}

\end{document}